\documentclass[aps,prc,amsfonts, reprint,longbibliography, superscriptaddress, amsmath,amssymb,longbibliography,
reprint
]{revtex4-1}

\usepackage{hyperref}
\hypersetup{colorlinks=true, linkcolor=blue, citecolor=red, urlcolor=blue}
\usepackage{tikz}
\usetikzlibrary{arrows}
\usetikzlibrary{decorations.pathmorphing}
\usepackage{hyperref}
\usepackage{amssymb}
\usepackage{amsmath}
\usepackage{color}
\usepackage{xcolor}
\usepackage{graphicx}
\usepackage{soul}
\usepackage{mathtools}

\newcommand{\comment}[1]{}

\newcommand{\RN}[1]{
}

\usetikzlibrary{patterns}

\begin{document}

\title{Feynman-Vernon influence functional approach to quantum transport in interacting nanojunctions: An analytical  hierarchical study}

\author{Luca Magazz\`u}
\affiliation{Institute for Theoretical Physics, University of
Regensburg, 93040 Regensburg, Germany}
\author{Milena Grifoni}
\affiliation{Institute for Theoretical Physics, University of
Regensburg, 93040 Regensburg, Germany}
\date{\today}

\begin{abstract}
{
We present a nonperturbative and formally exact approach for the charge transport in interacting nanojunctions based on a real-time path integral formulation of the reduced system dynamics.  For reservoirs of noninteracting fermions,  the exact trace over the leads' degrees of freedom results in the time-nonlocal Feynman-Vernon influence functional, a functional of the Grassmann-valued paths of the nanojunction, which induces correlations among the tunneling transitions in and out of the nanojunction. 
An expansion of the influence functional in terms of the number of tunneling transitions, and integration of the Grassmann variables between the tunneling times, allows us to obtain  a still exact generalized master equation (GME)  for the populations of the reduced density matrix (RDM) in the occupation number representation, as well as a formally exact expression for the current. 
By borrowing the nomenclature of the famous spin-boson model, we parametrize the two-state dynamics of each single-particle fermionic degree of freedom, in the occupation number representation, in terms of blips and sojourns.\\
\indent We apply our formalism to the exactly solvable resonant level model (RLM) and to the single-impurity Anderson model (SIAM), the latter being a prototype system for studying  strong correlations.
For both systems, we demonstrate a hierarchical  diagrammatic structure. While the hierarchy closes at the second-tier for the RLM, this is not the case for the interacting SIAM.  Upon inspection of the current kernel, known results from various perturbative and nonperturbative approximation schemes to quantum transport in the SIAM are recovered. 
Finally, a novel noncrossing approximation for the hierarchical kernel is developed, which enables us to  systematically decrease temperature at each next level of the approximation. Analytical results for a simplified fourth-tier scheme are presented both in equilibrium and nonequilibrium and with an applied magnetic field.
}
\end{abstract}
\maketitle


\section{Introduction}
\label{intro}
The qualitative understanding and quantitative description of transport properties of interacting nanojunctions is one of the core problems of nonequilibrium condensed matter physics. Interacting nanojunctions describe the general class of open systems whereby a quantum system S of interest (a molecule, a quantum wire, a set of quantum dots etc.) is coupled via tunneling to two or more fermionic reservoirs held at different chemical potential and/or temperature, see Fig. 1.
Relevant observables of interest are typically the average current flowing through the junctions, or its higher-order cumulants, which result from a nonequilibrium configuration in the leads.\\ 
\indent The presence of many-body electronic interactions in the central system, in combination with the large number of degrees of freedom in the fermionic reservoirs, renders the solution of the transport problem  a challenge.   Not even for the archetypal single-impurity Anderson model (SIAM) \cite{Anderson1961}, where the central system is a single orbital which can accommodate two electrons of opposite spin, the current-voltage characteristics for this model has yet been  obtained in closed analytic form in the whole regime of parameters.  The SIAM is a prototypical example to investigate the interplay between strong correlations in the central system and a continuum of degrees of freedom provided by the leads electrons. Below a critical temperature, known as Kondo temperature $T_{\rm K}$, this interplay gives rise to the emergence of the Kondo singlet, a bound state at the Fermi level signaling the screening of  the unpaired impurity spin by the conduction electrons~\cite{Ng1988, Glazman1988,Hewson1993, vanderWiel2000, Kouwenhoven2001}. Importantly, the Kondo temperature depends exponentially on the  tunneling coupling, showing the need of nonperturbative approaches in the tunneling to capture this effect~\cite{Hewson1993}.\\ 
\indent The necessity to develop approximation schemes enabling the treatment of tunneling and interactions on the same footing is at the core of various approaches to interacting quantum transport which have been proposed in the literature.
 In order to understand the novelty of the method proposed in this work, and to put it in a proper context, we shortly summarize the very essence of the main approaches available so far. We start by distinguishing between numerically exact methods and analytical or semi-analytical schemes. 
 In equilibrium, numerically exact methods such as the numerical renormalization group (NRG) \cite{Wilson1975,Bulla2008} or the density matrix renormalization group (DM-NRG) \cite{White1992,Schollwoeck2011} are well-established to evaluate the linear conductance through nanojunctions with only few degrees of freedom of the central system. In this work we shall use results from DM-NRG simulations to benchmark various approximation routes for the SIAM. 
 Numerical schemes also applicable in nonequilibrium situations are being developed and involve, among others, time-dependent DM-NRG methods \cite{Anders2005,Heidrich-Meisner2009},  iterative~\cite{Weiss2008} and Monte Carlo~\cite{Muehlbacher2008} path integral schemes, and hierarchical equation of motion approaches \cite{Jin2008,Schinabeck2016}, also based on the path integral approach, or auxiliary function methods~\cite{Arrigoni2013,Arrigoni2014, Fugger2018}. See also~\cite{Eckel2010} for a more detailed review and comparative study of some of these approaches.
 The computational effort however grows exponentially with the number of degrees of freedom of the central system, which renders  numerical approaches impractical for interacting  nanojunctions with more than a few degrees of freedom.\\ 
\indent For this reason, also semi-analytical and analytical schemes have attracted much interest to address the transport problem.
 These encompass frameworks where the expression for the current involves the calculation of  nonequilibrium Greens' functions  and associated self-energies~\cite{Meir1992, Haug2008, Scheer2010, Ryndyk2016,Thoss2018, Cohen2020, Evers2020} to ones where the current results from a statistical average, and thus the central quantities are the density operator or the reduced density matrix (RDM) of the open system, see e.g. \cite{Blum2012,Schoeller1997,Timm2008, Timm2011}. Given the large variety of methods and their different range of applicability, it is rather difficult to keep focus of the enormous amount of literature by now available, so that comparison between various schemes and short topical reviews become very valuable \cite{Andergassen2010, Levy2013, Koller2010, deSouza2019, Reimer2019, Lindner2019, Ferguson2020}.\\ 
\indent From the perspective of this work, it is convenient to separate the available methods in two main groups. 
 In the first one, and by far the most popular, starting point are dynamical equations for the relevant quantities, which  are solved by truncating a hierarchy of equations,  or by systematic perturbation schemes.  The most known dynamical schemes involve equations of motion for the Green's functions \cite{Meir1991, Kashcheyevs2006, VanRoermund2010, Levy2013, Lavagna2015,Eckern2020}, kinetic equations for the density matrix \cite{Pedersen2005, Karlstrom2013} or the reduced density operator \cite{Schoeller1994, Koenig1996, Schoeller1997, Leijnse2008, Kern2013, Jin2014, Saptsov2012, Saptsov2014}, and perturbative RG-schemes \cite{Schoeller2000, Rosch2003,Paaske2004a, Paaske2004b,Rosch2005, Karrasch2006,Pletyukhov2012, Nestmann2021}.
 In the second and much less explored one,  starting point  are formally exact expressions for generating functions or for the reduced density matrix obtained with field integral methods.  Here the relevant information on the time-evolution of the  open system is captured e.g. by Keldysh effective action \cite{Kamenev2005,Altland2010, Bock2016}  or double-path Feynman-Vernon influence functionals~\cite{Weiss2012, Jin2010}, resulting from an exact trace over the reservoir degrees of freedom. 
The advantage of these approaches is to enable analytical solutions being intrinsically nonperturbative in both the tunneling and interaction. For example, generating functional methods have been used to treat zero-bias anomalies in metallic islands~\cite{Altland2009},   and  the nonequilibrium Kondo effect in the SIAM~\cite{Wingreen1994, Smirnov2013} and in carbon nanotube-based quantum dots~\cite{Schmid2015,Niklas2016, Lahiri2020}. However, a treatment of interacting nanojunctions based on an exact path integral expression for the junction's RDM has not been discussed yet.    
In this work we wish to bridge this gap.\\ 
\indent Here we propose an analytical  method, based on the Feynman-Vernon influence functional approach for fermionic reservoirs. This approach provides an exact expression for the RDM in the fermionic coherent-state representation~\cite{Negele1988, Cahill1999} and has been used to investigate transient and stationary transport in noninteracting nanojunctions~\cite{Tu2008,Jin2010, Zhang2012, Yang2016}.\\
\indent Here we show that the influence functional is also a powerful tool to treat interaction effects all the way down to low temperatures for a generic nanojunction linearly coupled to nonequilibrium fermionic reservoirs. Starting from the exact formal expression for the system's RDM, we derive a still exact quantum master equation for the same object. Importantly, and one major result of this work, a nested hierarchical structure of the quantum master equation kernel is recognized which allows for devising systematic, non perturbative schemes in the calculation of the kernel.  Similarly, a path-integral expression for the current through the nanojunction is obtained and its relation to current formulae in terms of nonequilibrium Green's functions~\cite{Meir1992} elucidated.   We apply then our formalism to two archetypal examples. The first one, the exactly solvable resonant level model (RLM)~\cite{Reimer2019}, is used to show that the nesting in the hierarchical structure is finite for  noninteracting models, and thus a closed analytical form for the current can be obtained. The second is the SIAM, where the combined effect of interactions and tunnel coupling imply an infinite, hierarchical structure. On the one hand, by performing an expansion of the kernel in powers of the tunneling coupling,  Coulomb blockade physics, single electron tunneling and cotunneling effects occurring in the weak coupling limit~\cite{Grabert1992} can be described, in agreement with established diagrammatic perturbative schemes, see e.g.~\cite{Koenig1996,Koller2010}. 
On the other hand, by truncating the hierarchy to the second tier, the diagrammatic resonant tunneling approximation (RTA)~\cite{Koenig1996,Karlstrom2013} is recovered. By neglecting crossed diagrams in the RTA, a second-tier noncrossing approximation (NCA2) is obtained. Noticeably, the NCA2 reproduces the famous result for the SIAM Green's function as obtained in~\cite{Meir1993} using the equation of motion (EOM) method. Finally, by only including charge fluctuations, the NCA2 reduces to the dressed second order (DSO) approximation discussed in~\cite{Kern2013}. Such schemes foresee the onset of the Kondo zero-bias anomaly, but are plagued by a pinning problem of the self-energy at the particle-hole symmetry point when decreasing the temperature, see e.g.~\cite{Kashcheyevs2006}. Further, the temperature for the onset of the anomaly differs from the proper Kondo temperature, see e.g.~\cite{Kern2013}. Thus higher-oder tiers are required. In  this work,  we develop an infinite-tier scheme named dressed bubble approximation (DBA), and a simplified NCA version of it,  whereby the problem of finding the self-energies, and thus the retarded Green's function, is reduced to a geometrical problem involving the inversion of matrices of dimension $4\times 4$, at most, for the SIAM. Exemplarily, we discuss a fourth-tier scheme, the NCA4, whose simplified version allows for an easy-to-handle analytical solution that improves over the second-tier schemes by lifting the pinning problem. We test the predictions for the linear conductance of this simplified NCA4 scheme against numerically exact DM-NRG results. Moreover, we study the transport properties in a nonequilibrium situation and in the presence of an applied magnetic field.\\
\indent The paper is structured as follows. 
In Sec.~\ref{PI_representation} we introduce the generic model for  interacting nanojunctions and derive formally exact path integral expressions in the coherent-state representation~\cite{Negele1988} for both the RDM and for the current at a given lead~\cite{Tu2008,Jin2010}. Here, as a result of the trace over the fermionic reservoirs, tunneling events in and out of the central system become correlated through the action of the time-nonlocal Feynman-Vernon influence functional. How to then obtain  an exact master equation for the RDM in the case of noninteracting nanojunctions is further discussed in~\cite{Tu2008,Jin2010}.  
 Since our focus is on the interplay of interactions and tunneling, we perform a first crucial step by  expressing  the exact propagator in the {\em{occupation number representation}} starting from the coherent-state picture. This transformation paves the route for the  expansion of the influence functional in series of tunneling transitions discussed in Sec.~\ref{tunneling_expansion}, and for the  diagrammatic representation of the propagator in terms of {\em {blips}} and {\em{sojourns}} illustrated in Sec.~\ref{state_conserving_and_DR}. Here, borrowing the nomenclature from the famous spin-boson problem~\cite{Leggett1987}, we show that by expanding the influence functional and integrating out the Grassmann variables we can view a path as a sequence of blips/sojourns as for the two-state system in the spin-boson problem; the two states of the spin  correspond here to (fermionic) degrees of freedom of the central system  being empty or singly occupied. In Sec.~\ref{GME_pop_current} this knowledge is used to obtain an exact generalized master equation (GME) for the diagonal elements (populations) of the RDM,  as well as an integral equation for the current. The hierarchical structure of the populations kernel in Laplace space  
and the Dyson equation for its propagator are derived in the central Sec.~\ref{exact_kernel}. Specializing to the case of proportional coupling, the connection between the current kernel and the retarded Green's function is established in Sec.~\ref{prop_coupl_and_GF}. In such case, the Meir-Wingreen formula for the current is recovered.
There follow two sections where we apply our general formalism to the exactly solvable RLM, Sec.~\ref{RLM}, and to the SIAM, Sec.~\ref{SIAM}. Since the RLM accommodates at most one electron, interactions effects play no role here, and   the Dyson equation for the propagator is solved exactly at the second-tier level. In the SIAM in contrast, the hierarchy of equations for the propagator does not close, and approximation schemes are required. We show how to recover within our formalism various common approximation schemes for the SIAM and discuss further the novel infinite-tier DBA scheme. Analytical results, which include the temperature dependence of the linear conductance and the differential conductance in the presence of an applied magnetic field, are then exemplarily obtained within the NCA4, a truncation of this scheme to the fourth tier, with some additional simplifications.  Finally, conclusions are drawn in Sec.~\ref{conclusions}.
Some of the detailed derivations are deferred to the appendices.

\section{Path integral representation for the reduced density matrix and the current} 
\label{PI_representation}
\begin{figure}[ht]
\begin{center}
\includegraphics[width=5cm,angle=0]{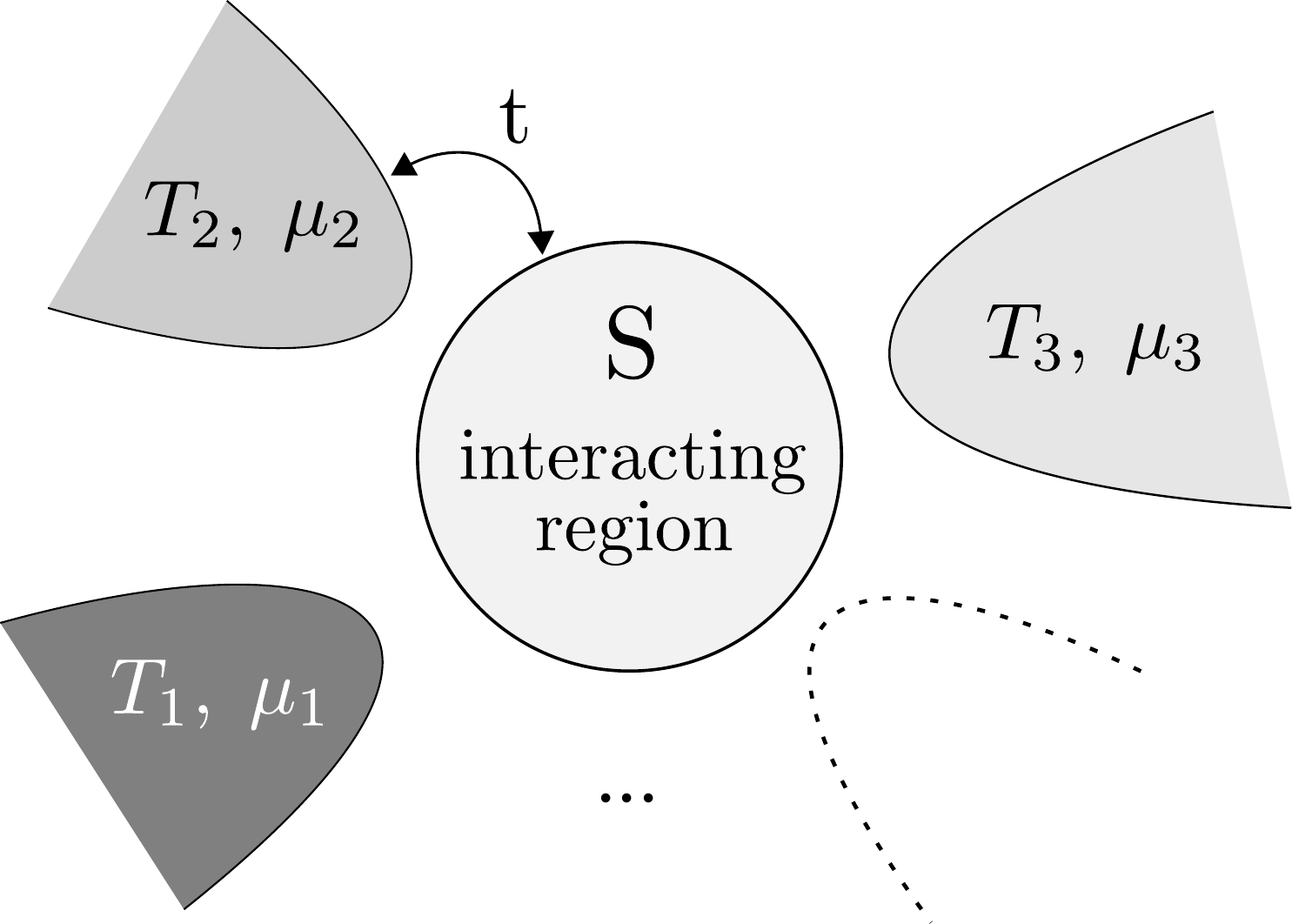}
\caption{\small{General transport setting where a central interacting region, the system $\rm S$, is tunnel-coupled to  several noninteracting fermionic leads with given temperature and chemical potential.}}
\label{scheme}
\end{center}
\end{figure}
 We consider the general transport setting depicted in Fig.~\ref{scheme}, where a central interacting region, with a number $N$ of available electron states, indexed by $i$ or $j$ in what follows, is connected via tunnel coupling to noninteracting fermionic leads, held in general at different chemical potentials and/or different temperatures. 
This general setting can describe molecular junctions~\cite{Cohen2020,Evers2020},   manufactured nanostructures, such as lateral quantum dots~\cite{Hanson2007}, or other complex junctions~\cite{Laird2015}.\\
\indent The Hamiltonian of this transport setup consists of three terms, corresponding to the partition in central system (S) plus leads coupled via a particle exchange term, and reads     
\begin{equation}
\begin{aligned}
\label{H_general}
H=&H_{\rm S}+\sum_{\alpha k \sigma}\epsilon_{\alpha k}c^{\dag}_{\alpha k \sigma}c_{\alpha k \sigma}\\
&+\sum_{i \alpha k \sigma}\left[ {\rm t}_{i\alpha k \sigma} a_{i}^{\dag}c_{\alpha k \sigma}+ {\rm t}^{*}_{i\alpha k \sigma} c^{\dag}_{\alpha k \sigma}a_{i}\right]\;.
\end{aligned}
\end{equation}
The central system part is left unspecified at the present stage, being some function of the fermionic creation and annihilation operators $a_i^\dag$ and $a_i$ relative to the single-particle basis $\{|i\rangle\}$ in S. It contains in principle interaction terms which are quartic in these operators. Further, for simplicity, $H_{\rm S}$ is assumed to be time-independent, although the inclusion of time-dependent terms in a real-time path integral formalism is straightforward~\cite{Grifoni1998, Weiss2012}.
The second term is the free leads part with creation and annihilation operators $c_{\alpha k \sigma}^\dag$ and $c_{\alpha k \sigma}$, where $\alpha$ runs over the leads, $\sigma$ is the spin degree of freedom, and $k$ denotes the $k$th electronic state in the lead $\alpha$. The third term in Eq.~\eqref{H_general} describes the exchange of particles between dot and leads, with the energies ${\rm t}_{i\alpha k \sigma}$ giving the amplitude  of the tunnel coupling. In the continuum limit, denoting with $\varrho_{\alpha \sigma}(\epsilon)$ the density of states of lead $\alpha$ in energy space, we set $\sum_{\alpha k \sigma}\rightarrow \sum_{\alpha \sigma}\int d\epsilon \varrho_{\alpha \sigma}(\epsilon)$. Then, the tunnel coupling is characterized by the energy-dependent hybridization matrix
$\boldsymbol{\Gamma}(\epsilon)=\sum_\alpha\boldsymbol{\Gamma}_{\alpha}(\epsilon)$ whose elements are 
\begin{equation}
\label{Gamma_matrix}
[\boldsymbol{\Gamma}_\alpha(\epsilon)]_{ij}:=2\pi\sum_\sigma \varrho_{\alpha \sigma}(\epsilon){\rm t}_{i \alpha \sigma}(\epsilon){\rm t}_{j \alpha \sigma}^*(\epsilon) \;.
\end{equation}

\subsection{Reduced density matrix and current} 
\indent Let us denote with $\rho$ the reduced density matrix (RDM) of the central system.
The RDM is obtained from $\rho_{\rm tot}(t)$, the total density matrix, by tracing out the leads degrees of freedom $\rho(t)={\rm Tr}_{\rm leads}[\rho_{\rm tot}(t)]$, with the time evolution of $\rho_{\rm tot}$ being governed by the evolution operator associated to the Hamiltonian~\eqref{H_general}. Since we have assumed noninteracting leads, this trace can be performed exactly in the coherent-state representation using standard path integral techniques~\cite{Negele1988, Tu2008}. We consider for simplicity an initially factorized density matrix  $\rho_{\rm tot}(t_0)=\rho(t_0)\otimes\rho_{\rm leads}^{\rm th}$, where $\rho_{\rm leads}^{\rm th}=\bigotimes_\alpha\rho_{\alpha}^{\rm th}$, with the lead $\alpha$ in the grand-canonical equilibrium state  at a given temperature $T_\alpha$ and chemical potential $\mu_\alpha$, see Fig.~\ref{scheme}. 
The propagator $\mathcal{J}$ yields the matrix elements of the RDM in the coherent-state representation at time $t$ according to 
\begin{equation}
\begin{aligned}
\label{rho_ab}
\langle\boldsymbol{\xi}_{a}|\rho(t)|\boldsymbol{\xi}_{b}\rangle=\int d^2\boldsymbol{\xi}_0 d^2\bar{\boldsymbol{\xi}}_0 \mathcal{J}(\boldsymbol{\xi}^{*}_{a},\boldsymbol{\xi}_{b},t;\boldsymbol{\xi}_{0},\bar{\boldsymbol{\xi}}^{*}_{0},t_{0})\rho_{\boldsymbol\xi_0\bar{\boldsymbol{\xi}}_{0}}(t_{0})\;,
\end{aligned}
\end{equation}
where $\rho_{\boldsymbol\xi_0\bar{\boldsymbol{\xi}}_{0}}(t_{0})=\langle \boldsymbol\xi_0|\rho(t_{0})|\bar{\boldsymbol{\xi}}_{0}\rangle$. The Grassmann variables $\boldsymbol{\xi}=(\dots,\xi^i,\dots)$ and $\boldsymbol{\xi}^{*}=(\dots,\xi^{i*},\dots)$ have one component for each electronic state which is defined by $\hat{a}_{i}|\boldsymbol{\xi}\rangle=\xi^i|\boldsymbol{\xi}\rangle$ and $\langle\boldsymbol{\xi}|\hat{a}_{i}^{\dag}=\xi^{i*}\langle\boldsymbol{\xi}|$.
Following the procedure outlined in Appendix~\ref{propagator_PI}, the propagator acquires the formal, exact path integral expression in the coherent-state representation
\begin{equation}
\begin{aligned}
\label{propagator_rho}
\mathcal{J}(\boldsymbol{\xi}^{*}_{a},\boldsymbol{\xi}_{b},t;&\boldsymbol{\xi}_{0},\bar{\boldsymbol{\xi}}^{*}_{0},t_{0})=\int_{\boldsymbol{\xi}_{0}}^{\boldsymbol{\xi}^{*}_{a}} D\boldsymbol{\xi}\int_{\bar{\boldsymbol{\xi}}^{*}_{0}}^{\boldsymbol{\xi}_{b}} D \bar{\boldsymbol{\xi}}\\
&\times e^{\frac{\rm i}{\hbar}[S_{\rm S}(\boldsymbol{\xi}^{*},\boldsymbol{\xi})-S_{\rm S}^*(\bar{\boldsymbol{\xi}}^{*},\bar{\boldsymbol{\xi}})]}\mathcal{F}(\boldsymbol{\xi}^{*},\boldsymbol{\xi},\bar{\boldsymbol{\xi}}^{*},\bar{\boldsymbol{\xi}})\;,
\end{aligned}
\end{equation}
where $\int D\boldsymbol{\xi}= \int\prod_{k=1}^{K} d\boldsymbol{\xi}(t_k)^*d\boldsymbol{\xi}(t_k)$ and  $\int D\bar{\boldsymbol{\xi}}= \int\prod_{k=1}^K d\bar{\boldsymbol{\xi}}(t_k)^*d\bar{\boldsymbol{\xi}}(t_k)$ denote the sums over paths in the forward and backward time branches, respectively, with fixed end-points and $K\rightarrow \infty$. The action of the central system is given by the time-discretized expression ($\boldsymbol\xi_k\equiv \boldsymbol\xi(t_k)$ and $t_{k+1}=t_k+\delta t$)
\begin{equation}
\begin{aligned}
\label{action_S}
e^{\frac{\rm i}{\hbar}S_{\rm S}(\boldsymbol{\xi}^{*},\boldsymbol{\xi})}=&\prod_{k=0}^{K} e^{-\boldsymbol{\xi}_k^* \cdot\boldsymbol{\xi}_k+\boldsymbol{\xi}^*_{k+1}\cdot\boldsymbol{\xi}_k-\frac{\rm i}{\hbar}H_{\rm S}(\boldsymbol{\xi}^*_{k+1},\boldsymbol{\xi}_k)\delta t}\;,\\
e^{\frac{\rm i}{\hbar}S_{\rm S}^*(\bar{\boldsymbol{\xi}}^{*},\bar{\boldsymbol{\xi}})}=&\prod_{k=0}^{K} e^{-\bar{\boldsymbol{\xi}}^*_k\cdot \bar{\boldsymbol{\xi}}_k+\bar{\boldsymbol{\xi}}^*_k\cdot\bar{\boldsymbol{\xi}}_{k+1}+\frac{\rm i}{\hbar}H_{\rm S}(\bar{\boldsymbol{\xi}}^*_k,\bar{\boldsymbol{\xi}}_{k+1})\delta t}\;,
\end{aligned}
\end{equation}
with $\boldsymbol{\xi}^*(t_{K+1})\equiv \boldsymbol{\xi}_a^*$ and $\bar{\boldsymbol{\xi}}(t_{K+1})\equiv \boldsymbol{\xi}_b$.
 Due to the trace over the leads, these paths are coupled by the Feynman-Vernon influence functional~\cite{Weiss2012} $\mathcal{F}(\boldsymbol{\xi}^{*},\boldsymbol{\xi},\bar{\boldsymbol{\xi}}^{*},\bar{\boldsymbol{\xi}})=\exp[\varPhi(\boldsymbol{\xi}^{*},\boldsymbol{\xi},\bar{\boldsymbol{\xi}}^{*},\bar{\boldsymbol{\xi}})]$ whose phase can be given in the following symmetric form, see Appendix~\ref{Influence_phase},
\begin{equation}
\begin{aligned}
\label{phase_IF}
\varPhi(&\boldsymbol{\xi}^{*}, \boldsymbol{\xi},\bar{\boldsymbol{\xi}}^{*},\bar{\boldsymbol{\xi}})=
-\int_{t_{0}}^{t}dt'\int_{t_{0}}^{t'}d t''\Big[\\
&\boldsymbol{\xi}^{*}(t')\cdot  \mathbf{g}_-(t'-t'')\cdot\boldsymbol{\xi}(t'') + \boldsymbol{\xi}(t') \cdot \mathbf{g}_+^*(t'-t'')\cdot\boldsymbol{\xi}^{*}(t'')\\
-&\bar{\boldsymbol{\xi}}(t')\cdot  \mathbf{g}_-^*(t'-t'')\cdot\bar{\boldsymbol{\xi}}^{*}(t'')-\bar{\boldsymbol{\xi}}^{*}(t')\cdot \mathbf{g}_+(t'-t'')\cdot\bar{\boldsymbol{\xi}}(t'')\\
+&\bar{\boldsymbol{\xi}}^{*}(t')\cdot  \mathbf{g}_-(t'-t'')\cdot\boldsymbol{\xi}(t'') + \bar{\boldsymbol{\xi}}(t')\cdot \mathbf{g}_+^*(t'-t'')\cdot\boldsymbol{\xi}^{*}(t'')\\
-&\boldsymbol{\xi}(t')\cdot \mathbf{g}_-^*(t'-t'')\cdot\bar{\boldsymbol{\xi}}^{*}(t'')-\boldsymbol{\xi}^{*}(t')\cdot \mathbf{g}_+(t'-t'')\cdot\bar{\boldsymbol{\xi}}(t'')\Big]\;.
\end{aligned}
\end{equation}
The correlation matrices $\mathbf{g}_{\pm}$ have elements 
\begin{equation}
\begin{aligned}
\label{corr_matrices}
{\rm g}_{ij,\pm}(t)=&\frac{1}{\hbar^2}\sum_{\alpha k \sigma} {\rm t}_{i \alpha k \sigma}{\rm t}^{*}_{j \alpha k \sigma}f_{\pm}^\alpha(\epsilon_{k})e^{-\frac{\rm i}{\hbar}\epsilon_{\alpha k}t}
\;,
\end{aligned}
\end{equation}
where $f_+^\alpha(\epsilon_{k})=[1+e^{\beta_{\alpha}(\epsilon_{\alpha k}-\mu_{\alpha})}]^{-1}$ is the Fermi function  of lead $\alpha$ and $f_-^\alpha(\epsilon_k):=1-f_+^\alpha(\epsilon_k)$. As shown in Appendix~\ref{leads_corr_func}, these matrix elements are the correlation functions of the leads' force operator.\\
\begin{figure}[ht]
\begin{center}
\includegraphics[width=6.5cm,angle=0]{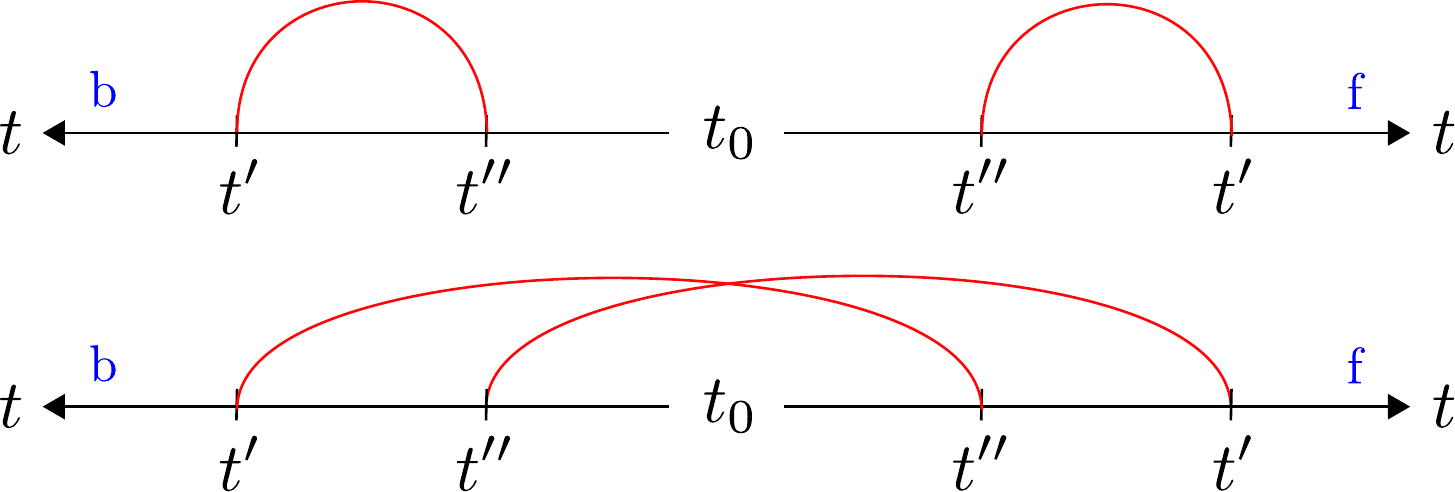}
\caption{\small{The elementary processes displayed in the phase of the influence functional, Eq.~\eqref{phase_IF}: Each line joining a pair of tunneling transitions represents either a  creation/annihilation or annihilation/creation process, giving a total of eight elementary processes. In the lower panel, the paths in the forward (f) and backward (b) time branch are coupled by the influence phase $\varPhi$.}}
\label{processes}
\end{center}
\end{figure}
\indent The phase of the influence functional, Eq.~\eqref{phase_IF}, displays the eight fundamental processes involved in the transport setup consisting of pairs of tunneling events, each creating or annihilating one electron in the central system, connected by a fermion line. These processes are shown in Fig.~\ref{processes}. A fermion line is mathematically represented by the time-dependent part of the correlation function calculated at the difference between the times of the two events, see Eq.~\eqref{corr_matrices}. To the two tunneling transitions coupled by a correlation matrix $\mathbf{g}$ we attributed the product of the two tunnel amplitudes with the appropriate Fermi function, as given by the prefactors of $\mathbf{g}$. As an example of process in the phase of the influence functional, consider $\xi^{i*}(t')  {\rm g}_{ij,-}(t'-t'')\xi^j(t'')$, which is one of the terms generated by the scalar product in the first term of the sum in Eq.~\eqref{phase_IF}. This is the forward process depicted in the upper-right part of Fig.~\ref{processes} and consists in the destruction of one electron in the state $j$ of the central system at time $t''$ followed by the creation of one electron at a later time $t'$ in the state $i$. Note that this is actually a collection of processes, as there is a sum over the leads and their states in the correlation function. Likewise, $\bar{\xi}^i(t'){\rm g}_{ij,-}^*(t'-t'')\bar{\xi}^{j*}(t'')$ gives the creation of one electron in the dot at time $t''$ followed by the annihilation of an electron at a later time $t'$ in the backward time branch, see the upper-left part of Fig.~\ref{processes}.\\
\indent As can be seen from Eq.~\eqref{phase_IF}, in the influence functional, forward and backward paths of the individual degrees of freedom are self-interacting and also coupled to each other in a time-nonlocal fashion. The latter feature ensures that the Feynman-Vernon approach takes fully into account the back-action due to the leads in the system evolution.

\subsection{Current}

We define the particle current in lead $l$ as the expectation value of $\dot{\hat{N}}_l(t)$, the time derivative of the particle number operator of lead $l$. In the Heisenberg picture $\hat{N}_l(t)=\sum_{k \sigma} c_{l k \sigma}^{\dag}(t)c_{l k \sigma}(t)$ so that, with the general Hamiltonian in Eq. (\ref{H_general}),
\begin{equation}
\begin{aligned}
\dot{\hat{N}}_l(t)=&-\frac{\rm i}{\hbar}\left[\hat{N}_l(t),H(t)\right]\\
=&-\frac{\rm i}{\hbar}\sum_{i k \sigma}\left[{\rm t}_{i l k \sigma}^*c_{l k \sigma}^{\dag}(t)a_i(t)-{\rm t}_{i l k \sigma}a_i^{\dag}(t)c_{l k \sigma}(t)\right]\;.
\end{aligned}\end{equation}
The electron current $I_l(t)=-e\langle \dot{\hat{N}}_l (t)\rangle$, where $\langle \dot{\hat{N}}_l (t)\rangle={\rm Tr}[\dot{\hat{N}}_l (t)\rho_{\rm tot}]$, assumes the form
\begin{equation}
\begin{aligned}
\label{current}
I_l (t)=&\;e\frac{\rm i}{\hbar}\sum_{i k \sigma}\left[{\rm t}_{i l  k \sigma}^*\langle c_{l  k \sigma}^{\dag}(t)a_i(t)\rangle -{\rm t}_{i l  k \sigma}\langle a_i^{\dag}(t)c_{l  k \sigma}(t)\rangle \right]\\
\equiv &\;e 2{\rm Re}\;{\rm Tr}_{\rm S}[\boldsymbol{\mathcal{A}}_l (t)]\;,
\end{aligned}
\end{equation}
where, using $\hat{O}_H(t)=U^\dag (t,t_0)\hat{O}_S U (t,t_0)$, we have defined the system operator $\mathcal{A}_l (t)$ as the following trace over the leads
\begin{equation}
\begin{aligned}
\label{A_alpha}
[\boldsymbol{\mathcal{A}}_l (t)]_{ii}:= -\frac{\rm i}{\hbar}\sum_{k \sigma}{\rm t}_{i l  k  \sigma}{\rm Tr}_{\rm leads}\left[a_i^\dag c_{l  k \sigma}\rho_{\rm tot}(t)\right]\;.
\end{aligned}
\end{equation}
The system operator $\boldsymbol{\mathcal{A}}_l (t)$ admits a path integral representation similar to the one carried out for the RDM, namely
\begin{equation}
\begin{aligned}
\label{A_alpha_PI}
\langle \boldsymbol\xi_a|\boldsymbol{\mathcal{A}}_l (t)| \boldsymbol\xi_b\rangle= \int d^2\boldsymbol{\xi}_0 d^2\bar{\boldsymbol{\xi}}_0 \mathcal{J}^I_l (\boldsymbol{\xi}^{*}_a,\boldsymbol{\xi}_b,t;\boldsymbol{\xi}_{0},\bar{\boldsymbol{\xi}}^{*}_{0},t_{0})\rho_{\boldsymbol\xi_0\bar{\boldsymbol{\xi}}_{0}}(t_{0})\;.
\end{aligned}
\end{equation}
The current propagator is given by~\cite{Tu2008,Jin2010} 
\begin{equation}
\begin{aligned}
\label{propagator_I}
\mathcal{J}^I_l  &(\boldsymbol{\xi}^{*}_{a},\boldsymbol{\xi}_{b},t;\boldsymbol{\xi}_{0},\bar{\boldsymbol{\xi}}^{*}_{0},t_{0})=\int_{\boldsymbol{\xi}_{0}}^{\boldsymbol{\xi}^{*}_{a}} D\boldsymbol{\xi}\int_{\bar{\boldsymbol{\xi}}^{*}_{0}}^{\boldsymbol{\xi}_{b}} D \bar{\boldsymbol{\xi}}\\
&\times e^{\frac{\rm i}{\hbar}[S_{\rm S}(\boldsymbol{\xi}^{*},\boldsymbol{\xi})-S_{\rm S}^*(\bar{\boldsymbol{\xi}}^{*},\bar{\boldsymbol{\xi}})]}\mathcal{I}_l (\boldsymbol{\xi}^{*},\boldsymbol{\xi},\bar{\boldsymbol{\xi}})\mathcal{F}(\boldsymbol{\xi}^{*},\boldsymbol{\xi},\bar{\boldsymbol{\xi}}^{*},\bar{\boldsymbol{\xi}})\;.
\end{aligned}
\end{equation}
This expression is similar to that of the propagator for the system RDM, Eq.~\eqref{propagator_rho}, the difference being the multiplicative current functional 
\begin{equation}
\begin{aligned}
\label{current_functional}
\mathcal{I}_l (\boldsymbol{\xi}^{*},\boldsymbol{\xi},\bar{\boldsymbol{\xi}})=-\int_{t_0}^t  dt'\;  \boldsymbol{\xi}^*(t)\big[&\mathbf{g}_{l ,-}(t-t')\boldsymbol{\xi}(t')\\
- & \mathbf{g}_{l ,+}(t-t')\bar{\boldsymbol{\xi}}(t') \big]\;.
\end{aligned}
\end{equation}
Here, the correlation matrices bear the index $l $ (which is not summed over) of the lead considered for the calculation of the current, with $\mathbf{g}_{\pm}(t)=\sum_{\alpha}\mathbf{g}_{\alpha,\pm}(t)$.
Moreover, there is one single time integral and the last Grassmann variable has the time argument fixed at the final time $t$ while the argument of the first runs from $t_0$ to $t$. Finally, the structure of the integrand in $\mathcal{I}$ is similar to that of $\varPhi$, the exponent of the influence functional $\mathcal{F}$ given in Eq.~\eqref{phase_IF}, except for the two constraints that fix the nature of the last Grassmann variable, reflecting the fact that the operator $a_i^\dag$ is fixed in the calculation of $\mathcal{A}_l $, see Eq.~\eqref{A_alpha}. In Appendix~\ref{Greens_functions-PI}, we show the connection between the path integral expression for the current and the Green's functions.

\subsection{Propagators in the occupation number representation}
\label{occupation_rep}

For a system with $N$ electronic states $i=1,\dots,N$, we introduce the composite index $\boldsymbol{n}=(n_1,\dots,n_N)$ collecting the occupations of the states in the occupation number representation, with $n_i=0,1$ for state $i$. The anticommutation relation obeyed by any two Grassmann variables yields the property~\cite{Negele1988}
\begin{equation}
\label{Grassmann_integrals}
\int d\xi^* d\xi  \Big\{
\xi^* \xi\;,\;\xi^*\;,\;\xi\;,\;1\Big\}
=\Big\{-1\;,\;0\;,\;0\;,\;0\Big\}\;.
\end{equation}
Note that $\xi^*$ and $\xi$ are independent Grassmann variables.
The definition of coherent states 
\[
|\boldsymbol{\xi}\rangle=\prod_{i=1}^N (1-\xi^i a_i^\dag)|0_i\rangle
\]
 and the property of the Grassmann integrals, Eq.~\eqref{Grassmann_integrals}, allow us to define the projectors
that map the system state from the coherent-state to the occupation number representation.
For example, in the case of a single electron state ($N=1$) an element of the RDM reads in the coherent-state representation
\begin{equation}
\begin{aligned}\label{RDM1state}
\langle\xi|\rho(t)|\bar\xi\rangle
=&\rho_{00}(t)+\rho_{01}(t)\bar\xi+\rho_{10}(t)\xi^*+\rho_{11}(t)\xi^*\bar\xi\;,
\end{aligned}
\end{equation}
where we have used $\langle 0|\xi\rangle=\langle0|(1-\xi a^\dag)|0\rangle=1-\langle0|\xi|1\rangle=1-\xi\langle0|1\rangle=1$ and $\langle 1|\xi\rangle=\langle 0|a|\xi\rangle=\langle 0|\xi|\xi\rangle=\xi\langle0|\xi\rangle=\xi$.
The elements of the RDM in the occupation number representation are then recovered by performing the Grassmann integrals
\begin{equation}
\begin{aligned}
\rho_{nn'}(t)=\Pi(n')\Pi^*(n)\;\langle\xi|\rho(t)|\bar\xi\rangle\;,
\end{aligned}
\end{equation}
where the projectors $\Pi^*(n)$ and $\Pi(n)$ integrate out the Grassmann variables to the left and to the right of the operator $|0\rangle\langle 0|$, respectively.  Their definitions are 
\begin{equation}
\begin{aligned}
\Pi^*(0) &= \int d\xi^* \xi^*\;,\quad \Pi^*(1)=\int d\xi^* \;,\\ 
\Pi(0) &= \int d\bar\xi \;\bar\xi\;,\qquad \Pi(1)=\int d\bar\xi \;,
\end{aligned}
\end{equation}
as can be checked by applying the rules in Eq.~\eqref{Grassmann_integrals} for the Grassmann integrals.
In the general case, the populations, identified by the occupations $n_1,\dots,n_N$, are given by ${\rm P}_{\boldsymbol{n}}(t)=\rho_{\boldsymbol{n}\boldsymbol{n}}(t)=\Pi_b(\boldsymbol{n})\Pi_a^*(\boldsymbol{n}) \;\rho_{ab}(t)$, where
\begin{equation}
\begin{aligned}
\label{def_projector}
\Pi(\boldsymbol{n})=\prod_{i=1}^N\Pi^{i}(n_i)\;,\qquad \Pi^*(\boldsymbol{n})=\prod_{i=N}^1\Pi^{i*}(n_i) \;.
\end{aligned}
\end{equation}
The propagator for the populations in the occupation number representation gives the population vector at time $t$  according to 
\begin{equation}
\label{Populations}
{\rm P}_{\boldsymbol{n}'}(t)=\sum_{\boldsymbol{n}''}J_{\boldsymbol{n}'\boldsymbol{n}''}(t,t_0){\rm P}_{\boldsymbol{n}''}(t_0)\;.
\end{equation}
Thus, the matrix element $(\boldsymbol{n}',\boldsymbol{n})$ of the propagator is obtained by fixing the initial state to $\rho(t_0)=|\boldsymbol{n}\rangle\langle \boldsymbol{n}|$ so that  ${\rm P}_{\boldsymbol{n}''}(t_0)=\delta_{\boldsymbol{n}''\boldsymbol{n}}$. Then
\begin{equation}
\begin{aligned}
\label{Propagator_general}
J_{\boldsymbol{n}'\boldsymbol{n}}(t,t_0)=&\Pi_b(\boldsymbol{n}')\Pi_a^*(\boldsymbol{n}')\int d^2\boldsymbol{\xi}_0 d^2\bar{\boldsymbol{\xi}}_0 \mathcal{J}(\boldsymbol{\xi}^{*}_a,\bar{\boldsymbol{\xi}}_b,t;\boldsymbol{\xi}_0,\bar{\boldsymbol{\xi}}^{*}_0,t_0)\\
&\times \langle \boldsymbol\xi_0|\boldsymbol{n}\rangle\langle \boldsymbol{n}|\bar{\boldsymbol{\xi}}_0\rangle\;,
\end{aligned}
\end{equation}
where we use Eq.~\eqref{RDM1state} to calculate the matrix element of the RDM at $t_0$ in the coherent-state representation.
Equation~\eqref{Propagator_general} provides the recipe to obtain the propagator in the occupation number representation starting from the coherent-state path integral picture.\\
\indent Likewise, the diagonal elements of the current propagator in Eq.~\eqref{A_alpha_PI} are obtained as
\begin{equation}
\begin{aligned}
\label{}
J^I_{l ,\boldsymbol{n}'\boldsymbol{n}}(t,t_0)=&\Pi_b(\boldsymbol{n}')\Pi_a^*(\boldsymbol{n}')\int d^2\boldsymbol{\xi}_0 d^2\bar{\boldsymbol{\xi}}_0 \mathcal{J}^I_l (\boldsymbol{\xi}^{*}_a,\bar{\boldsymbol{\xi}}_b,t;\boldsymbol{\xi}_0,\bar{\boldsymbol{\xi}}^{*}_0,t_0)\\
&\times \langle \boldsymbol\xi_0|\boldsymbol{n}\rangle\langle \boldsymbol{n}|\bar{\boldsymbol{\xi}}_0\rangle\;.
\end{aligned}
\end{equation}
\section{Tunneling expansion of the influence functional}
\label{tunneling_expansion}
The influence functional couples the processes within and between the forward and backward time branches of the propagators. To move forward in the actual calculations, first we unify the two time branches in a single one and then discretize the paths of the central system in this unique time branch by expanding the influence functional in the number of processes, namely in powers of $\Gamma$, see Eq.~\eqref{Gamma_matrix}. This expansion gives rise to a diagrammatic unraveling of the propagator. The peculiarity of the present approach is the parametrization of the paths of S in terms of $N$ paths of reduced density matrices of individual two-state systems, one for each electron state of the central system.

\subsection{Diagrammatic unravelling of the propagator from the expansion of the influence functional}

\begin{figure}[ht]
\includegraphics[width=7cm,angle=0]{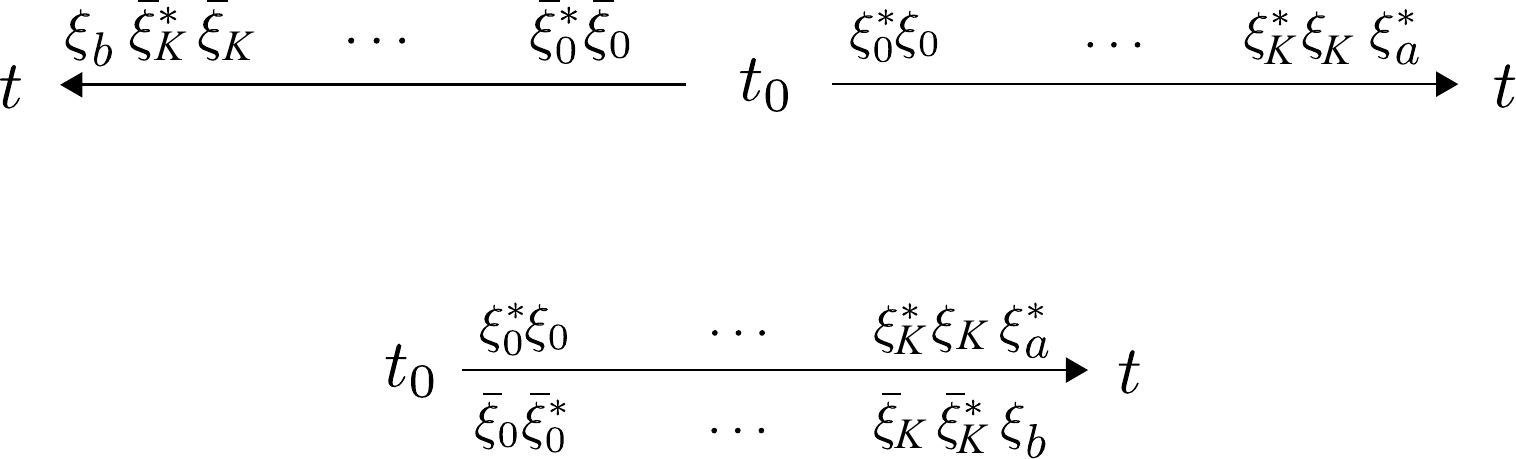}
\caption{\small{Forward (non-barred) and backward (barred) Grassmann variables arranged in a single (forward) time axis with reversal of the time direction for the backward branch. This requires associating to $\bar{\xi}^*$ the annihilation and to $\bar{\xi}$ the creation of an electron in the central system.}}
\label{path_collapsed}
\end{figure}

\begin{figure}[ht]
\includegraphics[width=7cm,angle=0]{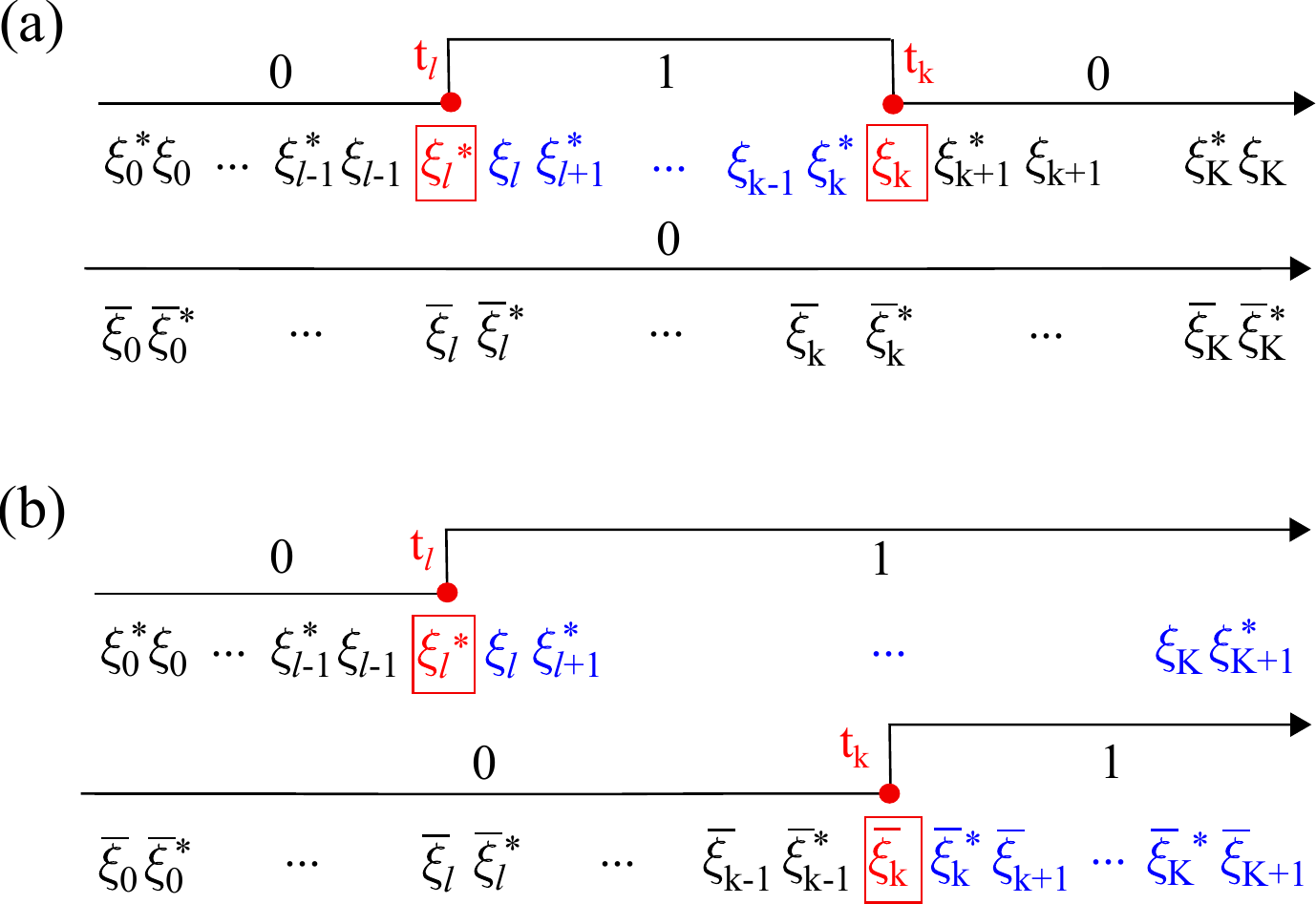}
\caption{\small{Two examples of paths of an individual electron state in the central system. The transition times and the corresponding Grassmann variables $\xi_k\equiv\xi(t_k)$ contained in the influence functions, Eq.~\eqref{Fkl_general}, are highlighted by red boxes. Zeros and ones indicate the occupation of the state in the forward and backward branches.}}
\label{path_detailed}
\end{figure}

Adopting the following notation    
\begin{equation}
\begin{aligned}\label{symbols}
\boldsymbol{\xi}_{+1}^{+1}&=\boldsymbol{\xi}\;,\quad\boldsymbol{\xi}_{+1}^{-1}=\boldsymbol{\xi}^*,\quad\boldsymbol{\xi}_{-1}^{+1}=\bar{\boldsymbol{\xi}},\quad\boldsymbol{\xi}_{-1}^{-1}=\bar{\boldsymbol{\xi}}^*,\\
\mathbf{g}_{+1}^{+1}&=\mathbf{g}_+,\quad \mathbf{g}_{+1}^{-1}=\mathbf{g}_+^*,\quad  \mathbf{g}_{-1}^{+1}=\mathbf{g}_-,\quad \mathbf{g}_{-1}^{-1}=\mathbf{g}_-^*\;,
\end{aligned}\end{equation}
where the lower index identifies the time branch (sign of the Fermi function) and the upper index performs the complex (Hermitian) conjugation for the Grassmann-valued paths (correlation matrices),
the phase of the influence functional, Eq.~\eqref{phase_IF}, can be expressed in the compact form $\varPhi(\boldsymbol{\xi}^{*},\boldsymbol{\xi},\bar{\boldsymbol{\xi}}^{*},\bar{\boldsymbol{\xi}})=\int_{t_{0}}^{t}dt'\int_{t_{0}}^{t'}d t'' \;F(t',t'')$, where
\begin{equation}
\begin{aligned}\label{IF_compact}
F(t',t'')=-\sum_{x,y,z=\pm 1}x\;\boldsymbol{\xi}_{y}^{z}(t')\mathbf{g}_{xz}^{-z}(t'-t'')\boldsymbol{\xi}_{x}^{-z}(t'')\;.
\end{aligned}\end{equation}
The eight elementary processes comprised by the phase of the influence functional are rendered by the sum over the three binary indexes $x,y,$ and $z$.
As these processes consist of couples of tunneling transitions, the expansion in the tunnel coupling is given by the sum over the number $m$ of pairs of  transitions 
\[
\mathcal{J}(\boldsymbol{\xi}^{*}_{a},\boldsymbol{\xi}_{b},t;\boldsymbol{\xi}_{0},\bar{\boldsymbol{\xi}}^{*}_{0},t_{0})=\sum_{m=0}^{\infty}\mathcal{J}^{(m)}(\boldsymbol{\xi}^{*}_{a},\boldsymbol{\xi}_{b},t;\boldsymbol{\xi}_{0},\bar{\boldsymbol{\xi}}^{*}_{0},t_{0})\;.
\]
The term with $2m$ transitions (order $m$ in $\Gamma$, cf. Eq.~\eqref{Gamma_matrix}) reads
\begin{equation}
\begin{aligned}\label{J_n}
\mathcal{J}^{(m)}(\boldsymbol{\xi}^{*}_{a},\boldsymbol{\xi}_{b},t; & \;\boldsymbol{\xi}_{0},\bar{\boldsymbol{\xi}}^{*}_{0},t_{0})=\int \mathcal{D}\{t\}_{m}\int_{\boldsymbol{\xi}_{0}}^{\boldsymbol{\xi}^{*}_{a}} D\boldsymbol{\xi}\int_{\bar{\boldsymbol{\xi}}^{*}_{0}}^{\boldsymbol{\xi}_{b}} D \bar{\boldsymbol{\xi}}\\
&\times e^{\frac{\rm i}{\hbar}[S_{\rm S}(\boldsymbol{\xi}^{*},\boldsymbol{\xi})-S_{\rm S}^*(\bar{\boldsymbol{\xi}}^{*},\bar{\boldsymbol{\xi}})]}\sum_{\mathcal{P}_m}\prod_{p=1}^m {\rm F}_{k_p,l_p}\;,
\end{aligned}\end{equation}
where $\mathcal{P}_m$ denotes one of the $(2m)!/(2^m m!)$ possible arrangements of $2m$ time indexes in groups of $2$ with no repetitions, meaning that two transitions at the same time instant are not allowed. The symbol $\int\mathcal{D}\{t\}_{m}$ comprises the nested time integrations over the $2m$ transition times. Explicitly 
\begin{equation}\label{Dt}
\int \mathcal{D}\{t\}_{m}:= \int_{t_0}^t dt_{2m}\int_{t_0}^{t_{2m}}dt_{2m-1}\dots\int_{t_0}^{t_2}dt_1\;.
\end{equation}
In Eq.~\eqref{J_n}, we have introduced the Grassmann-valued functions
\begin{equation}
\begin{aligned}\label{Fkl_general}
{\rm F}_{kl}=&-\sum_{i,j=1}^N\sum_{x,y,z=\pm 1}x\;[{\xi}^i_k]_{y}^{z}\;[{\rm g}_{ij}(t_k-t_l)]_{xz}^{-z}\; [{\xi}^j_l]_{x}^{-z}\;.
\end{aligned}\end{equation}	
Thus, in the present \emph{time-discretized} picture of the influence functional, the Grassmann-valued paths, which are expressed in terms of a set of Grassmann numbers associated to the specific time instants of the tunneling transitions, consist of individual transition at specific times (the sequence of times being ordered). The nested time integrals in Eq.~\eqref{Dt} reproduce all the possibilities for the sequences of processes. Finally, the sum over the set of coefficients $x,y,$ and $z$ in Eq.~\eqref{Fkl_general} and the sum over the system states, implicit in the scalar products with the correlation matrix, produce the sum-over-paths.\\
\indent To deal with the forward and backward paths with a single parametrization, it is convenient to use a single time direction for the two time branches depicted in Fig.~\ref{processes}.  In order to do so, it is necessary to make the associations
\begin{equation}
\begin{aligned}\label{convention}
\xi^*/\xi&\qquad \text{creation/annihilation in S}\\
 \bar{\xi}^*/\bar{\xi}&\qquad \text{annihilation/creation in S}\;,
\end{aligned}\end{equation}
as suggested by Fig.~\ref{path_collapsed}.\\ 
The resulting diagrammatic notation is more compact and the topology of the diagrams is different with respect to the Keldysh formalism. For example, diagrams displaying crossings in the Keldysh contour, as e.g. in the seminal work by K\"{o}nig \emph{et al.}~\cite{Koenig1996}, can be crossing-free when the time branches are collapsed in a single one, as in the present treatment, see also e.g.~\cite{Lindner2019,Schiro2019}.\\
\indent Two examples of paths, comprising two tunneling transitions each, are detailed in Fig.~\ref{path_detailed} where the Grassmann variables associated to the transition times $t_k$ and $t_l$ and contained in the functions ${\rm F}_{kl}$ are highlighted.
Both processes in Fig.~\ref{path_detailed} fall in the class succinctly represented by the only diagram generated by the first order term in the expansion of the influence function, namely
\begin{equation}
F_{21}
\rightarrow \begin{gathered}
\resizebox{1.5cm}{!}{
\begin{tikzpicture}[] 
\draw[thick] (-0.5,0) -- (1.5,0); 
\draw[black,thick] (0,0) arc (180:0:0.5cm  and 0.75cm);
 \end{tikzpicture}
}
\end{gathered}\;.
\end{equation}
As a further example, the $2$nd order term in the expansion of the influence functional gives
\begin{equation}
\begin{aligned}\label{exampleIF}
\sum_{\mathcal{P}_2}&\prod_{p=1}^2{\rm F}_{k_p\,l_p}= {\rm F}_{21}{\rm F}_{43}+{\rm F}_{32}{\rm F}_{41}+{\rm F}_{31}{\rm F}_{42}\\
\rightarrow & \begin{gathered}
\resizebox{2.5cm}{!}{
\begin{tikzpicture}[] 
\draw[thick] (-0.5,0) -- (0,0); 
\draw[thick] (0,0) -- (1,0); 
\draw[thick] (1,0) -- (2,0);
\draw[thick] (2,0) -- (3,0); 
\draw[thick] (3,0) -- (3.5,0);
\draw[black,thick] (0,0) arc (180:0:0.5cm  and 0.75cm);
\draw[black,thick] (2,0) arc (180:0:0.5cm  and 0.75cm) ;
 \end{tikzpicture}
 }
\end{gathered}
+
\begin{gathered}
\resizebox{2.2cm}{!}{
\begin{tikzpicture}[] 
\draw[thick] (-0.5,0) -- (0,0); 
\draw[thick] (0,0) -- (1,0); 
\draw[thick] (1,0) -- (2,0);
\draw[thick] (2,0) -- (3,0); 
\draw[thick] (3,0) -- (3.5,0);
\draw[black,thick] (0,0) arc (180:0:1.5cm  and 1cm);
\draw[black,thick] (1,0) arc (180:0:0.5cm  and 0.5cm) ;
 \end{tikzpicture}
 }
\end{gathered}
+
\begin{gathered}
\resizebox{2.2cm}{!}{
\begin{tikzpicture}[] 
\draw[thick] (-0.5,0) -- (0,0); 
\draw[thick] (0,0) -- (1,0); 
\draw[thick] (1,0) -- (2,0);
\draw[thick] (2,0) -- (3,0); 
\draw[thick] (3,0) -- (3.5,0);
\draw[black,thick] (0,0) arc (180:0:1cm  and 1cm);
\draw[black,thick] (1,0) arc (180:0:1cm  and 1cm) ;
 \end{tikzpicture}
 }
\end{gathered}\;,\nonumber
\end{aligned}
\end{equation}
where we provided a diagrammatic picture of the three terms resulting from the sum over the permutations.
Note that each function ${\rm F}_{kl}$ contains a pair of Grassmann-valued (vector) variables and therefore the functions ${\rm F}$ commute with each other.  

\section{State-conserving tunneling and diagrammatic rules}
\label{state_conserving_and_DR}

\subsection{Diagonal hybridization matrix}\label{diagonal_Gamma}

The expansion in terms of the influence functions $F_{kl}$ given in Eq.~\eqref{J_n} is very general and only relies on the properties of Grassmann numbers. In the remaining of this work we focus for simplicity on the case in which the correlation matrices are diagonal in the basis $\{|i\rangle\}$ of single-electron states of the central system S, namely
\begin{equation}
\label{corr_matrices_diagonal}
 [\mathbf{g}_{\alpha,\pm}(t)]_{i j}=\frac{1}{\hbar^2}\sum_{k \sigma} |{\rm t}_{i \alpha k \sigma}|^2f_{\pm}^\alpha(\epsilon_{k})e^{-\frac{\rm i}{\hbar}\epsilon_{\alpha k}t}\delta_{ij}\;,
\end{equation}
with $\mathbf{g}_{\pm}(t)=\sum_\alpha\mathbf{g}_{\alpha,\pm}(t)$. This implies that the paths of the different electron states in S are correlated only via the interaction term at the level of the system Hamiltonian and are otherwise independent. 
As a result, the energy-dependent hybridization matrix of lead $\alpha$, Eq.~\eqref{Gamma_matrix}, specializes to  
\begin{equation}
\label{Gamma_matrix_diagonal}
[\boldsymbol{\Gamma}_\alpha(\epsilon)]_{ij}=2\pi 
\sum_\sigma\varrho_{\alpha \sigma}(\epsilon) |{\rm t}_{i \alpha \sigma}(\epsilon)|^2\delta_{ij}\;, 
\end{equation}
which is still state-dependent, i.e. not proportional to the identity. Note that the hybridization matrices of the different leads are simultaneously diagonal in the occupation basis $\{|n_1,\dots,n_N\rangle\}$. Archetype examples of systems to which Eq.~\eqref{corr_matrices_diagonal} applies are the resonant level model (RLM) and the single-impurity Anderson model (SIAM) discussed in Secs.~\ref{RLM} and~\ref{SIAM}, respectively.\\
\indent Due to the diagonal correlation matrices, Eq.~\eqref{corr_matrices_diagonal}, the influence functional is factorized  and the fermion lines only connect transitions which change the occupation of individual states. As a result, if no coherences are present at the initial time $t_0$, none will be produced at later times. This is not true for non-diagonal correlation matrices, where coherences can develop and couple to the populations. This aspect is crucial for example in the so-called spin-valve setup~\cite{Koenig2003, Braun2006, Hornberger2008, Rezaei2020, Rohrmeier2021} and for interacting nanojunctions displaying interference effects~\cite{Pedersen2005, Darau2009, Donarini2019}.\\
\indent With the correlation matrices given by Eq~\eqref{corr_matrices_diagonal}, the Grassmann-valued functions in Eq.~\eqref{Fkl_general} specialize to ${\rm F}_{kl}=\sum_{i=1}^N {\rm F}_{kl}^i$, where
\begin{equation}
\begin{aligned}\label{FklSIAM}
{\rm F}_{kl}^i=&\sum_{x,y,z=\pm 1}-x\;[\xi^i_k]_{y}^{z} [ {\rm g}_{ii}(t_k-t_l)]_{xz}^{-z} [\xi^i_l]_{x}^{-z}
\;.
\end{aligned}\end{equation}
\indent The assumption of diagonal correlation matrices allows us to establish diagrammatic rules for the paths of individual fermionic single-particle states $i$ and to express the contribution of a composite diagram, involving different electron states, in terms of the individual diagrammatic contributions and of a \emph{common} phase factor accounting for the interactions.\\

\subsection{Parametrization for a single degree of freedom (resonant level model)}
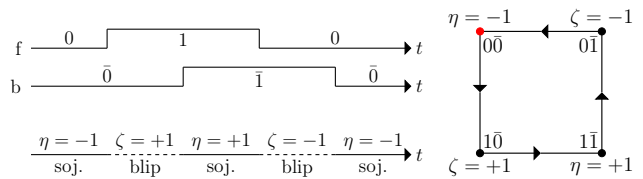
\begin{figure}[ht]
\resizebox{5.7cm}{!}{
\begin{tikzpicture}[]
\draw[thick,-] (0,1) node[left] {\Large{f\;}} -- node[above] {\Large{$0$}} (2,1)  ; 
\draw[thick] (2,1) -- (2,1.5); 
\draw[thick] (6,1) -- (6,1.5);
\draw[thick] (2,1.5) -- node[below] {\Large{$1$}} (6,1.5);
\draw[thick,-triangle 90,fill=black] (6,1) -- node[above] {\Large{$0$}} (10,1) node[right] {\Large{$t$}};
\draw[thick,-] (0,0) node[left] {\Large{b\;}} -- node[above] {\Large{$\bar{0}$}} (4,0);
\draw[thick] (4,0) -- (4,0.5);
\draw[thick] (4,0.5) -- node[below] {\Large{$\bar{1}$}} (8,0.5);
\draw[thick,-triangle 90,fill=black] (8,0) -- node[above] {\Large{$\bar{0}$}} (10,0) node[right] {\Large{$t$}}; 
\draw[thick] (8,0) -- (8,0.5);
\draw[thick] (0,-1.8)  -- node[above] {\Large{$\eta=-1$}} node[below] {\Large{soj.}}  (2,-1.8);
\draw[dashed,thick] (2,-1.8) -- node[above] {\Large{$\zeta=+1$}}  node[below] {\Large{blip}} (4,-1.8);
\draw[thick] (4,-1.8) -- node[above] {\Large{$\eta=+1$}} node[below] {\Large{soj.}} (6,-1.8);
\draw[dashed,thick] (6,-1.8) -- node[above] {\Large{$\zeta=-1$}}  node[below] {\Large{blip}} (8,-1.8);
\draw[thick,-triangle 90,fill=black] (8,-1.8) -- node[above] {\Large{$\eta=-1$}} node[below] {\Large{soj.}} (10,-1.8) node[right] {\Large{$t$}} ;
\end{tikzpicture}
}
\resizebox{2.7cm}{!}{
\begin{tikzpicture}[]
\draw[thick,-triangle 90,fill=black] (1.5,1.5) -- (0,1.5);
\draw[thick] (0,1.5)--(-1.5,1.5);
\draw[thick,-triangle 90,fill=black] (-1.5,1.5) -- (-1.5,0);
\draw[thick](-1.5,0) -- (-1.5,-1.5);
\draw[thick,-triangle 90,fill=black] (-1.5,-1.5) -- (0,-1.5);
\draw[thick] (0,-1.5) -- (1.5,-1.5);
\draw[thick,-triangle 90,fill=black] (1.5,-1.5) -- (1.5,0);
\draw[thick] (1.5,0) -- (1.5,1.5);
\filldraw[red]
(-1.5,1.5) circle (2.5pt) node[align=left,   below,black] {\Large{$\quad\; 0\bar{0}$}} node[above,black] {\Large{$\eta=-1$}};
\filldraw
(1.5,1.5) circle (2.5pt) node[align=center, below] {\Large{$0\bar{1}\quad\;$}} node[above] {\Large{$\zeta=-1$}}
(1.5,-1.5) circle (2.5pt) node[align=center, above] {\Large{$1\bar{1}\quad\;$}} node[below] {\Large{$\eta=+1$}}
(-1.5,-1.5) circle (2.5pt) node[align=center, above] {\Large{$\quad\; 1\bar{0}$}} node[below] {\Large{$\zeta=+1$}};
\end{tikzpicture}
}
\caption{\small{Blip/sojourn parametrization. Left -- Forward and backward paths associated to the occupation of an individual electron state with the corresponding collapsed, single-branch path parametrized in terms of blip and sojourns (below). Right -- The time evolution now occurs along the four sides of the square, whose corners define the four elements of the density matrix of a two-state system. The red full dot on the top-left corner denotes the starting (and final) state of the path.}}
\label{parametrization_paths}
\end{figure}

\begin{figure}[ht]
\begin{center}
\resizebox{7cm}{!}{
\begin{tikzpicture}[]
\draw (8.8,1.5) node[] {$\eta_0$};
\draw[thick] (8,1) -- (9.5,1); 
\draw (10.3,1.5) node[] {$\zeta_1$};
\draw[dashed,thick] (9.5,1) -- (11,1);
\draw (11.8,1.5) node[] {$\eta_1$}; 
\draw[thick] (11,1) -- (12.5,1);
\draw (13.3,1.5) node[] {$\zeta_2$};
\draw[dashed,thick] (12.5,1) -- (14,1);
\draw (14.8,1.5) node[] {$\eta_2$}; 
\draw[thick] (14,1) -- (15.5,1);
\draw[thick,->] (14,1) -- (15.5,1);
\draw[thick] (9.5,0.9) -- node[align=left,   below] {$\xi_{-\eta_0\zeta_1}^{-\zeta_1}$} node[align=left, above] {$t_1$}  (9.5,1.1);
\draw[thick] (11,0.9) --
 node[align=left,   below] {$\xi_{-\eta_1\zeta_1}^{\zeta_1}$} node[align=left, above] {$t_2$}
(11,1.1);
\draw[thick] (12.5,0.9) --
 node[align=left,   below] {$\xi_{-\eta_1\zeta_2}^{-\zeta_2}$} node[align=left, above] {$t_3$}
(12.5,1.1);
\draw[thick] (14,0.9) -- node[align=left,   below] {$\xi_{-\eta_2\zeta_2}^{\zeta_2}$} node[align=left, above] {$t_4$}
(14,1.1);
\end{tikzpicture}
}
\caption{\small{Sequence of Grassmann variables associated to the transition times of an individual degree of freedom of the central system in the blip/sojourn parametrization, see Eqs.~\eqref{symbols} and~\eqref{convention}.}}
\label{parametrization}
\end{center}
\end{figure}
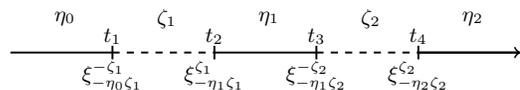

\indent Before considering the general case, we focus our attention on  an individual degree of freedom of the central system S or, equivalently, on the simplest case of a \emph{spinless level} coupled to electronic reservoirs, the so-called resonant level model, see Fig.~\ref{scheme_spinless} below.\\
\indent To this extent we notice that a single fermionic degree of freedom is characterized, in the occupation number representation, by the two values $0,1$, for the state being empty or occupied, respectively. Thus, one can consider the degree of freedom as a two-state system; the corresponding time evolution of generic forward and backward paths for such pseudo-spin, or qubit, is shown in Fig.~\ref{parametrization_paths}. Borrowing ideas from the path integral formulation of the famous spin-boson problem~\cite{Weiss2012}, we conveniently collapse the two-state paths on the forward and backward branches into a single four-state path. As shown in Fig.~\ref{parametrization_paths}, in analogy to the spin-boson nomenclature, we call sojourns the states $(0,\bar{0})$ and $(1,\bar{1})$, and blips the combinations $(0,\bar{1})$ and $(1,\bar{0})$.
This is done under the assumption of instantaneous tunneling events that change the occupation number of the electron states in S. A sojourn state corresponds to having the same occupation of the electron state both in the forward and in the backward path, meaning that the state is either empty ($\eta=-1$) or occupied ($\eta=+1$) in both branches. On the contrary, a blip state refers to different occupation of the two branches, namely occupied in the forward and empty in the backward ($\zeta=+1$) or \emph{vice-versa} ($\zeta=-1$), see the right panel of Fig.~\ref{parametrization_paths}. In the present case, since $N=1$, we disregard the state index $i$, and  Eq.~\eqref{FklSIAM} reduces to ${\rm F}_{kl}^i\rightarrow {\rm F}_{kl}$.\\
\indent To proceed, we perform the integration of the Grassmann variables associated to the time instants \emph{between} transitions, as done explicitly in Appendix.~\ref{integration_Grassmann_var}. This results in the phase factors related to the central system S which are discussed below. The residual Grassmann variables are the ones associated to the transition times, as shown in Fig.~\ref{parametrization} for a collapsed path  which comprises four tunneling transitions.\\ 

\indent A path of the electron state with $2m$ transitions starting and ending in a sojourn, has $m+1$ sojourn intervals ($\eta_0,\dots,\eta_m$) and $m$ blip intervals ($\zeta_1,\dots,\zeta_m$). 
The path is thus uniquely identified by the corresponding sequence 
\[
\eta_0,\zeta_1,\eta_1,\dots,\zeta_k,\eta_k,\dots,\zeta_m,\eta_m.
\]
With the associations made in Eqs.~\eqref{symbols} and~\eqref{convention}, to this sequence of blip/sojourn indexes there correspond the following sequence of Grassmann variables
\[
 \xi_{-\eta_0\zeta_1}^{-\zeta_1},\xi_{-\eta_1\zeta_1}^{\zeta_1},\dots, \xi_{-\eta_{k-1}\zeta_k}^{-\zeta_k},\xi_{-\eta_k\zeta_k}^{\zeta_k},\dots ,\xi_{-\eta_m\zeta_m}^{\zeta_m}\;,
\]
 as shown  in Fig \ref{parametrization}. To each transition to a blip state (odd 
transition times $t_{2k-1}$) is associated the Grassmann variable $ [\xi_{2k-1}]_{-\eta_{k-1}\zeta_k}^{-\zeta_k}\equiv
 \xi_{-\eta_{k-1}\zeta_k}^{-\zeta_k} $ and to a transition to a sojourn (even transition times $t_{2k}$) is associated $[\xi_{2k}]_{-\eta_k\zeta_k}^{\zeta_k} \equiv \xi_{-\eta_k\zeta_k}^{\zeta_k}$. Grassmann variables at different times are independent, otherwise a path with two Grassmann variables of the same type (e.g. creation in the forward path) would yield a vanishing contribution due to the property $(\xi^*)^2=\xi^2=0$.
Note that to a given path there  correspond different arrangements of the functions ${\rm F}_{kl}$ (fermion lines) attached to couples of transitions.\\
\indent From Eq.~(\ref{FklSIAM}), and using $(\zeta_k)^2=1$, according to the type of transitions being involved at times $t_k$ and $t_l$ (blip-sojourn, sojourn-blip, blip-blip, or sojourn-sojourn), the function ${\rm F}_{kl}$ acquires one of the forms 
\begin{equation}\begin{aligned}\label{Fkl}
&({\rm b} \rightarrow {\rm s})  \qquad {\rm F}_{2k\, 2l-1} =\xi_{-\eta_k\zeta_k}^{\zeta_k} {\rm f}_{2k\, 2l-1} \;\xi_{-\eta_{l-1}\zeta_l}^{-\zeta_l}
\\
&({\rm s} \rightarrow {\rm b})\qquad{\rm F}_{2k-1\,2l} =
\xi_{-\eta_{k-1}\zeta_k}^{-\zeta_k}  {\rm f}_{2k-1\,2l}\; \xi_{-\eta_l\zeta_l}^{\zeta_l}
\\
&({\rm b} \rightarrow {\rm b})\quad{\rm F}_{2k-1\,2l-1} =\xi_{-\eta_{k-1}\zeta_k}^{-\zeta_k}{\rm f}_{2k-1\,2l-1}\;\xi_{-\eta_{l-1}\zeta_l}^{-\zeta_l}
\\
&({\rm s} \rightarrow {\rm s})\qquad\quad\;{\rm F}_{2k\,2l}=\xi_{-\eta_k\zeta_k}^{\zeta_k} {\rm f}_{2k\,2l}\; \xi_{-\eta_l\zeta_l}^{\zeta_l}
\;,
\end{aligned}\end{equation}
where
\begin{equation}\begin{aligned}\label{fkl}
{\rm f}_{2k\, 2l-1} =&\;\eta_{l-1}\zeta_l\;{\rm g}^{-\zeta_l}_{-\eta_{l-1}}(t_{2k}-t_{2l-1}) \delta_{\zeta_k,\zeta_l} \\
{\rm f}_{2k-1\,2l} =&\;\eta_l\zeta_l\;{\rm g}^{\zeta_l}_{\eta_l}(t_{2k-1}-t_{2l}) \delta_{\zeta_k,\zeta_l}\\
{\rm f}_{2k-1\,2l-1} =&\;\eta_{l-1}\zeta_l\;{\rm g}^{-\zeta_l}_{-\eta_{l-1}}(t_{2k-1}-t_{2l-1}) \delta_{\zeta_k,-\zeta_l}\\
{\rm f}_{2k\,2l}=&\;\eta_l\zeta_l\;{\rm g}^{\zeta_l}_{\eta_l}(t_{2k}-t_{2l}) \delta_{\zeta_k,-\zeta_l}\;.
\end{aligned}\end{equation}

 Equation~\eqref{Fkl} is the combined result of the parametrization of the paths shown in Figs.~\ref{parametrization_paths} and~\ref{parametrization} and the form of the influence functions in Eq.~\eqref{FklSIAM}. This result is essential to establish the diagrammatic rules discussed in the next section once the residual Grassmann variables are integrated out. The following scheme clarifies the associations in Eq.~\eqref{Fkl} in the case $m=2$
\begin{equation}\label{b-s_scheme}
\begin{gathered}
\resizebox{2.6cm}{!}{
\begin{tikzpicture}[] 
\draw[thick] (-0.5,0) -- (0,0); 
\draw[dashed,thick] (0,0) -- (1,0); 
\draw[thick] (1,0) -- (2,0);
\draw[dashed,thick] (2,0) -- (3,0); 
\draw[thick] (3,0) -- (3.5,0);
\draw[](0.5,1) node[] {\large{b $\rightarrow$ s}};
\draw[black,thick] (0,0)  node[below] {\large{$t_1$}} arc (180:0:0.5cm  and 0.75cm) node[below] {\large{$t_2$}};
\draw[](2.5,1) node[] {\large{b $\rightarrow$ s}};
\draw[black,thick] (2,0)  node[below] {\large{$t_3$}} arc (180:0:0.5cm  and 0.75cm)  node[below] {\large{$t_4$}} ;
\end{tikzpicture}
}
\end{gathered}\\
\begin{gathered}
\resizebox{2.5cm}{!}{
\begin{tikzpicture}[] 
\draw[thick] (-0.5,0) -- (0,0); 
\draw[dashed,thick] (0,0) -- (1,0); 
\draw[thick] (1,0) -- (2,0);
\draw[dashed,thick] (2,0) -- (3,0); 
\draw[thick] (3,0) -- (3.5,0);
\draw[](1.5,1.3) node[] {\large{b $\rightarrow$ s}};
\draw[black,thick] node[below] {\large{$t_1$}} (0,0) arc (180:0:1.5cm  and 1cm)  node[below] {\large{$t_4$}};
\draw[](1.5,0.7) node[] {\large{s $\rightarrow$ b}};
\draw[black,thick]  (1,0)  node[below] {\large{$t_2$}} arc (180:0:0.5cm  and 0.5cm)  node[below] {\large{$t_3$}};
\end{tikzpicture}
}
\end{gathered}\\
\begin{gathered}
\resizebox{2.5cm}{!}{
\begin{tikzpicture}[] 
\draw[thick] (-0.5,0) -- (0,0); 
\draw[dashed,thick] (0,0) -- (1,0); 
\draw[thick] (1,0) -- (2,0);
\draw[dashed,thick] (2,0) -- (3,0); 
\draw[thick] (3,0) -- (3.5,0);
\draw[](0.8,1.3) node[] {\large{b $\rightarrow$ b}};
\draw[black,thick] (0,0)  node[below] {\large{$t_1$}} arc (180:0:1cm  and 1cm)  node[below] {\large{$t_3$}};
\draw[](2.2,1.3) node[] {\large{s $\rightarrow$ s}};
\draw[black,thick] (1,0)  node[below] {\large{$t_2$}} arc (180:0:1cm  and 1cm)  node[below] {\large{$t_4$}};
\end{tikzpicture}
}
\end{gathered}\;.
\end{equation}
Besides depending on the influence functions ${\rm F}^i_{kl}$, the propagator also depends on the action of the central system $S_{\rm S}(\boldsymbol{\xi}^{*},\boldsymbol{\xi})$, see Eq.~\eqref{propagator_rho}. Upon integrating out the Grassmann variables between transitions, this action produces the phase factors ${\rm b}_{kl}$ associated to the influence functions in Eq.~\eqref{Fkl}. 
For a central system with a single spinless level of energy $\epsilon$ these phase factors are schematized by  
\begin{equation}\begin{aligned}\label{Bm_spinless}
&({\rm b} \rightarrow {\rm s})  \quad e^{-\frac{\rm i}{\hbar}\zeta_l \epsilon \; (t_{2k}-t_{2l-1})}\\
&({\rm s} \rightarrow {\rm b})\quad e^{+\frac{\rm i}{\hbar}\zeta_l \epsilon \; (t_{2k-1}-t_{2l})}\\
&({\rm b} \rightarrow {\rm b})\quad e^{-\frac{\rm i}{\hbar}\zeta_l \epsilon \; (t_{2k-1}-t_{2l-1})}\\
&({\rm s} \rightarrow {\rm s})\quad e^{+\frac{\rm i}{\hbar}\zeta_l \epsilon \; (t_{2k}-t_{2l})}\;.
\end{aligned}\end{equation}
Note that  the sign of the exponent is determined by the state (blip/sojourn) from which the fermion line departs, see the scheme~\eqref{b-s_scheme}. 
Applying Eq.~\eqref{Propagator_general} to the  propagator order by order, Eq.~\eqref{J_n}, we are left with the following expansion of the propagator for the populations in the resonant level model:  $J_{{\eta}'{\eta}}(t;t_{0})=\sum_{m=0}^\infty J_{{\eta}'{\eta}}^{(m)}(t;t_{0})$, where
\begin{equation}\label{Jm_RLM}
J^{(m)}_{{\eta}'{\eta}}(t;t_0)=\sum_{{\rm paths}_{m}} \int \mathcal{D}\{t\}_{m} \sum_{\mathcal{P}}
 \mathcal{B}_m(\mathcal{P}) {\Phi}_m (\mathcal{P})\;.
\end{equation}
Here, the sum over the permutations $\mathcal{P}$ accounts for the different ways in which the fermion lines can connect $m$ pairs of tunneling transitions within the path joining the two sojourns $\eta$ and $\eta'$, see Eq.~\eqref{b-s_scheme}.
The central system and influence functional part are given by
\begin{equation}\begin{aligned}\label{Phim_RLM}
 \mathcal{B}_m(\mathcal{P})=&\prod_{p=1}^{m} {\rm b}_{k_p\,l_p} \;,\\
{\Phi}_m (\mathcal{P})=& \int \mathcal{D}\{\xi\}_{m}\prod_{p=1}^{m} {\rm F}_{k_p\,l_p}\;,
\end{aligned}\end{equation}
respectively.\\
\indent The sum over paths in Eq.~\eqref{Jm_RLM} amounts to summing over the possible values of the intermediate blip and sojourn state with $2m$ tunneling transitions. For example, in the case $m=2$
\begin{equation}
\begin{aligned}
\sum_{{\rm paths}_{2}}=\sum_{\zeta_1,\eta_1,\zeta_2}
\begin{gathered}
\resizebox{6cm}{!}{
\begin{tikzpicture}[]
\draw[thick] (0,0.5) -- node[above] {\Large{$\eta$}} (2,0.5)  ; 
\draw[dashed,thick] (2,0.5) -- node[above] {\Large{$\zeta_1$}} (4,0.5); 
\draw[thick] (4,0.5) -- node[above] {\Large{$\eta_1$}} (6,0.5);
\draw[dashed,thick] (6,0.5) -- node[above] {\Large{$\zeta_2$}} (8,0.5); 
\draw[thick] (8,0.5) -- node[above] {\Large{$\eta'$}}(10,0.5);
\end{tikzpicture}
}
\end{gathered}\;.
\end{aligned}
\end{equation}
\indent Finally, the symbol $\int\mathcal{D}\{\xi\}_{m}$ performs the integration over the residual Grassmann variables associated to the $2m$ transition times
\begin{equation}
\begin{aligned}\label{DxiRLM} 
\int \mathcal{D}\{\xi\}_{m}:=\int \prod_{k=1}^{m} (-\zeta_k\eta_k) d\xi_{-\eta_{k-1}\zeta_k}^{-\zeta_k}d\xi_{-\eta_k\zeta_k}^{\zeta_k}\;.
\end{aligned}
\end{equation}
The factors $-\eta_k\zeta_k$ in the above integration measure reflect the non-commuting nature of the symbols $d\xi$ and are introduced to keep track of the order in which the Grassmann-valued coordinates appear originally in the integration measure, i.e. $\prod_k d^2\xi(t_k) d^2\bar{\xi}(t_k)=\prod_k d\xi^*(t_k)  d\xi(t_k) d\bar{\xi}^*(t_k) d\bar{\xi}(t_k)$,  with the $*$-numbers to the left within the two classes of forward and backward variables, and with the backward variables to the right of the forward. This is exemplified in Appendix~\ref{integration_measure_Dxi}.\\

\subsection{Parametrization for $N$ degrees of freedom}

\begin{figure}[ht]
\begin{center}
\resizebox{8cm}{!}{
\begin{tikzpicture}[]
\draw[blue,thick] (0,2.5) node[blue,left] {\Large{$\uparrow$\;}} -- node[above] {\Large{$\eta^\uparrow_0$}} (1,2.5)  ; 
\draw[blue,thick] (1,2.5) arc (180:0:2.5cm  and 1.5cm);
\draw[blue,dashed,thick] (1,2.5) -- node[above] {\Large{$\zeta^\uparrow_1$}} (6,2.5); 
\draw[blue,thick] (6,2.5) -- node[above] {\Large{$\eta^\uparrow_1$}}(10,2.5);
\draw (4,2) node[left] {\large{$\epsilon_\uparrow+\frac{U}{2}(1+\eta_0^\downarrow)$}};
\draw (5.7,2) node[left] {\large{$\epsilon_\uparrow+\frac{U}{2}$}};
\draw[red,thick] (0,0) node[left] {\Large{$\downarrow$\;}} -- node[above] {\Large{$\eta^\downarrow_0\qquad$}} (4,0)  ; 
\draw[red,thick] (4,0) arc (180:0:2.5cm  and 1.5cm);
\draw[red,dashed,thick] (4,0) -- node[above] {\Large{$\qquad\zeta^\downarrow_1\qquad$}} (9,0); 
\draw[red,thick] (9,0) -- node[above] {\Large{$\eta^\downarrow_1$}}(10,0);
\draw[gray,dotted,thick] (1,3) -- (1,-1);
\draw[gray,dotted,thick] (4,3) -- (4,-1);
\draw[gray,dotted,thick] (6,3) -- (6,-1);
\draw[gray,dotted,thick] (9,3) -- (9,-1);
\draw (5.7,-.5) node[left] {\large{$\epsilon_\downarrow+\frac{U}{2}$}};
\draw (9,-.5) node[left] {\large{$\epsilon_\downarrow+\frac{U}{2}(1+\eta_1^\uparrow)$}};
\end{tikzpicture}
}
\caption{\small{Path-dependent energies in the SIAM. The energies $E_\sigma$ in the phase factors, Eq.~\eqref{Bm}, associated to the overlap of blip and sojourn of the two electron states $\sigma=\;\uparrow,\downarrow$ are $E_\sigma(\eta )=\epsilon_\sigma+(1+\eta) U/2$ for blip-sojourn overlap and $E_\sigma=\epsilon_\sigma+U/2$ for blip-blip overlap.
The time axis is divided in $5$ intervals $\tau_k$ separated by $4$ tunneling transitions ($m=2$) distributed between the two paths. The phase factors descend from the action of the central system, see Eqs.~\eqref{propagator_rho} and~\eqref{action_S}, 
after integrating-out the Grassmann variables between the tunneling transitions.}}
\label{scheme_interaction}
\end{center}
\end{figure}
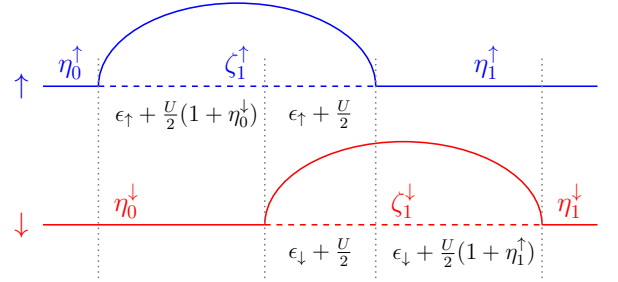

Due to the diagonal hybridization matrices introduced in Sec.~\ref{diagonal_Gamma}, the influence functional factorizes in the product of functionals for the individual  
electron states $i$, or, equivalently, the phase of the influence functional, Eq.~\eqref{phase_IF}, is the sum over the electron states $i$. As a result, the above description of the resonant level model generalizes in a straightforward fashion to $N$ electronic states.
In this case, the populations are identified with the string of sojourns associated to the different electron states via the vector index $\boldsymbol{\eta}=\{\eta^i\}$, with the correspondences $n_i=0\leftrightarrow\eta^i=-1$ and  $n_i=1\leftrightarrow\eta^i=+1$.
 The propagator for the populations now reads   $J_{\boldsymbol{\eta}'\boldsymbol{\eta}}(t;t_{0})=\sum_{m=0}^\infty J_{\boldsymbol{\eta}'\boldsymbol{\eta}}^{(m)}(t;t_{0})$, where
\begin{equation}\begin{aligned}\label{Jm}
J^{(m)}_{\boldsymbol{\eta}'\boldsymbol{\eta}}(t;t_0)=
\sum_{{\rm paths}_{m}} \int \mathcal{D}\{t\}_{m} \prod
_i\sum_{\mathcal{P}_i}
 \mathcal{B}_{m_i}^i(\mathcal{P}_i) {\Phi}_{m_i}^i (\mathcal{P}_i)\;,
\end{aligned}\end{equation}
with $\sum_i m_i = m$, and where
\begin{equation}\begin{aligned}\label{phim}
 \mathcal{B}_{m_i}^i(\mathcal{P})=&\prod_{p=1}^{m_i} {\rm b}^i_{k_p\,l_p} \;,\\
{\Phi}_{m_i}^i (\mathcal{P}_i)=& \int \mathcal{D}\{\xi\}_{m_i}\prod_{p=1}^{m_i} {\rm F}^i_{k_p\,l_p}\;.
\end{aligned}\end{equation}
The sum over the permutations $\mathcal{P}_i$ accounts now for the different ways in which the fermion lines can connect $m_i$ pairs of tunneling transitions within the same path, that of the electron state $i$ joining the two sojourns $\eta^i$ and ${\eta^i}'$. \\
\indent While the influence functions ${\Phi}_{m_i}^i$ depend exclusively on the path of the individual state $i$, the phase factors in $\mathcal{B}_{m_i}^i$ couple the paths of the different states via the interaction. Specifically, the constant single-particle energies $\epsilon_i$ turn into the path-dependent energies $E_i$; they depend on the instantaneous states of all degrees of freedom $\{ \eta^j\}_n$ during the time interval $\tau_n$ between consecutive transitions, not necessarily of the same electron state. For example, assume that a fermion line associated to the state $i$ departs form a blip state at time $t_l$ and encompasses $W$ intervals with $\sum_n^W\tau_n=t_k-t_l$. Then the corresponding phase factor reads  
\begin{equation}\label{Bm}
{\rm b}^i_{k l}:=\prod_{n=1}^W e^{-\frac{\rm i}{\hbar}\zeta_l E_i(\{\eta\}_n)\tau_n}\;,
\end{equation}
which reduces for a noninteracting system to
\begin{equation}
{\rm b}^i_{k l}=\prod_{n=1}^W e^{-\frac{\rm i}{\hbar}\zeta_l \epsilon_i\tau_n}=e^{-\frac{\rm i}{\hbar}\zeta_l \epsilon_i(t_k-t_l)} \;.
\end{equation}
Thus, in the noninteracting case, the integrand in Eq.~\eqref{Jm} is actually factorized in the system's degrees on freedom.\\
\indent Finally, the sum over paths in Eq.~\eqref{Jm} now takes into account the different possibilities to distribute $2m$ transitions among the paths of the $N$  individual states $i$ connecting the initial and final sojourn states $\boldsymbol{\eta}$ and $\boldsymbol{\eta}'$ with $\sum_i m_i = m$.\\
\indent To exemplify how these phase factors and the sum-over-paths work in the case of a multi-state system ($N>1$), consider the case of the SIAM. As it describes an interacting central system which is a single, spinful level, we have  $N=2$ and $i\equiv \sigma=\;\uparrow,\downarrow$, cf. Eq.~\eqref{H_SIAM} below.
In this specific case, denoting with $\bar\sigma$ the opposite spin state with respect to $\sigma$, 
the energies associated to the spin $\sigma$ in the phase factors read
\begin{equation}\label{E_sigma}
\begin{aligned}
E_\sigma(\eta )=&\epsilon_\sigma+(1+\eta) U/2\qquad({\rm blip-sojourn})\\
E_\sigma=&\epsilon_\sigma+U/2\qquad\qquad\qquad({\rm blip-blip})\;,
\end{aligned}
\end{equation}
Thus, for example, if the path of $\bar\sigma$ is in a sojourn state with $\eta =+1$, then $E_\sigma=\epsilon_\sigma+U$: This is the addition energy to be payed for adding a further electron to the dot. The presence of the term $U/2$ in the second line of Eq.~\eqref{E_sigma} implies that overlap of different fermion lines can produce the energy $U$ 
according to the relative sign of the index $\zeta$,  
 see Appendix~\ref{integration_Grassmann_var} for details. In Fig.~\ref{scheme_interaction} an example which shows the energies $E_\sigma$ is provided for a path with two transition for each spin path.\\
 The sum over paths with $4$ transitions connecting the populations $\boldsymbol{\eta}=(\eta^\uparrow,\eta^\downarrow)$ and $\boldsymbol{\eta}'=({\eta'}^\uparrow,{\eta'}^\downarrow)$ is given by
\begin{equation}
\begin{aligned}\label{sum-over-paths}
\sum_{{\rm paths}_{2}}=&\sum_{\zeta^\uparrow_1,\eta^\uparrow_1,\zeta^\uparrow_2}\begin{gathered}
\resizebox{6cm}{!}{
\begin{tikzpicture}[]
\draw[blue,thick] (0,0.5) -- node[blue,above] {\Large{$\eta^\uparrow$}} (2,0.5)  ; 
\draw[blue,dashed,thick] (2,0.5) -- node[blue,above] {\Large{$\zeta^\uparrow_1$}} (4,0.5); 
\draw[blue,thick] (4,0.5) -- node[blue,above] {\Large{$\eta^\uparrow_1$}} (6,0.5);
\draw[blue,dashed,thick] (6,0.5) -- node[blue,above] {\Large{$\zeta^\uparrow_2$}} (8,0.5); 
\draw[blue,thick] (8,0.5) -- node[blue,above] {\Large{${\eta'}^\uparrow$}}(10,0.5);
\draw[red,thick] (0,0) -- node[below] {\Large{$\eta^\downarrow={\eta'}^\downarrow$}} (10,0);
\end{tikzpicture}
}
\end{gathered}\\
&+\sum_{\zeta^\uparrow_1,\zeta^\downarrow_1}
\begin{gathered}
\resizebox{6cm}{!}{
\begin{tikzpicture}[]
\draw[blue,thick] (0,0.5)  -- node[blue,above] {\Large{$\eta^\uparrow$}} (2,0.5)  ; 
\draw[blue,dashed,thick] (2,0.5) -- node[blue,above] {\Large{$\zeta^\uparrow_1$}} (6,0.5); 
\draw[blue,thick] (6,0.5) -- node[blue,above] {\Large{${\eta'}^\uparrow$}}(10,0.5);
\draw[red,thick] (0,0) -- node[below] {\Large{$\eta^\downarrow$}} (4,0)  ; 
\draw[red,dashed,thick] (4,0) -- node[below] {\Large{$\zeta^\downarrow_1\qquad$}} (8,0); 
\draw[red,thick] (8,0) -- node[below] {\Large{${\eta'}^\downarrow$}}(10,0);
\end{tikzpicture}
}
\end{gathered}\\
+&\sum_{\zeta^\downarrow_1,\eta^\downarrow_1,\zeta^\downarrow_2}
\begin{gathered}
\resizebox{6cm}{!}{
\begin{tikzpicture}[]
\draw[red,thick] (0,0) -- node[below] {\Large{$\eta^\downarrow$}} (2,0)  ; 
\draw[red,dashed,thick] (2,0) -- node[below] {\Large{$\zeta^\downarrow_1$}} (4,0); 
\draw[red,thick] (4,0) -- node[below] {\Large{$\eta^\downarrow_1$}} (6,0);
\draw[red,dashed,thick] (6,0) -- node[below] {\Large{$\zeta^\downarrow_2$}} (8,0); 
\draw[red,thick] (8,0) -- node[below] {\Large{${\eta'}^\downarrow$}}(10,0);
\draw[blue,thick] (0,0.5) -- node[blue,above] {\Large{$\eta^\uparrow={\eta'}^\uparrow$}} (10,0.5);
\end{tikzpicture}
}
\end{gathered}\;.
\end{aligned}
\end{equation}
Note that if ${\eta'}^\downarrow\neq\eta^\downarrow$, then the uppermost line  of Eq.~\eqref{sum-over-paths} does not contribute to the sum -over-paths. The same holds for $\sigma=\,\uparrow$ and the bottom line. If initial and final sojourns, $\boldsymbol{\eta}$ and $\boldsymbol{\eta}'$, differ for both spin states, then only the central line contributes to the sum.\\
\indent Finally, the Grassmann variables and blip/sojourn indexes in the symbol $\int\mathcal{D}\{\xi\}_{m_i}$ acquire the state index $i$
\begin{equation}
\begin{aligned}\label{Dxi} 
\int \mathcal{D}\{\xi\}_{m_i}:=\int \prod_{k=1}^{m_i} (-\zeta_k^i\eta_k^i) d\xi_{-\eta_{k-1}^i\zeta_k^i}^{-\zeta_k^i}d\xi_{-\eta_k^i\zeta_k^i}^{\zeta_k^i}\;,
\end{aligned}
\end{equation}
cf. Eq.~\eqref{DxiRLM}.\\
\indent Using the parametrization of the paths in Fig.~\ref{parametrization} for the individual degrees of freedom $i$ and the integration measure in Eq.~\eqref{Dxi} for the residual Grassmann variables associated to the transition times, we are able to automatically carry out the integrations over these variables. The result is simply an overall sign given by the anticommutation property of the $\xi$'s, as detailed in Appendix~\ref{integration_Grassmann_var}. In other words, we find from Eq.~\eqref{Fkl} the simple form
\begin{equation}\begin{aligned}\label{phim3}
{\Phi}_{m_i}^i (\mathcal{P}_i)=& {\rm sgn}_{\mathcal{P}_i}\prod_{k=1}^{m_i} \zeta_k^i\eta_k^i\prod_{p=1}^{m_i} {\rm f}^i_{k_p\,l_p}\;,
\end{aligned}\end{equation}
where ${\rm sgn}_{\mathcal{P}_i}$ is an overall sign given by the integration over the Grassmann 
variables associated  to the transitions. This sign depends on the order of the transitions, and thus on the permutation $\mathcal{P}$, due to the non-commuting character of the Grassmann variables. Importantly, Eq.~\eqref{phim3} allows us to establish diagrammatic rules, whereby the explicit form of the functions ${\rm f}_{kl}$ is obtained by just looking at the arrangement of the fermion lines in the associated diagram, cf. Eq.~\eqref{fkl}. The diagrammatic rules are summarized below.

\subsection{Diagrammatic rules in the time domain for the individual electron states}
\label{diagrammatic_rules}

\indent Once specified a path of the full system with $2m$ transitions, the individual influence functions ${\Phi}_{m_i}^i$, Eq.~\eqref{phim}, are the sums over the different arrangements of fermion lines ${\rm f}_{kl}^i$ connecting $2 m_i$ transitions, where $\sum_i m_i =m$. Each of these arrangements of fermion lines constitutes a diagram relative to the state $i$.  
In this section we establish diagrammatic rules individually for each state. This is convenient because, since the Pauli exclusion principle applies separately to the different  states in the central system, the overlap of fermion lines yields different diagrammatic contributions according to whether the lines involve the same or different electron states.  Diagrams relative to different states are then coupled by the phase factors in $\mathcal{B}_{m_i}^i$, see Eqs.~\eqref{phim} and~\eqref{Bm}. Each diagrammatic contribution $\mathcal{B}_{m_i}^i(\mathcal{P}_i) {\Phi}_{m_i}^i (\mathcal{P}_i)$ to Eq.~\eqref{Jm} consists of 
(from here on the state index $i$ is understood)
\begin{itemize}
\item The overall sign $(-1)^{\rm n.\; crossings}$ due to the integration of the Grassmann variables in $\Phi_m$, see Eqs.~\eqref{Fkl},~\eqref{phim}, and~Eq.~\eqref{phim3}. 

\item The product $\prod_{k=1}^m (-\zeta_k\eta_k) $,  from the normal ordering of the Grassmann integration measure, see Eqs.~\eqref{DxiRLM} and~\eqref{Dxi}.
\item The product of the functions ${\rm f}_{kl}$, Eq. (\ref{fkl}), representing the fermion lines which connect two tunneling transitions, times the corresponding phase factor ${\rm b}_{kl}$ of the central system, Eqs.~(\ref{Bm_spinless}) and~\eqref{Bm}. To each fermion line is associated the constraint on the $\zeta$'s connected by the line, according to the scheme in Eq. (\ref{fkl}).
\end{itemize}

Below we show examples with $m$ from $0$ to $2$. More examples, with higher order diagrams, are shown in Appendix~\ref{examples_diagrams}. For $m=0$ there are no tunneling transitions. Hence
\begin{equation}\begin{aligned}\label{0th}
(0)
\begin{gathered}
\resizebox{1.6cm}{!}{
\begin{tikzpicture}[] 
\draw[thick] (-0.5,0) -- (1.5,0);
 \end{tikzpicture}
 }
\end{gathered}
\quad\delta_{\eta',\eta}\\
\end{aligned}\end{equation}
For $m=1$ there is only one fermion line connecting two tunneling times
\begin{equation}\begin{aligned}\label{1st}
(1)
\begin{gathered}
\resizebox{1.6cm}{!}{
\begin{tikzpicture}[] 
\draw[thick] (-0.5,0) -- (0,0); 
\draw[dashed,thick] (0,0) -- (1,0); 
\draw[thick] (1,0) -- (1.5,0);
\draw[black,thick] (0,0) arc (180:0:0.5cm  and 0.75cm);
\filldraw 
(0,0) circle (2pt) node[align=left,   below] {$\eta_0\zeta_1$} ;
\end{tikzpicture}
 }
\end{gathered}
\quad &
(+1)(-\zeta_1\eta_1)\;{\rm f}_{21}{\rm b}_{21}\\
=&(-\zeta_1\eta_1)\;\eta_0\zeta_1 {\rm g}_{-\eta_0}^{-\zeta_1}(t_2-t_1){\rm b}_{21}\\
=&\;\eta'\eta\;[-{\rm g}_{-\eta}^{-\zeta_1}(t_2-t_1)]{\rm b}_{21}\;.
\end{aligned}\end{equation}
Here we used $(\zeta_1)^2=1$ and identified $\eta_0=\eta$ and $\eta_1=\eta'$. The full dot in the above diagram indicates the vertex, here associated to the transition from which the fermion line departs. Analogously, with $m=2$
\begin{equation}\begin{aligned}\label{2nda}
&(2a)
\begin{gathered}
\resizebox{3cm}{!}{
\begin{tikzpicture}[] 
\draw[thick] (-0.5,0) -- (0,0); 
\draw[dashed,thick] (0,0) -- (1,0); 
\draw[thick] (1,0) -- (2,0);
\draw[dashed,thick] (2,0) -- (3,0); 
\draw[thick] (3,0) -- (3.5,0);
\draw[black,thick] (0,0) arc (180:0:0.5cm  and 0.75cm);
\draw[black,thick] (2,0) arc (180:0:0.5cm  and 0.75cm) ;
\filldraw 
(0,0) circle (2pt) node[align=left,   below] {$\eta_0\zeta_1$} 
(2,0) circle (2pt) node[align=center, below] {$\eta_1\zeta_2$} ;
\end{tikzpicture}
 }
\end{gathered}\\
&(+1)\zeta_1\eta_1\zeta_2\eta_2\;{\rm f}_{21}{\rm b}_{21}{\rm f}_{43}{\rm b}_{43}\\
&=\zeta_1\eta_1\zeta_2\eta_2\;\eta_0\zeta_1{\rm g}_{-\eta_0}^{-\zeta_1}(t_2-t_1){\rm b}_{21}\\
&\quad\times\eta_1\zeta_2\; {\rm g}_{-\eta_1}^{-\zeta_2}(t_4-t_3){\rm b}_{43}\\
&=\eta'\eta\;[-{\rm g}_{-\eta}^{-\zeta_1}(t_2-t_1)]{\rm b}_{21}[-{\rm g}_{-\eta_1}^{-\zeta_2}(t_4-t_3)]{\rm b}_{43}\;,
\end{aligned}\end{equation}
\begin{equation}\begin{aligned}\label{2ndb}
&(2b)
\begin{gathered}
\resizebox{3cm}{!}{
\begin{tikzpicture}[] 
\draw[thick] (-0.5,0) -- (0,0); 
\draw[dashed,thick] (0,0) -- (1,0); 
\draw[thick] (1,0) -- (2,0);
\draw[dashed,thick] (2,0) -- (3,0); 
\draw[thick] (3,0) -- (3.5,0);
\draw[black,thick] (0,0) arc (180:0:1.5cm  and 1cm);
\draw[black,thick] (1,0) arc (180:0:0.5cm  and 0.5cm) ;
\filldraw 
(0,0) circle (2pt) node[align=left,   below] {$\eta_0\zeta_1$} 
(1,0) circle (2pt) node[align=center, below] {$\zeta_1\eta_1$} ;
 \end{tikzpicture}
 }
\end{gathered}\\
&(+1)\zeta_1\eta_1\zeta_2\eta_2\;{\rm f}_{41}{\rm b}_{41}{\rm f}_{32}{\rm b}_{32}\\
&=\zeta_1\eta_1\zeta_2\eta_2\;\eta_0\zeta_1\; {\rm g}_{-\eta_0}^{-\zeta_1}(t_4-t_1){\rm b}_{41}\\
&\quad\times\zeta_1\eta_1\;{\rm g}_{\eta_1}^{\zeta_1}(t_3-t_2){\rm b}_{32}\delta_{\zeta_2,\zeta_1}\\
&=\eta'\eta\delta_{\zeta_2,\zeta_1}\;[-{\rm g}_{-\eta}^{-\zeta_1}(t_4-t_1)]{\rm b}_{41}\;[-{\rm g}_{\eta_1}^{\zeta_1}(t_3-t_2)]{\rm b}_{32}\;,
\end{aligned}\end{equation}

\begin{equation}\begin{aligned}\label{2ndc}
&(2c)
\begin{gathered}
\resizebox{3cm}{!}{
\begin{tikzpicture}[] 
\draw[thick] (-0.5,0) -- (0,0); 
\draw[dashed,thick] (0,0) -- (1,0); 
\draw[thick] (1,0) -- (2,0);
\draw[dashed,thick] (2,0) -- (3,0); 
\draw[thick] (3,0) -- (3.5,0);
\draw[black,thick] (0,0) arc (180:0:1cm  and 1cm);
\draw[black,thick] (1,0) arc (180:0:1cm  and 1cm) ;
\filldraw 
(0,0) circle (2pt) node[align=left,   below] {$\eta_0\zeta_1$} 
(1,0) circle (2pt) node[align=center, below] {$\zeta_1\eta_1$} ;
 \end{tikzpicture}
 }
\end{gathered}\\
&(-1)\zeta_1\eta_1\zeta_2\eta_2\;{\rm f}_{31}{\rm b}_{31}{\rm f}_{42}{\rm b}_{42}\\
&=\zeta_1\eta_1\zeta_2\eta_2\;\eta_0\zeta_1 \; {\rm g}_{-\eta_0}^{-\zeta_1}(t_3-t_1){\rm b}_{31}\\
&\quad\times\zeta_1\eta_1\;{\rm g}_{\eta_1}^{\zeta_1}(t_4-t_2){\rm b}_{42}\delta_{\zeta_2,-\zeta_1}\\
&=\eta'\eta \;[-{\rm g}_{-\eta}^{-\zeta_1}(t_3-t_1)]{\rm b}_{31}\;[-{\rm g}_{\eta_1}^{\zeta_1}(t_4-t_2)]{\rm b}_{42}\delta_{\zeta_2,-\zeta_1}\;,
\end{aligned}\end{equation}
where we used $\zeta_1\zeta_2\delta_{\zeta_2,-\zeta_1}=-\delta_{\zeta_2,-\zeta_1}$.

Noticeably, for all second-order diagrams, the product of the inner sojourns and blips results in a factor $+1$. 
Multiplication by internal sojourn indexes emerges as we go to higher orders and overlap of more than two fermion lines. This is exemplified by the following diagram of order $m=3$ (see also the complete list in Appendix~\ref{examples_diagrams})
\begin{equation}\begin{aligned}\label{3rda}
&(3)
\begin{gathered}
\resizebox{4.5cm}{!}{
\begin{tikzpicture}[] 
\draw[thick] (-0.5,0) -- (0,0); 
\draw[dashed,thick] (0,0) -- (1,0); 
\draw[thick] (1,0) -- (2,0);
\draw[dashed,thick] (2,0) -- (3,0); 
\draw[thick] (3,0) -- (4,0);
\draw[dashed,thick] (4,0) -- (5,0); 
\draw[thick] (5,0) -- (5.5,0);
\draw[black,thick] (0,0) arc (180:0:2.5cm  and 1.5cm);
\draw[black,thick] (1,0) arc (180:0:1.5cm  and 1cm) ;
\draw[black,thick] (2,0) arc (180:0:0.5cm  and 0.5cm) ;
\filldraw 
(0,0) circle (2pt) node[align=left,   below] {$\eta_0\zeta_1$} 
(1,0) circle (2pt) node[align=center, below] {$\zeta_1\eta_1$} 
(2,0) circle (2pt) node[align=center, below] {$\eta_1\zeta_2$} ;
 \end{tikzpicture}
 }
\end{gathered}\\
&\prod_{k=1}^3(-\zeta_k\eta_k)\eta_0\zeta_1\;{\rm g}_{-\eta_0}^{-\zeta_1}(6,1){\rm b}_{61}\;\zeta_1\eta_1{\rm g}_{\eta_1}^{\zeta_1}(5,2){\rm b}_{52}\\
&\quad\times\;\eta_1\zeta_2{\rm g}_{-\eta_1}^{-\zeta_2}(4,3){\rm b}_{43}\delta_{\zeta_3,\zeta_1}\\
&=\eta'\eta \eta_1\eta_2 \;[-{\rm g}_{-\eta}^{-\zeta_1}(6,1)]{\rm b}_{61}\;[-{\rm g}_{\eta_1}^{\zeta_1}(5,2)]{\rm b}_{52}\\
&\quad\times[-{\rm g}_{-\eta_1}^{-\zeta_2}(4,3)]{\rm b}_{43}\delta_{\zeta_3,\zeta_1}\;,
\end{aligned}
\end{equation}
where, for the sake of compactness, we set ${\rm g}(t_k-t_l)\equiv {\rm g}(k,l)$.

By applying the above rules we notice that the multiplicative factors $\zeta$ are always compensated by the product $\prod_{k=1}^m (-\zeta_k\eta_k) $ and by the constraints $\delta_{\zeta_k,\pm\zeta_l}$ contained in the functions ${\rm f}_{kl}$, Eq.~\eqref{fkl}. The multiplicative sojourn indexes are instead compensated solely by the products $\zeta_{l}\eta_l$ or $\eta_{l-1}\zeta_l$ associated to each departing line (i.e. to the vertexes). As a result, each diagram presents, as a multiplicative factor, the product $\eta'\eta$ of the last and first sojourn indexes times the product of the $\eta_k$'s of the internal sojourns which are not compensated, i.e. the ones with zero or two departing lines. 

We are now in the position to set the diagrammatic rules for the individual states (to a full diagram will correspond the product of the diagrammatic contributions from each state, see Eq.~\eqref{Jm} and the examples in Sec.~\ref{diagrams_N_states}). To each diagram we associate 
\begin{enumerate}
\item An overall sign $(-1)^{\rm n.\; crossings}$. 

\item A sign given by the products of the non-compensated $\zeta$ indexes (namely the ones of the blip states with zero or two vertexes) times the corresponding constraints. For example $\zeta_k\zeta_l\delta_{\zeta_k,-\zeta_l}=-\delta_{\zeta_k,-\zeta_l}$ and $\zeta_k\zeta_l\delta_{\zeta_k,\zeta_l}=\delta_{\zeta_k,\zeta_l}$, (because $\zeta=\pm 1$). These constraints make the corresponding sums in the sum-over-paths collapse.

\item The product $\eta'\eta$ times the product of the non-compensated $\eta$ indexes, namely the ones of the sojourn states with zero or two vertexes.

\item The products of the correlators $-{\rm g}_x^y(t_k-t_l)$ and the associated phase factors ${\rm b}_{kl}$ of the central system for each fermion line.

\end{enumerate}

To exemplify this, we consider the following $3$rd-order diagram
\begin{equation}\begin{aligned}\label{3rdb}
&
\begin{gathered}
\resizebox{5.5cm}{!}{
\begin{tikzpicture}[] 
\draw[thick] (-0.5,0) -- (0,0); 
\draw[dashed,thick] (0,0) --  node[above]{$\zeta_1$} (1,0); 
\draw[thick] (1,0) -- node[above]{$\eta_1$} (2,0);
\draw[dashed,thick] (2,0) --node[above]{$\zeta_2$}  (3,0); 
\draw[thick] (3,0) -- node[above]{$\eta_2$} (4,0);
\draw[dashed,thick] (4,0) --node[above]{$\zeta_3$}  (5,0); 
\draw[thick] (5,0) -- node[above]{$\eta'$} (5.5,0);
\draw[black,thick] (1,0) arc (180:0:2cm  and 1.5cm);
\draw[] (1.3,1.5) node[above]{$\delta_{\zeta_2\zeta_1}$};
\draw[black,thick] (2,0) arc (180:0:1cm  and 1cm) ;
\draw[] (3.5,1.5) node[above]{$\delta_{\zeta_3,-\zeta_1}$};
\draw[black,thick] (0,0) arc (180:0:1.5cm  and 1.5cm) ;
\filldraw 
(0,0) circle (2pt) node[align=left,   below] {$\eta\zeta_1$} 
(1,0) circle (2pt) node[align=center, below] {$\zeta_1\eta_1$} 
(2,0) circle (2pt) node[align=center, below] {$\eta_1\zeta_2$} ;
\end{tikzpicture}
 }
\end{gathered}\\
&-\eta'\eta\eta_1\eta_2 [-{\rm g}_{-\eta}^{-\zeta_1}(4,1)]{\rm b}_{41}\;[-{\rm g}_{\eta_1}^{\zeta_1}(6,2)]{\rm b}_{62}\\
&\quad\times[-{\rm g}_{-\eta_1}^{-\zeta_1}(5,3)]{\rm b}_{53}\delta_{\zeta_3,-\zeta_1}\delta_{\zeta_2,\zeta_1}\;,
\end{aligned}\end{equation}
where, as in Eq.~\eqref{2ndc}, we used $\zeta_1\zeta_3\delta_{\zeta_3,-\zeta_1}=-\delta_{\zeta_3,-\zeta_1}$.\\
Another example is given by
\begin{equation}\begin{aligned}\label{3rdc}
&
   \begin{gathered}
\resizebox{5.5cm}{!}{
\begin{tikzpicture}[] 
\draw[thick] (-0.5,0) -- (0,0); 
\draw[dashed,thick] (0,0) -- node[above]{$\zeta_1$} (1,0); 
\draw[thick] (1,0) -- node[above]{$\eta_1$} (2,0);
\draw[dashed,thick] (2,0) -- node[above]{$\zeta_2$}(3,0); 
\draw[thick] (3,0) --node[above]{$\eta_2$} (4,0);
\draw[dashed,thick] (4,0) -- node[above]{$\zeta_3$}(5,0); 
\draw[thick] (5,0) -- node[above]{$\eta'$}(5.5,0);
\draw[black,thick] (1,0) arc (180:0:1.5cm  and 1.5cm);
\draw[] (1.3,1.5) node[above]{$\delta_{\zeta_2\zeta_1}$};
\draw[black,thick] (2,0) arc (180:0:1.5cm  and 1.5cm) ;
\draw[] (2.5,1.5) node[above]{$\delta_{\zeta_3\zeta_1}$};
\draw[black,thick] (0,0) arc (180:0:1.5cm  and 1.5cm) ;
\draw[] (3.7,1.5) node[above]{$\delta_{\zeta_3\zeta_2}$};
\filldraw 
(0,0) circle (2pt) node[align=left,   below] {$\eta\zeta_1$} 
(1,0) circle (2pt) node[align=center, below] {$\zeta_1\eta_1$} 
(2,0) circle (2pt) node[align=center, below] {$\eta_1\zeta_2$} ;
\end{tikzpicture}
 }
\end{gathered} \\
&-\eta'\eta\eta_1\eta_2 [-{\rm g}_{-\eta}^{-\zeta_1}(4,1)]{\rm b}_{41}\;[-{\rm g}_{\eta_1}^{\zeta_1}(5,2)]{\rm b}_{52}\\
&\quad\times[-{\rm g}_{-\eta_1}^{-\zeta_1}(6,3)]{\rm b}_{63}\delta_{\zeta_3,\zeta_1}\delta_{\zeta_2,\zeta_1}\;,
\end{aligned}\end{equation}
where we used $\zeta_1\zeta_3\delta_{\zeta_3,\zeta_1}=\delta_{\zeta_3,\zeta_1}$.\\
\indent By inspection of the diagrams in Eqs.~\eqref{3rda}-\eqref{3rdc}, one sees that none of the correlators ${\rm g}$ bears the index $\eta_2$. If also the phase factors ${\rm b}_{kl}$  do not depend on $\eta_2$, as e.g. in the case of the resonant level model, or in the non-interacting case, see Eq.~\eqref{Bm}, then by performing the sum over paths the diagrams with this topology \emph{vanish collectively} due to the sum over $\eta_2$. A similar argument holds for all the diagrams with more than two overlapping fermion lines of the same state $i$ because overlap of more than two fermion lines entails the presence of sojourns with no vertexes. This means, in particular, that:
\begin{itemize}
\item The exact propagator for the resonant level model is reproduced by the diagrams with at most two overlapping fermion lines.
\item For the  non-interacting spinful level the exact propagator is reproduced by the diagrams with at most four overlapping fermion lines, of which no more that two can belong to the same spin.
\end{itemize}
These results are in agreement with what has been proved using a Liouville space approach in~\cite{Karlstrom2013}.\\
\subsection{Diagrams for $N$-state systems}
\label{diagrams_N_states}
A full diagram for $N>1$ and with diagonal hybridization matrices is given by the arrangements of fermion lines, with some fixed topology, connecting the transitions within the paths of the individual system states $i$.
As an example, consider again the SIAM. A so-called crossing diagram  ($m=2$, with crossing fermion lines) 
\begin{equation}
\begin{aligned}
&\begin{gathered}
\resizebox{6cm}{!}{
\begin{tikzpicture}[] 
\draw[thick] (0.5,0) -- (9.5,0); 
\draw[black,thick] (2,0) arc (180:0:2.cm  and 1.2cm) ;
\draw[thick] (4,0) arc (180:0:2.cm  and 1.2cm);
\filldraw 
(2,0) circle (2pt)
(4,0) circle (2pt);
 \end{tikzpicture}
}
\end{gathered}
\end{aligned}
\end{equation}
can be obtained in the following ways
\begin{equation}
\begin{aligned}
&\begin{gathered}
\resizebox{6cm}{!}{
\begin{tikzpicture}[]
\draw[blue,thick] (0.5,0.5) node[left]{\Large{$\uparrow$}} -- (2,0.5)  ; 
\draw[blue,dashed,thick] (2,0.5) -- (4,0.5); 
\draw[blue,thick] (4,0.5) -- (6,0.5);
\draw[blue,dashed,thick] (6,0.5) -- (8,0.5); 
\draw[blue,thick] (8,0.5) -- (9.5,0.5);
\draw[blue,thick] (2,0.5) arc (180:0:2.cm  and 1.2cm) ;
\draw[blue,thick] (4,0.5) arc (180:0:2.cm  and 1.2cm);
\filldraw [blue]
(2,0.5) circle (2pt)
(4,0.5) circle (2pt);
\draw[red,thick] (0.5,0)node[left]{\Large{$\downarrow$}} --  (9.5,0);
\end{tikzpicture}
}
\end{gathered}\\
&\begin{gathered}
\resizebox{6cm}{!}{
\begin{tikzpicture}[]
\draw[blue,thick] (0.5,0.5) node[left]{\Large{$\uparrow$}}  -- (2,0.5)  ; 
\draw[blue,dashed,thick] (2,0.5) --  (6,0.5); 
\draw[blue,thick] (6,0.5) -- (9.5,0.5);
\draw[blue,thick] (2,0.5) arc (180:0:2.cm  and 1.2cm) ;
\filldraw[blue] 
(2,0.5) circle (2pt);
\draw[red,thick] (0.5,0)node[left]{\Large{$\downarrow$}}  -- (4,0)  ; 
\draw[red,dashed,thick] (4,0) --  (8,0); 
\draw[red,thick] (8,0) -- (9.5,0);
\draw[red,thick] (4,0.5) arc (180:0:2.cm  and 1.2cm);
\draw[red,thick] (4,0) --  (4,0.5); 
\draw[red,thick] (8,0) --  (8,0.5); 
\filldraw[red] 
(4,0) circle (2pt);
\end{tikzpicture}
}
\end{gathered}\\
&\begin{gathered}
\resizebox{6cm}{!}{
\begin{tikzpicture}[]
\draw[red,thick] (0.5,0)node[left]{\Large{$\downarrow$}}   -- (2,0)  ; 
\draw[red,dashed,thick] (2,0) --  (6,0); 
\draw[red,thick] (6,0) -- (9.5,0);
\draw[red,thick] (2,0.5) arc (180:0:2.cm  and 1.2cm) ;
\draw[blue,thick] (0.5,0.5)node[left]{\Large{$\uparrow$}}  -- (4,0.5)  ; 
\draw[blue,dashed,thick] (4,0.5) --  (8,0.5); 
\draw[blue,thick] (8,0.5) -- (9.5,0.5);
\draw[blue,thick] (4,0.5) arc (180:0:2.cm  and 1.2cm);
\filldraw[blue] 
(4,0.5) circle (2pt);
\draw[red,thick] (2,0) --  (2,0.5); 
\draw[red,thick] (6,0) --  (6,0.5);
\filldraw[red] 
(2,0) circle (2pt);
; 
\end{tikzpicture}
}
\end{gathered}\\
&\begin{gathered}
\resizebox{6cm}{!}{
\begin{tikzpicture}[]
\draw[red,thick] (0.5,0)node[left]{\Large{$\downarrow$}}  --  (2,0)  ; 
\draw[red,dashed,thick] (2,0) --  (4,0); 
\draw[red,thick] (4,0) -- (6,0);
\draw[red,dashed,thick] (6,0) -- (8,0); 
\draw[red,thick] (8,0) -- (9.5,0);
\draw[red,thick] (2,0.5) arc (180:0:2.cm  and 1.2cm) ;
\draw[red,thick] (4,0.5) arc (180:0:2.cm  and 1.2cm);
\draw[red,thick] (2,0) --  (2,0.5); 
\draw[red,thick] (6,0) --  (6,0.5); 
\draw[red,thick] (4,0) --  (4,0.5); 
\draw[red,thick] (8,0) --  (8,0.5); 
\filldraw[red] 
(2,0) circle (2pt)
(4,0) circle (2pt);
\draw[blue,thick] (0.5,0.5)node[left]{\Large{$\uparrow$}}  --  (9.5,0.5);
\end{tikzpicture}
}
\end{gathered}\;.
\end{aligned}
\end{equation}
The diagrammatic rules developed in the previous section are thus applied to the two states $\sigma=\;\{\uparrow,\downarrow\}$, according to how the fermion lines are distributed among these electron states. The resulting integrand in Eq.~\eqref{Jm} is given by the product of the contributions from each state.  Also, in the interacting case, the phase factors ${\rm b}^\sigma_{kl}$, see Eqs.~\eqref{Bm} and~\eqref{E_sigma} and Fig.~\ref{scheme_interaction}, depend on the details of the paths of both states simultaneously. For example, the second diagrammatic contribution above is evaluated as
\begin{equation}
\begin{aligned}
&\begin{gathered}
\resizebox{6cm}{!}{
\begin{tikzpicture}[]
\draw[blue,thick] (0.5,0.5) node[left]{\Large{$\uparrow$}}  -- (2,0.5)  ; 
\draw[blue,dashed,thick] (2,0.5) --  (6,0.5); 
\draw[blue,thick] (6,0.5) -- (9.5,0.5);
\draw[blue,thick] (2,0.5) arc (180:0:2.cm  and 1.2cm) ;
\filldraw[blue] 
(2,0.5) circle (2pt);
\draw[red,thick] (0.5,0)node[left]{\Large{$\downarrow$}}  -- (4,0)  ; 
\draw[red,dashed,thick] (4,0) --  (8,0); 
\draw[red,thick] (8,0) -- (9.5,0);
\draw[red,thick] (4,0.5) arc (180:0:2.cm  and 1.2cm);
\draw[red,thick] (4,0) --  (4,0.5); 
\draw[red,thick] (8,0) --  (8,0.5); 
\filldraw[red] 
(4,0) circle (2pt);
\end{tikzpicture}
}
\end{gathered}\\
&\;{\eta^\uparrow}'\eta^\uparrow{\eta^\downarrow}'\eta^\downarrow\;[-{\rm g}_{-\eta^\uparrow}^{-\zeta^\uparrow_1}(t_3-t_1)]{\rm b}_{31}^\uparrow\;[-{\rm g}_{-\eta^\downarrow}^{-\zeta^\downarrow_1}(t_4-t_2)]{\rm b}_{42}^\downarrow\;,
\end{aligned}
\end{equation}
with
\begin{equation}
\begin{aligned}
{\rm b}_{31}^\uparrow=&e^{-\frac{\rm i}{\hbar}\zeta^\uparrow_1 E_\uparrow(\eta^\downarrow)(t_2-t_1)}e^{-\frac{\rm i}{\hbar}\zeta^\uparrow_1 E_\uparrow(t_3-t_2)}\;,\\
{\rm b}_{42}^\downarrow=&e^{-\frac{\rm i}{\hbar}\zeta^\downarrow_1 E_\downarrow(t_3-t_2)}e^{-\frac{\rm i}{\hbar}\zeta^\downarrow_1 E_\downarrow({\eta^\uparrow}')(t_4-t_3)}\;,
\end{aligned}
\end{equation}
with $E_\sigma=\epsilon_\sigma+U/2$ and $E_\sigma(\eta )=\epsilon_\sigma+U/2(1+\eta )$, see the scheme in Fig.~\ref{scheme_interaction}. In the absence of interactions $E_\sigma,E_\sigma(\eta )\rightarrow \epsilon_\sigma$ and therefore ${\rm b}_{31}^\uparrow\rightarrow \exp[-{\rm i}\zeta^\uparrow_1 \epsilon
_\uparrow(t_3-t_1)/{\hbar}]$ and ${\rm b}_{42}^\downarrow\rightarrow\exp[-{\rm i}\zeta^\downarrow_1 \epsilon
_\downarrow(t_4-t_2)/{\hbar}]$, as in Eq.~\eqref{Bm_spinless}. 
\indent As an application of the diagrammatic rules developed in Sec.~\ref{diagrammatic_rules} to the SIAM, let us consider the three diagrams whose fermion lines involve both spin states
\begin{equation}\begin{aligned}\label{examples8th}
(A)&
\begin{gathered}
\resizebox{5cm}{!}{
\begin{tikzpicture}[] 
\draw[blue,thick] (1,-1) arc (180:0:2.5cm  and 1.5cm);
\draw[red,thick] (2,-1) arc (180:0:1.5cm  and 1cm) ;
\draw[red,thick] (0,-1) arc (180:0:2cm  and 1.25cm) ;
\draw[blue,thick] (3,-1) arc (180:0:2cm  and 1.25cm) ;
\draw[blue,thick] (-0.5,-1)node[left]{\Large{$\uparrow$}}  -- (1,-1); 
\draw[blue,dashed,thick] (1,-1) -- (3,-1); 
\draw[blue,thick] (3,-1) --  (6,-1); 
\draw[blue,dashed,thick] (6,-1) -- (7,-1)[]; 
\draw[blue,thick] (7,-1) -- (7.5,-1); 
\draw[red,thick] (-0.5,-2)node[left]{\Large{$\downarrow$}}  -- (0,-2); 
\draw[red,dashed,thick] (0,-2) -- (2,-2); 
\draw[red,thick] (2,-2) -- (4,-2); 
\draw[red,dashed,thick] (4,-2) -- (5,-2); 
\draw[red,thick] (5,-2) -- (7.5,-2); 
\draw[red] (0,-1) -- (0,-2); 
\draw[red] (2,-1) -- (2,-2); 
\draw[red] (4,-1) -- (4,-2); 
\draw[red] (5,-1) -- (5,-2); 
\filldraw[blue] 
(3,-1) circle (2pt)
(1,-1) circle (2pt); 
\filldraw[red] 
(0,-2) circle (2pt)
(2,-2) circle (2pt);
\end{tikzpicture}
 }
\end{gathered}\\
(B)&
\begin{gathered}
\resizebox{5cm}{!}{
\begin{tikzpicture}[] 
\draw[blue,thick] (1,-1) arc (180:0:2.5cm  and 1.5cm);
\draw[blue,thick] (2,-1) arc (180:0:1.5cm  and 1cm) ;
\draw[red,thick] (0,-1) arc (180:0:2cm  and 1.25cm) ;
\draw[blue,thick] (3,-1) arc (180:0:2cm  and 1.25cm) ;
\draw[blue,thick] (-0.5,-1)node[left]{\Large{$\uparrow$}}  -- (1,-1); 
\draw[blue,dashed,thick] (1,-1) -- (2,-1); 
\draw[blue,thick] (2,-1) -- (3,-1); 
\draw[blue,dashed,thick] (3,-1) -- (5,-1); 
\draw[blue,thick] (5,-1) --node[above] {$\eta^\uparrow_2$} (6,-1) ; 
\draw[blue,dashed,thick] (6,-1) -- (7,-1); 
\draw[blue,thick] (7,-1) -- (7.5,-1); 
\draw[red,thick] (-0.5,-2) node[left]{\Large{$\downarrow$}} -- (0,-2); 
\draw[red,dashed,thick] (0,-2) -- (4,-2); 
\draw[red,thick] (4,-2) -- (7.5,-2); 
\draw[red] (0,-1) -- (0,-2); 
\draw[red] (4,-1) -- (4,-2); 
\filldraw[blue] 
(3,-1) circle (2pt)
(2,-1) circle (2pt)
(1,-1) circle (2pt); 
\filldraw[red] 
(0,-2) circle (2pt); 
 \end{tikzpicture}
 }
\end{gathered} = 0\\
(C)&
\begin{gathered}
\resizebox{5cm}{!}{
\begin{tikzpicture}[] 
\draw[blue,thick] (1,-1) arc (180:0:2.5cm  and 1.5cm);
\draw[blue,thick] (2,-1) arc (180:0:1.5cm  and 1cm) ;
\draw[blue,thick] (0,-1) arc (180:0:2cm  and 1.25cm) ;
\draw[red,thick] (3,-1) arc (180:0:2cm  and 1.25cm) ;
\draw[blue,thick] (-0.5,-1) node[left]{\Large{$\uparrow$}}  -- (0,-1); 
\draw[blue,dashed,thick] (0,-1) -- (1,-1); 
\draw[blue,thick] (1,-1) -- (2,-1); 
\draw[blue,dashed,thick] (2,-1) -- (4,-1); 
\draw[blue,thick] (4,-1) -- node[above] {$\eta^\uparrow_2$} (5,-1); 
\draw[blue,dashed,thick] (5,-1) -- (6,-1); 
\draw[blue,thick] (6,-1) -- (7.5,-1); 
\draw[red,thick] (-0.5,-2) node[left]{\Large{$\downarrow$}}  -- (3,-2); 
\draw[red,dashed,thick] (3,-2) -- (7,-2); 
\draw[red,thick] (7,-2) -- (7.5,-2); 
\draw[red] (3,-1) -- (3,-2); 
\draw[red] (7,-1) -- (7,-2); 
\filldraw[blue] 
(0,-1) circle (2pt)
(2,-1) circle (2pt)
(1,-1) circle (2pt); 
\filldraw[red] 
(3,-2) circle (2pt); \end{tikzpicture}\;.
 }
\end{gathered}\neq 0
\end{aligned}\end{equation}
The multiplicative factor $\eta^\uparrow_2$ is brought by the overlap of more than two fermion lines of the same spin state. Consider the noninteracting case. While $(A)$ contributes, the diagrams $(B)$ and $(C)$ vanish collectively once the sum over paths (specifically over $\eta^\uparrow_2$) is performed, as the phase factor $\mathcal{B}$ in the propagator is independent of the sojourn indexes, see Eq.~\eqref{Bm} and Fig.~\ref{scheme_interaction}. In the interacting case,  $(C)$ contributes because there is a phase factor associated to $\eta^\uparrow_2$ due to the sojourn-blip overlap, while $(B)$ is still vanishing, as the sojourn-sojourn overlap does not contribute to the phase factors, see also Ref.~\cite{Karlstrom2013}.\\

\section{Generalized master equation for the populations and the current}
\label{GME_pop_current}
In the absence of time-dependent driving,  the propagator has time-translation symmetry. It is therefore convenient to Laplace-transform the population propagator order by order and obtain a generalized master equation (GME) for the populations and an integral equation for the current. The kernels of these equations are related to each other and, in turn, connected to the Green's functions. This connection will be elaborated in Sec.~\ref{prop_coupl_and_GF}.
In the following, we indicate as $\hat{f}(\lambda)=\int_0^\infty dt \exp(-\lambda t)f(t)$ the Laplace transform of a function $f(t)$.

\subsection{GME for the populations}\label{GME_populations}

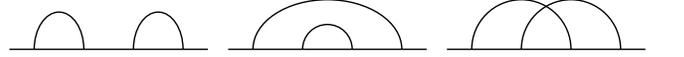
\begin{figure}[ht]
\resizebox{2.8cm}{!}{
\begin{tikzpicture}[] 
\draw[thick] (-0.5,0) -- (3.5,0); 
\draw[black,thick] (0,0) arc (180:0:0.5cm  and 0.75cm);
\draw[black,thick] (2,0) arc (180:0:0.5cm  and 0.75cm) ;
 \end{tikzpicture}
 }
\resizebox{2.8cm}{!}{
\begin{tikzpicture}[] 
\draw[thick] (-0.5,0) -- (3.5,0); 
\draw[black,thick] (0,0) arc (180:0:1.5cm  and 1cm);
\draw[black,thick] (1,0) arc (180:0:0.5cm  and 0.5cm) ;
 \end{tikzpicture}
 }
\resizebox{2.8cm}{!}{
\begin{tikzpicture}[] 
\draw[thick] (-0.5,0) -- (3.5,0); 
\draw[black,thick] (0,0) arc (180:0:1cm  and 1cm);
\draw[black,thick] (1,0) arc (180:0:1cm  and 1cm) ;
 \end{tikzpicture}
}
\caption{\small{The three topologies of $2$nd-order diagrams. Each fermion line can belong to each of the $N$ states of the central system S. The first diagram is reducible while the second and third are irreducible.}}
\label{diagrams_2}
\end{figure}

 Due to the nested time integrals in the definition of $J^{(m)}$, Eq.~\eqref{Jm}, the reducible contributions, i.e. the ones that can be cut by a vertical line not crossing any fermion line, factorize in Laplace space. For this reason, the populations and current kernels collect the so-called irreducible diagrammatic contributions, the ones that cannot be cut by a vertical line not crossing any fermion line.\\
\indent Using Eq.~\eqref{0th}, the zeroth-order contribution to the propagator is 
\begin{equation}\label{J0}
J^{(0)}_{\boldsymbol{\eta}'\boldsymbol{\eta}}(t;0)=\delta_{\boldsymbol{\eta}'\boldsymbol{\eta}}\;,
\end{equation}
and its Laplace transform reads 
\begin{equation}\label{J0lambda}
\hat{J}^{(0)}_{\boldsymbol{\eta}'\boldsymbol{\eta}}(\lambda)=\frac{1}{\lambda}\delta_{\boldsymbol{\eta}'\boldsymbol{\eta}}\;.
\end{equation}
\indent Let us denote with $\boldsymbol{\eta}^{(i)}$ the set of sojourn indexes associated to all states except $i$, namely
$$
\boldsymbol{\eta}^{(i)}=(\dots,\eta^{i-1},\eta^{i+1},\dots)
$$
and set the initial time $t_0=0$. 
The term $m=1$ contains two tunneling transitions. This implies the change in the occupation of one state at most. Thus, the first-order propagator has composite indexes $\boldsymbol{\eta}'$ and $\boldsymbol{\eta}$ that differ for at most one entry.  
The resulting first-order propagator  can be readily calculated according to the definition, Eq.~\eqref{Jm}, and the diagrammatic rules set up in Sec.~\ref{diagrammatic_rules}, yielding 
\begin{equation}\begin{aligned}\label{J1}
J^{(1)}_{\boldsymbol{\eta}'\boldsymbol{\eta}}(t;0)=&\sum_i\int_0^t dt_2\int_0^{t_2} dt_1
\sum_{\zeta^i}e^{-\frac{\rm i}{\hbar} \zeta^i E_i(\boldsymbol{\eta}^{(i)})(t_2-t_1)}\\
&\qquad\times {\eta'}^i\eta^i\;[-{\rm g}_{-\eta^i}^{-\zeta^i}(t_2-t_1)]\delta_{{{\boldsymbol{\eta}'}^{(i)}}\boldsymbol{\eta}^{(i)}}\;.
\end{aligned}\end{equation}
If, for some $i$,  ${\eta'}^i\neq\eta^i$, we have ${\eta'}^i\eta^i=-1$ because $\eta=\pm 1$. Moreover, the sum over $i$ collapses to a single term due to the constraint $\delta_{{\boldsymbol{\eta}^{(i)}}'\boldsymbol{\eta}^{(i)}}$.
Taking into account Eq.~\eqref{J0}, this implies that, up to first order, 
$$
J_{\boldsymbol{\eta}\boldsymbol{\eta}}(t;0)=1 - \sum_i J^{({\eta'}^i\neq\eta^i)}_{\boldsymbol{\eta}'\boldsymbol{\eta}}(t;0)\;,
$$ 
in agreement with the conservation of the total probability.\\
\indent Let us introduce the irreducible kernel of order $1$ using the explicit form for the correlator, Eq.~\eqref{corr_matrices_diagonal},
\begin{equation}\begin{aligned}\label{K1_t}
\mathcal{K}^{(1)}_{\boldsymbol{\eta}'\boldsymbol{\eta}}(\tau)
=&\prod_{j=1}^N {\eta'}^j\eta^j
\sum_{i \zeta^i\alpha k \sigma}e^{-\frac{\rm i}{\hbar} \zeta^i [ E_i(\boldsymbol{\eta}^{(i)})-\epsilon_k]\tau} \\
&\times\frac{-|{\rm t}_{i \alpha k \sigma}|^2}{\hbar^2} f^\alpha_{-\eta^i}(\epsilon_k)\delta_{{\boldsymbol{\eta}'}^{(i)}\boldsymbol{\eta}^{(i)}}\;,
\end{aligned}\end{equation}
where we singled out the prefactor $\prod_j {\eta'}^j  \eta^j=\pm 1$, common to all orders (see the diagrammatic rules), by exploiting the property that when two sojourn indexes are the same they contribute as $(\eta^i)^2=1$.
In Laplace space, the first-order propagator acquires then the form
\begin{equation}\begin{aligned}\label{J1Laplace}
\hat{J}^{(1)}_{\boldsymbol{\eta}'\boldsymbol{\eta}}(\lambda)
=&\frac{1}{\lambda}\hat{\mathcal{K}}^{(1)}_{\boldsymbol{\eta}'\boldsymbol{\eta}}(\lambda)\frac{1}{\lambda}\;,
\end{aligned}\end{equation}
with
\begin{equation}\begin{aligned}\label{K1_lambda}
\hat{\mathcal{K}}^{(1)}_{\boldsymbol{\eta}'\boldsymbol{\eta}}(\lambda)=&\prod_{j=1}^{N} {\eta'}^j\eta^j
\sum_{i \zeta^i\alpha k \sigma}\frac{-(|{\rm t}_{i \alpha k \sigma}|^2/\hbar^2)\;f^\alpha_{-\eta^i}(\epsilon_k)}{\lambda+{\rm i}\zeta^i[E_i(\boldsymbol{\eta}^{(i)})-\epsilon_k]/\hbar}\delta_{{\boldsymbol{\eta}'}^{(i)}\boldsymbol{\eta}^{(i)}}
\;.
\end{aligned}
\end{equation}
If, for some $i$, ${\eta'}^i\neq \eta^i$, then the prefactor is $-1$ and the sum over $i$ collapses to a single term, as for the first-order propagator. This entails that,
from Eqs.~\eqref{J1} and~\eqref{J1Laplace}, the diagonal elements of the irreducible kernel are related to the off-diagonal ones by 
\begin{equation}\label{K1_lambda_diagonal}
\hat{\mathcal{K}}^{(1)}_{\boldsymbol{\eta}\boldsymbol{\eta}}(\lambda)=-\sum_i \hat{\mathcal{K}}^{(1),{\eta'}^i\neq\eta^i}_{\boldsymbol{\eta}'\boldsymbol{\eta}}(\lambda)\;.
\end{equation}
Since, as we show below, the rates $\hat{\mathcal{K}}^{(1)}_{\boldsymbol{\eta}'\boldsymbol{\eta}}(0)$ are the steady-state rates of the master equation for the populations in the sequential tunneling approximation~\cite{Koller2010}, Eq.~\eqref{K1_lambda_diagonal} is consistent with the conservation of the total probability.\\
\indent Higher-order contributions can be calculated as well according to the diagrammatic rules given in Sec.~\ref{diagrammatic_rules}. The $2$nd-order contribution to the propagator  for the populations is the sum of the three classes of diagrams in Fig.~\ref{diagrams_2}.
The first is a reducible diagram and in Laplace space is the product of two  lower-order diagrams
\begin{equation}\begin{aligned}\label{}
\hat J_{\includegraphics[height=0.2cm,angle=0]{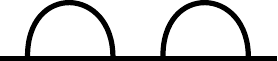},\boldsymbol{\eta}'\boldsymbol{\eta}}^{(2)}(\lambda)=\frac{1}{\lambda}\sum_{\boldsymbol{\eta}''}\hat{\mathcal{K}}^{(1)}_{\boldsymbol{\eta}'\boldsymbol{\eta}''}(\lambda)\frac{1}{\lambda}\hat{\mathcal{K}}^{(1)}_{\boldsymbol{\eta}''\boldsymbol{\eta}}(\lambda)\frac{1}{\lambda}\;.
\end{aligned}\end{equation}
Here, the internal sum over $\boldsymbol{\eta}$ has been singled-out, allowing for the use of the matrix notation
\begin{equation}\begin{aligned}\label{}
\hat{\boldsymbol{J}}_{\includegraphics[height=0.2cm,angle=0]{diagram_2nd-red.pdf}}^{(2)}(\lambda)=\frac{1}{\lambda}\frac{\hat{\boldsymbol{\mathcal{K}}}^{(1)}(\lambda)}{\lambda}\cdot\frac{\hat{\boldsymbol{\mathcal{K}}}^{(1)}(\lambda)}{\lambda}\;.
\end{aligned}\end{equation}
The full propagating function $\hat{\boldsymbol{J}}^{(2)}$ in Laplace space, expressed as the sum of the three diagrammatic contributions shown in Fig.~\ref{diagrams_2}, is given by
\begin{equation}\begin{aligned}\label{J2lambda}
\hat{\boldsymbol{J}}^{(2)}(\lambda)=\frac{1}{\lambda}\left[\frac{\hat{\boldsymbol{\mathcal{K}}}^{(1)}(\lambda)}{\lambda}\cdot\frac{\hat{\boldsymbol{\mathcal{K}}}^{(1)}(\lambda)}{\lambda}+\frac{\hat{\boldsymbol{\mathcal{K}}}^{(2)}(\lambda)}{\lambda}\right]\;,
\end{aligned}\end{equation}
where the irreducible kernel of $2$nd order $\hat{\boldsymbol{\mathcal{K}}}^{(2)}(\lambda)$ is the sum of the two $2$nd-order irreducible diagrams in Fig.~\ref{diagrams_2} (the second and the third) in Laplace space. As discussed in Sec.~\ref{exact_kernel} below, these contributions can be written as the contraction of a matrix block with a vertex, as in Eq.~\eqref{K1_lambda}, with the difference that the block has now internal processes. The same applies to higher-order irreducible kernels leading to the final Eq.~\eqref{Kexact} below.\\
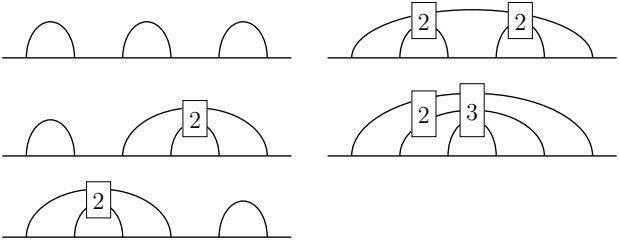
\begin{figure}[ht]
\begin{center}
\begin{equation}\begin{aligned}\nonumber
&\resizebox{4cm}{!}{
\begin{tikzpicture}[] 
\draw[thick] (-0.5,0) -- (5.5,0); 
\draw[black,thick] (0,0) arc (180:0:0.5cm  and 0.75cm);
\draw[black,thick] (2,0) arc (180:0:0.5cm  and 0.75cm) ;
\draw[black,thick] (4,0) arc (180:0:0.5cm  and 0.75cm) ;
 \end{tikzpicture}
 }\quad
 \resizebox{4cm}{!}{
\begin{tikzpicture}[] 
\draw[thick] (-0.5,0) -- (5.5,0); 
\draw[black,thick] (0,0) arc (180:0:2.5cm  and 1.cm);
\draw[black,thick] (1,0) arc (180:0:0.5cm  and 0.7cm) ;
\fill[white] (1.25,0.4) rectangle (1.75,1.15);
\draw [] (1.25,0.4) rectangle (1.75,1.15);
\draw[](1.5,0.75) node[] {\Large{$2$}};
\draw[black,thick] (3,0) arc (180:0:0.5cm  and 0.75cm) ;
\fill[white] (3.75,0.4) rectangle (3.25,1.15);
\draw [] (3.75,0.4) rectangle (3.25,1.15);
\draw[](3.5,0.75) node[] {\Large{$2$}};
 \end{tikzpicture}
 }\\
&\resizebox{4cm}{!}{
\begin{tikzpicture}[] 
\draw[thick] (-2.5,0) -- (3.5,0); 
\draw[black,thick] (0,0) arc (180:0:1.5cm  and 1cm);
\draw[black,thick] (1,0) arc (180:0:0.5cm  and 0.7cm) ;
\fill[white] (1.25,0.4) rectangle (1.75,1.15);
\draw [] (1.25,0.4) rectangle (1.75,1.15);
\draw[](1.5,0.75) node[] {\Large{$2$}};
\draw[black,thick] (-2,0) arc (180:0:0.5cm  and 0.75cm) ;
 \end{tikzpicture}
 }\quad
 \resizebox{4cm}{!}{
\begin{tikzpicture}[] 
\draw[thick] (-0.5,0) -- (5.5,0); 
\draw[black,thick] (0,0) arc (180:0:2.5cm  and 1.3cm);
\draw[black,thick] (1,0) arc (180:0:1.5cm  and .95cm) ;
\fill[white] (1.25,0.5) rectangle (1.75,1.35);
\draw [] (1.25,0.4) rectangle (1.75,1.35);
\draw[](1.5,0.85) node[] {\Large{$2$}};
\draw[black,thick] (2,0) arc (180:0:0.5cm  and 0.75cm) ;
\fill[white] (2.25,0.4) rectangle (2.75,1.5);
\draw [] (2.25,0.4) rectangle (2.75,1.5);
\draw[](2.5,0.9) node[] {\Large{$3$}};
\end{tikzpicture}
}\\
&\resizebox{4cm}{!}{
\begin{tikzpicture}[] 
\draw[thick] (-0.5,0) -- (5.5,0); 
\draw[black,thick] (0,0) arc (180:0:1.5cm  and 1cm);
\draw[black,thick] (1,0) arc (180:0:0.5cm  and 0.7cm) ;
\fill[white] (1.25,0.4) rectangle (1.75,1.15);
\draw [] (1.25,0.4) rectangle (1.75,1.15);
\draw[](1.5,0.75) node[] {\Large{$2$}};
\draw[black,thick] (4,0) arc (180:0:0.5cm  and 0.75cm) ;
 \end{tikzpicture}
 }
\end{aligned}\end{equation}
\caption{\small{Compact representation for the 15 third-order diagrams. The five diagrams in the left column are reducible and can be seen as combinations of the first- and second-order ones, see Fig.~\ref{diagrams_2}. The ten diagrams on the right are irreducible and their explicit forms are the same as the ones listed in Appendix~\ref{examples_diagrams} for an individual degree of freedom of S.
 The lower-right class of diagrams vanishes when each of the three fermion lines belongs to the same state.}}
\label{diagrams3compact}
\end{center}
\end{figure}
\indent Going to the $3$rd-order propagator, it collects the contributions from the 15 diagrams listed in Fig.~\ref{diagrams3compact} where, in order to give a compact visualization, we introduce the symbols 
\begin{equation}\begin{aligned}\label{box2}
\begin{gathered}
\resizebox{0.8cm}{!}{
\begin{tikzpicture}[] 
\draw[] (0.4,0.8) node[below] {\Large{$2$}};
\draw[thick] (0,0) -- (0.8,0) ;
\draw[thick] (0.8,0) -- (0.8,1) ; 
\draw[thick] (0.8,1) -- (0,1)  ;
\draw[thick] (0,1) -- (0,0) ;
 \end{tikzpicture}
}
\end{gathered}
=&
\begin{gathered}
\resizebox{0.8cm}{!}{
\begin{tikzpicture}[] 
\draw[thick] (0,0.2) -- (0.8,0.2) ;
\draw[thick] (0.8,0.8) -- (0,0.8)  ;
 \end{tikzpicture}
}
\end{gathered}
+
\begin{gathered}
\resizebox{0.8cm}{!}{
\begin{tikzpicture}[] 
\draw[thick] (0,0.2) -- (0.8,0.8) ;
\draw[thick] (0.8,0.2) -- (0,0.8)  ;
 \end{tikzpicture}
}
\end{gathered}
\end{aligned}\end{equation}
and
\begin{equation}\begin{aligned}\label{box3}
\begin{gathered}
\resizebox{0.8cm}{!}{
\begin{tikzpicture}[] 
\draw[] (0.4,0.8) node[below] {\Large{$3$}};
\draw[thick] (0,0) -- (0.8,0) ;
\draw[thick] (0.8,0) -- (0.8,1) ; 
\draw[thick] (0.8,1) -- (0,1)  ;
\draw[thick] (0,1) -- (0,0) ;
 \end{tikzpicture}
}
\end{gathered}
=&
\begin{gathered}
\resizebox{0.8cm}{!}{
\begin{tikzpicture}[] 
\draw[thick] (0,0.2) -- (0.8,0.2) ;
\draw[thick] (0,0.8) -- (0.8,0.8)  ;
\draw[thick] (0,1.4) -- (0.8,1.4)  ;
 \end{tikzpicture}
}
\end{gathered}
+
\begin{gathered}
\resizebox{0.8cm}{!}{
\begin{tikzpicture}[] 
\draw[thick] (0,0.2) -- (0.8,0.8) ;
\draw[thick] (0.8,0.2) -- (0,0.8)  ;
\draw[thick] (0,1.4) -- (0.8,1.4)  ;
 \end{tikzpicture}
}
\end{gathered}
+
\begin{gathered}
\resizebox{0.8cm}{!}{
\begin{tikzpicture}[] 
\draw[thick] (0,0.2) -- (0.8,1.4) ;
\draw[thick] (0,0.8) -- (0.8,0.8)  ;
\draw[thick] (0,1.4) -- (0.8,0.2)  ;
 \end{tikzpicture}
}
\end{gathered}
\end{aligned}\end{equation}
The crosses have the role of exchanging the fermion lines to produce the different topologies of diagrams.
In the first column of Fig.~\ref{diagrams3compact} are listed the reducible $3$rd-order diagrams that can be obtained by combining the two $2$nd-order irreducible diagrams in Fig.~\ref{diagrams_2} and a $1$st-order one. The second column of Fig.~\ref{diagrams3compact} lists the irreducible diagrams divided for convenience in the two classes with overlap of two and three fermion lines. Irreducible diagrams containing $n$ overlapping fermion lines are called $n$-tier diagrams.\\
\indent Along the same lines as with the $2$nd order, we can write the $3$rd order propagating function in Laplace space as the sum of products of lower-order irreducible kernels plus the irreducible $3$rd order kernel $\hat{\boldsymbol{\mathcal{K}}}^{(3)}(\lambda)$, i.e., 
\begin{equation}\begin{aligned}\label{}
\hat{\boldsymbol{J}}^{(3)}(\lambda)=&\frac{1}{\lambda}\Bigg[\left(\frac{\hat{\boldsymbol{\mathcal{K}}}^{(1)}(\lambda)}{\lambda}\right)^3+
\frac{\hat{\boldsymbol{\mathcal{K}}}^{(1)}(\lambda)}{\lambda}\cdot\frac{\hat{\boldsymbol{\mathcal{K}}}^{(2)}(\lambda)}{\lambda}\\
&+
\frac{\hat{\boldsymbol{\mathcal{K}}}^{(2)}(\lambda)}{\lambda}\cdot\frac{\hat{\boldsymbol{\mathcal{K}}}^{(1)}(\lambda)}{\lambda}+
\frac{\hat{\boldsymbol{\mathcal{K}}}^{(3)}(\lambda)}{\lambda}\Bigg]\;,
\end{aligned}\end{equation}
where $\hat{\boldsymbol{\mathcal{K}}}^{(3)}(\lambda)$ collects the irreducible $3$rd order diagrams in  Fig.~\ref{diagrams3compact} (the ones in the second column).\\
\indent At this point we are in the position to derive the formally exact GME for the populations and the current.  
The exact propagator is obtained by summing over all orders $m$ the $m$th-order propagators as follows
\begin{equation}\begin{aligned}\label{JspinfulLaplace}
\hat{\boldsymbol{J}}&(\lambda)=\sum_{m=0}^\infty \hat{\boldsymbol{J}}^{(m)}(\lambda)\\
=&\frac{1}{\lambda}\Bigg[\mathbf{1}+\frac{\hat{\boldsymbol{\mathcal{K}}}^{(1)}(\lambda)}{\lambda}+\left(\frac{\hat{\boldsymbol{\mathcal{K}}}^{(1)}(\lambda)}{\lambda}\right)^2+\frac{\hat{\boldsymbol{\mathcal{K}}}^{(2)}(\lambda)}{\lambda}\\
&+\left(\frac{\hat{\boldsymbol{\mathcal{K}}}^{(1)}(\lambda)}{\lambda}\right)^3+
2\frac{\hat{\boldsymbol{\mathcal{K}}}^{(1)}(\lambda)}{\lambda}\cdot\frac{\hat{\boldsymbol{\mathcal{K}}}^{(2)}(\lambda)}{\lambda}+
\frac{\hat{\boldsymbol{\mathcal{K}}}^{(3)}(\lambda)}{\lambda}+\dots\Bigg]\\
=&\frac{1}{\lambda}\sum_{m=0}^\infty\left(\frac{\hat{\boldsymbol{\mathcal{K}}}^{(1)}(\lambda)}{\lambda}+\frac{\hat{\boldsymbol{\mathcal{K}}}^{(2)}(\lambda)}{\lambda}+\dots\right)^m\\
=&\frac{1}{\lambda}\sum_{m=0}^\infty\left(\frac{\hat{\boldsymbol{\mathcal{K}}}(\lambda)}{\lambda}\right)^m\\
=&\left[\lambda\mathbf{1}-\hat{\boldsymbol{\mathcal{K}}}(\lambda)\right]^{-1}\;,
\end{aligned}\end{equation}
where we introduced the kernel
\begin{equation}\label{irreducible_kernel}
\hat{\boldsymbol{\mathcal{K}}}(\lambda)=\sum_{m=1}^\infty\hat{\boldsymbol{\mathcal{K}}}^{(m)}(\lambda)\;
\end{equation}
which collects all the irreducible contributions to $\hat{\boldsymbol{J}}$. 
From Eq.~\eqref{JspinfulLaplace}, $\lambda \hat{\boldsymbol{J}}(\lambda)-\mathbf{1}=\hat{\boldsymbol{\mathcal{K}}}(\lambda)\cdot\hat{\boldsymbol{J}}(\lambda)$.
By transforming back to the time domain, this implies that $\boldsymbol{J}$ is the solution of the following GME
\begin{equation}\begin{aligned}\label{GME-J}
\frac{d}{dt}\boldsymbol{J}(t)=\int_0^t dt'\;\boldsymbol{\mathcal{K}}(t-t')\cdot \boldsymbol{J}(t')\;.
\end{aligned}\end{equation}
According to Eq.~\eqref{Populations}, the populations are obtained by multiplying the above matrix equation by $\mathbf{P}(0)$, the  population vector at the initial  time $t=0$, which results in
\begin{equation}\begin{aligned}\label{GME-P}
\frac{d}{dt}\mathbf{P}(t)=\int_0^t dt'\;\boldsymbol{\mathcal{K}}(t-t')\cdot \mathbf{P}(t')\;.
\end{aligned}\end{equation}
The asymptotic populations are the solution of the equation $\mathbf{0}=\hat{\boldsymbol{\mathcal{K}}}(0)\cdot\mathbf{P}^\infty$, which is obtained upon applying to Eq.~\eqref{GME-P} the final value theorem $f(t\rightarrow\infty)=\lim_{\lambda\to 0}\lambda \hat{f}(\lambda)$.

\subsection{Integral equation for the current} 
\label{integral_Eq_current}

\begin{figure}[ht]
\begin{center}
\resizebox{2.4cm}{!}{
\begin{tikzpicture}[] 
\draw[cyan] (1,0.6) node[] {$i,l $};
\draw[thick] (-0.5,0) -- (1,0); 
\draw[cyan,thick] (1,0) -- node[below] {${\eta'}^i=+1$} (1.5,0); 
\draw[cyan,thick] (0,0) arc (180:0:0.5cm  and 0.5cm);
 \end{tikzpicture}
 }
\resizebox{4.2cm}{!}{
\begin{tikzpicture}[] 
\draw[cyan] (2.8,0.9) node[] {$i,l $};
\draw[thick] (-0.5,0) -- (3,0); 
\draw[cyan,thick] (3,0) -- node[below] {${\eta'}^i=+1$} (3.5,0);
\draw[cyan,thick] (0,0) arc (180:0:1.5cm  and 1cm);
\draw[black,thick] (1,0) arc (180:0:0.5cm  and 0.5cm) ;
 \end{tikzpicture}
 }
\resizebox{3.8cm}{!}{
\begin{tikzpicture}[] 
\draw[cyan](2.7,0.5) node[] {$i,l $};
\draw[thick] (-0.5,0) -- (2.5,0); 
\draw[cyan,thick] (2.5,0) -- node[below] {${\eta'}^i=+1$}(3.,0);
\draw[black,thick] (0,0) arc (180:0:0.5cm  and 0.6cm) ;
\draw[cyan,thick] (1.5,0) arc (180:0:0.5cm  and 0.6cm);
 \end{tikzpicture}
 } 
\resizebox{4.2cm}{!}{
\begin{tikzpicture}[] 
\draw[cyan](2.8,0.9) node[] {$i,l $};
\draw[thick] (-0.5,0) -- (3,0); 
\draw[cyan,thick] (3,0) -- node[below] {${\eta'}^i=+1$}(3.5,0);
\draw[black,thick] (0,0) arc (180:0:1.cm  and 0.8cm) ;
\draw[cyan,thick] (1,0) arc (180:0:1.cm  and 0.8cm);
 \end{tikzpicture}
 } 
\caption{\small{Examples of diagrams contributing to $\boldsymbol{\mathcal{A}}_l $. The sojourn of the degree of freedom $i$ associated to the last transition is constrained to be ${\eta^i}'=+1$.}}
\label{diagramsA}
\end{center}
\end{figure}
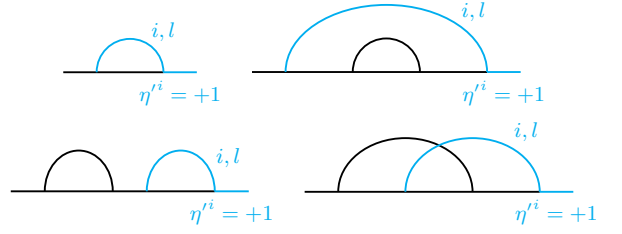

In the present situation of diagonal hybridization matrices, the current functional $\mathcal{I}_l $, Eq.~\eqref{current_functional}, entering the expression for the current through the lead $l $ via Eq.~\eqref{propagator_I}, specializes to
\begin{equation}\begin{aligned}\label{mISIAM}
\mathcal{I}_l (\boldsymbol{\xi}^{*},\boldsymbol{\xi},\bar{\boldsymbol{\xi}})
=&-\int_{t_0}^{t}dt'\sum_{i,x,y,z}\left[x\;\xi_{y}^{z}(t) {\rm g}_{l ,xz}^{-z}(t-t')\xi_{x}^{-z}(t')\right]_i\\
&\qquad\qquad\qquad\times\delta_{z,-1}\delta_{y,+1}\;.
\end{aligned}\end{equation}
The current functional has thus the same form as the phase of the influence functional except that there is no integration over the time of the last tunneling transition and there are constrains on the contributing processes. This entails that the diagrammatic unraveling of the current propagator, obtained by expanding the influence functional as a series in the tunneling transitions, goes along the same lines as the one for the populations. The differences consist in the \emph{last} fermion line of the diagrams bearing the constraints associated with the current and the nested time integrals missing the integration over the last tunneling time. 
The expansion of $J^I_{l ,\boldsymbol{\eta}'\boldsymbol{\eta}}(t,t_{0})$ starts from $m=1$ because, not being at the exponent, the current functional adds two additional transitions to the ones generated by expanding the influence functional. This also implies that, if there are no coherences in the initial state of the system, then also the paths contributing to the current start and end in sojourn states.
The current propagator, in the discretized picture and in the occupation number representation, is then
$J^I_{l ,\boldsymbol{\eta}'\boldsymbol{\eta}}(t;0)=\sum_{m=1}^\infty J^{I(m)}_{l ,\boldsymbol{\eta}'\boldsymbol{\eta}}(t;0)$, where
\begin{equation}\begin{aligned}\label{J_I_n_SIAM}
J^{I(m)}_{l ,\boldsymbol{\eta}'\boldsymbol{\eta}}(t;0)=&\sum_{{\rm paths}_{m}} \int \mathcal{D}^I\{t\}_{m} 
 \prod
_i\sum_{\mathcal{P}_i}
 \mathcal{B}_{m_i}^i(\mathcal{P}_i) {\Phi}_{m_i}^i (\mathcal{P}_i)\\
&\qquad\times {\rm constraints}\;,
\end{aligned}\end{equation} 
with $\sum_i m_i=m$ and $\mathcal{B}_{m_i}^i$ and $\Phi_{m_i}^i$ defined in Eq.~\eqref{phim}. 
To order $m$ there are $2m$ tunneling transitions, as for the populations, however in the case of the current the last transition occurs at the final time $t$, yielding the definition
\begin{equation}
\begin{aligned}\label{D_I} 
\int \mathcal{D}^I\{t\}_{m}:= & \int_{t_0}^t dt_{2m-1}\int_{t_0}^{t_{2m-1}}dt_{2m-2}\dots\int_{t_0}^{t_2}dt_1\;.
\end{aligned}
\end{equation}
Going to the details of the constrains in the current calculations, they can be read-off from Eq.~\eqref{mISIAM}. The first is that the correlator of the last fermion line is not summed over the leads but has the index $l $ of the considered lead. Further,  
according to Eq.~\eqref{Fkl} the Grassmann variable associated to the last transition (the paths start from and land in a sojourn) is of the type $\xi^{\zeta'}_{-\eta'\zeta'}$. Then, the constraints in Eq.~\eqref{mISIAM} translate into
$\zeta'=-1$ and $-\eta'\zeta'=+1$, which imply $\delta_{\eta',+1}$, namely the last sojourn of the degree of freedom $i$  associated to the last transition has to be $+1$. 
Summarizing, the current constraints on the last fermion line are 
\begin{itemize}
\item The last fermion line is specific to the lead $l $ so that there is no contraction over the lead index $\alpha'$.

\item The index $\zeta'$ of the last fermion line is constrained to be $\zeta'=-1$.

\item The final sojourn of the state $i$ associated to the fermion line making the last transition must be ${\eta'}^i=+1$ (the sum over $i$ accounts for all possible processes).
\end{itemize}
Examples of paths that contribute to $\boldsymbol{\mathcal{A}}_l $ with associated fermion lines are shown in Fig. \ref{diagramsA}.\\

\indent The explicit expression for the  term $m=1$  in the expansion of the current propagator is 
\begin{equation}\begin{aligned}\label{J1_I}
J^{I(1)}_{l ,\boldsymbol{\eta}'\boldsymbol{\eta}}(t;0)=&\prod_{j=1}^{N} {\eta'}^j\eta^j\sum_i \int_0^t dt_1 \sum_{\zeta^i}
e^{-\frac{\rm i}{\hbar} \zeta^i E_i(\boldsymbol{\eta}^{(i)})(t_2-t_1)}\\
&\times \;[-{\rm g}_{l ,-\eta^i}^{-\zeta^i}(t_2-t_1)]\delta_{{\boldsymbol{\eta}'}^{(i)}\boldsymbol{\eta}^{(i)}}\delta_{\zeta^i,-1}\delta_{{\eta'}^i,+1}
\end{aligned}\end{equation}
(cf. Eq.~\eqref{J1}). In Laplace space
\begin{equation}\begin{aligned}\label{J1_I_Laplace}
\hat{\boldsymbol{J}}^{I(1)}_l (\lambda)
=&\hat{\boldsymbol{\mathcal{K}}}^{I(1)}_l (\lambda)\frac{1}{\lambda}\;.
\end{aligned}\end{equation}
\indent Let us denote with $\hat{\boldsymbol{\mathcal{K}}}_l ^{I}(\lambda)=\sum_{m}\hat{\boldsymbol{\mathcal{K}}}_l ^{I(m)}(\lambda)$ the sum of all the irreducible diagrammatic contributions to $\hat{\boldsymbol{\mathcal{A}}}_l (\lambda)$ with the last fermion line satisfying the constraints given by the current functional. Note that the reducible contributions to $\hat{\boldsymbol{\mathcal{A}}}_l (\lambda)$ are products of ordinary irreducible kernels $\hat{\boldsymbol{\mathcal{K}}}(\lambda)$ with only the last factor of the type $\hat{\boldsymbol{\mathcal{K}}}_l ^I(\lambda)$. This is because only the last fermion line bears the constraints of the current calculation. Then we find that the exact current propagator is the following sum over all orders $m$ of the $m$-th order propagators 
\begin{equation}\begin{aligned}\label{A_alpha_Laplace}
\hat{\boldsymbol{\mathcal{A}}}_l (\lambda)=&\frac{\hat{\boldsymbol{\mathcal{K}}}_l ^I(\lambda)}{\lambda}\cdot\Bigg[\mathbf{1}+\frac{\hat{\boldsymbol{\mathcal{K}}}^{(1)}(\lambda)}{\lambda}+\left(\frac{\hat{\boldsymbol{\mathcal{K}}}^{(1)}(\lambda)}{\lambda}\right)^2+\frac{\hat{\boldsymbol{\mathcal{K}}}^{(2)}(\lambda)}{\lambda}\\
&+\left(\frac{\hat{\boldsymbol{\mathcal{K}}}^{(1)}(\lambda)}{\lambda}\right)^3+
\frac{\hat{\boldsymbol{\mathcal{K}}}^{(1)}(\lambda)}{\lambda}\cdot\frac{\hat{\boldsymbol{\mathcal{K}}}^{(2)}(\lambda)}{\lambda}+\dots\Bigg]\\
=&\hat{\boldsymbol{\mathcal{K}}}_l ^I(\lambda)\cdot\left[\lambda\mathbf{1}-\hat{\boldsymbol{\mathcal{K}}}(\lambda)\right]^{-1}\\
=&\hat{\boldsymbol{\mathcal{K}}}_l ^I(\lambda)\cdot\hat{\boldsymbol{J}}(\lambda)\;,
\end{aligned}\end{equation}
or, in the time domain
\begin{equation}\begin{aligned}\label{GME-A}
\boldsymbol{\mathcal{A}}_l (t)=\int_0^t dt'\;\boldsymbol{\mathcal{K}}_l ^I(t-t')\cdot\boldsymbol{J}(t')\;.
\end{aligned}\end{equation}
\indent Similarly to the steady state-populations, the steady-state current is found by applying to Eq.~\eqref{A_alpha_Laplace} the final value theorem, which results in
\begin{equation}\begin{aligned}\label{current1}
I_l ^\infty
=&\;e 2{\rm Re}\;{\rm Tr}_{\rm S}[\boldsymbol{\mathcal{A}}_l ^\infty]\\
=&\lim_{\lambda\to 0}\; e 2{\rm Re}\;{\rm Tr}_{\rm S}[\hat{\boldsymbol{\mathcal{K}}}_l ^I(\lambda)\cdot\lambda\hat{\boldsymbol{J}}(\lambda)]\\
=&\;e 2{\rm Re}\;{\rm Tr}_{\rm S}[\hat{\boldsymbol{\mathcal{K}}}_l ^I(0)\cdot\boldsymbol{J}^\infty ]\\
=&\;e2{\rm Re}\;\sum_{\boldsymbol{\eta}'\boldsymbol{\eta}}\hat{\mathcal{K}}_{l ,\boldsymbol{\eta}'\boldsymbol{\eta}}^I(0){\rm P}_{\boldsymbol{\eta}}^\infty
\;,
\end{aligned}\end{equation}
where we assumed that the matrix elements of the asymptotic propagator $\boldsymbol{J}^\infty$ are independent of their column index, i.e. that the steady-state populations are independent of their initial values. In other words, in the asymptotic propagator matrix each column is equal to the asymptotic population vector.

\section{Exact formal expression for the kernel}
\label{exact_kernel}

\subsection{Block structure of the irreducible kernel}
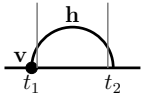
\begin{figure}[ht]
\begin{center}
\resizebox{2cm}{!}{
\begin{tikzpicture}[] 
\draw[line width=0.5mm] (-0.5,0) -- (2,0); 
\draw[](0.75,1) node[] {\large{$\mathbf{h}$}};
\draw[line width=0.5mm] (0,0)  node[below] {\large{$t_1$}} arc (180:0:0.75cm  and 0.75cm) node[below] {\large{$t_2$}};
\filldraw 
(0,0) circle (3pt) node[above] {\large{$\mathbf{v}\quad$}};
\draw[gray,thick] (0.1,0) -- (0.1,1.2);
\draw[gray,thick] (1.4,0) -- (1.4,1.2);
\end{tikzpicture}
}
\caption{\small{Irreducible diagram of order $1$. The vertex $\mathbf{v}$ is denoted by the full dot. The two ends of the fermion line are contracted, namely the indexes $\zeta^i,\alpha$, $k$, and $\sigma$ are summed over. The time interval $t_2-t_1$ is a blip only for the state associated to the fermion line. Note that the states $i$ are summed over in contracting the fermion line, see Eq.~\eqref{contraction}.}}
\label{scheme1st}
\end{center}
\end{figure}

\indent The diagrammatic contributions to the populations and current propagators display an exponential dependence on time, cf. Eqs.~\eqref{corr_matrices_diagonal} and~\eqref{Bm}. This feature and the nested time integrals enable one to express the irreducible kernels in Laplace space as the \emph{contraction} of products of blocks - each equipped with a matrix structure - with an initial vertex. 
The simplest example is provided by the first-order irreducible kernel $\hat{\boldsymbol{\mathcal{K}}}^{(1)}(\lambda)$ in Eq.~\eqref{K1_lambda}, which can be rendered by the contraction of the product of two matrices, associated to an initial vertex $\mathbf{v}$ and a block $\mathbf{h}(\lambda)$ which encompasses a free fermion line, respectively, see Fig.~\ref{scheme1st}.\\

\indent The matrix blocks are indexed by the state index $i$ and the associated collective index   
\begin{equation}\begin{aligned}\label{chi_index}
\boldsymbol\chi:=(\underbrace{\zeta^i,\alpha, k,\sigma}_{\boldsymbol\kappa},\boldsymbol{\eta}^{(i)})\;
\end{aligned}\end{equation}
which includes path and leads variables (note that the components of $\chi$ depend on the state $i$).
The scalar product between two generic  blocks $ \mathbf{A}$ and $ \mathbf{B}$ is given by
\begin{equation}\label{}
[\mathbf{A}\cdot \mathbf{B}]^{i'i}_{\boldsymbol{\chi}'\boldsymbol{\chi}}=\sum_{i'' \boldsymbol\chi''} [\mathbf{A}]^{i'i''}_{\boldsymbol{\chi'}\boldsymbol{\chi''}} [\mathbf{B}]^{i''i}_{\boldsymbol{\chi''}\boldsymbol{\chi}}\;.
\end{equation}
We denote by the symbol $\langle \cdot\rangle$ the contraction of a matrix block $\mathbf{C}$ with an initial vertex ${\rm v}_{-\eta^i}$. The contraction consists in summing over the initial and final index $\boldsymbol\kappa$ and $\boldsymbol\kappa'$, cf. Eq.~\eqref{chi_index}, namely
\begin{equation}\label{}
\langle\mathbf{C}\cdot\mathbf{v}_{- \eta^i}\rangle^{i'i}_{{\boldsymbol{\eta}'}^{(i')}\boldsymbol{\eta}^{(i)}}=\sum_{\boldsymbol{\kappa}'\boldsymbol{\kappa}} [\mathbf{C}]^{i'i}_{\boldsymbol{\chi'}\boldsymbol{\chi}}{\rm v}^{i}_{- \eta^i}(\boldsymbol\kappa)\;.
\end{equation}
As a result, the first-order kernel in Eq.~\eqref{K1_lambda} can be written as 
\begin{equation}\label{K1_lambda_blocks}
\hat{\mathcal{K}}^{(1)}_{\boldsymbol{\eta}'\boldsymbol{\eta}}(\lambda)=\prod_{j=1}^{N} {\eta'}^j\eta^j
\sum_{i'i}\langle \mathbf{h}(\lambda)\cdot \mathbf{v}_{- \eta^i}\rangle^{i'i}_{{\boldsymbol{\eta}'}^{(i')}\boldsymbol{\eta}^{(i)}}\;,
\end{equation}
which is the contraction of the scalar product
\begin{equation}\begin{aligned}\label{contraction}
\langle \mathbf{h}(\lambda)\cdot \mathbf{v}_{- \eta^i}\rangle^{i'i}_{{\boldsymbol{\eta}'}^{(i')}\boldsymbol{\eta}^{(i)}}=&\sum_{{\boldsymbol{\kappa}}',\boldsymbol{\kappa}}\sum_{i''\boldsymbol\chi''} [\mathbf{h}(\lambda)]^{i'i''}_{\boldsymbol{\chi}'\boldsymbol{\chi''}} [\mathbf{v}_{- \eta^i}]^{i'' i}_{\boldsymbol{\chi''}\boldsymbol{\chi}}\;.
\end{aligned}\end{equation}
Here, we have defined the matrix blocks of elements
\begin{equation}\begin{aligned}\label{h_v}
[\mathbf{h}(\lambda)]^{i'i}_{\boldsymbol{\chi'}\boldsymbol{\chi}}:=&\frac{1}{\lambda+{\rm i}\zeta^i[E_i(\boldsymbol{\eta}^{(i)})-\epsilon_k]/\hbar}\delta_{i'i}\delta_{\boldsymbol{\chi'}\boldsymbol{\chi}}\;,\\
[\mathbf{v}_{\pm \eta^i}]^{i'i}_{\boldsymbol{\chi'}\boldsymbol{\chi}}:=&-\frac{|{\rm t}_{i \alpha \sigma}(\epsilon_k)|^2}{\hbar^2} f^\alpha_{\pm\eta^i}(\epsilon_k)\delta_{i'i}\delta_{\boldsymbol{\chi'}\boldsymbol{\chi}}\\
&\equiv {\rm v}^{i}_{\pm \eta^i}(\boldsymbol\kappa)\delta_{i'i}\delta_{\boldsymbol{\chi'}\boldsymbol{\chi}}
\;.
\end{aligned}\end{equation}
Graphically, the vertexes ${\rm v}_{\pm}$ are associated to the two processes
\begin{equation}\begin{aligned}\label{vertexes}
\begin{gathered}
\resizebox{1.5cm}{!}{
\begin{tikzpicture}[] 
\draw[thick] (-1,0) node[black,below] {\Large{$\eta^i$}} -- (0,0); 
\draw[dashed,thick] (0,0) -- (1,0); 
\draw[black,thick] (1,1) arc[start angle=90, end angle=180, radius=1cm];
\filldraw 
(0,0) circle (2pt);
 \end{tikzpicture}
 }
\end{gathered}
\quad{\rm v}_{-\eta^i}\;,\qquad
\begin{gathered}
\resizebox{1.5cm}{!}{
\begin{tikzpicture}[] 
\draw[dashed,thick] (-1,0) -- (0,0); 
\draw[thick] (0,0) -- (1,0) node[black,below] {\Large{$\eta^i$}} ; 
\draw[black,thick] (1,1) arc[start angle=90, end angle=180, radius=1cm];
\filldraw 
(0,0) circle (2pt);
 \end{tikzpicture}
 }
\end{gathered}
\;{\rm v}_{+\eta^i}\;,
\end{aligned}\end{equation}
with the $\pm$ sign of ${\rm v}$ being established directly by the form of the influence functional. Note that, since we deal with the population propagator, the paths start and end in sojourns, thus the initial vertex is always of the type ${\rm v}_{-\eta}$.
 In all diagrams, the two ends of each fermion line are contracted in this way. 
Analogously, the matrix element of the  irreducible current kernel of first order $\boldsymbol{\mathcal{K}}^{I(1)}_l (\lambda)$ reads
\begin{equation}\label{K1_I_lambda}
[\hat{\boldsymbol{\mathcal{K}}}^{I(1)}_l (\lambda)]_{\boldsymbol{\eta}' \boldsymbol{\eta}}
=\prod_{j=1}^{N} {\eta'}^j\eta^j\sum_{i'i}\delta_{{\eta'}^i,+1}
\langle {\rm c}'_l \mathbf{h}(\lambda)\cdot \mathbf{v}_{- \eta^i}\rangle^{i'i}_{{\boldsymbol{\eta}'}^{(i')}\boldsymbol{\eta}^{(i)}}\;,
\end{equation}
where ${\rm c}'_l :=\delta_{{\zeta'}^i,-1}\delta_{\alpha',l }$, so that
\begin{equation}\label{h_I}
 [{\rm c}'_l \mathbf{h}(\lambda)]^{i'i}_{\boldsymbol{\chi'}\boldsymbol{\chi}}:=\frac{	\delta_{\zeta^i,-1}\delta_{\alpha,l }}{\lambda+{\rm i}\zeta^i[E_i(\boldsymbol{\eta}^{(i)})-\epsilon_k]/\hbar}\delta_{i'i}\delta_{\boldsymbol{\chi'}\boldsymbol{\chi}}
\;,
\end{equation}
cf. Eq.~\eqref{h_v}.\\
\begin{figure}[ht]
\begin{center}
\resizebox{3.cm}{!}{
\begin{tikzpicture}[] 
\draw[line width=0.5mm] (-0.5,0) -- (3,0); 
\filldraw (0,0) node[above] {\large{$\mathbf{v}\quad$}} circle (2pt);
\filldraw (1,0) circle (2pt);
\draw[gray,thick] (0.1,0) -- (0.1,1.4);
\draw[gray,thick] (0.9,0) -- (0.9,1.4);
\draw[gray,thick] (2.1,0) -- (2.1,1.4);
\draw[gray,thick] (2.9,0) -- (2.9,1.4);
\draw[line width=0.5mm] (3,0) --  (3.5,0);
\draw[line width=0.5mm] (0,0)node[below] {\large{$t_1$}}  arc (180:0:1.5cm  and 1cm)node[below] {\large{$t_4$}} ;
\draw[line width=0.5mm] (1,0)node[below] {\large{$t_2$}}  arc (180:0:0.5cm  and 0.5cm)node[below] {\large{$t_3$}};
\draw[](0.5,1.4) node[] {\large{$\mathbf{h}$}};
\draw[](1.5,1.4) node[] {\large{$\mathbf{B}$}};
\draw[](2.5,1.4) node[] {\large{$\mathbf{h}$}};
\end{tikzpicture}
 }
 \qquad
 \resizebox{3.cm}{!}{
\begin{tikzpicture}[] 
\draw[line width=0.5mm] (-0.5,0) -- (3,0); 
\filldraw (0,0) node[above] {\large{$\mathbf{v}\quad$}} circle (2pt);
\filldraw (1,0) circle (2pt);
\draw[gray,thick] (0.1,0) -- (0.1,1.4);
\draw[gray,thick] (0.9,0) -- (0.9,1.4);
\draw[gray,thick] (2.1,0) -- (2.1,1.4);
\draw[gray,thick] (2.9,0) -- (2.9,1.4);
\draw[line width=0.5mm] (3,0) --  (3.5,0);
\draw[line width=0.5mm] (0,0)node[below] {\large{$t_1$}}  arc (180:0:1cm  and 1cm)node[below] {\large{$t_4$}} ;
\draw[line width=0.5mm] (1,0)node[below] {\large{$t_2$}}  arc (180:0:1cm  and 1cm)node[below] {\large{$t_3$}};
\draw[](0.5,1.4) node[] {\large{$\mathbf{h}$}};
\draw[](1.5,1.4) node[] {\large{$\mathbf{X}$}};
\draw[](2.5,1.4) node[] {\large{$\mathbf{h}$}};
\end{tikzpicture}
 }
\caption{\small{Irreducible diagrams contributing to $J^{(2)}$. A vertex is denoted by a full dot. The two ends of the lines connecting two blocks carry  path indexes that are summed over in the connection. The first block is contracted with the vertex $\mathbf{v}$ and the right end of the last block is also contracted.}}
\label{scheme2nd}
\end{center}
\end{figure}
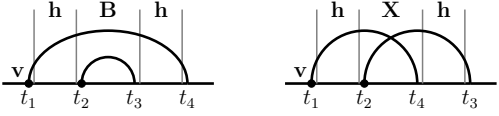
\begin{figure}[ht]
\begin{center}
\resizebox{!}{2.cm}{
\begin{tikzpicture}[] 
\draw[line width=0.5mm] (0.5,1.3)node[left] {\LARGE{$i\boldsymbol\kappa$}}  arc (120:60:2.5cm  and 2.5cm)node[right] {\LARGE{$i\boldsymbol\kappa$}};
\draw[line width=0.5mm] (0.5,-0.5) node[left] {\LARGE{$\boldsymbol{\eta}^{(i)}$}} -- (3,-0.5) node[right] {\LARGE{$\boldsymbol{\eta}^{(i)}$}}; 
\draw [green,line width=0.5mm] (2.8,-0.8) rectangle (0.8,1.9);
\draw[](1.8,2.4) node[] {\LARGE{$\mathbf{h}$}};
\end{tikzpicture}
}
\resizebox{!}{2.cm}{
\begin{tikzpicture}[] 
\draw[line width=0.5mm] (0.5,1.3)node[left] {\LARGE{$i\boldsymbol\kappa$}}  arc (120:60:2.5cm  and 2.5cm)node[right] {\LARGE{$i\boldsymbol\kappa$}};
\draw[line width=0.5mm] (0.5,-0.5) node[left] {\LARGE{$\boldsymbol{\eta}^{(i)}$}} -- (3,-0.5) node[right] {\LARGE{${\boldsymbol{\eta}'}^{(i)}$}}; 
\filldraw 
(1,-0.5) circle (3pt);
\draw[line width=0.5mm] (1,-0.5) arc (180:0:0.8cm  and 1.cm);
\draw [green,line width=0.5mm] (2.8,-0.8) rectangle (0.8,1.9);
\draw[](1.8,2.4) node[] {\LARGE{$\mathbf{B}$}};
\end{tikzpicture}
}
\resizebox{!}{2.cm}{
\begin{tikzpicture}[] 
\draw[line width=0.5mm] (0.5,1.3)node[left] {\LARGE{$i\boldsymbol\kappa$}}  arc (100:0:1.8cm  and 1.8cm);
\draw[line width=0.5mm]
(1,-0.5) arc (0:100:-1.8cm  and 1.8cm)node[right] {\LARGE{$i'\boldsymbol\kappa'$}} ;
\draw[line width=0.5mm] (0.5,-0.5) node[left] {\LARGE{$\boldsymbol{\eta}^{(i)}$}} -- (3,-0.5) node[right] {\LARGE{${\boldsymbol{\eta}'}^{(i)}$}}; 
\filldraw 
(1,-0.5) circle (3pt);
\draw [green,line width=0.5mm] (2.8,-0.8) rectangle (0.8,1.9);
\draw[](1.8,2.4) node[] {\LARGE{$\mathbf{X}$}};
\end{tikzpicture}
}
\caption{\small{The free propagator, bubble, and crossing blocks involved in the diagrams of Figs.~\ref{scheme1st} and~\ref{scheme2nd}.}}
\label{scheme_blocks_general}
\end{center}
\end{figure}
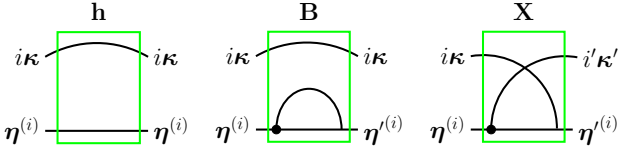
\indent Going to orders higher than the first, consider the four-transition diagrams of Fig.~\ref{scheme2nd}: The time slicing yields a block-product structure in Laplace space.
The blocks are shown in Fig.~\ref{scheme_blocks_general}. Each of them is a matrix with state indexes $i,j$ and the two collective indexes $\boldsymbol\chi'$ and $\boldsymbol\chi$ that specify the path and lead variables according to Eq.~\eqref{chi_index}.\\
\indent The sum over paths is performed automatically by the matrix multiplications implied by forming the diagrams from blocks which, in turn, possibly contain internal processes. The simplest examples of the latter are given by the bubble and crossing blocks, $\mathbf{B}$ and $\mathbf{X}$, of Fig.~\ref{scheme_blocks_general}.  
The irreducible kernel of order $2$ 
 in Laplace space acquires the following expressions in terms of the blocks defined above
\begin{equation}\begin{aligned}\label{K2_blocks}
\hat{\mathcal{K}}^{(2)}_{\boldsymbol{\eta}'\boldsymbol{\eta}}=\prod_{j=1}^{N}{\eta'}^j\eta^j\sum_{i'i}\langle & \mathbf{h} \cdot \Big[\mathbf{B}+\mathbf{X}\Big]\cdot\mathbf{h}\cdot \mathbf{v}_{- \eta^i}\rangle^{i'i}_{{\boldsymbol{\eta}'}^{(i')}\boldsymbol{\eta}^{(i)}}\;,
\end{aligned}\end{equation}
where the dependences on $\lambda$ are understood.\\
\indent The bubble and the crossing shown in Fig.~\ref{scheme_blocks_general} constitute the building blocks of the important resonant tunneling approximation (RTA)~\cite{Koenig1996} in which diagrams with overlap of more than two overlapping fermion lines are neglected, see Eq.~\eqref{phi_RTA} below.
The $3$rd order irreducible kernel in Laplace space reads, within the RTA,
\begin{equation}\begin{aligned}\label{K2_blocks_RTA}
\hat{\mathcal{K}}^{(3)}_{{\rm  RTA}\boldsymbol{\eta}'\boldsymbol{\eta}}=&\prod_{j=1}^{N}{\eta'}^j\eta^j \sum_{i'i}\langle \mathbf{h}\cdot\Big[\mathbf{B}+\mathbf{X}\Big]\cdot\mathbf{h}\\
&\quad\cdot\Big[\mathbf{B}+\mathbf{X}\Big]\cdot\mathbf{h}
\cdot \mathbf{v}_{- \eta^i}\rangle^{i'i}_{{\boldsymbol{\eta}'}^{(i')}\boldsymbol{\eta}^{(i)}}\;.
\end{aligned}\end{equation}
Including in a formal way the contributions beyond-RTA and the higher-order irreducible diagrams with overlap of arbitrarily many fermion lines, we obtain a picture where the kernel results from contraction of a dressed block, the irreducible propagator $\boldsymbol{\phi}$, whose definition is the object of the next section.

\subsection{Diagrammatic unraveling of the kernel}
\label{diagrammatic_unravelling}
\indent The sum of all irreducible diagrams can be obtained by contracting with an initial vertex the dressed block $\boldsymbol{\phi}$  which gives the exact kernel according to 
\begin{equation}\label{Kexact}
\hat{\mathcal{K}}_{\boldsymbol{\eta}'\boldsymbol{\eta}}(\lambda)=  \prod_{j=1}^{N}{\eta'}^j\eta^j \sum_{i'i}\langle\boldsymbol{\phi}(\lambda)\cdot\mathbf{v}_{- \eta^i}\rangle^{i'i}_{{\boldsymbol{\eta}'}^{(i')}\boldsymbol{\eta}^{(i)}}\;.
\end{equation}
The irreducible propagator $\boldsymbol{\phi}$ can be seen as a block with an incoming and an outgoing fermion line, similar to the ones shown in Fig.~\ref{scheme_blocks_general}, dressed by processes of all orders. Using the notation introduced in Sec.~\ref{GME_pop_current}, with the boxes that permute the fermion lines numbered according to the number of overlapping lines,  we can give the following (symbolic) exact expression 
\vspace{1.5cm}
\begin{widetext}
\begin{equation}\begin{aligned}\label{phi_exact_symbolic}
\begin{gathered}
\resizebox{!}{1.8cm}{
\begin{tikzpicture} 
\draw[line width=0.5mm] (0.8,-0.5) node[left] {\Large{$\boldsymbol{\eta}^{(i)}$}} -- (3.2,-0.5) 
node[right] {\Large{${\boldsymbol{\eta}'}^{(i')}$}}; 
\draw[line width=0.5mm] (0.8,0.9) node[left] {\Large{$i\boldsymbol\kappa$}} -- (3.2,0.9) node[right] {\Large{$i'\boldsymbol\kappa'$}}; 
\fill[white] (2.8,-0.6) rectangle (1.2,1.15);
\draw [pattern=north west lines, pattern color=black] (2.8,-0.6) rectangle (1.2,1.15);
\draw[](2,1.5) node[] {\Large{$\boldsymbol{\phi}$}};
 \end{tikzpicture}
}
\end{gathered}
=
\begin{gathered}
\resizebox{!}{1.8cm}{
\begin{tikzpicture} 
\draw[line width=0.5mm] (0.8,-0.5) -- (1.5,-0.5); 
\draw[line width=0.5mm] (0.8,0.9)  -- (1.5,0.9); 
\draw[](1.2,1.6) node[] {\Large{$\mathbf{h}_1$}};
\draw[](2.5,0.3) node[] {\LARGE{$\begin{aligned}
\sum_{n=0}^\infty\Bigg\{
\end{aligned}$}};
\draw[line width=0.5mm] (3.5,-0.5) -- (5,-0.5); 
\draw[line width=0.5mm] (3.5,0.9)  -- (5,0.9); 
\draw[line width=0.5mm] (3.5,-0.2) arc (180:0:0.75cm  and 0.9cm);
\draw[line width=0.5mm] (3.5,-0.5)  -- (3.5,-0.2); 
\draw[line width=0.5mm] (5,-0.5)  -- (5,-0.2); 
\filldraw[](3.5,-0.5) circle (2.5pt); 
\fill[white] (4,0.4) rectangle (4.5,1.1);
\draw [] (4,0.4) rectangle (4.5,1.1);
\draw[](4.2,1.5) node[] {\Large{$\langle{\rm v}\mathbf{S}_2\rangle$}};
\draw[](4.25,0.75) node[] {\Large{$2$}};
\draw[line width=0.5mm] (5.2,-0.5) -- (6,-0.5); 
\draw[line width=0.5mm] (5.2,0.9)  -- (6,0.9); 
\draw[](5.6,1.6) node[] {\Large{$\mathbf{h}_1$}};
\draw[](6.5,0.3) node[] {\LARGE{$\begin{aligned}
+
\end{aligned}$}};
\draw[line width=0.5mm] (7,-0.5) -- (8.5,-0.5); 
\draw[line width=0.5mm] (7,0.9)  -- (8.5,0.9); 
\draw[line width=0.5mm] (8,0.65)  -- (8.5,0.65);
\draw[line width=0.5mm] (7,-0.2) arc (180:95:0.75cm  and 0.9cm);
\draw[line width=0.5mm] (7,-0.5)  -- (7,-0.2); 
\draw[line width=0.5mm] (8.5,-0.5)  -- (8.5,-0.2); 
\filldraw[](7,-0.5) circle (2.5pt); 
\fill[white] (7.5,0.4) rectangle (8,1.1);
\draw [] (7.5,0.4) rectangle (8,1.1);
\draw[](7.7,1.5) node[] {\Large{$\langle{\rm v}\mathbf{S}_2$}};
\draw[](7.75,0.75) node[] {\Large{$2$}};
\fill[white] (8.2,-0.7) rectangle (8.8,1.1);
\draw[](9.2,0.3) node[] {\LARGE{$\begin{aligned}
\sum_{m=1}^\infty\Bigg[
\end{aligned}$}};
\draw[line width=0.5mm] (10,-0.5) -- (11.5,-0.5); 
\draw[line width=0.5mm] (10,0.9)  -- (11.5,0.9); 
\draw[line width=0.5mm] (10,0.65)  -- (11.5,0.65);
\draw[line width=0.5mm] (10,-0.2) arc (180:0:0.75cm  and 0.675cm);
\draw[line width=0.5mm] (10,-0.5)  -- (10,-0.2); 
\draw[line width=0.5mm] (11.5,-0.5)  -- (11.5,-0.2); 
\filldraw[](10,-0.5) circle (2.5pt); 
\fill[white] (10.5,0.3) rectangle (11,1.1);
\draw [] (10.5,0.3) rectangle (11,1.1);
\draw[](10.7,1.5) node[] {\Large{$\langle{\rm v}\mathbf{S}_3\rangle$}};
\draw[](10.75,0.7) node[] {\Large{$3$}};
\draw[line width=0.5mm] (11.7,-0.5) -- (12.5,-0.5); 
\draw[line width=0.5mm] (11.7,0.9)  -- (12.5,0.9); 
\draw[line width=0.5mm] (11.7,0.65)  -- (12.5,0.65);
\draw[](12.1,1.6) node[] {\Large{$\mathbf{h}_2$}};
\draw[](13,0.3) node[] {\LARGE{$\begin{aligned}
+
\end{aligned}$}};
\draw[line width=0.5mm] (13.5,-0.5) -- (15,-0.5); 
\draw[line width=0.5mm] (13.5,0.9)  -- (15,0.9); 
\draw[line width=0.5mm] (13.5,0.65)  -- (15,0.65);
\draw[line width=0.5mm] (13.5,-0.2) arc (180:90:0.75cm  and 0.675cm);
\draw[line width=0.5mm] (13.5,-0.5)  -- (13.5,-0.2); 
\draw[line width=0.5mm] (15,-0.5)  -- (15,-0.2); 
\draw[line width=0.5mm] (14.5,0.45)  -- (15,0.45); 
\filldraw[](13.5,-0.5) circle (2.5pt); 
\fill[white] (14,0.3) rectangle (14.5,1.1);
\draw [] (14,0.3) rectangle (14.5,1.1);
\draw[](14.2,1.5) node[] {\Large{$\langle{\rm v}\mathbf{S}_3$}};
\draw[](14.25,0.7) node[] {\Large{$3$}};
\fill[white] (14.7,-0.7) rectangle (16,1.1);
\draw[](16.5,0.3) node[] {\LARGE{$\begin{aligned}
\sum_{k=1}^\infty\Bigg(\dots\Bigg)^k
\end{aligned}$}};
\draw[line width=0.5mm] (18.25,-0.5) -- (19,-0.5); 
\draw[line width=0.5mm] (18.25,0.9)  -- (19,0.9); 
\draw[line width=0.5mm] (18.25,0.65)  -- (19,0.65);
\draw[line width=0.5mm] (19,-0.2) arc (0:90:0.75cm  and 0.675cm);
\draw[line width=0.5mm] (19,-0.5)  -- (19,-0.2); 
\draw[line width=0.5mm] (19.2,-0.5) -- (20,-0.5); 
\draw[line width=0.5mm] (19.2,0.9)  -- (20,0.9); 
\draw[line width=0.5mm] (19.2,0.65)  -- (20,0.65);
\draw[](19.2,1.6) node[] {\Large{$\rangle\quad\mathbf{h}_2$}};
\draw[](20.5,0.3) node[] {\LARGE{$\begin{aligned}
\Bigg]^m
\end{aligned}$}};
\draw[line width=0.5mm] (21.25,-0.5) -- (22,-0.5); 
\draw[line width=0.5mm] (21.25,0.9)  -- (22,0.9); 
\draw[line width=0.5mm] (22,-0.2) arc (0:90:0.75cm  and 0.85cm);
\draw[line width=0.5mm] (22,-0.5)  -- (22,-0.2); 
\draw[line width=0.5mm] (22.2,-0.5) -- (23,-0.5); 
\draw[line width=0.5mm] (22.2,0.9)  -- (23,0.9); 
\draw[](22.2,1.6) node[] {\Large{$\rangle\quad\mathbf{h}_1$}};
\draw[](23.5,0.3) node[] {\LARGE{$\begin{aligned}
\Bigg\}^n
\end{aligned}$}};
\end{tikzpicture}
}
\end{gathered}
\end{aligned}\end{equation}
\end{widetext}
Notice that, for simplicity,
the diagrams and the corresponding formulas above them are both ordered from left to right here, at variance with the convention used throughout the text, where formulas are ordered from right to left.
This formal unraveling of the propagator constitutes a main result of the present work. It systematically encompasses the truncation schemes based on the order in $\Gamma$ (e.g. sequential tunneling and cotunneling) and on the depth of the hierarchy of overlap of fermion lines (e.g. RTA), as we exemplify below. In Eq.~\eqref{phi_exact_symbolic} we have introduced $\mathbf{h}_1\equiv \mathbf{h}$ and similarly, for the  blocks with higher overlap of noncrossing fermion lines, $\mathbf{h}_n$. 
 As in the definition of the first-order irreducible kernel in Eq.~\eqref{contraction}, the symbol $\langle \cdot \rangle$ implies the contraction with an initial vertex of a fermion line, in this case the most internal. The boxes $\boxed{2}$ and $\boxed{3}$ are defined in Eqs.~\eqref{box2} and~\eqref{box3}, respectively. The box $\boxed{4}$ reads
\begin{equation}\begin{aligned}\label{box4}
\begin{gathered}
\resizebox{0.8cm}{!}{
\begin{tikzpicture}[] 
\draw[] (0.4,0.8) node[below] {\Large{$4$}};
\draw[thick] (0,0) -- (0.8,0) ;
\draw[thick] (0.8,0) -- (0.8,1) ; 
\draw[thick] (0.8,1) -- (0,1)  ;
\draw[thick] (0,1) -- (0,0) ;
 \end{tikzpicture}
}
\end{gathered}
=&
\begin{gathered}
\resizebox{0.8cm}{!}{
\begin{tikzpicture}[] 
\draw[thick] (0,0.2) -- (0.8,0.2) ;
\draw[thick] (0,0.6) -- (0.8,0.6)  ;
\draw[thick] (0,1.0) -- (0.8,1.0)  ;
\draw[thick] (0,1.4) -- (0.8,1.4)  ;
 \end{tikzpicture}
}
\end{gathered}
+
\begin{gathered}
\resizebox{0.8cm}{!}{
\begin{tikzpicture}[] 
\draw[thick] (0,0.2) -- (0.8,0.6) ;
\draw[thick] (0,0.6) -- (0.8,0.2)  ;
\draw[thick] (0,1.0) -- (0.8,1.0)  ;
\draw[thick] (0,1.4) -- (0.8,1.4)  ;
 \end{tikzpicture}
}
\end{gathered}
+
\begin{gathered}
\resizebox{0.8cm}{!}{
\begin{tikzpicture}[] 
\draw[thick] (0,0.2) -- (0.8,1.0) ;
\draw[thick] (0,0.6) -- (0.8,0.6)  ;
\draw[thick] (0,1.0) -- (0.8,0.2)  ;
\draw[thick] (0,1.4) -- (0.8,1.4)  ;
 \end{tikzpicture}
}
\end{gathered}
+
\begin{gathered}
\resizebox{0.8cm}{!}{
\begin{tikzpicture}[] 
\draw[thick] (0,0.2) -- (0.8,1.4) ;
\draw[thick] (0,0.6) -- (0.8,0.6)  ;
\draw[thick] (0,1.0) -- (0.8,1.0)  ;
\draw[thick] (0,1.4) -- (0.8,0.2)  ;
 \end{tikzpicture}
}
\end{gathered}\;,
\end{aligned}\end{equation}
and analogous definitions hold for the higher-order boxes.  Finally, $\mathbf{S}_n$  contains the box $\boxed{n}$ that operates on $n$ fermion lines by exchanging them in pairs and thus represent a class of blocks.\\ 
\indent From the schematic expression in Eq.~\eqref{phi_exact_symbolic} we find in Laplace space
\begin{equation}\label{phi_exact}
\boldsymbol{\phi}=
\sum_{n=0}^\infty\left(\mathbf{h}\langle \tilde{\mathbf{S}}_2\mathbf{v}\rangle\right)^n\mathbf{h} 
\end{equation}
with $\langle\mathbf{S}_2\mathbf{v}\rangle\equiv\mathbf{B}+\mathbf{X}$, the RTA self-energy, and $\mathbf{h}_1\equiv\mathbf{h}$, see Fig.~\ref{scheme_blocks_general}.
In Eq.~\eqref{phi_exact}, we have defined the dressed self-energies \emph{iteratively} as 
\begin{equation}\begin{aligned}\label{hierarchy_S}
\tilde{\mathbf{S}}_{n-1}:=\sum_{k=0}^\infty\Bigg(\mathbf{h}_{n-1}\langle \tilde{\mathbf{S}}_{n}\mathbf{v}\rangle\Bigg)^k\mathbf{S}_{n-1}\;.
\end{aligned}\end{equation}
Summing the geometrical series in Eq.~\eqref{phi_exact} we find the exact result
\begin{equation}\begin{aligned}\label{phi_exact_Dyson}
\boldsymbol{\phi}=[\mathbf{h}^{-1}-
\langle \tilde{\mathbf{S}}_2\mathbf{v}\rangle
]^{-1}\;.
\end{aligned}\end{equation}
Thus, the function $\boldsymbol{\phi}$ is the solution of the Dyson equation 
\begin{equation}\begin{aligned}\label{Dyson_phi}
\boldsymbol{\phi}=\mathbf{h}+\mathbf{h}\langle \tilde{\mathbf{S}}_2\mathbf{v}\rangle\boldsymbol{\phi}\;.
\end{aligned}\end{equation}
The formal exact expression for $\boldsymbol{\phi}$ generates the different approximations schemes used in literature by suitably truncating the series expression, Eq.~\eqref{phi_exact}, and/or the depth of the  hierarchy in Eq.~\eqref{hierarchy_S}. For example, the sequential tunneling (ST) scheme, first order in $\Gamma$, is recovered by truncating Eq.~\eqref{phi_exact} to $n=0$,
\begin{equation}\label{phi_ST}
\boldsymbol{\phi}_{\rm ST}=\mathbf{h}\;.
\end{equation}
The next-order perturbative scheme includes the cotunneling (CT) diagrams, with $n =0, 1$ in  Eq.~\eqref{phi_exact} and $\tilde{\mathbf{S}}_2=\mathbf{S}_2$, so that $\langle \mathbf{S}_2\mathbf{v}\rangle=\mathbf{B}+\mathbf{X}$. Then the irreducible propagator reads
\begin{equation}\label{phi_CT}
\boldsymbol{\phi}_{\rm CT}=\mathbf{h}+\mathbf{h}(\mathbf{B}+\mathbf{X})\mathbf{h}\;.
\end{equation}
Retaining all orders in the series expansion for $\boldsymbol{\phi}$ and truncating the hierarchy in Eq.~\eqref{hierarchy_S} to two overlapping fermion lines, i.e. to the second tier, where $\tilde{\mathbf{S}}_2=\mathbf{S}_2$, we obtain the Dyson equation for the RTA propagator 
\begin{equation}\begin{aligned}\label{phi_RTA}
\boldsymbol{\phi}_{\rm RTA}=\mathbf{h}+\mathbf{h}(\mathbf{B}+\mathbf{X})\boldsymbol{\phi}_{\rm RTA}
\;.
\end{aligned}\end{equation}
This approximation scheme is nonperturbative in the tunnel coupling $\Gamma$ and corresponds to the scheme proposed in~\cite{Koenig1996,Karlstrom2013}. Note that the propagator $\boldsymbol\phi_{\rm CT}$ is readily recovered by expanding $\boldsymbol\phi_{\rm RTA}$ to order $\Gamma^2$. From the RTA, neglecting the crossing block, we obtain the second-tier noncrossing approximation  (NCA2), to be discussed below, which is the solution of 
\begin{equation}\begin{aligned}\label{phi_gDSO}
\boldsymbol{\phi}_{\rm NCA2}=\mathbf{h}+\mathbf{h}\mathbf{B}\boldsymbol{\phi}_{\rm NCA2}
\end{aligned}\end{equation}
and reproduces the results obtained by Meir, Wingreen, and Lee in~\cite{Meir1991} with the equation of motion method. 

\subsection{Dyson equation for the kernel in terms of dressed bubbles and crossings}
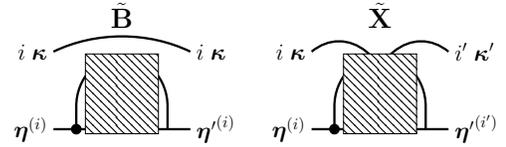
\begin{figure}[ht]
\begin{center}
\resizebox{!}{2cm}{
\begin{tikzpicture}[] 
\draw[line width=0.5mm] (0.5,1.2) node[left] {\Large{$\;\;i\;\boldsymbol\kappa$}} arc (120:60:3cm  and 2.5cm) node[right] {\Large{$i\;\boldsymbol\kappa$}};
\draw[line width=0.5mm]  (1,0) arc (180:0:1.cm  and 1.1cm); 
\draw[line width=0.5mm] (0.5,-0.5) 
node[left] {\Large{$\boldsymbol{\eta}^{(i)}$}} -- (3.5,-0.5) node[right] {\Large{${\boldsymbol{\eta}'}^{(i)}$}}; 
\draw[line width=0.5mm] (1,-0.5) -- (1,0); 
\draw[line width=0.5mm] (3,-0.5) -- (3,0); 
\filldraw(1,-0.5) circle (3pt); 
\fill[white] (2.8,-0.6) rectangle (1.2,1.15);
\draw [pattern=north west lines, pattern color=black] (2.8,-0.6) rectangle (1.2,1.15);
\draw[](2,2) node[] {\LARGE{$\tilde{\mathbf{B}}$}};
 \end{tikzpicture}
}
\resizebox{!}{2cm}{
\begin{tikzpicture}[] 
\draw[line width=0.5mm] (0.5,1.2) node[left] {\Large{$\;\;i\;\boldsymbol\kappa$}} arc (120:10:1.2cm  and 1.7cm) ;
\draw[line width=0.5mm] (3.5,1.2) node[right] {\Large{$i'\;\boldsymbol\kappa'$}}  arc (120:10:-1.2cm  and 1.7cm); 
\draw[line width=0.5mm]  (1,0) arc (180:0:1.cm  and 1.1cm); 
\draw[line width=0.5mm] (0.5,-0.5) 
node[left] {\Large{$\boldsymbol{\eta}^{(i)}$}} -- (3.5,-0.5)node[right] {\Large{${\boldsymbol{\eta}'}^{(i')}$}}; 
(3.5,-0.5); 
\draw[line width=0.5mm] (1,-0.5) -- (1,0); 
\draw[line width=0.5mm] (3,-0.5) -- (3,0);
\filldraw(1,-0.5) circle (3pt); 
\fill[white] (2.8,-0.6) rectangle (1.2,1.15);
\draw [pattern=north west lines, pattern color=black] (2.8,-0.6) rectangle (1.2,1.15);
\draw[](2,2) node[] {\LARGE{$\tilde{\mathbf{X}}$}};
 \end{tikzpicture}
 }
\caption{\small{Dressed, irreducible bubble and crossing blocks that generalize the ones shown in  Fig.~\ref{scheme_blocks_general}}. The sum $\tilde{\mathbf{B}}+\tilde{\mathbf{X}}$ is the self-energy for the exact propagator $\boldsymbol{\phi}$. Within the dressed bubble approximation (DBA), the irreducible bubble diagram $\tilde{\mathbf{B}}$ is the self-energy of the propagator $\boldsymbol{\phi}_{\rm DBA}$.}
\label{dressed_BX}
\end{center}
\end{figure}

We split the exact, dressed self-energy of Eqs.~\eqref{phi_exact_Dyson} and~\eqref{Dyson_phi} as $\langle \tilde{\mathbf{S}}_2\mathbf{v}\rangle=\tilde{\mathbf{B}}+\tilde{\mathbf{X}}$, according to whether the incoming and outgoing fermion lines are the same or not. 
This yields
\begin{equation}
\boldsymbol{\phi}=\left[\mathbf{h}^{-1}-(\tilde{\mathbf{B}}+\tilde{\mathbf{X}})\right]^{-1}\;.
\end{equation}
Hence, the sum $\tilde{\mathbf{B}}+\tilde{\mathbf{X}}$ is the self-energy for the exact propagator $\boldsymbol{\phi}$.
The irreducible block $\tilde{\mathbf{B}}$ is a fermion line dressed by processes of all orders, including crossings. In the dressed, irreducible  crossing block $\tilde{\mathbf{X}}$, the incoming and outgoing fermion lines are not the same, which implies the presence of crossings involving these main fermion lines. These two dressed blocks are shown in Fig.~\ref{dressed_BX}.\\
\indent Let us introduce the dressed propagator in the dressed bubble approximation (DBA), $\boldsymbol{\phi}_{\rm DBA}$, obtained by setting to zero the crossing block $\tilde{\mathbf{X}}$. It satisfies the same Dyson equation as  $\boldsymbol{\phi}$, but with the dressed bubble self-energy alone; namely, in Laplace space,
\begin{equation}\begin{aligned}\label{dressed_line}
\boldsymbol{\phi}_{\rm DBA}=&\left[\mathbf{h}^{-1}-\tilde{\mathbf{B}}\right]^{-1}
\;,
\end{aligned}\end{equation}
or equivalently
\begin{equation}\begin{aligned}\label{Dyson_phiD}
\boldsymbol{\phi}_{\rm DBA}=&\mathbf{h}+\mathbf{h}\tilde{\mathbf{B}}\boldsymbol{\phi}_{\rm DBA}\;.
\end{aligned}\end{equation}
Truncation of the hierarchy to the second tier yields the NCA2, Eq.~\eqref{phi_gDSO}, where $\tilde{\mathbf{B}}\equiv \mathbf{B}$.
In terms of $\boldsymbol{\phi}_{\rm DBA}$, we can cast the exact equation for $\boldsymbol{\phi}$, Eq.~\eqref{Dyson_phi}, in the form
\begin{equation}
\label{Dyson_phi_a}
\boldsymbol{\phi}=\boldsymbol{\phi}_{\rm DBA}+\boldsymbol{\phi}_{\rm DBA}\tilde{\mathbf{X}}\boldsymbol{\phi}\;.
\end{equation}
Due to the lack of crossings on the main fermion line, $\boldsymbol\phi_{\rm DBA}$ is diagonal in $\boldsymbol\kappa$ and in $i$  (but not in $\boldsymbol{\eta}^{(i)}$)  namely 
\begin{equation}\label{varphiB}
[\boldsymbol\phi_{\rm DBA}]^{i'i}_{\boldsymbol\chi'\boldsymbol\chi}=\varphi^{ii}
_{{\rm DBA},{\boldsymbol{\eta}'}^{(i)}\boldsymbol{\eta}^{(i)}}(\boldsymbol\kappa)\delta_{i'i}\delta_{\boldsymbol\kappa'\boldsymbol\kappa}\;.
\end{equation}
Component-wise in $\boldsymbol\kappa$, retaining the matrix structure induced by $\boldsymbol{\eta}^{(i)}=(\dots,\eta^{i-1},\eta^{i+1},\dots)$ and with the dependence on $\lambda$ left implicit, Eq.~\eqref{Dyson_phi_a} reads
\begin{equation}\begin{aligned}\label{Dyson_phi_component_a}
\boldsymbol\phi^{i'i}(\boldsymbol\kappa',&\boldsymbol\kappa)=\boldsymbol\varphi_{\rm DBA}^{ii}(\boldsymbol\kappa')\delta_{i'i}\delta_{\boldsymbol\kappa' \boldsymbol\kappa}\\
&+\boldsymbol\varphi_{\rm DBA}^{i'i'}(\boldsymbol\kappa')\cdot\sum_{i''\boldsymbol\kappa''}\tilde{\mathbf{X}}^{i'i''}(\boldsymbol\kappa',\boldsymbol\kappa'')\cdot\boldsymbol\phi^{i''i}(\boldsymbol\kappa'',\boldsymbol\kappa)\;,
\end{aligned}\end{equation}
where $\boldsymbol\varphi_{\rm DBA}^{ii}(\boldsymbol\kappa)$ is the left-contracted propagator given by
\begin{equation}\label{varphi_B}
\boldsymbol\varphi_{\rm DBA}^{ii}(\boldsymbol\kappa)=\sum_{i'\boldsymbol\kappa'}\boldsymbol\phi^{i'i}_{\rm DBA}(\boldsymbol\kappa',\boldsymbol\kappa)\;.
\end{equation}

The steady-state population kernel can then be written as
\begin{equation}\begin{aligned}\label{SS_Kcurrent}
\hat{\mathcal{K}}_{\boldsymbol{\eta}'\boldsymbol{\eta}}(0)
=&\prod_{j=1}^N{\eta'}^j\eta^j\sum_{i'i}
\langle \boldsymbol\phi \cdot\mathbf{v}_{- \eta^i}\rangle^{i'i}_{{\boldsymbol{\eta}'}^{(i')}\boldsymbol{\eta}^{(i)}}\\
=&\prod_{j=1}^N{\eta'}^j\eta^j \sum_{i' i \boldsymbol{\kappa}'\boldsymbol{\kappa}}
\phi^{i'i}_{{\boldsymbol{\eta}'}^{(i')}\boldsymbol{\eta}^{(i)}}(\boldsymbol{\kappa}',\boldsymbol{\kappa}){\rm v}^{i}_{-\eta^i}(\boldsymbol{\kappa})\;.
\end{aligned}\end{equation}
As anticipated in Sec.~\ref{GME_populations}, the steady-state populations ${\rm P}^\infty_{\boldsymbol{\eta}}$ are the solutions of the matrix equation 
\begin{equation}\label{P_steady}
\mathbf{0}=\sum_{\boldsymbol{\eta}}\hat{\mathcal{K}}_{\boldsymbol{\eta}'\boldsymbol{\eta}}(0){\rm P}^\infty_{\boldsymbol{\eta}}\;.
\end{equation}

\subsection{Current kernel}
\label{current_calculation}

As shown in Sec.~\ref{integral_Eq_current}, the current kernel can be calculated along the same lines as the population kernel provided that the additional constraints 
$$
\delta_{{\eta'}^{i'},+1}{\rm c}'_l  \equiv\delta_{{\eta'}^{i'},+1}\delta_{\alpha',l }\delta_{\zeta',-1}$$ 
are introduced for the last fermion line.

According to Eq.~\eqref{current1}, the general formula for the steady-state current on lead $l $ is 
\begin{equation}\begin{aligned}\label{current3}
I^\infty_l =\; e2{\rm Re}\sum_{\boldsymbol{\eta}'\boldsymbol{\eta}} \hat{\mathcal{K}}^{I}_{l ,\boldsymbol{\eta}'\boldsymbol{\eta}}(0){\rm P}^\infty_{\boldsymbol{\eta}}\;,
\end{aligned}\end{equation}
where the kernel is in Laplace space and calculated at $\lambda=0$. The  current kernel formally reads 
\begin{equation}\begin{aligned}\label{Kcurrent}
\hat{\mathcal{K}}&^{I}_{l ,\boldsymbol{\eta}'\boldsymbol{\eta}}(0)
=\prod_{j=1}^N{\eta'}^j\eta^j\sum_{i'i}\delta_{{\eta'}^{i'},+1}
\langle{\rm c}'_l  \boldsymbol\phi \cdot\mathbf{v}_{- \eta^i}\rangle^{i'i}_{{\boldsymbol{\eta}'}^{(i')}\boldsymbol{\eta}^{(i)}}\\
=&\prod_{j=1}^N{\eta'}^j\eta^j \sum_{i' i \boldsymbol{\kappa}'\boldsymbol{\kappa}}\delta_{{\eta'}^{i'},+1}{\rm c}'_l 
\phi^{i'i}_{{\boldsymbol{\eta}'}^{(i')}\boldsymbol{\eta}^{(i)}}(\boldsymbol{\kappa}',\boldsymbol{\kappa}){\rm v}^{i}_{-\eta^i}(\boldsymbol{\kappa})\;,
\end{aligned}\end{equation}
cf. Eq.~\eqref{SS_Kcurrent}.\\
\indent Summarizing, the steady-state populations and currents of the $N$-state system coupled to multiple leads can be obtained via  Eqs.~\eqref{P_steady} and~\eqref{current3}, respectively. Both the population and the current kernels are in turn directly given by the irreducible propagator $\boldsymbol{\phi}$ which is calculated via the Dyson equations~\eqref{Dyson_phiD} and~\eqref{Dyson_phi_a}.

\section{Proportional coupling and connection with the Green's functions}
\label{prop_coupl_and_GF}

Consider the situation in which the central system is connected to two leads ($L$ and $R$) and the tunneling amplitudes are related by 
$$
 |{\rm t}_{i R \sigma}(\epsilon)|^2=\gamma_{i R} |{\rm t}_{i L \sigma}(\epsilon)|^2/\gamma_{i L} 
$$
with $\gamma_{i R}+\gamma_{i L}=1$ (proportional coupling).
Current conservation at the steady state $I^\infty=I_L^\infty=-I_R^\infty$ implies,  for proportional coupling, $I^\infty=\sum_i [\gamma_{i R} I_{i L}^\infty-\gamma_{i L}I_{i R}^\infty]$, where $I_{i'\alpha}^\infty=e2{\rm Re}\sum_{\boldsymbol{\eta}'\boldsymbol{\eta}} \hat{\mathcal{K}}^{I}_{i' l ,\boldsymbol{\eta}'\boldsymbol{\eta}}(0){\rm P}^\infty_{\boldsymbol{\eta}}$ is the state-resolved steady-state current and $\hat{\mathcal{K}}^{I}_{i' l ,\boldsymbol{\eta}'\boldsymbol{\eta}}(0)$ is given by Eq.~\eqref{Kcurrent} without the sum over the final state $i'$. Correspondingly, we introduce the current kernel
\begin{equation}\begin{aligned}\label{Kcurrent_PC}
\hat{\mathcal{K}}^{I}_{\boldsymbol{\eta}'\boldsymbol{\eta}}
(0):=&\sum_{i'}\left[\gamma_{i' R}\hat{\mathcal{K}}^{I}_{i' L,\boldsymbol{\eta}'\boldsymbol{\eta}}(0)-\gamma_{i' L}\hat{\mathcal{K}}^{I}_{i' R,\boldsymbol{\eta}'\boldsymbol{\eta}}(0)\right]\;,
\end{aligned}\end{equation} 
with the steady-state current obtained as
\begin{equation}\begin{aligned}\label{current4}
I^\infty=\;e 2{\rm Re}\sum_{\boldsymbol{\eta}'\boldsymbol{\eta}} \hat{\mathcal{K}}^{I}_{\boldsymbol{\eta}'\boldsymbol{\eta}}(0){\rm P}^\infty_{\boldsymbol{\eta}}\;.
\end{aligned}\end{equation}
Consider now the Dyson equation for $\boldsymbol\phi$, Eq.~\eqref{Dyson_phi_component_a}. To obtain the current kernel for the current on lead $l$ we make a contraction with the vertex as in Eq.~\eqref{Kcurrent}, which gives
\begin{equation}\begin{aligned}\label{Dyson_phi_contracted}
\langle{\rm c}'_l  \boldsymbol\phi\cdot\mathbf{v}_{- \eta^i}\rangle
=&\langle{\rm c}'_l \left[\boldsymbol\phi_{\rm DBA}+\boldsymbol\phi_{\rm DBA}\cdot\tilde{\mathbf{X}}\cdot\boldsymbol\phi\right]\cdot\mathbf{v}_{- \eta^i}\rangle\;,
\end{aligned}\end{equation}
or, in symbols,
\begin{equation}\begin{aligned}\label{Dyson_phi_symbolic}
\begin{gathered}
\resizebox{!}{1.1cm}{
\begin{tikzpicture} 
\draw[line width=0.5mm] (0.4,-0.5)-- (3.6,-0.5) ; 
\draw[red,line width=0.5mm] (2.2,0.9) arc (90:0:1.4cm  and 1.4cm) ;
\draw[red] (3.5,0.7) node[]{\LARGE{$l$}};
\draw[line width=0.5mm] (0.4,-0.5) arc (180:90:1.4cm  and 1.4cm);
\filldraw[] (0.4,-0.5)  circle (3pt); 
\fill[white] (2.8,-0.6) rectangle (1.2,1.15);
\draw [pattern=north west lines, pattern color=black] (2.8,-0.6) rectangle (1.2,1.15);
\draw[](2,1.6) node[] {\huge{$\boldsymbol{\phi}$}};
 \end{tikzpicture}
}
\end{gathered}
=&
\begin{gathered}
\resizebox{!}{1.1cm}{
\begin{tikzpicture} 
\draw[line width=0.5mm] (0.4,-0.5)-- (3.6,-0.5) ; 
\draw[red,line width=0.5mm] (2.2,0.9) arc (90:0:1.4cm  and 1.4cm); 
\draw[red] (3.5,0.7) node[]{\LARGE{$l$}};
\draw[red,line width=0.5mm] (0.4,-0.5) arc (180:90:1.4cm  and 1.4cm);
\filldraw[red] (0.4,-0.5)  circle (3.5pt); 
\draw[red,line width=0.5mm] (1.8,0.9) -- (2.2,0.9); 
\fill[white] (2.8,-0.6) rectangle (1.2,.6);
\draw [pattern=north west lines, pattern color=black] (2.8,-0.6) rectangle (1.2,.6);
\draw[](2,1.6) node[] {\huge{$\boldsymbol{\phi}_{\rm DBA}$}};
 \end{tikzpicture}
}
\end{gathered}
+
\begin{gathered}
\resizebox{!}{1.1cm}{
\begin{tikzpicture} 
\draw[line width=0.5mm] (0.4,-0.5)-- (3.6,-0.5) ; 
\draw[line width=0.5mm] (0.4,-0.5) arc (180:90:1.4cm  and 1.4cm);
\draw[line width=0.5mm] (1.8,0.9) -- (3.6,0.9); 
\filldraw[] (0.4,-0.5)  circle (3pt); 
\fill[white] (2.8,-0.6) rectangle (1.2,1.15);
\draw [pattern=north west lines, pattern color=black] (2.8,-0.6) rectangle (1.2,1.15);
\draw[](2,1.6) node[] {\huge{$\boldsymbol{\phi}$}};
 \end{tikzpicture}
}
\end{gathered}
\hspace{-0.25cm}
\begin{gathered}
\resizebox{!}{1.1cm}{
\begin{tikzpicture} 
\draw[line width=0.5mm] (0.8,-0.5)-- (3.2,-0.5) ; 
\draw[line width=0.5mm] (0.8,0.9) arc (90:0:1.4cm  and 1.4cm); 
\fill[white] (2.8,-0.6) rectangle (1.2,.6);
\draw [pattern=north west lines, pattern color=black] (2.8,-0.6) rectangle (1.2,.6);
\draw[red,line width=0.5mm] (1.8,-0.5) arc (180:90:1.4cm  and 1.4cm);
\filldraw[red] (1.8,-0.5)  circle (3pt); 
\draw[](2,1.6) node[] {\huge{$\tilde{\mathbf{X}}$}};
 \end{tikzpicture}
}
\end{gathered}
\hspace{-0.25cm}
\begin{gathered}
\resizebox{!}{1.1cm}{
\begin{tikzpicture} 
\draw[line width=0.5mm] (0.4,-0.5)-- (3.6,-0.5) ; 
\draw[red,line width=0.5mm] (2.2,0.9) arc (90:0:1.4cm  and 1.4cm); 
\draw[red] (3.5,0.7) node[]{\LARGE{$l$}};
\draw[red,line width=0.5mm] (0.4,0.9) -- (2.2,0.9); 
\fill[white] (2.8,-0.6) rectangle (1.2,.6);
\draw [pattern=north west lines, pattern color=black] (2.8,-0.6) rectangle (1.2,.6);
\draw[](2,1.6) node[] {\huge{$\boldsymbol{\phi}_{\rm DBA}$}};
 \end{tikzpicture}
}
\end{gathered}\;,
\end{aligned}\end{equation}
where we highlighted the last fermion line which bears the current constraints of the lead $l$.
In the first term on the right-hand side, the vertex from which the last fermion line departs is the first vertex, the one explicitly appearing in the contraction. On the other hand, in the second term, this vertex is inside the dressed block with crossing of the main fermion line $\tilde{\mathbf{X}}$. Using the relation $
f^\alpha_{-\eta}(\epsilon_k)=\delta_{\eta,+1}-\eta f^\alpha_+(\epsilon_k)
$, both these vertexes (denoted with red full dots in Eq.~\eqref{Dyson_phi_symbolic}) can be split as
\begin{equation}\label{vertex_split}
{\rm v}_{-\eta^i}^{i}(\boldsymbol{\kappa})=\delta_{\eta^i,+1}{\rm v}^{i}(\boldsymbol{\kappa})-\eta^i {\rm v}_+^{i}(\boldsymbol{\kappa})\;,
\end{equation} 
where ${\rm v}^{i}(\boldsymbol{\kappa})=-|{\rm t}_{i \alpha\sigma}(\epsilon_{k})|^2 /\hbar^2$ does not contain the Fermi function.
Then, we can write the dressed crossing block by singling-out this internal vertex as follows
\begin{equation}\label{X_PC}
\tilde{\mathbf{X}}^{i'i}(\boldsymbol{\kappa}' \boldsymbol{\kappa})={\rm v}^{i'}(\boldsymbol{\kappa}')\tilde{\mathbf{x}}^{i'i}(\boldsymbol{\kappa}' \boldsymbol{\kappa})-{\rm v}_+^{i'}(\boldsymbol{\kappa}')\tilde{\mathbf{x}}_+^{i'i}(\boldsymbol{\kappa}' \boldsymbol{\kappa})\;.
\end{equation} 
Taking the difference in Eq.~\eqref{Kcurrent_PC}, the terms not containing the Fermi function, i.e. the ones with ${\rm v}^{i'}$, cancel out for proportional coupling, so that from \eqref{Kcurrent},~\eqref{Dyson_phi_contracted}, and~\eqref{X_PC} we obtain 
\begin{equation}\begin{aligned}\label{current_kernel_prop_coupling}
\hat{\mathcal{K}}^{I}_{\boldsymbol{\eta}'\boldsymbol{\eta}}(0)
= & \prod_{j=1}^N{\eta'}^j\eta^j\sum_{i'i}\delta_{{\eta'}^{i'},+1}\Big[\gamma_{i' R}
\langle{\rm c}'_L  \boldsymbol\phi \cdot\mathbf{v}_{- \eta^i}\rangle^{i'i}_{{\boldsymbol{\eta}'}^{(i')}\boldsymbol{\eta}^{(i)}}\\
&\qquad\qquad\qquad\qquad-\gamma_{i' L}
\langle{\rm c}'_R  \boldsymbol\phi \cdot\mathbf{v}_{- \eta^i}\rangle^{i'i}_{{\boldsymbol{\eta}'}^{(i')}\boldsymbol{\eta}^{(i)}}\Big]\\
= - \sum_{ i' \boldsymbol{\kappa}'}&
[\gamma_{i' R}{\rm v}_{+}^{i'L}(\boldsymbol{\kappa}')-\gamma_{i' L}{\rm v}_{+}^{i'R}(\boldsymbol{\kappa}')]\Omega^{i'}_{{\boldsymbol{\eta}'}\boldsymbol{\eta}}(\boldsymbol{\kappa}')\delta_{\zeta',-1}\;,
\end{aligned}\end{equation}
with ${\rm v}_{+}^{i l}(\boldsymbol{\kappa})={\rm v}_{+}^{i}(\boldsymbol{\kappa})\delta_{\alpha,l}$ and
\begin{equation}\begin{aligned}\label{omega}
\Omega^{i'}_{{\boldsymbol{\eta}'}\boldsymbol{\eta}}(\boldsymbol{\kappa}')&:=\prod_{j=1}^N{\eta'}^j\eta^j\delta_{{\eta'}^{i'},+1}\sum_i\Big[ \eta^i \boldsymbol\varphi^{ii}_{\rm DBA}(\boldsymbol{\kappa}')\delta_{i'i}\\
+\boldsymbol\varphi^{i'i'}_{\rm DBA}&(\boldsymbol{\kappa}')\sum_{i''\boldsymbol{\kappa}''\boldsymbol{\kappa}}\tilde{\mathbf{x}}_+^{i' i''}(\boldsymbol{\kappa}',\boldsymbol{\kappa}'')\boldsymbol\phi^{i''i}(\boldsymbol{\kappa}'',\boldsymbol{\kappa}){\rm v}^i_{-\eta^i}(\boldsymbol{\kappa})\Big]_{{\boldsymbol{\eta}'}^{(i')}\boldsymbol{\eta}^{(i)}}\;,
\end{aligned}\end{equation}
where matrix multiplication with respect to the composite sojourn indexes $\boldsymbol{\eta}^{(i)}$ is understood in the second line.\\
\indent
 Using the above definitions and Appendix~\ref{prop_coupling}, the stationary current with \emph{proportional coupling} reads $I^\infty=-2{\rm Re}{\rm Tr}_{\rm S}[\boldsymbol{\mathcal{A}}^\infty]$. 
From Eqs.~\eqref{current4} and~\eqref{current_kernel_prop_coupling}
\begin{equation}\begin{aligned}\label{I_steady_prop_coupl_general}
{\rm Tr}_{\rm S}[\boldsymbol{\mathcal{A}}^\infty]
=&\sum_{\boldsymbol{\eta}'\boldsymbol{\eta}} \hat{\mathcal{K}}^{I}_{\boldsymbol{\eta}'\boldsymbol{\eta}}(0){\rm P}^\infty_{\boldsymbol{\eta}}\\
=&-\sum_i\frac{\gamma_{i L}\gamma_{i R}}{2\pi\hbar^2}\int d\epsilon[f^{L}_+(\epsilon)-f^{R}_+(\epsilon)]\Gamma_{ii}(\epsilon)\\
&\times \sum_{\zeta}
\sum_{{\boldsymbol{\eta}}'\boldsymbol{\eta}}\Omega^{i}_{\boldsymbol{\eta}'\boldsymbol{\eta}}(\zeta,\epsilon){\rm P}^\infty_{\boldsymbol{\eta}}\delta_{\zeta,-1}\;,
\end{aligned}\end{equation}
where we used the property $\gamma_{i L}
{\Gamma}_{ii R}(\epsilon)=\gamma_{i R}{\Gamma}_{ii L}(\epsilon)=\gamma_{i L}\gamma_{i R}{\Gamma}_{ii}(\epsilon)$. 
On the other hand, from Eqs.~\eqref{IPropCoupling} and~\eqref{IPropCouplingLandauer2}
\begin{equation}\begin{aligned}\label{}
{\rm Tr}_{\rm S}[\boldsymbol{\mathcal{A}}^\infty]
=&
\sum_i\frac{-{\rm i}\gamma_{i L}\gamma_{i R}}{2\pi\hbar} \int d\epsilon \left[f_+^L(\epsilon)-f_+^R(\epsilon)\right]\Gamma_{ii}(\epsilon)\mathcal{G}^a_{ii}(\epsilon)\;,
\end{aligned}\end{equation}
where $\mathcal{G}^a$ is the advanced Green's function of the central system. 
Note that the blip index $\zeta$ always multiplies the imaginary unit, as can be seen in the definition of the correlation functions ${\rm g}^\zeta$ which enter the diagrammatic contributions to the propagator, see Eqs~\eqref{0th}-\eqref{3rdc}, along with the phase factors ${\rm b}_{kl}$ defined Eq.~\eqref{Bm}. This means that, in the time domain, $\zeta$ establishes the sign of the time variable and the real part can be obtained by removing the constraint $\delta_{\zeta,-1}$ and summing over $\zeta$. We can thus make the following identification with the retarded/advanced Green's functions 
\begin{equation}\begin{aligned}\label{GF_general}
\mathcal{G}^{(\zeta)}_{i i}(\epsilon)
=- &\frac{{\rm i}\zeta}{\hbar}\sum_{{\boldsymbol{\eta}}'\boldsymbol{\eta}}\Omega^{i}_{\boldsymbol{\eta}'\boldsymbol{\eta}}(\zeta,\epsilon){\rm P}^\infty_{\boldsymbol{\eta}}\;,
\end{aligned}\end{equation}
where $\zeta=+1$ ($-1$) gives the retarded (advanced) Green's function.  
 The exact steady state current acquires, in the continuum limit, the form of the Meir-Wingreen formula~\cite{Meir1992}
\begin{equation}\begin{aligned}\label{I_MW}
I^\infty
=&\frac{e}{\hbar}\sum_i\int d \epsilon\left[\frac{\Gamma_L(\epsilon)\Gamma_R(\epsilon)}{\Gamma_L(\epsilon)+\Gamma_R(\epsilon)}\right]_{ii}[f^{L}_+(\epsilon)-f^{R}_+(\epsilon)]\\
&\times\left[-\frac{1}{\pi}{\rm Im}\; \mathcal{G}^r_{i i}(\epsilon)\right]\;,
\end{aligned}\end{equation}
where $\Gamma_{ii \alpha}(\epsilon)=2\pi\sum_\sigma\varrho_{\alpha \sigma} (\epsilon) |{\rm t}_{i \alpha \sigma}(\epsilon)|^2 $ and where we used ${\rm Re}[-{\rm i} \mathcal{G}^a_{i i}(\epsilon)]={\rm Im} \mathcal{G}^a_{i i}(\epsilon)=-{\rm Im} \mathcal{G}^r_{i i}(\epsilon)$. The function $-(1/\pi){\rm Im}\; \mathcal{G}^r_{i i}(\epsilon)$ is the system's density of states in the presence of tunnel-coupling to the leads.\\
\indent A quantity used to characterize the transport properties of the setup in a nonequilibrium setting, namely in the presence of a voltage bias $eV:=\mu_L-\mu_R$, is the differential conductance $\partial I/\partial V$. 
At equilibrium, $\mu_L=\mu_R=\mu$, the behavior of the transport setup is described by the linear conductance $G$, defined as the limiting value 
of the differential conductance for vanishing bias. 
Setting $\mu_L=\mu+eV/2$ and $\mu_R=\mu-eV/2$, and using $\partial f_+^L/\partial V=-(e/2)\partial f_+^L/\partial \epsilon$ and $\partial f_+^R/\partial V=\;(e/2)\partial f_+^R/\partial \epsilon$, the linear conductance assumes the form~\cite{Wingreen1994}
\begin{equation}\begin{aligned}\label{G}
G
=&-\pi G_0\sum_i\int d \epsilon\left[\frac{\Gamma_L(\epsilon)\Gamma_R(\epsilon)}{\Gamma_L(\epsilon)+\Gamma_R(\epsilon)}\right]_{ii}\frac{\partial f_+(\epsilon)}{\partial \epsilon}\\
&\times\left[-\frac{1}{\pi}{\rm Im}\; \mathcal{G}^r_{i i}(\epsilon)\right]\;,
\end{aligned}\end{equation}
where, due to the vanishing bias, $ f_+^L(\epsilon)=f_+^R(\epsilon)= f_+(\epsilon)$. Here, $G_0:=2e^2/h$ is the conductance quantum.
 At $T=0$, the derivative of the Fermi function is $-\delta(\epsilon-\mu)$ so that
\begin{equation}\begin{aligned}\label{GT0}
G_{T=0}=\pi G_0\sum_i\left[\frac{\Gamma_L(\mu)\Gamma_R(\mu)}{\Gamma_L(\mu)+\Gamma_R(\mu)}\right]_{ii}\left[-\frac{1}{\pi}{\rm Im}\; \mathcal{G}^r_{i i}(\mu)\right]\;.
\end{aligned}\end{equation}
\indent In the following, we shall apply the general formalism developed here to two archetypal models, the resonant level model and the SIAM.

\section{Resonant level model}
\label{RLM}

\begin{figure}[ht]
\begin{center}
\includegraphics[width=4.5cm,angle=0]{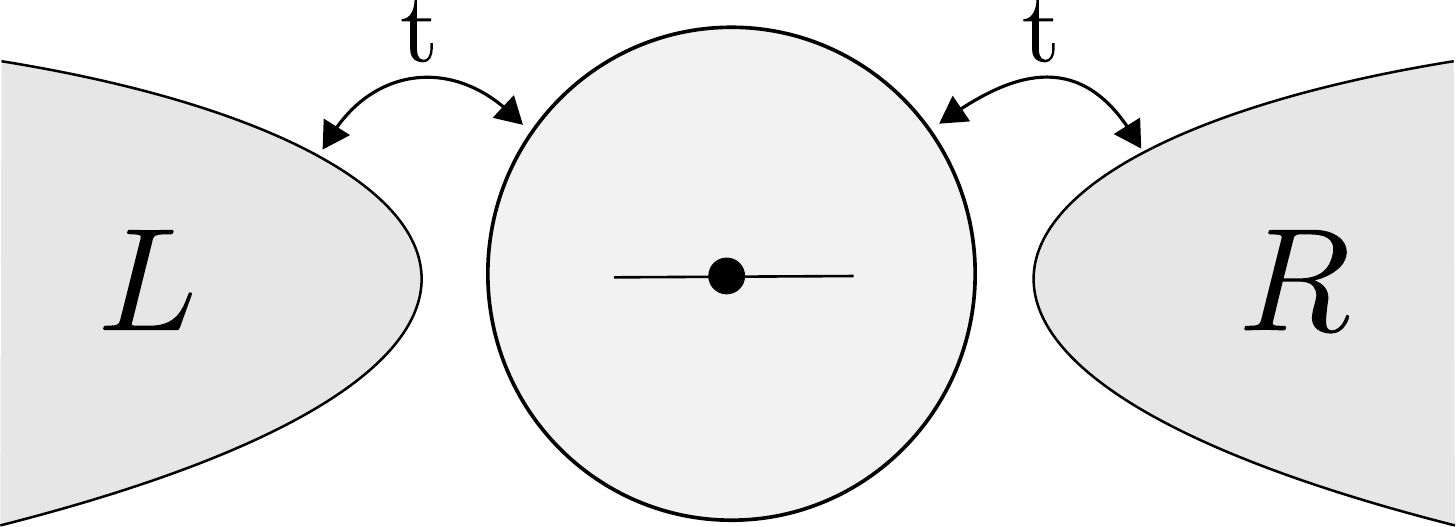}
\caption{\small{Scheme of a spinless level of energy $\epsilon_0$ tunnel-coupled to two noninteracting leads.}}
\label{scheme_spinless}
\end{center}
\end{figure}

Up to here, we have considered a general interacting central system connected to a number of noninteracting leads, with possibly energy- and state-dependent tunnel coupling. The only constraint has been given on the correlation matrices in the form of Eq.~\eqref{corr_matrices_diagonal}. In Sec.~\ref{prop_coupl_and_GF}, we have specialized the discussion to the case of two leads and proportional coupling.\\
\indent To exemplify the construction carried out so far, we consider in this section the resonant level model (RLM). This model describes a single, spinless level of energy $\epsilon_0$ coupled to two noninteracting leads as shown in Fig.~\ref{scheme_spinless}. 

\begin{figure}[ht]
\begin{center}
\resizebox{!}{2cm}{
\begin{tikzpicture}[] 
\draw[line width=0.5mm] (0.5,1.3)node[left] {\LARGE{$\boldsymbol\kappa$}}  arc (120:60:2.5cm  and 2.5cm)node[right] {\LARGE{$\boldsymbol\kappa$}};
\draw[dashed,line width=0.5mm] (0.3,-0.5)  -- (3.2,-0.5); 
\draw [green,line width=0.5mm] (2.8,-0.8) rectangle (0.8,1.9);
\draw[](1.8,2.4) node[] {\LARGE{$\mathbf{h}$}};
\end{tikzpicture}
}
\resizebox{!}{2cm}{
\begin{tikzpicture}[] 
\draw[line width=0.5mm] (0.5,1.3)node[left] {\LARGE{$\boldsymbol\kappa$}}  arc (120:60:2.5cm  and 2.5cm)node[right] {\LARGE{$\boldsymbol\kappa$}};
\draw[dashed,line width=0.5mm] (0.3,-0.5)-- (3.2,-0.5); 
\draw[line width=0.5mm] (1,-0.5)--  node[above] {\LARGE{$\eta$}} (2.6,-0.5); 
\filldraw 
(1,-0.5) circle (3pt);
\draw[line width=0.5mm] (1,-0.5) arc (180:0:0.8cm  and 1.cm);
\draw[]  (2,.5)  node[above] {\LARGE{$\boldsymbol\kappa_{1}$}}  ;
\draw [green,line width=0.5mm] (2.8,-0.8) rectangle (0.8,1.9);
\draw[](1.8,2.4) node[] {\LARGE{$\mathbf{B}$}};
\end{tikzpicture}
}
\resizebox{!}{2cm}{
\begin{tikzpicture}[] 
\draw[line width=0.5mm] (0.5,1.3)node[left] {\LARGE{$\boldsymbol\kappa$}}  arc (100:0:1.8cm  and 1.8cm);
\draw[line width=0.5mm]
(1,-0.5) arc (0:100:-1.8cm  and 1.8cm)node[right] {\LARGE{$\boldsymbol\kappa'$}} ;
\draw[dashed,line width=0.5mm] (0.3,-0.5)  -- node[above] {\LARGE{$\eta$}} (3.2,-0.5); 
\draw[line width=0.5mm] (1,-0.5)  --  (2.6,-0.5); 
\filldraw 
(1,-0.5) circle (3pt);
\draw [green,line width=0.5mm] (2.8,-0.8) rectangle (0.8,1.9);
\draw[](1.8,2.4) node[] {\LARGE{$\mathbf{X}$}};
\end{tikzpicture}
}
\caption{\small{Blocks involved in the propagator of the resonant level model. Full dots represent the vertexes defined in Eqs.~\eqref{h_v} and~\eqref{vertexes}. Dashed and solid lines at the bottom denote blip and sojourn states, respectively. The sojourn index $\eta$ assumes the values $+1$ and $-1$ for occupied and empty dot, respectively. These \emph{internal} sojourns are summed over, the sum being included in the definitions of the blocks. }}
\label{scheme_blocks_spinless}
\end{center}
\end{figure}
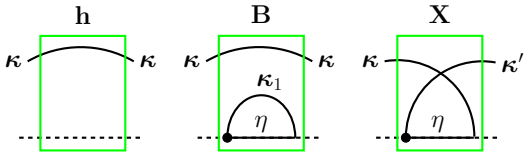

The model Hamiltonian is
\begin{equation}
\begin{aligned}
\label{H_RLM}
H=&\;\epsilon_0 \;a^\dag a
+\sum_{\alpha k \sigma}\epsilon_{\alpha k}c^{\dag}_{\alpha k \sigma}c_{\alpha k \sigma}\\
&+\sum_{\alpha k \sigma}\big[ {\rm t}_{\alpha k} a^{\dag}c_{\alpha k \sigma} + {\rm t}^{*}_{\alpha k} c^{\dag}_{\alpha k \sigma}a\big]\;.
\end{aligned}
\end{equation}
Due to the lack of interactions in the dot, this model admits an exact solution and has been analyzed with a variety of methods in Ref.~\cite{Reimer2019}, see also~\cite{Ferguson2020}. According to the diagrammatic rules set up in Sec.~\ref{diagrammatic_rules}, the contributing diagrams for a noninteracting system can have at most two overlapping fermion lines for the same electron state. Since the dot is equipped with a single state, the RTA, discussed in Sec.~\ref{diagrammatic_unravelling}, provides the exact description of the resonant level. Hence, the exact equation~\eqref{Dyson_phi_a} reduces to the RTA, where the crossing block ${\rm X}$ is not dressed by internal processes and the dressed propagator $\phi_{\rm B}$ is given by the bare propagator $\rm h$ dressed by the bubbles ${\rm B}$, see Fig.~\ref{scheme_blocks_spinless}. 
As a result, the exact Dyson equation~\eqref{Dyson_phi_component_a} specializes to     
\begin{equation}\begin{aligned}\label{Dyson_phi_RLM}
\phi(\kappa',\kappa)=&\varphi_{\rm NCA2}(\kappa)\delta_{\kappa'\kappa}
+\varphi_{\rm NCA2}(\kappa')\sum_{\kappa''}{\rm X}(\kappa',\kappa'')\phi(\kappa'',\kappa)\;,
\end{aligned}\end{equation}
where the matrix structure in the sojourn index $\boldsymbol{\eta}^{(i)}$ is lost due the fact that there is a single electron state so that $\boldsymbol\chi\equiv \boldsymbol\kappa$ with 
$$
 \boldsymbol\kappa=(\zeta,\alpha,k)\;.$$ 
The dressed propagator $\phi_{\rm NCA2}(\kappa',\kappa)=\varphi_{\rm NCA2}(\kappa')\delta_{\kappa'\kappa}$ is in turn given by
\begin{equation}\begin{aligned}\label{varphiB_RLM}
\boldsymbol{\phi}_{\rm NCA2}=&\left[\mathbf{h}^{-1}-\mathbf{B}\right]^{-1}
\;,
\end{aligned}\end{equation}
according to Eq.~\eqref{dressed_line}.\\
\indent In Laplace space the bare propagator $\mathbf{h}(\lambda=0^+)$ reads
\begin{equation}\label{h_RLM}
[\mathbf{h}]_{\boldsymbol\kappa'\boldsymbol\kappa}={\rm i}\hbar \frac{1}{\zeta(\epsilon_k-\epsilon_0)+{\rm i} 0^+}\delta_{\boldsymbol\kappa'\boldsymbol\kappa}\;,
\end{equation}
see Eq.~\eqref{h_v}.\\
\indent The bubble block, the central diagram in Fig.~\ref{scheme_blocks_spinless}, is the contraction with a vertex of the internal fermion line (indexed with $1$) of the
free propagator with two overlapping fermion lines 
\begin{equation}
\begin{aligned}
\label{h2_RLM}
[\mathbf{h}_{2}]_{\boldsymbol\kappa'\boldsymbol\kappa}=
\begin{gathered}
\resizebox{!}{1cm}{
\begin{tikzpicture}[] 
\draw[line width=0.7mm] (1,1.5) 
arc (120:60:2cm  and 2.1cm)  node[right] {\LARGE{$\boldsymbol\kappa$}};
\draw[line width=0.7mm] (1,0.8) 
arc (120:60:2cm  and 2.8cm) node[right] {\LARGE{$\boldsymbol\kappa_{1}$}} ; 
\draw[line width=0.7mm] (1,0) -- (3,0)node[right] {\LARGE{$\eta$}} ; 
 \end{tikzpicture}
 }
\end{gathered}
={\rm i} \hbar\frac{\delta_{\zeta_{1},-\zeta}}{\zeta(\epsilon_k-\epsilon_{k_{1}})+{\rm i}  0^+}\delta_{\boldsymbol\kappa'\boldsymbol\kappa}
\;.
\end{aligned}
\end{equation}
Here, the index $\zeta_{1}$ is constrained to be equal to $-\zeta$ by the diagrammatic rules, see Eq.~\eqref{2ndb} where the upper indexes of the correlation functions ${\rm g}^\zeta_\eta$ have opposite signs.
Note that the block ${\rm h}_2$ bears no dependence on the dot energy as the system is in a sojourn state, denoted with $\eta$, contrary to the block ${\rm h}$ in Eq.~\eqref{h_RLM} where the system is in a blip state (dashed line, see the left diagram in Fig.~\ref{scheme_blocks_spinless}).  
 Including the sum over the internal sojourn $\eta$ due to the sum-over-paths, the bubble block is evaluated as follows 
\begin{equation}
\begin{aligned}
\label{bubble_RLM}
[\mathbf{B}]_{\boldsymbol\kappa'\boldsymbol\kappa}=&\begin{gathered}
\resizebox{!}{1cm}{
\begin{tikzpicture}[] 
\draw[line width=0.7mm] (0.5,1.5) arc (120:60:3cm  and 2.1cm) node[right] {\LARGE{$\boldsymbol\kappa$}} ;
\draw[line width=0.7mm] (1,0) arc (180:0:1cm  and 1.2cm); 
\draw[]  (2,1.1)  node[above] {\LARGE{$\boldsymbol\kappa_{1}$}}  ;
\draw[dashed,line width=0.8mm]  (0.5,0) -- (1,0); 
\draw[line width=0.7mm] (1,0) -- node[above] {\LARGE{$\eta$}} (3,0); 
\draw[dashed,line width=0.7mm] (3,0) -- (3.5,0); 
\filldraw 
(1,0) circle (5pt);
 \end{tikzpicture}
 }
\end{gathered}
=\sum_{\eta}\langle {\rm h}_{2}{\rm v}_\eta\rangle\\
=&{\rm i}\zeta\hbar\sum_{\boldsymbol\kappa_{1}}\frac{\sum_{\eta}{\rm v}_{\eta}(\boldsymbol{\kappa}_1)\delta_{\zeta_{1},-\zeta}}{\epsilon_k-\epsilon_{k_{1}}+{\rm i}\zeta 0^+}\delta_{\boldsymbol\kappa'\boldsymbol\kappa}\\
=&-\frac{{\rm i}\zeta}{\hbar}\sum_{\alpha_{1},k_{1}}\frac{|{\rm t}_{\alpha_{1}}(\epsilon_{k_1})|^2 }{\epsilon_k-\epsilon_{k_{1}}+{\rm i}\zeta 0^+}\delta_{\boldsymbol\kappa'\boldsymbol\kappa}\;,
\end{aligned}
\end{equation}
where we used $\sum_\eta f_\eta(\epsilon_k)=1$ in the vertex
\begin{equation}\label{v_RLM}
{\rm v}_{\eta}(\boldsymbol{\kappa}):=-\frac{|{\rm t}_{\alpha}(\epsilon_k)|^2}{\hbar^2} f^\alpha_{\eta}(\epsilon_k)\;.
\end{equation}
In the wide-band limit (WBL), i.e. for energy independent tunneling amplitudes, using the result of Eq.~\eqref{I0} we obtain
\begin{equation}
\begin{aligned}
\label{B+bare}
[\mathbf{B}]_{\boldsymbol\kappa'\boldsymbol\kappa}=&\zeta\frac{\rm i}{\hbar}\sum_{\alpha_{1}}\varrho_{\alpha_{1}}|{\rm t}_{\alpha_{1}}|^2 \int d\epsilon_{1}\; \frac{\delta_{\boldsymbol\kappa'\boldsymbol\kappa}}{\epsilon_{1}-\epsilon_k-{\rm i}\zeta 0^+}\\
=&-\frac{1}{\hbar}\pi\sum_{\alpha_{1}}\varrho_{\alpha_{1}}|{\rm t}_{\alpha_{1}}|^2\delta_{\boldsymbol\kappa'\boldsymbol\kappa}\\
=&-\frac{\Gamma}{2\hbar}\delta_{\boldsymbol\kappa'\boldsymbol\kappa}\;,
\end{aligned}
\end{equation}
where $\Gamma=2\pi\sum_\alpha\varrho_\alpha |{\rm t}_\alpha|^2$. As a result, from Eq.~\eqref{varphiB_RLM}, 
\begin{equation}\label{phiB_RLM}
\varphi_{\rm NCA2}(\zeta,k)={\rm i}\zeta\hbar \frac{1}{\epsilon_k-\epsilon_0+{\rm i}\zeta \Gamma/2}\;.
\end{equation}
The block $\mathbf{X}$ is given by attaching the vertex ${\rm v}_\eta$  to the outgoing fermion line (with index $\boldsymbol\kappa'$) in the propagator 
\begin{equation}
\begin{aligned}
\label{h2X_RLM}
[\mathbf{h}^{\rm X}_{2}]_{\boldsymbol\kappa'\boldsymbol\kappa}=&{\rm i} \hbar\frac{\delta_{\zeta',-\zeta}}{\zeta(\epsilon_k-\epsilon_{k'})+{\rm i} 0^+}\;,
\end{aligned}
\end{equation}
resulting in
\begin{equation}
\begin{aligned}
\label{X_RLM}
[\mathbf{X}]_{\boldsymbol\kappa'\boldsymbol\kappa}=&\begin{gathered}
\resizebox{!}{0.7cm}{
\begin{tikzpicture}[] 
\draw[line width=0.7mm] (0.5,1.5)node[left] {\LARGE{$\boldsymbol\kappa$}} arc (100:2:2cm  and 1.6cm) ;
\draw[line width=0.7mm] (1,0) arc (2:100:-2cm  and 1.6cm)node[right] {\LARGE{$\boldsymbol\kappa'$}}; 
\draw[dashed,line width=0.8mm]  (0.5,0) -- (1,0); 
\draw[line width=0.7mm] (1,0) -- node[above] {\LARGE{$\eta$}} (2.8,0); 
\draw[dashed,line width=0.7mm] (2.8,0) -- (3.3,0); 
\filldraw 
(1,0) circle (5pt);
 \end{tikzpicture}
 }
\end{gathered}\\
=&\frac{{\rm i}\zeta\hbar\;\sum_{\eta}{\rm v}_{\eta}(\boldsymbol{\kappa}')}{\epsilon_k-\epsilon_{k'} + {\rm i}\zeta0^+}\delta_{\zeta',-\zeta}\\
=&-\frac{{\rm i}\zeta}{\hbar}\frac{\;|{\rm t}_{\alpha'}(\epsilon_{k'})|^2}{\epsilon_k-\epsilon_{k'} + {\rm i}\zeta0^+}\delta_{\zeta',-\zeta}
\;.
\end{aligned}
\end{equation}
Here, as for $\mathbf{B}$, the internal sojourn index $\eta$ is summed over due to the sum-over-path with the result that $\mathbf{X}$ does not contain the Fermi function.\\
\indent The retarded/advanced Green's function are given by Eq.~\eqref{GF_general} via the function $\Omega_{\eta'\eta}$ defined in Eq.~\eqref{omega} which, in the RLM, adapts to
\begin{equation}\begin{aligned}\label{omega_RLM}
\Omega_{{\eta}'\eta}(\boldsymbol{\kappa}')=&\;\eta'\eta\delta_{\eta',+1}\Big[ \eta \varphi_{\rm NCA2}(\boldsymbol{\kappa}')\\
+&\;\varphi_{\rm NCA2}(\boldsymbol{\kappa}')\sum_{\boldsymbol{\kappa}\boldsymbol{\kappa}''}{\rm x}_+(\boldsymbol{\kappa}',\boldsymbol{\kappa})\phi(\boldsymbol{\kappa},\boldsymbol{\kappa}''){\rm v}_{-\eta}(\boldsymbol{\kappa}'')\Big]\;.
\end{aligned}\end{equation}
The block ${\rm x}_+$, see Eqs.~\eqref{vertex_split}-\eqref{X_PC} is easy to evaluate and from Eq.~\eqref{X_RLM} reads
\[
{\rm x}_+=\frac{{\rm i}\zeta\hbar\;\sum_{\eta}\eta}{\epsilon_k-\epsilon_{k'} + {\rm i}\zeta0^+}\delta_{\zeta',-\zeta}=0\;.
\]
Thus,
\[
\Omega_{{\eta}'\eta}(\zeta,k)=\delta_{\eta',+1}\varphi_{\rm NCA2}(\zeta,k)\;,
\]
and the resulting expression for the Green's function is
\begin{equation}\begin{aligned}\label{GF_RLM}
\mathcal{G}^{(\zeta)}(\epsilon_k)
=&-\frac{{\rm i}\zeta}{\hbar}\sum_{\eta'\eta}\Omega_{\eta'\eta}(\zeta,k){\rm P}^\infty_{\eta}\\
=&-\frac{{\rm i}\zeta}{\hbar}\varphi_{\rm NCA2}(\zeta,k)\sum_{\eta}{\rm P}^\infty_{\eta}\\
=&\frac{1}{\epsilon_k-\epsilon_0+{\rm i}\zeta \Gamma/2}\;,
\end{aligned}\end{equation}
where we used Eq.~\eqref{phiB_RLM}. Thus, as expected, the single-particle Green's function  acquires a broadening $\Gamma/2$ due to the coupling of the resonant level to the leads.\\
\indent Substituting $\mathcal{G}^{(\zeta=+1)}(\epsilon)$ in Eq.~\eqref{I_MW}, the  current for proportional coupling, $\gamma_L\Gamma_R=\gamma_R\Gamma_L$ (with $\gamma_L+\gamma_R=1$),  in the WBL reads~\footnote{This integral can be solved by noting that the denominator in the integral splits as $\left\{[(\epsilon-\epsilon_0)-{\rm i}\Gamma/2]^{-1}-[(\epsilon-\epsilon_0)+{\rm i}\Gamma/2]^{-1}\right\}/{\rm i}\Gamma$ and by applying Eq.~\eqref{Isummary}.} 
\begin{equation}\begin{aligned}\label{I_MW_spinless}
I^\infty
=&\frac{e}{h}\int d \epsilon\; [f^{L}_+(\epsilon)-f^{R}_+(\epsilon)]
\frac{\Gamma_L\Gamma_R}{(\epsilon-\epsilon_0)^2+\Gamma^2/4}\\
=&\;\frac{e}{h}\frac{\Gamma_L\Gamma_R}{\Gamma}r(\epsilon_0)\;,
\end{aligned}\end{equation}
where $\Gamma=\Gamma_L+\Gamma_R$ and where we defined
\begin{equation}\begin{aligned}\label{r}
r(x)=&{\rm Im}\psi\left(\frac{1}{2}+{\rm i}\frac{x-\mu_L-{\rm i}\frac{\Gamma}{2}}{2\pi k_{\rm B}T}\right)\\
&- {\rm Im}\psi\left(\frac{1}{2}+{\rm i}\frac{x-\mu_R-{\rm i}\frac{\Gamma}{2}}{2\pi k_{\rm B}T}\right) \;,
\end{aligned}\end{equation} 
with $\Psi(x)$ the digamma function. Equation~\eqref{I_MW_spinless} provides the exact, finite-temperature expression for the current in the RLM.\\
\indent Summarizing, while the RTA gives the exact irreducible propagator $\phi$ and consequently  the exact density matrix and current for the RLM, the crossing diagrams do not contribute to the retarded/advanced Green's functions (and therefore to the current for proportional coupling).
Note that, as the dot can host at most one electron, the system is necessarily noninteracting and consequently inelastic processes are absent. This gives the Landauer formula Eq.~\eqref{I_MW_spinless}, where the temperature  dependence is exclusively in the Fermi functions. Using Eq.~\eqref{G} for the conductance, with the imaginary part of Eq.~\eqref{GF_RLM}, we readily obtain the analytical expression for the conductance at $T=0$, where $-\partial f_+/\partial \epsilon=\delta(\epsilon-\mu)$, which is of the Breit-Wigner form~\cite{Haug2008, Scheer2010}
\begin{equation}\label{G_RLM}
G=\frac{2e^2}{h}\frac{\Gamma_L\Gamma_R/2}{(\mu-\epsilon_0)^2+\Gamma^2/4}\;.
\end{equation}
For $\Gamma_\alpha=\Gamma/2$ the conductance saturates to half of the quantum of conductance $G_0=2e^2/h$ at resonance, i.e. for $\mu=\epsilon_0$. In the noninteracting \emph{spinful} model, where the dot can host two electrons with opposite spin,  the sum over $\sigma$ yields the maximum  
$G_{\rm max}=G_0$.

\section{Single-impurity Anderson model (SIAM)}
\label{SIAM}

\begin{figure}[ht]
\begin{center}
\includegraphics[width=4.7cm,angle=0]{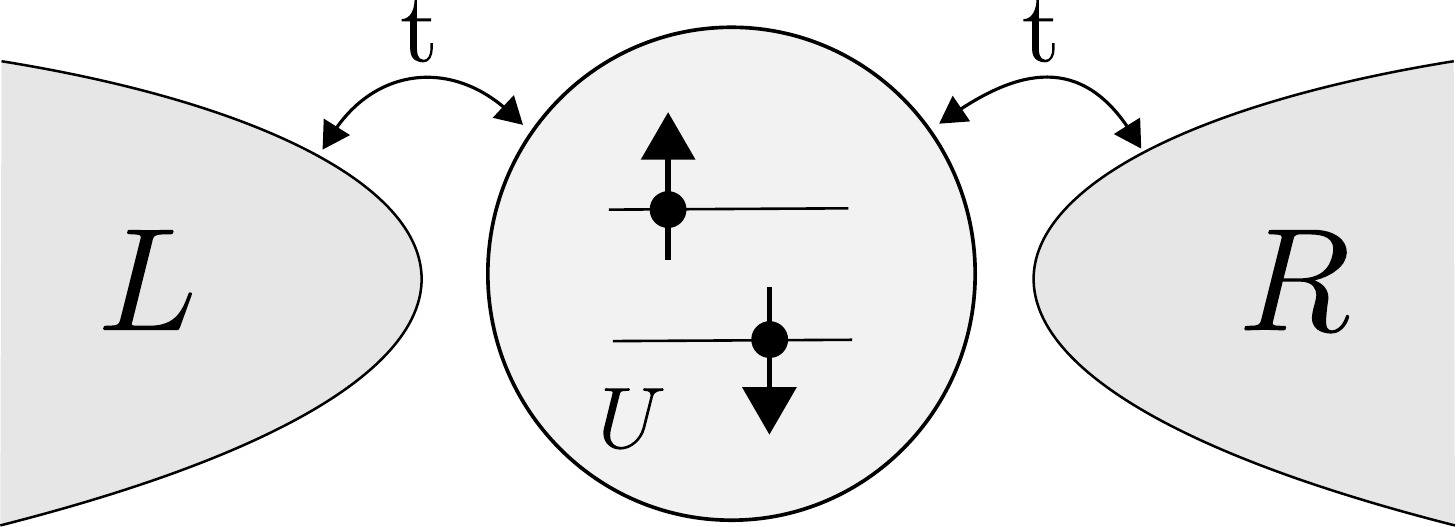}
\caption{\small{Single-impurity Anderson model realized by a quantum dot tunnel-coupled to  two noninteracting leads.}}
\label{scheme_SIAM}
\end{center}
\end{figure}
We now specialize the discussion to the simplest, yet highly nontrivial, instance of the general model of interacting nanostructure described by Eq.~\eqref{H_general}, the single-impurity Anderson model (SIAM). In the SIAM, the central system is a quantum dot that can host at most two electrons with opposite spin states,  the latter being denoted by $\sigma=\;\uparrow,\downarrow$. The dot is connected to two leads, $L$ and $R$,  via a tunnel coupling which we assume here to be spin-independent. A scheme of the model is provided in Fig.~\ref{scheme_SIAM}. The resulting four many-body dot states are given by $|0\rangle$, $|\uparrow\rangle$, $|\downarrow\rangle$, and $|2\rangle$, or, in terms of the sojourn indexes $(\eta^\uparrow,\eta^\downarrow)$, by $|-1-1\rangle$, $|+1-1\rangle$, $|-1+1\rangle$, and $|+1+1\rangle$, respectively. 
 The difference in chemical potentials produces at long times a stationary current which is essentially determined by three energy scales: The interaction energy $U$ between the electrons in the dot, the tunnel coupling $\Gamma$, and the thermal energy $k_{\rm B}T$.
Despite its simplicity, this model already displays a variety of interesting physical effects and transport regimes  arising from the interplay of these energy scales, see Fig.~\ref{scheme_approximations} below. For example, the Kondo effect~\cite{Hewson1993,vanderWiel2000, Kouwenhoven2001}, an exquisitely nonperturbative many-body phenomenon due to the correlation between electrons in the dot and the leads, becomes relevant for temperatures around and below a certain value $T_{\rm K}$ which depends exclusively on $\Gamma$, $U$, and the single-particle energies $\epsilon_\sigma$.\\
\indent The full Hamiltonian of the setup reads
\begin{equation}
\begin{aligned}
\label{H_SIAM}
H=&\sum_\sigma\epsilon_\sigma \hat{n}_\sigma+U \hat{n}_\uparrow \hat{n}_\downarrow
+\sum_{\alpha k \sigma}\epsilon_{\alpha k}c^{\dag}_{\alpha k \sigma}c_{\alpha k \sigma}\\
&+\sum_{\alpha k \sigma}\big[ {\rm t}_{\alpha k} a_{\sigma}^{\dag}c_{\alpha k \sigma} + {\rm t}^{*}_{\alpha k} c^{\dag}_{\alpha k \sigma}a_{\sigma}\big]\;,
\end{aligned}
\end{equation}
where $\alpha=L,R$, and $\hat{n}_\sigma= a_\sigma^{\dag}a_\sigma$. The single-particle energies $\epsilon_\sigma$ are possibly split by an externally applied magnetic field, $\epsilon_\uparrow-\epsilon_\downarrow=\Delta_B$. Note that spin-independent coupling constants $\rm t_{\alpha k}$ (and non-magnetic leads) imply that the correlation matrices $\mathbf{g}_{\pm}$, Eq.~\eqref{corr_matrices}, are not only diagonal, but also proportional to the identity, namely $[\mathbf{g}_{\pm}(t)]_{\sigma'\sigma}= \delta_{\sigma'\sigma}\; {\rm g}_{\pm}(t)$ with
\begin{equation}
\label{corr_matrices_SIAM}
 {\rm g}_{\pm}(t)=\frac{1}{\hbar^2}\sum_{\alpha k} |{\rm t}_{\alpha k}|^2f_{\pm}^\alpha(\epsilon_{k})e^{-\frac{\rm i}{\hbar}\epsilon_{\alpha k}t}\;.
\end{equation}
In what follows we use the compact notation of Eq.~\eqref{symbols} which, in the case considered here, adapts to 
\begin{equation}
\begin{aligned}\label{symbols_SIAM}
{\rm g}_{+1}^{+1}&={\rm g}_+,\quad {\rm g}_{+1}^{-1}={\rm g}_+^*,\quad  {\rm g}_{-1}^{+1}={\rm g}_-,\quad {\rm g}_{-1}^{-1}={\rm g}_-^*\;.
\end{aligned}\end{equation}
In the WBL, the tunnel coupling is quantified by $
\Gamma:=2\pi \sum_{\alpha}\varrho_\alpha |{\rm t}_{\alpha}|^2$. 
The collective index $\boldsymbol\chi$, defined for the general case in Eq.~\eqref{chi_index}, specializes in the SIAM to 
$$
\boldsymbol\chi:=(\boldsymbol\kappa,\eta^{\bar\sigma}),\quad{\rm 
where}\quad \boldsymbol\kappa=(\zeta,\alpha,k)
$$
and where $\bar{\sigma}$ denotes the opposite spin with respect to $\sigma$.
In order to simplify the notation, from here on we make the identifications 
$$
\eta^{\sigma}\equiv\nu\quad{\rm and}\quad \eta^{\bar\sigma}\equiv\eta\;,
$$
with $\bar\nu\equiv -\nu$ and $\bar\eta\equiv -\eta$, so that $\boldsymbol\chi:=(\boldsymbol\kappa,\eta)$.
The system energies $E_i$ in the phase factors of Eq.~\eqref{Bm} 
are in this case the dot energies shown in Fig.~\ref{scheme_interaction}. For the fermion lines associated to the state $\sigma$ these energies read
\begin{equation}\begin{aligned}\label{energies_SIAM}
&E_\sigma=\epsilon_\sigma+U/2\quad\qquad\qquad\qquad \bar\sigma\text{ in a blip state}\\
&E_\sigma(\eta)=\epsilon_{\sigma}+(1+\eta)U/2\;\;\qquad\bar\sigma\text{ in the sojourn $\eta$}\;,
\end{aligned}\end{equation}
with similar expressions for $E_{\bar\sigma}$ and $E_{\bar\sigma}(\nu)$.\\
\indent The retarded ($\zeta=+1$) and advanced ($\zeta=-1$) dot Green's functions, Eq.~\eqref{GF_general}, adopt in the SIAM the following form 
\begin{equation}\begin{aligned}\label{GF_SIAM}
\mathcal{G}^{(\zeta)}_{\sigma\sigma}(\epsilon_k)
=- &\frac{{\rm i}\zeta}{\hbar}\sum_{{\boldsymbol{\eta}}' \boldsymbol{\eta}}\Omega^\sigma_{\boldsymbol{\eta}'\boldsymbol{\eta}}(\zeta,k){\rm P}^\infty_{\boldsymbol{\eta}}\;,
\end{aligned}\end{equation}
with
\begin{equation}\begin{aligned}\label{omega_SIAM}
\Omega^\sigma_{{\boldsymbol{\eta}'}\boldsymbol{\eta}}&(\boldsymbol{\kappa})
=\eta'\eta\delta_{\nu' ,+1}\sum_{s=\uparrow,\downarrow}\Big[ \boldsymbol\varphi_{\rm B}^{\sigma}(\boldsymbol{\kappa})\delta_{\sigma s}+\nu\boldsymbol\varphi_{\rm B}^{\sigma}(\boldsymbol{\kappa})\\
&\times\sum_{\sigma''\boldsymbol{\kappa}''\boldsymbol{\kappa}'}\tilde{\mathbf{x}}_+^{\sigma \sigma''}(\boldsymbol{\kappa} ,\boldsymbol{\kappa}'')\boldsymbol\phi^{\sigma''s}(\boldsymbol{\kappa}'',\boldsymbol{\kappa}'){\rm v}_{-\eta^{s}}(\boldsymbol{\kappa}')\Big]_{\eta' \eta^{\bar{s}}}\;,
\end{aligned}\end{equation}
where we used $(\nu)^2=1$. 
Here, the blocks in parenthesis have a $2\times 2$ matrix structure induced by the sojourn indexes $\eta'\eta $, and the vertex is given by
\begin{equation}\label{v_SIAM}
{\rm v}_{-\eta}(\boldsymbol{\kappa}):=-\frac{|{\rm t}_{\alpha}(\epsilon_k)|^2}{\hbar^2} f^\alpha_{-\eta}(\epsilon_k)\;.
\end{equation}
The Meir-Wingreen formula, which gives the current for a general system in the case of proportional coupling with the leads, Eq.~\eqref{I_MW}, adapts for the SIAM to
\begin{equation}\begin{aligned}\label{I_MW_SIAM}
I^\infty
=&\frac{e}{\hbar}\frac{\Gamma_L\Gamma_R}{\Gamma}\sum_\sigma\int d \epsilon\; [f^{L}_+(\epsilon)-f^{R}_+(\epsilon)]\left[-\frac{1}{\pi}{\rm Im} \mathcal{G}^r_{\sigma\sigma}(\epsilon)\right]\;.
\end{aligned}\end{equation}
The asymptotic population of the spin states $\sigma$ is the trace over the occupation states of $\bar\sigma$ of the SIAM populations ${\rm P}_{\boldsymbol{\eta}}^\infty$, namely ${\rm p}^{\sigma}_{\nu}=\sum_{\eta}{\rm P}_{\boldsymbol{\eta}}^\infty$. In terms of the expectation value of the number operator $\hat{n}_\sigma$
\begin{equation}\begin{aligned}\label{p_sigma}
{\rm p}^\sigma_+=\langle \hat{n}_\sigma\rangle\;,\qquad {\rm p}^\sigma_-=1- \langle \hat{n}_\sigma\rangle\;.
\end{aligned}\end{equation}
These expectation values can be calculated either by solving the master equation for ${\rm P}_{\boldsymbol{\eta}}^\infty$, see Eq.~\eqref{P_steady}, or self-consistently, via the equations of motion technique~\cite{Lavagna2015}, where, in the wide-band limit, 
\begin{equation}\begin{aligned}\label{n_barsigma_SIAM}
\langle \hat{n}_\sigma \rangle
=\frac{1}{\Gamma}\int d\epsilon\left[\Gamma_L f_+^L(\epsilon)+\Gamma_R f_+^R(\epsilon) \right]\left[-\frac{1}{\pi}{\rm Im}\mathcal{G}^r_{\sigma\sigma}(\epsilon)\right]\;,
\end{aligned}\end{equation}
with $\Gamma=\Gamma_L+\Gamma_R$.\\

\subsection{Approximation schemes}
\begin{figure}[ht]
\resizebox{7cm}{!}{
\begin{tikzpicture}[]
\draw[line width=0.5mm,->] (-1,0) --(10,0) node[right] 
{\Large{$T$}};
\draw[line width=0.5mm] (-1,0.2)--  (-1,-0.2)node[below] {\Large{$0$}};
\draw[line width=0.5mm] (1.5,0.2) --  (1.5,-0.2) node[below] {\Large{$T_{\rm K}$}};
\draw[line width=0.5mm] (5,0.2) --  (5,-0.2) node[below] {\Large{$\Gamma$}};
\draw[dashed] (10,0) --  (10,2.8);
\draw[dashed] (-1,0) --  (-1,2.8);
\draw[dashed] (2.2,0) --  (2.2,2);
\draw[dashed] (3.5,0) --  (3.5,1.2);
\draw[dashed] (6.5,0) --  (6.5,0.4);
\draw[gray,line width=0.5mm,<->] (6.5,0.4) --(10,0.4);
\draw[]  (8.2,0.4)  node[above] {\Large{CT, ST\; (pert.)}};
\draw[gray,line width=0.5mm,<->] (3.5,1.2) --(10,1.2);
\draw[]  (6.5,1.2)  node[above] {\Large{RTA, NCA2, DSO\quad ($2$nd-tier)}};
\draw[gray,line width=0.5mm,->] (2.2,2) -- (10,2);
\draw[gray,dashed,line width=0.5mm,-] (0.5,2) -- (2.2,2);
\draw[]  (3.2,2)  node[above] {\Large{NCA4, sNCA4  ($4$th-tier)}};
\draw[gray,line width=0.5mm,->] (2,2.8) --(10,2.8);
\draw[gray,dashed,line width=0.5mm,<-] (-1,2.8) -- (2,2.8);
\draw[]  (1.8,2.8)  node[above] {\Large{DBA,\;NCA\quad ($\infty$-tier)}};
\end{tikzpicture}
}
\caption{\small{The different approximation schemes for the SIAM discussed in this work, in their range of validity ($k_{\rm B}=1$). Dashed lines indicate the expected regime of validity. Sequential tunneling (ST) and cotunneling (CT) are perturbative in $\Gamma$ (first and second order, respectively). The second-tier noncrossing approximation (NCA2) and the resonant tunneling approximation (RTA) are nonperturbative in $\Gamma$ and neglect diagrams with overlap of more than two fermion lines (second-tier schemes). The first is obtained from the second by neglecting the crossing diagrams and yields the Meir-Wingreen-Lee result~\cite{Meir1991} for the retarded Green's function. Higher-tier approximation schemes deepen the hierarchy of fermion lines. We propose novel infinite- and fourth-tier schemes which neglect the crossings at the first (DBA) or at all levels (NCA, NCA4, sNCA4) and can be seen as dressed versions of the NCA2.}}
\label{scheme_approximations}
\end{figure}
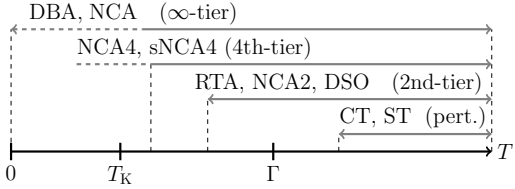

Despite the existing rich literature on the SIAM, an exact analytical solution for the dot Green's function, Eq.~\eqref{GF_SIAM}, encompassing the whole regime of parameters $U$, $\Gamma$, and $T$, is not known so far. In the forthcoming sections, we show how known perturbative schemes in $\Gamma$, as well as some nonperturbative ones, are recovered within our approach. Furthermore, a novel infinite-tier approximation scheme is discussed.\\ 
\indent In Fig.~\ref{scheme_approximations} we sketch the regime of validity of different analytical approaches derived from the diagrammatic unravelling of the exact irreducible propagator in Eq.~\eqref{phi_exact_symbolic}. The sequential tunneling and cotunneling schemes are perturbative in $\Gamma$ and thus valid when $\Gamma$ is the smallest energy scale, namely $\Gamma\ll k_{\rm B}T$ for $U=0$ and $\Gamma\ll k_{\rm B}T, U$ for $U\neq 0$.  In order to access the regime $\Gamma\sim k_{\rm B}T$ one needs to include processes of all orders. The simplest way to do this is to truncate the hierarchy of diagrams discussed in Sec.~\ref{diagrammatic_unravelling} to a maximum overlap of two fermion lines. We call the resulting schemes second-tier. Iterating the insertion of the bare bubble and crossing blocks defined above in the bare propagator $\mathbf{h}$ and summing the geometrical series results in the RTA~\cite{Koenig1996,Karlstrom2013}, where the propagator  $\boldsymbol\phi_{\rm RTA}$ is the solution of the Dyson equation~\eqref{phi_RTA}. Neglecting the crossing blocks of the RTA, one is left with a main fermion line dressed by bubbles, a scheme which we call NCA2. It generalizes the dressed second order (DSO), which accounts only for charge fluctuations internal to the main fermion line~\cite{Kern2013}. In our language the DSO considers the bubble blocks diagonal in the index $\eta$. On the contrary, the  
NCA2 takes into account the full matrix structure of the bubbles and, along with the RTA, recovers the noninteracting Green's functions. 
These second-tier schemes display artifacts such as the pinning of the density of states at the particle-hole symmetry point, due to the temperature-independent self-energy at this symmetry point, and do not predict the correct Kondo temperature $T_{\rm K}$. Crossing diagrams contribute to inelastic processes but are not expected to be relevant for investigating the zero-bias anomaly. For these reasons, in Sec.~\ref{NCA-like}, we discuss the infinite-tier approximation DBA which allows to recover the NCA2 form for the  Green's function but with dressed self-energies, see Eqs.~\eqref{Gzeta} and~\eqref{relationB-Sigma} below. In Sec.~\ref{gDSO4}, we explicitly evaluate such self-energies in a simplified version of the fourth-tier scheme NCA4 (sNCA4). In concluding this sub-section, we notice that the NCA scheme discussed here is different from the slave-boson NCA for the SIAM reviewed in~\cite{Bickers1987}. We nevertheless use the same name since we similarly neglect the crossing diagrams at all levels. 

\subsection{Perturbative schemes}

\subsubsection{Sequential tunneling}
\label{seq_tunneling}

The simplest approximation, valid for $k_{\rm B}T,U\gg\Gamma$ (or $k_{\rm B}T\gg\Gamma$ if $U=0$), consists in truncating to the lowest order in $\Gamma$ the Dyson equation for the propagator $\boldsymbol\phi$, Eq.~\eqref{Dyson_phi}. This results in $\boldsymbol{\phi}_{\rm ST}=\mathbf{h}$, where the bare propagator with a single fermion line reads for the SIAM $[\mathbf{h}]^{\sigma'\sigma}_{\boldsymbol\chi'\boldsymbol\chi}={\rm h}^\sigma_{\eta\eta}\delta_{\sigma'\sigma}\delta_{\boldsymbol\chi'\boldsymbol\chi}$, where
\begin{equation}\begin{aligned}\label{h_SIAM}
{\rm h}^\sigma_{\eta\eta}=&
\begin{gathered}
\resizebox{!}{1.2cm}{
\begin{tikzpicture}[] 
\draw[blue,line width=0.7mm] (0.5,1.3) arc (120:60:2.5cm  and 2.5cm) node[right] {\LARGE{$\sigma\boldsymbol\kappa$}} ;
\draw[blue,dashed,line width=0.7mm]  (0.5,0) -- (3,0); 
\draw[red,line width=0.7mm] (0.5,-0.5)   -- (3,-0.5) node[right] {\huge{$\eta$}}; 
 \end{tikzpicture}
 }
\end{gathered}
=
\;{\rm i}\hbar \frac{1}{\zeta[\epsilon_k-E_\sigma(\eta)]+{\rm i} 0^+}\;.
\end{aligned}\end{equation}
This propagator yields the current to the lowest order, namely the second order in the tunneling amplitude ${\rm t}$ (or first order in $\Gamma$). Note that the propagator $\boldsymbol{\phi}_{\rm ST}$ is diagonal both in the spin and in the remaining variables $\boldsymbol\chi$. Using Eqs.~\eqref{GF_SIAM} and~\eqref{omega_SIAM}, the Green's functions in the ST approximation is 
\begin{equation}\begin{aligned}\label{GF_ST}
 \mathcal{G}^{(\zeta)}_{\sigma\sigma,{\rm ST}}(\epsilon_k)=&-\frac{{\rm i}\zeta}{\hbar}\sum_{\eta'\eta}\eta'\eta\varphi^{\sigma\sigma}_{{\rm ST},\eta'\eta}(\zeta,k){\rm p}^{\bar\sigma}_\eta\\
 =&\sum_{\eta}\frac{{\rm p}^{\bar\sigma}_\eta}{\epsilon_k-E_\sigma(\eta)+{\rm i}\zeta 0^+}\;,
\end{aligned}\end{equation}
where ${\rm p}^{\bar\sigma}_{\eta}=\sum_{\nu}{\rm P}_{\boldsymbol{\eta}}^\infty$. Recall that $\varphi^{\sigma\sigma}_{{\rm ST},\eta'\eta}(\kappa)=\sum_{\sigma'\kappa'}\phi^{\sigma'\sigma}_{{\rm ST},\eta'\eta}(\kappa',\kappa)$, see Eq.~\eqref{varphi_B}. 
\begin{figure}[t!]
\begin{center}
\includegraphics[width=9cm,angle=0]{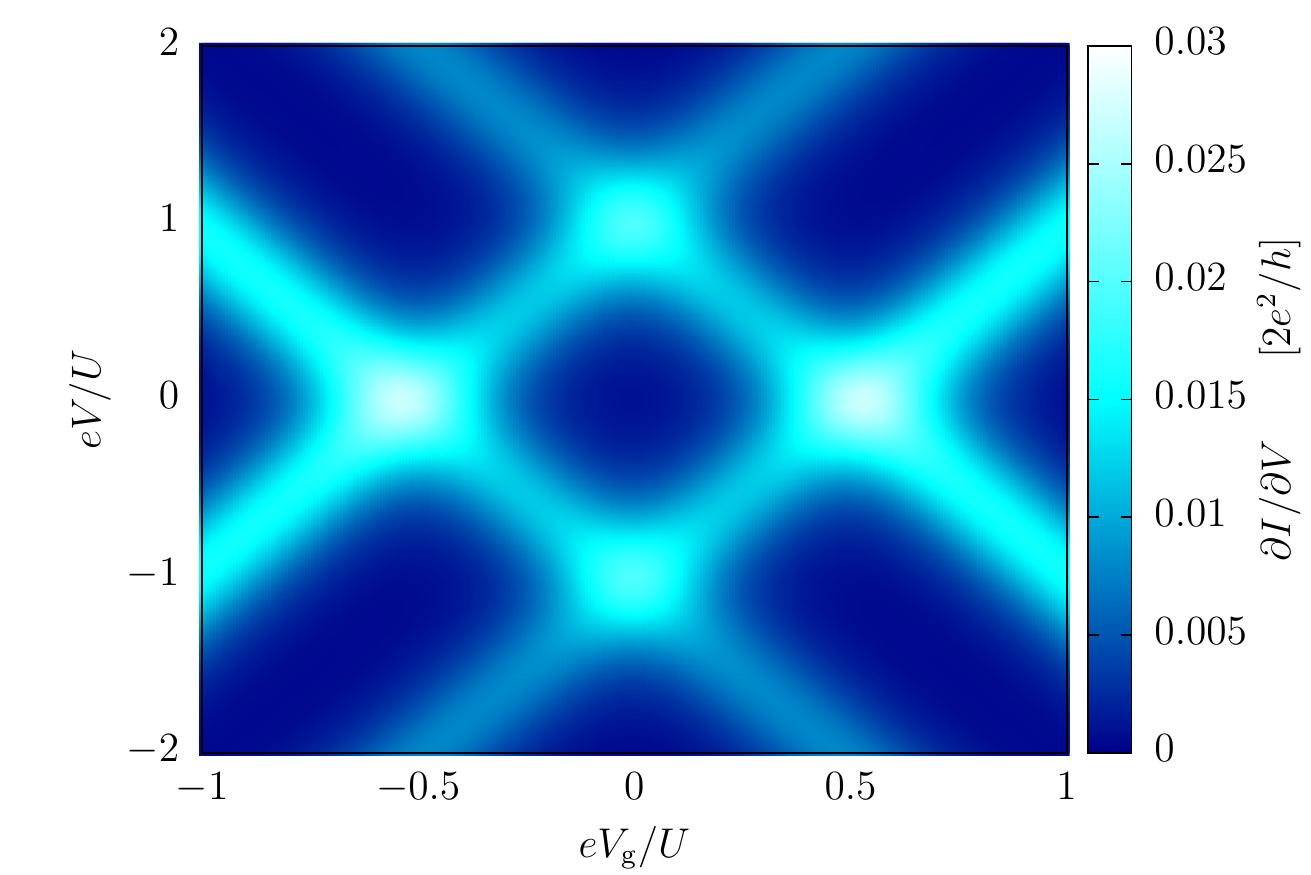}
\caption{\small{Differential conductance  \emph{vs} the gate voltage $V_g$ and bias voltage $V$ in the sequential tunneling (ST) approximation. 
At low bias voltages, the current is strongly  suppressed inside the regions enclosed by so-called Coulomb diamonds due to charging effects. The ST approximation does not account for virtual processes (of higher order in $\Gamma$) which enable transport of charge also inside the Coulomb diamonds.    
Degenerate case, $\epsilon_\sigma=\epsilon_0=-U/2+eV_{\rm g}$, with temperature $k_{\rm B}T=0.1~U$ and tunneling rates $\Gamma_L=\Gamma_R=0.005~U$.}}
\label{diff_cond_ST}
\end{center}
\end{figure}
In the ST approximation, from Eqs.~\eqref{p_sigma} and~\eqref{GF_ST}, 
\begin{equation}\begin{aligned}
-\frac{1}{\pi}{\rm Im}\mathcal{G}^r_{\sigma\sigma}(\epsilon)
=&\delta(\epsilon-\epsilon_\sigma)(1-\langle \hat{n}_{\bar\sigma}\rangle)+\delta(\epsilon-\epsilon_\sigma-U)\langle \hat{n}_{\bar\sigma}\rangle\;,
\end{aligned}\end{equation}
where we have used $\lim_{\varepsilon \to 0^+}\varepsilon/(x^2+\varepsilon^2)=\pi\delta(x)$, 
and the general formula~\eqref{I_MW_SIAM} gives for the curent 
\begin{equation}\begin{aligned}\label{I_MW_ST}
I_{\rm ST}^\infty
=&\frac{e}{\hbar}\frac{\Gamma_L\Gamma_R}{\Gamma}\sum_\sigma \Big\{ [f^{L}_+(\epsilon_\sigma)-f^{R}_+(\epsilon_\sigma)](1-\langle \hat{n}_{\bar\sigma}\rangle)\\
&+[f^{L}_+(\epsilon_\sigma+U)-f^{R}_+(\epsilon_\sigma+U)]\langle \hat{n}_{\bar\sigma}\rangle\Big\}\;.
\end{aligned}\end{equation}
The levels's populations are obtained by solving Eq.~\eqref{n_barsigma_SIAM} which yields in this case
\begin{equation}\label{n_sigma_ST}
\langle \hat{n}_{\bar\sigma}\rangle=\frac{\sum_\alpha\Gamma_\alpha f_+^\alpha(\epsilon_\sigma)}{\Gamma+\sum_\alpha\Gamma_\alpha f_+^\alpha(\epsilon_\sigma)-\sum_\alpha\Gamma_\alpha f_+^\alpha(\epsilon_\sigma+U)}\;.
\end{equation}
\indent In Fig.~\ref{diff_cond_ST}, we show the differential conductance $\partial I_{\rm ST}^\infty/\partial V$ in the degenerate case,  obtained from Eqs.~\eqref{I_MW_ST} and~\eqref{n_sigma_ST} by using $\partial f_+^L/\partial V=-(e/2)\partial f_+^L/\partial \epsilon$ and $\partial f_+^R/\partial V=\;(e/2)\partial f_+^R/\partial \epsilon$. The differential conductance is shown as a function of the bias voltage $eV=\mu_L-\mu_R$, and the gate voltage, which shifts the position of the (degenerate) level via $\epsilon_0=-U/2+eV_{\rm g}$. Such plot, called stability diagram, highlights the resonances which form diamond-shaped regions where the differential conductance is zero. In the central regions called Coulomb diamonds,  the dot populations are 0, 1, and 2, from left to right, and the current is suppressed. This effect is called    
Coulomb blockade and appears in the present regime of weak tunnel coupling , where the Coulomb interaction dominates and $\Gamma/k_{\rm B}T\ll 1$.\\
\indent A horizontal cut ($V=0$) of the stability diagram gives the linear conductance $G=\partial I/\partial V|_{V=0}$, see Eq.~\eqref{G}. The linear conductance, which is suppressed at the center of the Coulomb diamond, shows two peaks separated by the energy $U$, see Fig.~\ref{fig_gDSO} below. 
A straightforward extension of the ST which accounts for a $\Gamma$-broadening in the Green's function is discussed in Sec.~\ref{Gamma_broadening} below.

\subsubsection{Cotunneling}
\label{co_tunneling}

The next improvement over the ST, also perturbative in $\Gamma$, is the cotunneling approximation. It allows charge transfer across the dot also in the parameter regimes where ST processes are exponentially suppressed due to the Coulomb blockade~\cite{Koller2010}, i.e. around the center of the Coulomb diamond, $V_{\rm g}\sim 0$, see  Fig.~\ref{diff_cond_ST}. In our diagrammatic approach, this occurs via virtual processes encoded in the bubble and crossing blocks $\mathbf{B}$ and $\mathbf{X}$, i.e., according to Eq.~\eqref{phi_CT},
\begin{equation}\label{phi_SIAM_CT}
\boldsymbol{\phi}_{\rm CT}=\mathbf{h}+\mathbf{h}(\mathbf{B}+\mathbf{X})\mathbf{h}\;.
\end{equation}
As the sequential tunneling approximation, also this scheme is valid when $\Gamma$ is the smallest energy scale of the problem.\\ 
\indent As already noticed in the case of the RLM, the diagrammatic rules in Sec.~\ref{diagrammatic_rules}
 and in Appendix~\ref{examples_diagrams} imply that when there are at most two overlapping fermion lines associated to the same state (in the present case to the same spin state), the fermion line of a bubble has the index $\zeta$ opposite to the one of the main fermion line. Similarly, in a crossing, the outgoing line has the index $\zeta$ opposite to the one of the incoming line. These constraints do not apply when the spin states involved in a bubble or in a crossing block are different.    
It is therefore natural to distinguish these two cases.\\ 
\indent The bubble block is obtained by contracting with a vertex the internal fermion line (indexed with $1$) of the bare propagator with two overlapping fermion lines which can be of the same [$\sigma(\sigma)$], see Eq.~\eqref{h2_+_SIAM}, or of opposite [$\sigma(\bar\sigma)$] spin.
The first is given by $[\mathbf{h}_2^{(\sigma)}]^{\sigma'\sigma}_{\boldsymbol\chi'\boldsymbol\chi,\nu'\nu}={\rm h}_{2}^{\sigma(\sigma)}\delta_{\sigma'\sigma}\delta_{\boldsymbol\kappa'\boldsymbol\kappa}\delta_{\boldsymbol{\eta}'\boldsymbol{\eta}}$, where
\begin{equation}\begin{aligned}\label{h2_+_SIAM}
{\rm h}_2^{\sigma(\sigma)}=
\begin{gathered}
\resizebox{!}{1.3cm}{
\begin{tikzpicture}[] 
\draw[blue,line width=0.7mm] (0.5,2)  arc (120:60:2.5cm  and 1.cm)node[right] {\LARGE{$\sigma\boldsymbol\kappa$}};
\draw[blue,line width=0.7mm] (0.5,1.3)  arc (120:60:2.5cm  and 2.5cm)node[right] {\LARGE{$\sigma\boldsymbol\kappa_{1}$}};
\draw[blue,line width=0.7mm]  (0.5,0) -- (3,0)node[right] {\huge{$\nu$}}; 
\draw[red,line width=0.7mm] (0.5,-0.7)  -- (3,-0.7) node[right] {\huge{$\eta$}}; 
 \end{tikzpicture}
 }
\end{gathered}
=
{\rm i} \hbar\frac{ \delta_{\zeta_{1},-\zeta}}{\zeta(\epsilon_k-\epsilon_{k_{1}})+{\rm i}  0^+}\;,
\end{aligned}\end{equation}
(cf. Eq.~\eqref{h2_RLM}).
Note that the block $\mathbf{h}_2^{(\sigma)}$ has a $4\times 4$ structure in the composite index $(\nu,\eta)\equiv (\eta^\sigma,\eta^{\bar\sigma})$.
The propagator with overlap of two fermion lines with opposite spin is given by the block $[\mathbf{h}_2^{(\bar\sigma)}]^{\sigma'\sigma}_{\boldsymbol\kappa'\boldsymbol\kappa}={\rm h}_{2}^{\sigma(\bar\sigma)}\delta_{\sigma'\sigma}\delta_{\boldsymbol\kappa'\boldsymbol\kappa}$, with
\begin{equation}\begin{aligned}\label{h2_-_SIAM}
{\rm h}_{2}^{\sigma(\bar\sigma)}=
\begin{gathered}
\resizebox{!}{1.3cm}{
\begin{tikzpicture}[] 
\draw[blue,line width=0.7mm] (0.5,2)  arc (120:60:2.5cm  and 1.cm)node[right] {\LARGE{$\sigma\boldsymbol\kappa$}};
\draw[red,line width=0.7mm] (0.5,1.3)  arc (120:60:2.5cm  and 2.5cm)node[right] {\LARGE{$\bar\sigma\boldsymbol\kappa_1$}};
\draw[blue,dashed,line width=0.7mm]  (0.5,0) -- (3,0); 
\draw[red,dashed,line width=0.7mm] (0.5,-0.7) -- (3,-0.7) ; 
 \end{tikzpicture}
 }
\end{gathered}
=
{\rm i}\hbar\frac{1}{ \zeta(\epsilon_k-E_\sigma)+ \zeta_{1}(\epsilon_{k_{1}}-E_{\bar\sigma})+{\rm i}0^+}
\end{aligned}\end{equation}
which has no structure in the sojourn indexes.
The bubble block resulting from the contraction of the internal fermion lines of the two above propagators with the vertexes
\begin{equation}\label{}
{\rm v}_{\pm\eta}(\boldsymbol\kappa):=-\frac{|{\rm t}_{\alpha}(\epsilon_k)|^2}{\hbar^2} f^\alpha_{\pm\eta}(\epsilon_k)\;,
\end{equation}
is given by  
$[\mathbf{B}]^{\sigma'\sigma}_{\boldsymbol\chi'\boldsymbol\chi}=
{\rm B}^{\sigma}_{\eta'\eta}\delta_{\sigma'\sigma}\delta_{\boldsymbol\kappa'\boldsymbol\kappa}$. The function ${\rm B}^{\sigma}_{\eta'\eta}$ is the sum
\begin{equation}\begin{aligned}\label{B_SIAM}
{\rm B}^{\sigma}_{\eta'\eta}=&
{\rm B}^{\sigma(\sigma)}_{\eta'\eta}+{\rm B}^{\sigma(\bar\sigma)}_{\eta'\eta}\\
=&
\begin{gathered}
\resizebox{!}{1.3cm}{
\begin{tikzpicture}[] 
\draw[blue,line width=0.7mm] (0.5,1.3) arc  (120:60:3cm  and 2.6cm) node[right] {\LARGE{$\sigma\boldsymbol\kappa$}}  ;
\draw[blue,line width=0.7mm] (1,0) arc (180:0:1.cm  and 1.cm); 
\draw[blue]  (2,.9)  node[above] {\LARGE{$\sigma\boldsymbol\kappa_{1}$}}  ;
\draw[blue,dashed,line width=0.7mm]  (0.5,0) -- (1,0); 
\draw[blue,line width=0.7mm] (1,0) --node[above] {\LARGE{$\nu$}}  (3,0); 
\draw[blue,dashed,line width=0.7mm] (3,0) -- (3.5,0); 
\draw[red,line width=0.7mm] (0.5,-0.5)-- (3.5,-0.5) node[right] {\huge{$\eta$}}; 
\filldraw[blue] (1,0) circle (4pt); 
 \end{tikzpicture}
 }
\end{gathered}
+
\begin{gathered}
\resizebox{!}{1.3cm}{
\begin{tikzpicture}[] 
\draw[blue,line width=0.7mm] (0.5,1.3) arc (120:60:3cm  and 2.6cm) node[right] {\LARGE{$\sigma\boldsymbol\kappa$}};
\draw[red,line width=0.7mm]  (1,0) arc (180:0:1.cm  and 1.cm)  ;
\draw[red]  (2,.9)  node[above] {\LARGE{$\bar\sigma\boldsymbol\kappa_{1}$}}  ;
\draw[blue,dashed,line width=0.7mm]  (0.5,0) -- (3.5,0); 
\draw[red,line width=0.7mm] (0.5,-0.5)node[left] {\huge{$\eta$}} -- (1,-0.5); 
\draw[red,line width=0.7mm] (1,-0.5) -- (1,0); 
\draw[red,dashed,line width=0.7mm] (1,-0.5) -- (3,-0.5); 
\draw[red,line width=0.7mm] (3,-0.5) -- (3.5,-0.5)node[right] {\huge{$\eta'$}}; 
\draw[red,line width=0.7mm] (3,-0.5) -- (3,-0) ; 
\filldraw[red](1,-0.5) circle (4pt);
 \end{tikzpicture}
 }
\end{gathered}\\
=&\sum_{\boldsymbol\kappa_{1}}\frac{{\rm i}\hbar\sum_{\nu}{\rm v}_{\nu}(\boldsymbol\kappa_{1})\delta_{\zeta_{1},-\zeta}}{\zeta(\epsilon_k-\epsilon_{k_{1}}) + {\rm i} 0^+}\delta_{\eta'\eta}\\
&+\sum_{\boldsymbol\kappa_{1}}\frac{{\rm i}\hbar\;{\rm v}_{-\eta}(\boldsymbol\kappa_{1})}{\zeta(\epsilon_{k} - E_\sigma) +\zeta_{1}(\epsilon_{k_{1}} - E_{\bar\sigma})+ {\rm i} 0^+ }\;.
\end{aligned}\end{equation}
The first contribution, which is diagonal in all the indexes, is calculated as in Eqs.~\eqref{bubble_RLM}-\eqref{B+bare}, and reads ${\rm B}^{\sigma(\sigma)}_{\eta'\eta}=-\Gamma/(2\hbar)\delta_{\eta'\eta}$ in the WBL. 
The second term in Eq.~\eqref{B_SIAM} depends on $\eta$ via the vertex and can be evaluated as well in the WBL  where $\sum_{\boldsymbol\kappa_{1}}\rightarrow\sum_{\zeta_{1}\alpha_{1}}\varrho_{\alpha_{1}}\int d\epsilon_{1}$. We obtain, see Appendix \ref{contraction_integrals},
\begin{equation}\begin{aligned}\label{}
-\frac{\rm i}{\hbar}&\sum_{\zeta_{1}}\zeta_{1} \sum_{\alpha_{1}}\frac{\Gamma_{\alpha_{1}}}{2\pi}\int_{-W}^{W} d\epsilon_{1}\;  \frac{f^{\alpha_{1}}_{-\eta}(\epsilon_{1})}{\epsilon_{1}-\mathcal{E}_{\zeta_{1}}+{\rm i}\zeta_{1} 0^+}\\
=&{\rm i}\eta \sum_{\alpha}\frac{\Gamma_{\alpha}}{2\pi\hbar} \Bigg[{\rm Re}\psi\left(\frac{1}{2}+{\rm i}\frac{\mathcal{E}_{+1}-\mu_{\alpha}}{2\pi k_{\rm B} T}\right)\\
&\qquad\qquad\;-{\rm Re}\psi\left(\frac{1}{2}+{\rm i}\frac{\mathcal{E}_{-1}-\mu_{\alpha}}{2\pi k_{\rm B} T}\right)\Bigg]\\
&-\sum_{\alpha}\frac{\Gamma_{\alpha}}{2 \hbar}\left[f^{\alpha}_{-\eta}(\mathcal{E}_{+1})+f^{\alpha}_{-\eta}(\mathcal{E}_{-1})\right]\;,
\end{aligned}\end{equation}
where $\mathcal{E}_{\zeta_{1}}:=E_{\bar\sigma}+\zeta\zeta_{1}(E_\sigma-\epsilon)$.
All in all, the WBL expression for the bubble block reads
\begin{equation}\begin{aligned}\label{B_SIAM_WBL}
{\rm B}^{\sigma}_{\eta'\eta}&=
-\frac{\Gamma}{2\hbar}\delta_{\eta'\eta}\\
-&\sum_{\alpha_{1},\zeta_{1}}\frac{\Gamma_{\alpha_{1}}}{2\hbar}\Bigg[f^{\alpha_{1}}_{-\eta}(\mathcal{E}_{\zeta_{1}})
-{\rm i}\frac{\eta\zeta_{1}}{\pi}{\rm Re}\psi\left(\frac{1}{2}+{\rm i}\frac{\mathcal{E}_{\zeta_{1}}-\mu_{\alpha_{1}}}{2\pi k_{\rm B} T}\right)\Bigg]\;.
\end{aligned}\end{equation}
Note that the second term, while depending on the sojourn index $\eta$, is independent of $\eta'$.
Thus, in the WBL, the matrix elements of the bubble can be written as the sum
\begin{equation}\label{B_bare_SIAM_WBL}
{\rm B}^{\sigma}_{\eta'\eta}=
-\frac{\Gamma}{2\hbar}\delta_{\eta'\eta}+{\rm B}^{\sigma}_{\bar\eta\eta}\;.
\end{equation}
\indent The crossing blocks, with overlap of fermion lines with same and opposite spin, read
\begin{equation}\begin{aligned}\label{X_SIAM}
[\mathbf{X}]^{\sigma\sigma}_{\boldsymbol\chi'\boldsymbol\chi}=&
\begin{gathered}
\resizebox{!}{1cm}{
\begin{tikzpicture}[] 
\draw[blue,line width=0.7mm] (0.5,1.3) node[left] {\LARGE{$\sigma,\kappa$}} arc (100:10:1.7cm  and 1.6cm) ;
\draw[blue,line width=0.7mm] (3,1.3)node[right] {\LARGE{$\sigma,\kappa'$}} arc (100:10:-1.7cm  and 1.6cm); 
\draw[blue,line width=0.7mm] (1,0) --node[above] {\huge{$\nu$}} (2.5,0); 
\draw[blue,dashed,line width=0.7mm] (0.5,0) -- (3,0); 
\draw[red,line width=0.7mm] (0.5,-0.5) node[left] {\huge{$\eta$}}-- (3,-0.5)node[right] {\huge{$\eta$}}; 
\filldraw[blue] 
(1,0) circle (4pt); 
 \end{tikzpicture}
 }
\end{gathered}=\frac{{\rm i}\hbar\;\sum_{\nu}{\rm v}_{\nu}(\boldsymbol\kappa')}{\zeta(\epsilon_k-\epsilon_{k'}) + {\rm i} 0^+}\delta_{\zeta',-\zeta}\delta_{{\eta }'\eta }\;,\\
[\mathbf{X}]^{\bar\sigma\sigma}_{\boldsymbol\chi'\boldsymbol\chi}
=&\begin{gathered}
\resizebox{!}{1cm}{
\begin{tikzpicture}[] 
\draw[blue,line width=0.7mm] (0.5,1.3)node[left] {\LARGE{$\sigma,\kappa$}} arc (100:10:1.7cm  and 1.6cm) ;
\draw[red,line width=0.7mm] (3,1.3)node[right] {\LARGE{$\bar\sigma,\kappa'$}} arc (100:10:-1.7cm  and 1.6cm); 
\draw[blue,dashed,line width=0.7mm]  (0.5,0) -- (2.5,0); 
\draw[red,line width=0.7mm] (0.5,-0.5) node[left] {\huge{$\eta$}}-- (1,-0.5); 
\draw[red,line width=0.7mm] (1,-0.5) -- (1,0); 
\draw[red,dashed,line width=0.7mm] (1,-0.5) -- (3,-0.5); 
\draw[blue,line width=0.7mm] (2.5,-0) -- (3,0)node[right] {\huge{$\nu'$}}; 
\filldraw[red](1,-0.5) circle (4pt);
 \end{tikzpicture}
 }
\end{gathered}
=\frac{{\rm i}\hbar\;{\rm v}_{-\eta}(\boldsymbol\kappa')}{\zeta(\epsilon_{k} - E_\sigma) +\zeta'(\epsilon_{k'} - E_{\bar\sigma})+ {\rm i} 0^+ }\;,
\end{aligned}\end{equation}
respectively.
Here, in the spin-diagonal crossing block,  we include the sum over the internal sojourn $\nu$ in the definition, as done for the bubble with overlap of same-spin fermion lines, see Eq.~\eqref{B_SIAM}.\\
\indent In order to write the Green's function, we use the cotunneling  irreducible propagator in Eq.~\eqref{phi_SIAM_CT}. This yields 
\begin{equation}\begin{aligned}\label{}
\Omega^{\sigma,{\rm CT}}_{ {\boldsymbol{\eta}'}\boldsymbol{\eta}}&(\boldsymbol{\kappa})
=\eta'\eta\delta_{\nu' ,+1}\Big\{ {\rm h}^{\sigma}_{\eta\eta}(\boldsymbol{\kappa})\delta_{\eta'\eta}\\
+&\Big[\mathbf{h}^{\sigma}(\boldsymbol{\kappa})\mathbf{B}^{\sigma}(\boldsymbol{\kappa})\mathbf{h}^{\sigma}(\boldsymbol{\kappa})]_{\eta'\eta}\\
+&\sum_{\sigma'\boldsymbol{\kappa}'}\Big[\nu\mathbf{h}^{\sigma}(\boldsymbol{\kappa})\mathbf{x}_+^{\sigma \sigma'}(\boldsymbol{\kappa} ,\boldsymbol{\kappa}')\mathbf{h}^{\sigma'}(\boldsymbol{\kappa}'){\rm v}_{-\eta^{\sigma'}}(\boldsymbol{\kappa}')\Big]_{\eta' \eta^{\bar\sigma'}}\Big\}\;,
\end{aligned}\end{equation}
where, according to Eqs.~\eqref{vertex_split} and~\eqref{X_PC}, we split the crossing block $\mathbf{X}^{\sigma'\sigma}$ as
\begin{equation}\label{X_PC_SIAM}
{\mathbf{X}}^{\sigma\sigma'}(\boldsymbol{\kappa} \boldsymbol{\kappa}')={\rm v}(\boldsymbol{\kappa}){\mathbf{x}}^{\sigma\sigma'}(\boldsymbol{\kappa} \boldsymbol{\kappa}')-{\rm v}_+(\boldsymbol{\kappa}){\mathbf{x}}_+^{\sigma\sigma'}(\boldsymbol{\kappa} \boldsymbol{\kappa}')\;.
\end{equation}
Now, from inspection of Eq.~\eqref{X_SIAM} we find that $\mathbf{x}_+^{\sigma \sigma}=0$ due to the sum over $\nu$ (and in analogy to the RLM case) because nothing else depends on $\nu$.
We are then left with
\begin{equation}\begin{aligned}\label{omega_CT}
\Omega^{\sigma,{\rm CT}}_{{\boldsymbol{\eta}'}\boldsymbol{\eta}}(\boldsymbol{\kappa})
=&\eta'\eta\delta_{\nu' ,+1} \Big[{\rm h}_{\eta\eta}^{\sigma}(\boldsymbol{\kappa})\delta_{\eta'\eta}\\
&+{\rm h}_{\eta'\eta'}^{\sigma}(\boldsymbol{\kappa}){\rm B}^{\sigma}_{\eta'\eta}(\boldsymbol{\kappa}){\rm h}_{\eta\eta}^{\sigma}(\boldsymbol{\kappa})\\
&+\nu\sum_{\boldsymbol{\kappa}'}{\rm h}_{\eta'\eta'}^{\sigma}(\boldsymbol{\kappa}){\rm x}_{+,\eta'\nu}^{\sigma \bar\sigma}(\boldsymbol{\kappa} ,\boldsymbol{\kappa}'){\rm h}_{\nu\nu}^{\bar\sigma}(\boldsymbol{\kappa}'){\rm v}_{-\eta}(\boldsymbol{\kappa}')\Big]\;,
\end{aligned}\end{equation}
where
\begin{equation}\begin{aligned}
\mathbf{x}_{+,\eta'\nu}^{\sigma\bar\sigma}(\boldsymbol\kappa,\boldsymbol\kappa')
=&\frac{{\rm i}\hbar\;\eta'}{\zeta(\epsilon_{k} - E_\sigma) +\zeta'(\epsilon_{k'} - E_{\bar\sigma})+ {\rm i} 0^+ }
\;.
\end{aligned}\end{equation}
Using the above result and Eq.~\eqref{B_SIAM} we can cast $\Omega^\sigma_{\rm CT}$ in the form
\begin{equation}\begin{aligned}\label{omega_CT_2}
\Omega^{\sigma,{\rm CT}}_{\boldsymbol{\eta}' \boldsymbol{\eta}}(\boldsymbol{\kappa})
=&\eta'\eta\delta_{\nu' ,+1} \Bigg\{{\rm h}_{\eta\eta}^{\sigma}(\boldsymbol{\kappa})\delta_{\eta'\eta}+{\rm h}_{\eta'\eta'}^{\sigma}(\boldsymbol{\kappa}){\rm B}^{\sigma(\sigma)}_{\eta'\eta}(\boldsymbol{\kappa}){\rm h}_{\eta\eta}^{\sigma}(\boldsymbol{\kappa})\\
&+{\rm i}\hbar\sum_{\boldsymbol{\kappa}'}
\frac{{\rm h}_{\eta'\eta'}^{\sigma}(\boldsymbol{\kappa})[{\rm h}_{\eta\eta}^{\sigma}(\boldsymbol{\kappa})+ \nu\eta'{\rm h}_{\nu\nu}^{\bar\sigma}(\boldsymbol{\kappa}')]{\rm v}_{-\eta}(\boldsymbol{\kappa}')}{\zeta(\epsilon_{k} - E_\sigma) +\zeta'(\epsilon_{k'} - E_{\bar\sigma})+ {\rm i} 0^+ }\Bigg\}
\;.
\end{aligned}\end{equation}
As a result, the Green's function is the sum of the ST contribution, given by the first term in Eq.~\eqref{omega_CT_2}, plus the terms of fourth order in the tunneling amplitude (second order in $\Gamma$) and reads  
\begin{equation}\begin{aligned}
\mathcal{G}^{(\zeta)}_{\sigma\sigma,{\rm CT}}(\epsilon_k)
=- &\frac{{\rm i}\zeta}{\hbar}\sum_{{\boldsymbol{\eta}}' \boldsymbol{\eta}}\Omega^\sigma_{\boldsymbol{\eta}' \boldsymbol{\eta},{\rm CT}}(\zeta,k){\rm P}^\infty_{\boldsymbol{\eta}}\\
=&\mathcal{G}^{(\zeta)}_{\sigma\sigma,{\rm ST}}(\epsilon_k)+\mathcal{G}^{(\zeta)}_{\sigma\sigma,{\rm 4th}}(\epsilon_k)\;.
\end{aligned}\end{equation}
For a comprehensive diagrammatic analysis of cotunneling effects we refer to~\cite{Koller2010}. In the recent article~\cite{Rohrmeier2021}, interference phenomena at the cotunneling level, where one needs to go beyond the assumption of state-conserving tunneling, are discussed for interacting double quantum dots. In this work, we rather focus on nonperturbative schemes.

\subsection{Nonperturbative, second-tier schemes}

\subsubsection{Resonant tunneling approximation (RTA)}
\label{RTA}
Iterating the insertion of the cotunneling blocks $\mathbf{B}$ and $\mathbf{X}$ in the bare propagator $\mathbf{h}$, one obtains the nonperturbative RTA. The Dyson equation for the irreducible propagator $\boldsymbol\phi_{\rm RTA}=[\mathbf{h}^{-1}-\mathbf{B}-\mathbf{X}]^{-1}$, Eq.~\eqref{phi_RTA}, can be given in terms of the NCA2 propagator $\boldsymbol\phi_{\rm NCA2}=[\mathbf{h}^{-1}-\mathbf{B}]^{-1}$(see the next section),  as follows
\begin{equation}\begin{aligned}
\label{}
\boldsymbol{\phi}_{\rm RTA}=&\boldsymbol{\phi}_{\rm NCA2}+\boldsymbol{\phi}_{\rm NCA2}{\mathbf{X}}\boldsymbol{\phi}_{\rm RTA}\;.
\end{aligned}\end{equation}
Component-wise in $\boldsymbol\kappa$, with the  $2\times 2$ matrix structure induced by the sojourn indexes $\eta^\uparrow,\eta^\downarrow$ left implicit, this equation reads
\begin{equation}\begin{aligned}\label{dyson_phi_RTA}
\boldsymbol\phi_{\rm RTA}^{\sigma'\sigma}(\boldsymbol\kappa',&\boldsymbol\kappa)=\boldsymbol\varphi_{\rm NCA2}^{\sigma'\sigma'}(\boldsymbol\kappa')\delta_{\boldsymbol\kappa' \boldsymbol\kappa} \delta_{\sigma'\sigma}\\
&+
\boldsymbol\varphi_{\rm NCA2}^{\sigma'\sigma'}(\boldsymbol\kappa')\cdot\sum_{\sigma''\boldsymbol\kappa''}{\mathbf{X}}^{\sigma'\sigma''}(\boldsymbol\kappa',\boldsymbol\kappa'')\cdot\boldsymbol\phi_{\rm RTA}^{\sigma''\sigma}(\boldsymbol\kappa'',\boldsymbol\kappa)
\;.
\end{aligned}\end{equation}
The RTA is equivalent to the second order von Neumann approach~\cite{Karlstrom2013} and in the noninteracting case reproduces the current for the SIAM, but not the full density matrix, contrary to the case of the RLM which is fully described by the RTA. Indeed, as discussed in Sec.~\ref{diagrammatic_rules}, for $U=0$ the contributing diagrams have at most four overlapping fermion lines, of which at most two with the same spin.\\
\indent In the infinite-$U$ limit, the RTA admits an analytical solution to be found along the lines of~\cite{KoenigDiploma1995,Koenig1996}. In this limit, the blocks  specialize to (consider $\zeta=+1$)
\begin{equation}\begin{aligned}\label{matrix_functions_SIAM_Uinfty}
[\mathbf{h}]^{\sigma'\sigma}_{\boldsymbol\chi'\boldsymbol\chi}=&\;{\rm i}\hbar \frac{1}{\epsilon_k-\epsilon_\sigma+{\rm i} 0^+}\delta_{\sigma'\sigma}\delta_{\boldsymbol\chi'\boldsymbol\chi}\delta_{\eta,-1}\;,\\
[\mathbf{X}]^{\sigma\sigma}_{\boldsymbol\chi'\boldsymbol\chi}
=&\;{\rm i}\hbar\frac{\sum_{\nu}{\rm v}_{\nu}^{\alpha'}(k')}{\epsilon_k-\epsilon_{k'} + {\rm i} 0^+}\delta_{\zeta',-\zeta}\;,\\
[\mathbf{X}]^{\bar\sigma\sigma}_{\boldsymbol\chi'\boldsymbol\chi}
=&\;{\rm i}\hbar\frac{{\rm v}_{+}^{\alpha'}(k')}{\epsilon_{k} - \epsilon_{k'} + {\rm i}0^+ }\delta_{\zeta',-\zeta}\;,\\
[\mathbf{B}]^{\sigma'\sigma}_{\boldsymbol\chi'\boldsymbol\chi}=&\;{\rm i}\hbar\sum_{\alpha''k''}\frac{\sum_{\nu}{\rm v}_{\nu}^{\alpha''}(k'')+{\rm v}_{-\eta }^{\alpha''}(k'')}{\epsilon_k-\epsilon_{k''} + {\rm i} 0^+}\delta_{\sigma'\sigma}\delta_{\boldsymbol\chi'\boldsymbol\chi}\;.
\end{aligned}\end{equation}
The $2\times 2$ matrix structure of Eq.~\eqref{dyson_phi_RTA} is lost because $\eta$ can only assume the value $-1$, as induced by the interaction energy appearing at the denominator of ${\rm h}$ for $\eta=+1$, namely the dot can be occupied at most by one electron. As a result, we get the following two equations for the diagonal and off-diagonal elements in the spin index (we omit the labels RTA and NCA2)
\begin{equation}\begin{aligned}\label{}
\phi^{\sigma\sigma}(\epsilon,\epsilon')=&\varphi^{\sigma\sigma}(\epsilon)\delta(\epsilon-\epsilon')\\
&+
\varphi^{\sigma\sigma}(\epsilon)\int d\epsilon''{\rm X}^{\sigma\sigma}(\epsilon,\epsilon'')\phi^{\sigma\sigma}(\epsilon'',\epsilon)\\
&+
\varphi^{\sigma\sigma}(\epsilon)\int d\epsilon''{\rm X}^{\sigma\bar\sigma}(\epsilon,\epsilon'')\phi^{\bar\sigma\sigma}(\epsilon'',\epsilon')\;,\\
\phi^{\bar\sigma\sigma}(\epsilon,\epsilon')=&
\varphi^{\bar\sigma\bar\sigma}(\epsilon)\int d\epsilon''{\rm X}^{\bar\sigma\sigma}(\epsilon,\epsilon'')\phi^{\sigma\sigma}(\epsilon'',\epsilon)\\
&+
\varphi^{\bar\sigma\bar\sigma}(\epsilon)\int d\epsilon''{\rm X}^{\bar\sigma\bar\sigma}(\epsilon,\epsilon'')\phi^{\bar\sigma\sigma}(\epsilon'',\epsilon)\;,
\end{aligned}\end{equation}
where ${\rm X}^{\bar\sigma\bar\sigma}={\rm X}^{\sigma\sigma}$ and ${\rm X}^{\bar\sigma\sigma}={\rm X}^{\sigma\bar\sigma}$, see Eq.~\eqref{matrix_functions_SIAM_Uinfty}.
Consider now the degenerate system, $\epsilon_\uparrow=\epsilon_\downarrow=\epsilon_0$. In this case $\varphi^{\sigma\sigma}=\varphi^{\bar\sigma\bar\sigma}=\varphi^\sigma$ and, by summing the above two equations, we obtain
\begin{equation}\begin{aligned}\label{RTA_equation_degenerate}
\phi^\sigma(\epsilon,\epsilon')=&\varphi^\sigma(\epsilon)\delta(\epsilon-\epsilon')+
\varphi^\sigma(\epsilon)\int d \epsilon''{\rm X}(\epsilon,\epsilon'')\phi^\sigma(\epsilon'',\epsilon)\;,
\end{aligned}\end{equation}
where
\begin{equation}\begin{aligned}\label{}
\phi^\sigma(\epsilon,\epsilon')=&\sum_{\sigma'}\phi^{\sigma'\sigma}(\epsilon,\epsilon')\;,\\
{\rm X}(\epsilon,\epsilon')=&{\rm X}^{\sigma\sigma}(\epsilon,\epsilon')+{\rm X}^{\bar\sigma\sigma}(\epsilon,\epsilon')\\
=&\;{\rm i}\hbar\sum_\alpha\frac{2{\rm v}_+^{\alpha}(\epsilon)+{\rm v}_-^{\alpha}(\epsilon)}{\epsilon - \epsilon' + {\rm i}0^+ }\;,
\end{aligned}\end{equation}
see Eq.~\eqref{matrix_functions_SIAM_Uinfty}.
Solving Eq.~\eqref{RTA_equation_degenerate}, it is found that the real part of $\phi$, the one which enters the current, as in Eqs.~\eqref{current3} and~\eqref{Kcurrent}, reads
\begin{equation}
{\rm Re}\;\int d\epsilon'\phi^\sigma(\epsilon,\epsilon'){\rm v}_{-\eta}(\epsilon')= C_{\eta} \sum_\alpha[2{\rm v}_+^{\alpha}(\epsilon)+{\rm v}_-^{\alpha}(\epsilon)]|\varphi(\epsilon)|^2\;,
\end{equation}
where $C_{\eta}$ is a constant depending on the initial vertex ${\rm v}_{-\eta}$.
As shown in~\cite{Kern2013}, in this regime of large interaction $U$, the RTA and the DSO (see Sec.~\ref{sec_DSO}) display a qualitatively similar behavior of the linear conductance, though the RTA predicts higher peak-conductance. Moreover, the two schemes share the same prediction for the zero-bias anomaly temperature scale, which is given in Eq.~\eqref{TKgDSO} below.

\subsubsection{Second-tier noncrossing approximation (NCA2)}
\label{gDSO}

\begin{figure}[t!]
\begin{center}
\includegraphics[width=9cm,angle=0]{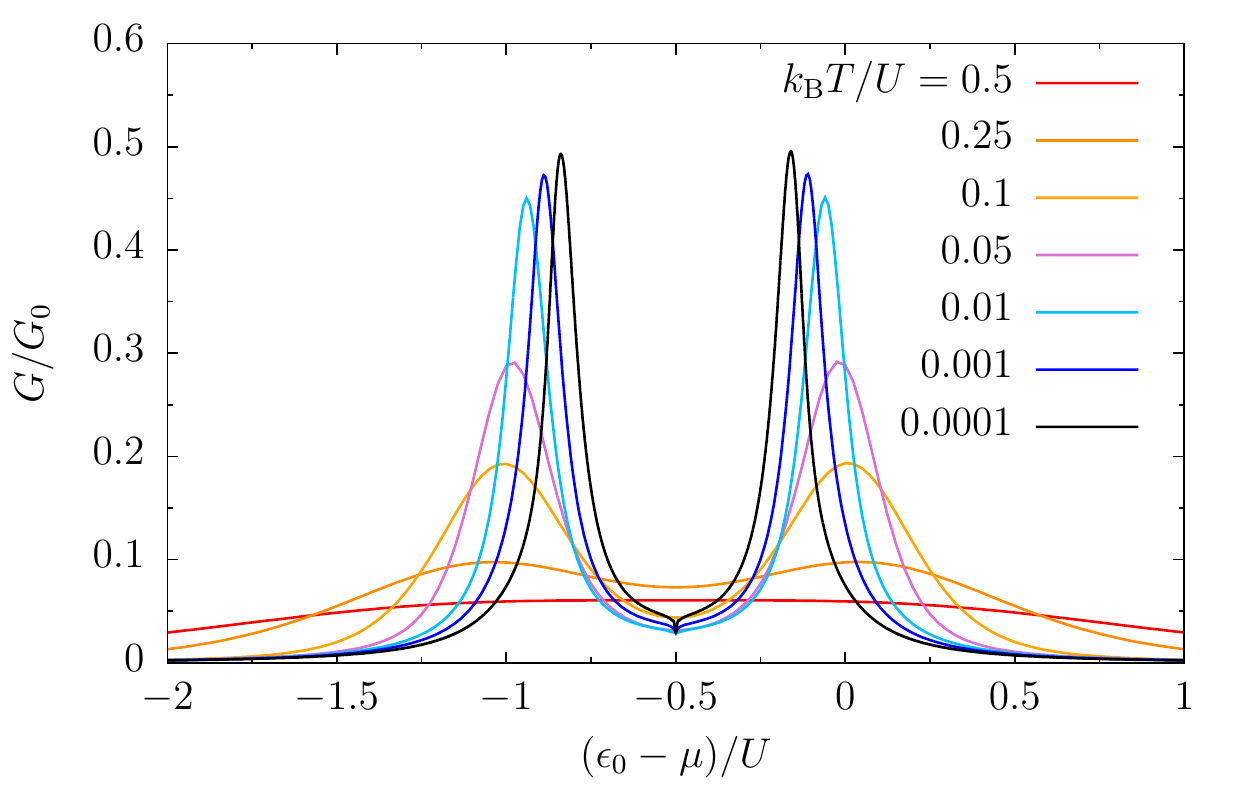}
\caption{\small{Linear conductance within the NCA2, in units of the conductance quantum $G_0=2e^2/h$, \emph{vs} the gate voltage $\epsilon_0-\mu$ in the degenerate case $\epsilon_\sigma=\epsilon_0$. 
The peaks shrink and move towards the center as temperature is decreased. The curves at the lowest temperatures display an unphysical pinning at the particle-hole symmetry point.
The tunnel coupling is $\Gamma=0.1~U$ with $\Gamma_L=\Gamma_R=\Gamma/2$.}}
\label{fig_gDSO}
\end{center}
\end{figure}

In the NCA2, the propagator $\boldsymbol\phi_{\rm NCA2}$ is given by the Dyson equation~\eqref{phi_gDSO}. As a result, the bare ST propagator is dressed by the bubble diagrams $\mathbf{B}$, namely
\begin{equation}\begin{aligned}\label{phiB_gDSO}
\boldsymbol{\phi}_{\rm NCA2}=&\left[\mathbf{h}^{-1}-{\mathbf{B}}\right]^{-1}\;.
\end{aligned}\end{equation}
While being diagonal in $\sigma$ and $\boldsymbol\kappa$, the bubble blocks $\mathbf{B}$ have a nontrivial $2\times 2$ matrix structure in terms of the sojourn index $\eta$, see Eq.~\eqref{B_bare_SIAM_WBL}, a feature which accounts for charge transfer by  processes internal to the main fermion line. As a result, they induce a $2\times 2$ structure to the contracted (matrix) function $\boldsymbol\varphi_{\rm NCA2}^{\sigma\sigma}(\kappa)$, defined by $\boldsymbol{\phi}_{\rm NCA2}^{\sigma'\sigma}(\kappa',\kappa)=\boldsymbol{\varphi}_{\rm NCA2}^{\sigma\sigma}(\kappa)\delta_{\kappa'\kappa}\delta_{\sigma'\sigma}$.\\
\indent Neglecting the crossings in the main fermion lines results in the Green's function, see Eqs.~\eqref{GF_SIAM} and~\eqref{omega_SIAM},
\begin{equation}\begin{aligned}\label{GF_gDSO}
 \mathcal{G}^{(\zeta)}_{\sigma\sigma}(\epsilon_k)=&-\frac{{\rm i}\zeta}{\hbar}\sum_{\eta'\eta}\eta'\eta\boldsymbol\varphi^{\sigma\sigma}_{{\rm NCA2},\eta'\eta}(\zeta,k){\rm p}^{\bar\sigma}_\eta\\
=&-\frac{{\rm i}\zeta}{\hbar}\sum_{\eta}\left[\boldsymbol\varphi^{\sigma\sigma}_{{\rm NCA2},\eta\eta}(\zeta,k)-\boldsymbol\varphi^{\sigma\sigma}_{{\rm NCA2},\bar\eta\eta}(\zeta,k)\right]{\rm p}^{\bar\sigma}_\eta\;.
\end{aligned}\end{equation}
The matrix elements of $\boldsymbol\varphi_{\rm NCA2}$, from Eq.~\eqref{phiB_gDSO}, read
\begin{equation}\begin{aligned}\label{GD-solution}
\boldsymbol{\varphi}^{\sigma\sigma}_{{\rm NCA2},\eta\eta}=&\frac{({\rm h}^{\sigma}_{\bar\eta\bar\eta})^{-1}-{\rm B}^{\sigma}_{\bar\eta\bar\eta}}{[({\rm h}^{\sigma}_{\eta\eta})^{-1}-{\rm B}^{\sigma}_{\eta\eta}][({\rm h}^{\sigma}_{\bar\eta\bar\eta})^{-1}-{\rm B}^{\sigma}_{\bar\eta\bar\eta}]-{\rm B}^{\sigma}_{\eta\bar\eta}{\rm B}^{\sigma}_{\bar\eta\eta}}\;,\\
\boldsymbol{\varphi}^{\sigma\sigma}_{{\rm NCA2},\bar\eta\eta}=&\frac{{\rm B}^{\sigma}_{\bar\eta\eta}}{[({\rm h}^{\sigma}_{\eta\eta})^{-1}-{\rm B}^{\sigma}_{\eta\eta}][({\rm h}^{\sigma}_{\bar\eta\bar\eta})^{-1}-{\rm B}^{\sigma}_{\bar\eta\bar\eta}]-{\rm B}^{\sigma}_{\eta\bar\eta}{\rm B}^{\sigma}_{\bar\eta\eta}}\;,
\end{aligned}\end{equation}
where ${\rm h}^{\sigma}_{\eta'\eta}$ is defined in Eq.~\eqref{h_SIAM} and where the dependence on $\boldsymbol\kappa$ of the block functions is understood. 
Note that the denominators are independent of $\eta$.
Plugging these results in Eq.~\eqref{GF_gDSO}, with ${\rm p}^\sigma_\eta$ given by Eq.~\eqref{p_sigma}, the dot Green's function reads
\begin{equation}\begin{aligned}\label{Gzeta_gDSO}
 \mathcal{G}^{(\zeta)}_{\sigma\sigma}
&=\frac{1-\langle \hat{n}_{\bar\sigma} \rangle}{\epsilon-\epsilon_\sigma-{\rm i}\zeta\hbar{\rm B}^{\sigma}_{--}+{\rm i}\zeta\hbar{\rm B}^{\sigma}_{+-}\frac{\epsilon-\epsilon_\sigma-{\rm i}\zeta\hbar[{\rm B}^{\sigma}_{--}+{\rm B}^{\sigma}_{-+}]}{\epsilon-\epsilon_\sigma-U-{\rm i}\zeta\hbar[{\rm B}^{\sigma}_{++}+{\rm B}^{\sigma}_{+-}]}}\\
+&\frac{\langle \hat{n}_{\bar\sigma} \rangle }{\epsilon-\epsilon_\sigma-U-{\rm i}\zeta\hbar{\rm B}^{\sigma}_{++}+{\rm i}\zeta\hbar{\rm B}^{\sigma}_{-+}\frac{\epsilon-\epsilon_\sigma-U-{\rm i}\zeta\hbar[{\rm B}^{\sigma}_{++}+{\rm B}^{\sigma}_{+-}]}{\epsilon-\epsilon_\sigma-{\rm i}\zeta\hbar[{\rm B}^{\sigma}_{--}+{\rm B}^{\sigma}_{-+}]}}
\;.
\end{aligned}\end{equation}
In the WBL, we can simplify this expression by using the result in Eq.~\eqref{B_bare_SIAM_WBL} and the property
\begin{equation}\label{property_B_gDSO}
\sum_\eta {\rm B}^{\sigma}_{\bar\eta \eta}=-\Gamma/\hbar\;,
\end{equation}
which can be checked by inspection of Eq.~\eqref{B_SIAM_WBL}. The retarded ($\zeta=+1$) Green's function in the NCA2 reads
\begin{equation}\begin{aligned}\label{GRgDSO}
\mathcal{G}^r_{\sigma\sigma}(\epsilon)=&\frac{1-\langle \hat{n}_{\bar\sigma} \rangle}{\epsilon-\epsilon_\sigma+{\rm i}\Gamma/2+{\Sigma}_{\sigma-}(\epsilon)\frac{U}{\epsilon-\epsilon_\sigma-U+{\rm i}3\Gamma/2}}\\
+&\frac{\langle \hat{n}_{\bar\sigma} \rangle}{\epsilon-\epsilon_\sigma-U+{\rm i}\Gamma/2-{\Sigma}_{\sigma+}(\epsilon)\frac{U}{\epsilon-\epsilon_\sigma+{\rm i}3\Gamma/2}}\;,
\end{aligned}\end{equation}
where we have singled-out the constant broadening ${\rm i}\Gamma/2$ and identified the 
interaction-induced contributions with the off-diagonal matrix elements of the bubbles via
\begin{equation}\begin{aligned}\label{relationB-Sigma_gDSO}
\Sigma_{\sigma\eta}(\epsilon)&:={\rm i}\hbar{\rm B}^{\sigma}_{\bar\eta\eta}(\kappa)|_{\zeta=+1}\\
=-\sum_{\alpha_{1}}&\frac{\Gamma_{\alpha_{1}}}{2\pi}\sum_{\zeta_{1}}\Bigg[\eta\zeta_{1}{\rm Re}\psi\left(\frac{1}{2}+{\rm i}\frac{\mathcal{E}_{\zeta_{1}}-\mu_{\alpha_{1}}}{2\pi k_{\rm B} T}\right)+{\rm i}\pi f^{\alpha_{1}}_{-\eta}(\mathcal{E}_{\zeta_{1}})\Bigg]\;,
\end{aligned}\end{equation}
with $\mathcal{E}_{\zeta_{1}}:=E_{\bar\sigma}+\zeta_{1}(E_\sigma-\epsilon)$ and $E_\sigma=\epsilon_\sigma+U/2$.
 Note that, by virtue of the property~\eqref{property_B_gDSO}, which implies 
\begin{equation}\label{property_Sigma_gDSO}
 \Sigma_{\sigma\eta}(\epsilon)=-\Sigma_{\sigma\bar\eta}(\epsilon)-{\rm i}\Gamma\;,
\end{equation} 
we can express the retarded Green's function in terms of a single self energy, e.g. $\Sigma_{\sigma-}(\epsilon)$. Interestingly, the result in Eq.~\eqref{GRgDSO} is equivalent to the one obtained by Meir, Wingreen, and Lee in~\cite{Meir1991} with the equations of motion technique.\\
\indent We now summarize the properties and predictions of the NCA2.
For vanishing interaction, $U= 0$, we recover the correct WBL result for the noninteracting case 
\begin{equation}\begin{aligned}\label{GzetaGDSOWBL0}
\mathcal{G}^r_{\sigma\sigma}(\epsilon)=&\frac{1}{\epsilon-\epsilon_\sigma+{\rm i}\Gamma/2}\;.
\end{aligned}\end{equation}
As anticipated below Eq.~\eqref{G_RLM}, the zero-temperature conductance in the noninteracting limit 
\begin{equation}\label{G_SIAM_U0}
G=G_0\frac{\Gamma_L\Gamma_R}{(\mu-\epsilon_0)^2+\Gamma^2/4}
\end{equation}
saturates, at resonance and for $\Gamma_\alpha=\Gamma/2$, to the conductance quantum $G_0$ due to the sum over the spin degree of freedom. This can be seen by applying the general formula, Eq.~\eqref{G}, to the SIAM with the retarded Green's function given by Eq.~\eqref{GzetaGDSOWBL0}.\\
\indent For finite interaction, on the other hand, the NCA2 displays a zero-bias anomaly at a characteristic temperature $T^*=T_{\rm NCA2}$. 
This temperature is defined as the one at which the real parts of the denominators vanish, causing a peak in the density of states $-{\rm Im}[\mathcal{G}^r_{\sigma\sigma}(\mu)]/\pi$. By virtue of the property~\eqref{property_Sigma_gDSO} satisfied by the self-energies, this condition is the same for the two terms in Eq.~\eqref{GRgDSO} and reads
\begin{equation}\begin{aligned}\label{conditionTK}
(\mu-\epsilon_0)(\mu-\epsilon_0-U)-3\Gamma^2/4+U\;{\rm Re}[\Sigma_{\sigma-}(\mu)|_{T^*}]=0\;.
\end{aligned}\end{equation}
For sufficiently large interaction energy $U$, away from the condition $\mu-\epsilon_0=U/2$, we can approximate the real part of the self-energy in Eq.~\eqref{relationB-Sigma_gDSO} as
\begin{equation}\begin{aligned}\label{ReSigma_gDSO}
{\rm Re}\Sigma_{\sigma-}(\mu)|_{T^*}\simeq
\frac{\Gamma}{2\pi}\ln\left(\frac{U}{2\pi k_{\rm B} T^*}\right)\;.
\end{aligned}\end{equation}
Then, assuming $U\gg \Gamma$, solving Eq.~\eqref{conditionTK} for $T^*= T_{\rm NCA2}$, we find 
\begin{equation}\begin{aligned}\label{TKgDSO}
k_{\rm B} T_{\rm NCA2}\simeq \frac{U}{2\pi}e^{2\pi\frac{(\mu-\epsilon_0)(\mu-\epsilon_0-U)}{U\Gamma}}\;.
\end{aligned}\end{equation}
Notice that this result differs from the Kondo temperature $T_{\rm K}$, which has the prefactor $\pi$ in place of $2\pi$ at the exponent~\cite{vanderWiel2000,Kouwenhoven2001}, namely
\begin{equation}\begin{aligned}\label{TKondo}
k_{\rm B}T_{\rm K}= \;\frac{\sqrt{U\Gamma}}{2}e^{\pi\frac{(\mu-\epsilon_0)(\mu-\epsilon_0-U)}{U\Gamma}}\;,
\end{aligned}\end{equation}
with $\lim_{U\to \infty} k_{\rm B}T_{\rm K}\propto\exp[-\pi(\mu-\epsilon_0)/\Gamma]$.\\
\indent Another problem of this approximation scheme is the temperature-independent behavior at the particle-hole symmetry point (sp) $\mu-\epsilon_0=U/2$ at equilibrium in the degenerate case~\cite{Kashcheyevs2006}. Indeed, since $\mathcal{E}_\pm-\mu=\mp(\epsilon-\mu)$, we have $f_+(\mathcal{E}_+)=f_-(\mathcal{E}_-)$. Also ${\rm Re}\psi(x+{\rm i}y)={\rm Re}\psi(x-{\rm i}y)$. Thus, from Eq.~\eqref{relationB-Sigma_gDSO} we find that the self-energy is purely imaginary and temperature-independent
\begin{equation}\begin{aligned}\label{}
\Sigma_{\sigma\eta}^{\rm sp}(\epsilon) = -{\rm i}\Gamma/2\;.
\end{aligned}\end{equation}
Consequently, the retarded Green's function becomes
\begin{equation}\begin{aligned}\label{GRgDSO-Pinning}
\mathcal{G}^{r,{\rm sp}}_{\sigma\sigma}(\epsilon)=&\frac{1-\langle \hat{n}_{\bar\sigma} \rangle}{\epsilon+U/2+{\rm i}\Gamma/2-{\rm i}(\Gamma/2)\frac{U}{\epsilon-U/2+{\rm i}3\Gamma/2}}\\
+&\frac{\langle \hat{n}_{\bar\sigma} \rangle}{\epsilon-U/2+{\rm i}\Gamma/2+{\rm i}(\Gamma/2)\frac{U}{\epsilon+ U/2+{\rm i}3\Gamma/2}}\;.
\end{aligned}\end{equation}
This feature causes the onset of an artifact in the linear conductance $G$ when the temperature decreases below $\Gamma$: A pinning at a temperature-independent value of $G$ at the symmetry point.
Note that inclusion of the crossings does not lift this problem: This can be seen from Fig.~3 of Ref.~\cite{Pedersen2005} where a scheme equivalent to the RTA, which includes the crossings, is used.\\
\indent The linear conductance from the NCA2 is shown in Fig.~\ref{fig_gDSO}, for the degenerate case, as a function of the gate voltage $\epsilon_0-\mu$ for various temperatures. At  high temperatures, $k_{\rm B}T>\Gamma$, there are two temperature-broadened peaks separated by the energy $U$, in agreement with the ST result shown in Fig.~\ref{diff_cond_ST}. Upon decreasing $T$, the peaks get narrower and closer, witnessing the transition to $\Gamma$-broadened conductance peaks, where the dot energies are renormalized by the tunnel coupling to the leads. This is captured by the nonperturbative character of the NCA2. As anticipated above, at low temperatures, $k_{\rm B}T\ll\Gamma$, the pinning of $G$ at a temperature-independent value appears at the particle-hole symmetry point $\epsilon_0-\mu=-U/2$. The onset of this artifact signals the breakdown of the approximation scheme.
As shown in Sec.~\ref{gDSO4}, introducing higher-tier processes which dress the NCA2 bubble diagrams, this problem is lifted as the dependence of the self-energies on the temperature is restored.

\subsubsection{Charge fluctuations only: Dressed second order}\label{sec_DSO}

\begin{figure}[t!]
\begin{center}
\includegraphics[width=9cm,angle=0]{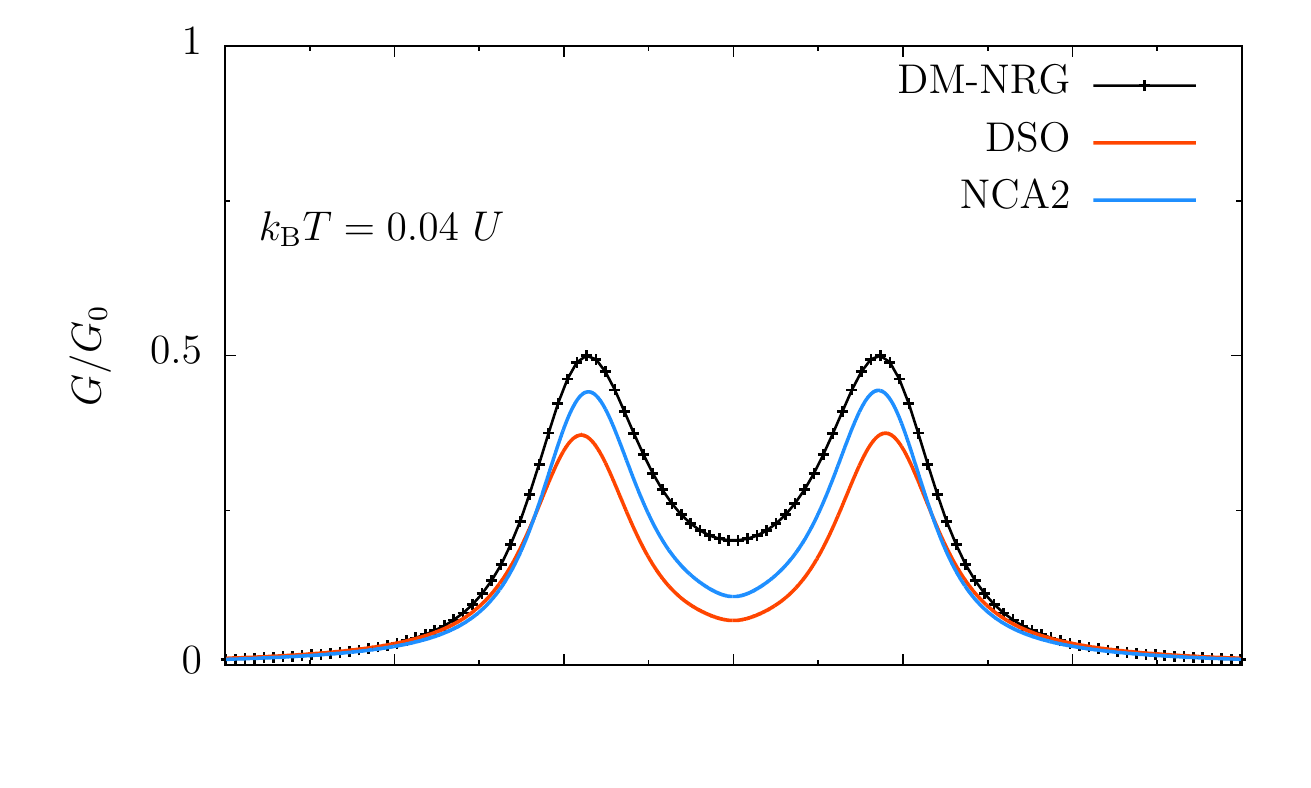}\\
\vspace{-0.85cm}
\includegraphics[width=9cm,angle=0]{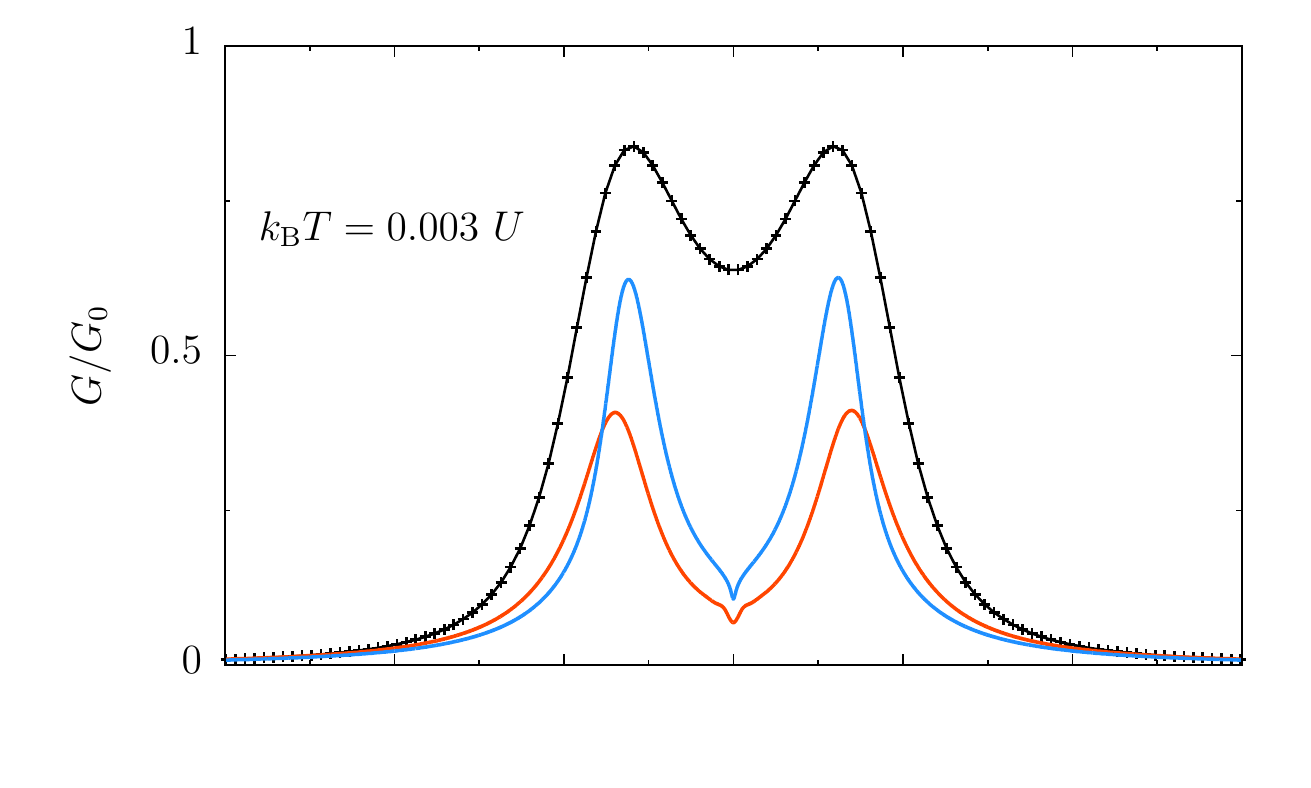}\\
\vspace{-0.85cm}
\includegraphics[width=9cm,angle=0]{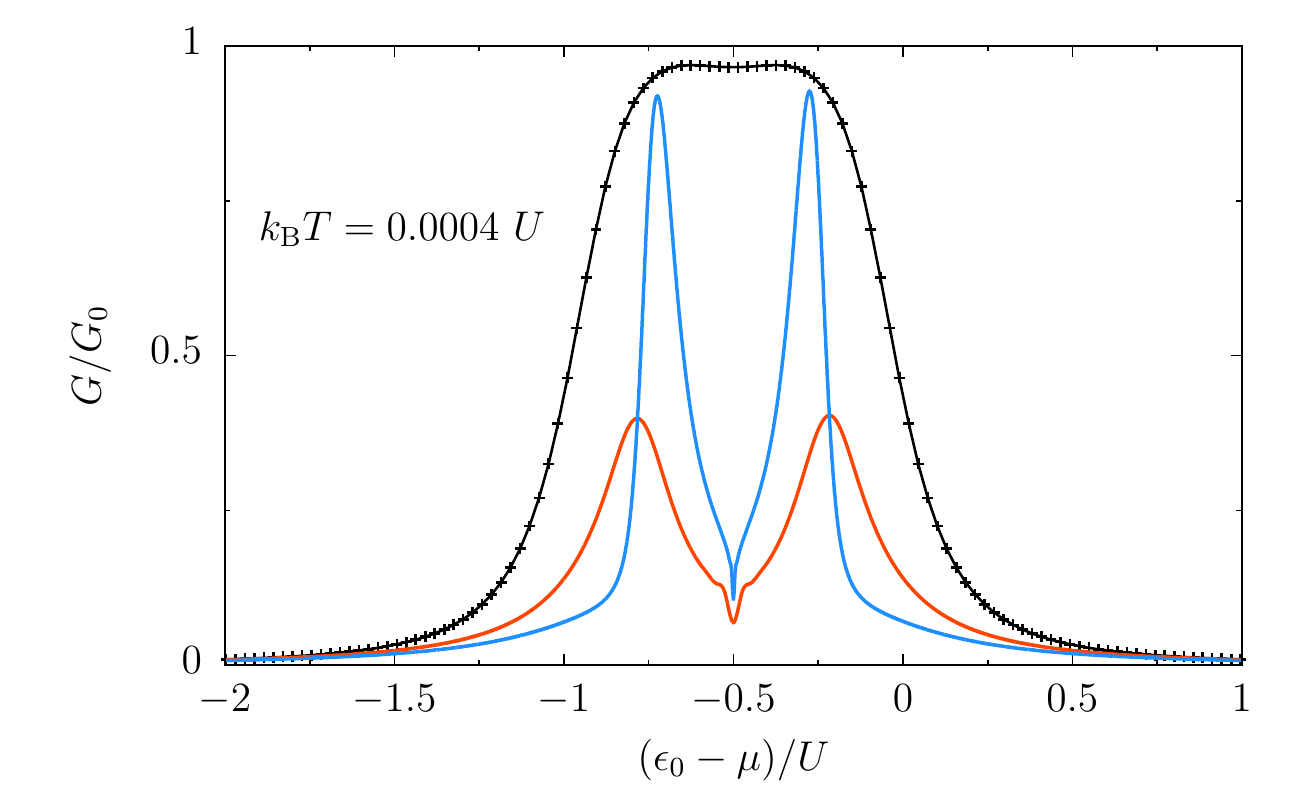}
\caption{\small{Linear conductance within the NCA2 and the DSO schemes compared with the curves from the numerically-exact DM-NRG. We consider the degenerate case $\epsilon_\sigma=\epsilon_0$ for three different temperatures.  The tunnel coupling is $\Gamma=0.2~U$ with $\Gamma_L=\Gamma_R=\Gamma/2$. The parameters are chosen to allow a direct comparison with the results of~\cite{deSouza2019}.}}
\label{fig_gDSO_vs_DSO}
\end{center}
\end{figure}
The DSO is the simplest, nontrivial  approximation scheme nonperturbative in $\Gamma$, being the dressed version of the sequential tunneling, see Eq.~\eqref{phi_ST}, where the main fermion line is dressed by charge fluctuations~\cite{Koller2012, Kern2013, Dirnaichner2015}. The  diagrams retained are formally similar to the ones of the NCA2, namely they consist in  dressing the main fermion lines with bare bubble diagrams. The difference is that the DSO only accounts for charge fluctuations of the main fermion line, meaning that the sojourns  states before and after a bubble are the same, $\eta=\eta_1=\dots=\eta'$. As a consequence, the charge in the dot does not vary by more than one unit between the two ends of the main fermion line, net charge transfers being operated solely by the latter. In this  scheme, the kernel connects the same states as those connected by the sequential tunneling, the states that differ by no more than one electron in occupancy. Note that the same is not true for the NCA2, which includes also processes that vary the dot charge also within the main line (pair tunneling). 
The fact that the internal processes leave the intermediate sojourns unchanged means that the bubbles in the Dyson equation~\eqref{DysonNCA} for the function $\varphi^{\sigma\sigma}$ have a diagonal structure yielding the solution  
\begin{equation}\begin{aligned}\label{phi_DSO}
\varphi^{\sigma\sigma}_{\eta'\eta,{\rm DSO}}(\kappa)=&\frac{1}{({\rm h}^{\sigma}_{\eta\eta})^{-1}-{\rm B}^{\sigma}_{\eta\eta}}\delta_{\eta'\eta}\;.
\end{aligned}\end{equation}
Thus, according to Eq.~\eqref{GF_gDSO}, the retarded DSO Green's function assumes, in the WBL, the form
\begin{equation}\begin{aligned}\label{GzetaDSO}
\mathcal{G}^r_{\sigma\sigma}(\epsilon)=&\frac{1-\langle \hat{n}_{\bar\sigma} \rangle}{\epsilon-\epsilon_\sigma+{\rm i}\Gamma/2-\Sigma_{\sigma-}(\epsilon)}\\
+&\frac{\langle \hat{n}_{\bar\sigma} \rangle}{\epsilon-\epsilon_\sigma-U+{\rm i}\Gamma/2-\Sigma_{\sigma+}(\epsilon)}
\;,
\end{aligned}\end{equation}
with  $\Sigma_{\sigma\eta}(\epsilon)$  given by Eq.~\eqref{relationB-Sigma_gDSO}.
Note that, contrary to the  NCA2, the  correct result in the noninteracting limit is \emph{not} recovered within the DSO. 
However, in the limit $U\rightarrow \infty$, the DSO reproduces the NCA2 result in the same limit, namely one obtains
\begin{equation}\begin{aligned}\label{GrDSOUinfty}
\lim_{U\to\infty}\mathcal{G}^r_{\sigma\sigma}(\epsilon)=&\frac{1-\langle \hat{n}_{\bar\sigma} \rangle}{\epsilon-\epsilon_\sigma+{\rm i}\Gamma/2-\Sigma_{\sigma-}(\epsilon)}\;.
\end{aligned}\end{equation}
The corresponding prediction for the Kondo-like temperature $T_{\rm DSO}$ is in this limit
\begin{equation}\label{TKDSOUinfty}
k_{\rm B} T_{\rm DSO}\simeq \frac{U}{2\pi}e^{-2\pi(\mu-\epsilon_0)/\Gamma}\;,
\end{equation}
and is the same as for the NCA2 and RTA, with the wrong prefactor in the exponent.
Nevertheless, the DSO is the simplest scheme, namely the one with the minimal collection of diagrams, which captures the emergence of a zero-bias anomaly at low temperatures, $k_{\rm B}T\leq\Gamma$.  
A comparison between the linear conductance calculated within the NCA2 and the one from the DSO is shown in Fig.~\ref{fig_gDSO_vs_DSO}, for $k_{\rm B}T/\Gamma=0.2$, where the linear conductance follows qualitatively the DM-NRG result, down to $k_{\rm B}T/\Gamma=0.002$, where both schemes break down. The parameters are chosen so as to allow for a direct comparison with Fig.~6 of Ref.~\cite{deSouza2019}, where the scheme EOM2 (equation of motion method) behaves similarly to the NCA2.

\subsubsection{$\Gamma$-broadening of the ST}\label{Gamma_broadening}
Dressing the main fermion line exclusively with the temperature-independent bubble, namely the first term of Eq.~\eqref{B_SIAM}, amounts to neglecting in the DSO retarded Green's function the temperature-dependent self-energies
$\Sigma_{\sigma\pm}(\epsilon)$, see Eq.~\eqref{GzetaDSO}. The only internal process retained here corresponds to the temperature-independent diagrammatic contributions in Ref.~\cite{Saptsov2014}. The resulting Green's function is that of a $\Gamma$-broadened version of the ST approximation, where the single-particle energies acquire a broadening $\Gamma$ due to the coupling to the leads and reads 
\begin{equation}\begin{aligned}\label{GF_dressed_ST}
 G^{r}_{\sigma\sigma}(\epsilon_k)=\sum_{\eta}\frac{{\rm p}^{\bar\sigma}_\eta}{\epsilon_k-E_\sigma(\eta)+{\rm i} \Gamma/2}\;.
\end{aligned}\end{equation}
This scheme is the same as the EOM0 reviewed in~\cite{deSouza2019}, which is derived with the EOM method.
With this result for the Green's function, the general formula Eq.~\eqref{I_MW}, gives for the current in the ST approximation
\begin{equation}\begin{aligned}\label{}
I^\infty
=&\frac{e}{h}\Gamma_L\Gamma_R \sum_\sigma\int d \epsilon\; [f^{L}_+(\epsilon)-f^{R}_+(\epsilon)]\\
&\left[\frac{1-\langle \hat{n}_{\bar\sigma}\rangle}{(\epsilon-\epsilon_\sigma)^2+\Gamma^2/4}+\frac{\langle \hat{n}_{\bar\sigma}\rangle}{(\epsilon-\epsilon_\sigma-U)^2+\Gamma^2/4}\right]\\
=&\;\frac{e}{h}\frac{\Gamma_L\Gamma_R}{\Gamma}\sum_\sigma\Big[(1-\langle \hat{n}_{\bar\sigma}\rangle)r(\epsilon_\sigma)+ \langle \hat{n}_{\bar\sigma}\rangle r(\epsilon_\sigma+U) \Big]\;,
\end{aligned}\end{equation}
where $r(x)$ is defined in Eq.~\eqref{r}. \\
\begin{figure}[t!]
\begin{center}
\includegraphics[width=8.5cm,angle=0]{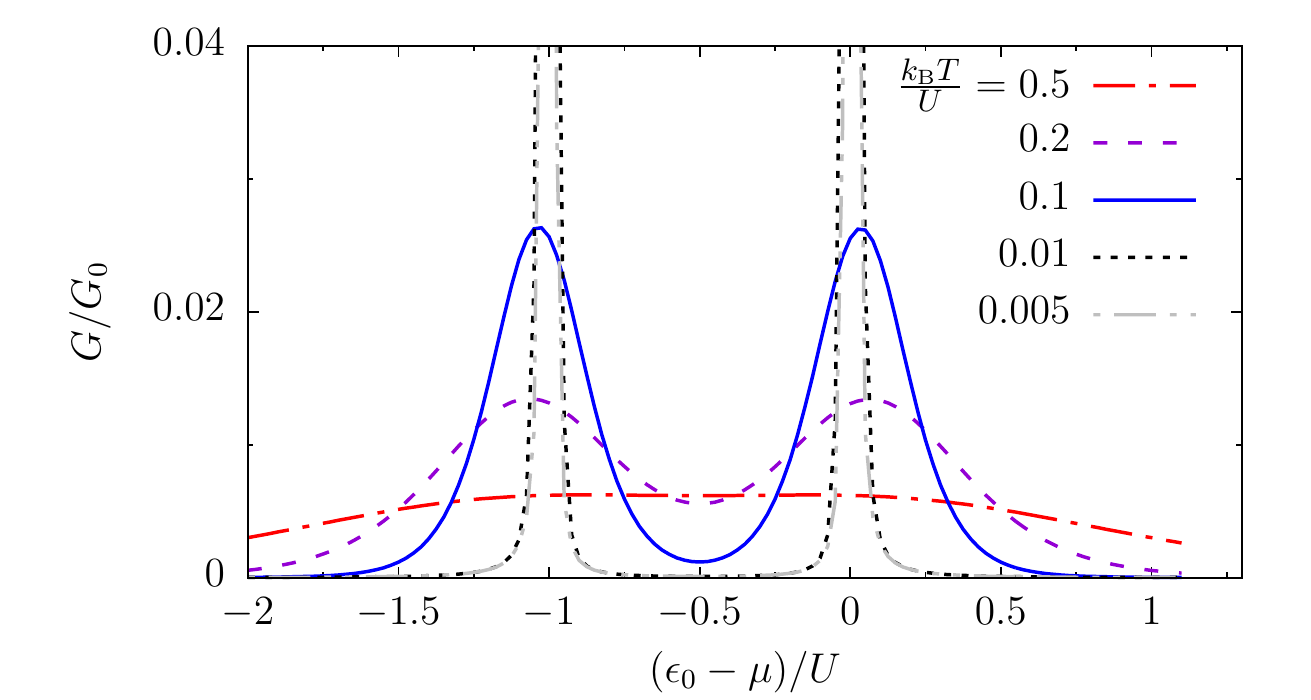}
\caption{\small{Linear conductance within the $\Gamma$-broadened ST, Eq.~\eqref{GF_dressed_ST} \emph{vs} the gate voltage $\epsilon_0-\mu$ in the degenerate case $\epsilon_\sigma=\epsilon_0$. Lowering the temperature, the peaks shrink until $k_{\rm B}T \simeq \Gamma$ but they do not move towards the center. Their height grows as $T^{-1}$. The tunnel coupling is $\Gamma=0.01~U$ with $\Gamma_L=\Gamma_R=\Gamma/2$.}}
\label{fig_dressed_ST}
\end{center}
\end{figure}
\indent In Fig.~\ref{fig_dressed_ST}, we show the linear conductance $G$, obtained by inserting the retarded Green's function ($\zeta=+1$) of Eq.~\eqref{GF_dressed_ST} into the general expression~\eqref{G}. The linear conductance is plotted, in the degenerate case, where $\epsilon_\uparrow=\epsilon_\downarrow=\epsilon_0$, against the gate voltage. Decreasing the temperature, $G$ is suppressed around the particle-hole symmetry point $\epsilon_0-\mu=-U/2$, the central region between the two peaks separated by the energy $U$, the separation being independent of the temperature, contrary to the NCA2, cf. Fig.~\ref{fig_gDSO}.

\subsection{Infinite-tier schemes I: Neglecting the crossings in the main fermion line (DBA)}
\label{NCA-like}
From the considerations above, the second-tier approximations well describe the behavior of the SIAM above temperatures of the order $k_{\rm B}T\sim \Gamma$. Also they capture the onset of a zero-bias anomaly at low temperature. However, artifacts occur when the temperature is lowered even further. As shown in the next sections, the problems with the Kondo temperature $T_{\rm K}$ and the pinning at the particle-hole symmetry point are mitigated by deepening the hierarchy of internal process including third- and fourth-tier bubble diagrams.\\
\indent Let us go to the exact formal expression for $\boldsymbol\phi$,  Eq.~\eqref{phi_exact_Dyson}, and assume that the main fermion line does not undergo crossings, which yields a diagonal irreducible propagator in $\sigma$ and  $\kappa$. 
Within this assumption, the equation for $\boldsymbol\phi\rightarrow \boldsymbol\phi_{\rm DBA}$ is
\begin{equation}\begin{aligned}\label{DysonNCA}
\boldsymbol{\phi}_{\rm DBA}=[\mathbf{h}^{-1}-\tilde{\mathbf{B}}]^{-1}\;,
\end{aligned}\end{equation}
see Eq.~\eqref{dressed_line}. 
The propagator is dressed by internal processes according to the hierarchy of Eq.~\eqref{hierarchy_S} with $\langle\tilde{\mathbf{S}}_2\mathbf{v}\rangle \rightarrow \tilde{\mathbf{B}}_2\equiv \tilde{\mathbf{B}}$, where the block $[\tilde{\mathbf{B}}]^{\sigma'\sigma}_{\boldsymbol\chi'\boldsymbol\chi}=\tilde{\rm B}^{\sigma}_{\eta'\eta}(\kappa)\delta_{\sigma'\sigma}\delta_{\kappa'\kappa}$ is given by the contraction of a free propagator dressed by all possible (irreducible) processes, including crossings. The diagrammatic of this dressing is in fact the same as the exact irreducible propagator $\boldsymbol\phi$ itself, see Fig.~\ref{dressed_BX}.\\
\indent As the bare bubbles of the  NCA2, the dressed bubbles are diagonal in all indexes except for $\eta$, inducing a $2\times 2$ structure to the contracted (matrix) function $\boldsymbol\varphi_{\rm DBA}(\kappa)$.
Along similar lines as in the NCA2, the matrix elements of $\boldsymbol\varphi_{\rm DBA}$ read
\begin{equation}\begin{aligned}\label{DBA_matrix_elements}
\boldsymbol{\varphi}^{\sigma\sigma}_{{\rm DBA},\eta\eta}=&\frac{({\rm h}^{\sigma}_{\bar\eta\bar\eta})^{-1}-\tilde{\rm B}^{\sigma}_{\bar\eta\bar\eta}}{[({\rm h}^{\sigma}_{\eta\eta})^{-1}-\tilde{\rm B}^{\sigma}_{\eta\eta}][({\rm h}^{\sigma}_{\bar\eta\bar\eta})^{-1}-\tilde{\rm B}^{\sigma}_{\bar\eta\bar\eta}]-\tilde{\rm B}^{\sigma}_{\eta\bar\eta}\tilde{\rm B}^{\sigma}_{\bar\eta\eta}}\\
\boldsymbol{\varphi}^{\sigma\sigma}_{{\rm DBA},\bar\eta\eta}=&\frac{\tilde{\rm B}^{\sigma}_{\bar\eta\eta}}{[({\rm h}^{\sigma}_{\eta\eta})^{-1}-\tilde{\rm B}^{\sigma}_{\eta\eta}][({\rm h}^{\sigma}_{\bar\eta\bar\eta})^{-1}-\tilde{\rm B}^{\sigma}_{\bar\eta\bar\eta}]-\tilde{\rm B}^{\sigma}_{\eta\bar\eta}\tilde{\rm B}^{\sigma}_{\bar\eta\eta}}\;,
\end{aligned}\end{equation}
where the bare block ${\rm B}$ has been replaced by the dressed one $\tilde{\rm B}$.
Similar to Eq.~\eqref{GF_gDSO}, the  dot Green's function in the $\infty$-tier scheme reads
\begin{equation}\begin{aligned}\label{Gzeta}
 \mathcal{G}^{(\zeta)}_{\sigma\sigma}
&=\frac{1-\langle \hat{n}_{\bar\sigma} \rangle}{\epsilon-\epsilon_\sigma-{\rm i}\zeta\hbar\tilde{\rm B}^{\sigma}_{--}+{\rm i}\zeta\hbar\tilde{\rm B}^{\sigma}_{+-}\frac{\epsilon-\epsilon_\sigma-{\rm i}\zeta\hbar[\tilde{\rm B}^{\sigma}_{--}+\tilde{\rm B}^{\sigma}_{-+}]}{\epsilon-\epsilon_\sigma-U-{\rm i}\zeta\hbar[\tilde{\rm B}^{\sigma}_{++}+\tilde{\rm B}^{\sigma}_{+-}]}}\\
+&\frac{\langle \hat{n}_{\bar\sigma} \rangle }{\epsilon-\epsilon_\sigma-U-{\rm i}\zeta\hbar\tilde{\rm B}^{\sigma}_{++}+{\rm i}\zeta\hbar\tilde{\rm B}^{\sigma}_{-+}\frac{\epsilon-\epsilon_\sigma-U-{\rm i}\zeta\hbar[\tilde{\rm B}^{\sigma}_{++}+\tilde{\rm B}^{\sigma}_{+-}]}{\epsilon-\epsilon_\sigma-{\rm i}\zeta\hbar[\tilde{\rm B}^{\sigma}_{--}+\tilde{\rm B}^{\sigma}_{-+}]}}
\;.
\end{aligned}\end{equation}
As above, the dependence on $\kappa$ is understood.
Thus, neglecting the crossing of the main fermion line, we obtain a general structure for the retarded Green's function by a simple $2\times 2$ (block) matrix inversion.
The matrix elements of the dressed bubbles are schematized by 
\begin{equation}\begin{aligned}\label{dressed_bubbles}
\tilde{\rm B}^{\sigma}_{\eta'\eta}=&
\begin{gathered}
\resizebox{!}{1.5cm}{
\begin{tikzpicture}[] 
\draw[blue,thick] (0.5,1.3) node[left] {\Large{$\sigma$}} arc (120:60:3cm  and 2.5cm) node[right] {\Large{$\sigma$}};
\draw[blue,thick] (1,0) arc (180:0:1.cm  and 1.1cm); 
\draw[blue,dashed,thick]  (0.5,0) -- (1,0); 
\draw[blue,thick] (1,0) -- (3,0); 
\draw[blue,dashed,thick] (3,0) -- (3.5,0); 
\draw[red,thick] (0.5,-0.5) node[left] {\Large{$\eta$}} -- node[right] {\Large{$\qquad\quad\eta'$}} (3.5,-0.5); 
\filldraw[blue] 
(1,0) circle (2pt);
\fill[white] (2.6,-0.6) rectangle (1.4,1.2);
\draw [pattern=north west lines, pattern color=black] (2.6,-0.6) rectangle (1.4,1.2);
 \end{tikzpicture}
 }
\end{gathered}
+
\begin{gathered}
\resizebox{!}{1.5cm}{
\begin{tikzpicture}[] 
\draw[blue,thick] (0.5,1.2) node[left] {\Large{$\sigma$}} arc (120:60:3cm  and 2.5cm) node[right] {\Large{$\sigma$}} ;
\draw[red,thick]  (1,0) arc (180:0:1.cm  and 1.1cm); 
\draw[blue,dashed,thick]  (0.5,0) -- (3.5,0); 
\draw[red,thick] (0.5,-0.5) node[left] {\Large{$\eta$}} -- (1,-0.5); 
\draw[red,thick] (1,-0.5) -- (1,0); 
\draw[red,dashed,thick] (1,-0.5) -- (3,-0.5); 
\draw[red,thick] (3,-0.5) -- node[right] {\Large{$\;\eta'$}}  (3.5,-0.5); 
\draw[red,thick] (3,-0.5) -- (3,-0); 
\filldraw[red](1,-0.5) circle (2pt); 
\fill[white] (2.6,-0.6) rectangle (1.4,1.2);
\draw [pattern=north west lines, pattern color=black] (2.6,-0.6) rectangle (1.4,1.2);
 \end{tikzpicture}
 }
\end{gathered}\;.
\end{aligned}\end{equation}
Here, the bare contribution (no internal processes) to the first bubble diagram is diagonal in $\eta'\eta$ and is also independent of the value of the sojourn $\eta$, as it does not include interactions. This bare bubble is the same as the one of the NCA2 and, in the WBL, is evaluated to be $-\mathbf{1}\Gamma/2\hbar$, see Eq.~\eqref{B_SIAM}. In the presence of internal processes, the first dressed bubble in Eq.~\eqref{dressed_bubbles} depends in principle on the sojourn $\eta'$. This occurs when the latter has overlap with a blip of the $\sigma$ path. However, in this case the block vanishes because the overlap implies overlap of three fermion lines of the state $\sigma$, resulting in a factor $\nu$ that makes the whole diagram vanish upon summing over $\nu$, as exemplified in the diagram $(B)$ of Eq.~\eqref{examples8th}. 
As a result, this block is independent of the value of the last sojourn $\eta'$. The same independence of  $\eta'$ holds for the second block, because the sojourn $\eta'$, although involved with interactions, lies outside the block itself. On the contrary, the initial sojourn is relevant for both bubbles because it determines the vertex of the fermion line of spin $\bar\sigma$ according to the definition in Eq.~\eqref{vertexes}. Summarizing, on the basis of the diagrammatic rules we conclude that the dressed bubble  $\tilde{\mathbf{B}}$ possess the same property of the bare bubble, namely  
\begin{equation}\begin{aligned}\label{symmetry_dressedB}
\tilde{\mathbf{B}}_{\eta'\eta}=&\;-\frac{\Gamma}{2\hbar}\mathbf{1}\delta_{\eta'\eta}+\tilde{\mathbf{B}}_{\bar\eta\eta}\;,
\end{aligned}\end{equation}
where, again, boldface objects indicate  diagonal matrices in the indexes $\sigma$ and $\boldsymbol\kappa$.
Exploiting the  symmetry in Eq.~\eqref{symmetry_dressedB}, we can simplify the form of the Green's function in Eq~\eqref{Gzeta}. In particular, the retarded ($\zeta=+1$) Green's function in the WBL reads
\begin{equation}\begin{aligned}\label{GRnoncrossing}
\mathcal{G}^r_{\sigma\sigma}(\epsilon)&=\frac{1-\langle \hat{n}_{\bar\sigma} \rangle }{\epsilon-\epsilon_\sigma+{\rm i}\frac{\Gamma}{2}+\tilde{\Sigma}_{\sigma-}(\epsilon)\frac{U}{\epsilon-\epsilon_\sigma-U+{\rm i}\frac{\Gamma}{2}-\tilde{\Sigma}_{\sigma}(\epsilon)}}\\
+&\frac{\langle \hat{n}_{\bar\sigma} \rangle}{\epsilon-\epsilon_\sigma-U+{\rm i}\frac{\Gamma}{2}-\tilde{\Sigma}_{\sigma+}(\epsilon)\frac{U}{\epsilon-\epsilon_\sigma+{\rm i}\frac{\Gamma}{2}-\tilde{\Sigma}_{\sigma}(\epsilon)}}
\;,
\end{aligned}\end{equation}
where
\begin{equation}
\tilde\Sigma_{\sigma}(\epsilon)=\sum_\eta \tilde\Sigma_{\sigma\eta}(\epsilon)\;.
\end{equation}
Equation~\eqref{GRnoncrossing} is one of the main results of the present work.
The problem of calculating the retarded Green's function, and thus the relevant physical properties for proportional coupling, reduces to that of determining the retarded, dressed self-energies $\tilde\Sigma_{\sigma\eta}$, here identified with the off-diagonal elements of the dressed bubbles via
\begin{equation}\begin{aligned}\label{relationB-Sigma}
\tilde{\Sigma}_{\sigma\eta}(\epsilon):={\rm i}\hbar\tilde{\rm B}^{\sigma}_{\bar\eta\eta}(\kappa)|_{\zeta=+1}\;.
\end{aligned}\end{equation}
The evaluation of these self-energies remains complicated due to the inner processes dressing the bubble $\tilde{\mathbf{B}}^{\sigma}$.\\
\indent The retarded Green's function in Eq.~\eqref{GRnoncrossing} has the same form of the one found in~\cite{Lavagna2015} with a self-consistent truncation of the equations of motion. As such, provided that the self-energies have the correct form, the DBA can in principle  reproduce the unitary limit at $T=0$. In the degenerate case, the latter is obtained if, at the particle-hole symmetry point $\epsilon_0-\mu=-U/2$, the dressed self-energies acquire the values
\begin{equation}\label{SE_T0}
\tilde{\Sigma}_{\sigma\pm}(\mu)=\pm \frac{\Gamma}{\pi}\ln\left(\frac{2k_{\rm B}T_{\rm K}}{\sqrt{U\Gamma}}\right)+{\rm i}\frac{{\rm Im}\tilde{\Sigma}_{\sigma}(\mu)}{2}
\end{equation}
in terms of the Kondo temperature, Eq.~\eqref{TKondo}, so that ${\rm Im}\mathcal{G}^r_{\sigma\sigma}(\mu)=-2/\Gamma$.
 We next introduce an approximation scheme, the noncrossing approximation (NCA), where also the internal crossings are neglected. Note that this scheme, obtained by systematically neglecting the crossings in the diagrammatic unravelling of the self-energies, \emph{does not coincide} with the NCA  well-known in the Green's functions literature~\cite{Wingreen1994}

\subsection{Infinite-tier schemes II: Neglecting all crossings (NCA)}
\label{NCA}

In the absence of crossings at all levels, the hierarchy in Eq.~\eqref{hierarchy_S} simplifies to
\begin{equation}\begin{aligned}\label{hierarchy_B}
\tilde{\mathbf{h}}_{n-1}=&\;\sum_{k=0}^\infty\Bigg(\mathbf{h}_{n-1}\langle \tilde{\mathbf{h}}_{n}\mathbf{v}\rangle\Bigg)^k\mathbf{h}_{n-1}\\
=&\;[\mathbf{h}_{n-1}^{-1}-\tilde{\mathbf{B}}_{n}^{\rm NCA}]^{-1}\;,
\end{aligned}\end{equation}
with $\mathbf{h}_n=\mathbf{1}{\rm h}_n$ denoting the bare propagator with $n$ overlapping fermion lines and $\tilde{\mathbf{h}}_n$ the corresponding  propagator  dressed by higher-tier bubbles $\tilde{\mathbf{B}}_{n}^{\rm NCA}=\langle \tilde{\mathbf{h}}_{n}\mathbf{v}\rangle$. In particular $\tilde{\mathbf{B}}_2^{\rm NCA}=\langle \tilde{\mathbf{h}}_2\mathbf{v}\rangle\equiv \tilde{\mathbf{B}}$, where $ \tilde{\mathbf{h}}_2=[\mathbf{h}_2^{-1}-\tilde{\mathbf{B}}_{3}^{\rm NCA}]^{-1}$. In the following we drop the indexes and make the identification $\tilde{\mathbf{B}}_2^{\rm NCA}\equiv\tilde{\mathbf{B}}$. The dressed bubble $\tilde{\mathbf{B}}$ in Eq.~\eqref{dressed_bubbles}  is then schematically described by the following sum of  two contributions
\begin{equation}\begin{aligned}\label{dressed_bubbles_gDSO4}
\tilde{\rm B}^{\sigma}_{\eta'\eta}=&\;
\tilde{\rm B}^{\sigma(\sigma)}_{\eta'\eta}+\tilde{\rm B}^{\sigma(\bar\sigma)}_{\eta'\eta}\\
=&
\begin{gathered}
\resizebox{!}{1.5cm}{
\begin{tikzpicture}[] 
\draw[blue,line width=0.5mm] (0.5,1.3) arc (120:60:3cm  and 2.5cm) ;
\draw[blue,line width=0.5mm] (1,0) arc (180:0:1.cm  and 1.2cm); 
\draw[blue,dashed,line width=0.5mm]  (0.5,0) -- (1,0); 
\draw[blue,line width=0.5mm] (1,0) -- (3,0); 
\draw[blue,dashed,line width=0.5mm] (3,0) -- (3.5,0); 
\draw[red,line width=0.5mm] (0.5,-0.5) node[left] {\Large{$\eta$}} -- node[right] {\Large{$\qquad\quad\eta'$}} (3.5,-0.5); 
\filldraw[blue] 
(1,0) circle (3pt);
\fill[white] (2.6,-0.8) rectangle (1.4,.4);
\draw [] (2.6,-0.8) rectangle (1.4,.4);
\draw[](2.,-0.2) node[] {\LARGE{$\tilde{\mathbf{h}}^{\sigma(\sigma)}_{2}$}}; 
\end{tikzpicture}
 }
\end{gathered}
+
\begin{gathered}
\resizebox{!}{1.5cm}{
\begin{tikzpicture}[] 
\draw[blue,line width=0.5mm] (0.5,1.2) arc (120:60:3cm  and 2.5cm) ;
\draw[red,line width=0.5mm]  (1,0) arc (180:0:1.cm  and 1.2cm); 
\draw[blue,dashed,line width=0.5mm]  (0.5,0) -- (3.5,0); 
\draw[red,line width=0.5mm] (0.5,-0.5) node[left] {\Large{$\eta$}} -- (1,-0.5); 
\draw[red,line width=0.5mm] (1,-0.5) -- (1,0); 
\draw[red,dashed,line width=0.5mm] (1,-0.5) -- (3,-0.5); 
\draw[red,line width=0.5mm] (3,-0.5) -- node[right] {\Large{$\;\eta'$}}  (3.5,-0.5); 
\draw[red,line width=0.5mm] (3,-0.5) -- (3,-0); 
\filldraw[red](1,-0.5) circle (3pt); 
\fill[white] (2.6,-0.8) rectangle (1.4,.4);
\draw [] (2.6,-0.8) rectangle (1.4,.4);
\draw[](2.,-0.2) node[] {\LARGE{$\tilde{\rm h}_2^{\sigma(\bar\sigma)}$}}; 
 \end{tikzpicture}
 }
\end{gathered}\;.
\end{aligned}\end{equation}
Here, the white rectangles indicate the dressing of the propagators $\mathbf{h}_2$ with two overlapping fermion lines by iteration of third-tier bubbles. Specifically, the dressed bubble of type $\sigma(\sigma)$ is given by the following contraction of a dressed propagator with $4\times 4$ matrix structure in $\boldsymbol{\eta}=(\nu,\eta)$
\begin{equation}\begin{aligned}\label{dressed_bubble+}
\tilde{\rm B}_{\eta'\eta}^{\sigma(\sigma)}=&\sum_{\nu}\langle\sum_{\nu'}[\tilde{\mathbf{h}}^{\sigma(\sigma)}_{2}]_{\boldsymbol{\eta}'\boldsymbol{\eta}}{\rm v}_{\nu}\rangle\;,
\end{aligned}\end{equation}
where $\tilde{\mathbf{h}}_2^{\sigma(\sigma)}$ is obtained by dressing the bare propagator  $\mathbf{1}{\rm h}_2^{\sigma(\sigma)}$, Eq.~\eqref{h2_+_SIAM}, with the third-tier bubble $\tilde{\mathbf{B}}_3^{\sigma(\sigma)}$, namely
\begin{equation}\begin{aligned}\label{tildeh2ss}
\tilde{\mathbf{h}}_2^{\sigma(\sigma)}=\left[[\mathbf{1}{\rm h}_2^{\sigma(\sigma)}]^{-1}-\tilde{\mathbf{B}}_3^{\sigma(\sigma)}\right]^{-1}\;.
\end{aligned}
\end{equation}
The third-tier bubble $\tilde{\mathbf{B}}_3^{\sigma(\sigma)}$ is in turn given by the sum
\begin{equation}\begin{aligned}\label{3rdtier_gDSO4_s}
\tilde{\rm B}_{3,\boldsymbol{\eta}'\boldsymbol{\eta}}^{\sigma(\sigma)}=&
\begin{gathered}
\resizebox{!}{1.9cm}{
\begin{tikzpicture}[] 
\draw[blue,line width=0.5mm] (0.5,1.5) arc (120:60:3cm  and 2.cm) ;
\draw[blue,line width=0.5mm] (0.5,.8) arc (120:60:3cm  and 5cm) ;
\draw[red,line width=0.5mm]  (1,0) arc (180:0:1.cm  and 1.2cm); 
\draw[blue,line width=0.5mm]  (0.5,0) node[left] {\Large{$\nu$}}   --  (3.5,0) node[right] {\Large{$\;\nu'$}}; 
\draw[red,line width=0.5mm] (0.5,-0.5) node[left] {\Large{$\eta$}} -- (1,-0.5); 
\draw[red,line width=0.5mm] (1,-0.5) -- (1,0); 
\draw[red,dashed,line width=0.5mm] (1,-0.5) -- (3,-0.5); 
\draw[red,line width=0.5mm] (3,-0.5) -- node[right] {\Large{$\;\eta'$}}  (3.5,-0.5); 
\draw[red,line width=0.5mm] (3,-0.5) -- (3,-0); 
\filldraw[red](1,-0.5) circle (3pt); 
\fill[white] (2.6,-0.8) rectangle (1.4,.4);
\draw [] (2.6,-0.8) rectangle (1.4,.4);
\draw[](2.,-0.2) node[] {\large{$\tilde{\mathbf{h}}_3^{\sigma(\sigma\bar\sigma)}$}}; 
 \end{tikzpicture}
 }
\end{gathered}
+
\begin{gathered}
\resizebox{!}{1.9cm}{
\begin{tikzpicture}[] 
\draw[blue,line width=0.5mm] (0.5,1.5) arc (120:60:3cm  and 2.cm) ;
\draw[blue,line width=0.5mm] (0.5,.8) arc (120:60:3cm  and 5cm) ;
\draw[blue,line width=0.5mm] (1,0) arc (180:0:1.cm  and 1.2cm); 
\draw[blue,line width=0.5mm]  (0.5,0)node[left] {\Large{$\nu$}}  -- (1,0); 
\draw[blue,dashed,line width=0.5mm] (1,0) -- (3,0); 
\draw[blue,line width=0.5mm] (3,0) -- (3.5,0)node[right] {\Large{$\;\nu'$}} ; 
\draw[red,line width=0.5mm] (0.5,-0.5) node[left] {\Large{$\eta$}} -- node[right] {\Large{$\qquad\quad\eta'$}} (3.5,-0.5); 
\filldraw[blue] 
(1,0) circle (3pt);
\fill[white] (2.6,-0.8) rectangle (1.4,.4);
\draw [] (2.6,-0.8) rectangle (1.4,.4);
\draw[](2.,-0.2) node[] {\large{$\tilde{\mathbf{h}}_3^{\sigma(\sigma\sigma)}$}};
 \end{tikzpicture}
 }
\end{gathered}\\
=&\;\nu'\nu\langle \tilde{\rm h}^{\sigma(\sigma\bar\sigma)}_{3,\nu'\nu}{\rm v}_{-\eta}\rangle+\nu'\nu\langle \tilde{\rm h}^{\sigma(\sigma\sigma)}_{3,\eta'\eta}{\rm v}_{-\nu}\rangle\\
\equiv&\;\nu'\nu\tilde{\rm B}_{3,\nu'\nu}^{A}(\eta) +\nu'\nu \tilde{\rm B}_{3,\eta'\eta}^{B}(\nu)\;.
\end{aligned}\end{equation}
  The prefactor $\nu'\nu$ stems from overlap of three fermion lines of spin $\sigma$, according to the diagrammatic rules, see Eqs.~\eqref{3rda}-\eqref{3rdc} and also Eq.~\eqref{4tha}. Note that these third-tier bubbles can in principle change the occupation state of both spin degrees of freedom. As a result, they capture spin-flip process, which are virtual processes by which the state of the dot with single occupation changes spin due to multiple (virtual) transitions as for example in
$$\uparrow\;\longrightarrow\;(\uparrow\downarrow)\;\longrightarrow\; \downarrow\;.$$
We anticipate that, since in the SIAM we deal with two degrees of freedom, the $4 \times 4$ structure of $\tilde{\mathbf{B}}_{3}^{\sigma(\sigma)}$ is the largest in the hierarchical analysis.\\   
\indent The other second-tier bubble in Eq.~\eqref{dressed_bubbles_gDSO4} is $\tilde{\rm B}^{\sigma(\bar\sigma)}$. It is calculated as the contraction of a dressed propagator which bears no structure in the sojourn indexes
\begin{equation}\begin{aligned}\label{dressed_bubble-}
\tilde{\rm B}_{\eta'\eta}^{\sigma(\bar\sigma)}=&\langle\tilde{\rm h}_2^{\sigma(\bar\sigma)}{\rm v}_{-\eta} \rangle\;,
\end{aligned}\end{equation}
where $\tilde{\rm h}_2^{\sigma(\bar\sigma)}$
is obtained by dressing the bare propagator ${\rm h}_2^{\sigma(\bar\sigma)}$ in Eq.~\eqref{h2_-_SIAM} with the third-tier bubble $\tilde{\rm B}_3^{\sigma(\bar\sigma)}$  according to
\begin{equation}\begin{aligned}\label{}
\tilde{\rm h}_2^{\sigma(\bar\sigma)}=\left[[{\rm h}_2^{\sigma(\bar\sigma)}]^{-1}-\tilde{\rm B}_3^{\sigma(\bar\sigma)}\right]^{-1}\;.
\end{aligned}
\end{equation}
The third-tier bubble entering this equation is
\begin{equation}\begin{aligned}\label{3rdtier_gDSO4_o}
\tilde{\rm B}_3^{\sigma(\bar\sigma)}=&
\begin{gathered}
\resizebox{!}{1.9cm}{
\begin{tikzpicture}[] 
\draw[blue,line width=0.5mm] (0.5,1.5) arc (120:60:3cm  and 2.cm) ;
\draw[red,line width=0.5mm] (0.5,.8) arc (120:60:3cm  and 5cm) ;
\draw[blue,line width=0.5mm] (1,0) arc (180:0:1.cm  and 1.2cm); 
\draw[blue,dashed,line width=0.5mm]  (0.5,0) -- (1,0); 
\draw[blue,line width=0.5mm] (1,0) -- (3,0); 
\draw[blue,dashed,line width=0.5mm] (3,0) -- (3.5,0); 
\draw[red,dashed,line width=0.5mm] (0.5,-0.5) node[left,white] {\Large{$\eta$}} -- node[right,white] {\Large{$\qquad\quad\eta'$}} (3.5,-0.5); 
\filldraw[blue] 
(1,0) circle (3pt);
\fill[white] (2.6,-0.8) rectangle (1.4,.4);
\draw [] (2.6,-0.8) rectangle (1.4,.4);
\draw[](2.,-0.2) node[] {\large{$\tilde{\mathbf{h}}_3^{\sigma(\sigma\bar\sigma)}$}};
 \end{tikzpicture}
 }
\end{gathered}
+
\begin{gathered}
\resizebox{!}{1.9cm}{
\begin{tikzpicture}[] 
\draw[blue,line width=0.5mm] (0.5,1.5) arc (120:60:3cm  and 2.cm) ;
\draw[red,line width=0.5mm] (0.5,.8) arc (120:60:3cm  and 5cm) ;
\draw[red,line width=0.5mm]  (1,0) arc (180:0:1.cm  and 1.2cm); 
\draw[blue,dashed,line width=0.5mm]  (0.5,0) -- (3.5,0); 
\draw[red,dashed,line width=0.5mm] (0.5,-0.5) node[left,white] {\Large{$\eta$}} -- (1,-0.5); 
\draw[red,line width=0.5mm] (1,-0.5) -- (1,0); 
\draw[red,line width=0.5mm] (1,-0.5) -- (3,-0.5); 
\draw[red,dashed,line width=0.5mm] (3,-0.5) -- node[right,white] {\Large{$\;\eta'$}}  (3.5,-0.5); 
\draw[red,line width=0.5mm] (3,-0.5) -- (3,-0); 
\filldraw[red](1,-0.5) circle (3pt); 
\fill[white] (2.6,-0.8) rectangle (1.4,.4);
\draw [] (2.6,-0.8) rectangle (1.4,.4);
\draw[](2.,-0.2) node[] {\large{$\tilde{\mathbf{h}}_3^{\sigma(\bar\sigma\bar\sigma)}$}};
 \end{tikzpicture}
 }
\end{gathered}\\
=&\sum_{\nu}\langle\sum_{\nu'} \tilde{\rm h}^{\sigma(\bar\sigma\sigma)}_{3,\nu'\nu}{\rm v}_{\nu}\rangle+\sum_{\eta}\langle\sum_{\eta'} \tilde{\rm h}^{\sigma(\bar\sigma\bar\sigma)}_{3,\eta'\eta}{\rm v}_{\eta}\rangle\;.
\end{aligned}\end{equation}
Note that also the bubble $\tilde{\mathbf{B}}_3^{\sigma(\bar\sigma)}$ has no structure in the sojourn indexes. Moreover, we have included the sums over the sojourns internal to the bubbles in the definitions. The corresponding self-energy $\tilde{\Sigma}_{\sigma\eta}^{(\bar\sigma)}(\epsilon):={\rm i}\hbar\tilde{\rm B}^{\sigma(\bar\sigma)}_{\bar\eta\eta}(\kappa)|_{\zeta=+1}$ is given by
\begin{equation}\begin{aligned}\label{SigmaB_sNCA4}
\tilde{\Sigma}_{\sigma\eta}^{(\bar\sigma)}(\epsilon)
=-\eta  \sum_{\alpha}\frac{\Gamma_\alpha}{2\pi} \Bigg[
&\psi\left(\frac{1}{2}+{\rm i}\frac{\tilde{\mathcal{E}}_{+}-\mu_\alpha}{2\pi k_{\rm B}T}\right)\\
-&\psi^* \left(\frac{1}{2}+{\rm i}\frac{\tilde{\mathcal{E}}_{-}-\mu_\alpha}{2\pi k_{\rm B}T}\right)
\Bigg]-{\rm i}\frac{\Gamma}{2}\;,
\end{aligned}\end{equation}
see Appendix~\ref{dressingB-}, where
\begin{equation}\begin{aligned}\label{}
\tilde{\mathcal{E}}_{\zeta_1}=\epsilon_{\bar\sigma}+\zeta_1(\epsilon_\sigma-\epsilon)+\delta_{\zeta_1,+1}U +\zeta_1{\rm Re}\;\tilde{\Sigma}_{3\sigma,\zeta_1}^{(\bar\sigma)}-{\rm i}|{\rm Im}\;\tilde{\Sigma}_{3\sigma,\zeta_1}^{(\bar\sigma)}|\;.
\end{aligned}\end{equation}
Note that for vanishing third-tier self-energies, given by Eq.~\eqref{B3d_2}, one recovers the NCA2 self-energies $\Sigma_{\sigma\eta}(\epsilon)$, see Eq.~\eqref{relationB-Sigma_gDSO}.\\
\indent The task of finding the dressed self-energies $\tilde{\Sigma}_{\sigma\eta}(\epsilon)$ has thus been reduced to the evaluation of the dressed bubbles $\tilde{\mathbf{B}}_3$ together with the inversion of the $4\times4$ matrix in Eq.~\eqref{tildeh2ss}.
In turn, the propagators  $\tilde{\mathbf{h}}_3$ in Eqs.~\eqref{3rdtier_gDSO4_s} and~\eqref{3rdtier_gDSO4_o} are given by dressing the bare propagators $\mathbf{h}_3$ with overlap of three fermion lines with the fourth-tier bubbles $\tilde{\mathbf{B}}_4$, namely 
\begin{equation}\begin{aligned}\label{}
\tilde{\mathbf{h}}_3=&\left[[\mathbf{h}_3]^{-1}-\tilde{\mathbf{B}}_4\right]^{-1}\;,
\end{aligned}\end{equation}
where we made no reference to the spin of the fermion lines. \\
\indent The hierarchy of internal processes proceeds similarly for higher overlaps of fermion lines. Note that the dimension associated to the matrix structure in the sojourn indexes varies between $0$ and $4$ (never exceeding this upper bound in the SIAM) according to the number and the spin of the overlapping fermion lines.

\subsection{Fourth-tier scheme: NCA4}
\label{gDSO4}

Up to this point, the description of the hierarchy of diagrammatic contributions to the second-tier bubbles, namely to the self-energies, see Eq.~\eqref{relationB-Sigma}, is exact, within the approximation of neglecting the crossings. The truncation of the hierarchy in Eq.~\eqref{hierarchy_B} to the level $n=4$ gives rise to a fourth-tier scheme where $\tilde{\mathbf{h}}_{4}=\mathbf{h}_{4}$, namely $\tilde{\mathbf{B}}_4\equiv \mathbf{B}_4$.\\ 
\indent The fourth-tier bubbles $\mathbf{B}_4$ are similar to the NCA2 second-tier bubbles ${\rm B}^{\sigma}_{\eta'\eta}$, see Eq.~\eqref{B_SIAM}, except for the additional layers of fermion lines, and the products of sojourn indexes associated to the overlap of three fermion lines of the same spin. The 4th-tier bubbles dressing the propagators $\tilde{\rm h}_3$ in Eq.~\eqref{3rdtier_gDSO4_s} are schematized as 
\begin{equation}\begin{aligned}\label{B4s}
{\rm B}^{\sigma(\sigma\bar\sigma)}_{4,\nu'\nu}=&
\begin{gathered}
\resizebox{!}{1.6cm}{
\begin{tikzpicture}[] 
\draw[blue,line width=0.5mm] (0.5,2) arc (120:60:3cm  and 1.5cm) ;
\draw[blue,line width=0.5mm] (0.5,1.6) arc (120:60:3cm  and 2.cm) ;
\draw[red,line width=0.5mm] (0.5,1.2) arc (120:60:3cm  and 2.5cm) ;
\draw[red,line width=0.5mm]  (1,0) arc (180:0:1.cm  and 1.2cm); 
\draw[blue,line width=0.5mm]  (0.5,0)node[left]{\Large{$\nu$}} -- node[right]{\Large{$\qquad\quad\nu$}} (3.5,0); 
\draw[red,dashed,line width=0.5mm] (0.5,-0.5) -- (1,-0.5); 
\draw[red,line width=0.5mm] (1,-0.5) -- (1,0); 
\draw[red,line width=0.5mm] (1,-0.5) -- (3,-0.5); 
\draw[red,dashed,line width=0.5mm] (3,-0.5) --   (3.5,-0.5); 
\draw[red,line width=0.5mm] (3,-0.5) -- (3,-0); 
\filldraw[red](1,-0.5) circle (3pt); 
 \end{tikzpicture}
 }
\end{gathered}
+
\begin{gathered}
\resizebox{!}{1.6cm}{
\begin{tikzpicture}[] 
\draw[blue,line width=0.5mm] (0.5,2) arc (120:60:3cm  and 1.5cm) ;
\draw[blue,line width=0.5mm] (0.5,1.6) arc (120:60:3cm  and 2.cm) ;
\draw[red,line width=0.5mm] (0.5,1.2) arc (120:60:3cm  and 2.5cm) ;
\draw[blue,line width=0.5mm] (1,0) arc (180:0:1.cm  and 1.2cm); 
\draw[blue,line width=0.5mm]  (0.5,0)node[left] {\Large{$\nu$}} --(1,0); 
\draw[blue,dashed,line width=0.5mm] (1,0) -- (3,0); 
\draw[blue,line width=0.5mm] (3,0)--  node[]{\Large{$\qquad\quad\nu'$}}(3.5,0) ; 
\draw[red,dashed,line width=0.5mm] (0.5,-0.5) --  (3.5,-0.5); 
\filldraw[blue] 
(1,0) circle (3pt);
\end{tikzpicture}
 }
\end{gathered}\\
=&\sum_\eta\langle {\rm h}^{\sigma(\sigma\bar\sigma\bar\sigma)}_{4,\nu'\nu}{\rm v}_{\eta}\rangle+\nu'\nu\langle \tilde{\rm h}^{\sigma(\sigma\bar\sigma\sigma)}_{4}{\rm v}_{-\nu}\rangle\\
{\rm B}^{\sigma(\sigma\sigma)}_{4,\eta'\eta}=&
\begin{gathered}
\resizebox{!}{1.6cm}{
\begin{tikzpicture}[] 
\draw[blue,line width=0.5mm] (0.5,2) arc (120:60:3cm  and 1.5cm) ;
\draw[blue,line width=0.5mm] (0.5,1.6) arc (120:60:3cm  and 2.cm) ;
\draw[blue,line width=0.5mm] (0.5,1.2) arc (120:60:3cm  and 2.5cm) ;
\draw[blue,line width=0.5mm] (1,0) arc (180:0:1.cm  and 1.2cm); 
\draw[blue,dashed,line width=0.5mm]  (0.5,0) -- (1,0); 
\draw[blue,line width=0.5mm] (1,0) -- (3,0); 
\draw[blue,dashed,line width=0.5mm] (3,0) -- (3.5,0); 
\draw[red,line width=0.5mm] (0.5,-0.5) node[left] {\Large{$\eta$}} -- node[right] {\Large{$\qquad\quad\eta$}} (3.5,-0.5); 
\filldraw[blue] 
(1,0) circle (3pt);
\end{tikzpicture}
 }
\end{gathered}
+
\begin{gathered}
\resizebox{!}{1.6cm}{
\begin{tikzpicture}[] 
\draw[blue,line width=0.5mm] (0.5,2) arc (120:60:3cm  and 1.5cm) ;
\draw[blue,line width=0.5mm] (0.5,1.6) arc (120:60:3cm  and 2.cm) ;
\draw[blue,line width=0.5mm] (0.5,1.2) arc (120:60:3cm  and 2.5cm) ;
\draw[red,line width=0.5mm]  (1,0) arc (180:0:1.cm  and 1.2cm); 
\draw[blue,dashed,line width=0.5mm]  (0.5,0) -- (3.5,0); 
\draw[red,line width=0.5mm] (0.5,-0.5) node[left] {\Large{$\eta$}} -- (1,-0.5); 
\draw[red,line width=0.5mm] (1,-0.5) -- (1,0); 
\draw[red,dashed,line width=0.5mm] (1,-0.5) -- (3,-0.5); 
\draw[red,line width=0.5mm] (3,-0.5) -- node[right] {\Large{$\;\eta'$}}  (3.5,-0.5); 
\draw[red,line width=0.5mm] (3,-0.5) -- (3,-0); 
\filldraw[red](1,-0.5) circle (3pt); 
\end{tikzpicture}
 }
\end{gathered}\\
=&\sum_{\nu}\langle {\rm h}^{\sigma(\sigma\sigma\sigma)}_{4,\eta'\eta}{\rm v}_{\nu}\rangle+\langle \tilde{\rm h}^{\sigma(\sigma\sigma\bar\sigma)}_{4}{\rm v}_{-\eta}\rangle
\;.
\end{aligned}\end{equation}
Analogously, the propagators $\tilde{\rm h}_3$ in Eq.~\eqref{3rdtier_gDSO4_o} are dressed by the fourth-tier bubbles
\begin{equation}\begin{aligned}\label{B4o}
{\rm B}^{\sigma(\bar\sigma\sigma)}_{4,\nu'\nu}=&
\begin{gathered}
\resizebox{!}{1.6cm}{
\begin{tikzpicture}[] 
\draw[blue,line width=0.5mm] (0.5,2) arc (120:60:3cm  and 1.5cm) ;
\draw[red,line width=0.5mm] (0.5,1.6) arc (120:60:3cm  and 2.cm) ;
\draw[blue,line width=0.5mm] (0.5,1.2) arc (120:60:3cm  and 2.5cm) ;
\draw[red,line width=0.5mm]  (1,0) arc (180:0:1.cm  and 1.2cm); 
\draw[blue,line width=0.5mm]  (0.5,0)node[left]{\Large{$\nu$}} -- node[right]{\Large{$\qquad\quad\nu$}} (3.5,0); 
\draw[red,dashed,line width=0.5mm] (0.5,-0.5) -- (1,-0.5); 
\draw[red,line width=0.5mm] (1,-0.5) -- (1,0); 
\draw[red,line width=0.5mm] (1,-0.5) -- (3,-0.5); 
\draw[red,dashed,line width=0.5mm] (3,-0.5) --   (3.5,-0.5); 
\draw[red,line width=0.5mm] (3,-0.5) -- (3,-0); 
\filldraw[red](1,-0.5) circle (3pt); 
 \end{tikzpicture}
 }
\end{gathered}
+
\begin{gathered}
\resizebox{!}{1.6cm}{
\begin{tikzpicture}[] 
\draw[blue,line width=0.5mm] (0.5,2) arc (120:60:3cm  and 1.5cm) ;
\draw[red,line width=0.5mm] (0.5,1.6) arc (120:60:3cm  and 2.cm) ;
\draw[blue,line width=0.5mm] (0.5,1.2) arc (120:60:3cm  and 2.5cm) ;
\draw[blue,line width=0.5mm] (1,0) arc (180:0:1.cm  and 1.2cm); 
\draw[blue,line width=0.5mm]  (0.5,0)node[left] {\Large{$\nu$}} --(1,0); 
\draw[blue,dashed,line width=0.5mm] (1,0) -- (3,0); 
\draw[blue,line width=0.5mm] (3,0)--  node[]{\Large{$\qquad\quad\nu'$}}(3.5,0) ; 
\draw[red,dashed,line width=0.5mm] (0.5,-0.5) --  (3.5,-0.5); 
\filldraw[blue] 
(1,0) circle (3pt);
\end{tikzpicture}
 }
\end{gathered}\\
=&\sum_\eta\langle {\rm h}^{\sigma(\bar\sigma\sigma\bar\sigma)}_{4,\nu'\nu}{\rm v}_{\eta}\rangle+\nu'\nu\langle {\rm h}^{\sigma(\bar\sigma\sigma\sigma)}_{4}{\rm v}_{-\nu}\rangle\\
{\rm B}^{\sigma(\bar\sigma\bar\sigma)}_{4,\eta'\eta}=&
\begin{gathered}
\resizebox{!}{1.6cm}{
\begin{tikzpicture}[] 
\draw[blue,line width=0.5mm] (0.5,2) arc (120:60:3cm  and 1.5cm) ;
\draw[red,line width=0.5mm] (0.5,1.6) arc (120:60:3cm  and 2.cm) ;
\draw[red,line width=0.5mm] (0.5,1.2) arc (120:60:3cm  and 2.5cm) ;
\draw[blue,line width=0.5mm] (1,0) arc (180:0:1.cm  and 1.2cm); 
\draw[blue,dashed,line width=0.5mm]  (0.5,0) -- (1,0); 
\draw[blue,line width=0.5mm] (1,0) -- (3,0); 
\draw[blue,dashed,line width=0.5mm] (3,0) -- (3.5,0); 
\draw[red,line width=0.5mm] (0.5,-0.5) node[left] {\Large{$\eta$}} -- node[right] {\Large{$\qquad\quad\eta$}} (3.5,-0.5); 
\filldraw[blue] 
(1,0) circle (3pt);
\end{tikzpicture}
 }
\end{gathered}
+
\begin{gathered}
\resizebox{!}{1.6cm}{
\begin{tikzpicture}[] 
\draw[blue,line width=0.5mm] (0.5,2) arc (120:60:3cm  and 1.5cm) ;
\draw[red,line width=0.5mm] (0.5,1.6) arc (120:60:3cm  and 2.cm) ;
\draw[red,line width=0.5mm] (0.5,1.2) arc (120:60:3cm  and 2.5cm) ;
\draw[red,line width=0.5mm]  (1,0) arc (180:0:1.cm  and 1.2cm); 
\draw[blue,dashed,line width=0.5mm]  (0.5,0) -- (3.5,0); 
\draw[red,line width=0.5mm] (0.5,-0.5) node[left] {\Large{$\eta$}} -- (1,-0.5); 
\draw[red,line width=0.5mm] (1,-0.5) -- (1,0); 
\draw[red,dashed,line width=0.5mm] (1,-0.5) -- (3,-0.5); 
\draw[red,line width=0.5mm] (3,-0.5) -- node[right] {\Large{$\;\eta'$}}  (3.5,-0.5); 
\draw[red,line width=0.5mm] (3,-0.5) -- (3,-0); 
\filldraw[red](1,-0.5) circle (3pt); 
\end{tikzpicture}
 }
\end{gathered}\\
=&\sum_{\nu}\langle {\rm h}^{\sigma(\bar\sigma\bar\sigma\sigma)}_{4,\eta'\eta}{\rm v}_{\nu}\rangle+\eta'\eta\langle {\rm h}^{\sigma(\bar\sigma\bar\sigma\bar\sigma)}_{4}{\rm v}_{-\eta}\rangle
\;.
\end{aligned}\end{equation}
Note the prefactor $\eta'\eta$ in the last line which is absent in Eq.~\eqref{B4s}.
The structure of the dressed propagators $\tilde{\mathbf{h}}_3^{\sigma}$ is the same as the one of $\boldsymbol\phi^{\sigma\sigma}_{\rm NCA2}$, see Eq.~\eqref{GD-solution} and of $\boldsymbol\phi^{\sigma\sigma}_{\rm DBA}$, Eq.~\eqref{DBA_matrix_elements}, and their matrix elements read (we omit any reference to the spin)
\begin{equation}\begin{aligned}\label{h3dressed}
\tilde{\rm h}_{3,\eta\eta}=&\frac{{\rm h}^{-1}_{3,\bar\eta\bar\eta}-{\rm B}_{4,\bar\eta\bar\eta}}{[{\rm h}^{-1}_{3,\eta\eta}-{\rm B}_{4,\eta\eta}][{\rm h}^{-1}_{3,\bar\eta\bar\eta}-{\rm B}_{4,\bar\eta\bar\eta}]-{\rm B}_{4,\eta\bar\eta}{\rm B}_{4,\bar\eta\eta}}\\
\tilde{\rm h}_{3,\bar\eta\eta}=&\frac{{\rm B}_{4,\bar\eta\eta}}{[{\rm h}^{-1}_{3,\eta\eta}-{\rm B}_{4,\eta\eta}][{\rm h}^{-1}_{3,\bar\eta\bar\eta}-{\rm B}_{4,\bar\eta\bar\eta}]-{\rm B}_{4,\eta\bar\eta}{\rm B}_{4,\bar\eta\eta}}\;.
\end{aligned}\end{equation}
Finally, the bare propagators with overlap of three fermion lines are diagonal $2\times 2$ matrices with elements
\begin{equation}\begin{aligned}\label{h3ss}
{\rm h}^{\sigma(\sigma\bar\sigma)}_{3,\nu'\nu}
&=
\begin{gathered}
\resizebox{!}{1.3cm}{
\begin{tikzpicture}[] 
\draw[blue,line width=0.7mm] (0.5,2.3)  arc (120:60:2.5cm  and 1.cm)  node[right] {\LARGE{$\sigma\kappa$}};
\draw[blue,line width=0.7mm] (0.5,1.7)  arc (120:60:2.5cm  and 1.8cm)  node[right] {\LARGE{$\sigma\kappa_1$}};
\draw[red,line width=0.7mm] (0.5,1.1)  arc (120:60:2.5cm  and 2.5cm)  node[right] {\LARGE{$\bar\sigma\kappa_2$}};
\draw[blue,line width=0.7mm]  (0.5,0)  -- (3,0)node[right] {\LARGE{$\nu$}} ; 
\draw[red,dashed,line width=0.7mm] (0.5,-0.7) -- (3,-0.7) ; 
 \end{tikzpicture}
 }
\end{gathered}
=
\frac{{\rm i}\hbar\delta_{\nu'\nu}}{\zeta(\epsilon_k-\epsilon_{k_{1}})+ \zeta_{2}[\epsilon_{k_{2}}-E_{\bar\sigma}(\nu)]+{\rm i}0^+}\\
{\rm h}^{\sigma(\sigma\sigma)}_{3,\eta'\eta}
&=
\begin{gathered}
\resizebox{!}{1.3cm}{
\begin{tikzpicture}[] 
\draw[blue,line width=0.7mm] (0.5,2.3)  arc (120:60:2.5cm  and 1.cm)  node[right] {\LARGE{$\sigma\kappa$}};
\draw[blue,line width=0.7mm] (0.5,1.7)  arc (120:60:2.5cm  and 1.8cm)  node[right] {\LARGE{$\sigma\kappa_1$}};
\draw[blue,line width=0.7mm] (0.5,1.1)  arc (120:60:2.5cm  and 2.5cm)  node[right] {\LARGE{$\sigma\kappa_2$}};
\draw[blue,dashed,line width=0.7mm]  (0.5,0)  -- (3,0); 
\draw[red,line width=0.7mm] (0.5,-0.7)  -- (3,-0.7)node[right] {\LARGE{$\eta$}}  ; 
 \end{tikzpicture}
 }
\end{gathered}
=
\frac{{\rm i}\hbar\delta_{\eta'\eta}}{\zeta(\epsilon_k-\epsilon_{k_{1}})+ \zeta_{2}[\epsilon_{k_{2}}-E_{\sigma}(\eta)]+{\rm i}0^+}\;,
\end{aligned}
\end{equation}
and
\begin{equation}\begin{aligned}\label{h3sbs}
{\rm h}^{\sigma(\bar\sigma\sigma)}_{3,\nu'\nu}
&=
\begin{gathered}
\resizebox{!}{1.3cm}{
\begin{tikzpicture}[] 
\draw[blue,line width=0.7mm] (0.5,2.3)  arc (120:60:2.5cm  and 1.cm)  node[right] {\LARGE{$\sigma\kappa$}};
\draw[red,line width=0.7mm] (0.5,1.7)  arc (120:60:2.5cm  and 1.8cm)  node[right] {\LARGE{$\bar\sigma\kappa_1$}};
\draw[blue,line width=0.7mm] (0.5,1.1)  arc (120:60:2.5cm  and 2.5cm)  node[right] {\LARGE{$\bar\sigma\kappa_2$}};
\draw[blue,line width=0.7mm]  (0.5,0)  -- (3,0)node[right] {\LARGE{$\nu$}} ; 
\draw[red,dashed,line width=0.7mm] (0.5,-0.7) -- (3,-0.7) ; 
 \end{tikzpicture}
 }
\end{gathered}
=
\frac{{\rm i}\hbar\delta_{\nu'\nu}}{\zeta(\epsilon_k-\epsilon_{k_2})+ \zeta_1[\epsilon_{k_1}-E_{\bar\sigma}(\nu)]+{\rm i}0^+}\\
{\rm h}^{\sigma(\bar\sigma\bar\sigma)}_{3,\eta'\eta}
&=
\begin{gathered}
\resizebox{!}{1.3cm}{
\begin{tikzpicture}[] 
\draw[blue,line width=0.7mm] (0.5,2.3)  arc (120:60:2.5cm  and 1.cm)  node[right] {\LARGE{$\sigma\kappa$}};
\draw[red,line width=0.7mm] (0.5,1.7)  arc (120:60:2.5cm  and 1.8cm)  node[right] {\LARGE{$\bar\sigma\kappa_1$}};
\draw[red,line width=0.7mm] (0.5,1.1)  arc (120:60:2.5cm  and 2.5cm)  node[right] {\LARGE{$\bar\sigma\kappa_2$}};
\draw[blue,dashed,line width=0.7mm]  (0.5,0)  -- (3,0); 
\draw[red,line width=0.7mm] (0.5,-0.7)  -- (3,-0.7)node[right] {\LARGE{$\eta$}}  ; 
 \end{tikzpicture}
 }
\end{gathered}
=
\frac{{\rm i}\hbar\delta_{\eta'\eta}}{\zeta_1(\epsilon_{k_1}-\epsilon_{k_2})+ \zeta[\epsilon_k-E_{\sigma}(\eta)]+{\rm i}0^+}\;,
\end{aligned}
\end{equation}
where $E_{\sigma}(\eta)=\epsilon_{\sigma}+(1+\eta) U/2$.\\
\indent The determination of the dressed second-tier bubble $\tilde{\mathbf{B}}^{\sigma(\bar\sigma)}$, Eq.~\eqref{dressed_bubble-}, and in turn of the self-energy $\tilde{\Sigma}_{\sigma\eta}^{(\bar\sigma)}(\epsilon)$, Eq.~\eqref{SigmaB_sNCA4}, relies on the calculation of the dressed NCA4 third-tier bubbles~\eqref{3rdtier_gDSO4_o}. On the other hand, evaluating the dressed second-tier bubble $\tilde{\mathbf{B}}^{\sigma(\sigma)}$, Eq.~\eqref{dressed_bubble+}, is more involved, as we need in principle to invert and contract a $4 \times 4$ matrix whose matrix structure is inherited by the one of the dressed third-tier bubbles~\eqref{3rdtier_gDSO4_s}.\\
\indent As shown in Appendix~\ref{dressingB+}, a closed formal expression can be found for the bubble $\tilde{\mathbf{B}}^{\sigma(\sigma)}$ with a two-stage procedure that yields
\begin{equation}\begin{aligned}\label{}
\tilde{\rm B}_{\eta'\eta}^{\sigma(\sigma)}=\begin{gathered}
\resizebox{!}{1.6cm}{
\begin{tikzpicture}[] 
\draw[blue,line width=0.5mm] (0.5,1.3) arc (120:60:3cm  and 2.5cm) ;
\draw[blue,line width=0.5mm] (1,0) arc (180:0:1.cm  and 1.2cm); 
\draw[blue,dashed,line width=0.5mm]  (0.5,0) -- (1,0); 
\draw[blue,line width=0.5mm] (1,0) -- (3,0); 
\draw[blue,dashed,line width=0.5mm] (3,0) -- (3.5,0); 
\draw[red,line width=0.5mm] (0.5,-0.5) node[left] {\Large{$\eta$}} -- node[right] {\Large{$\qquad\quad\eta'$}} (3.5,-0.5); 
\filldraw[blue] 
(1,0) circle (3pt);
\fill[white] (2.6,-0.8) rectangle (1.4,.4);
\draw [] (2.6,-0.8) rectangle (1.4,.4);
\draw[](2.,-0.2) node[] {\LARGE{$\tilde{\mathbf{h}}^{\sigma(\sigma)}_{2}$}}; 
 \end{tikzpicture}
 }
\end{gathered}
=&-\frac{\Gamma}{2\hbar}\delta_{\eta'\eta}+\langle{\rm K}_{\eta'\eta}^{\sigma(\sigma)}{\rm v}_{+}\rangle
\;,
\end{aligned}\end{equation}
where, as a key result,
\begin{equation}
\begin{aligned}
\label{K_main}
{\rm K}_{\eta'\eta}^{\sigma(\sigma)}=&\;\eta\frac{\Delta A^{\bar\sigma}_+}{([{\rm h}^{\sigma(\sigma)}_{2}]^{-1}+\Gamma/\hbar)^2-\Delta A^{\sigma}_+\Delta A^{\bar\sigma}_+}
\;.
\end{aligned}
\end{equation}
Here, the functions $\Delta A_+^{\sigma/\bar\sigma}$ are differences of the dressed propagators $\tilde{\rm h}^{\sigma(\sigma\sigma)}_{3,\eta\eta}$ or $\tilde{\rm h}^{\sigma(\sigma\bar\sigma)}_{3,\nu\nu}$, as seen in Eq.~\eqref{DeltaA_exact}. They lead to a nontrivial, temperature-dependent, renormalization of the self-energy $\tilde{\Sigma}_{\sigma\eta}^{(\sigma)}(\epsilon):={\rm i}\hbar\tilde{\rm B}^{\sigma(\sigma)}_{\bar\eta\eta}(\kappa)|_{\zeta=+1}\;$, see e.g. Eq.~\eqref{SigmaA_sNCA4} below.
We notice that $\sum_{\eta}{\rm K}_{\bar\eta\eta}^{\sigma(\sigma)}=0$, therefore the corresponding self-energy has the property $\sum_\eta \tilde{\Sigma}_{\sigma\eta}^{(\sigma)}(\epsilon)=0$. On the other hand, it is easy to see from Eq~\eqref{SigmaB_sNCA4} that $\sum_\eta \tilde{\Sigma}_{\sigma\eta}^{(\bar\sigma)}(\epsilon)=-{\rm i}\Gamma$.
The dressed self-energies have thus the property 
\begin{equation}\label{SigmaNCA4}
\sum_{\eta}\tilde\Sigma_{\sigma\eta}(\epsilon)=\tilde\Sigma_{\sigma}(\epsilon)=-{\rm i}\Gamma\;,
\end{equation}
where
$$\tilde\Sigma_{\sigma\eta}=\tilde\Sigma_{\sigma\eta}^{(\sigma)}+\tilde\Sigma_{\sigma\eta}^{(\bar\sigma)}\;.
$$
This is the same property as the one obeyed by the bare self-energies in the NCA2, Eq.~\eqref{property_B_gDSO}. Therefore, in the NCA4, the retarded Green's function, whose general noncrossing-approximated form is provided in Eq.~\eqref{GRnoncrossing}, simplifies to
\begin{equation}\begin{aligned}\label{GRgDSO4}
\mathcal{G}^r_{\sigma\sigma}(\epsilon)=&\frac{1-\langle \hat{n}_{\bar\sigma} \rangle}{\epsilon-\epsilon_\sigma+{\rm i}\Gamma/2+\tilde{\Sigma}_{\sigma-}(\epsilon)\frac{U}{\epsilon-\epsilon_\sigma-U+{\rm i}3\Gamma/2}}\\
+&\frac{\langle \hat{n}_{\bar\sigma} \rangle}{\epsilon-\epsilon_\sigma-U+{\rm i}\Gamma/2 -\tilde{\Sigma}_{\sigma+}(\epsilon)\frac{U}{\epsilon-\epsilon_\sigma+{\rm i}3\Gamma/2}}\;.
\end{aligned}\end{equation}
The important difference with the NCA2  is that the self-energies are now dressed by higher-level processes, and specifically by $3$rd-tier bubbles, see Eqs.~\eqref{dressed_bubble+} and~\eqref{dressed_bubble-}. This crucial feature lifts the pinning problem at the symmetry point $\mu-\epsilon_0=U/2$, as the self-energies remain temperature-dependent. Using the sum rule Eq.~\eqref{SigmaNCA4}, we can give the retarded Green's function solely in terms of the self-energy $\tilde{\Sigma}_{\sigma-}$.\\ 
\indent  Consider the degenerate case, $\epsilon_\uparrow=\epsilon_\downarrow=\epsilon_0$, at equilibrium, $\mu_L=\mu_R=\mu$. As for the NCA2 and the DSO, a zero-bias peak in the conductance appears for temperature below a certain value $T^*$ for which the real parts of the denominators vanish, causing a peak in the density of states $-{\rm Im}[\mathcal{G}^r_{\sigma\sigma}(\mu)]/\pi$.  This condition is 
\begin{equation}\begin{aligned}\label{conditionTK2}
(\mu-\epsilon_0)(\mu-\epsilon_0-U)-3\Gamma^2/4+U\;{\rm Re}[\tilde{\Sigma}_{\sigma,-}(\mu)|_{T^*}]=0\;,
\end{aligned}\end{equation}
which is formally the same as for the NCA2, except that here the self-energy is dressed. In the degenerate case, $\Delta A^{\bar\sigma}_+=\Delta A^{\sigma}_+=\Delta A_+$, and a decomposition of Eq.~\eqref{K_main} in partial fractions allows us to express
\begin{equation}
\begin{aligned}
\label{K_main_degenerate}
{\rm K}_{\eta'\eta}^{\sigma(\sigma)}=
\frac{\eta}{2}
&\Bigg[\frac{1}{[{\rm h}^{\sigma(\sigma)}_{2}]^{-1}+\Gamma/\hbar-\Delta A_+}\\
&-\frac{1}{[{\rm h}^{\sigma(\sigma)}_{2}]^{-1}+\Gamma/\hbar + \Delta A_+}\Bigg]\;.
\end{aligned}
\end{equation}
Notice that $\Delta A_+=(a_{\rm R}+{\rm i}a_{\rm I})/\hbar$ is a complex function. This leads to a temperature-dependent renormalization of the propagator ${\rm h}_2^{\sigma(\sigma)}$, cf. Eq.~\eqref{h2_+_SIAM}. Explicitly, we define $\Gamma_{\pm}(T)=\Gamma \pm a_{\rm R}(T)$; the imaginary part $a_{\rm I}(T)$ yields energy renormalization.  
As seen in Eq.~\eqref{DeltaA_exact}, $\Delta A_+$ shares with the Green's function, Eq.~\eqref{GRgDSO4}, the same  renormalization of the dot energies $E_{\bar\sigma}(\nu)$ ($E_{\sigma}(\eta)$) which thus occurs also at the level of the self-energy and in principle at all (even) levels of the  hierarchy. We find (not shown) a similar structure for the third-tier bubbles that renormalize the dot energy in the  self-energy $\tilde\Sigma_{\sigma\eta}^{(\bar\sigma)}(\epsilon)$.
Lastly, we notice that, at the particle-hole symmetry point $\epsilon_0-\mu=-U/2$, and in the degenerate case, the equilibrium NCA4 retarded Green's function in Eq.~\eqref{GRgDSO4} acquires the particularly simple expression 
\begin{equation}
\label{GNCA4PHsp}
\mathcal{G}^r_{\sigma\sigma}(\mu)=\frac{{\rm i}3\Gamma /2}{(U/2+{\rm i}\Gamma /2)(-U/2+{\rm i}3\Gamma /2)+U\tilde{\Sigma}_{\sigma,-}(\mu)}\;.
\end{equation}
\indent In what follows, by discarding the nontrivial, off-diagonal contributions from the fourth-tier bubbles, we obtain an approximate fourth-tier scheme easier to handle for analytical evaluations.

\subsubsection{Simplified NCA4 (sNCA4)}

To provide an easy-to-handle, analytical treatment that improves on the NCA2, we consider a simplified version of the NCA4 propagators $\tilde{\mathbf{h}}_3$. Specifically, we neglect the second terms in Eqs.~\eqref{B4s}-\eqref{B4o}. This approximation yields, for all the 4th-tier bubbles, the simple result $\mathbf{B}_4=-\Gamma/(2\hbar)\mathbf{1}$, where $\mathbf{1}$ is the two-dimensional identity in the index $\eta $ or $\nu$. As a consequence, the bare propagators in Eqs.~\eqref{h3ss}-\eqref{h3sbs} simply acquire a broadening $\Gamma/(2\hbar)$ and the dressed propagators in Eq.~\eqref{h3dressed} become diagonal, as the  nontrivial parts of $\mathbf{B}_4$ are disregarded. 
 Note that this treatment repeats what is done in the $\Gamma$-broadened sequential tunneling approximation, see Sec.~\ref{Gamma_broadening}, but at the fourth level of the hierarchy rather than the second. In this case $\Delta A^{\bar\sigma}_+$ and $\Delta A^{\sigma}_+$ are real. In addition, the third-tier bubbles $\tilde{\rm B}_3^{\sigma(\bar\sigma)}$, dressing the dot energy in the self-energy  $\tilde\Sigma_{\sigma\eta}^{(\bar\sigma)}(\epsilon)$, Eq.~\eqref{SigmaB_sNCA4}, are real at the symmetry point. Thus, contrary to the NCA4, the sNCA4 only accounts for the renormalization of the lifetime but not of the energy in the arguments of the self-energies.
In Appendices~\ref{third-level_bubbles} and \ref{dressingB+DSO4}, we give explicit expressions for the simplified third-tier bubbles of Eq.~\eqref{3rdtier_gDSO4_s}, and for the functions $\Delta A_+^{\sigma/\bar\sigma}$  defined in Eq.~\eqref{Delta_b}.
\\
\indent The resulting dressed second-tier bubbles are calculated in appendices~\ref{dressingB+DSO4} and~\ref{dressingB-}. They satisfy the properties
\begin{equation}
\begin{aligned}\label{prop_tildeB}
\tilde{\rm B}^{\sigma(\sigma)}_{\eta'\eta}
=-\frac{\Gamma}{2\hbar}\delta_{\eta'\eta}+\tilde{\rm B}^{\sigma(\sigma)}_{\bar\eta\eta}\quad{\rm and}
\quad
\tilde{\rm B}_{\eta'\eta}^{\sigma(\bar\sigma)}=\tilde{\rm B}_{\bar\eta\eta}^{\sigma(\bar\sigma)}\;.
\end{aligned}
\end{equation}
Thus, the sum $\tilde{\rm B}_{\eta'\eta}^{\sigma}=\tilde{\rm B}_{\eta'\eta}^{\sigma(\sigma)}+\tilde{\rm B}_{\eta'\eta}^{\sigma(\bar\sigma)}$ respects the property given in Eq.~\eqref{symmetry_dressedB} with the general, $\infty$-tier DBA scheme of Eq.~\eqref{DysonNCA}.\\
\noindent We find for the corresponding dressed, retarded self-energy $\tilde\Sigma_{\sigma\eta}^{(\sigma)}(\epsilon)={\rm i}\hbar\tilde{\rm B}^{\sigma(\sigma)}_{\bar\eta\eta}(\kappa)|_{\zeta=+1}$
\begin{equation}\begin{aligned}
\label{SigmaA_sNCA4}
\tilde{\Sigma}_{\sigma\eta}^{(\sigma)}&(\epsilon)=\\
-\frac{\eta}{2}&\sqrt{\frac{|\Delta A_+^{\bar\sigma}|}{|\Delta A_+^{\sigma}|}}\sum_{\alpha,p=\pm } p \frac{\Gamma_{\alpha}}{2\pi}
\Bigg[{\rm Re}\psi\left(\frac{1}{2}+\frac{ \Gamma_p(T)}{2\pi k_{\rm B}T }+{\rm i}\frac{\epsilon -\mu_{\alpha}}{2\pi k_{\rm B}T }\right)\\
&\qquad\qquad\quad\qquad-{\rm i}{\rm Im}\psi\left(\frac{1}{2}+\frac{\Gamma_p(T)}{2\pi k_{\rm B}T }+{\rm i}\frac{\epsilon -\mu_{\alpha}}{2\pi k_{\rm B}T }\right)\Bigg]\;,
\end{aligned}\end{equation}
where $\Gamma_\pm(T)= \Gamma \pm \hbar(\Delta A_+^{\sigma}\Delta A_+^{\bar\sigma})^{1/2}$. Further, $\tilde{\Sigma}_{\sigma\eta}^{(\bar\sigma)}(\epsilon)$ is given by Eq.~\eqref{SigmaB_sNCA4} with $\tilde{\rm B}^{\sigma(\bar\sigma)}_{3,\zeta_{1}}$ approximated as in Eq.~\eqref{B3d_2}.\\ 

\subsection{sNCA4 results at equilibrium}

In what follows we consider the symmetric coupling to the leads
$\Gamma_L=\Gamma_R=\Gamma/2$.
 Assuming that the parameters are such that the dot is close to the center of the Coulomb diamond, i.e. $\epsilon_0-\mu, U-\epsilon_0+\mu\gg\Gamma$, we can approximate the self-energy in order to obtain a simple expression for the Kondo-like temperature from the condition in Eq.~\eqref{conditionTK2}.
From Eq.~\eqref{SigmaR}, retaining the first term for sufficiently large $U$, away from the condition $\mu-\epsilon_0=U/2$, the real part of the self-energy $\tilde{\Sigma}_{\sigma,-}^{(\bar\sigma)}$ is the same as in the NCA2, cf. Eq~\eqref{ReSigma_gDSO},
 \begin{equation}\begin{aligned}\label{}
{\rm Re}[\tilde{\Sigma}_{\sigma,-}^{(\bar\sigma)}(\mu)]
\simeq & \frac{\Gamma}{2\pi}\log\left(\frac{U}{2\pi k_{\rm B}T}\right)\;.
 \end{aligned}\end{equation}
As shown in Appendix~\ref{dressingB+DSO4}, see Eq.~\eqref{Sigma+muEq_approx}, the retarded self-energy of type $(\sigma)$ is calculated to be
\begin{equation}
\tilde{\Sigma}_{\sigma,-}^{(\sigma)}(\mu)
\simeq \frac{\Gamma}{4\pi}
\log\left(\frac{ 2\Gamma}{2\pi k_{\rm B}T }\right)\;,
\end{equation}
and the resulting value for the Kondo-like temperature $T^*=T_{\rm sNCA4}$ is 
\begin{equation}\label{TKondogDSO4}
 k_{\rm B}T_{\rm sNCA4}=\frac{(2U^2\Gamma)^{1/3}}{2\pi e^{\pi\Gamma/(2U)}}e^{4\pi\frac{(\mu-\epsilon_0)(\mu-\epsilon_0-U)}{3U\Gamma}}\;,
\end{equation}
which essentially reproduces the one obtained in ~\cite{VanRoermund2010}. 
The result of the simplified NCA4 deviates from the true Kondo temperature $T_{\rm K}$, Eq.~\eqref{TKondo}. This deviation is ascribed to the fact that, in the dressing of the third-tier bubbles, only diagonal contributions were included, yielding a simple structure for the $4$th-tier bubbles. Diagrams describing spin fluctuations involve the fourth-tier bubbles in the second column of Eqs.~\eqref{B4s} and~\eqref{B4o} which are arguably relevant in the low temperature regimes.
In fact, as shown exemplarily in Eq.~\eqref{DeltaA_exact}, they lead to a temperature-dependent energy shift in the self-energy. \\
\begin{figure}[t!]
\begin{center}
\includegraphics[width=9cm,angle=0]{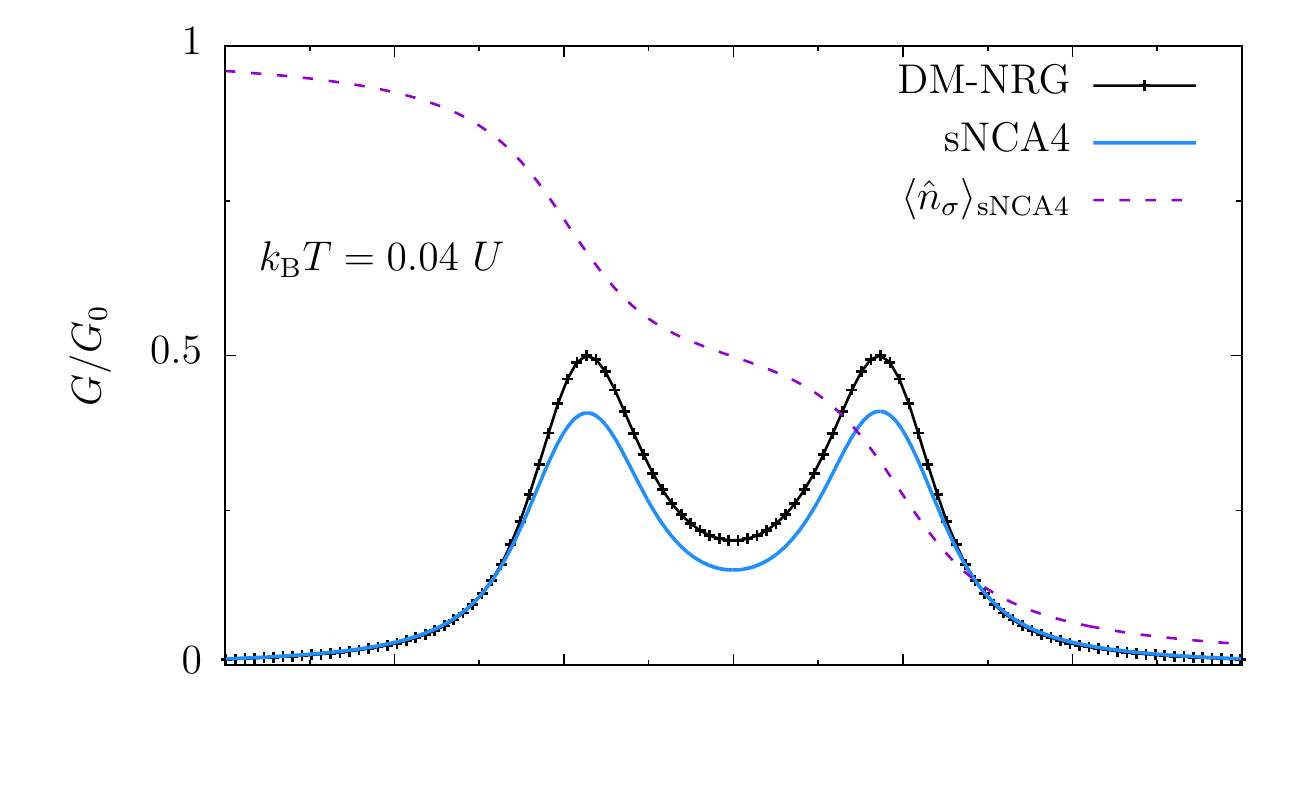}\\
\vspace{-0.85cm}
\includegraphics[width=9cm,angle=0]{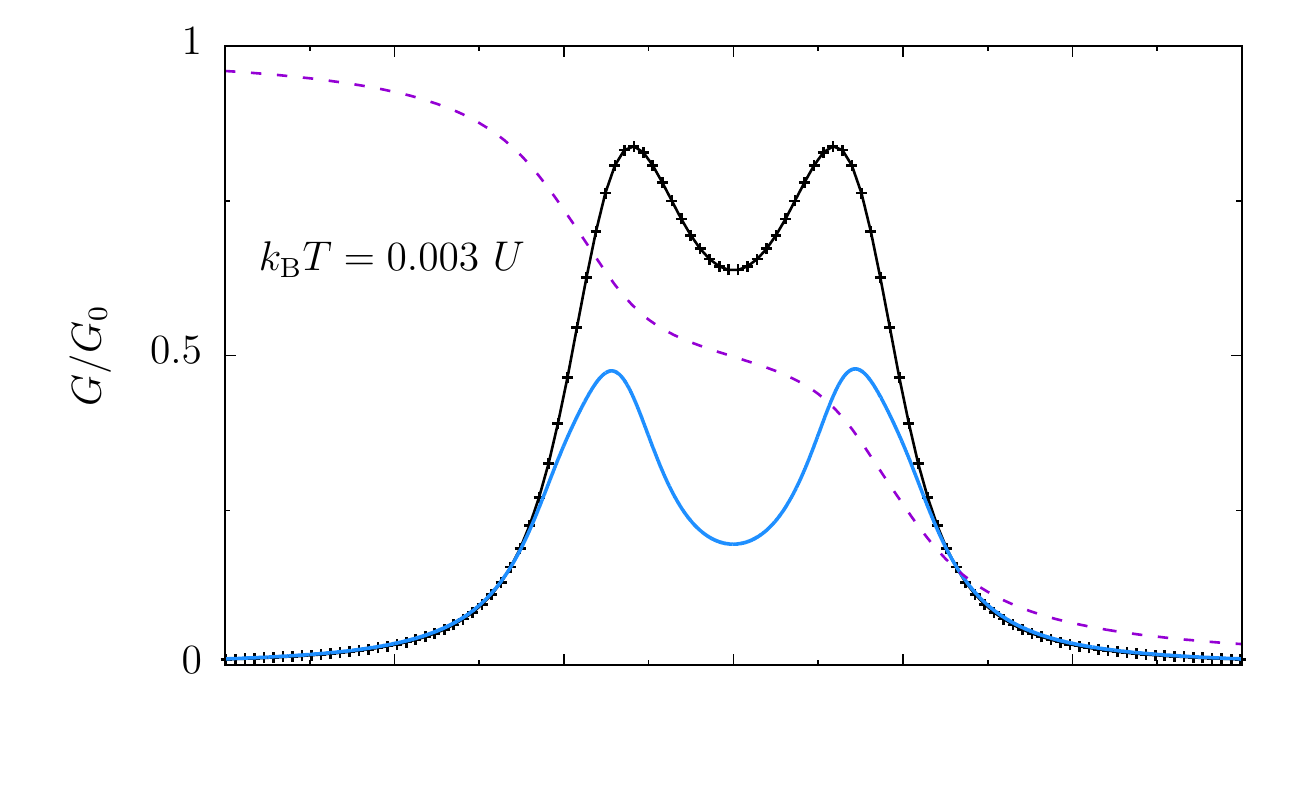}\\
\vspace{-0.85cm}
\includegraphics[width=9cm,angle=0]{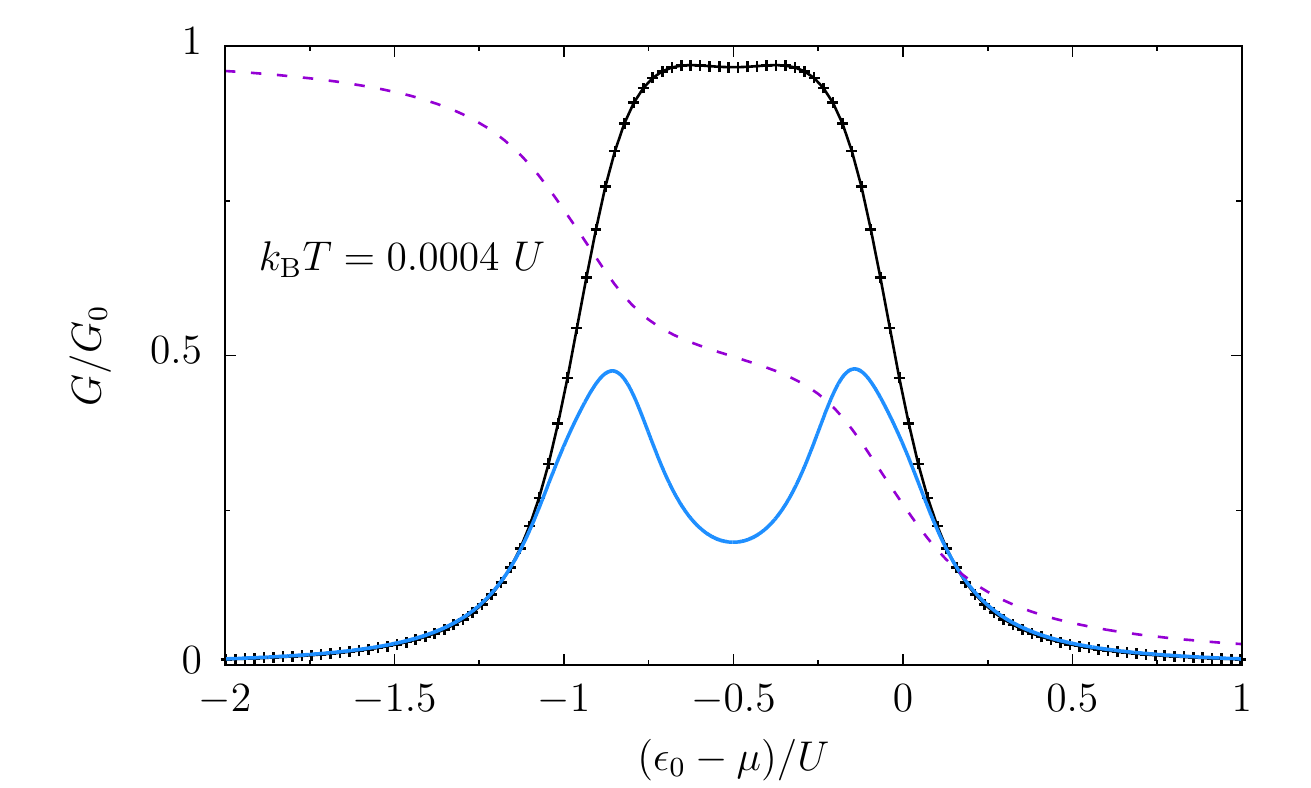}
\caption{\small{Linear conductance of the simplified NCA4 \emph{vs} the gate voltage for three different temperatures. Degenerate case, $\epsilon_\sigma=\epsilon_0$, with $\Gamma=0.2~U$. The dashed lines depict the expectation value $\langle\hat{n}_\uparrow\rangle=\langle\hat{n}_\downarrow\rangle$. The parameters are chosen to allow a direct comparison with the results of~\cite{deSouza2019}. The Kondo temperature, Eq.~\eqref{TKondo}, for this choice of parameters is $k_{\rm B}T_{\rm K}\simeq 0.004~U$. 
The sNCA4 conductance has a very weak dependence on temperature in the lowest two panels, a hint that the Fermi liquid regime is approached (see also Fig.~\ref{fig_gDSO4_GvsT}). However, in contrast to the DM-NRG curve, the unitary limit $G=G_0$ is not approached.}}
\label{fig_gDSO4}
\end{center}
\end{figure}
\begin{figure}[t!]
\begin{center}
\includegraphics[width=9cm,angle=0]{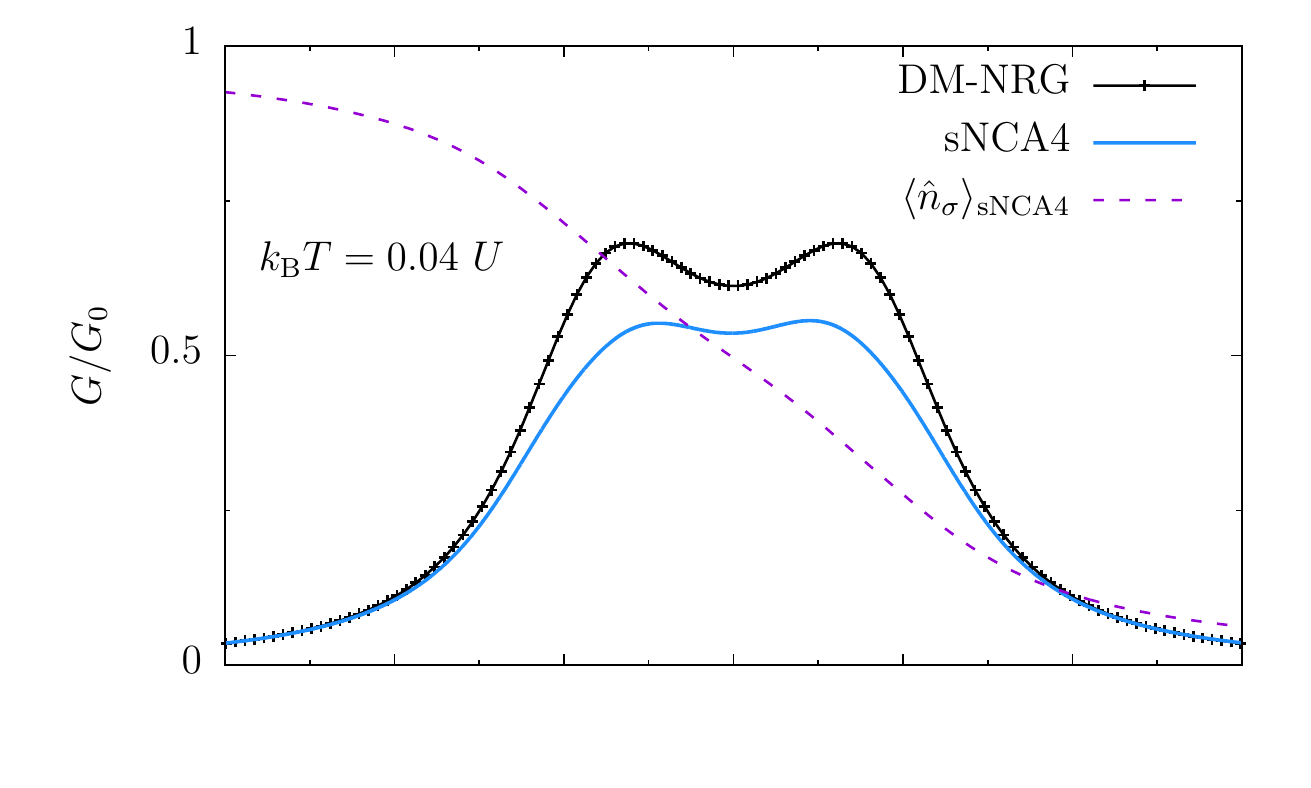}\\
\vspace{-0.85cm}
\includegraphics[width=9cm,angle=0]{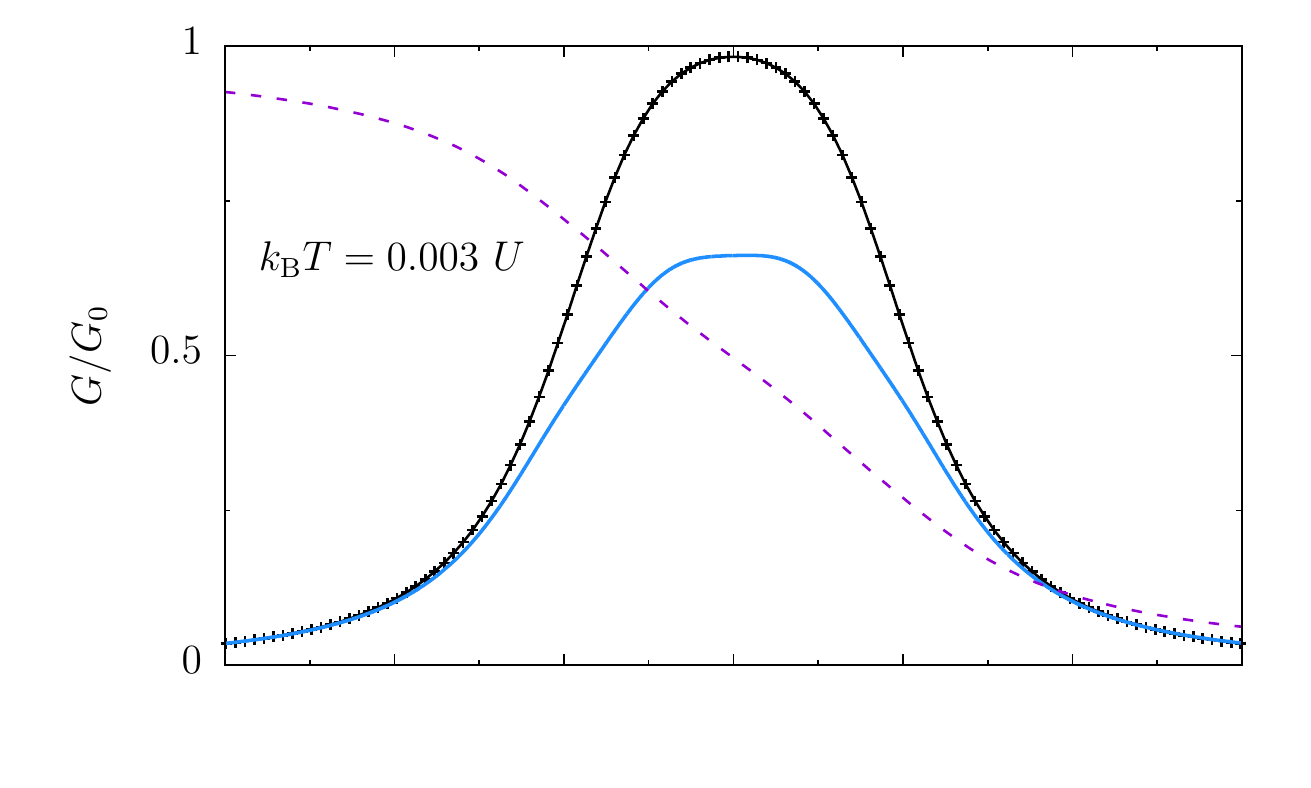}\\
\vspace{-0.85cm}
\includegraphics[width=9cm,angle=0]{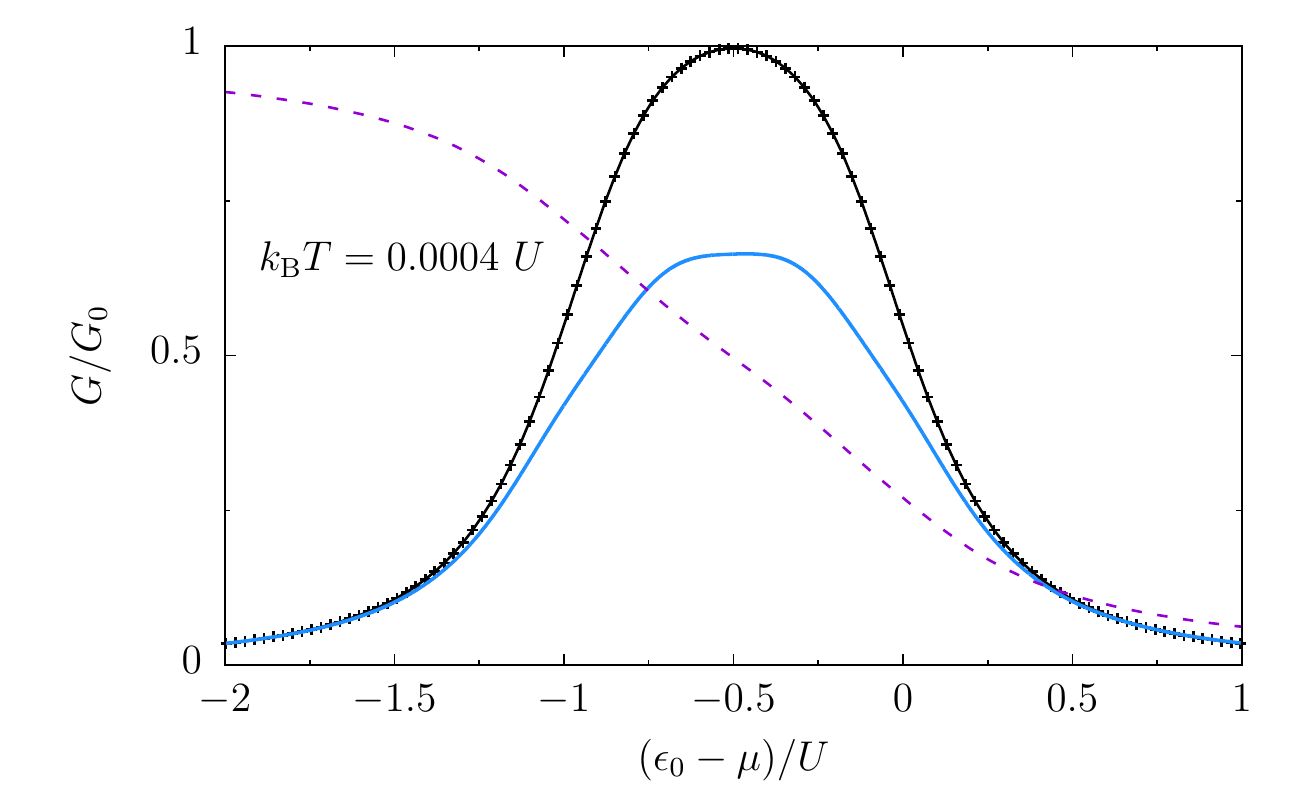}
\caption{\small{Linear conductance of the simplified NCA4 \emph{vs} the gate voltage for three different temperatures. Degenerate case, $\epsilon_\sigma=\epsilon_0$,  with $\Gamma=0.4~U$. The dashed lines depict the expectation value $\langle\hat{n}_\uparrow\rangle=\langle\hat{n}_\downarrow\rangle$. The Kondo temperature, Eq.~\eqref{TKondo}, for this choice of parameters is $k_{\rm B}T_{\rm K}\simeq 0.04~U$. Note that the sNCA4 performs better at low temperature with respect to the case $\Gamma=0.2~U$, cf. Fig.~\ref{fig_gDSO4}.}}
\label{fig_gDSO4_G04}
\end{center}
\end{figure}
\begin{figure}[t!]
\begin{center}
\includegraphics[width=9cm,angle=0]{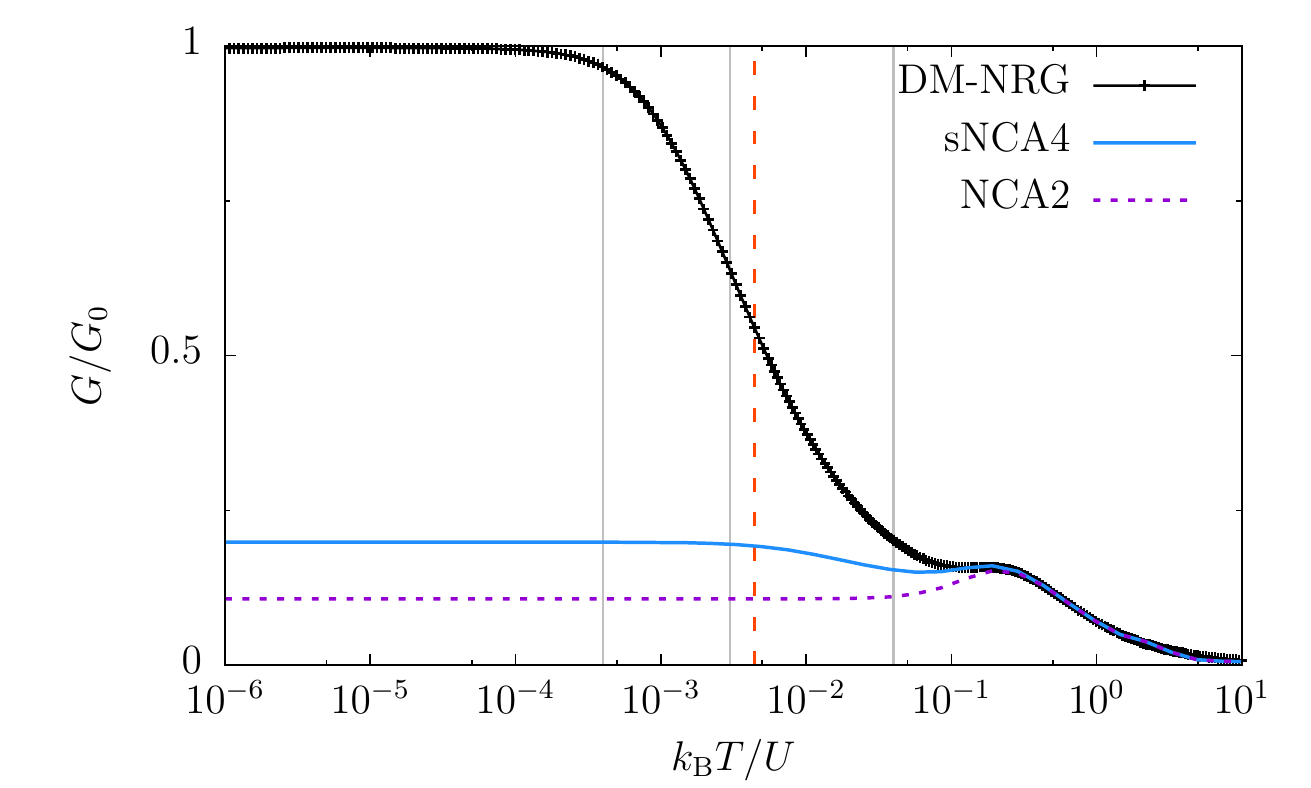}
\includegraphics[width=9cm,angle=0]{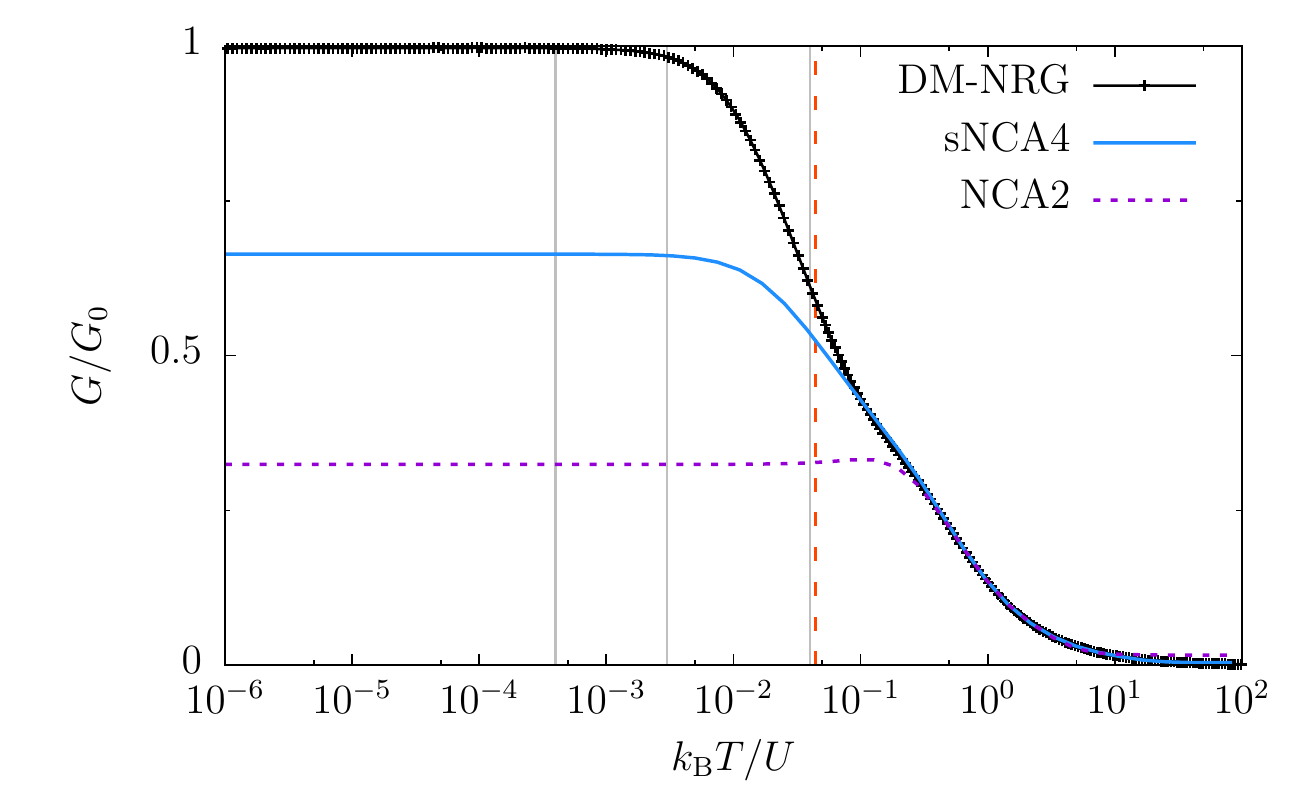}
\caption{\small{Temperature dependence of the  linear conductance in the simplified NCA4 and NCA2 at the particle-hole symmetry point $\epsilon_0-\mu=-U/2$. Degenerate case, $\epsilon_\sigma=\epsilon_0$,  with $\Gamma=0.2~U$ (upper panel) and $\Gamma=0.4~U$ (lower panel). The red dashed lines are at the values of the Kondo temperature $T_{\rm K}$ given by Eq.~\eqref{TKondo} while the vertical solid lines correspond to the three temperatures considered in in Figs.~\ref{fig_gDSO4} and~\ref{fig_gDSO4_G04}.
Consistently with the DM-NRG results, well below $T_{\rm K}$ the sNCA4 curves depend only weakly on the temperature. However, the unitary value $G_0$ is not attained.}}
\label{fig_gDSO4_GvsT}
\end{center}
\end{figure}
\indent In Figs.~\ref{fig_gDSO4} and~\ref{fig_gDSO4_G04}, we show the linear conductance calculated within the sNCA4 scheme considered here, for $\Gamma=0.2$ and $\Gamma=0.4~U$, respectively. In both cases we consider three values of the temperature. The   sNCA4 reproduces the DM-NRG curve quantitatively until temperatures slightly above $T_{\rm K}$ over the whole gate voltage range. After that, the conductance has only a weak temperature dependence, as expected in the Fermi liquid regime~\cite{Hewson1993}. However, it fails to reproduce the saturation  predicted for the SIAM to the  plateau value $G_0=2e^2/h$  for $T\rightarrow 0$. We notice that the sNCA4 has a better qualitative behavior at the larger value of $\Gamma$.  As in Fig.~\ref{fig_gDSO_vs_DSO}, the parameters of Fig.~\ref{fig_gDSO4} are chosen so as to allow for a direct comparison with Fig.~6 of Ref.~\cite{deSouza2019}, where different approximation schemes, some of which are derived with the EOM technique, are contrasted with the NRG results.

\subsubsection{Particle-hole symmetry point}

At the particle-hole symmetry point $\epsilon_0-\mu=-U/2$, and in the degenerate case, the retarded self-energy simplifies to $\tilde{\Sigma}_{\sigma,-}(\mu)=\tilde{\Sigma}_{\sigma,-}^{(\sigma)}(\mu)+\tilde{\Sigma}_{\sigma,-}^{(\bar\sigma)}(\mu)={\rm Re}\tilde{\Sigma}_{\sigma-}(\mu)-{\rm i}\Gamma/2$, 
 where, from Eqs.~\eqref{Sigma+PH} and~\eqref{Sigma-PH}, 
\begin{equation}\begin{aligned}\label{tildeSigma-PH}
{\rm Re}\tilde{\Sigma}_{\sigma,-}(\mu) = \frac{3\Gamma}{4\pi}
\sum_{p=\pm}p\;{\rm Re}\psi\left(\frac{1}{2}+\frac{ \Gamma_p}{2\pi k_{\rm B}T }\right)\;.
\end{aligned}
\end{equation}
In turn, the arguments $\Gamma_\pm$ read
$$\Gamma_\pm = \Gamma\Bigg[1 \pm\frac{2}{\pi}{\rm Im}\psi\left(\frac{1}{2}+\frac{\Gamma/2}{2\pi k_{\rm B}T}+{\rm i}\frac{U/2}{2\pi k_{\rm B}T}\right)\Bigg]\;.$$
From Eq.~\eqref{GNCA4PHsp}, this entails for the density of states
\begin{equation}\begin{aligned}\label{DOS_ph2}
-\frac{1}{\pi}{\rm Im}\mathcal{G}^r_{\sigma\sigma}(\mu)=\frac{3\Gamma/2\pi}{U^2/4+3\Gamma^2/4-U{\rm Re}\tilde{\Sigma}_{\sigma,-}(\mu)}\;.
\end{aligned}
\end{equation}
The linear conductance saturates at zero temperature and, as shown in Appendix~\ref{Fermi_liquid},  displays a Fermi liquid behavior at low $T$.\\
\indent The zero-temperature expression for the self-energy 
\begin{equation}\begin{aligned}\label{}
\tilde{\Sigma}_{\sigma,-}(\mu) =  \frac{3\Gamma}{4\pi}
\ln\left(\frac{1 + (2/\pi)\arctan(U/\Gamma)}{1 - (2/\pi)\arctan(U/\Gamma)}\right)-{\rm i}\frac{\Gamma}{2}\;,
\end{aligned}
\end{equation}
along with Eqs.~\eqref{GT0} and~\eqref{DOS_ph2}, provides the saturation value $G_{T=0}=-G_0 \Gamma\sum_{\sigma}{\rm Im}\mathcal{G}^r_{\sigma\sigma}(\mu)/4$ for the sNCA4 linear conductance.\\ 
\indent Figure~\ref{fig_gDSO4_GvsT} shows the linear conductance at the particle-hole symmetry point as a function of the temperature, for both values of $\Gamma$ considered in Figs.~\ref{fig_gDSO4} and~\ref{fig_gDSO4_G04}. In the limit $T\rightarrow 0$, the sNCA4 displays  saturation at values lower than the correct value $G_0$, which improves for larger $\Gamma$. In fact, saturation to $G_0$ is attained in the limit $\Gamma \gg U$, since the theory is exact in the noninteracting case. An improvement over the sNCA4 results is expected 
at the NCA4 level. Here, the inclusion of the off-diagonal contributions in the fourth-tier matrices $\mathbf{B}_4$ naturally leads to an energy shift also in the self-energies, see Eq.~\eqref{DeltaA_exact}, which becomes crucial below the Kondo temperature. The predictions of the full NCA4 will be the subject of future investigations.

\subsection{Nonequilibrium properties of  the sNCA4}
\begin{figure}[t!]
\begin{center}
\includegraphics[width=8cm,angle=0]{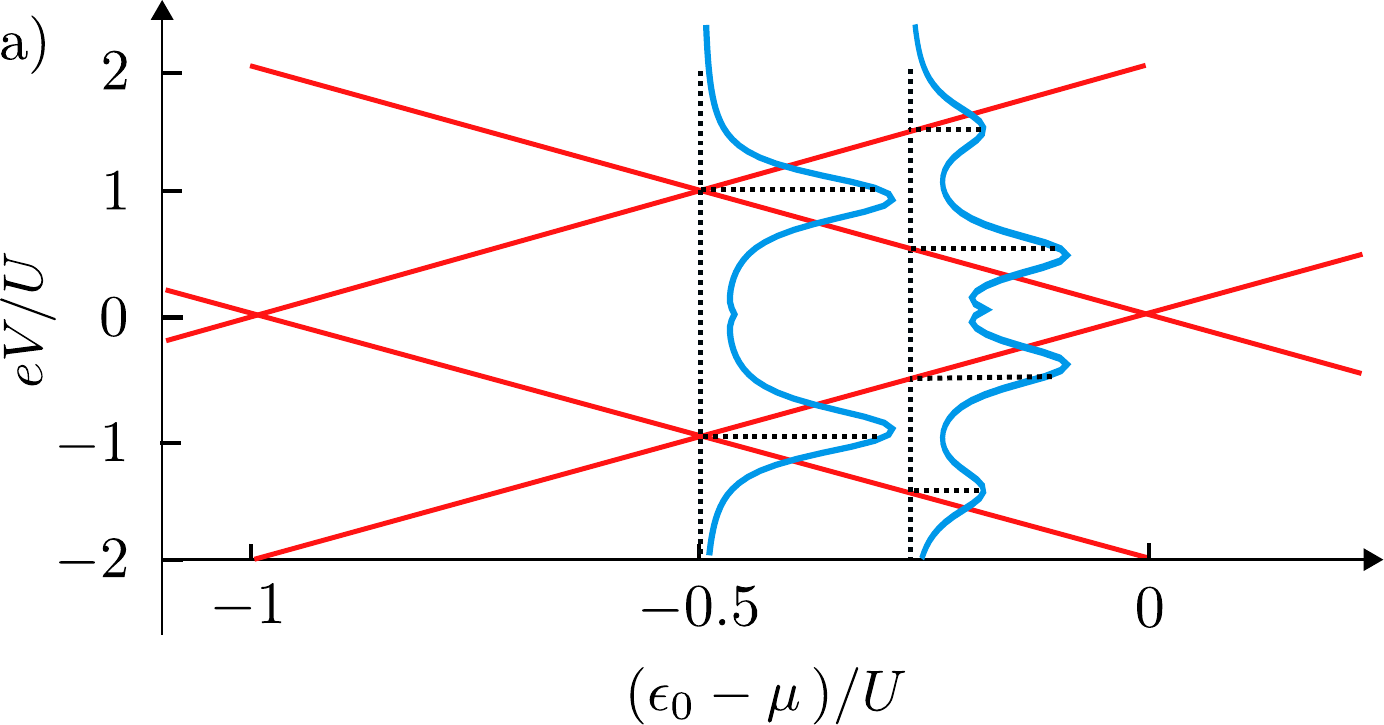}\\
\includegraphics[width=9.5cm,angle=0]{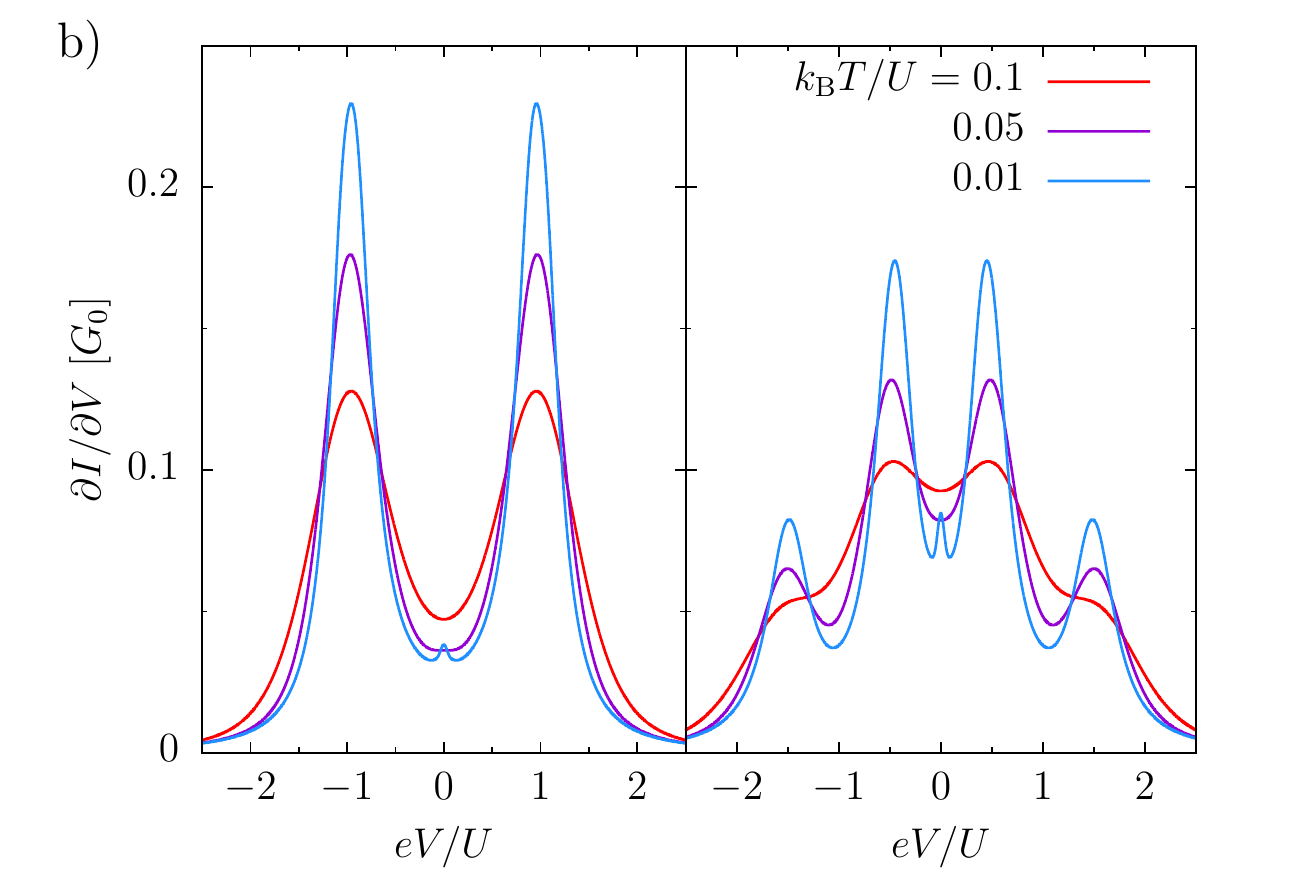}\\
\caption{\small{Effect of the temperature on the sNCA4 differential conductance for $\Gamma=0.1~U$, with $\Gamma_L=\Gamma_R=\Gamma/2$, in the degenerate case.
a) Scheme of the stability diagram for the SIAM, cf. Fig.~\ref{diff_cond_ST}. The curves are the differential conductance for two values of the gate voltage at $k_{\rm B}T=0.04~U$. b)  Differential conductance \emph{vs.} voltage bias $eV=\mu_L-\mu_R$, for different temperatures. The gate voltage is set to the values shown in panel a), namely $\epsilon_0-\mu=-U/2$ (left) and $\epsilon_0-\mu=-U/4$ (right). At the particle-hole symmetry point, $\epsilon_0-\mu=-U/2$, the Kondo temperature, Eq.~\eqref{TKondo}, is $k_{\rm B}T_{\rm K}\simeq 0.00006~U$.}}
\label{fig_Gdiff_vs_bias_Gamma01}
\end{center}
\end{figure}

\begin{figure}[t!]
\begin{center}
\includegraphics[width=9.5cm,angle=0]{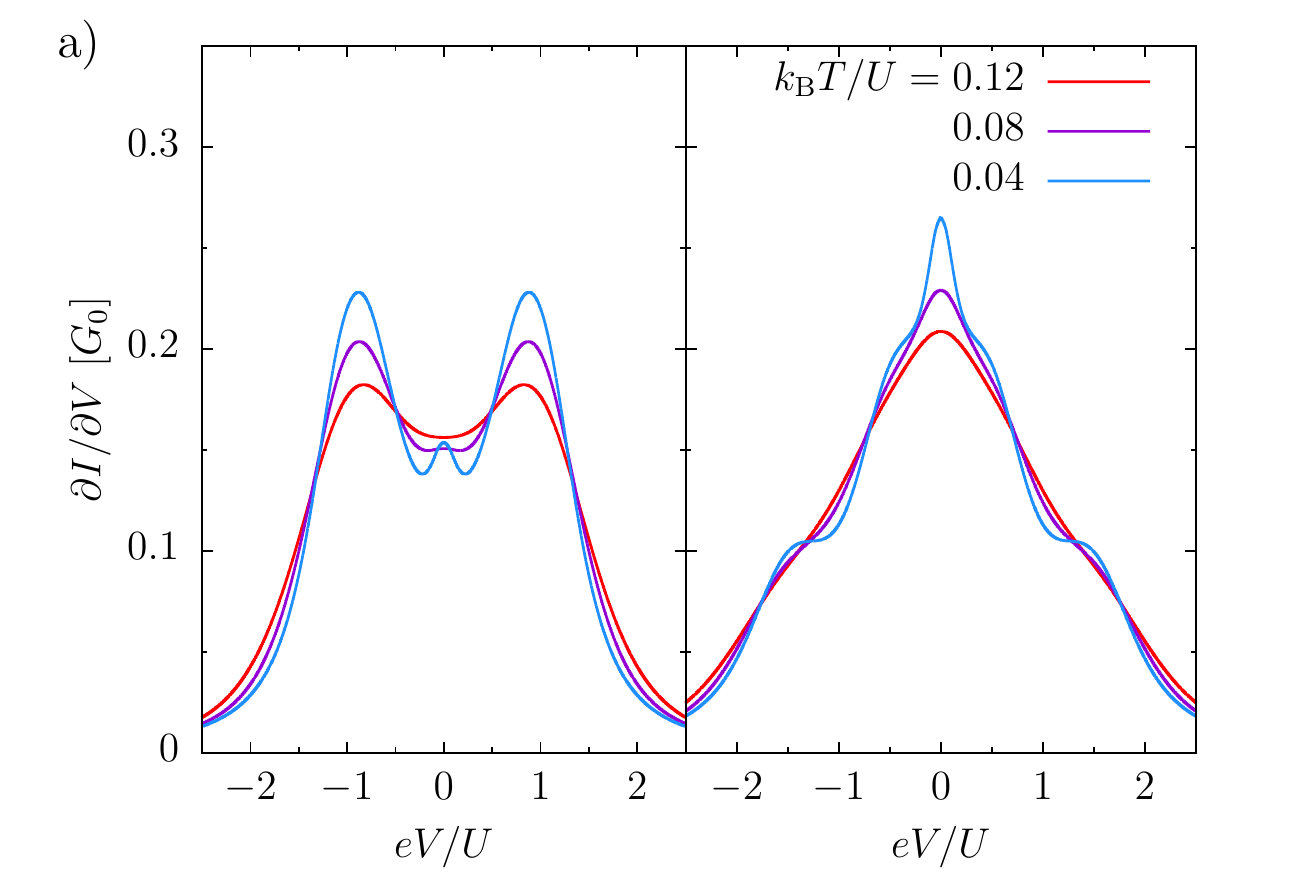}\\
\includegraphics[width=9.5cm,angle=0]{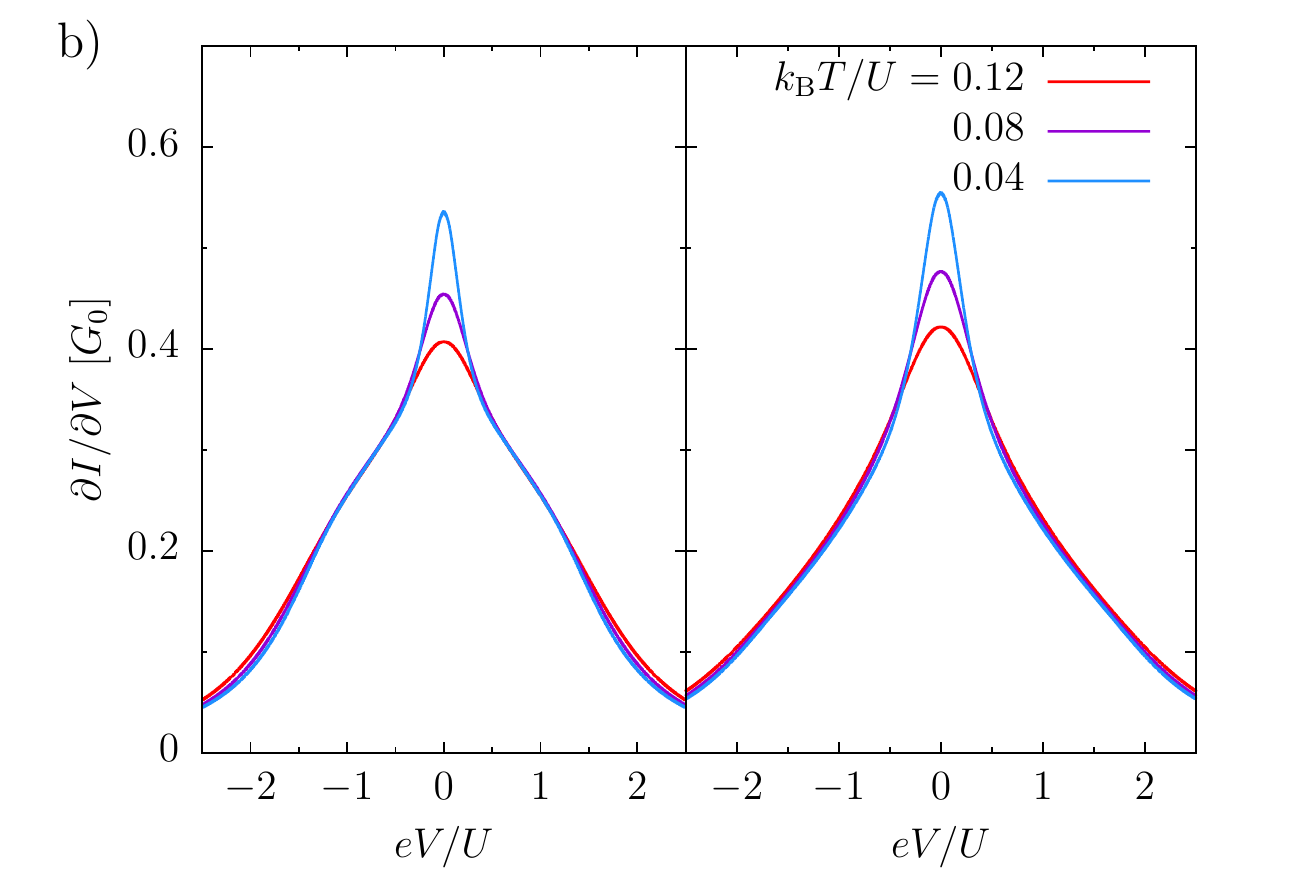}
\caption{\small{Effect of the temperature on the sNCA4 differential conductance \emph{vs.} voltage bias $eV=\mu_L-\mu_R$ in the degenerate case and for a) $\Gamma=0.2~U$ and b) $\Gamma=0.4~U$. The gate voltage is set to the same values as in Fig.~\ref{fig_Gdiff_vs_bias_Gamma01}, namely $\epsilon_0-\mu=-U/2$ (particle-hole symmetry point, left) and $\epsilon_0-\mu=-U/4$ (right). At the particle-hole symmetry point, the Kondo temperature, Eq.~\eqref{TKondo}, is $k_{\rm B}T_{\rm K}\simeq 0.004~U$ in panel a) and $k_{\rm B}T_{\rm K}\simeq 0.04~U$ in panel b).}}
\label{fig_Gdiff_vs_bias_Gamma02_04}
\end{center}
\end{figure}

\indent The simple form of the sNCA4 self-energies, Eqs.~\eqref{SigmaB_sNCA4} and~\eqref{SigmaA_sNCA4}, allows us to use the NCA4 Green's function, Eq.~\eqref{GRgDSO4}, to address the nonequilibrium situation. An insight on the peak structure of the differential conductance in the degenerate case, for different values of the tunnel coupling $\Gamma$, is given by Figs.~\ref{fig_Gdiff_vs_bias_Gamma01} and \ref{fig_Gdiff_vs_bias_Gamma02_04}. Specifically, we show the differential conductance as a function of the voltage bias $eV=\mu_L-\mu_R$ with $\mu_L=\mu+eV/2$ and $\mu_R=\mu-eV/2$ at the two gate voltages shown in Fig.~\ref{fig_Gdiff_vs_bias_Gamma01}a), see the cuts of the stability diagram. In Panel b) of Fig.~\ref{fig_Gdiff_vs_bias_Gamma01} the differential conductance \emph{vs.} bias voltage is shown at different temperatures for $\Gamma=0.1~U$. The same is done in  Fig.~\ref{fig_Gdiff_vs_bias_Gamma02_04}a) and b) for $\Gamma=0.2~U$ and $\Gamma=0.4~U$, respectively. As above, we  consider a symmetric coupling to the leads, $\Gamma_L=\Gamma_R=\Gamma/2$\\
\indent The Kondo temperature $T_{\rm K}$, Eq.~\eqref{TKondo}, is a function of $\epsilon_0-\mu$; for this reason to the two values of the gate voltage considered there correspond different Kondo temperatures at a given $\Gamma$. For $\Gamma=0.1~U$, Fig.~\ref{fig_Gdiff_vs_bias_Gamma01}b), the temperatures considered are larger than $T_{\rm K}$ for both gate voltages and the Kondo peak at zero bias is absent, although one can see the onset of the peak at the lowest value of $T$.   
The same holds for $\Gamma=0.2~U$, Fig.~\ref{fig_Gdiff_vs_bias_Gamma02_04}a), at the particle-hole symmetry point, $\epsilon_0-\mu=-U/2$, where $k_{\rm B}T_{\rm K}<0.04~U$. In contrast, the gate voltage $\epsilon_0-\mu=-U/4$ yields $k_{\rm B}T_{\rm K}>0.04~U$. This second situation is reflected by a zero-bias peak which develops fully as the temperature is decreased. For the largest value of the coupling, namely $\Gamma=0.4~U$, Fig.~\ref{fig_Gdiff_vs_bias_Gamma02_04}b),  at both  gate voltages the lowest value of $T$ used is lower than $T_{\rm K}$. This  entails that the two plots in panel b) show the same features, namely fully developed zero-bias peaks.

\subsubsection{Effect of an applied magnetic field}
\begin{figure}[t!]
\begin{center}
\includegraphics[width=8cm,angle=0]{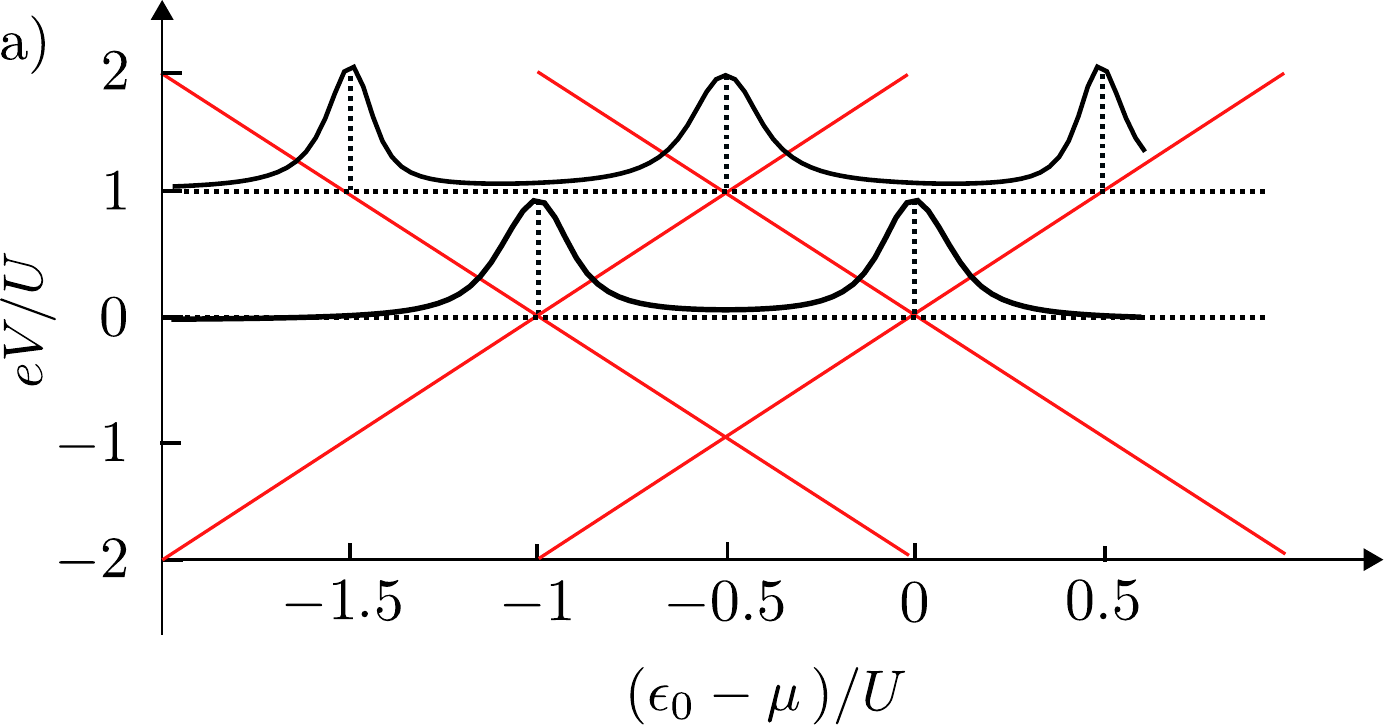}\\
\includegraphics[width=9cm,angle=0]{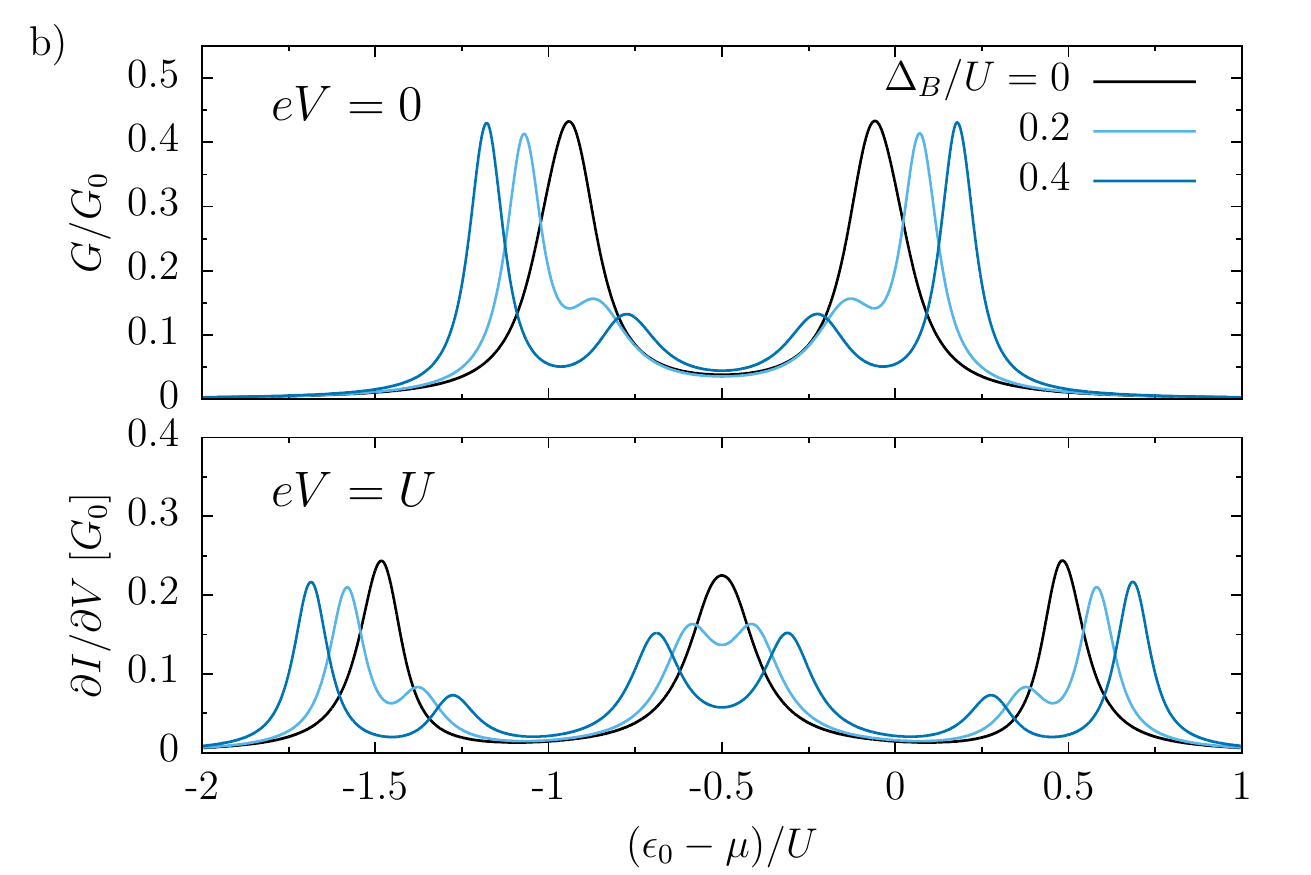}
\caption{\small{Effect of the magnetic field on the sNCA4 conductance. Symmetric coupling to the leads with $\Gamma=0.1~U$, and temperature $k_{\rm B}T=0.01~U$. a) Scheme of the stability diagram for the SIAM. The solid black curves are the differential and linear conductance at zero magnetic field and for two values of the bias voltage $eV=\mu_L-\mu_R$, namely $eV=0$ and $eV=U$. b) Linear and differential conductance \emph{vs.} the gate voltage, calculated for the two values of the bias voltage shown in panel a),
for different values of the Zeeman splitting $\Delta_B=\epsilon_\uparrow-\epsilon_\downarrow$.}}
\label{fig_Gdiff_vs_gate_Zeeman}
\end{center}
\end{figure}

In the presence of an applied magnetic field, the Zeeman splitting of the dot energies $\epsilon_\sigma$ is given by $\Delta_B=\epsilon_\uparrow-\epsilon_\downarrow$, with $\epsilon_\uparrow=\epsilon_0+\Delta_B/2$ and $\epsilon_\downarrow=\epsilon_0-\Delta_B/2$. We address the resulting non-degenerate situation in Fig.~\ref{fig_Gdiff_vs_gate_Zeeman}, where
the conductance is calculated as a function of the gate voltage both at equilibrium (zero voltage bias) and in nonequilibrium ($eV=U$). The same scheme of the stability diagram shown in Fig.~\ref{fig_Gdiff_vs_bias_Gamma01}a) is reproduced in Fig.~\ref{fig_Gdiff_vs_gate_Zeeman}a) and  presents the conductance at zero applied magnetic field for two fixed values of the voltage bias, corresponding to the horizontal cuts.
In panel b), we consider different values of $\Delta_B$ for the two values of the bias shown in panel a). According to the chosen bias voltage  $eV=\mu_L-\mu_R$, we obtain multiple-peak structures with different relative magnitudes. The peaks in the  conductance are split by the effect of the applied magnetic field.

\section{Conclusions}
\label{conclusions}
In summary, we have illustrated how the Feynman-Vernon approach, well-known in the study of the dissipative dynamics of quantum particles in bosonic environments \cite{Leggett1987,Grab88,Grifoni1998,Weiss2012}, is also a useful tool in the context of nonlinear transport in interacting nanojunctions. Integration over the reservoirs degrees of freedom enables one to obtain an exact 
path-integral representation for the reduced density matrix and the current for a general open system connected to several leads. Dealing with fermions, the path integral is given in terms of fermionic coherent states. Here, the Feynman-Vernon influence functional, a functional of the system paths, accounts  exactly for the effects of the leads on the system's dynamics. While the residual integration over the Grassmann variables can be easily performed for noninteracting systems \cite{Jin2008}, this is no longer the case when local interactions are present in the nanojunction, a situation which is the topic of our work.

In the first and general part, we show how this difficulty can be overcome by expressing the path integral for the propagator in the occupation number representation. This allows for a systematic expansion of the Feynman-Vernon influence functional in the system-leads tunneling amplitude  and its diagrammatic characterization.
The diagrammatic expansion is carried out for a general system provided that the tunneling matrices are diagonal in the system's states, meaning that the tunneling is state-preserving. In practice, we exclude from the discussion situations like those of non-collinearly polarized leads, or when orbital coherence is important  \cite{Koenig2003, Braun2006, Hornberger2008, Darau2009, Nilsson2010, Donarini2019, Rohrmeier2021}. This assumption enables us to consider exclusively the populations of the nanojunction. Since, due to Pauli exclusion principle, a single fermionic degree of freedom can only be empty or occupied, we exploit this "two-level" character to parametrize the propagator paths in terms of "blips" and "sojourns", in analogy  to the spin-boson model~\cite{Leggett1987}. 
The diagrammatic contributions to the propagating functions for the populations are summed to yield a formally exact generalized master equation (GME). Similarly, an integral equation for the current is derived. In the last part devoted to the general formalism, we give  hierarchical diagrammatic  expressions for both the kernels of the GME and of the integral equation for the current, which constitutes a major result of this general part. 
Due to the novelty of the approach, we have reported crucial steps of the derivations in numerous Appendices.
This allows non-expert readers to get acquainted with some mathematical intricacies. At the same time, readers not interested in the elimination of the Grassmann variables, can  start from the diagrammatic rules discussed in Sec.~\ref{state_conserving_and_DR} and continue with the derivation of the GME.\\ 
\indent In the second part of the work, the formalism is applied to  two important archetype models: the exactly solvable resonant level model and the single-impurity Anderson model (SIAM). Due to the vast literature on the topic,  we found it important to  show how seemingly different treatments or approximation schemes can be reconciled within our formalism. For example, the nonperturbative resonant tunneling approximation proposed in \cite{Koenig1996} is soon recovered by truncating the hierarchy in the kernel of the GME to the second tier. Also, the famous Meir-Wingreen formula for the SIAM retarded Green's function, derived with the equation of motion approach~\cite{Meir1991}, is obtained here within a selection of second-tier diagrams, which we call second-tier noncrossing approximation (NCA2). \\ 
\indent  While the resonant tunneling approximation and the NCA2  already capture the onset of the Kondo zero-bias anomaly upon decreasing temperature, they both have some drawbacks that can be overcome only by going to higher-tier treatments. To this aim, we develop first an infinite-tier approximation, the dressed bubble approximation (DBA). Then, we proceed with a simplified version of the DBA which neglects the crossing at all levels, the NCA.  Here, the evaluation of the SIAM retarded Green's function is formally reduced to the inversion of a $4\times 4$ self-energy matrix and the Green's function self-energies become dressed by virtual tunneling transitions.
Being interested in analytical solutions, we investigate the outcomes of our approximation  within fourth-tier schemes dubbed NCA4 and the  simplified NCA4 (sNCA4).
Here, like the Green's function, also the NCA4 self-energies acquire a finite lifetime and an energy shift.
While inclusion of the full NCA4 self-energy is still intricate,  its simplified version allows for a complete \emph{analytical} treatment of the SIAM. Exemplarily, we show that the conductance is well-reproduced from high temperatures down to the Kondo temperature for moderate interaction. The validity of the sNCA4 in this parameter range was checked for the equilibrium SIAM against exact numerical renormalization group simulations. While the  sNCA4 solves the pinning problem and displays a Fermi liquid behavior at low $T$, predicting a saturation of the conductance at zero temperature, it still does not yield the expected unitary value $G=2e^2/h$ for a Kondo impurity.
 A full NCA4 treatment at the level of the fourth tier is expected to improve the low-temperature predictions for the SIAM current-voltage characteristics. We defer the study of  this full fourth-tier scheme to future investigations.\\ 
\indent Finally, the purpose of this work is to introduce an analytical approach to interacting nanojunctions based on the Feynman-Vernon influence functional, and to apply it to archetypal models such as the resonant level model and the SIAM. Due to the generality of the method, more complex situations encompassing multilevel or multi-dot systems, state non-conserving tunneling, or junctions subject to time-dependent drive can be included in the theory. We hope that this potential will stimulate further investigations using the Feynman-Vernon approach.

\section*{Acknowledgements}

The authors thank D. Mantelli, for providing the DM-NRG data, and M. Wegewijs, for fruitful discussions and for pointing out an inconsistency in the diagrammatic expansion of the exact kernel. This work was supported by SFB 1277, project B02, and by the BMBF (German Ministry for Education and Research), project 13N15208, QuantERA  SiUCs.

\newpage

\onecolumngrid

\newpage

\appendix

\section{Path integral expression for the system propagator}
\label{propagator_PI}  
In the absence of an external time-dependent driving, the propagator for the quantum state of the full system, from the initial time $t_0$ to time $t$, reads
\begin{equation}
\begin{aligned}
\label{state_propagator-full}
U(t,t_0)=e^{-\frac{\rm i}{\hbar}H(t-t_0)}\;,
\end{aligned}
\end{equation}
with $H$ the complete Hamiltonian of the transport setup, Eq.~\eqref{H_general}.\\
\indent A path integral expression for the reduced density matrix of the central system can be obtained in the coherent-state representation~\cite{Negele1988}. The fermionic coherent states for the central system are defined as $|\boldsymbol{\xi}\rangle=\exp(-\sum_{i}\xi^i\hat{a}_{i}^{\dag})|\boldsymbol{0}\rangle=\prod_i\exp(-\xi^i\hat{a}_{i}^{\dag})|0_i\rangle=\prod_i(1-\xi^i\hat{a}_{i}^{\dag})|0_i\rangle=\prod_i (|0_i\rangle-\xi^i|1_i\rangle)$, where the Grassmann variables $\boldsymbol{\xi}=(\dots,\xi^i,\dots)$ and $\boldsymbol{\xi}^{*}=(\dots,\xi^{i*},\dots)$ have one component for each electronic state $i$ defined by $\hat{a}_{i}|\boldsymbol{\xi}\rangle=\xi^i|\boldsymbol{\xi}\rangle$ and $\langle\boldsymbol{\xi}|\hat{a}_{i}^{\dag}=\langle\boldsymbol{\xi}| \xi^{i*}$. The Grassmann variables obey the relations $\{\xi^i,\xi^j\}=\{\xi^i,\xi^{j*}\}=0$, meaning that $(\xi^i)^2=0$. 
Analogous definitions hold for the leads' states $|\boldsymbol{\phi}\rangle$ in the coherent-state representation.
Using the notation from Cahill and Glauber~\cite{Cahill1999}
\begin{eqnarray}\label{notationCG}
\int d^2\xi^i := \int d\xi^{i*}d\xi^i \;,\qquad
\int d^2\boldsymbol{\xi}:=\int\prod_{i}d^2\xi^i\;,\qquad\text{and}\qquad  \boldsymbol{\xi}^{*}\cdot\boldsymbol{\xi} = \sum_i\xi^{i*}\xi^i\;,
\end{eqnarray}
the identity in the Hilbert space of the  central system reads
\begin{equation}
\begin{aligned}
\label{id}
\hat{I}=\int d^2\boldsymbol{\xi} e^{-\boldsymbol{\xi}^{*}\cdot\boldsymbol{\xi}} |\boldsymbol{\xi}\rangle\langle\boldsymbol{\xi}| \;.
\end{aligned}
\end{equation}
The over completeness of the set of coherent states is manifest in the overlap between coherent states
\begin{equation}
\begin{aligned}
\label{overlap}
\langle \boldsymbol{\xi}_a|\boldsymbol{\xi}_b\rangle = e^{\;\boldsymbol{\xi}_a^{*}\cdot\boldsymbol{\xi}_b}\;.
\end{aligned}
\end{equation}
The trace of an operator in the coherent-state representation is
\begin{equation}
\begin{aligned}
\label{tr}
\text{Tr}\{\mathcal{A}\}=\int d^2\boldsymbol{\xi} e^{-\boldsymbol{\xi}^{*}\cdot\boldsymbol{\xi}}\langle-\boldsymbol{\xi}|\mathcal{A}|\boldsymbol{\xi}\rangle\;,
\end{aligned}
\end{equation}
and the Gaussian integrals are performed via
\begin{equation}
\begin{aligned}
\label{Gauss-int}
\int d^2\boldsymbol{\xi}
e^{-\boldsymbol{\xi}^{*}\cdot\mathcal{M}\cdot\boldsymbol{\xi}+\boldsymbol{\eta}^{*}\cdot\boldsymbol{\xi}+\boldsymbol{\xi}^{*}\cdot\boldsymbol{\psi}}=\text{det}[\mathcal{M}]e^{\boldsymbol{\eta}^{*}\cdot\mathcal{M}^{-1}\cdot\boldsymbol{\psi}}\;.
\end{aligned}
\end{equation}
Assuming the factorized initial condition $\rho_{\rm tot}(t_0)=\rho(t_{0})\otimes\rho_{\text{leads}}$ for the total density matrix, the matrix element $\langle\boldsymbol{\xi}_{a}|\rho(t)|\boldsymbol{\xi}_{b}\rangle$ in the coherent-state representation of the system RDM $\rho(t)$  is given by the following trace over the leads
\begin{equation}
\begin{aligned}
\label{rhoab}
\langle\boldsymbol{\xi}_{a}|\rho(t)|\boldsymbol{\xi}_{b}\rangle=&\langle\boldsymbol{\xi}_{a}|\text{Tr}_{\text{leads}}\{U(t,t_{0})\rho_{\rm tot}(t_{0})U^{\dag}(t,t_{0})\}|\boldsymbol{\xi}_{b}\rangle\\
=&\int d^2\boldsymbol{\phi} \;e^{-\boldsymbol{\phi}^{*}\cdot\boldsymbol{\phi}}\langle-\boldsymbol{\xi}_{a}\boldsymbol{\phi}|U(t,t_{0})\rho(t_{0})\otimes\rho_{\text{leads}}(t_{0})U^{\dag}(t,t_{0})|\boldsymbol{\phi}\boldsymbol{\xi}_{b}\rangle\:,
\end{aligned}
\end{equation}
where $|\boldsymbol{\phi}\rangle$ is the state of the leads in the coherent-state representation.
At this point we apply the standard procedure of dividing the time interval $t-t_0$ into $K$ small intervals of length $\delta t$ and introducing the identity for the composite system
\begin{equation}
\begin{aligned}
\label{}
I_k=\int d^2\boldsymbol{\xi}(t_k) d^2\boldsymbol{\phi}(t_k) \;e^{-\boldsymbol{\xi}^{*}(t_k)\cdot\boldsymbol{\xi}(t_k)} e^{-\boldsymbol{\phi}^{*}(t_k)\cdot\boldsymbol{\phi}(t_k)}
|\boldsymbol{\phi}(t_k)\boldsymbol{\xi}(t_k)\rangle\langle\boldsymbol{\xi}(t_k)\boldsymbol{\phi}(t_k)|
\end{aligned}
\end{equation}
at each time instant $t_k$, both in the forward and in the backward propagators $U(t,t_{0})$ and $U^\dag(t,t_{0})$. This results in
\begin{equation}
\begin{aligned}
\label{rhoab2}
\langle\boldsymbol{\xi}_{a}|\rho(t)|\boldsymbol{\xi}_{b}\rangle=&\int d^2\boldsymbol{\phi} \;e^{-\boldsymbol{\phi}^{*}\cdot\boldsymbol{\phi}}\int d^2\boldsymbol{\xi}_0 d^2\boldsymbol{\phi}_0 d^2\bar{\boldsymbol{\xi}}_0 d^2\bar{\boldsymbol{\phi}}_0\;e^{-\boldsymbol{\xi}^{*}_0\cdot\boldsymbol{\xi}_0-\bar{\boldsymbol{\xi}}^{*}_0\cdot\bar{\boldsymbol{\xi}}_0} \;e^{-\boldsymbol{\phi}^{*}_0\cdot\boldsymbol{\phi}_0-\bar{\boldsymbol{\phi}}^{*}_0\cdot\bar{\boldsymbol{\phi}}_0}\\
&\times \langle-\xi_{a}\boldsymbol{\phi}|U(t,t_{0})|\boldsymbol{\phi}_{0}\boldsymbol{\xi}_{0}\rangle\langle\boldsymbol{\xi}_{0}\boldsymbol{\phi}_{0}|\rho(t_{0})\otimes\rho_{\text{leads}}(t_0)|\bar{\boldsymbol{\phi}}_{0}\bar{\boldsymbol{\xi}}_{0}\rangle\langle\bar{\boldsymbol{\xi}}_{0}\bar{\boldsymbol{\phi}}_{0}|U^{\dag}(t,t_{0})|\boldsymbol{\phi}\boldsymbol{\xi}_{b}\rangle\;.
\end{aligned}
\end{equation}
Explicitly, setting $\boldsymbol{\xi}^*(t_{K+1})\equiv \boldsymbol{\xi}_a^*$, $\bar{\boldsymbol{\xi}}(t_{K+1})\equiv \boldsymbol{\xi}_b$, $\langle\boldsymbol{\phi}(t_{K+1})| \equiv \langle-\boldsymbol{\phi}|$, and $|\bar{\boldsymbol{\phi}}(t_{K+1})\rangle \equiv |\boldsymbol{\phi}\rangle$, the path integral expression for the matrix elements of the forward and backward propagator read
\begin{equation}
\begin{aligned}
\label{U}
\langle - \boldsymbol{\xi}_{a}\boldsymbol{\phi}| U(t,t_{0}) |\boldsymbol{\phi}_{0}\boldsymbol{\xi}_{0}\rangle=&\int \prod_{k=1}^{K}d^2\boldsymbol{\xi}(t_k)d^2\boldsymbol{\phi}(t_k) e^{-\boldsymbol{\xi}^{*}(t_k)\cdot\boldsymbol{\xi}(t_k)-\boldsymbol{\phi}^{*}(t_k)\cdot\boldsymbol{\phi}(t_k)}\\
&\times\prod_{k=1}^{K+1}e^{\boldsymbol{\xi}^{*}(t_k)\cdot \boldsymbol{\xi}(t_{k-1})+\boldsymbol{\phi}^{*}(t_k)\cdot \boldsymbol{\phi}(t_{k-1})}e^{-\frac{\rm i}{\hbar}H[\boldsymbol{\xi}^{*}(t_k),\boldsymbol{\phi}^{*}(t_k),\boldsymbol{\xi}(t_{k-1}),\boldsymbol{\phi}(t_{k-1})]\delta t}
\end{aligned}
\end{equation}
and
\begin{equation}
\begin{aligned}
\label{Udag}
\langle\bar{\boldsymbol{\xi}}_{0}\bar{\boldsymbol{\phi}}_{0}|U^{\dag}(t,t_{0})|\bar{\boldsymbol{\phi}}\bar{\boldsymbol{\xi}}_b\rangle=&\int \prod_{k=1}^{K}d^2\bar{\boldsymbol{\xi}}(t_k)d^2\bar{\boldsymbol{\phi}}(t_k) e^{-\bar{\boldsymbol{\xi}}^{*}(t_k)\cdot\bar{\boldsymbol{\xi}}(t_k)-\bar{\boldsymbol{\phi}}^{*}(t_k)\cdot\bar{\boldsymbol{\phi}}(t_k)}\\
&\times\prod_{k=1}^{K+1}e^{\bar{\boldsymbol{\xi}}^{*}(t_{k-1})\cdot \bar{\boldsymbol{\xi}}(t_k)+\bar{\boldsymbol{\phi}}^{*}(t_{k-1})\cdot \bar{\boldsymbol{\phi}}(t_k)}e^{\frac{\rm i}{\hbar}H[\bar{\boldsymbol{\xi}}^{*}(t_{k-1}),\bar{\boldsymbol{\phi}}^{*}(t_{k-1}),\bar{\boldsymbol{\xi}}(t_k),\boldsymbol{\phi}(t_k)]\delta t}\;,
\end{aligned}
\end{equation}
respectively. Collecting the above results, we obtain the matrix element of the RDM at time $t$
\begin{equation}
\begin{aligned}
\label{}
\langle\boldsymbol{\xi}_{a}|\rho(t)|\boldsymbol{\xi}_{b}\rangle=\int d^2\boldsymbol{\xi}_0 d^2\bar{\boldsymbol{\xi}}_0 \mathcal{J}(\boldsymbol{\xi}^{*}_{a},\boldsymbol{\xi}_{b},t;\boldsymbol{\xi}_{0},\bar{\boldsymbol{\xi}}^{*}_{0},t_{0})\langle\boldsymbol{\xi}_0|\rho(t_0)|\bar{\boldsymbol{\xi}}_0\rangle\;,
\end{aligned}
\end{equation}
where the propagator has the following path integral expression
\begin{equation}
\begin{aligned}
\label{path_int_propagator}
\mathcal{J}(\boldsymbol{\xi}^{*}_{a},\boldsymbol{\xi}_{b},t;\boldsymbol{\xi}_{0},\bar{\boldsymbol{\xi}}^{*}_{0},t_{0})=\int_{\boldsymbol{\xi}_{0}}^{\boldsymbol{\xi}^{*}_{a}} D\boldsymbol{\xi}\int_{\bar{\boldsymbol{\xi}}^{*}_{0}}^{\boldsymbol{\xi}_{b}} D \bar{\boldsymbol{\xi}}\; e^{\frac{\rm i}{\hbar}[S_{\rm S}(\boldsymbol{\xi}^{*},\boldsymbol{\xi})-S_{\rm S}^*(\bar{\boldsymbol{\xi}}^{*},\bar{\boldsymbol{\xi}})]}\mathcal{F}(\boldsymbol{\xi}^{*},\boldsymbol{\xi},\bar{\boldsymbol{\xi}}^{*},\bar{\boldsymbol{\xi}})\;.
\end{aligned}
\end{equation}
Here, the integration measures of the Grassmann-valued paths are defined by $\int D\boldsymbol{\xi}= \int\prod_{k=1}^K d\boldsymbol{\xi}_k^*d\boldsymbol{\xi}_k$ and  $\int D\bar{\boldsymbol{\xi}}= \int\prod_{k=1}^K d\bar{\boldsymbol{\xi}}_k^*d\bar{\boldsymbol{\xi}}_k$.
The functional containing the action of the central system is given by Eq.~\eqref{action_S} of the main text. The Feynman-Vernon influence functional $\mathcal{F}(\boldsymbol{\xi}^{*},\boldsymbol{\xi},\bar{\boldsymbol{\xi}}^{*},\bar{\boldsymbol{\xi}})=\exp[\varPhi(\boldsymbol{\xi}^{*},\boldsymbol{\xi},\bar{\boldsymbol{\xi}}^{*},\bar{\boldsymbol{\xi}})]$ is a functional of the Grassmann-valued paths of the central system which encapsulates the dissipative effect due to the coupling to the leads. Its phase reads~\cite{Tu2008,Jin2008,Jin2010} 
\begin{equation}
\begin{aligned}
\label{phase}
\varPhi&(\boldsymbol{\xi}^{*},\boldsymbol{\xi},\bar{\boldsymbol{\xi}}^{*},\bar{\boldsymbol{\xi}})=-\sum_{i j}\int_{t_{0}}^{t}dt'\int_{t_{0}}^{t'}d t''\left[ g_{i j}(t'-t'')\xi^{i*}(t')\xi^j(t'')+\mathcal{G}^{*}_{j i}(t'-t'') \bar{\xi}^{i*}(t'') \bar{\xi}^j(t')\right]\\
&-\sum_{i j}\int_{t_{0}}^{t}dt'\int_{t_0}^{t} d t''\left\{ g_{i j}(t'-t'')\bar{\xi}^{i*}(t')\xi^j(t'')-{g}_{+,i j}(t'-t'')\left[\xi^{i*}(t')+\bar{\xi}^{i*}(t')\right]	\left[\xi^j(t'')+\bar{\xi}^j(t'')\right]\right\},
\end{aligned}
\end{equation}
with the temperature-independent  and temperature-dependent correlation matrix of elements $g_{i j}(t)$ and ${g}_{+,i j}(t)$ defined as
\begin{equation}
\begin{aligned}
\label{corr-func}
g_{i j}(t)=&\frac{1}{\hbar^2}\sum_{\alpha  k \sigma}  {\rm t}_{i \alpha k }{\rm t}^{*}_{j \alpha k \sigma \sigma}e^{-\frac{\rm i}{\hbar}\epsilon_{\alpha k}t}\\
{g}_{+,i j}(t)=&\frac{1}{\hbar^2}\sum_{\alpha k \sigma} {\rm t}_{i \alpha k \sigma}{\rm t}^{*}_{j \alpha k \sigma}f_+^\alpha(\epsilon_{k})e^{-\frac{\rm i}{\hbar}\epsilon_{\alpha k}t}\;,
\end{aligned}
\end{equation}
respectively, where $f_+^\alpha(\epsilon_{k})=[1+e^{\beta_{\alpha}(\epsilon_{\alpha k}-\mu_{\alpha})}]^{-1}$ is the Fermi function  of lead $\alpha$. 
We also define
\begin{equation}
\begin{aligned}
\label{g-}
{g}_{-,i j}(t):=& g_{i j}(t)-{g}_{+,i j}(t)\\
=&\frac{1}{\hbar^2}\sum_{\alpha k \sigma} {\rm t}_{i \alpha k \sigma}{\rm t}^{*}_{j \alpha k \sigma}f_-^\alpha(\epsilon_{k})e^{-\frac{\rm i}{\hbar}\epsilon_{\alpha k}t}\;,
\end{aligned}
\end{equation}
where $f_-^\alpha(\epsilon_k):=1-f_+^\alpha(\epsilon_k)$.

\section{Phase of the influence functional}
\label{Influence_phase}

The terms in Eq.~\eqref{phase} can be rearranged in a convenient manner 
\begin{equation}
\begin{aligned}
\label{phase2}
\varPhi(\boldsymbol{\xi}^{*},\boldsymbol{\xi},\bar{\boldsymbol{\xi}}^{*},\bar{\boldsymbol{\xi}})=&-\sum_{i j}\int_{t_{0}}^{t}dt'\int_{t_{0}}^{t'}d t''\left[ g_{i j}(t'-t'')\xi^{i*}(t')\xi^j(t'')+\mathcal{G}^{*}_{j i}(t'-t'') \bar{\xi}^{i*}(t'') \bar{\xi}^j(t')\right]\\
+&\sum_{i j}\int_{t_{0}}^{t}dt'\int_{t_{0}}^{t}d t''\Big[ {g}_{+,i j}(t'-t'')\xi^{i*}(t')\xi^j(t'')+{g}_{+,i j}(t'-t'') \bar{\xi}^{i*}(t') \bar{\xi}^j(t'')\\
&\qquad\qquad\qquad\quad +{g}_{+,i j}(t'-t'')\xi^{i*}(t')\bar{\xi}^j(t'') - {g}_{-,i j}(t'-t'')\bar{\xi}^{i*}(t')\xi^j(t'')\Big]\;.
\end{aligned}
\end{equation}
\indent Further, exchanging the order of integration and using the relation $ g_{i j}(-t)= g^{*}_{j i}(t)$, see Eq.~\eqref{corr-func}, the influence phase in Eq.~\eqref{phase2} can be cast in the following compact form 
\begin{equation}
\begin{aligned}
\label{phase2b}
\varPhi(\boldsymbol{\xi}^{*},\boldsymbol{\xi},\bar{\boldsymbol{\xi}}^{*},\bar{\boldsymbol{\xi}})=&-\sum_{i j}\int_{t_{0}}^{t}dt'\int_{t_{0}}^{t'}d t''\left[ g_{i j}(t'-t'')\xi^{i*}(t')\xi^j (t'')+\mathcal{G}^{*}_{j i}(t'-t'') \bar{\xi}^{i*}(t'') \bar{\xi}^j (t')\right]\\
+&\sum_{i j}\int_{t_{0}}^{t}dt'\int_{t_{0}}^{t'}d t''\Big[ {g}_{+,i j}(t'-t'')\xi^{i*}(t')\xi^j (t'')+{g}_{+,i j}(t'-t'') \bar{\xi}^{i*}(t') \bar{\xi}^j (t'')\\
&\qquad\qquad\qquad\quad +{g}_{+,i j}(t'-t'')\xi^{i*}(t')\bar{\xi}^j (t'') - {g}_{-,i j}(t'-t'')\bar{\xi}^{i*}(t')\xi^j (t'')\\
&\qquad\qquad\qquad\quad +{g}_{+,j i}^{*}(t'-t'')\xi^{i*}(t'')\xi^j (t')+{g}_{+,j i}^{*}(t'-t'') \bar{\xi}^{i*}(t'') \bar{\xi}^j (t')\\
&\qquad\qquad\qquad\quad +{g}_{+,j i}^{*}(t'-t'')\xi^{i*}(t'')\bar{\xi}^j (t') - {g}_{-,j i}^{*}(t'-t'')\bar{\xi}^{i*}(t'')\xi^j (t')
\Big]\\
=&-\sum_{i j}\int_{t_{0}}^{t}dt'\int_{t_{0}}^{t'}d t''\Big[ {g}_{-,i j}(t'-t'')\xi^{i*}(t')\xi^j (t'')-{g}_{+,i j}(t'-t'') \bar{\xi}^{i*}(t') \bar{\xi}^j (t'')\\
&\qquad\qquad\qquad\qquad -{g}_{+,i j}(t'-t'')\xi^{i*}(t')\bar{\xi}^j (t'') + {g}_{-,i j}(t'-t'')\bar{\xi}^{i*}(t')\xi^j (t'')\\
&\qquad\qquad\qquad\qquad -{g}_{+,j i}^{*}(t'-t'')\xi^{i*}(t'')\xi^j (t')+{g}_{-,j i}^{*}(t'-t'') \bar{\xi}^{i*}(t'') \bar{\xi}^j (t')\\
&\qquad\qquad\qquad\qquad -{g}_{+,j i}^{*}(t'-t'')\xi^{i*}(t'')\bar{\xi}^j (t') + {g}_{-,j i}^{*}(t'-t'')\bar{\xi}^{i*}(t'')\xi^j (t')
\Big]\\
=&
-\int_{t_{0}}^{t}dt'\int_{t_{0}}^{t'}d t''\Big[\boldsymbol{\xi}^{*}(t')\cdot  \mathbf{g}_-(t'-t'')\cdot\boldsymbol{\xi}(t'') + \boldsymbol{\xi}(t')\cdot \mathbf{g}_+^*(t'-t'')\cdot\boldsymbol{\xi}^{*}(t'')\\
&\qquad\qquad\qquad\; -\bar{\boldsymbol{\xi}}(t')\cdot  \mathbf{g}_-^*(t'-t'')\cdot\bar{\boldsymbol{\xi}}^{*}(t'')-\bar{\boldsymbol{\xi}}^{*}(t')\cdot \mathbf{g}_+(t'-t'')\cdot\bar{\boldsymbol{\xi}}(t'')\\
&\qquad\qquad\qquad\;+\bar{\boldsymbol{\xi}}^{*}(t')\cdot  \mathbf{g}_-(t'-t'')\cdot\boldsymbol{\xi}(t'') + \bar{\boldsymbol{\xi}}(t')\cdot \mathbf{g}_+^*(t'-t'')\cdot\boldsymbol{\xi}^{*}(t'')\\
&\qquad\qquad\qquad\; - \boldsymbol{\xi}(t')\cdot\mathbf{g}_-^*(t'-t'')\cdot\bar{\boldsymbol{\xi}}^{*}(t'')-\boldsymbol{\xi}^{*}(t')\cdot \mathbf{g}_+(t'-t'')\cdot\bar{\boldsymbol{\xi}}(t'')\Big]\\
=&-\int_{t_{0}}^{t}dt'\int_{t_{0}}^{t'}d t''\sum_{x,y,z=\pm 1}x\;\boldsymbol{\xi}_{y}^{z}(t')\mathbf{g}_{xz}^{-z}(t'-t'')\boldsymbol{\xi}_{x}^{-z}(t'')\;,
\end{aligned}
\end{equation}
where we used the anticommutation property of the Grassmann variables and Eq.~\eqref{g-}. In the last line, we established the notation
\begin{equation}
\begin{aligned}
\label{}
\boldsymbol{\xi}_{+1}^{+1}&=\boldsymbol{\xi}\;,\quad\boldsymbol{\xi}_{+1}^{-1}=\boldsymbol{\xi}^*,\quad\boldsymbol{\xi}_{-1}^{+1}=\bar{\boldsymbol{\xi}},\quad\boldsymbol{\xi}_{-1}^{-1}=\bar{\boldsymbol{\xi}}^*,\\
\mathbf{g}_{+1}^{+1}&=\mathbf{g}_+,\quad \mathbf{g}_{+1}^{-1}=\mathbf{g}_+^*,\quad  \mathbf{g}_{-1}^{+1}=\mathbf{g}_-,\quad \mathbf{g}_{-1}^{-1}=\mathbf{g}_-^*\;.
\end{aligned}
\end{equation}
Equation~\eqref{phase2b} is the form of the influence phase used throughout this work.

\section{Leads' force operator correlation function}
\label{leads_corr_func}
The correlation functions $g_{\pm,i j}(t)$ are related to the correlation function of the (fermion) baths force operator which appears in the quantum Langevin equation for the dot operator $a_i(t)$. Indeed, given the full Hamiltonian Eq.~\eqref{H_general},
the Heisenberg equation of motion for the leads' operators $\dot{c}_{\alpha k \sigma}(t)={\rm i}[H,c_{\alpha k \sigma}(t)]/\hbar$ is solved by
\begin{equation}
\begin{aligned}
\label{solution_c}
c_{\alpha k \sigma}(t)=c_{\alpha k \sigma}(t_0)e^{-\frac{\rm i}{\hbar} \epsilon_{\alpha k}(t-t_0)}-\frac{\rm i}{\hbar}\sum_j {\rm t}_{j \alpha k \sigma}^* \int_{t_0}^t
dt'e^{-\frac{\rm i}{\hbar} \epsilon_{\alpha k}(t-t')} a_j(t')\;.
\end{aligned}
\end{equation}
Plugging this result in the Heisenberg equation  for the system operator $a_i(t)$
\begin{equation}
\begin{aligned}
\label{}
\dot{a}_i(t)=&\frac{\rm i}{\hbar}[H,a_i(t)]\\
=&\frac{\rm i}{\hbar}[H_{\rm S},a_i(t)]-\frac{\rm i}{\hbar}\sum_{\alpha k \sigma} {\rm t}_{i \alpha k \sigma}c_{\alpha k \sigma}(t)
\end{aligned}
\end{equation}
we obtain the quantum Langevin equation 
\begin{equation}
\begin{aligned}
\label{}
\dot{a}_i(t)=&\frac{\rm i}{\hbar}[H_{\rm S},a_i(t)]-\frac{1}{\hbar^2}\sum_{j \alpha k \sigma} {\rm t}_{i \alpha k \sigma}{\rm t}_{j \alpha k \sigma}^* \int_{t_0}^t
dt'e^{-\frac{\rm i}{\hbar} \epsilon_{\alpha k}(t-t')} a_j(t')+\hat\zeta_i(t)\;,
\end{aligned}
\end{equation}
where the baths force operator reads
\begin{equation}
\begin{aligned}
\label{}
\hat\zeta_i(t)=-\frac{\rm i}{\hbar}\sum_{\alpha k \sigma}{\rm t}_{i \alpha k \sigma}e^{-\frac{\rm i}{\hbar} \epsilon_{\alpha k}(t-t_0)}c_{\alpha k \sigma}(t_0)\;,
\end{aligned}
\end{equation}
see, e.g., Ref.~\cite{Kohler2004}. The correlation functions in Eq.~\eqref{corr-func} are thus related to the correlation function of the baths' force operators via
\begin{equation}
\begin{aligned}
\label{}
\langle\hat\zeta^\dag_i(t)\hat\zeta_j(t')\rangle =&{g}_{+,i j}(t-t')\\
\langle\hat\zeta_i(t)\hat\zeta^\dag_j(t')\rangle =&{g}_{-,i j}(t-t')\;.
\end{aligned}
\end{equation}

\section{Path integral representation of the current and the Green's functions}
\label{Greens_functions-PI}

Consider the current on lead $l$. Using the definition $f_-^l (\epsilon_k):=1-f_+^l (\epsilon_k)$, the current functional $\mathcal{I}$ in Eq.~\eqref{current_functional} can be rewritten as 
\begin{equation}
\begin{aligned}
\label{mI2}
\mathcal{I}_{l }(\boldsymbol{\xi}^{*},\boldsymbol{\xi},\bar{\boldsymbol{\xi}})=&-\int_{t_0}^t  dt'\; \left\{ \boldsymbol{\xi}^*(t)\mathbf{g}_{l }(t-t')\boldsymbol{\xi}(t')-\boldsymbol{\xi}^*(t)\tilde{\mathbf{g}}_{+,l }(t-t')[\boldsymbol{\xi}(t')+\bar{\boldsymbol{\xi}}(t')] \right\}\,,
\end{aligned}
\end{equation}
where
\begin{equation}
\begin{aligned}
\label{}
[\mathbf{g}_l (t)]_{i j}=[\mathbf{g}_{+,l }(t)+\mathbf{g}_{-,l }(t)]_{i j}=&\frac{1}{\hbar^2}\sum_{k \sigma} {\rm t}_{i l k \sigma}{\rm t}^{*}_{j l k \sigma}e^{-\frac{\rm i}{\hbar}\epsilon_{l k}t}\;.
\end{aligned}
\end{equation}
With this expressions, recalling the relation between the operators $a_i$ and the corresponding Grassmann variables $\xi^i$, the current $I_l (t)=- 2{\rm Re}\;{\rm Tr}_{\rm S}[\boldsymbol{\mathcal{A}}_l (t)]$, with the path integral representation of $\boldsymbol{\mathcal{A}}_l (t)$ given by eqs.~\eqref{A_alpha_PI} and~\eqref{propagator_I} , 
 can be seen as the path integral representation of the following trace over system and leads degrees of freedom
\begin{equation}
\begin{aligned}
\label{AalphaGF}
{\rm Tr}_{\rm S}[\boldsymbol{\mathcal{A}}_l (t)]=&-\sum_{ij}\int_{t_0}^t  dt'\;\Bigg\{ g_{l ij}(t-t'){\rm Tr}_{\rm tot}\left[a_i^\dag U(t,t') a_j U(t',t_0)\rho_{\rm tot}(t_0)U^\dag(t,t_0) \right]\\
&-{g}_{+,l ij}(t-t'){\rm Tr}_{\rm tot}\left[a_i^\dag U(t,t') a_j U(t',t_0)\rho_{\rm tot}(t_0)U^\dag(t,t_0)+a_i^\dag U(t,t_0)\rho_{\rm tot}(t_0)U(t',t_0)a_j U^\dag(t,t')\right]\Bigg\}\\
=&-\sum_{ij}\int_{t_0}^t  dt'\;\left[g_{l ij}(t-t')\langle a_i^\dag (t) a_j(t')\rangle -{g}_{+,l ij}(t-t') \langle \{a_i^\dag (t) , a_j(t')\}\rangle\right]\\
=&{\rm i}\hbar\sum_{ij}\int_{t_0}^t  dt'\;\left[g_{l ij}(t-t')\mathcal{G}^<_{ji}(t'-t)-{g}_{+,l ij}(t-t')\mathcal{G}^a_{ji}(t'-t)\right]\\
=&{\rm i}\hbar\int_{t_0}^t  dt'\;{\rm Tr}\left[\mathbf{g}_{l }(t-t')\cdot\boldsymbol{\mathcal{G}}^<(t'-t)-\mathbf{g}_{+,l }(t-t')\cdot\boldsymbol{\mathcal{G}}^a(t'-t)\right]
\;,
\end{aligned}
\end{equation}
where the last trace is in the matrix sense. The lesser, retarded, and advanced Green's functions are defined by
\begin{equation}
\begin{aligned}
\label{GFdef}
[\boldsymbol{\mathcal{G}}^<(t',t)]_{ij}=&{\rm i}\langle a_j^\dag (t) a_i(t') \rangle/\hbar\;,\\
[\boldsymbol{\mathcal{G}}^r(t,t')]_{ij}=&-{\rm i}\theta(t-t')\langle \{ a_i(t), a_j^\dag (t')\} \rangle/\hbar\;,\\
[\boldsymbol{\mathcal{G}}^a(t,t')]_{ij}=&[\boldsymbol{\mathcal{G}}^{r \dag}(t',t)]_{ij}\\
=&{\rm i}\theta(t'-t)\langle \{a_j^\dag(t'), a_i (t)\} \rangle/\hbar\;,
\end{aligned}
\end{equation}
respectively.
Note that the Heaviside function is already taken into account in the time integral that guarantees the ordering $t'< t$.

\section{Two leads and proportional coupling}
\label{prop_coupling}

Let us confine ourselves to the case of diagonal correlation matrices $\mathbf{g}_{+,\alpha}$. Having diagonal correlation matrices implies that, in  the continuum limit,
$$
[\boldsymbol{\Gamma}_{\alpha}(\epsilon)]_{ij}:=2\pi\sum_\sigma\varrho_{\alpha \sigma} (\epsilon) |{\rm t}_{i \alpha \sigma}(\epsilon)|^2\delta_{ij}\;.
$$
In a typical transport setting, the system is connected to two leads, $\alpha=L,R$. In the case of proportional coupling, the tunneling coefficients in the Hamiltonian are related by ${\rm t}_{i R \sigma}(\epsilon)=\sqrt{\gamma_{i R}/\gamma_{i L}}\;{\rm t}_{i L \sigma}(\epsilon)$ with $\gamma_{i L}+\gamma_{i R}=1$. Since $I_L^\infty=-I_R^\infty$, the current $I_L$ is asymptotically equal to the current $I(t)=\sum_i[\gamma_{i R} I_{i L}(t)-\gamma_{i L} I_{i R}(t)]$ which  we can directly write as 
\begin{equation}
\begin{aligned}
\label{IPropCoupling2}
I(t)=\;e 2{\rm Re}\; {\rm Tr}_{\rm S}[\boldsymbol{\mathcal{A}}(t)]\;.
\end{aligned}
\end{equation}
The path integral representation for the dot operator with diagonal elements $\mathcal{A}_{ii}(t):=\gamma_{i R} \mathcal{A}_{ii L}(t)-\gamma_{ii L} \mathcal{A}_{ii R}(t)$ is formally the same as the one in Eq.~\eqref{A_alpha_PI}. The current propagator $\mathcal{J}^I$ for $\mathcal{A}(t)$ is 
similar to $\mathcal{J}^I_l $, Eq.~\eqref{propagator_I}, the difference being the functional $\mathcal{I}(\boldsymbol{\xi}^{*},\boldsymbol{\xi},\bar{\boldsymbol{\xi}})$   in place of $\mathcal{I}_l (\boldsymbol{\xi}^{*},\boldsymbol{\xi},\bar{\boldsymbol{\xi}})$, where 
\begin{equation}
\label{CurrentFunctionalPropCoupl}
\mathcal{I}(\boldsymbol{\xi}^{*},\boldsymbol{\xi},\bar{\boldsymbol{\xi}})=\sum_i\int_{t_0}^t  dt'\; {\xi}^{i*}(t)\big[\gamma_{i R} {\rm g}_{+,iiL}(t-t')-\gamma_{i L}{\rm g}_{+,iiR}(t-t')\big][{\xi}^i(t')+\bar{\xi}^i(t')]\;.
\end{equation}
Here we used the property $f_-^{\alpha}(\epsilon)=1-f_+^{\alpha}(\epsilon)$ in the definition of the correlation matrices.
In the calculation of $I(t)$ for proportional coupling, the temperature-independent term involving the lesser Green's function in Eq.~\eqref{AalphaGF} drops and 
\begin{equation}
\begin{aligned}
\label{IPropCoupling}
{\rm Tr}_{\rm S}[\boldsymbol{\mathcal{A}}(t)]=&\sum_i\left[\gamma_{i R} \mathcal{A}_{ii L}(t)-\gamma_{i L} \mathcal{A}_{ii R}(t)\right]\\
=&-{\rm i}\hbar\sum_i\int_{t_0}^t  dt'\;
\left[\gamma_{i R} {\rm g}_{+,ii L}(t-t')-\gamma_{i L}{\rm g}_{+,ii R}(t-t')\right]\mathcal{G}_{ii}^a(t'-t)\;.
\end{aligned}
\end{equation}
In the continuum limit $\sum_{k \sigma}\rightarrow \sum_\sigma\int d\epsilon \varrho_{\alpha\sigma}(\epsilon)$, with $\varrho_{\alpha\sigma}(\epsilon)$ the density of states in energy space of lead $\alpha$. We define $\boldsymbol{\Gamma}(\epsilon)=\boldsymbol{\Gamma}_{L}(\epsilon)+\boldsymbol{\Gamma}_{R}(\epsilon)$, 
so that, for proportional coupling,  Eq.~\eqref{IPropCoupling} reads
\begin{equation}
\begin{aligned}
\label{IPropCouplingContinuum}
I(t)
=&\;e\sum_i\frac{\gamma_{i L}\gamma_{i R}}{\pi\hbar} \int d\epsilon \left[f_+^L(\epsilon)-f_+^R(\epsilon)\right] {\rm Im}\;\left[\Gamma_{ii}(\epsilon)\int_{t_0}^t dt'\; e^{-\frac{\rm i}{\hbar}\epsilon(t-t')}\mathcal{G}_{ii}^a(t'-t)\right]\;.
\end{aligned}
\end{equation}
\subsection*{Asymptotic limit}
In the limit $t-t_0\rightarrow\infty$, the time integral in Eq.~\eqref{IPropCouplingContinuum} yields the Fourier transform with
\begin{equation}
\begin{aligned}
\label{IPropCouplingLandauer}
I^\infty=&\;e\sum_i\frac{\gamma_{i L}\gamma_{i R}}{\pi\hbar} \int d\epsilon \left[f_+^L(\epsilon)-f_+^R(\epsilon)\right] {\rm Im}\;\left[{\Gamma}_{ii}(\epsilon)\mathcal{G}_{ii}^a(\epsilon)\right]\;.
\end{aligned}
\end{equation}
Taking into account the definition of the matrix $\boldsymbol{\Gamma}(\epsilon)$, the current formula~\eqref{IPropCouplingLandauer} coincides with the well-known result of Meir and Wingreen~\cite{Meir1992}.
We have
\begin{equation}
\begin{aligned}
\label{IPropCouplingLandauer2}
I^\infty=&\;e \sum_i\frac{\gamma_{i L}\gamma_{i R}}{\hbar}  \int d\epsilon \left[f_+^L(\epsilon)-f_+^R(\epsilon)\right]\Gamma_{ii}(\epsilon) \frac{1}{\pi}{\rm Im}\;\mathcal{G}^a_{ii}(\epsilon)\\
=&\frac{e}{\hbar}  \sum_i \int d\epsilon \left[f_+^L(\epsilon)-f_+^R(\epsilon)\right]\left[\frac{\boldsymbol\Gamma_{L}(\epsilon)\boldsymbol\Gamma_{R}(\epsilon)}{\boldsymbol\Gamma_{L}(\epsilon)+\boldsymbol\Gamma_{R}(\epsilon)}\right]_{ii}\left[-\frac{1}{\pi}{\rm Im}\;\mathcal{G}^r_{ii}(\epsilon)\right]\;,
\end{aligned}
\end{equation}
where we used the relation ${\rm Im}\;\mathcal{G}^a_{ii}(\epsilon)=-{\rm Im}\;\mathcal{G}^r_{ii}(\epsilon)$.

\newpage

\section{Integrating out the Grassmann variables  in the SIAM}
\label{integration_Grassmann_var}

In this appendix, we show how to trace over the Grassmann variables associated to the paths of the central system for specific instances of paths. This procedure yields ultimately the diagrammatic rules that can be traced back to the anticommutation property of  Grassmann numbers. First, we exemplify the procedure for the simplest case of a central system consisting of a single, spinless level, the resonant level model. Then, we make the calculations for the more involved case of the single-impurity Anderson model. Here, due to the Coulomb interaction, the phase associated to the action of the dot in the path integral expression for the propagator produces the phase factors that couple the diagrammatic contributions stemming from the individual spin paths.\\ 
\indent In order to perform specific calculations we employ the formula that connects the coherent-state representation of the propagator for the populations to a given order $m$ to the corresponding occupation number representation
\begin{equation}
\begin{aligned}
\label{Propagator_general_appendix}
J_{\boldsymbol{n}'\boldsymbol{n}}^{(m)}(t,t_0)=&\Pi_{b}(\boldsymbol{n}')\Pi_{a}^*(\boldsymbol{n}')\int d^2\boldsymbol{\xi}_0 d^2\bar{\boldsymbol{\xi}}_0 \mathcal{J}^{(m)}(\boldsymbol{\xi}^{*}_a,\bar{\boldsymbol{\xi}}_b,t;\boldsymbol{\xi}_0,\bar{\boldsymbol{\xi}}^{*}_0,t_0)\langle \boldsymbol\xi_0|\boldsymbol{n}\rangle\langle \boldsymbol{n}|\bar{\boldsymbol{\xi}}_0\rangle\;,
\end{aligned}
\end{equation}
where $\boldsymbol{n}=(\dots,n^i,\dots)$ with $n^i=0,1$. The projectors are defined by
\begin{equation}
\begin{aligned}
\label{def_projector_appendix}
\Pi(\boldsymbol{n})=\prod_{i=1}^N\Pi^i(n_i)\;,\qquad \Pi^*(\boldsymbol{n})=\prod_{i=N}^1\Pi^{i*}(n_i) \;,
\end{aligned}
\end{equation}
with
\begin{equation}
\begin{aligned}
\Pi^*(0) = \int d\xi^* \xi^*\;,\quad \Pi^*(1)=\int d\xi^* \;,\quad \Pi(0) = \int d\bar\xi \;\bar\xi\;,\quad \Pi(1)=\int d\bar\xi \;.
\end{aligned}
\end{equation}
\subsection*{Resonant level model}
 
In the RLM, the central system consists of a single, spinless level with energy $\epsilon$.
We start by considering in full detail specific instances of paths with low number of transitions, situated in the forward and backward paths. In this case the occupation of the level is the single degree of freedom of the central system. 
Let us use Eqs.~\eqref{Propagator_general_appendix} and~\eqref{def_projector_appendix}  to evaluate the contributions to the propagator at different orders in $\Gamma$ given by specific instances of paths. \\
\indent According to Eq.~\eqref{RDM1state} we have
$$
\langle \xi_0| 0 \rangle\langle 0 |\bar{\xi}_0\rangle=1\;\quad{\rm and}\quad \langle \xi_0| 1 \rangle\langle 1 |\bar{\xi}_0\rangle=\xi^*\bar\xi\;.
$$
\indent To order zero, using Eq.~\eqref{action_S} and defining $p:=-{\rm i}\epsilon\delta t/\hbar$
\begin{equation}\begin{aligned}\label{rho00discr0}
J^{(0)}_{00}(t;t_0)=&\Pi_{b}(0)\Pi_{a}^*(0)\int d^2\xi_0 d^2\bar{\xi}_0 \mathcal{J}^{(0)}(\xi^{*}_a,\bar{\xi}_b,t;\xi_0,\bar{\xi}^{*}_0,t_0)\langle\xi_0|0\rangle\langle 0|\bar{\xi}_0\rangle\\
=&\int  d\bar{\xi}_{N+1}  \bar{\xi}_{N+1}d\xi_{N+1}^*\xi_{N+1}^*\prod_{n=0}^{N}d^2\xi_n d^2\bar{\xi}_n e^{-\xi_n^* \xi_n}e^{-\bar{\xi}_n^* \bar{\xi}_n}\prod_{n=1}^{N+1}e^{\xi_n^*\xi_{n-1}p_n}e^{\bar{\xi}_{n-1}^*\bar{\xi}_n p_n^*}\\
=&\int  d\bar{\xi}_{N+1}  \bar{\xi}_{N+1}d\xi_{N+1}^*\xi_{N+1}^*\prod_{n=0}^{N}d^2\xi_n d^2\bar{\xi}_n e^{-\xi_n^* \xi_n}e^{-\bar{\xi}_n^* \bar{\xi}_n}\\
=&1\;,
\end{aligned}\end{equation}
with ${\xi}^*_{N+1}\equiv {\xi}_a^*$ and $\bar{{\xi}}_{N+1}\equiv {\xi}_b$.
Here we have used the property $\exp(\psi)=1+\psi$ and Eq. (\ref{Grassmann_integrals}). These properties imply that, for all $n$, the terms in the rightmost product contribute as $1$, otherwise there would be either products of same Grassmann numbers ($\psi^2=0$) or integrations not compensated by the corresponding Grassmann numbers in the integrand ($\int d\psi=0$), see Eq.~\eqref{Grassmann_integrals}.\\
\indent On the other hand, along the same lines one can see that a path with no transitions cannot join two states with different occupation, namely
\begin{equation}\begin{aligned}\label{rho11discr0}
J^{(0)}_{10}(t;t_0)=&\Pi_{b}(1)\Pi_{a}^*(1)\int d^2\xi_0 d^2\bar{\xi}_0 \mathcal{J}^{(0)}(\xi^{*}_a,\bar{\xi}_b,t;\xi_0,\bar{\xi}^{*}_0,t_0)\langle\xi_0|0\rangle\langle 0|\bar{\xi}_0\rangle\\
=&\int d\bar{\xi}_{N+1}  d\xi_{N+1}^* \prod_{n=0}^{N}d^2\xi_n d^2\bar{\xi}_n e^{-\xi_n^* \xi_n}e^{-\bar{\xi}_n^* \bar{\xi}_n}\prod_{n=1}^{N+1}e^{\xi_n^*\xi_{n-1}p_n}e^{\bar{\xi}_{n-1}^*\bar{\xi}_n p_n^*}\\
=&0\;,
\end{aligned}\end{equation}
where, again, we used the properties of the Grassmann integrals, Eq.~\eqref{Grassmann_integrals}.
In the following we associate the color red to the tunneling times and the colors black and blue to the  sojourn and blip times, respectively. These are the time intervals when the RDM is in a diagonal (resp. off-diagonal) state, see Fig.~\ref{parametrization_paths}.\\
\begin{figure}[ht]
\begin{center}
\includegraphics[width=14cm,angle=0]{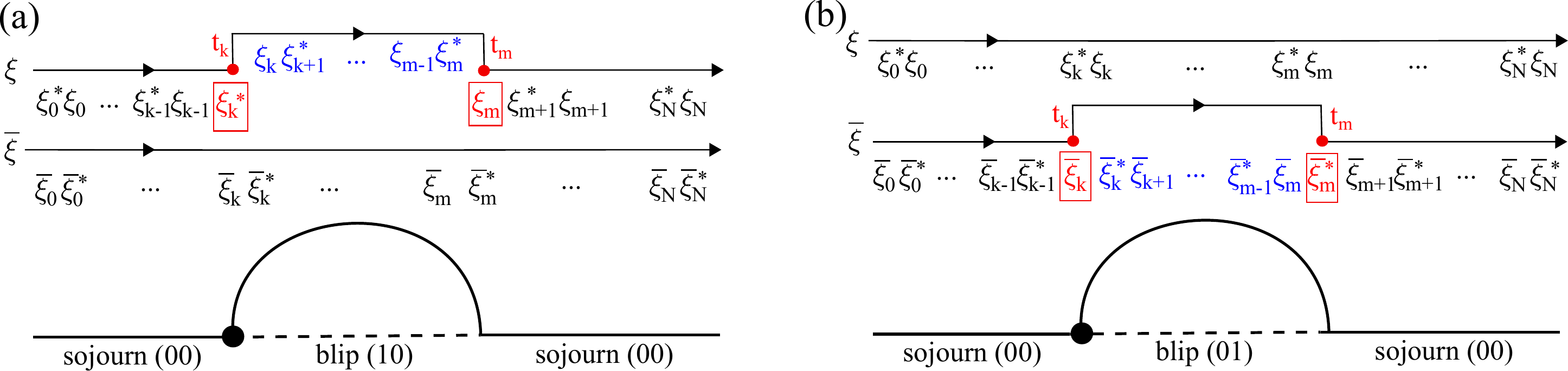}\\
\caption{\small{Two examples of paths with two transitions, either in the forward (a) or in the backward (b) time branch. The Grassmann variables boxed in red are the ones appearing in the influence functional for the examples of paths considered. Note that we use the same time direction for the two branches with the consequence that $\bar{\xi}$ and $\bar{\xi}^*$ creates and annihilates an electron in the dot, respectively.}}
\label{path2nd-a-b}
\end{center}
\end{figure}
\indent Going to first order in $\Gamma$,
 we first consider the path 
in Fig. \ref{path2nd-a-b}(a) with tunneling transitions in the forward branch. Specifically, an electron is created in the central system at time $t_k$ and subsequently annihilated at time $t_m$. The path is thus identified by the sequence $\xi_k^*, \xi_m$, see Fig.~\ref{parametrization}. 
Using the preliminary results
\begin{equation}\begin{aligned}\label{results_1}
\prod_{n=k}^m d^2\xi_n=&d\xi_k^*\left( \prod_{n=k+1}^m d\xi_{n-1}d\xi^*_n\right) d\xi_m\\
\prod_{n=k}^m d^2\bar{\xi}_n=&\prod_{n=m}^k d^2\bar{\xi}_n=d\bar{\xi}_m^*\left( \prod_{n=m}^{k+1} d\bar{\xi}_{n}d\bar{\xi}^*_{n-1}\right) d\bar{\xi}_k\;,
\end{aligned}\end{equation}
the contribution from path $(a)$ in Fig. \ref{path2nd-a-b} is obtained as 
\begin{equation}\begin{aligned}\label{rho00discr2a}
J_{(a),00}^{(1)}(t;t_0)=&\Pi_{b}(0)\Pi_{a}^*(0)\int d^2\xi_0 d^2\bar{\xi}_0 \mathcal{J}^{(1)}_{(a)}(\xi^{*}_a,\bar{\xi}_b,t;\xi_0,\bar{\xi}^{*}_0,t_0)\langle\xi_0|0\rangle\langle 0|\bar{\xi}_0\rangle\\
=&\int_{t_0}^t dt_m \int_{t_0}^{t_m} dt_k  \int d\bar{\xi}_{N+1} \bar{\xi}_{N+1}d\xi_{N+1}^* \xi_{N+1}^* \prod_{n=0}^{N}d^2\xi_n d^2\bar{\xi}_n e^{-\xi_n^* \xi_n}e^{-\bar{\xi}_n^* \bar{\xi}_n}\\
&\times\prod_{n=1}^{N+1}e^{\xi_n^*\xi_{n-1}p_n}e^{\bar{\xi}_{n-1}^*\bar{\xi}_n p_n^*}{\color{red}\xi_m}[- {\rm g}_+^*(t_m-t_k)]{\color{red}\xi_k^*}\\
=&\int_{t_0}^t dt_m \int_{t_0}^{t_m} dt_k\; [-{\rm g}_+^*(t_m-t_k)]\int d\xi_{N+1}^* d\bar{\xi}_{N+1} \bar{\xi}_{N+1}\xi_{N+1}^* \prod_{n=0}^{N}d^2\bar{\xi}_n e^{-\bar{\xi}_n^* \bar{\xi}_n}\prod_{n=0}^{k-1}d^2\xi_n e^{-\xi_n^* \xi_n}\prod_{n=m+1}^{N}d^2\xi_n e^{-\xi_n^* \xi_n}\\
&\times {\color{red}d\xi_k^*}\left({\color{blue}\prod_{n=k+1}^m d\xi_{n-1}d\xi^*_n}\right) {\color{red}d\xi_m}\left({\color{blue}\prod_{n=k+1}^m \xi^*_n\xi_{n-1}p_n}\right) {\color{red}\xi_m\xi_k^*}\\
=&\int_{t_0}^t dt_m \int_{t_0}^{t_m} dt_k\; [-{\rm g}_+^*(t_m-t_k)]\prod_{n=k+1}^m p_n\underbrace{\int d\xi_k^* d\xi_m \xi_m\xi_k^*}_{=1}\\
\longrightarrow & \; \int_{t_0}^t dt_2 \int_{t_0}^{t_2} dt_1\;[-{\rm g}_+^*(t_2-t_1) ]e^{-\frac{\rm i}{\hbar}\epsilon (t_2-t_1)}\;.
\end{aligned}\end{equation}
Similarly, the contribution from path $(b)$ in Fig. \ref{path2nd-a-b} is
\begin{equation}\begin{aligned}\label{rho00discr2b}
J_{(b),00}^{(1)}(t;t_0)=&\Pi_{b}(0)\Pi_{a}^*(0)\int d^2\xi_0 d^2\bar{\xi}_0 \mathcal{J}^{(1)}_{(b)}(\xi^{*}_a,\bar{\xi}_b,t;\xi_0,\bar{\xi}^{*}_0,t_0)\langle\xi_0|0\rangle\langle 0|\bar{\xi}_0\rangle\\
=&\int_{t_0}^t dt_m \int_{t_0}^{t_m} dt_k\int  d\bar{\xi}_{N+1} \bar{\xi}_{N+1}d\xi_{N+1}^*\xi_{N+1}^* \prod_{n=0}^{N}d^2\xi_n d^2\bar{\xi}_n e^{-\xi_n^* \xi_n}e^{-\bar{\xi}_n^* \bar{\xi}_n}\\
&\times\prod_{n=1}^{N+1}e^{\xi_n^*\xi_{n-1}p_n}e^{\bar{\xi}_{n-1}^*\bar{\xi}_n p_n^*}{\color{red}\bar{\xi}_m^*}\;{\rm g}_+(t_m-t_k){\color{red}\bar{\xi}_k}\\
=&\int_{t_0}^t dt_m \int_{t_0}^{t_m} dt_k \; {\rm g}_+(t_m-t_k) \int d\xi_{N+1}^* d\bar{\xi}_{N+1} \bar{\xi}_{N+1}\xi_{N+1}^* \prod_{n=0}^{N}d^2\xi_n e^{-\xi_n^* \xi_n}\\
&\times\prod_{n=0}^{k-1}d^2\bar{\xi}_n e^{-\bar{\xi}_n^* \bar{\xi}_n}\prod_{n=m+1}^{N}d^2\bar{\xi}_n e^{-\bar{\xi}_n^* \bar{\xi}_n} {\color{red}d\bar{\xi}_m^*}\left( {\color{blue}\prod_{n=m}^{k+1} d\bar{\xi}_{n}d\bar{\xi}^*_{n-1}}\right) {\color{red}d\bar{\xi}_k}\left( {\color{blue}\prod_{n=k+1}^m \bar{\xi}^*_{n-1}\bar{\xi}_{n}p^*_n}\right) {\color{red}\bar{\xi}^*_m\bar{\xi}_k}\\
=&\int_{t_0}^t dt_m \int_{t_0}^{t_m} dt_k \; {\rm g}_+(t_m-t_k)\prod_{n=k+1}^m p^*_n\underbrace{\int d\bar{\xi}_m^* d\bar{\xi}_k \bar{\xi}^*_m\bar{\xi}_k}_{=-1}\\
\longrightarrow &\int_{t_0}^t dt_2 \int_{t_0}^{t_2} dt_1\; [-{\rm g}_+(t_2-t_1)] e^{\frac{\rm i}{\hbar}\epsilon (t_2-t_1)}\;.
\end{aligned}\end{equation}
Hence, the two propagators for the specific paths (a) and (b) are the complex conjugated of each other.\\
\indent Next we consider the two paths contributing to $J^{(1)}(1,t;0,t_0)$ which are depicted in Fig. \ref{path2nd-c-d}.\\
\begin{figure}[ht]
\begin{center}
\includegraphics[width=14cm,angle=0]{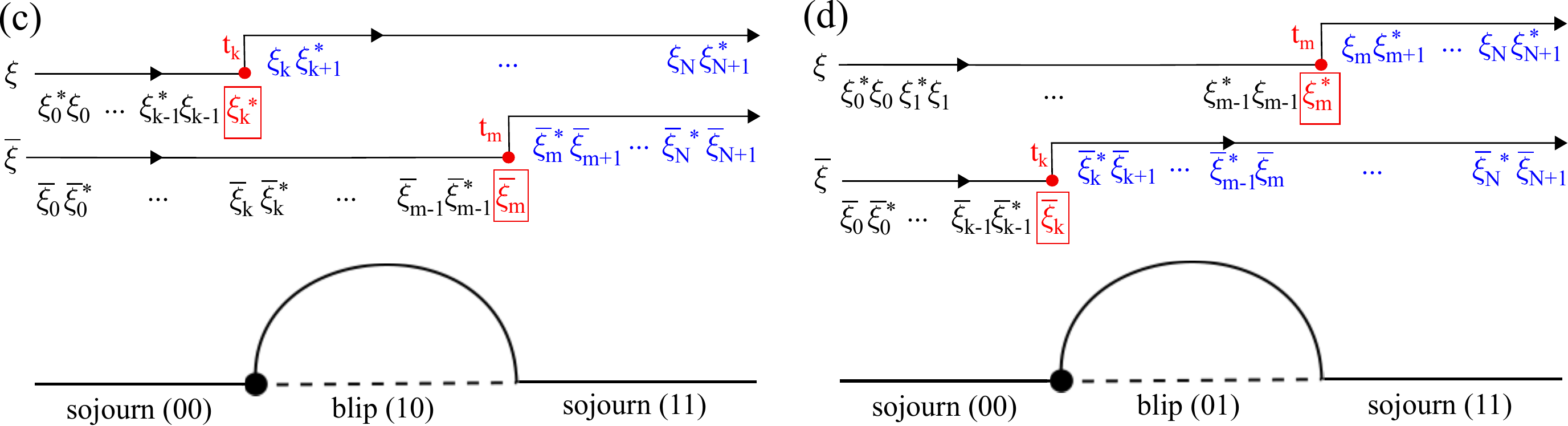}
\caption{\small{Paths with two transitions, one in the forward and the other in the backward branch. The boxed Grassmann variables in red are the ones appearing in the influence functional for the examples of paths considered.}}
\label{path2nd-c-d}
\end{center}
\end{figure}

\begin{figure}[ht]
\begin{center}
\includegraphics[width=12cm,angle=0]{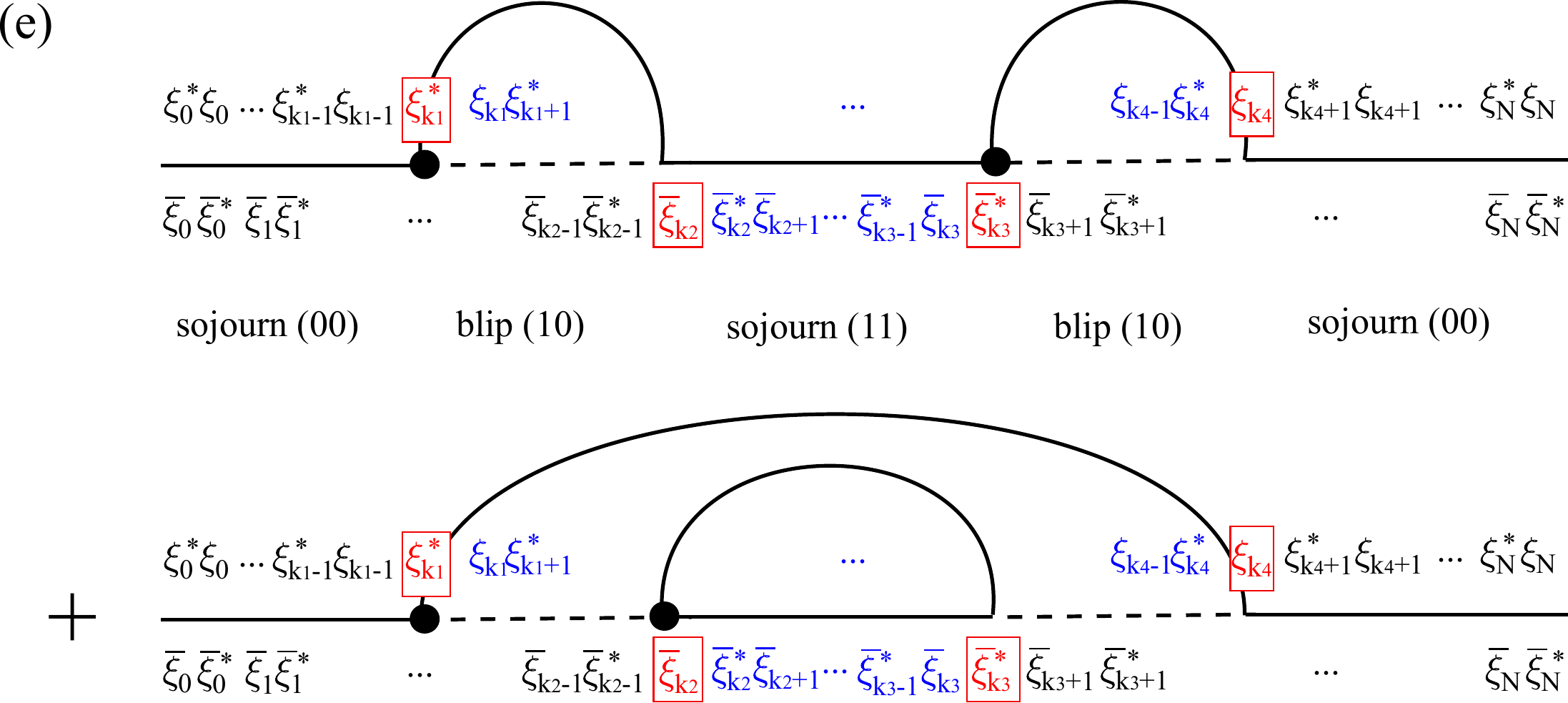}
\caption{\small{Path with four transitions, two in the forward and the others in the backward branch. Here, only the collective path is represented, with diagonal and off-diagonal states depicted with continuous and dashed lines, respectively. The boxed Grassmann variables in red are the ones appearing in the influence functional in this example.}}
\label{path4th}
\end{center}
\end{figure}

The contribution of path $(c)$ is evaluated similarly to those of paths $(a)$ and $(b)$. Explicitly
\begin{equation}\begin{aligned}
J_{(c),10}^{(1)}(t;t_0)=&\Pi_{b}(1)\Pi_{a}^*(1)\int d^2\xi_0 d^2\bar{\xi}_0 \mathcal{J}^{(1)}_{(c)}(\xi^{*}_a,\bar{\xi}_b,t;\xi_0,\bar{\xi}^{*}_0,t_0)\langle\xi_0|0\rangle\langle 0|\bar{\xi}_0\rangle\\
=&\int_{t_0}^t dt_m \int_{t_0}^{t_m} dt_k\int d\bar{\xi}_{N+1}d\xi_{N+1}^* \prod_{n=0}^{N}d^2\xi_n d^2\bar{\xi}_n e^{-\xi_n^* \xi_n}e^{-\bar{\xi}_n^* \bar{\xi}_n}\prod_{n=1}^{N+1}e^{\xi_n^*\xi_{n-1}p_n}e^{\bar{\xi}_{n-1}^*\bar{\xi}_n p_n^*}{\color{red}\bar{\xi}_m} [-{\rm g}_+^*(t_m-t_k)]{\color{red}\xi^*_k}\\
=&\int_{t_0}^t dt_m \int_{t_0}^{t_m} dt_k \;[-{\rm g}_+^*(t_m-t_k)] \int {\color{blue} d\bar{\xi}_{N+1}d\xi_{N+1}^*}\prod_{n=0}^{k-1}d^2\xi_n e^{-\xi_n^* \xi_n}\prod_{n=0}^{m-1}d^2\bar{\xi}_n e^{-\bar{\xi}_n^* \bar{\xi}_n}\\
&\times {\color{red}d\xi_k^*}\left( {\color{blue}\prod_{n=k+1}^N d\xi_{n-1}d\xi^*_n}\right) {\color{blue}d\xi_N}\left( {\color{blue}\prod_{n=k+1}^N \xi^*_n\xi_{n-1}p_n}\right) {\color{blue}\xi_{N+1}^*\xi_N p_{N+1}}\\
&\times {\color{blue}d\bar{\xi}_N^*}\left( {\color{blue}\prod_{n=N}^{m+1} d\bar{\xi}_{n}d\bar{\xi}^*_{n-1}}\right) {\color{red}d\bar{\xi}_m}\left( {\color{blue}\prod_{n=m+1}^N \bar{\xi}^*_{n-1}\bar{\xi}_{n}p^*_n}\right) {\color{blue}\bar{\xi}^*_N\bar{\xi}_{N+1}p_{N+1}^*}{\color{red}\bar{\xi}_m \xi^*_k}\nonumber
\end{aligned}\end{equation}

\begin{equation}\begin{aligned}\label{rho11discr2c}
=&-\int_{t_0}^t dt_m \int_{t_0}^{t_m} dt_k \;[-{\rm g}_+^*(t_m-t_k)] \int \prod_{n=0}^{k-1}d^2\xi_n e^{-\xi_n^* \xi_n}\prod_{n=0}^{m-1}d^2\bar{\xi}_n e^{-\bar{\xi}_n^* \bar{\xi}_n}\\
&\times \left( {\color{blue}\prod_{n=k+1}^N d\xi_{n-1}d\xi^*_n}\right) \left( {\color{blue}\prod_{n=k+1}^N \xi^*_n\xi_{n-1}p_n}\right) {\color{blue} p_{N+1}}\\
&\times \left( {\color{blue}\prod_{n=N}^{m+1} d\bar{\xi}_{n}d\bar{\xi}^*_{n-1}}\right) \left( {\color{blue}\prod_{n=m+1}^N \bar{\xi}^*_{n-1}\bar{\xi}_{n}p^*_n}\right) {\color{blue}p_{N+1}^*}\int{\color{red} d\xi_k^* d\bar{\xi}_m \bar{\xi}_m\xi_k^*}\\
=& - \int_{t_0}^t dt_m \int_{t_0}^{t_m} dt_k\;[-{\rm g}_+^*(t_m-t_k)]\prod_{n=k+1}^{N+1} p_n\prod_{n=m+1}^{N+1} p_n^*\int d\xi_k^* d\bar{\xi}_m \bar{\xi}_m\xi_k^*\\
=&- \int_{t_0}^t dt_m \int_{t_0}^{t_m} dt_k\;[-{\rm g}_+^*(t_m-t_k) ]\prod_{n=k+1}^{m} p_n\\
\longrightarrow & - \int_{t_0}^t dt_2 \int_{t_0}^{t_2} dt_1\;[-{\rm g}_+^*(t_2-t_1)]e^{-\frac{\rm i}{\hbar}\epsilon (t_2-t_1)}
\end{aligned}\end{equation}
Analogously, for the path $(d)$ in Fig. \ref{path2nd-c-d} we obtain
\begin{equation}\begin{aligned}
J_{(d)10}^{(1)}(t;,t_0)=&\Pi_{b}(1)\Pi_{a}^*(1)\int d^2\xi_0 d^2\bar{\xi}_0 \mathcal{J}^{(1)}_{(\bar\sigma)}(\xi^{*}_a,\bar{\xi}_b,t;\xi_0,\bar{\xi}^{*}_0,t_0)\langle\xi_0|0\rangle\langle 0|\bar{\xi}_0\rangle\\
=&\int_{t_0}^t dt_m \int_{t_0}^{t_m} dt_k\int d\bar{\xi}_{N+1} d\xi_{N+1}^* \prod_{n=0}^{N}d^2\xi_n d^2\bar{\xi}_n e^{-\xi_n^* \xi_n}e^{-\bar{\xi}_n^* \bar{\xi}_n}\prod_{n=1}^{N+1}e^{\xi_n^*\xi_{n-1}p_n}e^{\bar{\xi}_{n-1}^*\bar{\xi}_n p_n^*}{\color{red}\xi_m^*} {\rm g}_+(t_m-t_k){\color{red}\bar{\xi}_k}\\
=& \int_{t_0}^t dt_m \int_{t_0}^{t_m} dt_k\;{\rm g}_+(t_m-t_k)\int {\color{blue} d\bar{\xi}_{N+1}d\xi_{N+1}^*}\prod_{n=0}^{k-1}d^2\xi_n e^{-\xi_n^* \xi_n}\prod_{n=0}^{m-1}d^2\bar{\xi}_n e^{-\bar{\xi}_n^* \bar{\xi}_n}\\
&\times {\color{red}d\xi_m^*}\left( {\color{blue}\prod_{n=m+1}^N d\xi_{n-1}d\xi^*_n}\right) {\color{blue}d\xi_N}\left( {\color{blue}\prod_{n=m+1}^N \xi^*_n\xi_{n-1}p_n}\right) {\color{blue}\xi_{N+1}^*\xi_N p_{N+1}}\\
&\times {\color{blue}d\bar{\xi}_N^*}\left( {\color{blue}\prod_{n=N}^{k+1} d\bar{\xi}_{n}d\bar{\xi}^*_{n-1}}\right) {\color{red}d\bar{\xi}_k}\left( {\color{blue}\prod_{n=k+1}^N \bar{\xi}^*_{n-1}\bar{\xi}_{n}p^*_n}\right) {\color{blue}\bar{\xi}^*_N\bar{\xi}_{N+1}p_{N+1}^*}{\color{red}\xi_m^* \bar{\xi}_k}\nonumber
\end{aligned}\end{equation}

\begin{equation}\begin{aligned}\label{rho11discr2d}
=& - \int_{t_0}^t dt_m \int_{t_0}^{t_m} dt_k\;{\rm g}_+(t_m-t_k)\int {\color{blue} }\prod_{n=0}^{k-1}d^2\xi_n e^{-\xi_n^* \xi_n}\prod_{n=0}^{m-1}d^2\bar{\xi}_n e^{-\bar{\xi}_n^* \bar{\xi}_n}\\
&\times {\color{red}d\xi_m^*}\left( {\color{blue}\prod_{n=m+1}^N d\xi_{n-1}d\xi^*_n}\right) {\color{blue}d\xi_N d\xi_{N+1}^*}\left( {\color{blue}\prod_{n=m+1}^N \xi^*_n\xi_{n-1}p_n}\right) {\color{blue} \xi_{N+1}^*\xi_N p_{N+1}}\\
&\times {\color{blue}d\bar{\xi}_{N+1}d\bar{\xi}_N^*}\left( {\color{blue}\prod_{n=N}^{k+1} d\bar{\xi}_{n}d\bar{\xi}^*_{n-1}}\right) {\color{red}d\bar{\xi}_k}\left( {\color{blue}\prod_{n=k+1}^N \bar{\xi}^*_{n-1}\bar{\xi}_{n}p^*_n}\right) {\color{blue}\bar{\xi}^*_N\bar{\xi}_{N+1}p_{N+1}^*}{\color{red}\xi_m^* \bar{\xi}_k}\\
=&-  \int_{t_0}^t dt_m \int_{t_0}^{t_m} dt_k\;{\rm g}_+(t_m-t_k)\prod_{n=m+1}^{N+1} p_n\prod_{n=k+1}^{N+1} p_n^* \int {\color{red}{d\xi_m^* d\bar{\xi}_k \xi_m^*\bar{\xi}_k}}\\
=& -\int_{t_0}^t dt_m \int_{t_0}^{t_m} dt_k\;[-{\rm g}_+(t_m-t_k)]\prod_{n=k+1}^{m} p_n^*\\
\longrightarrow & - \int_{t_0}^t dt_2 \int_{t_0}^{t_2} dt_2\;[-{\rm g}_+(t_2-t_1)] e^{\frac{\rm i}{\hbar}\epsilon (t_2-t_1)}
\end{aligned}\end{equation}
We notice that Eqs.~\eqref{rho00discr2a} and~\eqref{rho11discr2c} only differ by a sign and the same holds for Eqs.~\eqref{rho00discr2b} and~\eqref{rho11discr2d}.\\
\indent Now we consider the specific four-transition path (second order in $\Gamma$) individuated by the ordered sequence of Grassmann variables $\{\xi_{k_1}^*\;,\bar{\xi}_{k_2}\;,\bar{\xi}_{k_3}^*\;,\xi_{k_4}\}$ and shown in Fig. \ref{path4th}.

The contribution of this path to $J^{(2)}_{00}(t;t_0)$ is evaluated along the same lines as above
\begin{equation}\begin{aligned}\label{rho00discr4}
J^{(2)}_{(e),00}(t;t_0)=&\Pi_{b}(0)\Pi_{a}^*(0)\int d^2\xi_0 d^2\bar{\xi}_0 \mathcal{J}^{(2)}_{(e)}(\xi^{*}_a,\bar{\xi}_b,t;\xi_0,\bar{\xi}^{*}_0,t_0)\langle\xi_0|0\rangle\langle 0|\bar{\xi}_0\rangle\\
=&\int_{t_0}^t dt_{k_4}\dots\int_{t_0}^{t_{k_2}} dt_{k_1}\;\int  d\bar{\xi}_{N+1} \bar{\xi}_{N+1}d\xi_{N+1}^*\xi_{N+1}^* 
\prod_{n=0}^{k_1-1}d^2\xi_n e^{-\xi_n^* \xi_n}\prod_{n=k_4+1}^{N}d^2\xi_n e^{-\xi_n^* \xi_n}\\
&\times\prod_{n=0}^{k_2-1}d^2\bar{\xi}_n e^{-\bar{\xi}_n^* \bar{\xi}_n}\prod_{n=k_3+1}^{N}d^2\bar{\xi}_n e^{-\bar{\xi}_n^* \bar{\xi}_n}\\
&\times {\color{red}d\xi_{k_1}^*}\left( {\color{blue}\prod_{n=k_1+1}^{k_4} d\xi_{n-1}d\xi^*_n}\right) {\color{red}d\xi_{k_4}}\left( {\color{blue}\prod_{n=k_1+1}^{k_4} \xi^*_n\xi_{n-1}p_n}\right)\\
&\times {\color{red}d\bar{\xi}_{k_3}^*}\left( {\color{blue}\prod_{n=k_3}^{k_2+1} d\bar{\xi}_{n}d\bar{\xi}^*_{n-1}}\right) {\color{red}d\bar{\xi}_{k_2}}\left( {\color{blue}\prod_{n=k_2+1}^{k_3} \bar{\xi}^*_{n-1}\bar{\xi}_{n}p^*_n}\right)\\
&\times\Big\{{\color{red}\bar{\xi}_{k_2}}[-{\rm g}_+^*(t_{k_2}-t_{k_1})] {\color{red}\xi_{k_1}^*}{\color{red}\xi_{k_4}}{\rm g}_-^*(t_{k_4}-t_{k_3}) {\color{red}\bar{\xi}_{k_3}^*}\\
&+{\color{red}\xi_{k_4}}[-{\rm g}_+^*(t_{k_4}-t_{k_1})] {\color{red}\xi_{k_1}^*}{\color{red}\bar{\xi}_{k_3}^*}{\rm g}_+(t_{k_3}-t_{k_2}) {\color{red}\bar{\xi}_{k_2}} \Big\}\\
=&\int_{t_0}^t dt_{k_4}\dots\int_{t_0}^{t_{k_2}} dt_{k_1}\int d\xi_{k_1}^*d\xi_4 d\bar{\xi}_{k_3}^*d\bar{\xi}_{k_2}\Big\{\bar{\xi}_{k_2}\xi_{k_1}^*\xi_{k_4}\bar{\xi}_{k_3}^* [-{\rm g}_+^*(t_{k_2}-t_{k_1})] {\rm g}_-^*(t_{k_4}-t_{k_3})\\
&+\xi_{k_4}\xi_{k_1}^*\bar{\xi}_{k_3}^*\bar{\xi}_{k_2} [-{\rm g}_+^*(t_{k_4}-t_{k_1})] {\rm g}_+(t_{k_3}-t_{k_2})\Big\}
\prod_{n=k_1+1}^{k_2} p_n\prod_{n=k_3+1}^{k_4} p_n\\
=&\int_{t_0}^t dt_{4}\dots\int_{t_0}^{t_{2}} dt_{1}\Big\{[-{\rm g}_+^*(t_{2}-t_{1}) ][-{\rm g}_-^*(t_{4}-t_{3})]+[-{\rm g}_+^*(t_{4}-t_{1}) ][-{\rm g}_+(t_{3}-t_{2})]\Big\}\\
&\times e^{-\frac{\rm i}{\hbar}\epsilon (t_{2}-t_{1})}e^{-\frac{\rm i}{\hbar}\epsilon (t_{4}-t_{3})}\;.
\end{aligned}\end{equation}
Note that a product of the type ${\rm g}(t_{4}-t_{2}){\rm g}(t_{3}-t_{1})$, implying a crossing of the fermion lines, is not present for this specific path because we have fixed the Grassmann variables at the transition times and the form of the influence functional prevents the fermion lines from joining two starred or two non-starred charges, see Eq.~\eqref{phase_IF}.\\
\indent From the examples above we can draw some conclusions
\begin{itemize}
\item Integrations over the \emph{sojourn} time intervals yield an overall phase factor $1$.
\item Integrations over the \emph{blip} time intervals yield the phase factors $\exp(-\zeta{\rm i}\epsilon \tau/\hbar)$, where $\tau$ is the blip length and $\zeta=\pm 1$, depending on the nature of the blip.
\end{itemize}
Once the trivial integrations over the sojourns/blip time intervals are performed, we are left with a final integration over the Grassmann variables associated to the transitions. As a result of this procedure, neither the Grassmann variables in the integration measures nor the ones in the integrands are time-ordered. Specifically, 
 \begin{itemize}
\item In the integration measure, the backward variables appear to the right of the forward and, within this two classes, starred variables are to the left of the non-starred ones. This reflects the original order of the integrations.
\item In the integrand, the Grassmann variables appear as a sequence of pairs whose order depends on how they are coupled by the functions ${\rm g}(t_j-t_i)$.  
\end{itemize}

\subsection*{SIAM}
 
\begin{figure}[ht]
\begin{center}
\includegraphics[width=9cm,angle=0]{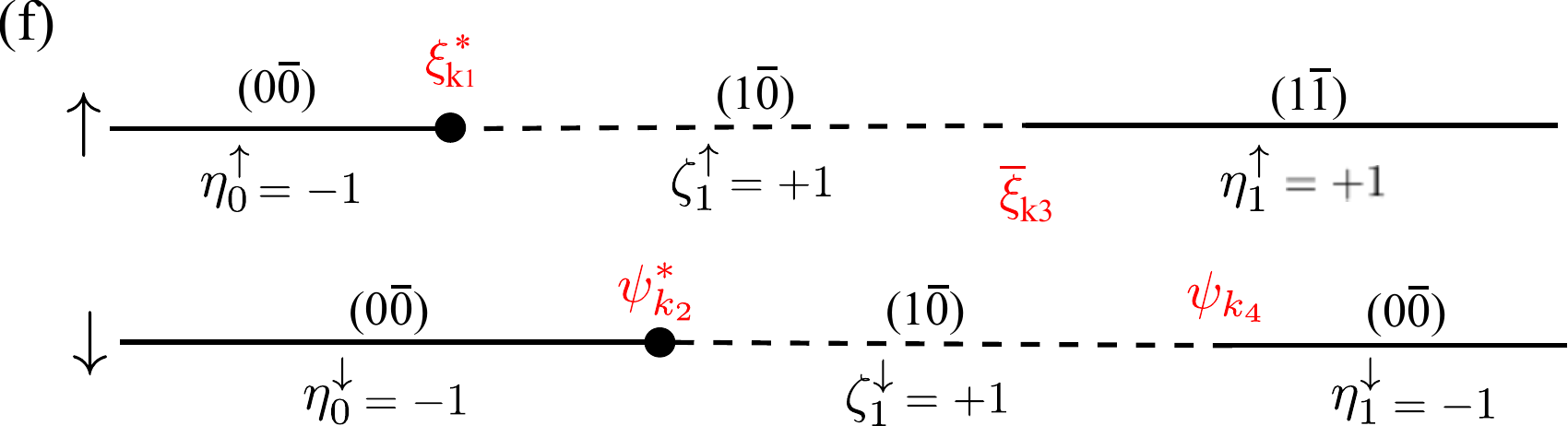}
\caption{\small{Path with four transitions distributed in the two sub-paths of the spin variables $\uparrow$ and $\downarrow$. Note that the path for $\sigma=~\uparrow$ is of the type (c) in Fig. \ref{path2nd-c-d}, while $\sigma=~\downarrow$ undergoes a sequence of the type (a) in Fig. \ref{path2nd-a-b}.}}
\label{path_spinful-A}
\end{center}
\end{figure}

\begin{figure}[ht]
\begin{center}
\includegraphics[width=9cm,angle=0]{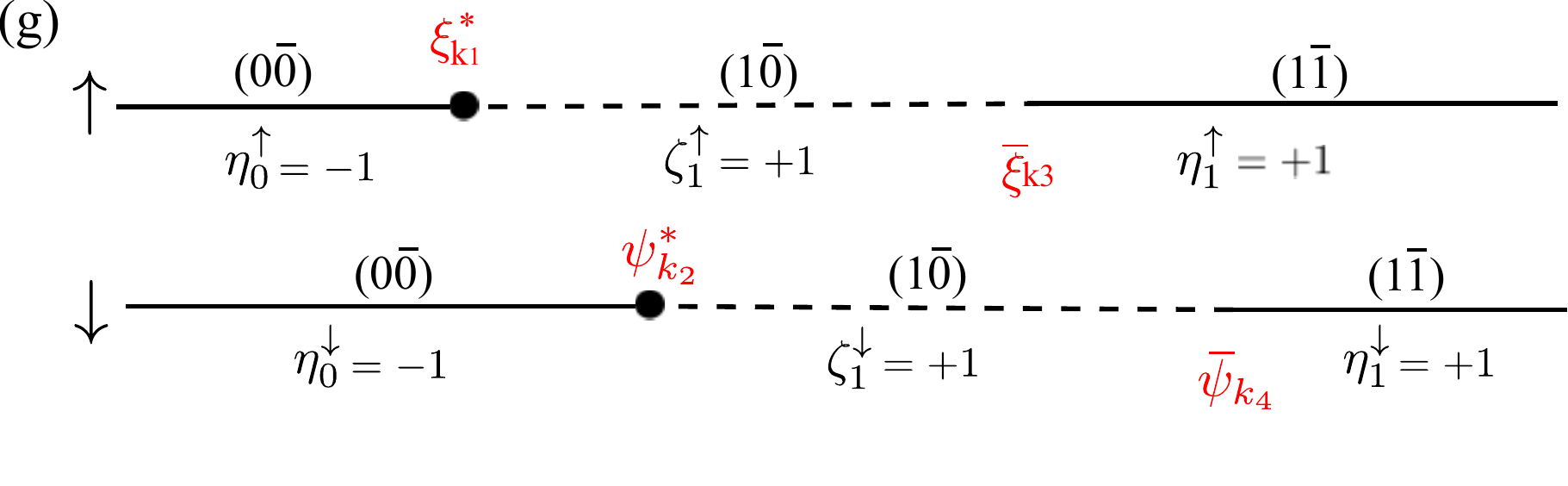}
\caption{\small{Path with four transitions distributed in the two sub-paths of the spin variables $\uparrow$ and $\downarrow$. 
Note that the sub-paths are both of the type (c) in Fig. \ref{path2nd-c-d}.}}
\label{path_spinful-B}
\end{center}
\end{figure}

We now generalize the procedure used for the RLM by analyzing specific instances of paths involving both spin states in the SIAM. To avoid adding further indexes we denote the Grassmann variables associated to $\sigma=\;\uparrow$ with the usual $\xi$ and the ones associated to $\sigma=\;\downarrow$ with the letter $\psi$. As for the RLM, instead of using the collective sojourn index $\boldsymbol{\eta}=(\eta^\uparrow,\eta^\downarrow)$, we indicate the initial and final occupation of the spin states of the dot in the argument of the propagator.
The integration measure has the property $d^2\boldsymbol{\xi}=d\xi^*d\psi^* d\xi d\psi=-d\xi^* d\xi d\psi^*  d\psi$, so that $d^2\boldsymbol{\xi} d^2\bar{\boldsymbol{\xi}}=d\xi^* d\xi d\psi^*  d\psi d\bar{\xi}^* d\bar{\xi} d\bar{\psi}^*  d\bar{\psi}=d^2\xi d^2 \bar{\xi} d^2\psi d^2 \bar{\psi}$. We use this property to factorize the integrations over the Grassmann variables for the two degrees of freedom.\\
\indent For the SIAM, the coherent states are expressed, in terms of occupation number states, as
\begin{equation}\begin{aligned}
|\boldsymbol{\xi}\rangle=&(1-\xi\hat{a}_{\uparrow}^{\dag})(1-\psi\hat{a}_{\downarrow}^{\dag})|0_\uparrow0_\downarrow\rangle\\
&=|0_\uparrow0_\downarrow\rangle+|1_\uparrow 0_\downarrow\rangle\xi+|0_\uparrow 1_\downarrow\rangle\psi+|1_\uparrow 1_\downarrow\rangle\psi\xi\\
\langle\boldsymbol{\xi}|=&\langle 0_\downarrow 0_\uparrow| (1-\hat{a}_{\downarrow}\psi^*) (1-\hat{a}_{\uparrow}\xi^*)\\
&=\langle 0_\downarrow 0_\uparrow|+\psi^*\langle 1_\downarrow 0_\uparrow|+\xi^*\langle 0_\downarrow 1_\uparrow| +\xi^*\psi^*\langle 1_\downarrow 1_\uparrow|\;.
\end{aligned}\end{equation}
As a result, the populations are found by calculating the matrix element of the impurity RDM
\begin{equation}\begin{aligned}
\langle\boldsymbol{\xi}| \rho |\bar{\boldsymbol{\xi}}\rangle={\rm P}_{00}+\dots+\psi^*\bar\psi\;{\rm P}_{01}+\dots+\xi^*\bar\xi\;{\rm P}_{10}+\dots+\xi^*\psi^*\bar\psi\bar\xi\;{\rm P}_{11}
\end{aligned}\end{equation}
and applying the projectors defined in Eq.~\eqref{def_projector_appendix} via ${\rm P}_{\boldsymbol{n}}\equiv\rho_{\boldsymbol{n}\boldsymbol{n}}=\Pi_b(\boldsymbol{n})\Pi_a^*(\boldsymbol{n})\langle\boldsymbol{\xi}| \rho |\bar{\boldsymbol{\xi}}\rangle$.\\
\indent Let us introduce the abbreviations 
\begin{equation}\begin{aligned}\label{abbreviations}
{\rm overlap}\quad 
O_n^{\uparrow}=&e^{-\xi_n^* \xi_n}\qquad\bar{O}_n^{\uparrow}=e^{-\bar{\xi}_n^* \bar{\xi}_n}\\
O_n^{\downarrow}=&e^{-\psi_n^* \psi_n}\qquad\bar{O}_n^{\downarrow}=e^{-\bar{\psi}_n^* \bar{\psi}_n}\\
 H_0\quad 
 \mathcal{P}_n^{\uparrow}=&e^{\xi_n^*\xi_{n-1}p_n^{\uparrow}}
\qquad \bar{\mathcal{P}}_n^{\uparrow}=e^{\bar{\xi}_{n-1}^*\bar{\xi}_{n}p_n^{*\uparrow}}\\
\mathcal{P}_n^{\downarrow}=&e^{\psi_n^*\psi_{n-1}p_n^{\downarrow}}
\qquad \bar{\mathcal{P}}_n^{\downarrow}=e^{\bar{\psi}_{n-1}^*\bar{\psi}_{n}p_n^{*\downarrow}}\\
{\rm interaction}\quad \mathcal{U}_n=&e^{\xi_n^*\xi_{n-1}\psi_n^*\psi_{n-1}u_n}\qquad\bar{\mathcal{U}}_n=e^{\bar{\xi}_{n-1}^*\bar{\xi}_{n}\bar{\psi}_{n-1}^*\bar{\psi}_{n}u_n^*}\;,
\end{aligned}\end{equation}
where $p_n^\sigma:=1-({\rm i}/\hbar)\epsilon_\sigma \delta t_n$ and $u_n:=-({\rm i}/\hbar)U \delta t_n$, with $U$ the interaction energy.
Consider the path in Fig.~\ref{path_spinful-A}. From Eqs.~\eqref{Propagator_general_appendix} and~\eqref{def_projector_appendix}, the contribution to $J^{(2)}_{10,00}(t;t_0)$ given by the path reads
\begin{equation}\begin{aligned}\label{}
J^{(2)}_{(f),10,00}(t;t_0)=&\Pi^\uparrow_{b}(1)\Pi^\downarrow_{b}(0)\Pi^{\downarrow*}_{a}(0)\Pi^{\uparrow*}_{a}(1)\int d^2\boldsymbol\xi_0 d^2\bar{\boldsymbol\xi}_0 \mathcal{J}^{(2)}_{(f)}(\boldsymbol\xi^{*}_a,\bar{\boldsymbol\xi}_b,t;\boldsymbol\xi_0,\bar{\boldsymbol\xi}^{*}_0,t_0)\langle\boldsymbol\xi_0|\mathbf{0}\rangle\langle \mathbf{0}|\bar{\boldsymbol\xi}_0\rangle\\
=&\int_{t_0}^t dt_{k_4}\dots\int_{t_0}^{t_{k_2}} dt_{k_1}\;\int d\bar{\xi}_{N+1} d\bar{\psi}_{N+1}\;  \bar{\psi}_{N+1}   d\psi^*_{N+1} \psi^*_{N+1}d\xi^*_{N+1}\prod_{n=0}^{N}d^2\xi_n  d^2\psi_n d^2\bar{\xi}_n d^2\bar{\psi}_n O_n^\uparrow O_n^\downarrow \bar{O}_n^{\uparrow} \bar{O}_n^{\downarrow}\\
&\times\prod_{n=1}^{N+1}\mathcal{P}_n^\uparrow \mathcal{P}_n^\downarrow\bar{\mathcal{P}}_n^{\uparrow}\bar{\mathcal{P}}_n^{\downarrow}\mathcal{U}_n\bar{\mathcal{U}}_n {\color{red}\bar{\xi}_{k_3}} [-{\rm g}_+^{\uparrow *}(t_{k_3}-t_{k_1})]{\color{red}\xi^*_{k_1}}{\color{red}\psi_{k_4}}  [-{\rm g}_+^{\downarrow *}(t_{k_4}-t_{k_2})]{\color{red}\psi_{k_2}^*}\\
=&\int_{t_0}^t dt_{k_4}\dots\int_{t_0}^{t_{k_2}} dt_{k_1}\;
\int d\bar{\xi}_{N+1}  d\xi^*_{N+1} \left(\prod_{n=0}^{k_1-1}d\xi_n^* d\xi_n O_n^\uparrow\right) d\xi_{k_1}^*\left(\prod_{n=k_1+1}^{N}d\xi_{n-1} d\xi_{n}^* \mathcal{P}_{n}^\uparrow\right) d\xi_N\mathcal{P}_{N+1}^\uparrow\\
&\times \bar{\mathcal{P}}_{N+1}^\uparrow d\bar{\xi}_N^*\left(\prod_{n=N}^{k_3+1}d\bar{\xi}_n d\bar{\xi}_{n-1}^* \bar{\mathcal{P}}_{n}^\uparrow\right)d\bar{\xi}_{k_3}\left(\prod_{n=k_3-1}^{0}d\bar{\xi}_n^* d\bar{\xi}_n \bar{O}_n^\uparrow\right){\color{red}\bar{\xi}_{k_3}} [-{\rm g}_+^{*\uparrow}(t_{k_3}-t_{k_1})]{\color{red}\xi^*_{k_1}}\\
&\times d\bar{\psi}_{N+1}\;  \bar{\psi}_{N+1}   d\psi^*_{N+1} \psi^*_{N+1} \left(\prod_{n=0}^{N}d\bar{\psi}_n^* d\bar{\psi}_n \bar{O_n^\downarrow}\right)\left(\prod_{n=0}^{k_2-1}d\psi_n^* d\psi_n O_n^\downarrow\right) d\psi_{k_2}^*\\
&\times\left(\prod_{n=k_2+1}^{k_4}d\psi_{n-1} d\psi_{n}^* \mathcal{P}_n^\downarrow\mathcal{U}_n\right)d\psi_{k_4}\left(\prod_{n=k_4+1}^{N}d\psi_n^* d\psi_n O_n^\downarrow\right){\color{red}\psi_{k_4}} [-{\rm g}_+^{*\downarrow}(t_{k_4}-t_{k_2})]{\color{red}\psi_{k_2}^*}\;.
\end{aligned}\end{equation}
Now, since (see Eq.~\eqref{abbreviations})
\begin{equation}\begin{aligned}
\prod_{n}d\xi_n^* d\xi_n O_n=&\prod_{n}d\xi_n^* d\xi_ne^{-\bar{\xi}_n^* \bar{\xi}_n}=1\\
{\rm and}\qquad \prod_{n}d\xi_{n-1} d\xi_{n}^* \mathcal{P}_{n}=&\prod_{n}d\xi_{n-1} d\xi_{n}^*e^{\xi_n^*\xi_{n-1}p_n}=\prod_{n}p_n\;,
\end{aligned}\end{equation}
we get
\begin{equation}\begin{aligned}\label{}
J^{(2)}_{(f),10,00}(t;t_0)=&\int_{t_0}^t dt_{k_4}\dots\int_{t_0}^{t_{k_2}} dt_{k_1}\;  \prod_{n=k_1+1}^{N}p_n^\uparrow\prod_{n=k_2+1}^{k_4}(p_n^\downarrow+u_n)\prod_{n=k_3+1}^{N}p_n^{\uparrow *}[-{\rm g}_+^{\uparrow *}(t_{k_3}-t_{k_1})][- {\rm g}_+^{\downarrow *}(t_{k_4}-t_{k_2})] \\
&\times \int d\bar{\xi}_{N+1}  d\xi^*_{N+1}  d\xi_{k_1}^* d\xi_N\mathcal{P}_{N+1}^\uparrow \bar{\mathcal{P}}_{N+1}^\uparrow d\bar{\xi}_N^*d\bar{\xi}_{k_3}\bar{\xi}_{k_3}\xi^*_{k_1}\\
&\times d\bar{\psi}_{N+1}\;  \bar{\psi}_{N+1}   d\psi^*_{N+1} \psi^*_{N+1}  d\psi_{k_2}^*d\psi_{k_4}\psi_{k_4} \psi_{k_2}^*\\
=&-\int_{t_0}^t dt_{k_4}\dots\int_{t_0}^{t_{k_2}} dt_{k_1}\;  \prod_{n=k_1+1}^{N+1}p_n^\uparrow\prod_{n=k_2+1}^{k_4}(p_n^\downarrow+u_n)\prod_{n=k_3+1}^{N+1}p_n^{\uparrow *}[-{\rm g}_+^{\uparrow *}(t_{k_3}-t_{k_1})][- {\rm g}_+^{\downarrow *}(t_{k_4}-t_{k_2})] \\
&\times \int  d\xi_{k_1}^*  d\bar{\xi}_{k_3}\bar{\xi}_{k_3}\xi^*_{k_1} d\psi_{k_2}^*d\psi_{k_4}\psi_{k_4} \psi_{k_2}^*\\
=&-\int_{t_0}^t dt_{4}\dots\int_{t_0}^{t_{2}} dt_{1}\;e^{-\frac{\rm i}{\hbar}\epsilon_\uparrow(t_{3}-t_{1})}e^{-\frac{\rm i}{\hbar}(\epsilon_\downarrow+U)(t_{4}-t_{2})}[-{\rm g}_+^{\uparrow *}(t_{3}-t_{1})][-{\rm g}_+^{\downarrow *}(t_{4}-t_{2})]\;,
\end{aligned}\end{equation}
where we have used Eq.~\eqref{results_1} and the properties that couples of Grassmann variables commute with other Grassmann variables and that variables belonging to different spins anticommute.\\
 \indent Next we consider the process depicted in Fig.~\ref{path_spinful-B}, which contributes to $\rho_{11,11}$. As above, we have
\begin{equation}\begin{aligned}\label{}
J^{(2)}_{(g)11,00}(t;t_0)=&\Pi^\uparrow_{b}(1)\Pi^\downarrow_{b}(1)\Pi^{\downarrow*}_{a}(1)\Pi^{\uparrow*}_{a}(1)\int d^2\boldsymbol\xi_0 d^2\bar{\boldsymbol\xi}_0 \mathcal{J}^{(2)}_{(g)}(\boldsymbol\xi^{*}_a,\bar{\boldsymbol\xi}_b,t;\boldsymbol\xi_0,\bar{\boldsymbol\xi}^{*}_0,t_0)\langle\boldsymbol\xi_0|\mathbf{0}\rangle\langle \mathbf{0}|\bar{\boldsymbol\xi}_0\rangle\\
=&\int_{t_0}^t dt_{k_4}\dots\int_{t_0}^{t_{k_2}} dt_{k_1}\;\int d\bar{\xi}_{N+1} d\bar{\psi}_{N+1}\;  d\psi^*_{N+1}d\xi^*_{N+1}
\prod_{n=0}^{N}d^2\xi_n  d^2\psi_n d^2\bar{\xi}_n d^2\bar{\psi}_n O_n^\uparrow O_n^\downarrow \bar{O}_n^{\uparrow} \bar{O}_n^{\downarrow}\\
&\times\prod_{n=1}^{N+1}\mathcal{P}_n^\uparrow \mathcal{P}_n^\downarrow\bar{\mathcal{P}}_n^{\uparrow}\bar{\mathcal{P}}_n^{\downarrow}\mathcal{U}_n\bar{\mathcal{U}}_n {\color{red}\bar{\xi}_{k_3}} [-{\rm g}_+^{\uparrow *}(t_{k_3}-t_{k_1})]{\color{red}\xi^*_{k_1}}{\color{red}\bar{\psi}_{k_4}}[-{\rm g}_+^{\downarrow *}(t_{k_4}-t_{k_2})]{\color{red}\psi_{k_2}^*}\\
=&\int_{t_0}^t dt_{k_4}\dots\int_{t_0}^{t_{k_2}} dt_{k_1}\; 
\int  d\bar{\xi}_{N+1}  d\xi^*_{N+1} \left(\prod_{n=0}^{k_1-1}d\xi_n^* d\xi_n O_n^\uparrow\right) d\xi_{k_1}^*\left(\prod_{n=k_1+1}^{N}d\xi_{n-1} d\xi_{n}^* \mathcal{P}_{n}^\uparrow\right) d\xi_N\mathcal{P}_{N+1}^\uparrow\\
&\times \bar{\mathcal{P}}_{N+1}^\uparrow d\bar{\xi}_N^*\left(\prod_{n=N}^{k_3+1}d\bar{\xi}_n d\bar{\xi}_{n-1}^* \bar{\mathcal{P}}_{n}^\uparrow\right)d\bar{\xi}_{k_3}\left(\prod_{n=k_3-1}^{0}d\bar{\xi}_n^* d\bar{\xi}_n \bar{O}_n^\uparrow\right){\color{red}\bar{\xi}_{k_3}} [-{\rm g}_+^{\uparrow *}(t_{k_3}-t_{k_1})]{\color{red}\xi^*_{k_1}}\\
&\times d\bar{\psi}_{N+1}d\psi^*_{N+1} \left(\prod_{n=0}^{k_2-1}d\psi_n^* d\psi_n O_n^\downarrow\right) d\psi_{k_2}^*\left(\prod_{n=k_2+1}^{N}d\psi_{n-1} d\psi_{n}^* \mathcal{P}_{n}^\downarrow\mathcal{U}_n\right) d\psi_N\mathcal{P}_{N+1}^\downarrow\mathcal{U}_{N+1}\\
&\times \bar{\mathcal{P}}_{N+1}^\downarrow\bar{\mathcal{U}}_{N+1} d\bar{\psi}_N^*\left(\prod_{n=N}^{k_4+1}d\bar{\psi}_n d\bar{\psi}_{n-1}^* \bar{\mathcal{P}}_{n}^\downarrow\bar{\mathcal{U}}_n\right)d\bar{\psi}_{k_4}\left(\prod_{n=k_4-1}^{0}d\bar{\psi}_n^* d\bar{\psi}_n \bar{O}_n^\downarrow\right){\color{red}\bar{\psi}_{k_4}} [-{\rm g}_+^{\downarrow *}(t_{k_4}-t_{k_2})]{\color{red}\psi_{k_2}^*}\;.
\end{aligned}\end{equation}
Again we use the definitions of $\mathcal{P}_{n}$ and $\mathcal{O}_{n}$ to integrate out the terms in parenthesis and get
\begin{equation}\begin{aligned}\label{}
J^{(2)}_{(g)11,00}(t;t_0)=& \int_{t_0}^t dt_{k_4}\dots\int_{t_0}^{t_{k_2}} dt_{k_1}\;  \prod_{n=k_1+1}^{N}p_n^\uparrow\prod_{n=k_2+1}^{N}(p_n^\downarrow+u_n)\prod_{n=k_3+1}^{N}p_n^{\uparrow *}\prod_{n=k_4+1}^{N}(p_n^{\downarrow *}+u_n^*)\\
&\times[-{\rm g}_+^{\uparrow *}(t_{k_3}-t_{k_1})][- {\rm g}_+^{\downarrow *}(t_{k_4}-t_{k_2})]\\
&\times\int  d\bar{\xi}_{N+1}  d\xi^*_{N+1}  d\xi_{k_1}^* d\xi_N\mathcal{P}_{N+1}^\uparrow \bar{\mathcal{P}}_{N+1}^\uparrow d\bar{\xi}_N^* d\bar{\xi}_{k_3} \bar{\xi}_{k_3}\xi^*_{k_1}\\
&\times d\bar{\psi}_{N+1}d\psi^*_{N+1} d\psi_{k_2}^* d\psi_N\mathcal{P}_{N+1}^\downarrow\mathcal{U}_{N+1} \bar{\mathcal{P}}_{N+1}^\downarrow\bar{\mathcal{U}}_{N+1} d\bar{\psi}_N^* d\bar{\psi}_{k_4}\bar{\psi}_{k_4}\psi_{k_2}^*\\
=& \int_{t_0}^t dt_{k_4}\dots\int_{t_0}^{t_{k_2}} dt_{k_1}\;  \prod_{n=k_1+1}^{N+1}p_n^\uparrow\prod_{n=k_2+1}^{N+1}(p_n^\downarrow+u_n)\prod_{n=k_3+1}^{N+1}p_n^{\uparrow *}\prod_{n=k_4+1}^{N+1}(p_n^{\downarrow *}+u_n^*)\\
&\times[-{\rm g}_+^{\uparrow *}(t_{k_3}-t_{k_1})][- {\rm g}_+^{\downarrow *}(t_{k_4}-t_{k_2})] \int d\xi_{k_1}^* d\bar{\xi}_{k_3} \bar{\xi}_{k_3}\xi^*_{k_1} d\psi_{k_2}^* d\bar{\psi}_{k_4} \bar{\psi}_{k_4}\psi_{k_2}^*\\
=&\int_{t_0}^t dt_{4}\dots\int_{t_0}^{t_{2}} dt_{1}\; e^{-\frac{\rm i}{\hbar}\epsilon_\uparrow(t_{3}-t_{1})}e^{-\frac{\rm i}{\hbar}(\epsilon_\downarrow+U)(t_{4}-t_{2})}[-{\rm g}_+^{\uparrow *}(t_{3}-t_{1})][- {\rm g}_+^{\downarrow *}(t_{4}-t_{2})]\;.
\end{aligned}\end{equation}
It is already apparent that the interaction is present in the time intervals where both spin states are occupied either in the forward or in the backward branch. Simultaneous double occupation in the two branches, as it is the case for the sojourn states (11,11), leads to a cancellation due to the sum of $u_n$ and $u_n^*$ at the exponent.
\begin{figure}[ht]
\begin{center}
\includegraphics[width=9cm,angle=0]{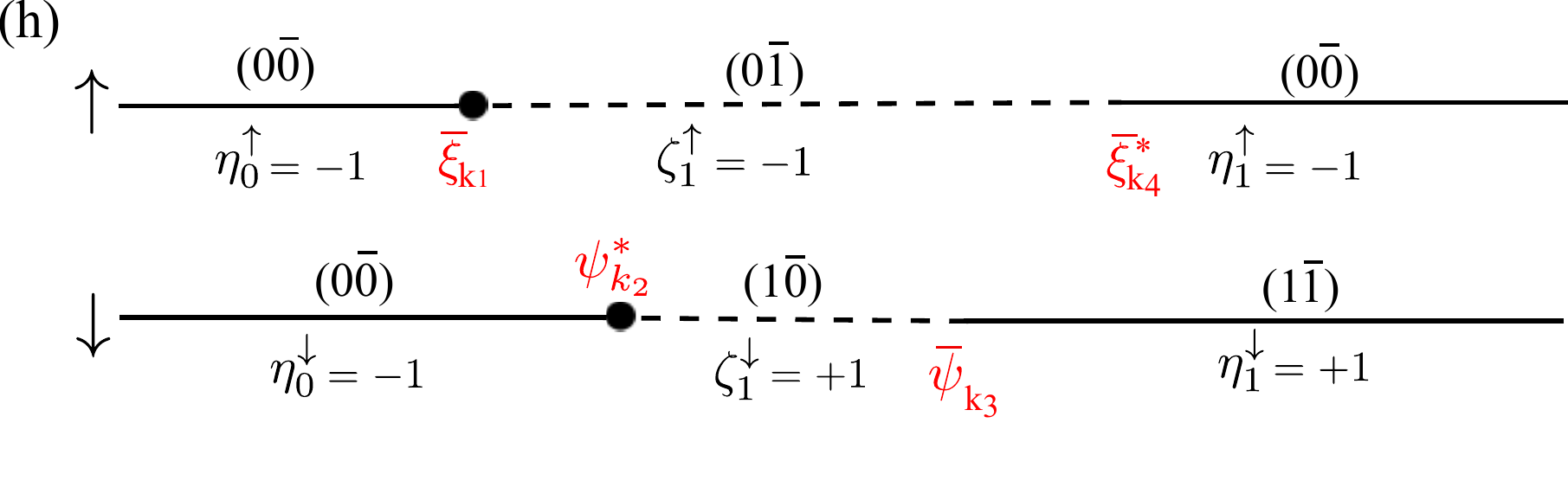}
\caption{\small{Path with four transitions distributed in the two sub-paths of the spin variables $\uparrow$ and $\downarrow$.
Note that the sub-path $\sigma=~\uparrow$ is of the type (b) in Fig. \ref{path2nd-a-b}, while $\sigma=~\downarrow$ undergoes a sequence of the type (c), see Fig. \ref{path2nd-c-d}.}}
\label{path_spinful-C}
\end{center}
\end{figure}
To better clarify this point, consider the example in Fig. (\ref{path_spinful-C}). This path contributes to $J^{(2)}_{01,00}(t;t_0)$, the contribution being
\begin{equation}\begin{aligned}\label{}
J^{(2)}_{(h)01,00}(t;t_0)=&\Pi^\uparrow_{b}(0)\Pi^\downarrow_{b}(1)\Pi^{\downarrow*}_{a}(1)\Pi^{\uparrow*}_{a}(0)\int d^2\boldsymbol\xi_0 d^2\bar{\boldsymbol\xi}_0 \mathcal{J}^{(2)}_{(h)}(\boldsymbol\xi^{*}_a,\bar{\boldsymbol\xi}_b,t;\boldsymbol\xi_0,\bar{\boldsymbol\xi}^{*}_0,t_0)\langle\boldsymbol\xi_0|\mathbf{0}\rangle\langle \mathbf{0}|\bar{\boldsymbol\xi}_0\rangle\\
=&\int_{t_0}^t dt_{k_4}\dots\int_{t_0}^{t_{k_2}} dt_{k_1}\;\int d\bar{\xi}_{N+1}\bar{\xi}_{N+1} d\bar{\psi}_{N+1}\;   d\psi^*_{N+1}  d\xi^*_{N+1} \xi^*_{N+1}\;\prod_{n=0}^{N}d^2\xi_n  d^2\psi_n d^2\bar{\xi}_n d^2\bar{\psi}_n O_n^\uparrow O_n^\downarrow \bar{O}_n^{\uparrow} \bar{O}_n^{\downarrow}\\
&\times\prod_{n=1}^{N+1}\mathcal{P}_n^\uparrow \mathcal{P}_n^\downarrow\bar{\mathcal{P}}_n^{\uparrow}\bar{\mathcal{P}}_n^{\downarrow}\mathcal{U}_n\bar{\mathcal{U}}_n {\color{red}\bar{\xi}_{k_4}^*} {\rm g}_+^{\uparrow}(t_{k_4}-t_{k_1}){\color{red}\bar{\xi}_{k_1}}{\color{red}\bar{\psi}_{k_3}} [-1{\rm g}_+^{\downarrow *}(t_{k_3}-t_{k_2})]{\color{red}\psi_{k_2}^*}\\
=& \int_{t_0}^t dt_{k_4}\dots\int_{t_0}^{t_{k_2}} dt_{k_1}\; 
\int d\xi^*_{N+1} d\bar{\xi}_{N+1} \bar{\xi}_{N+1} \xi^*_{N+1}\left(\prod_{n=0}^{N}d\xi_n^* d\xi_n O_n^\uparrow\right)\left(\prod_{n=N}^{k_4+1}d\bar{\xi}_n^* d\bar{\xi}_n \bar{O}_{n}^\uparrow\right)d\bar{\xi}_{k_4}^*\\
&\times\left(\prod_{n=k_4}^{k_1+1} d\bar{\xi}_n d\bar{\xi}_{n-1}^*\bar{\mathcal{P}}_n^\uparrow\right)d\bar{\xi}_1\left(\prod_{n=k_1-1}^{0}d\bar{\xi}_n^* d\bar{\xi}_n \bar{O}_{n}^\uparrow\right){\color{red}\bar{\xi}_{k_4}^*}[-{\rm g}_+^{\uparrow}(t_{k_4}-t_{k_1})]{\color{red}\bar{\xi}_{k_1}}\\
&\times d\psi^*_{N+1} d\bar{\psi}_{N+1}\left(\prod_{n=0}^{k_2-1}d\psi_n^* d\psi_n O_n^\downarrow\right) d\psi_{k_2}^*\left(\prod_{n=k_2+1}^{N}d\psi_{n-1} d\psi_{n}^* \mathcal{P}_n^\downarrow\right)d\psi_{N}\mathcal{P}_{N+1}^\downarrow\\
&\times  d\bar{\psi}_N^* \bar{\mathcal{P}}_{N+1}^\downarrow \left(\prod_{n=N}^{k_3+1}d\bar{\psi}_n d\bar{\psi}_{n-1}^* \bar{\mathcal{P}}_{n}^\downarrow\right)d\bar{\psi}_{k_3}\left(\prod_{n=k_3-1}^{0}d\bar{\psi}_n^* d\bar{\psi}_n \bar{O}_n^\downarrow\right)\prod_{n=k_3+1}^{k_4}\bar{\mathcal{U}}_n\\
&\times{\color{red}\bar{\psi}_{k_3}}[- {\rm g}_+^{\downarrow *}(t_{k_3}-t_{k_2})]{\color{red}\psi_{k_2}^*}\;.
\end{aligned}\end{equation}
Once the trivial integrations are carried out as before, we are left with
\begin{equation}\begin{aligned}\label{}
J^{(2)}_{(h)01,00}(t;t_0)=&  \int_{t_0}^t dt_{k_4}\dots\int_{t_0}^{t_{k_2}} dt_{k_1}\;  \prod_{n=k_2+1}^{N}p_n^{\downarrow}\prod_{n=k_1+1}^{k_4}p_n^{\uparrow*}\prod_{n=k_3+1}^{k_4}(p_n^{\downarrow*}+u_n^*)\prod_{n=k_4+1}^{N}p_n^{\downarrow*}[-{\rm g}_+^{\uparrow}(t_{k_4}-t_{k_1})][-{\rm g}_+^{\downarrow *}(t_{k_3}-t_{k_2})]\\
&\times \int d\xi^*_{N+1} d\bar{\xi}_{N+1} \bar{\xi}_{N+1} \xi^*_{N+1}d\bar{\xi}_{k_4}^*d\bar{\xi}_1\bar{\xi}_{k_4}^*\bar{\xi}_{k_1}d\psi^*_{N+1} d\bar{\psi}_{N+1} d\psi_{k_2}^*d\psi_{N}\mathcal{P}_{N+1}^\downarrow d\bar{\psi}_N^* \bar{\mathcal{P}}_{N+1}^\downarrow d\bar{\psi}_{k_3}\bar{\psi}_{k_3}\psi_{k_2}^*\\
=&  \int_{t_0}^t dt_{k_4}\dots\int_{t_0}^{t_{k_2}} dt_{k_1}\;  \prod_{n=k_2+1}^{N+1}p_n^{\downarrow}\prod_{n=k_1+1}^{k_4}p_n^{\uparrow*}\prod_{n=k_3+1}^{k_4}(p_n^{\downarrow*}+u_n^*)\prod_{n=k_4+1}^{N+1}p_n^{\downarrow*}[-{\rm g}_+^{\uparrow}(t_{k_4}-t_{k_1})][-{\rm g}_+^{\downarrow *}(t_{k_3}-t_{k_2})]\\
&\times \int d\bar{\xi}_{k_4}^*d\bar{\xi}_1\bar{\xi}_{k_4}^*\bar{\xi}_{k_1} d\psi_{k_2}^* d\bar{\psi}_{k_3}\bar{\psi}_{k_3}\psi_{k_2}^*\\
=&- \int_{t_0}^t dt_{4}\dots\int_{t_0}^{t_{2}} dt_{1}\; e^{\frac{\rm i}{\hbar}\epsilon_\uparrow(t_{4}-t_{1})}e^{-\frac{\rm i}{\hbar}\epsilon_\downarrow(t_{3}-t_{2})}e^{\frac{\rm i}{\hbar}U(t_{4}-t_{3})}[-{\rm g}_+^{\uparrow}(t_{4}-t_{1})][-{\rm g}_+^{\downarrow *}(t_{3}-t_{2})]\;.
\end{aligned}\end{equation}
In the above examples the phase factors stemming from the action of the dot have been factorized as $\prod_{j=1}^n \exp[-\frac{\rm i}{\hbar}E_j\tau_j]$ in order to reflect the time intervals $\tau_j$ between transitions. One can recognize that these phase factors are related to the blip/sojourn states of the underlying spin paths according to the example in Fig.~\ref{scheme_interaction}.

\section{Integration measure $\int\mathcal{D}\{\xi\}$}
\label{integration_measure_Dxi}

\begin{figure}[ht]
\begin{center}
\resizebox{12cm}{!}{
\begin{tikzpicture}[]
\draw (7,1.5) node[] {(a)};
\draw (8.8,1.5) node[] {$\eta_0$};
\draw[thick] (8,1) -- (9.5,1); 
\draw (10.3,1.5) node[] {$\zeta_1$};
\draw[dashed,thick] (9.5,1) -- (11,1);
\draw (11.8,1.5) node[] {$\eta_1$}; 
\draw[thick] (11,1) -- (12.5,1);
\draw (13.3,1.5) node[] {$\zeta_2$};
\draw[dashed,thick] (12.5,1) -- (14,1);
\draw (14.8,1.5) node[] {$\eta_2$}; 
\draw[thick] (14,1) -- (15.5,1);
\draw (16.3,1.5) node[] {$\zeta_3$}; 
\draw[dashed,thick] (15.5,1) -- (17,1);
\draw (17.8,1.5) node[] {$\eta_3$}; 
\draw[thick,->] (17,1) -- (18.5,1);
\draw[thick] (9.5,0.9) -- node[align=left,   below] {$[\xi_1]_{-\eta_0\zeta_1}^{-\zeta_1}$} node[align=left, above] {$t_1$}  (9.5,1.1);
\draw[thick] (11,0.9) --
 node[align=left,   below] {$[\xi_2]_{-\eta_1\zeta_1}^{\zeta_1}$} node[align=left, above] {$t_2$}
(11,1.1);
\draw[thick] (12.5,0.9) --
 node[align=left,   below] {$[\xi_3]_{-\eta_1\zeta_2}^{-\zeta_2}$} node[align=left, above] {$t_3$}
(12.5,1.1);
\draw[thick] (14,0.9) -- node[align=left,   below] {$[\xi_4]_{-\eta_2\zeta_2}^{\zeta_2}$} node[align=left, above] {$t_4$}
(14,1.1);
\draw[thick] (15.5,0.9) -- node[align=left,   below] {$[\xi_5]_{-\eta_2\zeta_3}^{-\zeta_3}$} node[align=left, above] {$t_5$}
(15.5,1.1);
\draw[thick] (17,0.9) -- node[align=left,   below] {$[\xi_6]_{-\eta_3\zeta_3}^{\zeta_3}$} node[align=left, above] {$t_6$}
(17,1.1);
\end{tikzpicture}
}\\
\vspace{0.5cm}
\resizebox{12cm}{!}{
\begin{tikzpicture}[]
\draw (7,1.5) node[] {(b)};
\draw (8.8,1.5) node[] {$-1$};
\draw[thick] (8,1) -- (9.5,1); 
\draw (10.3,1.5) node[] {$+1$};
\draw[dashed,thick] (9.5,1) -- (11,1);
\draw (11.8,1.5) node[] {$-1$}; 
\draw[thick] (11,1) -- (12.5,1);
\draw (13.3,1.5) node[] {$+1$};
\draw[dashed,thick] (12.5,1) -- (14,1);
\draw (14.8,1.5) node[] {$\eta_2$}; 
\draw[thick] (14,1) -- (15.5,1);
\draw (16.3,1.5) node[] {$+1$}; 
\draw[dashed,thick] (15.5,1) -- (17,1);
\draw (17.8,1.5) node[] {$-1$}; 
\draw[thick,->] (17,1) -- (18.5,1);
\draw[thick] (9.5,0.9) -- node[align=left,   below] {$\xi_1^*$} node[align=left, above] {$t_1$}  (9.5,1.1);
\draw[thick] (11,0.9) --
 node[align=left,   below] {$\xi_2$} node[align=left, above] {$t_2$}
(11,1.1);
\draw[thick] (12.5,0.9) --
 node[align=left,   below] {$\xi_3^*$} node[align=left, above] {$t_3$}
(12.5,1.1);
\draw[thick] (14,0.9) -- node[align=left,   below] {$[\xi_4]_{-\eta_2}$} node[align=left, above] {$t_4$}
(14,1.1);
\draw[thick] (15.5,0.9) -- node[align=left,   below] {$[\xi_5^*]_{-\eta_2}$} node[align=left, above] {$t_5$}
(15.5,1.1);
\draw[thick] (17,0.9) -- node[align=left,   below] {$\xi_6$} node[align=left, above] {$t_6$}
(17,1.1);
\end{tikzpicture}
}\\
\vspace{0.5cm}
\resizebox{12cm}{!}{
\begin{tikzpicture}[]
\draw (7,1.5) node[] {(c)};
\draw (8.8,1.5) node[] {$-1$};
\draw[thick] (8,1) -- (9.5,1); 
\draw (10.3,1.5) node[] {$+1$};
\draw[dashed,thick] (9.5,1) -- (11,1);
\draw (11.8,1.5) node[] {$-1$}; 
\draw[thick] (11,1) -- (12.5,1);
\draw (13.3,1.5) node[] {$+1$};
\draw[dashed,thick] (12.5,1) -- (14,1);
\draw (14.8,1.5) node[] {$\eta_2$}; 
\draw[thick] (14,1) -- (15.5,1);
\draw (16.3,1.5) node[] {$\zeta_3$}; 
\draw[dashed,thick] (15.5,1) -- (17,1);
\draw (17.8,1.5) node[] {$-1$}; 
\draw[thick,->] (17,1) -- (18.5,1);
\draw[thick] (9.5,0.9) -- node[align=left,   below] {$\xi_1^*$} node[align=left, above] {$t_1$}  (9.5,1.1);
\draw[thick] (11,0.9) --
 node[align=left,   below] {$\xi_2$} node[align=left, above] {$t_2$}
(11,1.1);
\draw[thick] (12.5,0.9) --
 node[align=left,   below] {$\xi_3^*$} node[align=left, above] {$t_3$}
(12.5,1.1);
\draw[thick] (14,0.9) -- node[align=left,   below] {$[\xi_4]_{-\eta_2}$} node[align=left, above] {$t_4$}
(14,1.1);
\draw[thick] (15.5,0.9) -- node[align=left,   below] {$[\xi_5^*]_{-\eta_2\zeta_3}^{-\zeta_3}$} node[align=left, above] {$t_5$}
(15.5,1.1);
\draw[thick] (17,0.9) -- node[align=left,   below] {$[\xi_6]_{\zeta_3}^{\zeta^3}$} node[align=left, above] {$t_6$}
(17,1.1);
\end{tikzpicture}
}
\caption{\small{Six-transition path. General case (a) and two specific examples: In the first $\eta_2$ is left unspecified while $\eta_0=\eta_1=\eta_3=-1$ and $\zeta_1=\zeta_2=\zeta_3=+1$ (b). In the second also $\zeta_3$ is left unspecified (c).}}
\label{path_2}
\end{center}
\end{figure}
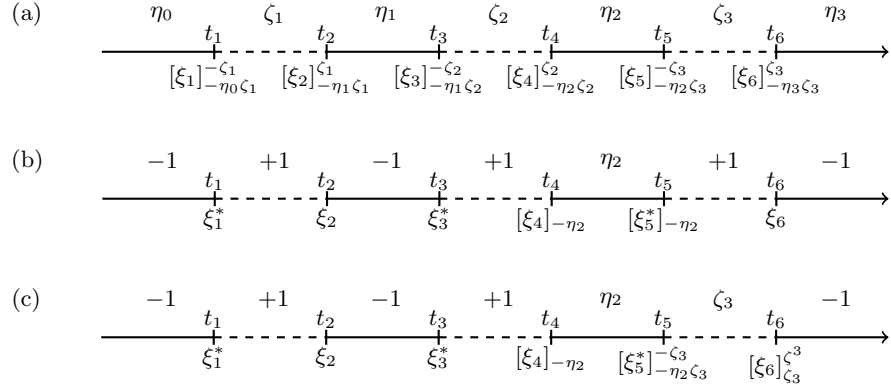

\indent  To see how the parametrization of the integrals over the Grassmann variables associated to the transition times works, consider the example of path shown in Fig.~\ref{path_2}(b).
Following Eq.~\eqref{Dxi}, the integration measure reads
\begin{equation}\begin{aligned}\label{}
&-\eta_2 d\xi_1^* d\xi_2 d\xi_3^* (d{\xi_4})_{-\eta_2}(d{\xi_5^*})_{-\eta_2} d\xi_6\\
&= \begin{cases}
 d\xi_1^* d\xi_2 d\xi_3^* d{\xi_4} d{\xi_5^*} d\xi_6\;, \quad \eta_2=-1\\
 d\xi_1^* d\xi_2 d\xi_3^* d\xi_6 d\bar{\xi}_5^* d\bar{\xi}_4\;, \quad \eta_2=+1
\end{cases}
\end{aligned}\end{equation}
As a further example, we leave also $\zeta_3$ unspecified, see Fig.~\ref{path_2}(c), so that the integration measure reads
\begin{equation}\begin{aligned}\label{}
&-\eta_2\zeta_3 d\xi_1^* d\xi_2 d\xi_3^* (d{\xi_4})_{-\eta_2}(d{\xi_5})^{-\xi_3}_{-\eta_2\zeta_3} (d{\xi_6})^{\zeta_3}_{\zeta_3}\\
&=\begin{cases}
d\xi_1^* d\xi_2 d\xi_3^* d{\xi_4} d{\xi_5^*} d\xi_6\;, \quad \eta_2=-1,\zeta_3=+1\\
 d\xi_1^* d\xi_2 d\xi_3^* d\xi_6 d\bar{\xi}_5^* d\bar{\xi}_4\;, \quad \eta_2=+1,\zeta_3=+1\\
   d\xi_1^* d\xi_2 d\xi_3^* d\xi_4 d\bar{\xi}_6^* d\bar{\xi}_5\;, \quad \eta_2=-1,\zeta_3=-1\\
  d\xi_1^* d\xi_2 d\xi_3^* d\xi_5 d\bar{\xi}_6^* d\bar{\xi}_4\;, \quad \eta_2=+1,\zeta_3=-1\\
\end{cases}
\end{aligned}\end{equation}
 \indent Carrying out an integration over the Grassmann variables associated to the tunneling transitions  is straightforward in the present derivation: The integral yields simply an overall sign, due to the order of the variables to be integrated, times the factors $(-\eta_k\zeta_k)$.

\section{Further examples of diagrammatic contributions from an individual state}
\label{examples_diagrams}

For the sake of compactness, in the following expressions we set $(t_j-t_i)\equiv (j,i)$. Then, the ten third-order irreducible diagrammatic  contributions $\mathcal{B}_{m_i}^i(\mathcal{P}_i) {\Phi}_{m_i}^i (\mathcal{P}_i)$, see Eq.~\eqref{Jm}, take the form
\begin{equation}\begin{aligned}\label{6th}
\begin{gathered}
\resizebox{4cm}{!}{
\begin{tikzpicture}[] 
\draw[thick] (-0.5,0) -- (0,0); 
\draw[dashed,thick] (0,0) -- (1,0); 
\draw[thick] (1,0) -- (2,0);
\draw[dashed,thick] (2,0) -- (3,0); 
\draw[thick] (3,0) -- (4,0);
\draw[dashed,thick] (4,0) -- (5,0); 
\draw[thick] (5,0) -- (5.5,0);
\draw[black,thick] (0,0) arc (180:0:2.5cm  and 1.25cm);
\draw[black,thick] (3,0) arc (180:0:0.5cm  and 0.75cm) ;
\draw[black,thick] (1,0) arc (180:0:0.5cm  and 0.75cm) ;
\filldraw 
(0,0) circle (2pt) node[align=left,   below] {$\eta_0\zeta_1$} 
(1,0) circle (2pt) node[align=center, below] {$\zeta_1\eta_1$} 
(3,0) circle (2pt) node[align=center, below] {$\zeta_2\eta_2$} ;
\end{tikzpicture}
 }
\end{gathered} &(+1)\prod_{n=1}^3\left(-\zeta_n\eta_n\right)\;[\eta_0\zeta_1{\rm g}_{-\eta_0}^{-\zeta_1}(6,1)]{\rm b}_{61}\;[\zeta_1\eta_1{\rm g}_{\eta_1}^{\zeta_1}(3,2)]{\rm b}_{32}\;[\zeta_2\eta_2{\rm g}_{\eta_2}^{\zeta_2}(5,4)]{\rm b}_{54}\;\delta_{\zeta_3,\zeta_1}\delta_{\zeta_2,\zeta_1}\\
&=\eta'\eta\;[-{\rm g}_{-\eta}^{-\zeta_1}(6,1)]{\rm b}_{61}\;[- {\rm g}_{\eta_1}^{\zeta_1}(3,2)]{\rm b}_{32}\;[-{\rm g}_{\eta_2}^{\zeta_1}(5,4)]{\rm b}_{54}\;\delta_{\zeta_3,\zeta_1}\delta_{\zeta_2,\zeta_1}\;,\\
\begin{gathered}
\resizebox{4cm}{!}{
\begin{tikzpicture}[] 
\draw[thick] (-0.5,0) -- (0,0); 
\draw[dashed,thick] (0,0) -- (1,0); 
\draw[thick] (1,0) -- (2,0);
\draw[dashed,thick] (2,0) -- (3,0); 
\draw[thick] (3,0) -- (4,0);
\draw[dashed,thick] (4,0) -- (5,0); 
\draw[thick] (5,0) -- (5.5,0);
\draw[black,thick] (0,0) arc (180:0:2cm  and 1.25cm);
\draw[black,thick] (3,0) arc (180:0:1cm  and 1cm) ;
\draw[black,thick] (1,0) arc (180:0:0.5cm  and 0.75cm) ;
\filldraw 
(0,0) circle (2pt) node[align=left,   below] {$\eta_0\zeta_1$} 
(1,0) circle (2pt) node[align=center, below] {$\zeta_1\eta_1$} 
(3,0) circle (2pt) node[align=center, below] {$\zeta_2\eta_2$} ;
\end{tikzpicture}
 }
\end{gathered} 
&(-1)\prod_{n=1}^3\left(-\zeta_n\eta_n\right)\;[\eta_0\zeta_1 {\rm g}_{-\eta_0}^{-\zeta_1}(5,1)]{\rm b}_{51}\;[\zeta_1\eta_1 {\rm g}_{\eta_1}^{\zeta_1}(3,2)]{\rm b}_{32}\;[\zeta_2\eta_2 {\rm g}_{\eta_2}^{\zeta_2}(6,4)]{\rm b}_{64}\;\delta_{\zeta_3,-\zeta_1}\delta_{\zeta_2,\zeta_1}\\
&=\eta'\eta\;[-{\rm g}_{-\eta}^{-\zeta_1}(5,1)]{\rm b}_{51}\;[- {\rm g}_{\eta_1}^{\zeta_1}(3,2)]{\rm b}_{32}\;[-{\rm g}_{\eta_2}^{\zeta_1}(6,4)]{\rm b}_{64}\;\delta_{\zeta_3,-\zeta_1}\delta_{\zeta_2,\zeta_1}\;,\\
\begin{gathered}
\resizebox{4cm}{!}{
\begin{tikzpicture}[] 
\draw[thick] (-0.5,0) -- (0,0); 
\draw[dashed,thick] (0,0) -- (1,0); 
\draw[thick] (1,0) -- (2,0);
\draw[dashed,thick] (2,0) -- (3,0); 
\draw[thick] (3,0) -- (4,0);
\draw[dashed,thick] (4,0) -- (5,0); 
\draw[thick] (5,0) -- (5.5,0);
\draw[black,thick] (1,0) arc (180:0:2cm  and 1.25cm);
\draw[black,thick] (3,0) arc (180:0:0.5cm  and 0.75cm) ;
\draw[black,thick] (0,0) arc (180:0:1cm  and 1cm) ;
\filldraw 
(0,0) circle (2pt) node[align=left,   below] {$\eta_0\zeta_1$} 
(1,0) circle (2pt) node[align=center, below] {$\zeta_1\eta_1$} 
(3,0) circle (2pt) node[align=center, below] {$\zeta_2\eta_2$} ;
\end{tikzpicture}
 }
\end{gathered}
&(-1)\prod_{n=1}^3\left(-\zeta_n\eta_n\right)\;[\eta_0\zeta_1 {\rm g}_{-\eta_0}^{-\zeta_1}(3,1)]{\rm b}_{31}\;[\zeta_1\eta_1 {\rm g}_{\eta_1}^{\zeta_1}(6,2)]{\rm b}_{62}\;[\zeta_2\eta_2 {\rm g}_{\eta_2}^{\zeta_2}(5,4)]{\rm b}_{54}\;\delta_{\zeta_3,-\zeta_1}\delta_{\zeta_2,\zeta_1}\\
&=\eta'\eta\;[-{\rm g}_{-\eta}^{-\zeta_1}(3,1)]{\rm b}_{31}\;[- {\rm g}_{\eta_1}^{\zeta_1}(6,2)]{\rm b}_{62}\;[-{\rm g}_{\eta_2}^{\zeta_1}(5,4)]{\rm b}_{54}\;\delta_{\zeta_3,-\zeta_1}\delta_{\zeta_2,\zeta_1}\;,\\
\end{aligned}
\end{equation}

\begin{equation}
\begin{aligned}\label{6th_pt2}
\begin{gathered}
\resizebox{4cm}{!}{
\begin{tikzpicture}[] 
\draw[thick] (-0.5,0) -- (0,0); 
\draw[dashed,thick] (0,0) -- (1,0); 
\draw[thick] (1,0) -- (2,0);
\draw[dashed,thick] (2,0) -- (3,0); 
\draw[thick] (3,0) -- (4,0);
\draw[dashed,thick] (4,0) -- (5,0); 
\draw[thick] (5,0) -- (5.5,0);
\draw[black,thick] (1,0) arc (180:0:1.5cm  and 1.25cm);
\draw[black,thick] (3,0) arc (180:0:1cm  and 1cm) ;
\draw[black,thick] (0,0) arc (180:0:1cm  and 1cm) ;
\filldraw 
(0,0) circle (2pt) node[align=left,   below] {$\eta_0\zeta_1$} 
(1,0) circle (2pt) node[align=center, below] {$\zeta_1\eta_1$} 
(3,0) circle (2pt) node[align=center, below] {$\zeta_2\eta_2$} ;
\end{tikzpicture}
 }
\end{gathered} 
&(+1)\prod_{n=1}^3\left(-\zeta_n\eta_n\right)\;[\eta_0\zeta_1 {\rm g}_{-\eta_0}^{-\zeta_1}(3,1)]{\rm b}_{31}\;[\zeta_1\eta_1 {\rm g}_{\eta_1}^{\zeta_1}(5,2)]{\rm b}_{52}\;[\zeta_2\eta_2 {\rm g}_{\eta_2}^{\zeta_2}(6,4)]{\rm b}_{64}\;\delta_{\zeta_3,\zeta_1}\delta_{\zeta_3,-\zeta_2}\\
&=\eta'\eta\;[-{\rm g}_{-\eta}^{-\zeta_1}(3,1)]{\rm b}_{31}\;[- {\rm g}_{\eta_1}^{\zeta_1}(5,2)]{\rm b}_{52}\;[-{\rm g}_{\eta_2}^{-\zeta_1}(6,4)]{\rm b}_{64}\;\delta_{\zeta_3,\zeta_1}\delta_{\zeta_2,-\zeta_1}\;,
\end{aligned}\end{equation}

\begin{equation}\begin{aligned}\label{}
\begin{gathered}
\resizebox{4.5cm}{!}{
\begin{tikzpicture}[] 
\draw[thick] (-0.5,0) -- (0,0); 
\draw[dashed,thick] (0,0) -- (1,0); 
\draw[thick] (1,0) -- (2,0);
\draw[dashed,thick] (2,0) -- (3,0); 
\draw[thick] (3,0) -- (4,0);
\draw[dashed,thick] (4,0) -- (5,0); 
\draw[thick] (5,0) -- (5.5,0);
\draw[black,thick] (0,0) arc (180:0:2.5cm  and 1.5cm);
\draw[black,thick] (1,0) arc (180:0:1.5cm  and 1cm) ;
\draw[black,thick] (2,0) arc (180:0:0.5cm  and 0.5cm) ;
\filldraw 
(0,0) circle (2pt) node[align=left,   below] {$\eta_0\zeta_1$} 
(1,0) circle (2pt) node[align=center, below] {$\zeta_1\eta_1$} 
(2,0) circle (2pt) node[align=center, below] {$\eta_1\zeta_2$} ;
 \end{tikzpicture}
 }
\end{gathered}
&\prod_{k=1}^3(-\zeta_k\eta_k)[\eta_0\zeta_1\;{\rm g}_{-\eta_0}^{-\zeta_1}(6,1)]{\rm b}_{61}\;[\zeta_1\eta_1{\rm g}_{\eta_1}^{\zeta_1}(5,2)]{\rm b}_{52}\;[\eta_1\zeta_2{\rm g}_{-\eta_1}^{-\zeta_2}(4,3)]{\rm b}_{43}\delta_{\zeta_3,\zeta_1}\\
&=\eta'\eta \eta_1\eta_2 \;[-{\rm g}_{-\eta}^{-\zeta_1}(6,1)]{\rm b}_{61}\;[-{\rm g}_{\eta_1}^{\zeta_1}(5,2)]{\rm b}_{52}[-{\rm g}_{-\eta_1}^{-\zeta_2}(4,3)]{\rm b}_{43}\delta_{\zeta_3,\zeta_1}\;,\\
\begin{gathered}
\resizebox{4.5cm}{!}{
\begin{tikzpicture}[] 
\draw[thick] (-0.5,0) -- (0,0); 
\draw[dashed,thick] (0,0) -- (1,0); 
\draw[thick] (1,0) -- (2,0);
\draw[dashed,thick] (2,0) -- (3,0); 
\draw[thick] (3,0) -- (4,0);
\draw[dashed,thick] (4,0) -- (5,0); 
\draw[thick] (5,0) -- (5.5,0);
\draw[black,thick] (0,0) arc (180:0:2.5cm  and 1.5cm);
\draw[black,thick] (1,0) arc (180:0:1cm  and 1cm) ;
\draw[black,thick] (2,0) arc (180:0:1cm  and 1cm) ;
\filldraw 
(0,0) circle (2pt) node[align=left,   below] {$\eta_0\zeta_1$} 
(1,0) circle (2pt) node[align=center, below] {$\zeta_1\eta_1$} 
(2,0) circle (2pt) node[align=center, below] {$\eta_1\zeta_2$} ;
\end{tikzpicture}
 }
\end{gathered} 
&(-1)\prod_{k=1}^3(-\zeta_k\eta_k)[\eta_0\zeta_1\;{\rm g}_{-\eta_0}^{-\zeta_1}(6,1)]{\rm b}_{61}\;[\zeta_1\eta_1{\rm g}_{\eta_1}^{\zeta_1}(4,2)]{\rm b}_{42}\;[\eta_1\zeta_2{\rm g}_{-\eta_1}^{-\zeta_2}(5,3)]{\rm b}_{53}\delta_{\zeta_3,-\zeta_2}\delta_{\zeta_2,-\zeta_1}\\
&=-\eta'\eta \eta_1\eta_2 \;[-{\rm g}_{-\eta}^{-\zeta_1}(6,1)]{\rm b}_{61}\;[-{\rm g}_{\eta_1}^{\zeta_1}(4,2)]{\rm b}_{42}[-{\rm g}_{-\eta_1}^{-\zeta_2}(5,3)]{\rm b}_{53}\;\delta_{\zeta_3,\zeta_1}\delta_{\zeta_2,-\zeta_1}\;,\\
 \begin{gathered}
\resizebox{4.5cm}{!}{
\begin{tikzpicture}[] 
\draw[thick] (-0.5,0) -- (0,0); 
\draw[dashed,thick] (0,0) -- (1,0); 
\draw[thick] (1,0) -- (2,0);
\draw[dashed,thick] (2,0) -- (3,0); 
\draw[thick] (3,0) -- (4,0);
\draw[dashed,thick] (4,0) -- (5,0); 
\draw[thick] (5,0) -- (5.5,0);
\draw[black,thick] (0,0) arc (180:0:2cm  and 1.5cm);
\draw[black,thick] (1,0) arc (180:0:2cm  and 1.5cm) ;
\draw[black,thick] (2,0) arc (180:0:0.5cm  and 0.75cm) ;
\filldraw 
(0,0) circle (2pt) node[align=left,   below] {$\eta_0\zeta_1$} 
(1,0) circle (2pt) node[align=center, below] {$\zeta_1\eta_1$} 
(2,0) circle (2pt) node[align=center, below] {$\eta_1\zeta_2$} ;
\end{tikzpicture}
 }
\end{gathered} 
&(-1)\prod_{k=1}^3(-\zeta_k\eta_k)[\eta_0\zeta_1\;{\rm g}_{-\eta_0}^{-\zeta_1}(5,1)]{\rm b}_{51}\;[\zeta_1\eta_1{\rm g}_{\eta_1}^{\zeta_1}(6,2)]{\rm b}_{62}\;[\eta_1\zeta_2{\rm g}_{-\eta_1}^{-\zeta_2}(4,3)]{\rm b}_{43}\delta_{\zeta_3,-\zeta_1}\\
&=\eta'\eta \eta_1\eta_2 \;[-{\rm g}_{-\eta}^{-\zeta_1}(5,1)]{\rm b}_{51}\;[-{\rm g}_{\eta_1}^{\zeta_1}(6,2)]{\rm b}_{62}[-{\rm g}_{-\eta_1}^{-\zeta_2}(4,3)]{\rm b}_{43}\;\delta_{\zeta_3,-\zeta_1}\;,
\end{aligned}\end{equation}

\begin{equation}\begin{aligned}\label{}
 \begin{gathered}
\resizebox{4.5cm}{!}{
\begin{tikzpicture}[] 
\draw[thick] (-0.5,0) -- (0,0); 
\draw[dashed,thick] (0,0) -- (1,0); 
\draw[thick] (1,0) -- (2,0);
\draw[dashed,thick] (2,0) -- (3,0); 
\draw[thick] (3,0) -- (4,0);
\draw[dashed,thick] (4,0) -- (5,0); 
\draw[thick] (5,0) -- (5.5,0);
\draw[black,thick] (0,0) arc (180:0:2cm  and 1.5cm);
\draw[black,thick] (1,0) arc (180:0:1cm  and 1cm) ;
\draw[black,thick] (2,0) arc (180:0:1.5cm  and 1.5cm) ;
\filldraw 
(0,0) circle (2pt) node[align=left,   below] {$\eta_0\zeta_1$} 
(1,0) circle (2pt) node[align=center, below] {$\zeta_1\eta_1$} 
(2,0) circle (2pt) node[align=center, below] {$\eta_1\zeta_2$} ;
\end{tikzpicture}
 }
\end{gathered} 
&(+1)\prod_{k=1}^3(-\zeta_k\eta_k)[\eta_0\zeta_1\;{\rm g}_{-\eta_0}^{-\zeta_1}(5,1)]{\rm b}_{51}\;[\zeta_1\eta_1{\rm g}_{\eta_1}^{\zeta_1}(4,2)]{\rm b}_{42}\;[\eta_1\zeta_2{\rm g}_{-\eta_1}^{-\zeta_2}(6,3)]{\rm b}_{63}\delta_{\zeta_3,-\zeta_1}\delta_{\zeta_2,-\zeta_1}\\
&=-\eta'\eta \eta_1\eta_2 \;[-{\rm g}_{-\eta}^{-\zeta_1}(5,1)]{\rm b}_{51}\;[-{\rm g}_{\eta_1}^{\zeta_1}(4,2)]{\rm b}_{42}[-{\rm g}_{-\eta_1}^{-\zeta_2}(6,3)]{\rm b}_{63}\;\delta_{\zeta_3,-\zeta_1}\;\delta_{\zeta_2,-\zeta_1}\;,\\
  \begin{gathered}
\resizebox{4.5cm}{!}{
\begin{tikzpicture}[] 
\draw[thick] (-0.5,0) -- (0,0); 
\draw[dashed,thick] (0,0) -- (1,0); 
\draw[thick] (1,0) -- (2,0);
\draw[dashed,thick] (2,0) -- (3,0); 
\draw[thick] (3,0) -- (4,0);
\draw[dashed,thick] (4,0) -- (5,0); 
\draw[thick] (5,0) -- (5.5,0);
\draw[black,thick] (1,0) arc (180:0:2cm  and 1.5cm);
\draw[black,thick] (2,0) arc (180:0:1cm  and 1cm) ;
\draw[black,thick] (0,0) arc (180:0:1.5cm  and 1.5cm) ;
\filldraw 
(0,0) circle (2pt) node[align=left,   below] {$\eta_0\zeta_1$} 
(1,0) circle (2pt) node[align=center, below] {$\zeta_1\eta_1$} 
(2,0) circle (2pt) node[align=center, below] {$\eta_1\zeta_2$} ;
\end{tikzpicture}
 }
\end{gathered} 
&(+1)\prod_{k=1}^3(-\zeta_k\eta_k)[\eta_0\zeta_1\;{\rm g}_{-\eta_0}^{-\zeta_1}(4,1)]{\rm b}_{41}\;[\zeta_1\eta_1{\rm g}_{\eta_1}^{\zeta_1}(6,2)]{\rm b}_{62}\;[\eta_1\zeta_2{\rm g}_{-\eta_1}^{-\zeta_2}(5,3)]{\rm b}_{53}\delta_{\zeta_3,-\zeta_1}\delta_{\zeta_2,\zeta_1}\\
&=-\eta'\eta \eta_1\eta_2 \;[-{\rm g}_{-\eta}^{-\zeta_1}(4,1)]{\rm b}_{41}\;[-{\rm g}_{\eta_1}^{\zeta_1}(6,2)]{\rm b}_{62}[-{\rm g}_{-\eta_1}^{-\zeta_1}(5,3)]{\rm b}_{53}\;\delta_{\zeta_3,-\zeta_1}\;\delta_{\zeta_2,\zeta_1}\;,\\
   \begin{gathered}
\resizebox{4.5cm}{!}{
\begin{tikzpicture}[] 
\draw[thick] (-0.5,0) -- (0,0); 
\draw[dashed,thick] (0,0) -- (1,0); 
\draw[thick] (1,0) -- (2,0);
\draw[dashed,thick] (2,0) -- (3,0); 
\draw[thick] (3,0) -- (4,0);
\draw[dashed,thick] (4,0) -- (5,0); 
\draw[thick] (5,0) -- (5.5,0);
\draw[black,thick] (1,0) arc (180:0:1.5cm  and 1.5cm);
\draw[black,thick] (2,0) arc (180:0:1.5cm  and 1.5cm) ;
\draw[black,thick] (0,0) arc (180:0:1.5cm  and 1.5cm) ;
\filldraw 
(0,0) circle (2pt) node[align=left,   below] {$\eta_0\zeta_1$} 
(1,0) circle (2pt) node[align=center, below] {$\zeta_1\eta_1$} 
(2,0) circle (2pt) node[align=center, below] {$\eta_1\zeta_2$} ;
\end{tikzpicture}
 }
\end{gathered} 
& (-1)\prod_{k=1}^3(-\zeta_k\eta_k)[\eta_0\zeta_1\;{\rm g}_{-\eta_0}^{-\zeta_1}(3,1)]{\rm b}_{31}\;[\zeta_1\eta_1{\rm g}_{\eta_1}^{\zeta_1}(4,2)]{\rm b}_{52}\;[\eta_1\zeta_2{\rm g}_{-\eta_1}^{-\zeta_2}(6,3)]{\rm b}_{63}\delta_{\zeta_3,\zeta_2}\delta_{\zeta_2,\zeta_1}\\
&=-\eta'\eta\eta_1\eta_2 [-{\rm g}_{-\eta}^{-\zeta_1}(4,1)]{\rm b}_{41}\;[-{\rm g}_{\eta_1}^{\zeta_1}(5,2)]{\rm b}_{52}\;[-{\rm g}_{-\eta_1}^{-\zeta_1}(6,3)]{\rm b}_{63}\delta_{\zeta_3,\zeta_1}\delta_{\zeta_2,\zeta_1}\;.
\end{aligned}\end{equation}
We also evaluate the following irreducible fourth-order diagram, which is relevant for the scheme NCA4 introduced in Sec.~\ref{gDSO4},
\begin{equation}\begin{aligned}\label{4tha}
&\begin{gathered}
\resizebox{6.5cm}{!}{
\begin{tikzpicture}[] 
\draw[thick] (-0.5,0) -- (0,0); 
\draw[dashed,thick] (0,0) -- (1,0); 
\draw[thick] (1,0) -- (2,0);
\draw[dashed,thick] (2,0) -- (3,0); 
\draw[thick] (3,0) -- (4,0);
\draw[dashed,thick] (4,0) -- (5,0); 
\draw[thick] (5,0) -- (6,0);
\draw[dashed,thick] (6,0) -- (7,0); 
\draw[thick] (7,0) -- (7.5,0);
\draw[black,thick] (0,0) arc (180:0:3.5cm  and 1.75cm);
\draw[black,thick] (1,0) arc (180:0:2.5cm  and 1.25cm) ;
\draw[black,thick] (2,0) arc (180:0:0.5cm  and 0.7cm) ;
\draw[black,thick] (4,0) arc (180:0:0.5cm  and 0.7cm) ;
\filldraw 
(0,0) circle (2pt) node[align=left,   below] {$\eta_0\zeta_1$} 
(1,0) circle (2pt) node[align=center, below] {$\zeta_1\eta_1$} 
(2,0) circle (2pt) node[align=center, below] {$\eta_1\zeta_2$} 
(4,0) circle (2pt) node[align=center, below] {$\eta_2\zeta_3$} ;
\end{tikzpicture}
 }
\end{gathered} \\
& (+1)\prod_{k=1}^4(-\zeta_k\eta_k)\;[\eta_0\zeta_1{\rm g}_{-\eta_0}^{-\zeta_1}(8,1)]{\rm b}_{81}\;
[\zeta_1\eta_1{\rm g}_{\eta_1}^{\zeta_1}(7,2)]{\rm b}_{72}\;[\eta_1\zeta_2{\rm g}_{-\eta_1}^{-\zeta_2}(4,3)]{\rm b}_{43}\;
[\eta_2\zeta_3{\rm g}_{-\eta_2}^{-\zeta_3}(6,5)]
{\rm b}_{65}\;\delta_{\zeta_4,\zeta_1}\\
&=\eta'\eta\eta_1\eta_3\;[-{\rm g}_{-\eta}^{-\zeta_1}(8,1)]{\rm b}_{81}\;
[-{\rm g}_{\eta_1}^{\zeta_1}(7,2)]{\rm b}_{72}\;[-{\rm g}_{-\eta_1}^{-\zeta_2}(4,3)]{\rm b}_{43}\;
[-{\rm g}_{-\eta_2}^{-\zeta_3}(6,5)]
{\rm b}_{65}\;\delta_{\zeta_4,\zeta_1}\;.
\end{aligned}\end{equation}

\newpage

\section{Contraction integrals in the wide-band limit}
\label{contraction_integrals}

\begin{itemize}

\item Integral not involving the Fermi function ($\zeta=\pm 1$)
\begin{equation}
\begin{aligned}
\label{I0}
I_0(\mathcal{E};\zeta)=&\int_{-W}^{W} dx\frac{1}{x-\mathcal{E}'+{\rm i}\zeta\;\mathcal{E}''}\\
\simeq &-{\rm i}\zeta\pi\qquad \qquad (W\gg \mathcal{E}')\,.
\end{aligned}
\end{equation}

\item Integral involving $f_{-\eta}(x)$,where $\eta=\pm 1$, with $f_+(x)=[e^{\beta(x-\mu)}+1]^{-1}$ the Fermi function and $f_-(x)=1-f_+(x)$. Assume $\mathcal{E}$ independent of $x$, with $\mathcal{E}''> 0$, and ${W}\gg {\mathcal{E}}',\mu$

\begin{equation}
\begin{aligned}
\label{I}
I_+(\mathcal{E})=&\int_{-W}^{W} d\epsilon \frac{f_{+}(\epsilon)}{\epsilon-\mathcal{E}'+{\rm i}\mathcal{E}''}\\
=&\int_{-\bar{W}}^{\bar{W}} dx \frac{\bar{f}_{+}(x)}{x-(\mathcal{E}'-\mu)/k_{\rm B}T+{\rm i}\mathcal{E}''/k_{\rm B}T}\qquad\qquad[\bar{f}_+(x)=(e^x+1)^{-1},\qquad \bar{W}\simeq W/{k_{\rm B} T}]\\
=&2\pi{\rm i}\sum_j{\rm Res}_j\left\{\frac{\bar{f}_{+}(z)}{z-(\mathcal{E}'-\mu)/k_{\rm B}T+{\rm i}\mathcal{E}''/k_{\rm B}T}\right\}\\
=&-2\pi{\rm i}\sum_{k=0}^{k_{\bar{W}}}\frac{1}{2\pi{\rm i}(k+1/2)-(\mathcal{E}'-\mu)/k_{\rm B}T+{\rm i}\mathcal{E}''/k_{\rm B}T}-{\rm i}\frac{\pi}{2}\\
=&-\sum_{k=0}^{k_{\bar{W}}}\frac{1}{k+1/2+\mathcal{E}''/(2\pi k_{\rm B}T)+{\rm i}(\mathcal{E}'-\mu)/(2\pi k_{\rm B}T)}-{\rm i}\frac{\pi}{2}\;.
\end{aligned}
\end{equation}
Now, since $W\rightarrow\infty$, the sum can be extended to infinity. Following Ref.~\cite{Leijnse2008},  if $\mathcal{E}$ is independent of $\epsilon$, we single out the $k=0$ term in the sum over $k$ and add and subtract the Euler-Mascheroni constant $\gamma_E=\lim_{K \to \infty}\sum_{k=1}^K 1/k-\ln(K)$. At this point, using the definition of digamma function $\psi(z)=-\gamma_E-1/z-\sum_{k=1}^\infty[1/(k+z)-1/k]$, we  obtain
\begin{equation}
\begin{aligned}
\label{}
I_+(\mathcal{E})=& {\rm Re}\psi\left(\frac{1}{2}+{\rm i}\frac{\mathcal{E}'-{\rm i}\mathcal{E}''-\mu}{2\pi k_{\rm B}T}\right)
-\ln\frac{W}{2\pi k_{\rm B}T}-{\rm i}\left[\frac{\pi}{2}-{\rm Im}\psi\left(\frac{1}{2}+{\rm i}\frac{\mathcal{E}'-{\rm i}\mathcal{E}''-\mu}{2\pi k_{\rm B}T}\right)\right]\\
I_-(\mathcal{E})=&\int_{-W}^{W} dx \frac{f_{-}(x)}{x-\mathcal{E}'+{\rm i}\mathcal{E}''}=\int_{-W}^{W} dx \frac{1-f_{+}(x)}{x-\mathcal{E}'+{\rm i}\mathcal{E}''}\\
=& -{\rm Re}\psi\left(\frac{1}{2}+{\rm i}\frac{\mathcal{E}'-{\rm i}\mathcal{E}''-\mu}{2\pi k_{\rm B}T}\right)
+\ln\frac{W}{2\pi k_{\rm B}T}-{\rm i}\left[\frac{\pi}{2}+{\rm Im}\psi\left(\frac{1}{2}+{\rm i}\frac{\mathcal{E}'-{\rm i}\mathcal{E}''-\mu}{2\pi k_{\rm B}T}\right)\right]
\;,
\end{aligned}
\end{equation}
where we used Eq.~\eqref{I0}. Thus
\begin{equation}
\begin{aligned}
\label{I+}
I(\mathcal{E};\eta)=&\int_{-W}^{W} dx \frac{f_{-\eta}(x)}{x-\mathcal{E}'+{\rm i}\mathcal{E}''}\\
=& -\eta\left[{\rm Re}\psi\left(\frac{1}{2}+{\rm i}\frac{\mathcal{E}'-{\rm i}\mathcal{E}''-\mu}{2\pi k_{\rm B}T}\right)
-\ln\frac{W}{2\pi k_{\rm B}T}\right]-{\rm i}\left[\frac{\pi}{2}+\eta{\rm Im}\psi\left(\frac{1}{2}+{\rm i}\frac{\mathcal{E}'-{\rm i}\mathcal{E}''-\mu}{2\pi k_{\rm B}T}\right)\right]
\end{aligned}
\end{equation}
Further
\begin{equation}
\begin{aligned}
\label{I-}
I^*(\mathcal{E};\eta)=&\int_{-W}^{W} dx \frac{f_{-\eta}(x)}{x-\mathcal{E}'-{\rm i}\mathcal{E}''}\\
=& -\eta\left[{\rm Re}\psi\left(\frac{1}{2}+{\rm i}\frac{\mathcal{E}'-{\rm i}\mathcal{E}''-\mu}{2\pi k_{\rm B}T}\right)
-\ln\frac{W}{2\pi k_{\rm B}T}\right]+{\rm i}\left[\frac{\pi}{2}+\eta{\rm Im}\psi\left(\frac{1}{2}+{\rm i}\frac{\mathcal{E}'-{\rm i}\mathcal{E}''-\mu}{2\pi k_{\rm B}T}\right)\right]
\end{aligned}
\end{equation}
Then, collecting the above results we can give the \textbf{compact expression}
\begin{equation}
\begin{aligned}
\label{Isummary}
I(\mathcal{E};\zeta,\eta)=&\int_{-W}^{W} dx \frac{f_{-\eta}(x)}{x-\mathcal{E}'+{\rm i}\zeta\mathcal{E}''}\\
=& -\eta\left[{\rm Re}\psi\left(\frac{1}{2}+{\rm i}\frac{\mathcal{E}'-{\rm i}\mathcal{E}''-\mu}{2\pi k_{\rm B}T}\right)
-\ln\frac{W}{2\pi k_{\rm B}T}\right]-{\rm i}\zeta\left[\frac{\pi}{2}+\eta{\rm Im}\psi\left(\frac{1}{2}+{\rm i}\frac{\mathcal{E}'-{\rm i}\mathcal{E}''-\mu}{2\pi k_{\rm B}T}\right)\right]
\end{aligned}
\end{equation}

\item Special case: $\mathcal{E}''=0^+$.
\begin{equation}
\begin{aligned}
\label{I1}
\int_{-W}^{W} dx\frac{f_{-\eta}(x)}{x-\mathcal{E}'+{\rm i}\zeta\; 0^+}
= -\eta\left[{\rm Re}\psi\left(\frac{1}{2}+{\rm i}\frac{\mathcal{E}'-\mu}{2\pi k_{\rm B}T}\right)-\ln\frac{W}{2\pi k_{\rm B}T}\right]-{\rm i}\zeta\pi f_{-\eta}(\mathcal{E}') 
\;,
\end{aligned}
\end{equation}
see also Eq.~(E1) of Ref.~\cite{Leijnse2008}. Here we used  the property
\begin{equation}
\begin{aligned}
\label{property_Fermi}
\frac{1}{2}\mp\frac{1}{\pi}{\rm Im}\psi\left(\frac{1}{2}+{\rm i}\frac{\mathcal{E}'-\mu}{2\pi k_{\rm B}T}\right)= f_\pm(\mathcal{E}')
\end{aligned}
\end{equation}
and also $f_{-\eta}(x)=\delta_{\eta,+1}-\eta f_+(x)$.
Note that 
\[
\int_{-W}^{W} dx\frac{1}{x-\mathcal{E}'+{\rm i}\zeta\; 0^+}=\sum_\eta\int_{-W}^{W} dx\frac{f_{-\eta}(x)}{x-\mathcal{E}'+{\rm i}\zeta\; 0^+}=-{\rm i}\zeta\pi\;,
\]
in agreement with Eq.~\eqref{I0}.

\item We assume that, in general, $\mathcal{E}=\mathcal{E}(\zeta,\eta)$ and consider the distinct cases $\mathcal{E}(\eta)=\mathcal{E}_\eta$ and $\mathcal{E}(\zeta)=\mathcal{E}_\zeta$. In the first case, summing over $\eta$

\begin{equation}
\begin{aligned}
\label{Prop1}
\sum_{\eta} I(\mathcal{E}_\eta;\zeta,\eta)=&-\left[{\rm Re}\psi\left(\frac{1}{2}+{\rm i}\frac{\mathcal{E}'_+-{\rm i}\mathcal{E}''_+-\mu}{2\pi k_{\rm B}T}\right)-{\rm Re}\psi\left(\frac{1}{2}+{\rm i}\frac{\mathcal{E}'_--{\rm i}\mathcal{E}''_--\mu}{2\pi k_{\rm B}T}\right)\right]\\
&-{\rm i}\zeta\left[\pi+{\rm Im}\psi\left(\frac{1}{2}+{\rm i}\frac{\mathcal{E}'_+-{\rm i}\mathcal{E}''_+-\mu}{2\pi k_{\rm B}T}\right)-{\rm Im}\psi\left(\frac{1}{2}+{\rm i}\frac{\mathcal{E}'_--{\rm i}\mathcal{E}''_--\mu}{2\pi k_{\rm B}T}\right) \right]\;.
\end{aligned}
\end{equation}
Note that if $\mathcal{E}$ is independent of $\eta$, then $\sum_{\eta} I(\mathcal{E};\zeta,\eta)=-{\rm i}\zeta\pi$. Likewise, when $\mathcal{E}=\mathcal{E}_\zeta$

\begin{equation}
\begin{aligned}
\label{Prop2}
\sum_{\zeta} \zeta\; I(\mathcal{E}_\zeta;\zeta,\eta)=&-\eta\left[{\rm Re}\psi\left(\frac{1}{2}+{\rm i}\frac{\mathcal{E}'_+-{\rm i}\mathcal{E}''_+-\mu}{2\pi k_{\rm B}T}\right)-{\rm Re}\psi\left(\frac{1}{2}+{\rm i}\frac{\mathcal{E}'_--{\rm i}\mathcal{E}''_--\mu}{2\pi k_{\rm B}T}\right)\right]\\
&-{\rm i}\left[\pi+\eta{\rm Im}\psi\left(\frac{1}{2}+{\rm i}\frac{\mathcal{E}'_+-{\rm i}\mathcal{E}''_+-\mu}{2\pi k_{\rm B}T}\right)+\eta{\rm Im}\psi\left(\frac{1}{2}+{\rm i}\frac{\mathcal{E}'_--{\rm i}\mathcal{E}''_--\mu}{2\pi k_{\rm B}T}\right) \right]\;.
\end{aligned}
\end{equation}
If $\mathcal{E}$ is independent of $\zeta$, then $\sum_{\zeta}\zeta I(\mathcal{E};\zeta,\eta)=2{\rm i}\zeta{\rm Im} I(\mathcal{E};\zeta,\eta)$.
\end{itemize}

\newpage

\section{Evaluation of the NCA4 fourth-tier bubbles $\mathbf{B}_4^{\sigma(\sigma\sigma)}$ and $\mathbf{B}_4^{\sigma(\sigma\bar\sigma)}$}
\label{gDSO4_4th_tier}

\indent The fourth-tier bubbles ${\rm B}_{4,\nu'\nu}^{\sigma(\sigma\bar\sigma)}$ and ${\rm B}_{4,\eta'\eta}^{\sigma(\sigma\sigma)}$ have the same structure as the NCA2 second-tier bubbles ${\rm B}^{\bar\sigma}_{\nu'\nu}$ and ${\rm B}^{\sigma}_{\eta'\eta}$ [Eq.~\eqref{B_SIAM}], respectively, except for the additional upper layers of fermion lines, and the products of sojourn indexes associated to the overlap of three fermion lines of the same spin. They are schematized as (in view of calculating the retarded self-energy we consider $\zeta=+1$)
\begin{equation}\begin{aligned}\label{B4s_a}
{\rm B}^{\sigma(\sigma\bar\sigma)}_{4,\nu'\nu}&|_{\zeta=+1}=
\begin{gathered}
\resizebox{!}{1.6cm}{
\begin{tikzpicture}[] 
\draw[blue,line width=0.5mm] (0.5,2) arc (120:60:3cm  and 1.5cm)  node[right] {\LARGE{$\sigma\kappa$}}; 
\draw[blue,line width=0.5mm] (0.5,1.6) arc (120:60:3cm  and 2.cm)  node[right]  {\LARGE{$\sigma\boldsymbol{\kappa}_{1}$}};
\draw[red,line width=0.5mm] (0.5,1.2) arc (120:60:3cm  and 2.5cm)  node[right] {\LARGE{$\bar\sigma\boldsymbol{\kappa}_{2}$}};
\draw[red,line width=0.5mm]  (1,0) arc (180:0:1.cm  and 1.2cm); 
\draw[blue,line width=0.5mm]  (0.5,0)node[left]{\Large{$\nu$}} -- node[right]{\Large{$\qquad\quad\nu'$}} (3.5,0); 
\draw[red,dashed,line width=0.5mm] (0.5,-0.5) -- (1,-0.5); 
\draw[red,line width=0.5mm] (1,-0.5) -- (1,0); 
\draw[red,line width=0.5mm] (1,-0.5) -- (3,-0.5); 
\draw[red,dashed,line width=0.5mm] (3,-0.5) --   (3.5,-0.5); 
\draw[red,line width=0.5mm] (3,-0.5) -- (3,-0); 
\filldraw[red](1,-0.5) circle (3pt); 
 \end{tikzpicture}
 }
\end{gathered}
+
\begin{gathered}
\resizebox{!}{1.6cm}{
\begin{tikzpicture}[] 
\draw[blue,line width=0.5mm] (0.5,2) arc (120:60:3cm  and 1.5cm)  node[right] {\LARGE{$\sigma\kappa$}}; 
\draw[blue,line width=0.5mm] (0.5,1.6) arc (120:60:3cm  and 2.cm) node[right]  {\LARGE{$\sigma\boldsymbol{\kappa}_{1}$}};
\draw[red,line width=0.5mm] (0.5,1.2) arc (120:60:3cm  and 2.5cm)  node[right] {\LARGE{$\bar\sigma\boldsymbol{\kappa}_{2}$}};
\draw[blue,line width=0.5mm] (1,0) arc (180:0:1.cm  and 1.2cm); 
\draw[blue,line width=0.5mm]  (0.5,0)node[left] {\Large{$\nu$}} --(1,0); 
\draw[blue,dashed,line width=0.5mm] (1,0) -- (3,0); 
\draw[blue,line width=0.5mm] (3,0)--  node[]{\Large{$\qquad\quad\nu'$}}(3.5,0) ; 
\draw[red,dashed,line width=0.5mm] (0.5,-0.5) --  (3.5,-0.5); 
\filldraw[blue] 
(1,0) circle (3pt);
\end{tikzpicture}
 }
\end{gathered}
=\sum_\eta\langle {\rm h}^{\sigma(\sigma\bar\sigma\bar\sigma)}_{4}{\rm v}_{\eta}\rangle_{\zeta=+1}\delta_{\nu'\nu}+\nu'\nu\langle \tilde{\rm h}^{\sigma(\sigma\bar\sigma\sigma)}_{4}{\rm v}_{-\nu}\rangle_{\zeta=+1}\\
=&\;\sum_{\boldsymbol{\kappa}_{3}}\frac{{\rm i} \hbar\, \sum_{\eta}{\rm v}_{\eta}(\boldsymbol{\kappa}_{3})\delta_{\zeta_{3},-\zeta_{2}}}{\zeta(\epsilon_k-\epsilon_{k_{1}})+\zeta_{2}(\epsilon_{k_{2}}-\epsilon_{k_{3}})+{\rm i}  0^+}\delta_{\nu'\nu}\Big|_{\zeta=+1}+\sum_{\boldsymbol{\kappa}_{3}}\frac{\nu'\nu\;{\rm i} \hbar\,  {\rm v}_{-\nu}(\boldsymbol{\kappa}_{3})}{\zeta(\epsilon_k-\epsilon_{k_{1}})+\zeta_{2}(\epsilon_{k_{2}}-E_{\bar\sigma})+\zeta_{3}(\epsilon_{k_{3}}-E_{\sigma})+{\rm i}  0^+}\Big|_{\zeta=+1}\\
=&\;\sum_{\boldsymbol{\kappa}_{3}}\frac{-{\rm i} \zeta_{2}\hbar\, {\rm v}(\boldsymbol{\kappa}_{3})\delta_{\zeta_{3},-\zeta_{2}}}{\epsilon_{k_{3}}-\epsilon_{k_{2}}-\zeta_{2}(\epsilon_k-\epsilon_{k_{1}})-{\rm i}\zeta_{2}  0^+}\delta_{\nu'\nu}+\sum_{\boldsymbol{\kappa}_{3}}\frac{\nu'\nu\;{\rm i} \zeta_{3}\hbar\,  {\rm v}_{-\nu}(\boldsymbol{\kappa}_{3})}{\epsilon_{k_{3}}-E_\sigma+\zeta_{3}(\epsilon_k-\epsilon_{k_{1}})+\zeta_{3}\zeta_{2}(\epsilon_{k_{2}}-E_{\bar\sigma})+{\rm i} \zeta_{3} 0^+}\\
=&\;\zeta_{2}\frac{\rm i}{\hbar}\sum_{\alpha}\varrho_{\alpha}|{\rm t}_{\alpha}|^2 \int d\epsilon_{3}\; \frac{\delta_{\zeta_{3},-\zeta_{2}}}{\epsilon_{3}-\mathcal{E}-{\rm i}\zeta_{2} 0^+}\delta_{\nu'\nu}-\nu'\nu\frac{\rm i}{\hbar}\sum_{\zeta_{3}}\zeta_{3} \sum_{\alpha}\frac{\Gamma_{\alpha}}{2\pi}\int_{-W}^{W} d\epsilon_{3}\;  \frac{f^{\alpha}_{-\nu}(\epsilon_{3})}{\epsilon_{3}-\mathcal{E}_{\sigma,\zeta_3}^{(4)}+{\rm i}\zeta_{3} 0^+}\\
=&\;-\frac{\Gamma}{2\hbar}\delta_{\nu'\nu}-\nu'\nu\sum_{\alpha}\frac{\Gamma_{\alpha}}{2\hbar}\left\{ f^{\alpha}_{-\nu}(\mathcal{E}_{\sigma,+}^{(4)})+ f^{\alpha}_{-\nu}(\mathcal{E}_{\sigma,-}^{(4)})-\frac{{\rm i}\nu}{\pi} \Bigg[{\rm Re}\psi\left(\frac{1}{2}+{\rm i}\frac{\mathcal{E}_{\sigma,+}^{(4)}-\mu_{\alpha}}{2\pi k_{\rm B} T}\right)-{\rm Re}\psi\left(\frac{1}{2}+{\rm i}\frac{\mathcal{E}_{\sigma,-}^{(4)}-\mu_{\alpha}}{2\pi k_{\rm B} T}\right)\Bigg]\right\}\\
\equiv &\;-\frac{\Gamma}{2\hbar}\delta_{\nu'\nu} - \nu'\nu\frac{\rm i}{\hbar}\Sigma_{4,\sigma\nu}^{(\sigma\bar\sigma)}\;,
\end{aligned}\end{equation}
where $E_\sigma=\epsilon_\sigma+U/2$ and ${\rm v}_{x}(\kappa):=-(|{\rm t}_{\alpha}(\epsilon_k)|^2/\hbar^2) f^\alpha_{x}(\epsilon)$ (note the product $\nu'\nu$ given by the overlap of three fermion lines with spin $\sigma$). Here we used the property~\eqref{property_Fermi} to define the fourth-tier self-energy
\begin{equation}\label{b4s}
\Sigma_{4,\sigma\nu}^{(\sigma\bar\sigma)}:=-\nu\sum_{\alpha}\frac{\Gamma_{\alpha}}{2\pi} \Bigg[\psi\left(\frac{1}{2}+{\rm i}\frac{\mathcal{E}_{\sigma,+}^{(4)}-\mu_{\alpha}}{2\pi k_{\rm B} T}\right)-\psi^*\left(\frac{1}{2}+{\rm i}\frac{\mathcal{E}_{\sigma,-}^{(4)}-\mu_{\alpha}}{2\pi k_{\rm B} T}\right)\Bigg]-{\rm i}\frac{\Gamma}{2}\;,
\end{equation}
where
$$
\mathcal{E}_{\sigma,\pm}^{(4)}:=\epsilon_\sigma\pm\zeta(\epsilon_1-\epsilon)\pm\zeta_{2}(\epsilon_{\bar\sigma}-\epsilon_{2})+\delta_{\zeta_2,\pm1}U\;.$$

Analogously,
\begin{equation}\begin{aligned}\label{B4s_b}
{\rm B}^{\sigma(\sigma\sigma)}_{4,\eta'\eta}=&
\begin{gathered}
\resizebox{!}{1.6cm}{
\begin{tikzpicture}[] 
\draw[blue,line width=0.5mm] (0.5,2) arc (120:60:3cm  and 1.5cm) node[right] {\LARGE{$\sigma\kappa$}}; 
\draw[blue,line width=0.5mm] (0.5,1.6) arc (120:60:3cm  and 2.cm)  node[right] {\LARGE{$\sigma\boldsymbol{\kappa}_{1}$}};
\draw[blue,line width=0.5mm] (0.5,1.2) arc (120:60:3cm  and 2.5cm)  node[right] {\LARGE{$\sigma\boldsymbol{\kappa}_{2}$}};
\draw[blue,line width=0.5mm] (1,0) arc (180:0:1.cm  and 1.2cm); 
\draw[blue,dashed,line width=0.5mm]  (0.5,0) -- (1,0); 
\draw[blue,line width=0.5mm] (1,0) -- (3,0); 
\draw[blue,dashed,line width=0.5mm] (3,0) -- (3.5,0); 
\draw[red,line width=0.5mm] (0.5,-0.5) node[left] {\Large{$\eta$}} -- node[right] {\Large{$\qquad\quad\eta'$}} (3.5,-0.5); 
\filldraw[blue] 
(1,0) circle (3pt);
\end{tikzpicture}
 }
\end{gathered}
+
\begin{gathered}
\resizebox{!}{1.6cm}{
\begin{tikzpicture}[] 
\draw[blue,line width=0.5mm] (0.5,2) arc (120:60:3cm  and 1.5cm) node[right] {\LARGE{$\sigma\kappa$}};
\draw[blue,line width=0.5mm] (0.5,1.6) arc (120:60:3cm  and 2.cm) node[right] {\LARGE{$\sigma\boldsymbol{\kappa}_{1}$}};
\draw[blue,line width=0.5mm] (0.5,1.2) arc (120:60:3cm  and 2.5cm) node[right] {\LARGE{$\sigma\boldsymbol{\kappa}_{2}$}};
\draw[red,line width=0.5mm]  (1,0) arc (180:0:1.cm  and 1.2cm); 
\draw[blue,dashed,line width=0.5mm]  (0.5,0) -- (3.5,0); 
\draw[red,line width=0.5mm] (0.5,-0.5) node[left] {\Large{$\eta$}} -- (1,-0.5); 
\draw[red,line width=0.5mm] (1,-0.5) -- (1,0); 
\draw[red,dashed,line width=0.5mm] (1,-0.5) -- (3,-0.5); 
\draw[red,line width=0.5mm] (3,-0.5) -- node[right] {\Large{$\;\eta'$}}  (3.5,-0.5); 
\draw[red,line width=0.5mm] (3,-0.5) -- (3,-0); 
\filldraw[red](1,-0.5) circle (3pt); 
\end{tikzpicture}
 }
\end{gathered}
=\sum_{\nu}\langle {\rm h}^{\sigma(\sigma\sigma\sigma)}_{4}{\rm v}_{\nu}\rangle\delta_{\eta'\eta}+\langle {\rm h}^{\sigma(\sigma\sigma\bar\sigma)}_{4}{\rm v}_{-\eta}\rangle\\
=&\;\sum_{\boldsymbol{\kappa}_{3}}\frac{{\rm i} \hbar\,\sum_{\nu}{\rm v}_{\nu}(\boldsymbol{\kappa}_{3})\delta_{\zeta_{3},-\zeta_{2}}}{\zeta(\epsilon_k-\epsilon_{k_{1}})+\zeta_{2}(\epsilon_{k_{2}}-\epsilon_{k_{3}})+{\rm i}  0^+}\delta_{\eta'\eta}+\sum_{\boldsymbol{\kappa}_{3}}\frac{{\rm i} \hbar\,  {\rm v}_{-\eta}(\boldsymbol{\kappa}_{3})}{\zeta(\epsilon_k-\epsilon_{k_{1}})+\zeta_{2}(\epsilon_{k_{2}}-E_\sigma)+\zeta_{3}(\epsilon_{k_{3}}-E_{\bar\sigma})+{\rm i}  0^+}\\
=&\;\sum_{\boldsymbol{\kappa}_{3}}\frac{-{\rm i} \zeta_{2}\hbar\,  {\rm v}(\boldsymbol{\kappa}_{3})\delta_{\zeta_{3},-\zeta_{2}}}{\epsilon_{k_{3}}-\epsilon_{k_{2}}-\zeta_{2}\zeta(\epsilon_k-\epsilon_{k_{1}})-{\rm i}\zeta_{2}  0^+}\delta_{\eta'\eta}+\sum_{\boldsymbol{\kappa}_{3}}\frac{{\rm i} \zeta_{3}\hbar\,  {\rm v}_{-\eta}(\boldsymbol{\kappa}_{3})}{\epsilon_{k_{3}}-E_{\bar\sigma}+\zeta_{3}\zeta(\epsilon_k-\epsilon_{k_{1}})+\zeta_{3}\zeta_{2}(\epsilon_{k_{2}}-E_\sigma)+{\rm i} \zeta_{3} 0^+}\\
=&\;\zeta_{2}\frac{\rm i}{\hbar}\sum_{\alpha}\varrho_{\alpha}|{\rm t}_{\alpha}|^2 \int d\epsilon_{3}\; \frac{\delta_{\zeta_{3},-\zeta_{2}}}{\epsilon_{3}-\mathcal{E}-{\rm i}\zeta_{2} 0^+}\delta_{\eta'\eta}-\frac{\rm i}{\hbar}\sum_{\zeta_{3}}\zeta_{3} \sum_{\alpha}\frac{\Gamma_{\alpha}}{2\pi}\int_{-W}^{W} d\epsilon_{3}\;  \frac{f^{\alpha}_{-\eta}(\epsilon_{3})}{\epsilon_{3}-\mathcal{E}_{\bar\sigma,\zeta_3}+{\rm i}\zeta_{3} 0^+}\\
=&\;-\frac{\Gamma}{2\hbar}\delta_{\eta'\eta}-\sum_{\alpha}\frac{\Gamma_{\alpha}}{2\hbar}\left\{ f^{\alpha}_{-\eta}(\mathcal{E}_{\bar\sigma,+})+ f^{\alpha}_{-\eta}(\mathcal{E}_{\bar\sigma,-})-\frac{{\rm i}\eta}{\pi} \Bigg[{\rm Re}\psi\left(\frac{1}{2}+{\rm i}\frac{\mathcal{E}_{\bar\sigma,+}-\mu_{\alpha}}{2\pi k_{\rm B} T}\right)-{\rm Re}\psi\left(\frac{1}{2}+{\rm i}\frac{\mathcal{E}_{\bar\sigma,-}-\mu_{\alpha}}{2\pi k_{\rm B} T}\right)\Bigg]\right\}\\
\equiv &-\frac{\Gamma}{2\hbar}\delta_{\eta'\eta} - \frac{\rm i}{\hbar}\Sigma_{4,\sigma\eta}^{(\sigma\sigma)}\;,
\end{aligned}\end{equation}
where $\Sigma_{4,\sigma\eta}^{(\sigma\sigma)}$ is defined as in Eq.~\eqref{b4s}, with $\nu\rightarrow \eta$ and $\sigma\rightarrow \bar\sigma$. Note that the 4th-tier self-energies $\Sigma_{4,\sigma}$ are formally identical to the 2nd-tier NCA2 self-energies, see Eq.~\eqref{relationB-Sigma_gDSO}.

\section{Dressing the bubble ${\rm B}^{\sigma(\sigma)}$ in the NCA4}
\label{dressingB+}

The dressed bubble $\tilde{\mathbf{B}}^{\sigma(\sigma)}$  is obtained by contracting the dressed propagator 
\begin{equation}\begin{aligned}\label{h2_dressed_bare}
\tilde{\mathbf{h}}_2^{\sigma(\sigma)}=\left[[\mathbf{1}{\rm h}_2^{\sigma(\sigma)}]^{-1}-\tilde{\mathbf{B}}_3^{\sigma(\sigma)}\right]^{-1}\;,\qquad{\rm where}\quad
{\rm h}_2^{\sigma(\sigma)}={\rm i} \hbar\frac{ \delta_{\zeta_{1},-\zeta}}{\zeta(\epsilon_k-\epsilon_{k_{1}})+{\rm i}  0^+}\;,
\end{aligned}\end{equation}
see Eq.~\eqref{tildeh2ss}, according to
\begin{equation}\begin{aligned}\label{B_sigma_sigma}
\tilde{\rm B}_{\eta'\eta}^{\sigma(\sigma)}=\begin{gathered}
\resizebox{!}{1.6cm}{
\begin{tikzpicture}[] 
\draw[blue,line width=0.5mm] (0.5,1.3) arc (120:60:3cm  and 2.5cm) ;
\draw[blue,line width=0.5mm] (1,0) arc (180:0:1.cm  and 1.2cm); 
\draw[blue,dashed,line width=0.5mm]  (0.5,0) -- (1,0); 
\draw[blue,line width=0.5mm] (1,0) -- (3,0); 
\draw[blue,dashed,line width=0.5mm] (3,0) -- (3.5,0); 
\draw[red,line width=0.5mm] (0.5,-0.5) node[left] {\Large{$\eta$}} -- node[right] {\Large{$\qquad\quad\eta'$}} (3.5,-0.5); 
\filldraw[blue] 
(1,0) circle (3pt);
\fill[white] (2.6,-0.8) rectangle (1.4,.4);
\draw [] (2.6,-0.8) rectangle (1.4,.4);
\draw[](2.,-0.2) node[] {\LARGE{$\tilde{\mathbf{h}}^{\sigma(\sigma)}_{2}$}}; 
 \end{tikzpicture}
 }
\end{gathered}
=&\;\sum_{\nu}\langle\sum_{\nu'}[\tilde{\mathbf{h}}^{\sigma(\sigma)}_{2}]_{\boldsymbol{\eta}'\boldsymbol{\eta}}{\rm v}_{\nu}\rangle\\
=&\;\sum_{\nu}\langle\sum_{\nu'}\tilde{\rm h }^{\sigma(\sigma)}_{2,\eta'\eta}(\nu',\nu)(\delta_{\nu,-1}{\rm v}+\nu{\rm v}_{+})\rangle\\
=&\;\langle\sum_{\nu'}\tilde{\rm h}^{\sigma(\sigma)}_{2,\eta'\eta}(\nu',-1){\rm v}\rangle+\sum_{\nu}\langle\sum_{\nu'}\nu\tilde{\rm h}^{\sigma(\sigma)}_{2,\eta'\eta}(\nu',\nu){\rm v}_{+}\rangle\\
=&\;-\frac{\Gamma}{2\hbar}\delta_{\eta'\eta}+\sum_{\nu}\langle\sum_{\nu'}\nu\tilde{\rm h}^{\sigma(\sigma)}_{2,\eta'\eta}(\nu',\nu){\rm v}_{+}\rangle\\
\equiv&-\frac{\Gamma}{2\hbar}\delta_{\eta'\eta}+\langle{\rm K}_{\eta'\eta}^{\sigma(\sigma)}{\rm v}_{+}\rangle
\;,
\end{aligned}\end{equation}
where we have introduced the $2\times 2$ matrix $\mathbf{K}^{\sigma(\sigma)}$ of elements ${\rm K}_{\eta'\eta}^{\sigma(\sigma)}:=\sum_{\nu}\sum_{\nu'}\nu\tilde{\rm h}^{\sigma(\sigma)}_{2,\eta'\eta}(\nu',\nu)$ and where the vertex reads
\begin{equation}\label{v_appendix}
{\rm v}_{\nu}=-\frac{|{\rm t}_{\alpha}(\epsilon_k)|^2}{\hbar^2} f^\alpha_{\nu}(\epsilon_k)\;.
\end{equation}
In Eq.~\eqref{B_sigma_sigma}, we have used the splitting of the Fermi function $f_\nu(x)=\delta_{\nu,-1}+\nu f_+(x) $ in the vertex. Also, we assumed that $\tilde{\rm h}^{\sigma(\sigma)}_{2,\eta'\eta}(\nu',-1)$ has no poles in the upper complex plane, so that the contraction with the temperature-independent vertex ${\rm v}$ simply yields $-\Gamma/2\hbar$, as in the non-dressed case, see Eqs.~\eqref{B+bare} and~\eqref{B_SIAM_WBL}.
The matrix $\tilde{\mathbf{h}}^{\sigma(\sigma)}_{2}$ has a $4\times 4$ structure in the collective sojourn index $\boldsymbol{\eta}=(\nu,\eta)$ induced by the third-tier bubble $\tilde{\mathbf{B}}_3^{\sigma(\sigma)}$.\\
\indent  In order to avoid the inversion of a four-dimensional matrix to evaluate $\tilde{\mathbf{h}}_2$, 
we exploit the specific form of the third-tier bubbles forming $\tilde{\mathbf{B}}_3^{\sigma(\sigma)}$ in the NCA4 and use a two-step procedure.
We start by writing the matrix element of the $4\times 4$ third-tier bubble as
\begin{equation}
\begin{aligned}
\label{B3s_appendix}
\tilde{\rm B}_{3,\boldsymbol{\eta}'\boldsymbol{\eta}}^{\sigma(\sigma)}= 
\begin{gathered}
\resizebox{!}{1.5cm}{
\begin{tikzpicture}[] 
\draw[blue,line width=0.5mm] (0.5,1.8) arc (120:60:3cm  and 2.cm) ;
\draw[blue,line width=0.5mm] (0.5,1.1) arc (120:60:3cm  and 5cm) ;
\draw[red,line width=0.5mm]  (1,0) arc (180:0:1.cm  and 1.2cm); 
\draw[blue,line width=0.5mm] (0.5,0.3) node[left] {\Large{$\nu$}}  -- (3.5,0.3) node[right] {\Large{$\;\nu'$}} ; 
\draw[red,line width=0.5mm] (0.5,-0.5) node[left] {\Large{$\eta$}} -- (1,-0.5); 
\draw[red,line width=0.5mm] (1,-0.5) -- (1,0); 
\draw[red,dashed,line width=0.5mm] (1,-0.5) -- (3,-0.5); 
\draw[red,line width=0.5mm] (3,-0.5) -- node[right] {\Large{$\;\eta'$}}  (3.5,-0.5); 
\draw[red,line width=0.5mm] (3,-0.5) -- (3,-0); 
\fill[white] (2.7,-0.8) rectangle (1.3,.6);
\draw [] (2.7,-0.8) rectangle (1.3,.6);
\filldraw[red] 
(1,-0.5) circle (3pt);
\draw[](2.,-0.2) node[] {\Large{${\rm B}^{\sigma(\sigma\bar\sigma)}_{4,\nu'\nu}$}}; 
 \end{tikzpicture}
 }
\end{gathered}
+
\begin{gathered}
\resizebox{!}{1.5cm}{
\begin{tikzpicture}[] 
\draw[blue,line width=0.5mm] (0.5,1.8) arc (120:60:3cm  and 2.cm) ;
\draw[blue,line width=0.5mm] (0.5,1.1) arc (120:60:3cm  and 5cm) ;
\draw[blue,line width=0.5mm] (1,0.3) arc (180:0:1.cm  and 1.cm); 
\draw[blue,line width=0.5mm]  (0.5,0.3)node[left] {\Large{$\nu$}}-- (1,0.3); 
\draw[blue,dashed,line width=0.5mm] (1,0.3) -- (3,0.3); 
\draw[blue,line width=0.5mm] (3,0.3) -- (3.5,0.3) node[right] {\Large{$\;\nu' $}} ; 
\draw[red,line width=0.5mm] (0.5,-0.5) node[left] {\Large{$\eta$}} -- node[right] {\Large{$\qquad\quad\eta'$}} (3.5,-0.5); 
\filldraw[blue] 
(1,0.3) circle (3pt);
\fill[white] (2.7,-0.8) rectangle (1.3,.6);
\draw [] (2.7,-0.8) rectangle (1.3,.6);
\draw[](2.,-0.2) node[] {\Large{${\rm B}^{\sigma(\sigma\sigma)}_{4,\eta'\eta}$}}; 
 \end{tikzpicture}
 }
\end{gathered}
 =\tilde{\rm B}_{3,\nu'\nu}^{A}(\eta) +\nu'\nu \tilde{\rm B}_{3,\eta'\eta}^{B}(\nu) \;.
\end{aligned}
\end{equation}
Note that $\tilde{\rm B}_{3,\eta'\eta}^{B}(\nu)$ is independent of $\nu'$ and $\tilde{\rm B}_{3,\nu'\nu}^{A}(\eta)$ is independent of $\eta'$, as shown in the above diagrams.
It is convenient to define separately the first and second contribution to $\tilde{\rm B}_{3,\boldsymbol{\eta}'\boldsymbol{\eta}}^{\sigma(\sigma)}$, in Eq.~\eqref{B3s_appendix} as two $2\times2$ matrices, denoted with $A$ and $B$ respectively. Let the first, $\tilde{\mathbf{B}}_3^{A}(\eta)$, be a matrix in the indexes $\nu'\nu$ with explicit dependence on the vertex index $\eta$. Specifically, the matrix elements of $\tilde{\mathbf{B}}_3^{A}(\eta)$ are given by the contraction of a  propagator dressed by 4th-tier bubbles
\begin{equation}
\begin{aligned}\label{B3A_contraction}
\tilde{\rm B}_{3,\nu'\nu}^{A}(\eta):=\langle \tilde{\rm h}^{\sigma(\sigma\bar\sigma)}_{3,\nu'\nu}{\rm v}_{-\eta}\rangle\;,
\end{aligned}
\end{equation}
where 
\begin{equation}\begin{aligned}\label{}
\tilde{\rm h}^{\sigma(\sigma\bar\sigma)}_{3,\nu\nu}=&\; \frac{({\rm h}^{\sigma(\sigma\bar\sigma)}_{3,\bar\nu\bar\nu})^{-1}-{\rm B}^{\sigma(\sigma\bar\sigma)}_{4,\bar\nu\bar\nu}}{D^{\sigma(\sigma\bar\sigma)}}\;,\qquad
\tilde{\rm h}^{\sigma(\sigma\bar\sigma)}_{3,\bar\nu\nu}=\frac{{\rm B}^{\sigma(\sigma\bar\sigma)}_{4,\bar\nu\nu}}{D^{\sigma(\sigma\bar\sigma)}}\;,
\end{aligned}\end{equation}
with ${\rm h}^{\sigma(\sigma\bar\sigma)}_{3,\nu'\nu}$ given in Eq.~\eqref{h3ss} and with  $D^{\sigma(\sigma\bar\sigma)}=[({\rm h}^{\sigma(\sigma\bar\sigma)}_{3,\nu\nu})^{-1}-{\rm B}^{\sigma(\sigma\bar\sigma)}_{4,\nu\nu}][({\rm h}^{\sigma(\sigma\bar\sigma)}_{3,\bar\nu\bar\nu})^{-1}-{\rm B}^{\sigma(\sigma\bar\sigma)}_{4,\bar\nu\bar\nu}]-{\rm B}^{\sigma(\sigma\bar\sigma)}_{4,\nu\bar\nu}{\rm B}^{\sigma(\sigma\bar\sigma)}_{4,\bar\nu\nu}$ a function of the energies, independent of the indexes $\nu'\nu$. As shown in Appendix~\ref{gDSO4_4th_tier}, the bare 4th-tier bubbles  ${\rm B}^{\sigma(\sigma\bar\sigma)}_{4,\nu'\nu}$, with NCA2 structure, are
\begin{equation}\begin{aligned}\label{F}
{\rm B}^{\sigma(\sigma\bar\sigma)}_{4,\nu'\nu}=-\frac{\Gamma}{2\hbar}\delta_{\nu'\nu}+\nu'\nu{\rm F}^{(\sigma)}_{\nu}\qquad{\rm where}\qquad {\rm F}^{(\sigma)}_{\nu}=-\frac{\rm i}{\hbar}\Sigma_{4,\sigma\nu}^{(\sigma\bar\sigma)}\;.
\end{aligned}\end{equation}
The symmetries of the fourth-tier bubble imply
\begin{equation}\begin{aligned}\label{h3A}
\tilde{\rm h}^{\sigma(\sigma\bar\sigma)}_{3,\nu\nu}=&\; \frac{({\rm h}^{\sigma(\sigma\bar\sigma)}_{3,\bar\nu\bar\nu})^{-1}+\Gamma/2\hbar}{D^{\sigma(\sigma\bar\sigma)}}
-\frac{{\rm F}^{(\sigma)}_{\bar\nu}}{D^{\sigma(\sigma\bar\sigma)}}\\
\tilde{\rm h}^{\sigma(\sigma\bar\sigma)}_{3,\bar\nu\nu}=&\;-\frac{{\rm F}^{(\sigma)}_{\nu}}{D^{\sigma(\sigma\bar\sigma)}}\;.
\end{aligned}\end{equation}
Thus, Eq.~\eqref{B3A_contraction} yields
\begin{equation}\begin{aligned}\label{B3A_contraction_2}
\tilde{\rm B}_{3,\nu\nu}^{A}(\eta)=&
\;\Biggl<\frac{({\rm h}^{\sigma(\sigma\bar\sigma)}_{3,\bar\nu\bar\nu})^{-1}+\Gamma/2\hbar}{D^{\sigma(\sigma\bar\sigma)}}{\rm v}_{-\eta}\Biggr>-\Biggl<\frac{{\rm F}^{(\sigma)}_{\bar\nu}}{D^{\sigma(\sigma\bar\sigma)}}{\rm v}_{-\eta}\Biggr>\\
\equiv&
\;A^{\bar\sigma}(\nu,\eta)-\Biggl<\frac{{\rm F}^{(\sigma)}_{\bar\nu}}{D^{\sigma(\sigma\bar\sigma)}}{\rm v}_{-\eta}\Biggr>\\
\tilde{\rm B}_{3,\bar\nu\nu}^{A}(\eta)=&
\;-\Biggl<\frac{{\rm F}^{(\sigma)}_\nu}{D^{\sigma(\sigma\bar\sigma)}}{\rm v}_{-\eta}\Biggr>\;.
\end{aligned}\end{equation}
For later purposes, let us define the matrix given by the sum over the vertex $\mathbf{S}_3^A:=\sum_\eta\tilde{\mathbf{B}}_3^{A}(\eta)$ and its off-diagonal elements 
\begin{equation}
\label{sA}
{\rm s}_{\nu}^A:=\langle [{\rm F}^{(\sigma)}_{\nu}/D^{\sigma(\sigma\bar\sigma)}]{\rm v}\rangle\;.
\end{equation}
We can express $\mathbf{S}_3^A$ in terms of its off-diagonal elements as
\begin{equation}\begin{aligned}
{\rm S}_{3, \nu\nu}^A:=&\sum_\eta\langle \tilde{\rm h}^{\sigma(\sigma\bar\sigma)}_{3,\nu\nu}{\rm v}_{-\eta}\rangle=\langle \tilde{\rm h}^{\sigma(\sigma\bar\sigma)}_{3,\nu\nu}{\rm v}\rangle=\sum_\eta A^{\bar\sigma}(\nu,\eta)-{\rm s}_{\bar\nu}^A=-\frac{\Gamma}{\hbar}+{\rm s}_{\nu}^A\\
{\rm S}_{3, \bar\nu\nu}^A:=&\sum_\eta\langle \tilde{\rm h}^{\sigma(\sigma\bar\sigma)}_{3,\bar\nu\nu}{\rm v}_{-\eta}\rangle=\langle \tilde{\rm h}^{\sigma(\sigma\bar\sigma)}_{3,\bar\nu\nu}{\rm v}\rangle=-{\rm s}_{\nu}^A\;,
\end{aligned}\end{equation}
where we used $\sum_\eta A^{\bar\sigma}(\nu,\eta)-\sum_\nu{\rm s}_{\nu}^A=\sum_\eta\langle (\tilde{\rm h}^{\sigma(\sigma\bar\sigma)}_{3,\nu\nu} + \tilde{\rm h}^{\sigma(\sigma\bar\sigma)}_{3,\bar\nu\nu}){\rm v}_{-\eta}\rangle=-\Gamma/\hbar$, assuming that the function 
\begin{equation}\begin{aligned}\label{B3A_diagonal}
\tilde{\rm h}^{\sigma(\sigma\bar\sigma)}_{3,\nu\nu} + \tilde{\rm h}^{\sigma(\sigma\bar\sigma)}_{3,\bar\nu\nu}
=&\;\frac{1}{({\rm h}^{\sigma(\sigma\bar\sigma)}_{3,\nu\nu})^{-1}  - {\rm B}^{\sigma(\sigma\bar\sigma)}_{4,\nu\nu} -{\rm B}^{\sigma(\sigma\bar\sigma)}_{4,\bar\nu\nu}
\frac{({\rm h}^{\sigma(\sigma\bar\sigma)}_{3,\nu\nu})^{-1} -{\rm B}^{\sigma(\sigma\bar\sigma)}_{4,\nu\nu}+{\rm B}^{\sigma(\sigma\bar\sigma)}_{4,\nu\bar\nu}}{({\rm h}^{\sigma(\sigma\bar\sigma)}_{3,\bar\nu\bar\nu})^{-1}-{\rm B}^{\sigma(\sigma\bar\sigma)}_{4,\bar\nu\bar\nu}+{\rm B}^{\sigma(\sigma\bar\sigma)}_{4,\bar\nu\nu}}}\\
=&\;\frac{{\rm i}\zeta\hbar}{\epsilon-\epsilon_{k_1}+\zeta\zeta_2[\epsilon_{k_2}-
E_{\bar\sigma}(\nu)]+{\rm i}\zeta\Gamma/2 - \zeta_2\nu \Sigma_{4,\sigma\nu}^{(\sigma\bar\sigma)}\frac{U}{\epsilon-\epsilon_{k_1}+\zeta\zeta_2[\epsilon_{k_2}-
E_{\bar\sigma}(\bar\nu)]+{\rm i}\zeta 3\Gamma/2}}\;
\end{aligned}\end{equation}
is analytical in the upper complex plane.
Here, we used Eqs.~\eqref{F}-\eqref{h3A} and~\eqref{h3ss}, and $E_{\bar\sigma}(\nu)-E_{\bar\sigma}(\bar\nu)=\nu U$. Note the formal similarity with the terms of the NCA2 Green's function, Eqs.~\eqref{Gzeta_gDSO} and~\eqref{GRgDSO}.
As a matrix with indexes $\nu'\nu$, $\mathbf{S}_3^A$ reads
\begin{equation}
\begin{aligned}\label{S3A}
\mathbf{S}_3^A=&-\frac{\Gamma}{\hbar}
\begin{pmatrix}
1&0\\
0&1
\end{pmatrix}
+
\begin{pmatrix}
{\rm s}_{+}^A&-{\rm s}_{-}^A\\
-{\rm s}_{+}^A&{\rm s}_{-}^A
\end{pmatrix}\\
=&-\frac{\Gamma}{\hbar}\mathbf{1}+\mathbf{s}^A\;.
\end{aligned}
\end{equation}
Analogously, we introduce the $2\times 2$ matrix $\tilde{\mathbf{B}}_3^{B}(\nu)$ in the indexes $\eta'\eta$  with explicit dependence on the vertex index $\nu$, see the second term in Eq.~\eqref{B3s_appendix}. The matrix element of $\tilde{\mathbf{B}}_3^{B}(\nu)$ is
\begin{equation}
\begin{aligned}\label{B3B}
\tilde{\rm B}_{3,\eta'\eta}^{B}(\nu):=\langle \tilde{\rm h}^{\sigma(\sigma\sigma)}_{3,\eta'\eta}{\rm v}_{-\nu}\rangle\;,
\end{aligned}
\end{equation}
where 
\begin{equation}\begin{aligned}\label{h3B}
\tilde{\rm h}^{\sigma(\sigma\sigma)}_{3,\eta\eta}=& \frac{({\rm h}^{\sigma(\sigma\sigma)}_{3,\bar\eta\bar\eta})^{-1}-{\rm B}^{\sigma(\sigma\sigma)}_{4,\bar\eta\bar\eta}}{D^{\sigma(\sigma\sigma)}}\;,\qquad
\tilde{\rm h}^{\sigma(\sigma\sigma)}_{3,\bar\eta\eta}=\frac{{\rm B}^{\sigma(\sigma\sigma)}_{4,\bar\eta\eta}}{D^{\sigma(\sigma\sigma)}}\;.
\end{aligned}\end{equation}
The bare 4th-tier bubbles  ${\rm B}^{\sigma(\sigma\sigma)}_{4,\eta'\eta}$, with NCA2 structure, can be written as
\begin{equation}\begin{aligned}\label{FB}
{\rm B}^{\sigma(\sigma\sigma)}_{4,\eta'\eta}=-\frac{\Gamma}{2\hbar}\delta_{\eta'\eta}+{\rm F}^{(\bar\sigma)}_{\eta}\qquad{\rm where}\qquad {\rm F}^{(\bar\sigma)}_{\eta}=-\frac{\rm i}{\hbar}\Sigma_{4,\sigma\eta}^{(\sigma\sigma)}\;,
\end{aligned}\end{equation}
see Appendix~\ref{gDSO4_4th_tier}.
As above, also in this case the diagonal elements can be expressed in terms of the off-diagonal ones, which results in
\begin{equation}\begin{aligned}\label{B3B_contraction_2}
\tilde{\rm B}_{3,\eta\eta}^{B}(\nu)=&
\;\Biggl<\frac{({\rm h}^{\sigma(\sigma\sigma)}_{3,\bar\eta\bar\eta})^{-1}+\Gamma/2\hbar}{D^{\sigma(\sigma\sigma)}}{\rm v}_{-\nu}\Biggr>-\Biggl<\frac{{\rm F}^{(\bar\sigma)}_{\bar\eta}}{D^{\sigma(\sigma\sigma)}}{\rm v}_{-\nu}\Biggr>\\
\equiv&
\;A^{\sigma}(\eta,\nu)-\Biggl<\frac{{\rm F}^{(\bar\sigma)}_{\bar\eta}}{D^{\sigma(\sigma\sigma)}}{\rm v}_{-\nu}\Biggr>\\
\tilde{\rm B}_{3,\bar\eta\eta}^{B}(\nu)=&
\;\Biggl<\frac{{\rm F}^{(\bar\sigma)}_\eta}{D^{\sigma(\sigma\sigma)}}{\rm v}_{-\nu}\Biggr>\;,
\end{aligned}\end{equation}
with  $D^{\sigma(\sigma\sigma)}=[({\rm h}^{\sigma(\sigma\sigma)}_{3,\eta\eta})^{-1}-{\rm B}^{\sigma(\sigma\sigma)}_{4,\eta\eta}][({\rm h}^{\sigma(\sigma\sigma)}_{3,\bar\eta\bar\eta})^{-1}-{\rm B}^{\sigma(\sigma\sigma)}_{4,\bar\eta\bar\eta}]-{\rm B}^{\sigma(\sigma\sigma)}_{4,\eta\bar\eta}{\rm B}^{\sigma(\sigma\sigma)}_{4,\bar\eta\eta}$. 
For the matrix  $\mathbf{S}_3^B:=\sum_\nu\tilde{\mathbf{B}}_3^{B}(\nu)$ of indexes $\eta'\eta$ we get 
\begin{equation}\begin{aligned}
{\rm S}_{3, \eta\eta}^B:=&\sum_\nu\langle \tilde{\rm h}^{\sigma(\sigma\sigma)}_{3,\eta\eta}{\rm v}_{-\nu}\rangle=\langle \tilde{\rm h}^{\sigma(\sigma\sigma)}_{3,\eta\eta}{\rm v}\rangle=\sum_\eta A^{\sigma}(\eta,\nu)-{\rm s}_{\bar\eta}^B=-\frac{\Gamma}{\hbar}+{\rm s}_{\eta}^B\\
{\rm S}_{3, \bar\eta\eta}^B:=&\sum_\nu\langle \tilde{\rm h}^{\sigma(\sigma\sigma)}_{3,\bar\eta\eta}{\rm v}_{-\nu}\rangle=\langle \tilde{\rm h}^{\sigma(\sigma\sigma)}_{3,\bar\eta\eta}{\rm v}\rangle={\rm s}_{\eta}^B\;,
\end{aligned}\end{equation}
where ${\rm s}_{\eta}^B:=\langle [{\rm F}^{(\bar\sigma)}_{\eta}/D^{\sigma(\sigma\sigma)}]{\rm v}\rangle$.
Similarly as for ${\rm S}_{3, \nu'\nu}^A$, we used $\sum_\nu A^{\sigma}(\eta,\nu)-\sum_\eta{\rm s}_{\eta}^B=\sum_\nu\langle (\tilde{\rm h}^{\sigma(\sigma\sigma)}_{3,\eta\eta} - \tilde{\rm h}^{\sigma(\sigma\sigma)}_{3,\bar\eta\eta}){\rm v}_{-\nu}\rangle=-\Gamma/\hbar$, assuming that the function 
\begin{equation}\begin{aligned}\label{B3B_diagonal}
\tilde{\rm h}^{\sigma(\sigma\sigma)}_{3,\eta\eta} - \tilde{\rm h}^{\sigma(\sigma\sigma)}_{3,\bar\eta\eta}
=&\;\frac{1}{({\rm h}^{\sigma(\sigma\sigma)}_{3,\eta\eta})^{-1} - {\rm B}^{\sigma(\sigma\sigma)}_{4,\eta\eta} +{\rm B}^{\sigma(\sigma\sigma)}_{4,\bar\eta\eta}
\frac{({\rm h}^{\sigma(\sigma\sigma)}_{3,\eta\eta})^{-1} -{\rm B}^{\sigma(\sigma\sigma)}_{4,\eta\eta}-{\rm B}^{\sigma(\sigma\sigma)}_{4,\eta\bar\eta}}{({\rm h}^{\sigma(\sigma\sigma)}_{3,\bar\eta\bar\eta})^{-1}-{\rm B}^{\sigma(\sigma\sigma)}_{4,\bar\eta\bar\eta}-{\rm B}^{\sigma(\sigma\sigma)}_{4,\bar\eta\eta}}}\\
=&\;\frac{{\rm i}\zeta\hbar}{\epsilon-\epsilon_{k_1}+\zeta\zeta_2[\epsilon_{k_2}-
E_{\sigma}(\eta)]+{\rm i}\zeta\Gamma/2 - \zeta_2\eta\Sigma_{4,\sigma\eta}^{(\sigma\sigma)}\frac{U}{\epsilon-\epsilon_{k_1}+\zeta\zeta_2[\epsilon_{k_2}-
E_{\sigma}(\bar\eta)]+{\rm i}\zeta 3\Gamma/2}}\;,
\end{aligned}\end{equation}
see Eqs.~\eqref{h3B}-\eqref{FB} and~\eqref{h3ss}, is analytical in the upper complex plane. 
 In matrix form, with indexes $\eta'\eta$,
\begin{equation}
\begin{aligned}\label{S3B}
\mathbf{S}_3^B=&-\frac{\Gamma}{\hbar}
\begin{pmatrix}
1&0\\
0&1
\end{pmatrix}
+
\begin{pmatrix}
{\rm s}_{+}^B&{\rm s}_{-}^B\\
{\rm s}_{+}^B&{\rm s}_{-}^B
\end{pmatrix}\\
=&-\frac{\Gamma}{\hbar}\mathbf{1}+\mathbf{s}^B\;.
\end{aligned}
\end{equation}

\indent The dressed bubble $\tilde{\mathbf{B}}^{\sigma(\sigma)}$, with $2\times 2$ structure in $\eta'\eta$, see Eq.~\eqref{B_sigma_sigma}, is then obtained as the contraction 
\begin{equation}
\begin{aligned}\label{Bs_appendix}
\tilde{\rm B}_{\eta'\eta}^{\sigma(\sigma)}
=&\begin{gathered}
\resizebox{!}{1.7cm}{
\begin{tikzpicture}[] 
\draw[blue,thick] (0.5,1.3) arc (120:60:3cm  and 2.5cm) ;
\draw[blue,thick] (1,0) arc (180:0:1.cm  and 1.1cm); 
\draw[blue,dashed,thick]  (0.5,0) -- (1,0); 
\draw[blue,thick] (1,0) -- (3,0); 
\draw[blue,dashed,thick] (3,0) -- (3.5,0); 
\draw[red,thick] (0.5,-0.5)  node[left] {\Large{$\eta$}} -- (3.5,-0.5)  node[right] {\Large{$\eta'$}}; 
\filldraw[blue] 
(1,0) circle (2pt); 
\fill[white] (2.6,-0.8) rectangle (1.4,.4);
\draw [] (2.6,-0.8) rectangle (1.4,.4);
\draw[](2.,-0.2) node[] {\LARGE{$\tilde{\mathbf{h}}^{A}_2$}}; 
\end{tikzpicture}
}
\end{gathered}
+
\begin{gathered}
\resizebox{!}{1.7cm}{
\begin{tikzpicture}[] 
\draw[blue,thick] (0.5,1.3) arc (120:60:6.5cm  and 2.5cm) ;
\draw[blue,thick] (1,0) arc (180:0:2.75cm  and 1.3cm); 
\draw[blue,thick] (3,0) arc (180:0:.75cm  and 1.1cm); 
\draw[blue,dashed,thick]  (0.5,0) -- (1,0); 
\draw[blue,thick] (1,0) -- (3,0); 
\draw[blue,dashed,thick] (3,0) -- (4.5,0); 
\draw[blue,thick] (4.5,0) -- (6.5,0); 
\draw[blue,dashed,thick]  (6.5,0) -- (7,0); 
\draw[red,thick] (0.5,-0.5) node[left] {\Large{$\eta$}} -- (7,-0.5) node[right] {\Large{$\eta'$}}; 
\filldraw[blue] (1,0) circle (2pt);
\filldraw[blue] (3,0) circle (2pt); 
\fill[white] (2.6,-0.8) rectangle (1.4,.4);
\draw [] (2.6,-0.8) rectangle (1.4,.4);
\draw[](2.,-0.2) node[] {\LARGE{$\tilde{\mathbf{h}}_2^{A}$}}; 
\fill[white] (6.1,-0.8) rectangle (4.9,.4);
\draw [] (6.1,-0.8) rectangle (4.9,.4);
\draw[](5.5,-0.2) node[] {\LARGE{$\tilde{\mathbf{h}}_2^{A}$}}; 
\fill[white] (3.3,-0.8) rectangle (4.2,.4);
\draw [] (3.3,-0.8) rectangle (4.2,.4);
\draw[](3.8,-0.2) node[] {\LARGE{$\tilde{\mathbf{B}}_3^B$}}; 
 \end{tikzpicture}
 }
\end{gathered}
+\dots\;,
\end{aligned}\end{equation}
where $\tilde{\mathbf{h}}_{2}^{A}(\eta)$ is the  propagator dressed solely by the $\tilde{\mathbf{B}}_3^A$ bubbles.

First we evaluate $\tilde{\mathbf{h}}_{2}^{A}(\eta)$, considering its matrix structure in $\nu$ and dependences on $\eta',\eta$ explicit. It is given by the series
\begin{equation}
\begin{aligned}
\label{tildeh2a}
\tilde{\mathbf{h}}_{2}^{A}(\eta',\eta)=&\;
\begin{gathered}
\resizebox{!}{1.55cm}{
\begin{tikzpicture}[] 
\fill[white] (1.3,-0.8) rectangle (2.2,.4);
\draw[blue,thick] (0.5,1.3) arc (120:60:3cm  and 2.5cm) ;
\draw[blue,thick] (0.5,0.6) arc (150:30:1.7cm  and 1.3cm); 
\draw[blue,thick]  (0.5,0) -- (3.5,0); 
\draw[red,thick] (0.5,-0.5) node[left] {\Large{$\eta$}} -- (3.5,-0.5); 
\end{tikzpicture}
 }
\end{gathered}
+
\begin{gathered}
\resizebox{!}{1.55cm}{
\begin{tikzpicture}[] 
\draw[blue,thick] (2,1.3) arc (120:60:3.5cm  and 2.5cm) ;
\draw[blue,thick] (2,0.6) arc (150:30:2cm  and 1.3cm); 
\draw[red,thick] (3,0) arc (180:0:.75cm  and .8cm); 
\draw[blue,thick]  (2,0) -- (5.5,0); 
\draw[red,thick] (2,-0.5) node[left] {\Large{$\eta$}} -- (3,-0.5);
\draw[red,thick,dashed] (3,-0.5) -- (4.5,-0.5);
\draw[red,thick] (3,-0.5) -- (3,0);
\draw[red,thick] (4.5,-0.5) -- (4.5,0); 
\draw[red,thick] (4.5,-0.5) -- (5.5,-0.5) node[right] {\Large{$\eta'$}}; 
\filldraw[red] (3,-0.5) circle (2pt); 
\fill[white] (3.3,-0.8) rectangle (4.2,.4);
\draw [] (3.3,-0.8) rectangle (4.2,.4);
\draw[](3.8,-0.2) node[] {\Large{${\rm B}_4$}};
\end{tikzpicture}
 }
\end{gathered}
+
\begin{gathered}
\resizebox{!}{1.55cm}{
\begin{tikzpicture}[] 
\draw[blue,thick] (0.5,1.3) arc (120:60:6.5cm  and 2.5cm) ;
\draw[blue,thick] (0.5,0.6) arc (150:30:3.7cm  and 1.3cm); 
\draw[red,thick] (1.5,0) arc (180:0:.75cm  and .8cm); 
\draw[red,thick] (4.5,0) arc (180:0:.75cm  and .8cm); 
\draw[blue,thick]  (0.5,0) -- (7,0); 
\draw[red,thick] (0.5,-0.5) node[left] {\Large{$\eta$}} -- (1.5,-0.5);
\draw[red,thick,dashed] (1.5,-0.5) -- (3,-0.5);
\draw[red,thick] (3,-0.5) -- node[below] {\Large{$\eta''$}} (4.5,-0.5);
\draw[red,dashed,thick] (4.5,-0.5) -- (6,-0.5);
\draw[red,thick] (6,-0.5) -- (7,-0.5) node[right] {\Large{$\eta'$}}; 
\draw[red,thick] (1.5,-0.5) -- (1.5,0);
\draw[red,thick] (3,-0.5) -- (3,0); 
\filldraw[red] (1.5,-0.5) circle (2pt); 
\fill[white] (1.8,-0.8) rectangle (2.7,.4);
\draw [] (1.8,-0.8) rectangle (2.7,.4);
\draw[](2.2,-0.2) node[] {\Large{${\rm B}_4$}};
\draw[red,thick] (6,-0.5) -- (6,0);
\draw[red,thick] (4.5,-0.5) -- (4.5,0); 
\filldraw[red] (4.5,-0.5) circle (2pt); 
\fill[white] (4.8,-0.8) rectangle (5.7,.4);
\draw [] (4.8,-0.8) rectangle (5.7,.4);
\draw[](5.2,-0.2) node[] {\Large{${\rm B}_4$}};
\end{tikzpicture}
 }
\end{gathered}+\dots\\
=&\;
\mathbf{1}{\rm h}^{\sigma(\sigma)}_{2}\delta_{\eta',\eta}+{\rm h}^{\sigma(\sigma)}_{2}\tilde{\mathbf{B}}_3^{A}(\eta){\rm h}^{\sigma(\sigma)}_{2}+{\rm h}^{\sigma(\sigma)}_{2}\sum_{\eta''}\tilde{\mathbf{B}}_3^{A}(\eta''){\rm h}^{\sigma(\sigma)}_{2}\tilde{\mathbf{B}}_3^{A}(\eta){\rm h}^{\sigma(\sigma)}_{2}+\dots\\
=&\;\mathbf{1}{\rm h}^{\sigma(\sigma)}_{2}\delta_{\eta',\eta}+{\rm h}^{\sigma(\sigma)}_{2}\tilde{\mathbf{B}}_3^{A}(\eta){\rm h}^{\sigma(\sigma)}_{2}+{\rm h}^{\sigma(\sigma)}_{2}\mathbf{S}^A_3{\rm h}^{\sigma(\sigma)}_{2}\tilde{\mathbf{B}}_3^{A}(\eta){\rm h}^{\sigma(\sigma)}_{2}+\dots\\
=&\;\mathbf{1}{\rm h}^{\sigma(\sigma)}_{2}\delta_{\eta',\eta}+\frac{{\rm h}^{\sigma(\sigma)}_{2}}{\mathbf{1}[{\rm h}^{\sigma(\sigma)}_{2}]^{-1}-\mathbf{S}^A_3}\tilde{\mathbf{B}}_3^{A}(\eta)\\
\equiv&\;\mathbf{1}{\rm h}^{\sigma(\sigma)}_{2}\delta_{\eta',\eta}+\tilde{\mathbf{C}}^A\tilde{\mathbf{B}}_3^{A}(\eta)
\;,
\end{aligned}
\end{equation}
where $\mathbf{S}^A_3$ has been introduced above and
\begin{equation}
\begin{aligned}
\label{SH}
\tilde{\mathbf{C}}^A:=\frac{{\rm h}^{\sigma(\sigma)}_{2}}{\mathbf{1}[{\rm h}^{\sigma(\sigma)}_{2}]^{-1}-\mathbf{S}^A_3}\;.
\end{aligned}
\end{equation}
Inversion of the matrix in the denominator of $\tilde{\mathbf{C}}^A$ then yields
\begin{equation}
\begin{aligned}\label{CA}
\tilde{\mathbf{C}}^A=&\;\frac{{\rm h}^{\sigma(\sigma)}_{2}}{([{\rm h}^{\sigma(\sigma)}_{2}]^{-1}+\Gamma/\hbar)([{\rm h}^{\sigma(\sigma)}_{2}]^{-1}+\Gamma/\hbar-{\rm s}_{+}^A-{\rm s}_{-}^A)}
\begin{pmatrix}
[{\rm h}^{\sigma(\sigma)}_{2}]^{-1}+\Gamma/\hbar - {\rm s}_{-}^A&\;-{\rm s}_{-}^A\\
-{\rm s}_{+}^A&\;[{\rm h}^{\sigma(\sigma)}_{2}]^{-1}+\Gamma/\hbar-{\rm s}_{+}^A
\end{pmatrix}\\
=&\;{\rm h}^{\sigma(\sigma)}_{2}\tilde{\rm h}^{\sigma(\sigma)}_{2}\left[\mathbf{1}+\tilde{\rm h}^{\sigma(\sigma)}_{A}
\begin{pmatrix}
{\rm s}_{+}^A&-{\rm s}_{-}^A\\
-{\rm s}_{+}^A&{\rm s}_{-}^A
\end{pmatrix}\right]
\;,
\end{aligned}
\end{equation}
where
\begin{equation}
\label{tilde_s_A}
\tilde{\rm h}^{\sigma(\sigma)}_{2}:=\frac{1}{[{\rm h}^{\sigma(\sigma)}_{2}]^{-1}+\Gamma/\hbar}\;, \qquad
\tilde{\rm h}^{\sigma(\sigma)}_{A}:=\frac{1}{[{\rm h}^{\sigma(\sigma)}_{2}]^{-1}+\Gamma/\hbar - {\rm s}^A}\;, \qquad{\rm and}\qquad{\rm s}^A:=\sum_{\nu}{\rm s}_{\nu}^A=-\frac{\Gamma}{\hbar}\langle [D^{\sigma(\sigma\bar\sigma)}]^{-1}{\rm v}\rangle\;.
\end{equation}
Now that we have a closed form for $\tilde{\mathbf{h}}_{2}^{A}(\eta',\eta)$, let us switch to representing every quantity as a matrix in $\eta'\eta$ while making explicit the dependencies on $\nu'$ and $\nu$. This is natural for $\tilde{\mathbf{B}}_3^B(\nu)$. On the other hand, as an intermediate step, we express $\tilde{\mathbf{h}}_{2}^{A}(\eta',\eta)$, Eq.~\eqref{tildeh2a}, as the matrix element

\begin{equation}
\begin{aligned}
\tilde{{\rm h}}_{2,\eta'\eta}^{A}(\nu',\nu)={\rm h}^{\sigma(\sigma)}_{2}\delta_{\eta',\eta}\delta_{\nu',\nu}+\sum_{\nu''}\tilde{\rm C}^A(\nu',\nu'')\tilde{\rm B}_{3,\eta'\eta}^{A}(\nu'',\nu)\;,
\end{aligned}
\end{equation}
where the product $\nu''\nu$ is associated to the overlap of three spin-$\sigma$ fermion lines in $\tilde{\rm B}_{3,\eta\eta}^{A}(\nu',\nu)$, see Eq.~\eqref{B3s_appendix}. Due to the sole dependency on the first-sojourn index $\eta$, the following property holds
\begin{equation}
\begin{aligned}\label{propB3A}
\tilde{\rm B}_{3,\eta\eta}^{A}(\nu',\nu)=\tilde{\rm B}_{3,\bar\eta\eta}^{A}(\nu',\nu)\;.
\end{aligned}
\end{equation}
Moreover, from Eqs.~\eqref{B3A_contraction_2} and~\eqref{B3B_contraction_2}, $\tilde{\rm B}_{3,\eta'\eta}^{A}(\nu',\nu)$ and $\tilde{\rm B}_{3,\eta'\eta}^{B}(\nu)$ have the symmetries
\begin{equation}
\begin{aligned}\label{propB3B}
\tilde{\rm B}_{3,\eta'\eta}^{A}(\nu,\nu)=&\;A^{\bar\sigma}(\nu,\eta)+\tilde{\rm B}_{3,\eta'\eta}^{A}(\nu,\bar\nu)\;,\\
\tilde{\rm B}_{3,\eta\eta}^{B}(\nu)=&\;A^{\sigma}(\eta,\nu)-\tilde{\rm B}_{3,\eta\bar\eta}^{B}(\nu)\;.
\end{aligned}
\end{equation}
From Eqs.~\eqref{CA} and~\eqref{tilde_s_A}, the function $\tilde{\rm C}^A(\nu',\nu)$, which does not depend on $\eta$, reads
\begin{equation}
\begin{aligned}\label{Hnu}
\tilde{\rm C}^A(\nu',\nu)
=&\;{\rm h}^{\sigma(\sigma)}_{2}\tilde{\rm h}^{\sigma(\sigma)}_{2}\left(\delta_{\nu'\nu}
+\tilde{\rm h}^{\sigma(\sigma)}_{A}\nu'\nu\;{\rm s}_{\nu}^A\right)\;.
\end{aligned}
\end{equation}
Therefore, as a matrix in $(\eta',\eta)$
\begin{equation}
\begin{aligned}\label{h2Amatrix}
\tilde{\mathbf{h}}_{2}^{A}(\nu',\nu)=&\;\mathbf{1}{\rm h}^{\sigma(\sigma)}_{2}\delta_{\nu',\nu}+\sum_{\nu''}\tilde{\rm C}^A(\nu',\nu'')\tilde{\mathbf{B}}_{3}^{A}(\nu'',\nu)\;.\\
\end{aligned}
\end{equation}
Finally, let us introduce the matrices
\begin{equation}
\begin{aligned}\label{KA}
\sum_{\nu}\nu\tilde{\mathbf{h}}_{2}^{A}(\nu',\nu):=&\;\mathbf{1}{\rm h}^{\sigma(\sigma)}_{2}\sum_{\nu}\nu\delta_{\nu',\nu}+\mathbf{K}^{A}(\nu')\;,\\
\mathbf{K}^{A}(\nu'):=&\;\sum_\nu\nu\sum_{\nu''}\tilde{\rm C}^A(\nu',\nu'')\tilde{\mathbf{B}}_{3}^{A}(\nu'',\nu)\;,\\
\mathbf{K}^{A}:=&\;\sum_{\nu'}\mathbf{K}^{A}(\nu')\;.
\end{aligned}
\end{equation}
Note that the matrices $\mathbf{K}^{A}(\nu)$ and $\mathbf{K}^{A}$ inherit the symmetry~\eqref{propB3A}.

\indent We are now in the position to iterate the third-tier bubbles $\tilde{\mathbf{B}}_3^B$ as outlined in Eq.~\eqref{Bs_appendix}. Accounting for the factors $\nu$ and $\nu'$ associated to overlap of three spin-$\sigma$ fermion lines in $\tilde{\mathbf{B}}_3^B$, the second term of Eq.~\eqref{B_sigma_sigma} yields the matrix in the indexes $\eta'\eta$ 
\begin{equation}
\begin{aligned}
\label{K}
\mathbf{K}^{\sigma(\sigma)}=&\;
\sum_{\nu'}\sum_{\nu}\nu\tilde{\mathbf{h}}_{2}^{\sigma(\sigma)}(\nu',\nu)\\
=&\;\sum_{\nu'}\sum_{\nu}\nu\tilde{\mathbf{h}}_{2}^{A}(\nu',\nu)+\sum_{\nu'} \sum_{\nu}\nu\sum_{\nu'''}  \tilde{\mathbf{h}}_{2}^{A}(\nu',\nu''')\nu'''\sum_{\nu''}\tilde{\mathbf{B}}_3^B(\nu'')\nu''\tilde{\mathbf{h}}_{2}^{A}(\nu'',\nu)\\
&+\sum_{\nu'} \sum_{\nu}\nu\sum_{\nu^{v}}  \tilde{\mathbf{h}}_{2}^{A}(\nu',\nu^{v})\nu^{v}\sum_{\nu^{iv}}\tilde{\mathbf{B}}_3^B(\nu^{iv})\nu^{iv}
\sum_{\nu'''}  \tilde{\mathbf{h}}_{2}^{A}(\nu^{iv},\nu''')\nu'''\sum_{\nu''}\tilde{\mathbf{B}}_3^B(\nu'')
\nu''\tilde{\mathbf{h}}_{2}^{A}(\nu'',\nu)+\dots\\
=&\;\mathbf{K}^{A}+\mathbf{K}^{A}\sum_{\nu''}\tilde{\mathbf{B}}_3^B(\nu'')\nu''\sum_\nu\tilde{\mathbf{h}}_{2}^{A}(\nu'',\nu)\nu\\
&+\mathbf{K}^{A}\sum_{\nu^{iv}}\tilde{\mathbf{B}}_3^B(\nu^{iv})\nu^{iv}
\sum_{\nu'''}  \tilde{\mathbf{h}}_{2}^{A}(\nu^{iv},\nu''')\nu'''\sum_{\nu''}\tilde{\mathbf{B}}_3^B(\nu'')
\nu''\sum_\nu\tilde{\mathbf{h}}_{2}^{A}(\nu'',\nu)	\nu+\dots\\
=&\;\mathbf{K}^{A}\frac{\mathbf{1}}{\mathbf{1}-\sum_{\nu'}\tilde{\mathbf{B}}_3^B(\nu')\nu'\sum_\nu\tilde{\mathbf{h}}_{2}^{A}(\nu',\nu) \nu}\\
[{\rm Eq.}~\eqref{KA}]\quad=&\;\mathbf{K}^{A}\frac{\mathbf{1}}{\mathbf{1}-{\rm h}^{\sigma(\sigma)}_{2}\mathbf{S}_3^B-\sum_{\nu}\tilde{\mathbf{B}}_3^B(\nu)\nu\mathbf{K}^{A}(\nu)}\\
[{\rm Eq.}~\eqref{S3B}]\quad=&\;\mathbf{K}^{A}\frac{\mathbf{1}}{\mathbf{1}(1+{\rm h}^{\sigma(\sigma)}_{2}\Gamma/\hbar)-{\rm h}^{\sigma(\sigma)}_{2}\mathbf{s}^B-\sum_{\nu}\tilde{\mathbf{B}}_3^B(\nu)\nu\mathbf{K}^{A}(\nu)}
\;.
\end{aligned}
\end{equation}
From the properties of the third-tier bubbles, Eqs.~\eqref{B3A_contraction_2},~\eqref{B3B_contraction_2},~\eqref{propB3A}-\eqref{propB3B} and the definitions of $\mathbf{K}^A$ and $\mathbf{K}^A(\nu)$, Eq.~\eqref{KA}, we have
\begin{equation}
\begin{aligned}
\label{}
\mathbf{K}^{A}_{\eta'\eta}=&\;\sum_{\nu',\nu}\tilde{\rm C}^A(\nu',\nu)\nu A^{\bar\sigma}(\nu,\eta)\\
\left[\sum_{\nu}\tilde{\mathbf{B}}_3^B(\nu)\nu\mathbf{K}^{A}(\nu)\right]_{\eta'\eta}=&\;\sum_{\nu',\nu}\nu'\nu \; \tilde{\rm C}^A(\nu',\nu)A^\sigma(\eta',\nu')A^{\bar\sigma}(\nu,\eta)
\;.
\end{aligned}
\end{equation}
After some lengthy manipulations one obtains the following results
\begin{equation}
\begin{aligned}
\label{intermediate_results}
\mathbf{K}^{A}_{\eta'\eta}=&\;\eta\;\Delta A^{\bar\sigma}_+\;{\rm h}^{\sigma(\sigma)}_{2}\tilde{\rm h}^{\sigma(\sigma)}_{2}\;,\\
\sum_{\nu}\tilde{\mathbf{B}}_3^B(\nu)\nu\mathbf{K}^{A}(\nu)=&\;\left(\mathbf{1}\Delta A^{\sigma}_+ \Delta A^{\bar\sigma}_+ + \mathbf{P}\right)\;{\rm h}^{\sigma(\sigma)}_{2}\tilde{\rm h}^{\sigma(\sigma)}_{2}\;,\\
{\rm where}\qquad\qquad\quad
\Delta A^{\sigma/\bar\sigma}_+=&\;A^{\sigma/\bar\sigma}(+,+)-A^{\sigma/\bar\sigma}(-,+)
\;,
\end{aligned}
\end{equation}
with $\mathbf{P}_{\eta\eta}=\mathbf{P}_{\bar\eta\eta}$. This latter property, along with $\mathbf{s}^B_{\eta\eta}=\mathbf{s}^B_{\bar\eta\eta}$, yields, upon inverting the matrix $\mathbf{K}$, the key result
\begin{equation}
\begin{aligned}\label{K_final}
{\rm K}_{\eta'\eta}^{\sigma(\sigma)}=&\;\eta\frac{\Delta A^{\bar\sigma}_+}{([{\rm h}^{\sigma(\sigma)}_{2}]^{-1}+\Gamma/\hbar)^{2}-\Delta A^{\sigma}_+\Delta A^{\bar\sigma}_+}\;.
\end{aligned}
\end{equation}
Note that the above expression implies for the retarded self-energy $\tilde{\Sigma}_{\sigma\eta}^{(\sigma)}(\epsilon)=\;{\rm i}\hbar\tilde{\rm B}^{\sigma(\sigma)}_{\bar\eta\eta}(\kappa)|_{\zeta=+1}$ the property
\begin{equation}\label{propertySigma+}
\sum_\eta\tilde{\Sigma}_{\sigma\eta}^{(\sigma)}(\epsilon) = 0\;.
\end{equation}
The explicit expression for $\Delta A^{\bar\sigma}_+ = A^{\bar\sigma}(+,+)-A^{\bar\sigma}(-,+)$ is found from the definition of $A^{\bar\sigma}(\nu,\eta)$, Eq.~\eqref{B3A_contraction_2}, adding  $(\Gamma/\hbar)/D^{\sigma(\sigma\bar\sigma)}$ to both the contracted  functions in the difference, and using Eq.~\eqref{B3A_diagonal}. The result is  
\begin{equation}
\begin{aligned}\label{DeltaA_exact}
\Delta A^{\bar\sigma}_+=&\;\Biggl<\left[\frac{({\rm h}^{\sigma(\sigma\bar\sigma)}_{3,--})^{-1}+\Gamma/2\hbar}{D^{\sigma(\sigma\bar\sigma)}}-\frac{({\rm h}^{\sigma(\sigma\bar\sigma)}_{3,++})^{-1}+\Gamma/2\hbar}{D^{\sigma(\sigma\bar\sigma)}}\right]{\rm v}_{-}\Biggr>\\
=&\;\Biggl<
\frac{{\rm i}\zeta\hbar}{\epsilon-\epsilon_{k_1}+\zeta\zeta_2[\epsilon_{k_2}-
E_{\bar\sigma}(+)]+{\rm i}\zeta\Gamma/2 - \zeta_2 \Sigma_{4,\sigma+}^{(\sigma\bar\sigma)}\frac{U}{\epsilon-\epsilon_{k_1}+\zeta\zeta_2[\epsilon_{k_2}-
E_{\bar\sigma}(-)]+{\rm i}\zeta 3\Gamma/2}}{\rm v}_{-}\Biggr>\\
-&\;
\Biggl<
\frac{{\rm i}\zeta\hbar}{\epsilon-\epsilon_{k_1}+\zeta\zeta_2[\epsilon_{k_2}-
E_{\bar\sigma}(-)]+{\rm i}\zeta\Gamma/2 + \zeta_2 \Sigma_{4,\sigma-}^{(\sigma\bar\sigma)}\frac{U}{\epsilon-\epsilon_{k_1}+\zeta\zeta_2[\epsilon_{k_2}-
E_{\bar\sigma}(+)]+{\rm i}\zeta 3\Gamma/2}}
{\rm v}_{-}\Biggr>\\
=&\;\sum_\nu \nu\Biggl<
\frac{{\rm i}\zeta_2\hbar}{\epsilon_{k_2}-
E_{\bar\sigma}(\nu)+\zeta\zeta_2(\epsilon-\epsilon_{k_1})+{\rm i}\zeta_2\Gamma/2 -\nu \zeta \Sigma_{4,\sigma\nu}^{(\sigma\bar\sigma)}\frac{U}{\epsilon_{k_2}-
E_{\bar\sigma}(\bar\nu)+\zeta\zeta_2(\epsilon-\epsilon_{k_1})+{\rm i}\zeta_2 3\Gamma/2}}{\rm v}_{-}\Biggr>\;,
\end{aligned}
\end{equation}
where the contraction involves the innermost fermion line indexed with $\boldsymbol{\kappa}_2$ and where we used $\zeta^2=1$. Here,  $E_{\bar\sigma}(\nu)=\epsilon_{\bar\sigma}+(1+\nu) U/2$. A similar result holds for $\Delta A^{\sigma}_+$.
Note that this expression displays the same structure as the one for the Green's function, Eq.~\eqref{GRgDSO4}; this makes evident that the renormalization of the dot energies $E_{\bar\sigma}(\nu)$ ($E_{\sigma}(\eta)$) occurs also at the level of the self-energy and in principle at all (even) levels of the  hierarchy.

\section{Evaluation of the dressed third-tier bubbles in the simplified NCA4}
\label{third-level_bubbles}

A simplification of the NCA4 is obtained by setting to zero the nontrivial parts of the 4th-tier bubbles, namely setting
$\Sigma_{4,\nu}^{(\sigma)}$ in Eq.~\eqref{DeltaA_exact}, or equivalently ${\rm F}_\nu^{(\sigma)}={\rm F}_\eta^{(\bar\sigma)}=0$ in Eqs.~\eqref{F} and~\eqref{FB}. 
As a result, the definition in Eqs.~\eqref{B3A_contraction_2} yields
\begin{equation}
\begin{aligned}\label{B3A_DSO4}
A^{\bar\sigma}(\nu,\eta)&\simeq
\begin{gathered}
\resizebox{!}{1.2cm}{
\begin{tikzpicture}[] 
\draw[blue,line width=0.5mm] (0.5,1.8) arc (120:60:3cm  and 2.cm)  node[right] {\LARGE{$\sigma\kappa$}};
\draw[blue,line width=0.5mm] (0.5,1.1) arc (120:60:3cm  and 5cm)  node[right] {\LARGE{$\sigma\kappa_1$}};
\draw[red,line width=0.5mm]  (1,0) arc (180:0:1.cm  and 1.2cm); 
\draw[blue,line width=0.5mm] (0.5,0.3) node[left] {\Large{$\nu$}}  -- (3.5,0.3) node[right] {\Large{$\;\nu$}} ; 
\draw[red,line width=0.5mm] (0.5,-0.5) node[left] {\Large{$\eta$}} -- (1,-0.5); 
\draw[red,line width=0.5mm] (1,-0.5) -- (1,0); 
\draw[red,dashed,line width=0.5mm] (1,-0.5) -- (3,-0.5); 
\draw[red,line width=0.5mm] (3,-0.5) -- node[right] {\Large{$\;\eta'$}}  (3.5,-0.5); 
\draw[red,line width=0.5mm] (3,-0.5) -- (3,-0); 
\filldraw[red](1,-0.5) circle (3pt); 
\draw [red,pattern=north west lines, pattern color=red] (2.4,-0.8) rectangle (1.6,-0.2);
 \end{tikzpicture}
 }
\end{gathered}=\Biggl<\frac{1}{({\rm h}^{\sigma(\sigma\bar\sigma)}_{3,\nu\nu})^{-1} + \Gamma/2\hbar}{\rm v}_{-\eta}\Biggr>
\\
&=\;\Biggl<
\frac{{\rm i}\zeta_2\hbar}{\epsilon_{k_2}-
E_{\bar\sigma}(\nu)+\zeta\zeta_2(\epsilon-\epsilon_{k_1})+{\rm i}\zeta_2\Gamma/2}{\rm v}_{-\eta}\Biggr>\\
&=-\frac{{\rm i}}{\hbar}\sum_{\alpha_{2}}\varrho_{\alpha_{2}} |{\rm t}_{\alpha_{2}}|^2\sum_{\zeta_{2}}\zeta_{2} \int_{-W}^{W} d\epsilon_{2}\; \frac{f^{\alpha_{2}}_{-\eta}(\epsilon_{2})}{\epsilon_{2}-[E_{\bar\sigma}(\nu)-\zeta_{2}\zeta(\epsilon-\epsilon_{1})]+{\rm i}\zeta_{2}\Gamma/2}\;,
\end{aligned}
\end{equation}
where $E_{\bar\sigma}(\nu)=\epsilon_{\bar\sigma}+(1+\nu) U/2$. Here, the vertex ${\rm v}_{\eta}(\boldsymbol{\kappa})$ is defined in Eq.~\eqref{v_SIAM}, the propagator ${\rm h}^{\sigma(\sigma\bar\sigma)}_{3,\nu\nu}$ is given in Eq.~\eqref{h3ss}, and the width $\Gamma/2\hbar$ is graphically rendered by the dashed box which is the sum of the geometrical series 
\begin{equation}\begin{aligned}
\begin{gathered}
\resizebox{!}{0.3cm}{
\begin{tikzpicture}[] 
\draw[dashed,line width=0.5mm] (1,0) -- (3,0); 
\draw [pattern=north west lines, pattern color=black] (2.4,-0.3) rectangle (1.6,0.3);
 \end{tikzpicture}
 }
\end{gathered}
=&
\begin{gathered}
\resizebox{!}{0.5cm}{
\begin{tikzpicture}[] 
\draw[white,line width=0.5mm] (1,0) arc (180:0:.5cm  and 1.2cm); 
\filldraw[white](1,0) circle (4pt); 
\draw[dashed,line width=0.5mm] (0,0) -- (2,0) ; 
\end{tikzpicture}
 }
\end{gathered}
+
\begin{gathered}
\resizebox{!}{0.5cm}{
\begin{tikzpicture}[] 
\draw[line width=0.5mm] (1,0) arc (180:0:.5cm  and 1.2cm); 
\draw[dashed,line width=0.5mm]  (0,0) -- (1,0); 
\draw[line width=0.5mm] (1,0) -- (2,0); 
\draw[dashed,line width=0.5mm] (2,0) -- (3,0); 
\filldraw[](1,0) circle (4pt); 
\end{tikzpicture}
 }
\end{gathered}
+
\begin{gathered}
\resizebox{!}{0.5cm}{
\begin{tikzpicture}[] 
\draw[line width=0.5mm] (1,0) arc (180:0:.5cm  and 1.2cm); 
\draw[line width=0.5mm] (3,0) arc (180:0:.5cm  and 1.2cm); 
\draw[dashed,line width=0.5mm]  (0,0) -- (1,0); 
\draw[line width=0.5mm] (1,0) -- (2,0); 
\draw[dashed,line width=0.5mm] (2,0) -- (3,0) ; 
\draw[line width=0.5mm] (3,0) -- (4,0); 
\draw[dashed,line width=0.5mm] (4,0) -- (5,0); 
\filldraw[](1,0) circle (4pt); 
\filldraw[](3,0) circle (4pt); 
\end{tikzpicture}
 }
\end{gathered}
+\quad\dots
\end{aligned}\end{equation}

Using  Eq.~\eqref{Isummary} to solve the integral we obtain
\begin{equation}
\begin{aligned}
\label{}
A^{\bar\sigma}(\nu,\eta)&\simeq\frac{\rm i}{\hbar}\sum_{\alpha}\frac{\Gamma_\alpha}{2\pi} \Bigg\{\eta \Bigg[{\rm Re}\psi\left(\frac{1}{2}+{\rm i}\frac{\zeta(\epsilon_1-\epsilon)+E_{\bar\sigma}(\nu)-{\rm i}\Gamma/2-\mu_{\alpha}}{2\pi k_{\rm B}T}\right)-{\rm Re}\psi\left(\frac{1}{2}+{\rm i}\frac{\zeta(\epsilon-\epsilon_1)+E_{\bar\sigma}(\nu)-{\rm i}\Gamma/2-\mu_{\alpha}}{2\pi k_{\rm B}T}\right)\Bigg]\\
&\qquad+{\rm i}\Bigg[\pi  +\eta {\rm Im}\psi\left(\frac{1}{2}+{\rm i}\frac{\zeta(\epsilon_1-\epsilon)+E_{\bar\sigma}(\nu)-{\rm i}\Gamma/2-\mu_{\alpha}}{2\pi k_{\rm B}T}\right)
+\eta{\rm Im}\psi\left(\frac{1}{2}+{\rm i}\frac{\zeta(\epsilon-\epsilon_1)+E_{\bar\sigma}(\nu)-{\rm i}\Gamma/2-\mu_{\alpha}}{2\pi k_{\rm B}T}\right)\Bigg]\Bigg\}\;,
\end{aligned}
\end{equation}
where $\Gamma_\alpha=2\pi \varrho_{\alpha} |{\rm t}_{\alpha}|^2$ and $\sum_\alpha\Gamma_\alpha=\Gamma$.
As a further approximation, we fix the argument $\epsilon_1=\epsilon$, which yields
\begin{equation}
\begin{aligned}
\label{B3A_DSO4_approx}
A^{\bar\sigma}(\nu,\eta)\simeq  - \sum_{\alpha} \frac{\Gamma_\alpha}{\hbar} \Bigg[\frac{1}{2}  +\eta \frac{1}{\pi}{\rm Im}\psi\left(\frac{1}{2}+\frac{\Gamma/2}{2\pi k_{\rm B}T}+{\rm i}\frac{E_{\bar\sigma}(\nu)-\mu_{\alpha}}{2\pi k_{\rm B}T}\right)\Bigg]\;.
\end{aligned}
\end{equation}
 A similar calculation for $A^{\sigma}(\eta,\nu)$, defined in Eq.~\eqref{B3B_contraction_2}, gives 
\begin{equation}
\begin{aligned}
\label{B3B_DSO4_approx}
A^{\sigma}(\eta,\nu)\simeq- \sum_{\alpha}\frac{\Gamma_\alpha}{\hbar} \Bigg[\frac{1}{2}  +\nu \frac{1}{\pi}{\rm Im}\psi\left(\frac{1}{2}+\frac{\Gamma/2}{2\pi k_{\rm B}T}+{\rm i}\frac{E_{\sigma}(\eta)-\mu_{\alpha}}{2\pi k_{\rm B}T}\right)\Bigg]\;.
\end{aligned}
\end{equation}

The 3rd-tier bubble of type $\sigma(\bar\sigma)$ does not have a matrix structure and is given by the contraction of the propagators in Eq.~\eqref{h3sbs}, dressed by the simplified fourth-tier bubble ${\rm B}_4=-\Gamma/2$, with the vertex ${\rm v}_{{\rm x}}=-|{\rm t}_{\alpha}(\epsilon_k)|^2  f^\alpha_{{\rm x}}(\epsilon_k)/\hbar^2$, where $\rm x$ stands for $\eta$ or $\nu$. It is the composite bubble
\begin{equation}
\begin{aligned}
\label{B3-}
\tilde{\rm B}_{3}^{\sigma(\bar\sigma)}\simeq &
\begin{gathered}
\resizebox{!}{1.4cm}{
\begin{tikzpicture}[] 
\draw[blue,line width=0.5mm] (0.5,1.5) arc (120:60:3cm  and 2.cm) node[right] {\LARGE{$\sigma\kappa$}};
\draw[red,line width=0.5mm] (0.5,.8) arc (120:60:3cm  and 5cm)node[right] {\LARGE{$\bar\sigma\kappa_1$}} ;
\draw[blue,line width=0.5mm] (1,0) arc (180:0:1.cm  and 1.2cm); 
\draw[blue,dashed,line width=0.5mm]  (0.5,0) -- (1,0); 
\draw[blue,line width=0.5mm] (1,0) -- (3,0); 
\draw[blue,dashed,line width=0.5mm] (3,0) -- (3.5,0); 
\draw[red,dashed,line width=0.5mm] (0.5,-0.5) node[left,white] {\Large{$\eta$}} -- node[right,white] {\Large{$\qquad\quad\eta'$}} (3.5,-0.5); 
\filldraw[blue] 
(1,0) circle (3pt);
\draw [red,pattern=north west lines, pattern color=red] (2.4,-0.7) rectangle (1.6,-0.3);
\end{tikzpicture}
 }
\end{gathered}
+
\begin{gathered}
\resizebox{!}{1.4cm}{
\begin{tikzpicture}[] 
\draw[blue,line width=0.5mm] (0.5,1.5) arc (120:60:3cm  and 2.cm) node[right] {\LARGE{$\sigma\kappa$}};
\draw[red,line width=0.5mm] (0.5,.8) arc (120:60:3cm  and 5cm) node[right] {\LARGE{$\bar\sigma\kappa_1$}};
\draw[red,line width=0.5mm]  (1,0) arc (180:0:1.cm  and 1.2cm); 
\draw[blue,dashed,line width=0.5mm]  (0.5,0) -- (3.5,0); 
\draw[red,dashed,line width=0.5mm] (0.5,-0.5) node[left,white] {\Large{$\eta$}} -- (1,-0.5); 
\draw[red,line width=0.5mm] (1,-0.5) -- (1,0); 
\draw[red,line width=0.5mm] (1,-0.5) -- (3,-0.5); 
\draw[red,dashed,line width=0.5mm] (3,-0.5) -- node[right,white] {\Large{$\;\eta'$}}  (3.5,-0.5); 
\draw[red,line width=0.5mm] (3,-0.5) -- (3,-0); 
\filldraw[red](1,-0.5) circle (3pt); 
\draw [blue,pattern=north west lines, pattern color=blue] (2.4,-0.2) rectangle (1.6,0.2);
 \end{tikzpicture}
 }
\end{gathered}\\
=&\hbar\sum_{\kappa_2}\left[
\sum_{\nu}\frac{{\rm v}_{\nu}^{\alpha_2}(\epsilon_{k_2})\delta_{\zeta_2,-\zeta}}{\Gamma/2-{\rm i} \zeta(\epsilon_k-\epsilon_{k_2})-{\rm i} \zeta_1[\epsilon_{k_1}-E_{\bar\sigma}(\nu)]}+\sum_{\eta}\frac{{\rm v}_{\eta}^{\alpha_2}(k_2)\delta_{\zeta_2,-\zeta_1}}{\Gamma/2-{\rm i} \zeta_1(\epsilon_{k_1}-\epsilon_{k_2})-{\rm i} \zeta[\epsilon_{k}-E_{\sigma}(\eta)]}\right]\\
=&\frac{{\rm i}\zeta}{\hbar}\sum_{\alpha_2}\varrho_{\alpha_2} |{\rm t}_{\alpha_2}|^2  \sum_{\nu} \int_{-W}^{W} d\epsilon_2\; \frac{f^{\alpha_2}_{\nu}(\epsilon_2)}{\epsilon_2-\epsilon+\zeta \zeta_1[E_{\bar\sigma}(\nu)-\epsilon_1]-{\rm i}\zeta \Gamma/2}\\
&+\frac{{\rm i}\zeta_1}{\hbar}\sum_{\alpha_2}\varrho_{\alpha_2}  |{\rm t}_{\alpha_2}|^2  \sum_{\eta} \int_{-W}^{W} d\epsilon_2\; \frac{f^{\alpha_2}_{\eta}(\epsilon_2)}{\epsilon_2-\epsilon_1+\zeta \zeta_1[E_{\sigma}(\eta)-\epsilon]-{\rm i}\zeta_1 \Gamma/2}\;,
\end{aligned}
\end{equation}
where $E_{\sigma}(\eta)=\epsilon_{\sigma}+(1+\eta) U/2$. From Eq.~\eqref{Isummary} we obtain
\begin{equation}
\begin{aligned}
\label{B3d}
\tilde{\rm B}_{3}^{\sigma(\bar\sigma)}\simeq\frac{{\rm i}}{\hbar}\sum_{\alpha}\frac{\Gamma_\alpha}{2\pi} \Bigg\{&\zeta{\rm Re}\psi\left(\frac{1}{2}+{\rm i}\frac{\epsilon - \zeta\zeta_1(\epsilon_{\bar\sigma}+U-\epsilon_1)
-{\rm i}\Gamma/2-\mu_{\alpha}}{2\pi k_{\rm B}T}\right)-\zeta{\rm Re}\psi\left(\frac{1}{2}+{\rm i}\frac{\epsilon - \zeta\zeta_1(\epsilon_{\bar\sigma}-\epsilon_1)-{\rm i}\Gamma/2-\mu_{\alpha}}{2\pi k_{\rm B}T}\right)\\
+\zeta_1 & {\rm Re}\psi\left(\frac{1}{2}+{\rm i}\frac{\epsilon_1 - \zeta\zeta_1(\epsilon_{\sigma}+U-\epsilon)-{\rm i}\Gamma/2-\mu_{\alpha}}{2\pi k_{\rm B}T}\right)-\zeta_1{\rm Re}\psi\left(\frac{1}{2}+{\rm i}\frac{\epsilon_1 - \zeta\zeta_1(\epsilon_{\sigma}-\epsilon)-{\rm i}\Gamma/2-\mu_{\alpha}}{2\pi k_{\rm B}T}\right)\\
+{\rm i}\Bigg[2\pi -&{\rm Im}\psi\left(\frac{1}{2}+{\rm i}\frac{\epsilon - \zeta\zeta_1(\epsilon_{\bar\sigma}+U-\epsilon_1)-{\rm i}\Gamma/2-\mu_{\alpha}}{2\pi k_{\rm B}T}\right)+{\rm Im}\psi\left(\frac{1}{2}+{\rm i}\frac{\epsilon - \zeta\zeta_1(\epsilon_{\bar\sigma}-\epsilon_1)-{\rm i}\Gamma/2-\mu_{\alpha}}{2\pi k_{\rm B}T}\right)\\
-&{\rm Im}\psi\left(\frac{1}{2}+{\rm i}\frac{\epsilon_1 - \zeta\zeta_1(\epsilon_{\sigma}+U-\epsilon)-{\rm i}\Gamma/2-\mu_{\alpha}}{2\pi k_{\rm B}T}\right)
+{\rm Im}\psi\left(\frac{1}{2}+{\rm i}\frac{\epsilon_1 - \zeta\zeta_1(\epsilon_{\sigma}-\epsilon)-{\rm i}\Gamma/2-\mu_{\alpha}}{2\pi k_{\rm B}T}\right)\Bigg]
\Bigg\}\;.
\end{aligned}
\end{equation}

Taking $\epsilon=\epsilon_1=\mu_\alpha$,
the retarded ($\zeta=+1$) third-tier self-energies $\tilde{\Sigma}_{3\sigma,\zeta_1}^{(\bar\sigma)}:={\rm i}\hbar\tilde{\rm B}_{3,\zeta_1}^{\sigma(\bar\sigma)}|_{\zeta=+1}$ read
\begin{equation}
\begin{aligned}
\label{B3d_2}
\tilde{\Sigma}_{3\sigma,\zeta_1}^{(\bar\sigma)}\simeq -\sum_{\alpha}\frac{\Gamma_\alpha}{2\pi} \Bigg\{&{\rm Re}\psi\left(\frac{1}{2}-{\rm i}\frac{  \zeta_1(\epsilon_{\bar\sigma}+U-\mu_\alpha)
+{\rm i}\Gamma/2}{2\pi k_{\rm B}T}\right)-{\rm Re}\psi\left(\frac{1}{2}-{\rm i}\frac{  \zeta_1(\epsilon_{\bar\sigma}-\mu_\alpha)+{\rm i}\Gamma/2}{2\pi k_{\rm B}T}\right)\\
+\zeta_1 & {\rm Re}\psi\left(\frac{1}{2}-{\rm i}\frac{ \zeta_1(\epsilon_{\sigma}+U-\mu_\alpha)+{\rm i}\Gamma/2}{2\pi k_{\rm B}T}\right)-\zeta_1{\rm Re}\psi\left(\frac{1}{2}-{\rm i}\frac{ \zeta_1(\epsilon_{\sigma}-\mu_\alpha)+{\rm i}\Gamma/2}{2\pi k_{\rm B}T}\right)\\
+{\rm i}\Bigg[2\pi -&{\rm Im}\psi\left(\frac{1}{2}-{\rm i}\frac{  \zeta_1(\epsilon_{\bar\sigma}+U-\mu_\alpha)
+{\rm i}\Gamma/2}{2\pi k_{\rm B}T}\right)+{\rm Im}\psi\left(\frac{1}{2}-{\rm i}\frac{  \zeta_1(\epsilon_{\bar\sigma}-\mu_\alpha)
+{\rm i}\Gamma/2}{2\pi k_{\rm B}T}\right)\\
-&{\rm Im}\psi\left(\frac{1}{2}-{\rm i}\frac{  \zeta_1(\epsilon_{\sigma}+U-\mu_\alpha)
+{\rm i}\Gamma/2}{2\pi k_{\rm B}T}\right)
+{\rm Im}\psi\left(\frac{1}{2}-{\rm i}\frac{  \zeta_1(\epsilon_{\sigma}-\mu_\alpha)
+{\rm i}\Gamma/2}{2\pi k_{\rm B}T}\right)\Bigg]
\Bigg\}\;.
\end{aligned}
\end{equation}
Consider the degenerate case, $\epsilon_\uparrow=\epsilon_\downarrow=\epsilon_0$, at equilibrium $\mu_\alpha=\mu$. At the particle-hole symmetry point $\mu-\epsilon_0=U/2$, Eq.~\eqref{B3d_2}   simplifies to
\begin{equation}
\begin{aligned}
\label{B3d_3}
\tilde{\Sigma}_{3\sigma,\pm}^{(\bar\sigma)}\simeq
-{\rm i}\Gamma\Bigg[1 \pm\frac{2}{\pi}{\rm Im}\psi\left(\frac{1}{2}+\frac{\Gamma/2}{2\pi k_{\rm B}T}+{\rm i}\frac{U/2}{2\pi k_{\rm B}T}\right)\Bigg]\;.
\end{aligned}
\end{equation}

\section{Evaluation of the dressed bubble $\tilde{\mathbf{B}}^{\sigma(\sigma)}$ within the  simplified NCA4}
\label{dressingB+DSO4}

Within the simplified NCA4, we can give an explicit expression for the functions $\Delta A_+^{\bar\sigma/\sigma}$ appearing in ${\rm K}_{\eta'\eta}^{\sigma(\sigma)}$, Eq.~\eqref{K_final}.
Using the simplified third-tier bubbles, Eqs.~\eqref{B3A_DSO4_approx}-\eqref{B3B_DSO4_approx}, we find 
\begin{equation}
\begin{aligned}\label{Delta_b}
\Delta A_+^{\bar\sigma}=&-\sum_{\alpha} \frac{\Gamma_\alpha}{\hbar}  \Bigg[\frac{1}{\pi}{\rm Im}\psi\left(\frac{1}{2}+\frac{\Gamma/2}{2\pi k_{\rm B}T}+{\rm i}\frac{\epsilon_{\bar\sigma}+U-\mu_\alpha}{2\pi k_{\rm B}T}\right) -  \frac{1}{\pi}{\rm Im}\psi\left(\frac{1}{2}+\frac{\Gamma/2}{2\pi k_{\rm B}T}+{\rm i}\frac{\epsilon_{\bar\sigma}-\mu_\alpha}{2\pi k_{\rm B}T}\right)\Bigg]\;,\\
\Delta A_+^{\sigma}=&-\sum_{\alpha} \frac{\Gamma_\alpha}{\hbar} \Bigg[\frac{1}{\pi}{\rm Im}\psi\left(\frac{1}{2}+\frac{\Gamma/2}{2\pi k_{\rm B}T}+{\rm i}\frac{\epsilon_{\sigma}+U-\mu_\alpha}{2\pi k_{\rm B}T}\right) -  \frac{1}{\pi}{\rm Im}\psi\left(\frac{1}{2}+\frac{\Gamma/2}{2\pi k_{\rm B}T}+{\rm i}\frac{\epsilon_{\sigma}-\mu_\alpha}{2\pi k_{\rm B}T}\right)\Bigg]\;.
\end{aligned}
\end{equation}
We can write ${\rm K}_{\eta'\eta}^{\sigma(\sigma)}$ as
\begin{equation}
\begin{aligned}
\label{K_simplified}
{\rm K}_{\eta'\eta}^{\sigma(\sigma)}=&\;-\eta
\frac{\Delta A_+^{\bar\sigma}}{2 (\Delta A_+^{\sigma}\Delta A_+^{\bar\sigma})^{1/2}}
\left[\frac{1}{[{\rm h}^{\sigma(\sigma)}_{2}]^{-1}+\Gamma/\hbar+(\Delta A_+^{\sigma}\Delta A_+^{\bar\sigma})^{1/2}}-\frac{1}{[{\rm h}^{\sigma(\sigma)}_{2}]^{-1}+\Gamma/\hbar - (\Delta A_+^{\sigma}\Delta A_+^{\bar\sigma})^{1/2}}\right]\;.
\end{aligned}
\end{equation}
Since the above expression is independent of $\eta'$, the bubble $\tilde{\mathbf{B}}^{\sigma(\sigma)}=-(\Gamma/2\hbar)\mathbf{1}+\langle\mathbf{K}^{\sigma(\sigma)}{\rm v}_{+}\rangle$ has the property, see Eq.~\eqref{B_sigma_sigma},
\begin{equation}\label{prp_tildeBs}
\tilde{\rm B}^{\sigma(\sigma)}_{\eta'\eta}
=-\frac{\Gamma}{2\hbar}\delta_{\eta'\eta}+\tilde{\rm B}^{\sigma(\sigma)}_{\bar\eta\eta}\;.
\end{equation}
Noting that $\Delta A_+^{\bar\sigma/\sigma}=-|\Delta A_+^{\bar\sigma/\sigma}|$, the off-diagonal elements are
\begin{equation}\begin{aligned}
\label{tildeB+Approx}
\tilde{\rm B}_{\bar\eta\eta}^{\sigma(\sigma)}
&=\langle{\rm K}_{\bar\eta\eta}^{\sigma(\sigma)}{\rm v}_+ \rangle\\
&= \frac{{\rm i}\zeta}{\hbar}\frac{\eta}{2}\sqrt{\frac{|\Delta A_+^{\bar\sigma}|}{|\Delta A_+^{\sigma}|}}\sum_{\alpha_1} \varrho_{\alpha_1}|{\rm t}_{\alpha_1}|^2\int_{-W}^W d\epsilon_1 \left[\frac{f^{\alpha_1}_+(\epsilon_1)}{\epsilon_1-\epsilon-{\rm i} \zeta \Gamma_+}-\frac{f^{\alpha_1}_+(\epsilon_1)}{\epsilon_1-\epsilon-{\rm i} \zeta \Gamma_-}\right]\;,
\end{aligned}
\end{equation}
where $\Gamma_\pm = \Gamma \pm \hbar(\Delta A_+^{\sigma}\Delta A_+^{\bar\sigma})^{1/2}$. 
Using Eq.~\eqref{I-} to perform the integrations we obtain for the corresponding retarded self-energies
\begin{equation}\begin{aligned}
\label{Sigma+}
\tilde{\Sigma}_{\sigma\eta}^{(\sigma)}(\epsilon)=&\;{\rm i}\hbar\tilde{\rm B}^{\sigma(\sigma)}_{\bar\eta\eta}(\kappa)|_{\zeta=+1}\\
=&\; -\frac{\eta}{2}\sqrt{\frac{|\Delta A_+^{\bar\sigma}|}{|\Delta A_+^{\sigma}|}}\sum_{\alpha}\frac{\Gamma_{\alpha}}{2\pi}
\Bigg\{{\rm Re}\psi\left(\frac{1}{2}+\frac{ \Gamma_+}{2\pi k_{\rm B}T }+{\rm i}\frac{\epsilon -\mu_{\alpha}}{2\pi k_{\rm B}T }\right)-{\rm Re}\psi\left(\frac{1}{2}+\frac{ \Gamma_-}{2\pi k_{\rm B}T }+{\rm i}\frac{\epsilon -\mu_{\alpha}}{2\pi k_{\rm B}T }\right)\\
&\quad\qquad\qquad\qquad-{\rm i} \Bigg[{\rm Im}\psi\left(\frac{1}{2}+\frac{\Gamma_+}{2\pi k_{\rm B}T }+{\rm i}\frac{\epsilon -\mu_{\alpha}}{2\pi k_{\rm B}T }\right)-{\rm Im}\psi\left(\frac{1}{2}+\frac{ \Gamma_-}{2\pi k_{\rm B}T }+{\rm i}\frac{\epsilon -\mu_{\alpha}}{2\pi k_{\rm B}T }\right)\Bigg]\Bigg\}\;.
\end{aligned}\end{equation}

Note that these self-energies maintain the  property of the full NCA4 result in Eq.~\eqref{propertySigma+}.
At equilibrium, $\mu_L=\mu_R=\mu$, and in the degenerate case, $\epsilon_\uparrow=\epsilon_\downarrow=\epsilon_0$,
\begin{equation}\begin{aligned}
\label{Sigma+muEq}
\tilde{\Sigma}_{\sigma\eta}^{(\sigma)}(\epsilon)\simeq &\; -\frac{\eta}{2}\frac{\Gamma}{2\pi}
\Bigg\{\psi^*\left(\frac{1}{2}+\frac{ \Gamma_+}{2\pi k_{\rm B}T }+{\rm i}\frac{\epsilon -\mu}{2\pi k_{\rm B}T }\right)-\psi^*\left(\frac{1}{2}+\frac{ \Gamma_-}{2\pi k_{\rm B}T }+{\rm i}\frac{\epsilon -\mu}{2\pi k_{\rm B}T }\right)\Bigg\}\;.
\end{aligned}\end{equation}
The self-energies are well-behaved in the limit $T\rightarrow 0$, though they do not yield the correct unitary limit for the conductance. Moreover, the correct behavior for the exponent of the expression for the Kondo temperature is not captured by this scheme. Indeed, for $\mu-\epsilon_0, U-\mu+\epsilon_0\gg\Gamma$, from Eq.~\eqref{Delta_b} $\Delta A_+^{\bar\sigma}=\Delta A_+^{\sigma}\sim -\Gamma/\hbar$ and the dressed self-energy $\tilde{\Sigma}_{\sigma-}^{(\sigma)}(\mu)$ is approximated, at low temperature, by
\begin{equation}\begin{aligned}
\label{Sigma+muEq_approx}
\tilde{\Sigma}_{\sigma,-}^{(\sigma)}(\mu)
\simeq &\; \frac{\Gamma}{4\pi}
\left[{\rm Re}\psi\left(\frac{1}{2}+\frac{ 2\Gamma}{2\pi k_{\rm B}T }\right)-{\rm Re}\psi\left(\frac{1}{2}\right)\right]\\
\simeq &\; \frac{\Gamma}{4\pi}
{\rm Re}\psi\left(\frac{1}{2}+\frac{ 2\Gamma}{2\pi k_{\rm B}T }\right)\;.
\end{aligned}\end{equation}
As shown in the main text, the prefactor yields an incorrect exponent in the Kondo temperature. \\
At the particle-hole symmetry point $\mu-\epsilon_0=U/2$,
\begin{equation}\begin{aligned}\label{Sigma+PH}
\tilde{\Sigma}_{\sigma\eta}^{(\sigma)}(\mu) = -\eta\frac{\Gamma}{4\pi}
\Bigg\{{\rm Re}\psi\left(\frac{1}{2}+\frac{ \Gamma_+}{2\pi k_{\rm B}T }\right)-{\rm Re}\psi\left(\frac{1}{2}+\frac{ \Gamma_-}{2\pi k_{\rm B}T }\right)\Bigg\}\;,
\end{aligned}
\end{equation}
where, since $\Delta A_+^{\bar\sigma}=\Delta A_+^{\sigma}=\Delta A_+$, the arguments of the digamma function read  
$$\Gamma_\pm = \Gamma \pm \hbar \Delta A_+ = \Gamma\Bigg[1 \pm\frac{2}{\pi}{\rm Im}\psi\left(\frac{1}{2}+\frac{\Gamma/2}{2\pi k_{\rm B}T}+{\rm i}\frac{U/2}{2\pi k_{\rm B}T}\right)\Bigg]\;.$$
Note that, according to Eq.~\eqref{B3d_3}, 
\begin{equation}
\label{b3PH}    
\tilde{\Sigma}_{3\sigma,\pm }^{(\bar\sigma)} =-{\rm i} \Gamma_\pm\;.
\end{equation}


\section{Evaluation of the dressed bubble $\tilde{\mathbf{B}}^{\sigma(\bar\sigma)}$}
\label{dressingB-}

In the NCA, the dressed bubble $\tilde{\mathbf{B}}^{\sigma(\bar\sigma)}$ is given by iteratively inserting the composite 3rd-level bubble $\tilde{\rm B}_{3}^{\sigma(\bar\sigma)}$ in Eq~\eqref{3rdtier_gDSO4_o} which results in the geometrical series
\begin{equation}\begin{aligned}\label{tildeB2-}
\tilde{\rm B}^{\sigma(\bar\sigma)}_{\eta'\eta}
=&\Biggl< \frac{1}{[{\rm h}_{2}^{\sigma(\bar\sigma)}]^{-1}-\tilde{\rm B}_{3}^{\sigma(\bar\sigma)}}{\rm v}_{-\eta} \Biggr>\;,
\end{aligned}\end{equation}
where the  bare propagator ${\rm h}_{2}^{\sigma(\bar\sigma)}$ is 
given by 
\begin{equation}\label{}
{\rm h}_{2}^{\sigma(\bar\sigma)}={\rm i}\hbar\frac{1}{ \zeta(\epsilon_k-E_\sigma)+ \zeta_{1}(\epsilon_{k_{1}}-E_{\bar\sigma})+{\rm i}0^+}
\end{equation}
and $E_\sigma=\epsilon_\sigma-U/2$.
From Eq.~\eqref{tildeB2-}, the retarded ($\zeta=+1$), dressed self-energy of type $(\bar\sigma)$  reads 
\begin{equation}\begin{aligned}\label{Sigma-WBL-gDSO4-R}
\tilde{\Sigma}_{\sigma,\eta}^{(\bar\sigma)}(\epsilon)
\equiv &{\rm i}\zeta\hbar\tilde{\rm B}^{\sigma(\bar\sigma)}_{\bar\eta\eta}(\kappa)|_{\zeta=+1}\\
=&{\rm i}\zeta\hbar^2\sum_{\boldsymbol{\kappa}_1}\frac{-(|{\rm t}_{\alpha_1}(\epsilon_{k_1})|^2/\hbar^2) f^{\alpha_1}_{-\eta}(\epsilon_{k_1})}{-{\rm i} \zeta(\epsilon_k-\epsilon_\sigma-U/2\hbar)-{\rm i} \zeta_1(\epsilon_{k_1}-\epsilon_{\bar\sigma}-U/2\hbar)-\hbar\tilde{\rm B}_{3}^{\sigma(\bar\sigma)}|_{\zeta=+1}}\\
=&\sum_{\alpha_1}\varrho_{\alpha_1}|{\rm t}_{\alpha_1}|^2 \sum_{\zeta_1}\zeta_1 \int_{-W}^{W} d\epsilon_1\;  \frac{f^{\alpha_1}_{-\eta}(\epsilon_1)}{\epsilon_1-[\epsilon_{\bar\sigma}+\zeta_1(\epsilon_\sigma-\epsilon)+\delta_{\zeta_1,+1}U]-{\rm i}\zeta_1\hbar\tilde{\rm B}_{3}^{\sigma(\bar\sigma)}|_{\zeta=+1}}\\
=&\sum_{\alpha_1}\varrho_{\alpha_1}|{\rm t}_{\alpha_1}|^2 \sum_{\zeta_1}\zeta_1 \int_{-W}^{W} d\epsilon_1\;  \frac{f^{\alpha_1}_{-\eta}(\epsilon_1)}{\epsilon_1-[\epsilon_{\bar\sigma}+\zeta_1(\epsilon_\sigma-\epsilon)+\delta_{\zeta_1,+1}U]-\zeta_1\tilde{\Sigma}_{3\sigma,\zeta_1}^{(\bar\sigma)}}
\;.
\end{aligned}\end{equation}
By summing over $\eta$ and applying Eq.~\eqref{I0}, one finds in the NCA (and in the lower-tier schemes)
\begin{equation}\begin{aligned}\label{SymmetryTildeSigma-}
\sum_\eta \tilde{\Sigma}_{\sigma\eta}^{(\bar\sigma)}(\epsilon)=-{\rm i}\Gamma\;.
\end{aligned}\end{equation}
Within the NCA4, and further assuming the $3$rd-tier self-energies  to be energy-independent, i.e. given by  Eq.~\eqref{B3d_2}, the integral in Eq.~\eqref{Sigma-WBL-gDSO4-R} is readily solved as
\begin{equation}\begin{aligned}\label{SigmaR}
\tilde{\Sigma}_{\sigma\eta}^{(\bar\sigma)}(\epsilon)
&\simeq  \sum_{\alpha}\varrho_{\alpha}|{\rm t}_{\alpha}|^2  \sum_{\zeta_1}\zeta_1  \int_{-W}^{W} d\epsilon_1\;  \frac{f^{\alpha}_{-\eta}(\epsilon_1)}{\epsilon_1-[\epsilon_{\bar\sigma}+\zeta_1(\epsilon_\sigma-\epsilon)+\delta_{\zeta_1,+1}U+\zeta_1{\rm Re}\;\tilde{\Sigma}_{3\sigma,\zeta_1}^{(\bar\sigma)}]-{\rm i}\zeta_1 {\rm Im}\;\tilde{\Sigma}_{3\sigma,\zeta_1}^{(\bar\sigma)}}\\
&=\sum_{\alpha}\varrho_{\alpha}|{\rm t}_{\alpha}|^2  \sum_{\zeta_1}\zeta_1  \int_{-W}^{W} d\epsilon_1\;  \frac{f^{\alpha}_{-\eta}(\epsilon_1)}{\epsilon_1-[\epsilon_{\bar\sigma}+\zeta_1(\epsilon_\sigma-\epsilon)+\delta_{\zeta_1,+1}U+\zeta_1{\rm Re}\;\tilde{\Sigma}_{3\sigma,\zeta_1}^{(\bar\sigma)}]+{\rm i}\zeta_1 |{\rm Im}\;\tilde{\Sigma}_{3\sigma,\zeta_1}^{(\bar\sigma)}|}\\
&=-\eta\sum_{\alpha}\frac{\Gamma_\alpha}{2\pi}\left[\psi\left(\frac{1}{2}+{\rm i}\frac{\tilde{\mathcal{E}}_{+1}-\mu_\alpha}{2\pi k_{\rm B}T}\right)-\psi^*\left(\frac{1}{2}+{\rm i}\frac{\tilde{\mathcal{E}}_{-1}-\mu_\alpha}{2\pi k_{\rm B}T}\right)\right] -{\rm i} \frac{\Gamma}{2}\;,
\end{aligned}\end{equation}
where, in the second line, we used explicitly the fact that the imaginary part of $\tilde{\Sigma}_{3\sigma,\zeta_1}^{(\bar\sigma)}$ is negative, see Eq.~\eqref{B3d}.   
Here, the energies in the arguments of the digamma functions read
\begin{equation}\begin{aligned}\label{Eprimed}
\tilde{\mathcal{E}}_{\zeta_1}=\epsilon_{\bar\sigma}+\zeta_1(\epsilon_\sigma-\epsilon)+\delta_{\zeta_1,+1}U +\zeta_1{\rm Re}\;\tilde{\Sigma}_{3\sigma,\zeta_1}^{(\bar\sigma)}-{\rm i}|{\rm Im}\;\tilde{\Sigma}_{3\sigma,\zeta_1}^{(\bar\sigma)}|\;.
\end{aligned}\end{equation}
In the degenerate case, $\epsilon_\uparrow=\epsilon_\downarrow=\epsilon_0$, at equilibrium,  $\mu_\alpha=\mu$
\begin{equation}
\begin{aligned}\label{SigmaB_sNCA4}
\tilde{\Sigma}_{\sigma\eta}^{(\bar\sigma)}(\epsilon)
=-\eta  \frac{\Gamma}{2\pi} 
\Bigg[
\psi\left(\frac{1}{2}+{\rm i}\frac{2\epsilon_0+U-\epsilon+\tilde{\Sigma}_{3\sigma,+}^{(\bar\sigma)}-\mu}{2\pi k_{\rm B}T}\right)
-\psi^* \left(\frac{1}{2}+{\rm i}\frac{\epsilon - \tilde{\Sigma}_{3\sigma,-}^{(\bar\sigma)*}-\mu}{2\pi k_{\rm B}T}\right)
\Bigg]-{\rm i}\frac{\Gamma}{2}\;.
\end{aligned}
\end{equation}
At the particle-hole symmetry point $\mu-\epsilon_0=U/2$
\begin{equation}\begin{aligned}\label{Sigma-PH}
\tilde{\Sigma}_{\sigma\eta}^{(\bar\sigma)}(\mu)
  &=-\eta\frac{\Gamma}{2\pi} \left[{\rm Re}\psi\left(\frac{1}{2}+\frac{\Gamma_+}{2\pi k_{\rm B}T}\right)-{\rm Re}\psi\left(\frac{1}{2}+\frac{\Gamma_-}{2\pi k_{\rm B}T}\right)\right]-{\rm i}\frac{\Gamma}{2}\;,
\end{aligned}\end{equation}
where we used Eq.~\eqref{b3PH}.

\section{Fermi liquid behavior within the sNCA4}
\label{Fermi_liquid}

Assuming a symmetric coupling,  $\Gamma_L=\Gamma_R=\Gamma/2$, in the degenerate case, $\epsilon_\sigma=\epsilon_0$, Eq.~\eqref{G} yields for the SIAM linear conductance 
\begin{equation}\begin{aligned}\label{GT}
G(T)=G_0\frac{\pi}{2}\int d\epsilon\left(-\frac{\partial f_+(\epsilon)}{\partial\epsilon}\right)g(\epsilon, T)\;,
\end{aligned}\end{equation}
where $G_0=2e^2/h$ is the unit of conductance and  $g(\epsilon,T):=\Gamma [-{\rm Im}\mathcal{G}^r_{\sigma\sigma}(\epsilon,T)/\pi]$. Integrating by parts Eq.~\eqref{GT}, using the Sommerfeld expansion for the resulting integral, and expanding $g(\mu,T)$ to second order in $T$, we obtain the following low-temperature expression for the linear conductance~\cite{Smirnov2013PRB}
\begin{equation}\begin{aligned}\label{Sommerfeld}
G(T)\simeq G_0\frac{\pi}{2}\left[g(\mu,0)+\partial_{T} g(\mu,0)T+\partial_{T}^2g(\mu,0)T^2/2+\pi^2k_{\rm B}^2\partial_{\epsilon}^2 g(\mu,0)T^2/6\right]\;.
\end{aligned}\end{equation}
From Eq.~\eqref{DOS_ph2}, we find for the temperature derivatives of the function $g(\epsilon,T)$, calculated  at the particle-hole symmetry point,
\begin{equation}\begin{aligned}\label{derivatives}
\partial_T g(\mu,0)
&= \frac{3U\Gamma^2}{2\pi}
\frac{\partial_T {\rm Re}\tilde{\Sigma}_{-}(\mu,0)}{[D(\mu,0)]^2}\;,\\
\partial^2_T g(\mu,0) &= \frac{3U\Gamma^2}{2\pi}
\frac{[D(\mu,0)]^2\partial^2_T {\rm Re}\tilde{\Sigma}_{-}(\mu,0)+2UD(\mu,0)[\partial_T {\rm Re}\tilde{\Sigma}_{-}(\mu,0)]^2}{[D(\mu,0)]^4}\;,
\end{aligned}
\end{equation}
where $D(\epsilon,T);= U^2/4+3\Gamma^2/4-U{\rm Re}\tilde{\Sigma}_{\sigma,-}(\epsilon,T)$.
The relation $\psi(1/2+z)=2\psi(2z)-\psi(z)-2\ln(2)$ and the asymptotic expansion of the digamma function $\psi(z)\sim \ln(z)+1/2z-1/12z^2$~\cite{Wolfram_digamma}, give, in the low-temperature limit, with $Z=X+{\rm i}Y$, the approximation
\begin{equation}\begin{aligned}\label{limit_psi}
\psi\left(\frac{1}{2}+\frac{Z}{2\pi k_{\rm B}T}\right)\sim \ln\left(\frac{|Z|}{2\pi k_{\rm B}T}\right) + {\rm i}\arctan\left(\frac{Y}{X}\right)+
\frac{1}{24}\left(\frac{2\pi k_{\rm B}T}{Z}\right)^2\;.
\end{aligned}
\end{equation}
Allowing for a temperature-dependent argument $Z=Z(T)$ and using the above expression we obtain
\begin{equation}\begin{aligned}
\label{derivatives_psi_T}
\partial_T\; \psi\left(\frac{1}{2}+\frac{Z(T)}{2\pi k_{\rm B}T}\right)&\sim -\frac{1}{T}+\frac{(\pi k_{\rm B})^2}{3[Z(T)]^2}T+\left[
\frac{1}{Z(T)}-\frac{(\pi k_{\rm B}T)^2}{3[Z(T)]^3}\right]\partial_T\;Z(T)\;,\\
\partial_T^2\; \psi\left(\frac{1}{2}+\frac{Z(T)}{2\pi k_{\rm B}T}\right)&\sim \frac{1}{T^2}+\frac{(\pi k_{\rm B})^2}{3[Z(T)]^2}+T\partial_T\frac{(\pi k_{\rm B})^2}{3[Z(T)]^2}+\partial_T\left[
\frac{1}{Z(T)}-\frac{(\pi k_{\rm B}T)^2}{3[Z(T)]^3}\right]\partial_T\;Z(T)+\left[
\frac{1}{Z(T)}-\frac{(\pi k_{\rm B}T)^2}{3[Z(T)]^3}\right]\partial_T^2\;Z(T)\;.
\end{aligned}
\end{equation}
These results can be applied to the self-energy $\tilde{\Sigma}_{\sigma,-}(\mu)=\tilde{\Sigma}_{\sigma,-}^{(\sigma)}(\mu)+\tilde{\Sigma}_{\sigma,-}^{(\bar\sigma)}(\mu)$ at the particle-hole symmetry point 
\begin{equation}\begin{aligned}\label{}
\tilde{\Sigma}_{\sigma,-}(\mu,T) =  \frac{3\Gamma}{4\pi}
\Bigg\{{\rm Re}\psi\left(\frac{1}{2}+\frac{ \Gamma_+(T)}{2\pi k_{\rm B}T }\right)-{\rm Re}\psi\left(\frac{1}{2}+\frac{ \Gamma_-(T)}{2\pi k_{\rm B}T }\right)\Bigg\}-{\rm i}\frac{\Gamma}{2}\;,
\end{aligned}
\end{equation}
with 
$$\Gamma_\pm (T) = \Gamma\Bigg[1 \pm\frac{2}{\pi}{\rm Im}\psi\left(\frac{1}{2}+\frac{\Gamma/2}{2\pi k_{\rm B}T}+{\rm i}\frac{U/2}{2\pi k_{\rm B}T}\right)\Bigg]\;.$$
Equation~\eqref{limit_psi} also yields 
\begin{equation}\begin{aligned}
\label{}
\partial_T\Gamma_\pm(T)  = \mp\frac{16 \pi k_{\rm B}^2}{3}\frac{\Gamma^2 U}{(\Gamma^2+U^2)^2}T
\;.
\end{aligned}
\end{equation}
As a result, 
\begin{equation}\begin{aligned}\label{}
\partial_T\tilde{\Sigma}_{\sigma,-}(\mu,0)&=0\;,\\
{\rm and}\qquad\partial_T^2\tilde{\Sigma}_{\sigma,-}(\mu,0) &=  \frac{\Gamma \pi k_{\rm B}^2}{4}
\Bigg\{
\left(\frac{1}{[\Gamma_+(0)]^2}-\frac{1}{[\Gamma_-(0)]^2}\right)-\left(
\frac{1}{\Gamma_+(0)}+\frac{1}{\Gamma_-(0)}\right)\frac{16 \Gamma^2 U}{\pi(\Gamma^2+U^2)^2}
\Bigg\}\;.
\end{aligned}
\end{equation}
From 
Eq.~\eqref{derivatives} 
\begin{equation}\begin{aligned}
\partial_T g(\mu,0)&=0\;,\qquad
\partial_T^2 g(\mu,0)=\frac{3U\Gamma^2}{2\pi}
\frac{\partial^2_T {\rm Re}\tilde{\Sigma}_{\sigma,-}(\mu,0)}{[U^2/4+3\Gamma^2/4-U{\rm Re}\tilde{\Sigma}_{\sigma,-}(\mu,0)]^2}\neq 0\;.
\end{aligned}
\end{equation}
The general expression for the NCA4 Green's function at equilibrium, conveniently rewritten as
\begin{equation}
\label{GRNCA4}
\mathcal{G}^{R}_{\sigma\sigma}(\epsilon)=\frac{\epsilon- \epsilon_\sigma-U+{\rm i}3\Gamma /2+U\langle n_{\bar\sigma}\rangle}{(\epsilon- \epsilon_\sigma+{\rm i}\Gamma /2)(\epsilon- \epsilon_\sigma-U+{\rm i}3\Gamma /2)+U\tilde{\Sigma}_{\sigma,-}(\epsilon)}\;,
\end{equation}
yields for the derivative with respect to $\epsilon$ of the function $g(\epsilon,T)$ calculated at the particle-hole symmetry point and in the degenerate case
\begin{equation}\begin{aligned}\label{}
\partial^2_\epsilon g(\mu,0) =&\frac{\Gamma/\pi}{[D(\mu,0)]^2}\left[7\Gamma + 2U\partial_\epsilon{\rm Re}\tilde{\Sigma}_{\sigma,-}(\mu,0)+(3/2)\Gamma U\partial^2_\epsilon{\rm Re}\tilde{\Sigma}_{\sigma,-}(\mu,0)\right]\\
&+\frac{\Gamma/\pi}{[D(\mu,0)]^3}\left\{12\Gamma^3-3\Gamma U^2\left[\partial_\epsilon{\rm Re}\tilde{\Sigma}_{\sigma,-}(\mu,0)\right]^2\right\}\;.
\end{aligned}
\end{equation}
For an energy-independent argument $Z$ we have for the derivatives with respect to the energy~\cite{Lahiri2020}
\begin{equation}\begin{aligned}
\label{derivatives_psi_e}
\partial_\epsilon\; \psi\left(\frac{1}{2}+\frac{Z}{2\pi k_{\rm B}T}+{\rm i}\frac{\epsilon-\mu}{2\pi k_{\rm B}T}\right)\Bigg|_{\epsilon=\mu,\; T=0}&=\frac{\sin(\varphi)+{\rm i}\cos(\varphi)}{|Z|}\;,\\
\
\partial_\epsilon^2\; \psi\left(\frac{1}{2}+\frac{Z}{2\pi k_{\rm B}T}+{\rm i}\frac{\epsilon-\mu}{2\pi k_{\rm B}T}\right)\Bigg|_{\epsilon=\mu,\; T=0}&=\frac{\cos(2\varphi)-{\rm i}\sin(2\varphi)}{|Z|^2}\;,
\end{aligned}
\end{equation}
where $Z=|Z|\exp({\rm i}\varphi)$. Using Eqs.~\eqref{Sigma+muEq} and~\eqref{SigmaB_sNCA4},  this entails 
\begin{equation}\begin{aligned}
\label{}
\partial_\epsilon{\rm Re}\tilde{\Sigma}_{\sigma,-}(\mu,0) =& -\frac{\Gamma}{2\pi}\frac{\sin(\varphi_+)}{|Z_+|}\;,\\
\partial^2_\epsilon{\rm Re}\tilde{\Sigma}_{\sigma,-}(\mu,0) =& \frac{\Gamma}{2\pi}\left[\frac{\cos(2\varphi_+)}{|Z_+|}-\frac{1}{\Gamma_-}\right]+\frac{\Gamma}{4\pi}\left[\frac{1}{\Gamma_+^2}-\frac{1}{\Gamma_-^2}\right]
\;,
\end{aligned}
\end{equation}
where $\varphi_+=\arctan(2\mu/\Gamma_+)$ and $|Z_+|=\sqrt{(2\mu)^2+\Gamma_+^2}$. 
The vanishing linear term in the low-temperature expansion of the linear conductance, Eq.~\eqref{Sommerfeld}, and the resulting quadratic dependence on $T$ indicate that the sNCA4 displays at low temperature a Fermi liquid behavior. Nevertheless, the saturation value at $T=0$ differs in the sNCA4 from the correct value $g(\mu, 0)=2/\pi$.

\newpage

\twocolumngrid


\begin{thebibliography}{95}%
\makeatletter
\providecommand \@ifxundefined [1]{%
 \@ifx{#1\undefined}
}%
\providecommand \@ifnum [1]{%
 \ifnum #1\expandafter \@firstoftwo
 \else \expandafter \@secondoftwo
 \fi
}%
\providecommand \@ifx [1]{%
 \ifx #1\expandafter \@firstoftwo
 \else \expandafter \@secondoftwo
 \fi
}%
\providecommand \natexlab [1]{#1}%
\providecommand \enquote  [1]{``#1''}%
\providecommand \bibnamefont  [1]{#1}%
\providecommand \bibfnamefont [1]{#1}%
\providecommand \citenamefont [1]{#1}%
\providecommand \href@noop [0]{\@secondoftwo}%
\providecommand \href [0]{\begingroup \@sanitize@url \@href}%
\providecommand \@href[1]{\@@startlink{#1}\@@href}%
\providecommand \@@href[1]{\endgroup#1\@@endlink}%
\providecommand \@sanitize@url [0]{\catcode `\\12\catcode `\$12\catcode
  `\&12\catcode `\#12\catcode `\^12\catcode `\_12\catcode `\%12\relax}%
\providecommand \@@startlink[1]{}%
\providecommand \@@endlink[0]{}%
\providecommand \url  [0]{\begingroup\@sanitize@url \@url }%
\providecommand \@url [1]{\endgroup\@href {#1}{\urlprefix }}%
\providecommand \urlprefix  [0]{URL }%
\providecommand \Eprint [0]{\href }%
\providecommand \doibase [0]{http://dx.doi.org/}%
\providecommand \selectlanguage [0]{\@gobble}%
\providecommand \bibinfo  [0]{\@secondoftwo}%
\providecommand \bibfield  [0]{\@secondoftwo}%
\providecommand \translation [1]{[#1]}%
\providecommand \BibitemOpen [0]{}%
\providecommand \bibitemStop [0]{}%
\providecommand \bibitemNoStop [0]{.\EOS\space}%
\providecommand \EOS [0]{\spacefactor3000\relax}%
\providecommand \BibitemShut  [1]{\csname bibitem#1\endcsname}%
\let\auto@bib@innerbib\@empty
\bibitem [{\citenamefont {Anderson}(1961)}]{Anderson1961}%
  \BibitemOpen
  \bibfield  {author} {\bibinfo {author} {\bibfnamefont {P.~W.}\ \bibnamefont
  {Anderson}},\ }\bibfield  {title} {\enquote {\bibinfo {title} {{Localized
  Magnetic States in Metals}},}\ }\href {\doibase 10.1103/PhysRev.124.41}
  {\bibfield  {journal} {\bibinfo  {journal} {Phys. Rev.}\ }\textbf {\bibinfo
  {volume} {124}},\ \bibinfo {pages} {41--53} (\bibinfo {year}
  {1961})}\BibitemShut {NoStop}%
\bibitem [{\citenamefont {Ng}\ and\ \citenamefont {Lee}(1988)}]{Ng1988}%
  \BibitemOpen
  \bibfield  {author} {\bibinfo {author} {\bibfnamefont {T.~K.}\ \bibnamefont
  {Ng}}\ and\ \bibinfo {author} {\bibfnamefont {P.~A.}\ \bibnamefont {Lee}},\
  }\bibfield  {title} {\enquote {\bibinfo {title} {{On-Site Coulomb Repulsion
  and Resonant Tunneling}},}\ }\href {\doibase 10.1103/PhysRevLett.61.1768}
  {\bibfield  {journal} {\bibinfo  {journal} {Phys. Rev. Lett.}\ }\textbf
  {\bibinfo {volume} {61}},\ \bibinfo {pages} {1768--1771} (\bibinfo {year}
  {1988})}\BibitemShut {NoStop}%
\bibitem [{\citenamefont {Glazman}\ and\ \citenamefont
  {Rahik}(1988)}]{Glazman1988}%
  \BibitemOpen
  \bibfield  {author} {\bibinfo {author} {\bibfnamefont {L.}~\bibnamefont
  {Glazman}}\ and\ \bibinfo {author} {\bibfnamefont {M.}~\bibnamefont
  {Rahik}},\ }\bibfield  {title} {\enquote {\bibinfo {title} {{Resonant Kondo
  transparency of a barrier with quasilocal impurity states}},}\ }\href@noop {}
  {\bibfield  {journal} {\bibinfo  {journal} {JETP Lett.}\ }\textbf {\bibinfo
  {volume} {11}},\ \bibinfo {pages} {2389} (\bibinfo {year}
  {1988})}\BibitemShut {NoStop}%
\bibitem [{\citenamefont {Hewson}(1993)}]{Hewson1993}%
  \BibitemOpen
  \bibfield  {author} {\bibinfo {author} {\bibfnamefont {A.P.}\ \bibnamefont
  {Hewson}},\ }\href@noop {} {\emph {\bibinfo {title} {{The Kondo problem to
  heavy fermions}}}}\ (\bibinfo  {publisher} {Cambridge University Press,
  Cambridge},\ \bibinfo {year} {1993})\BibitemShut {NoStop}%
\bibitem [{\citenamefont {van~der Wiel}\ \emph {et~al.}(2000)\citenamefont
  {van~der Wiel}, \citenamefont {{De Franceschi}}, \citenamefont {Fujisawa},
  \citenamefont {Elzerman}, \citenamefont {Tarucha},\ and\ \citenamefont
  {Kouwenhoven}}]{vanderWiel2000}%
  \BibitemOpen
  \bibfield  {author} {\bibinfo {author} {\bibfnamefont {W.~G.}\ \bibnamefont
  {van~der Wiel}}, \bibinfo {author} {\bibfnamefont {S.}~\bibnamefont {{De
  Franceschi}}}, \bibinfo {author} {\bibfnamefont {T.}~\bibnamefont
  {Fujisawa}}, \bibinfo {author} {\bibfnamefont {J.~M.}\ \bibnamefont
  {Elzerman}}, \bibinfo {author} {\bibfnamefont {S.}~\bibnamefont {Tarucha}}, \
  and\ \bibinfo {author} {\bibfnamefont {L.~P.}\ \bibnamefont {Kouwenhoven}},\
  }\bibfield  {title} {\enquote {\bibinfo {title} {{The Kondo Effect in the
  Unitary Limit}},}\ }\href {\doibase 10.1126/science.289.5487.2105} {\bibfield
   {journal} {\bibinfo  {journal} {Science}\ }\textbf {\bibinfo {volume}
  {289}},\ \bibinfo {pages} {2105} (\bibinfo {year} {2000})}\BibitemShut
  {NoStop}%
\bibitem [{\citenamefont {Kouwenhoven}\ and\ \citenamefont
  {Glazman}(2001)}]{Kouwenhoven2001}%
  \BibitemOpen
  \bibfield  {author} {\bibinfo {author} {\bibfnamefont {L.}~\bibnamefont
  {Kouwenhoven}}\ and\ \bibinfo {author} {\bibfnamefont {L.}~\bibnamefont
  {Glazman}},\ }\bibfield  {title} {\enquote {\bibinfo {title} {{Revival of the
  Kondo effect}},}\ }\href {\doibase 10.1088/2058-7058/14/1/28} {\bibfield
  {journal} {\bibinfo  {journal} {Physics World}\ }\textbf {\bibinfo {volume}
  {14}},\ \bibinfo {pages} {33--38} (\bibinfo {year} {2001})}\BibitemShut
  {NoStop}%
\bibitem [{\citenamefont {Wilson}(1975)}]{Wilson1975}%
  \BibitemOpen
  \bibfield  {author} {\bibinfo {author} {\bibfnamefont {K.~G.}\ \bibnamefont
  {Wilson}},\ }\bibfield  {title} {\enquote {\bibinfo {title} {{The
  renormalization group: Critical phenomena and the Kondo problem}},}\ }\href
  {\doibase 10.1103/RevModPhys.47.773} {\bibfield  {journal} {\bibinfo
  {journal} {Rev. Mod. Phys.}\ }\textbf {\bibinfo {volume} {47}},\ \bibinfo
  {pages} {773--840} (\bibinfo {year} {1975})}\BibitemShut {NoStop}%
\bibitem [{\citenamefont {Bulla}\ \emph {et~al.}(2008)\citenamefont {Bulla},
  \citenamefont {Costi},\ and\ \citenamefont {Pruschke}}]{Bulla2008}%
  \BibitemOpen
  \bibfield  {author} {\bibinfo {author} {\bibfnamefont {R.}~\bibnamefont
  {Bulla}}, \bibinfo {author} {\bibfnamefont {T.~A.}\ \bibnamefont {Costi}}, \
  and\ \bibinfo {author} {\bibfnamefont {T.}~\bibnamefont {Pruschke}},\
  }\bibfield  {title} {\enquote {\bibinfo {title} {{Numerical renormalization
  group method for quantum impurity systems}},}\ }\href {\doibase
  10.1103/RevModPhys.80.395} {\bibfield  {journal} {\bibinfo  {journal} {Rev.
  Mod. Phys.}\ }\textbf {\bibinfo {volume} {80}},\ \bibinfo {pages} {395--450}
  (\bibinfo {year} {2008})}\BibitemShut {NoStop}%
\bibitem [{\citenamefont {White}(1992)}]{White1992}%
  \BibitemOpen
  \bibfield  {author} {\bibinfo {author} {\bibfnamefont {S.~R.}\ \bibnamefont
  {White}},\ }\bibfield  {title} {\enquote {\bibinfo {title} {{Density matrix
  formulation for quantum renormalization groups}},}\ }\href {\doibase
  10.1103/PhysRevLett.69.2863} {\bibfield  {journal} {\bibinfo  {journal}
  {Phys. Rev. Lett.}\ }\textbf {\bibinfo {volume} {69}},\ \bibinfo {pages}
  {2863--2866} (\bibinfo {year} {1992})}\BibitemShut {NoStop}%
\bibitem [{\citenamefont {Schollw{\"o}ck}(2011)}]{Schollwoeck2011}%
  \BibitemOpen
  \bibfield  {author} {\bibinfo {author} {\bibfnamefont {U.}~\bibnamefont
  {Schollw{\"o}ck}},\ }\bibfield  {title} {\enquote {\bibinfo {title} {{The
  density-matrix renormalization group in the age of matrix product states}},}\
  }\href {\doibase 10.1016/j.aop.2010.09.012} {\bibfield  {journal} {\bibinfo
  {journal} {Ann. Phys.}\ }\textbf {\bibinfo {volume} {326}},\ \bibinfo {pages}
  {96--192} (\bibinfo {year} {2011})}\BibitemShut {NoStop}%
\bibitem [{\citenamefont {Anders}\ and\ \citenamefont
  {Schiller}(2005)}]{Anders2005}%
  \BibitemOpen
  \bibfield  {author} {\bibinfo {author} {\bibfnamefont {F.~B.}\ \bibnamefont
  {Anders}}\ and\ \bibinfo {author} {\bibfnamefont {A.}~\bibnamefont
  {Schiller}},\ }\bibfield  {title} {\enquote {\bibinfo {title} {{Real-Time
  Dynamics in Quantum-Impurity Systems: A Time-Dependent Numerical
  Renormalization-Group Approach}},}\ }\href {\doibase
  10.1103/PhysRevLett.95.196801} {\bibfield  {journal} {\bibinfo  {journal}
  {Phys. Rev. Lett.}\ }\textbf {\bibinfo {volume} {95}},\ \bibinfo {pages}
  {196801} (\bibinfo {year} {2005})}\BibitemShut {NoStop}%
\bibitem [{\citenamefont {Heidrich-Meisner}\ \emph {et~al.}(2009)\citenamefont
  {Heidrich-Meisner}, \citenamefont {Feiguin},\ and\ \citenamefont
  {Dagotto}}]{Heidrich-Meisner2009}%
  \BibitemOpen
  \bibfield  {author} {\bibinfo {author} {\bibfnamefont {F.}~\bibnamefont
  {Heidrich-Meisner}}, \bibinfo {author} {\bibfnamefont {A.~E.}\ \bibnamefont
  {Feiguin}}, \ and\ \bibinfo {author} {\bibfnamefont {E.}~\bibnamefont
  {Dagotto}},\ }\bibfield  {title} {\enquote {\bibinfo {title} {{Real-time
  simulations of nonequilibrium transport in the single-impurity Anderson
  model}},}\ }\href {\doibase 10.1103/PhysRevB.79.235336} {\bibfield  {journal}
  {\bibinfo  {journal} {Phys. Rev. B}\ }\textbf {\bibinfo {volume} {79}},\
  \bibinfo {pages} {235336} (\bibinfo {year} {2009})}\BibitemShut {NoStop}%
\bibitem [{\citenamefont {Weiss}\ \emph {et~al.}(2008)\citenamefont {Weiss},
  \citenamefont {Eckel}, \citenamefont {Thorwart},\ and\ \citenamefont
  {Egger}}]{Weiss2008}%
  \BibitemOpen
  \bibfield  {author} {\bibinfo {author} {\bibfnamefont {S.}~\bibnamefont
  {Weiss}}, \bibinfo {author} {\bibfnamefont {J.}~\bibnamefont {Eckel}},
  \bibinfo {author} {\bibfnamefont {M.}~\bibnamefont {Thorwart}}, \ and\
  \bibinfo {author} {\bibfnamefont {R.}~\bibnamefont {Egger}},\ }\bibfield
  {title} {\enquote {\bibinfo {title} {{Iterative real-time path integral
  approach to nonequilibrium quantum transport}},}\ }\href {\doibase
  10.1103/PhysRevB.77.195316} {\bibfield  {journal} {\bibinfo  {journal} {Phys.
  Rev. B}\ }\textbf {\bibinfo {volume} {77}},\ \bibinfo {pages} {195316}
  (\bibinfo {year} {2008})}\BibitemShut {NoStop}%
\bibitem [{\citenamefont {M{\"u}hlbacher}\ and\ \citenamefont
  {Rabani}(2008)}]{Muehlbacher2008}%
  \BibitemOpen
  \bibfield  {author} {\bibinfo {author} {\bibfnamefont {L.}~\bibnamefont
  {M{\"u}hlbacher}}\ and\ \bibinfo {author} {\bibfnamefont {E.}~\bibnamefont
  {Rabani}},\ }\bibfield  {title} {\enquote {\bibinfo {title} {{Real-Time Path
  Integral Approach to Nonequilibrium Many-Body Quantum Systems}},}\ }\href
  {\doibase 10.1103/PhysRevLett.100.176403} {\bibfield  {journal} {\bibinfo
  {journal} {Phys. Rev. Lett.}\ }\textbf {\bibinfo {volume} {100}},\ \bibinfo
  {pages} {176403} (\bibinfo {year} {2008})}\BibitemShut {NoStop}%
\bibitem [{\citenamefont {Jin}\ \emph {et~al.}(2008)\citenamefont {Jin},
  \citenamefont {Zheng},\ and\ \citenamefont {Yan}}]{Jin2008}%
  \BibitemOpen
  \bibfield  {author} {\bibinfo {author} {\bibfnamefont {J.}~\bibnamefont
  {Jin}}, \bibinfo {author} {\bibfnamefont {X.}~\bibnamefont {Zheng}}, \ and\
  \bibinfo {author} {\bibfnamefont {Y.~J.}\ \bibnamefont {Yan}},\ }\bibfield
  {title} {\enquote {\bibinfo {title} {{Exact dynamics of dissipative
  electronic systems and quantum transport: Hierarchical equations of motion
  approach}},}\ }\href {\doibase 10.1063/1.2938087} {\bibfield  {journal}
  {\bibinfo  {journal} {J. Chem. Phys.}\ }\textbf {\bibinfo {volume} {128}},\
  \bibinfo {pages} {234703} (\bibinfo {year} {2008})}\BibitemShut {NoStop}%
\bibitem [{\citenamefont {Schinabeck}\ \emph {et~al.}(2016)\citenamefont
  {Schinabeck}, \citenamefont {Erpenbeck}, \citenamefont {H{\"a}rtle},\ and\
  \citenamefont {Thoss}}]{Schinabeck2016}%
  \BibitemOpen
  \bibfield  {author} {\bibinfo {author} {\bibfnamefont {C.}~\bibnamefont
  {Schinabeck}}, \bibinfo {author} {\bibfnamefont {A.}~\bibnamefont
  {Erpenbeck}}, \bibinfo {author} {\bibfnamefont {R.}~\bibnamefont
  {H{\"a}rtle}}, \ and\ \bibinfo {author} {\bibfnamefont {M.}~\bibnamefont
  {Thoss}},\ }\bibfield  {title} {\enquote {\bibinfo {title} {{Hierarchical
  quantum master equation approach to electronic-vibrational coupling in
  nonequilibrium transport through nanosystems}},}\ }\href {\doibase
  10.1103/PhysRevB.94.201407} {\bibfield  {journal} {\bibinfo  {journal} {Phys.
  Rev. B}\ }\textbf {\bibinfo {volume} {94}},\ \bibinfo {pages} {201407}
  (\bibinfo {year} {2016})}\BibitemShut {NoStop}%
\bibitem [{\citenamefont {Arrigoni}\ \emph {et~al.}(2013)\citenamefont
  {Arrigoni}, \citenamefont {Knap},\ and\ \citenamefont {von~der
  Linden}}]{Arrigoni2013}%
  \BibitemOpen
  \bibfield  {author} {\bibinfo {author} {\bibfnamefont {E.}~\bibnamefont
  {Arrigoni}}, \bibinfo {author} {\bibfnamefont {M.}~\bibnamefont {Knap}}, \
  and\ \bibinfo {author} {\bibfnamefont {W.}~\bibnamefont {von~der Linden}},\
  }\bibfield  {title} {\enquote {\bibinfo {title} {{Nonequilibrium Dynamical
  Mean-Field Theory: An Auxiliary Quantum Master Equation Approach}},}\ }\href
  {\doibase 10.1103/PhysRevLett.110.086403} {\bibfield  {journal} {\bibinfo
  {journal} {Phys. Rev. Lett.}\ }\textbf {\bibinfo {volume} {110}},\ \bibinfo
  {pages} {086403--} (\bibinfo {year} {2013})}\BibitemShut {NoStop}%
\bibitem [{\citenamefont {Dorda}\ \emph {et~al.}(2014)\citenamefont {Dorda},
  \citenamefont {Nuss}, \citenamefont {von~der Linden},\ and\ \citenamefont
  {Arrigoni}}]{Arrigoni2014}%
  \BibitemOpen
  \bibfield  {author} {\bibinfo {author} {\bibfnamefont {A.}~\bibnamefont
  {Dorda}}, \bibinfo {author} {\bibfnamefont {M.}~\bibnamefont {Nuss}},
  \bibinfo {author} {\bibfnamefont {W.}~\bibnamefont {von~der Linden}}, \ and\
  \bibinfo {author} {\bibfnamefont {E.}~\bibnamefont {Arrigoni}},\ }\bibfield
  {title} {\enquote {\bibinfo {title} {{Auxiliary master equation approach to
  nonequilibrium correlated impurities}},}\ }\href {\doibase
  10.1103/PhysRevB.89.165105} {\bibfield  {journal} {\bibinfo  {journal} {Phys.
  Rev. B}\ }\textbf {\bibinfo {volume} {89}},\ \bibinfo {pages} {165105}
  (\bibinfo {year} {2014})}\BibitemShut {NoStop}%
\bibitem [{\citenamefont {Fugger}\ \emph {et~al.}(2018)\citenamefont {Fugger},
  \citenamefont {Dorda}, \citenamefont {Schwarz}, \citenamefont {von Delft},\
  and\ \citenamefont {Arrigoni}}]{Fugger2018}%
  \BibitemOpen
  \bibfield  {author} {\bibinfo {author} {\bibfnamefont {D.~M}\ \bibnamefont
  {Fugger}}, \bibinfo {author} {\bibfnamefont {A.}~\bibnamefont {Dorda}},
  \bibinfo {author} {\bibfnamefont {F.}~\bibnamefont {Schwarz}}, \bibinfo
  {author} {\bibfnamefont {J.}~\bibnamefont {von Delft}}, \ and\ \bibinfo
  {author} {\bibfnamefont {E.}~\bibnamefont {Arrigoni}},\ }\bibfield  {title}
  {\enquote {\bibinfo {title} {{Nonequilibrium Kondo effect in a magnetic
  field: auxiliary master equation approach}},}\ }\href {\doibase
  10.1088/1367-2630/aa9fdc} {\bibfield  {journal} {\bibinfo  {journal} {New J.
  Phys.}\ }\textbf {\bibinfo {volume} {20}},\ \bibinfo {pages} {013030}
  (\bibinfo {year} {2018})}\BibitemShut {NoStop}%
\bibitem [{\citenamefont {Eckel}\ \emph {et~al.}(2010)\citenamefont {Eckel},
  \citenamefont {Heidrich-Meisner}, \citenamefont {Jakobs}, \citenamefont
  {Thorwart}, \citenamefont {Pletyukhov},\ and\ \citenamefont
  {Egger}}]{Eckel2010}%
  \BibitemOpen
  \bibfield  {author} {\bibinfo {author} {\bibfnamefont {J.}~\bibnamefont
  {Eckel}}, \bibinfo {author} {\bibfnamefont {F.}~\bibnamefont
  {Heidrich-Meisner}}, \bibinfo {author} {\bibfnamefont {S.~G.}\ \bibnamefont
  {Jakobs}}, \bibinfo {author} {\bibfnamefont {M.}~\bibnamefont {Thorwart}},
  \bibinfo {author} {\bibfnamefont {M.}~\bibnamefont {Pletyukhov}}, \ and\
  \bibinfo {author} {\bibfnamefont {R.}~\bibnamefont {Egger}},\ }\bibfield
  {title} {\enquote {\bibinfo {title} {{Comparative study of theoretical
  methods for non-equilibrium quantum transport}},}\ }\href {\doibase
  10.1088/1367-2630/12/4/043042} {\bibfield  {journal} {\bibinfo  {journal}
  {New J. Phys.}\ }\textbf {\bibinfo {volume} {12}},\ \bibinfo {pages} {043042}
  (\bibinfo {year} {2010})}\BibitemShut {NoStop}%
\bibitem [{\citenamefont {Meir}\ and\ \citenamefont
  {Wingreen}(1992)}]{Meir1992}%
  \BibitemOpen
  \bibfield  {author} {\bibinfo {author} {\bibfnamefont {Y.}~\bibnamefont
  {Meir}}\ and\ \bibinfo {author} {\bibfnamefont {N.~S.}\ \bibnamefont
  {Wingreen}},\ }\bibfield  {title} {\enquote {\bibinfo {title} {{Landauer
  formula for the current through an interacting electron region}},}\ }\href
  {\doibase 10.1103/PhysRevLett.68.2512} {\bibfield  {journal} {\bibinfo
  {journal} {Phys. Rev. Lett.}\ }\textbf {\bibinfo {volume} {68}},\ \bibinfo
  {pages} {2512--2515} (\bibinfo {year} {1992})}\BibitemShut {NoStop}%
\bibitem [{\citenamefont {Haug}\ and\ \citenamefont {Jahuo}(2008)}]{Haug2008}%
  \BibitemOpen
  \bibfield  {author} {\bibinfo {author} {\bibfnamefont {H.~J.~W.}\
  \bibnamefont {Haug}}\ and\ \bibinfo {author} {\bibfnamefont {A.-P.}\
  \bibnamefont {Jahuo}},\ }\href@noop {} {\emph {\bibinfo {title} {{Quantum
  Kinetics in transport and optics of semiconductors, Springer Series in
  Solid-State Science}}}}\ (\bibinfo  {publisher} {Springer-Verlag, Berlin 2nd
  Ed.},\ \bibinfo {year} {2008})\BibitemShut {NoStop}%
\bibitem [{\citenamefont {Scheer}\ and\ \citenamefont
  {Cuevas}(2010)}]{Scheer2010}%
  \BibitemOpen
  \bibfield  {author} {\bibinfo {author} {\bibfnamefont {E.}~\bibnamefont
  {Scheer}}\ and\ \bibinfo {author} {\bibfnamefont {J.~C.}\ \bibnamefont
  {Cuevas}},\ }\href@noop {} {\emph {\bibinfo {title} {{Molecular electronics:
  an introduction to theory and experiment}}}}\ (\bibinfo  {publisher} {World
  Scientific, Singapore},\ \bibinfo {year} {2010})\BibitemShut {NoStop}%
\bibitem [{\citenamefont {Ryndyk}(2016)}]{Ryndyk2016}%
  \BibitemOpen
  \bibfield  {author} {\bibinfo {author} {\bibfnamefont {D.}~\bibnamefont
  {Ryndyk}},\ }\href@noop {} {\emph {\bibinfo {title} {{Theory of Quantum
  Transport at Nanoscale}}}}\ (\bibinfo  {publisher} {Springer, New York},\
  \bibinfo {year} {2016})\BibitemShut {NoStop}%
\bibitem [{\citenamefont {Thoss}\ and\ \citenamefont
  {Evers}(2018)}]{Thoss2018}%
  \BibitemOpen
  \bibfield  {author} {\bibinfo {author} {\bibfnamefont {M.}~\bibnamefont
  {Thoss}}\ and\ \bibinfo {author} {\bibfnamefont {F.}~\bibnamefont {Evers}},\
  }\bibfield  {title} {\enquote {\bibinfo {title} {{Perspective: Theory of
  quantum transport in molecular junctions}},}\ }\href {\doibase
  10.1063/1.5003306} {\bibfield  {journal} {\bibinfo  {journal} {J. Chem.
  Phys.}\ }\textbf {\bibinfo {volume} {148}},\ \bibinfo {pages} {030901}
  (\bibinfo {year} {2018})}\BibitemShut {NoStop}%
\bibitem [{\citenamefont {Cohen}\ and\ \citenamefont
  {Galperin}(2020)}]{Cohen2020}%
  \BibitemOpen
  \bibfield  {author} {\bibinfo {author} {\bibfnamefont {G.}~\bibnamefont
  {Cohen}}\ and\ \bibinfo {author} {\bibfnamefont {M.}~\bibnamefont
  {Galperin}},\ }\bibfield  {title} {\enquote {\bibinfo {title} {{Green's
  function methods for single molecule junctions}},}\ }\href {\doibase
  10.1063/1.5145210} {\bibfield  {journal} {\bibinfo  {journal} {J. Chem.
  Phys.}\ }\textbf {\bibinfo {volume} {152}},\ \bibinfo {pages} {090901}
  (\bibinfo {year} {2020})}\BibitemShut {NoStop}%
\bibitem [{\citenamefont {Evers}\ \emph {et~al.}(2020)\citenamefont {Evers},
  \citenamefont {Koryt{\'a}r}, \citenamefont {Tewari},\ and\ \citenamefont {van
  Ruitenbeek}}]{Evers2020}%
  \BibitemOpen
  \bibfield  {author} {\bibinfo {author} {\bibfnamefont {F.}~\bibnamefont
  {Evers}}, \bibinfo {author} {\bibfnamefont {R.}~\bibnamefont {Koryt{\'a}r}},
  \bibinfo {author} {\bibfnamefont {S.}~\bibnamefont {Tewari}}, \ and\ \bibinfo
  {author} {\bibfnamefont {J.~M.}\ \bibnamefont {van Ruitenbeek}},\ }\bibfield
  {title} {\enquote {\bibinfo {title} {{Advances and challenges in
  single-molecule electron transport}},}\ }\href {\doibase
  10.1103/RevModPhys.92.035001} {\bibfield  {journal} {\bibinfo  {journal}
  {Rev. Mod. Phys.}\ }\textbf {\bibinfo {volume} {92}},\ \bibinfo {pages}
  {035001} (\bibinfo {year} {2020})}\BibitemShut {NoStop}%
\bibitem [{\citenamefont {Blum}(2012)}]{Blum2012}%
  \BibitemOpen
  \bibfield  {author} {\bibinfo {author} {\bibfnamefont {K.}~\bibnamefont
  {Blum}},\ }\href@noop {} {\emph {\bibinfo {title} {{Density matrix theory and
  applications}}}},\ Vol.~\bibinfo {volume} {64}\ (\bibinfo  {publisher}
  {Springer, Berlin},\ \bibinfo {year} {2012})\BibitemShut {NoStop}%
\bibitem [{\citenamefont {Schoeller}(1997)}]{Schoeller1997}%
  \BibitemOpen
  \bibfield  {author} {\bibinfo {author} {\bibfnamefont {H.}~\bibnamefont
  {Schoeller}},\ }\enquote {\bibinfo {title} {{Transport Theory of Interacting
  Quantum Dots}},}\ in\ \href@noop {} {\emph {\bibinfo {booktitle} {{Mesoscopic
  Electron Transport}}}},\ \bibinfo {series and number} {{Nato Science Series
  E}}\ (\bibinfo  {publisher} {Kluwer, Dordrecht},\ \bibinfo {year} {1997})\
  pp.\ \bibinfo {pages} {291--330}\BibitemShut {NoStop}%
\bibitem [{\citenamefont {Timm}(2008)}]{Timm2008}%
  \BibitemOpen
  \bibfield  {author} {\bibinfo {author} {\bibfnamefont {C.}~\bibnamefont
  {Timm}},\ }\bibfield  {title} {\enquote {\bibinfo {title} {{Tunneling through
  molecules and quantum dots: Master-equation approaches}},}\ }\href {\doibase
  10.1103/PhysRevB.77.195416} {\bibfield  {journal} {\bibinfo  {journal} {Phys.
  Rev. B}\ }\textbf {\bibinfo {volume} {77}},\ \bibinfo {pages} {195416}
  (\bibinfo {year} {2008})}\BibitemShut {NoStop}%
\bibitem [{\citenamefont {Timm}(2011)}]{Timm2011}%
  \BibitemOpen
  \bibfield  {author} {\bibinfo {author} {\bibfnamefont {C.}~\bibnamefont
  {Timm}},\ }\bibfield  {title} {\enquote {\bibinfo {title}
  {{Time-convolutionless master equation for quantum dots: Perturbative
  expansion to arbitrary order}},}\ }\href {\doibase
  10.1103/PhysRevB.83.115416} {\bibfield  {journal} {\bibinfo  {journal} {Phys.
  Rev. B}\ }\textbf {\bibinfo {volume} {83}},\ \bibinfo {pages} {115416}
  (\bibinfo {year} {2011})}\BibitemShut {NoStop}%
\bibitem [{\citenamefont {Andergassen}\ \emph {et~al.}(2010)\citenamefont
  {Andergassen}, \citenamefont {Meden}, \citenamefont {Schoeller},
  \citenamefont {Splettstoesser},\ and\ \citenamefont
  {Wegewijs}}]{Andergassen2010}%
  \BibitemOpen
  \bibfield  {author} {\bibinfo {author} {\bibfnamefont {S.}~\bibnamefont
  {Andergassen}}, \bibinfo {author} {\bibfnamefont {V.}~\bibnamefont {Meden}},
  \bibinfo {author} {\bibfnamefont {H.}~\bibnamefont {Schoeller}}, \bibinfo
  {author} {\bibfnamefont {J.}~\bibnamefont {Splettstoesser}}, \ and\ \bibinfo
  {author} {\bibfnamefont {M.~R.}\ \bibnamefont {Wegewijs}},\ }\bibfield
  {title} {\enquote {\bibinfo {title} {{Charge transport through single
  molecules, quantum dots and quantum wires}},}\ }\href {\doibase
  10.1088/0957-4484/21/27/272001} {\bibfield  {journal} {\bibinfo  {journal}
  {Nanotechnology}\ }\textbf {\bibinfo {volume} {21}},\ \bibinfo {pages}
  {272001} (\bibinfo {year} {2010})}\BibitemShut {NoStop}%
\bibitem [{\citenamefont {Levy}\ and\ \citenamefont {Rabani}(2013)}]{Levy2013}%
  \BibitemOpen
  \bibfield  {author} {\bibinfo {author} {\bibfnamefont {T.~J.}\ \bibnamefont
  {Levy}}\ and\ \bibinfo {author} {\bibfnamefont {E.}~\bibnamefont {Rabani}},\
  }\bibfield  {title} {\enquote {\bibinfo {title} {{Symmetry breaking and
  restoration using the equation-of-motion technique for nonequilibrium quantum
  impurity models}},}\ }\href {\doibase 10.1088/0953-8984/25/11/115302}
  {\bibfield  {journal} {\bibinfo  {journal} {J. Phys.: Condens. Matter}\
  }\textbf {\bibinfo {volume} {25}},\ \bibinfo {pages} {115302} (\bibinfo
  {year} {2013})}\BibitemShut {NoStop}%
\bibitem [{\citenamefont {Koller}\ \emph {et~al.}(2010)\citenamefont {Koller},
  \citenamefont {Grifoni}, \citenamefont {Leijnse},\ and\ \citenamefont
  {Wegewijs}}]{Koller2010}%
  \BibitemOpen
  \bibfield  {author} {\bibinfo {author} {\bibfnamefont {S.}~\bibnamefont
  {Koller}}, \bibinfo {author} {\bibfnamefont {M.}~\bibnamefont {Grifoni}},
  \bibinfo {author} {\bibfnamefont {M.}~\bibnamefont {Leijnse}}, \ and\
  \bibinfo {author} {\bibfnamefont {M.~R.}\ \bibnamefont {Wegewijs}},\
  }\bibfield  {title} {\enquote {\bibinfo {title} {{Density-operator approaches
  to transport through interacting quantum dots: Simplifications in
  fourth-order perturbation theory}},}\ }\href {\doibase
  10.1103/PhysRevB.82.235307} {\bibfield  {journal} {\bibinfo  {journal} {Phys.
  Rev. B}\ }\textbf {\bibinfo {volume} {82}},\ \bibinfo {pages} {235307}
  (\bibinfo {year} {2010})}\BibitemShut {NoStop}%
\bibitem [{\citenamefont {de~Souza~Melo}\ \emph {et~al.}(2019)\citenamefont
  {de~Souza~Melo}, \citenamefont {{Dias da Silva}}, \citenamefont {Rocha},\
  and\ \citenamefont {Lewenkopf}}]{deSouza2019}%
  \BibitemOpen
  \bibfield  {author} {\bibinfo {author} {\bibfnamefont {B.}~\bibnamefont
  {de~Souza~Melo}}, \bibinfo {author} {\bibfnamefont {L.~G. G.~V.}\
  \bibnamefont {{Dias da Silva}}}, \bibinfo {author} {\bibfnamefont {A.~R.}\
  \bibnamefont {Rocha}}, \ and\ \bibinfo {author} {\bibfnamefont
  {C.}~\bibnamefont {Lewenkopf}},\ }\bibfield  {title} {\enquote {\bibinfo
  {title} {{Quantitative comparison of Anderson impurity solvers applied to
  transport in quantum dots}},}\ }\href {\doibase 10.1088/1361-648X/ab5773}
  {\bibfield  {journal} {\bibinfo  {journal} {J. Phys.: Condens. Matter}\
  }\textbf {\bibinfo {volume} {32}},\ \bibinfo {pages} {095602} (\bibinfo
  {year} {2019})}\BibitemShut {NoStop}%
\bibitem [{\citenamefont {Reimer}\ \emph {et~al.}(2019)\citenamefont {Reimer},
  \citenamefont {Wegewijs}, \citenamefont {Nestmann},\ and\ \citenamefont
  {Pletyukhov}}]{Reimer2019}%
  \BibitemOpen
  \bibfield  {author} {\bibinfo {author} {\bibfnamefont {V.}~\bibnamefont
  {Reimer}}, \bibinfo {author} {\bibfnamefont {M.~R.}\ \bibnamefont
  {Wegewijs}}, \bibinfo {author} {\bibfnamefont {K.}~\bibnamefont {Nestmann}},
  \ and\ \bibinfo {author} {\bibfnamefont {M.}~\bibnamefont {Pletyukhov}},\
  }\bibfield  {title} {\enquote {\bibinfo {title} {{Five approaches to exact
  open-system dynamics: Complete positivity, divisibility, and time-dependent
  observables}},}\ }\href {\doibase 10.1063/1.5094412} {\bibfield  {journal}
  {\bibinfo  {journal} {J. Chem. Phys.}\ }\textbf {\bibinfo {volume} {151}},\
  \bibinfo {pages} {044101} (\bibinfo {year} {2019})}\BibitemShut {NoStop}%
\bibitem [{\citenamefont {Lindner}\ \emph {et~al.}(2019)\citenamefont
  {Lindner}, \citenamefont {Kugler}, \citenamefont {Meden},\ and\ \citenamefont
  {Schoeller}}]{Lindner2019}%
  \BibitemOpen
  \bibfield  {author} {\bibinfo {author} {\bibfnamefont {C~J.}\ \bibnamefont
  {Lindner}}, \bibinfo {author} {\bibfnamefont {F.~B.}\ \bibnamefont {Kugler}},
  \bibinfo {author} {\bibfnamefont {V.}~\bibnamefont {Meden}}, \ and\ \bibinfo
  {author} {\bibfnamefont {H.}~\bibnamefont {Schoeller}},\ }\bibfield  {title}
  {\enquote {\bibinfo {title} {{Renormalization group transport theory for open
  quantum systems: Charge fluctuations in multilevel quantum dots in and out of
  equilibrium}},}\ }\href {\doibase 10.1103/PhysRevB.99.205142} {\bibfield
  {journal} {\bibinfo  {journal} {Phys. Rev. B}\ }\textbf {\bibinfo {volume}
  {99}},\ \bibinfo {pages} {205142} (\bibinfo {year} {2019})}\BibitemShut
  {NoStop}%
\bibitem [{\citenamefont {Ferguson}\ \emph {et~al.}(2020)\citenamefont
  {Ferguson}, \citenamefont {Zilberberg},\ and\ \citenamefont
  {Blatter}}]{Ferguson2020}%
  \BibitemOpen
  \bibfield  {author} {\bibinfo {author} {\bibfnamefont {M.~S.}\ \bibnamefont
  {Ferguson}}, \bibinfo {author} {\bibfnamefont {O.}~\bibnamefont
  {Zilberberg}}, \ and\ \bibinfo {author} {\bibfnamefont {G.}~\bibnamefont
  {Blatter}},\ }\bibfield  {title} {\enquote {\bibinfo {title} {{Open quantum
  systems beyond Fermi's golden rule: Diagrammatic expansion of the
  steady-state time-convolutionless master equation}},}\ }\href@noop {}
  {\bibfield  {journal} {\bibinfo  {journal} {arXiv:2010.09838}\ } (\bibinfo
  {year} {2020})}\BibitemShut {NoStop}%
\bibitem [{\citenamefont {Meir}\ \emph {et~al.}(1991)\citenamefont {Meir},
  \citenamefont {Wingreen},\ and\ \citenamefont {Lee}}]{Meir1991}%
  \BibitemOpen
  \bibfield  {author} {\bibinfo {author} {\bibfnamefont {Y.}~\bibnamefont
  {Meir}}, \bibinfo {author} {\bibfnamefont {N.~S.}\ \bibnamefont {Wingreen}},
  \ and\ \bibinfo {author} {\bibfnamefont {P.~A.}\ \bibnamefont {Lee}},\
  }\bibfield  {title} {\enquote {\bibinfo {title} {{Transport through a
  strongly interacting electron system: Theory of periodic conductance
  oscillations}},}\ }\href {\doibase 10.1103/PhysRevLett.66.3048} {\bibfield
  {journal} {\bibinfo  {journal} {Phys. Rev. Lett.}\ }\textbf {\bibinfo
  {volume} {66}},\ \bibinfo {pages} {3048--3051} (\bibinfo {year}
  {1991})}\BibitemShut {NoStop}%
\bibitem [{\citenamefont {Kashcheyevs}\ \emph {et~al.}(2006)\citenamefont
  {Kashcheyevs}, \citenamefont {Aharony},\ and\ \citenamefont
  {Entin-Wohlman}}]{Kashcheyevs2006}%
  \BibitemOpen
  \bibfield  {author} {\bibinfo {author} {\bibfnamefont {V.}~\bibnamefont
  {Kashcheyevs}}, \bibinfo {author} {\bibfnamefont {A.}~\bibnamefont
  {Aharony}}, \ and\ \bibinfo {author} {\bibfnamefont {O.}~\bibnamefont
  {Entin-Wohlman}},\ }\bibfield  {title} {\enquote {\bibinfo {title}
  {{Applicability of the equations-of-motion technique for quantum dots}},}\
  }\href {\doibase 10.1103/PhysRevB.73.125338} {\bibfield  {journal} {\bibinfo
  {journal} {Phys. Rev. B}\ }\textbf {\bibinfo {volume} {73}},\ \bibinfo
  {pages} {125338} (\bibinfo {year} {2006})}\BibitemShut {NoStop}%
\bibitem [{\citenamefont {Roermund}\ \emph {et~al.}(2010)\citenamefont
  {Roermund}, \citenamefont {Shiau},\ and\ \citenamefont
  {Lavagna}}]{VanRoermund2010}%
  \BibitemOpen
  \bibfield  {author} {\bibinfo {author} {\bibfnamefont {R.~Van}\ \bibnamefont
  {Roermund}}, \bibinfo {author} {\bibfnamefont {S-Y.}\ \bibnamefont {Shiau}},
  \ and\ \bibinfo {author} {\bibfnamefont {M.}~\bibnamefont {Lavagna}},\
  }\bibfield  {title} {\enquote {\bibinfo {title} {{Anderson model out of
  equilibrium: Decoherence effects in transport through a quantum dot}},}\
  }\href {\doibase 10.1103/PhysRevB.81.165115} {\bibfield  {journal} {\bibinfo
  {journal} {Phys. Rev. B}\ }\textbf {\bibinfo {volume} {81}},\ \bibinfo
  {pages} {165115} (\bibinfo {year} {2010})}\BibitemShut {NoStop}%
\bibitem [{\citenamefont {Lavagna}(2015)}]{Lavagna2015}%
  \BibitemOpen
\bibfield  {author} {\bibinfo {author} {\bibfnamefont {M.}\ \bibnamefont
  {Lavagna}},\ }\bibfield  {title} {\enquote {\bibinfo {title} {{Transport through an interacting quantum dot driven out-of-equilibrium}},}\ }\href {\doibase
  10.1088/1742-6596/592/1/012141} {\bibfield  {journal} {\bibinfo  {journal}
  {J. Phys.: Conf. Ser.}\ }\textbf {\bibinfo {volume} {592}},\ \bibinfo {pages}
  {012141} (\bibinfo {year} {2015})}\BibitemShut {NoStop}%
\bibitem [{\citenamefont {Eckern}\ and\ \citenamefont
  {Wysoki{\'n}ski}(2020)}]{Eckern2020}%
  \BibitemOpen
  \bibfield  {author} {\bibinfo {author} {\bibfnamefont {U.}~\bibnamefont
  {Eckern}}\ and\ \bibinfo {author} {\bibfnamefont {K.~I}\ \bibnamefont
  {Wysoki{\'n}ski}},\ }\bibfield  {title} {\enquote {\bibinfo {title} {{Two-
  and three-terminal far-from-equilibrium thermoelectric nano-devices in the
  Kondo regime}},}\ }\href {\doibase 10.1088/1367-2630/ab6874} {\bibfield
  {journal} {\bibinfo  {journal} {New J. Phys.}\ }\textbf {\bibinfo {volume}
  {22}},\ \bibinfo {pages} {013045} (\bibinfo {year} {2020})}\BibitemShut
  {NoStop}%
\bibitem [{\citenamefont {Pedersen}\ and\ \citenamefont
  {Wacker}(2005)}]{Pedersen2005}%
  \BibitemOpen
  \bibfield  {author} {\bibinfo {author} {\bibfnamefont {J.~N.}\ \bibnamefont
  {Pedersen}}\ and\ \bibinfo {author} {\bibfnamefont {A.}~\bibnamefont
  {Wacker}},\ }\bibfield  {title} {\enquote {\bibinfo {title} {{Tunneling
  through nanosystems: Combining broadening with many-particle states}},}\
  }\href {\doibase 10.1103/PhysRevB.72.195330} {\bibfield  {journal} {\bibinfo
  {journal} {Phys. Rev. B}\ }\textbf {\bibinfo {volume} {72}},\ \bibinfo
  {pages} {195330} (\bibinfo {year} {2005})}\BibitemShut {NoStop}%
\bibitem [{\citenamefont {{O. Karlstr{\"o}m}}\ \emph
  {et~al.}(2013)\citenamefont {{O. Karlstr{\"o}m}}, \citenamefont {{C. Emary}},
  \citenamefont {{P. Zedler}}, \citenamefont {{ J. N. Pedersen}}, \citenamefont
  {{C. Bergenfeldt}}, \citenamefont {{P. Samuelsson}}, \citenamefont {{T.
  Brandes}},\ and\ \citenamefont {{A. Wacker}}}]{Karlstrom2013}%
  \BibitemOpen
  \bibfield  {author} {\bibinfo {author} {\bibnamefont {{O. Karlstr{\"o}m}}},
  \bibinfo {author} {\bibnamefont {{C. Emary}}}, \bibinfo {author}
  {\bibnamefont {{P. Zedler}}}, \bibinfo {author} {\bibnamefont {{ J. N.
  Pedersen}}}, \bibinfo {author} {\bibnamefont {{C. Bergenfeldt}}}, \bibinfo
  {author} {\bibnamefont {{P. Samuelsson}}}, \bibinfo {author} {\bibnamefont
  {{T. Brandes}}}, \ and\ \bibinfo {author} {\bibnamefont {{A. Wacker}}},\
  }\bibfield  {title} {\enquote {\bibinfo {title} {{A diagrammatic description
  of the equations of motion, current and noise within the second-order von
  Neumann approach}},}\ }\href {\doibase 10.1088/1751-8113/46/6/065301}
  {\bibfield  {journal} {\bibinfo  {journal} {J. Phys. A: Math. Theor.}\
  }\textbf {\bibinfo {volume} {46}},\ \bibinfo {pages} {065301} (\bibinfo
  {year} {2013})}\BibitemShut {NoStop}%
\bibitem [{\citenamefont {Schoeller}\ and\ \citenamefont
  {Sch{\"o}n}(1994)}]{Schoeller1994}%
  \BibitemOpen
  \bibfield  {author} {\bibinfo {author} {\bibfnamefont {H.}~\bibnamefont
  {Schoeller}}\ and\ \bibinfo {author} {\bibfnamefont {G.}~\bibnamefont
  {Sch{\"o}n}},\ }\bibfield  {title} {\enquote {\bibinfo {title} {{Mesoscopic
  quantum transport: Resonant tunneling in the presence of a strong Coulomb
  interaction}},}\ }\href {\doibase 10.1103/PhysRevB.50.18436} {\bibfield
  {journal} {\bibinfo  {journal} {Phys. Rev. B}\ }\textbf {\bibinfo {volume}
  {50}},\ \bibinfo {pages} {18436--18452} (\bibinfo {year} {1994})}\BibitemShut
  {NoStop}%
\bibitem [{\citenamefont {{J. K{\"o}nig}}\ \emph {et~al.}(1996)\citenamefont
  {{J. K{\"o}nig}}, \citenamefont {{J. Schmid}}, \citenamefont {{H.
  Schoeller}},\ and\ \citenamefont {{G. Sch{\"o}n}}}]{Koenig1996}%
  \BibitemOpen
  \bibfield  {author} {\bibinfo {author} {\bibnamefont {{J. K{\"o}nig}}},
  \bibinfo {author} {\bibnamefont {{J. Schmid}}}, \bibinfo {author}
  {\bibnamefont {{H. Schoeller}}}, \ and\ \bibinfo {author} {\bibnamefont {{G.
  Sch{\"o}n}}},\ }\bibfield  {title} {\enquote {\bibinfo {title} {{Resonant
  tunneling through ultrasmall quantum dots: Zero-bias anomalies,
  magnetic-field dependence, and boson-assisted transport}},}\ }\href {\doibase
  10.1103/PhysRevB.54.16820} {\bibfield  {journal} {\bibinfo  {journal} {Phys.
  Rev. B}\ }\textbf {\bibinfo {volume} {54}},\ \bibinfo {pages} {16820--16837}
  (\bibinfo {year} {1996})}\BibitemShut {NoStop}%
\bibitem [{\citenamefont {Leijnse}\ and\ \citenamefont
  {Wegewijs}(2008)}]{Leijnse2008}%
  \BibitemOpen
  \bibfield  {author} {\bibinfo {author} {\bibfnamefont {M.}~\bibnamefont
  {Leijnse}}\ and\ \bibinfo {author} {\bibfnamefont {M.~R.}\ \bibnamefont
  {Wegewijs}},\ }\bibfield  {title} {\enquote {\bibinfo {title} {{Kinetic
  equations for transport through single-molecule transistors}},}\ }\href
  {\doibase 10.1103/PhysRevB.78.235424} {\bibfield  {journal} {\bibinfo
  {journal} {Phys. Rev. B}\ }\textbf {\bibinfo {volume} {78}},\ \bibinfo
  {pages} {235424} (\bibinfo {year} {2008})}\BibitemShut {NoStop}%
\bibitem [{\citenamefont {{J. Kern}}\ and\ \citenamefont {{M.
  Grifoni}}(2013)}]{Kern2013}%
  \BibitemOpen
  \bibfield  {author} {\bibinfo {author} {\bibnamefont {{J. Kern}}}\ and\
  \bibinfo {author} {\bibnamefont {{M. Grifoni}}},\ }\bibfield  {title}
  {\enquote {\bibinfo {title} {{Transport across an Anderson quantum dot in the
  intermediate coupling regime}},}\ }\href {\doibase
  10.1140/epjb/e2013-40618-9} {\bibfield  {journal} {\bibinfo  {journal}
  {EPJB}\ }\textbf {\bibinfo {volume} {86}},\ \bibinfo {pages} {384} (\bibinfo
  {year} {2013})}\BibitemShut {NoStop}%
\bibitem [{\citenamefont {Jin}\ \emph {et~al.}(2014)\citenamefont {Jin},
  \citenamefont {Li}, \citenamefont {Liu}, \citenamefont {Li},\ and\
  \citenamefont {Yan}}]{Jin2014}%
  \BibitemOpen
  \bibfield  {author} {\bibinfo {author} {\bibfnamefont {J.}~\bibnamefont
  {Jin}}, \bibinfo {author} {\bibfnamefont {J.}~\bibnamefont {Li}}, \bibinfo
  {author} {\bibfnamefont {Y.}~\bibnamefont {Liu}}, \bibinfo {author}
  {\bibfnamefont {X-Q}\ \bibnamefont {Li}}, \ and\ \bibinfo {author}
  {\bibfnamefont {YJ}~\bibnamefont {Yan}},\ }\bibfield  {title} {\enquote
  {\bibinfo {title} {{Improved master equation approach to quantum transport:
  From Born to self-consistent Born approximation}},}\ }\href {\doibase
  10.1063/1.4884390} {\bibfield  {journal} {\bibinfo  {journal} {J. Chem.
  Phys}\ }\textbf {\bibinfo {volume} {140}},\ \bibinfo {pages} {244111}
  (\bibinfo {year} {2014})}\BibitemShut {NoStop}%
\bibitem [{\citenamefont {Saptsov}\ and\ \citenamefont
  {Wegewijs}(2012)}]{Saptsov2012}%
  \BibitemOpen
  \bibfield  {author} {\bibinfo {author} {\bibfnamefont {R.~B.}\ \bibnamefont
  {Saptsov}}\ and\ \bibinfo {author} {\bibfnamefont {M.~R.}\ \bibnamefont
  {Wegewijs}},\ }\bibfield  {title} {\enquote {\bibinfo {title} {{Fermionic
  superoperators for zero-temperature nonlinear transport: Real-time
  perturbation theory and renormalization group for Anderson quantum dots}},}\
  }\href {\doibase 10.1103/PhysRevB.86.235432} {\bibfield  {journal} {\bibinfo
  {journal} {Phys. Rev. B}\ }\textbf {\bibinfo {volume} {86}},\ \bibinfo
  {pages} {235432} (\bibinfo {year} {2012})}\BibitemShut {NoStop}%
\bibitem [{\citenamefont {Saptsov}\ and\ \citenamefont
  {Wegewijs}(2014)}]{Saptsov2014}%
  \BibitemOpen
  \bibfield  {author} {\bibinfo {author} {\bibfnamefont {R.~B.}\ \bibnamefont
  {Saptsov}}\ and\ \bibinfo {author} {\bibfnamefont {M.~R.}\ \bibnamefont
  {Wegewijs}},\ }\bibfield  {title} {\enquote {\bibinfo {title}
  {{Time-dependent quantum transport: Causal superfermions, exact
  fermion-parity protected decay modes, and Pauli exclusion principle for mixed
  quantum states}},}\ }\href {\doibase 10.1103/PhysRevB.90.045407} {\bibfield
  {journal} {\bibinfo  {journal} {Phys. Rev. B}\ }\textbf {\bibinfo {volume}
  {90}},\ \bibinfo {pages} {045407} (\bibinfo {year} {2014})}\BibitemShut
  {NoStop}%
\bibitem [{\citenamefont {Schoeller}\ and\ \citenamefont
  {K{\"o}nig}(2000)}]{Schoeller2000}%
  \BibitemOpen
  \bibfield  {author} {\bibinfo {author} {\bibfnamefont {H.}~\bibnamefont
  {Schoeller}}\ and\ \bibinfo {author} {\bibfnamefont {J.}~\bibnamefont
  {K{\"o}nig}},\ }\bibfield  {title} {\enquote {\bibinfo {title} {{Real-Time
  Renormalization Group and Charge Fluctuations in Quantum Dots}},}\ }\href
  {\doibase 10.1103/PhysRevLett.84.3686} {\bibfield  {journal} {\bibinfo
  {journal} {Phys. Rev. Lett.}\ }\textbf {\bibinfo {volume} {84}},\ \bibinfo
  {pages} {3686--3689} (\bibinfo {year} {2000})}\BibitemShut {NoStop}%
\bibitem [{\citenamefont {Rosch}\ \emph {et~al.}(2003)\citenamefont {Rosch},
  \citenamefont {Paaske}, \citenamefont {Kroha},\ and\ \citenamefont
  {W{\"o}lfle}}]{Rosch2003}%
  \BibitemOpen
  \bibfield  {author} {\bibinfo {author} {\bibfnamefont {A.}~\bibnamefont
  {Rosch}}, \bibinfo {author} {\bibfnamefont {J.}~\bibnamefont {Paaske}},
  \bibinfo {author} {\bibfnamefont {J.}~\bibnamefont {Kroha}}, \ and\ \bibinfo
  {author} {\bibfnamefont {P.}~\bibnamefont {W{\"o}lfle}},\ }\bibfield  {title}
  {\enquote {\bibinfo {title} {{Nonequilibrium Transport through a Kondo Dot in
  a Magnetic Field: Perturbation Theory and Poor Man's Scaling}},}\ }\href
  {\doibase 10.1103/PhysRevLett.90.076804} {\bibfield  {journal} {\bibinfo
  {journal} {Phys. Rev. Lett.}\ }\textbf {\bibinfo {volume} {90}},\ \bibinfo
  {pages} {076804} (\bibinfo {year} {2003})}\BibitemShut {NoStop}%
\bibitem [{\citenamefont {Paaske}\ \emph
  {et~al.}(2004{\natexlab{a}})\citenamefont {Paaske}, \citenamefont {Rosch},\
  and\ \citenamefont {W{\"o}lfle}}]{Paaske2004a}%
  \BibitemOpen
  \bibfield  {author} {\bibinfo {author} {\bibfnamefont {J.}~\bibnamefont
  {Paaske}}, \bibinfo {author} {\bibfnamefont {A.}~\bibnamefont {Rosch}}, \
  and\ \bibinfo {author} {\bibfnamefont {P.}~\bibnamefont {W{\"o}lfle}},\
  }\bibfield  {title} {\enquote {\bibinfo {title} {{Nonequilibrium transport
  through a Kondo dot in a magnetic field: Perturbation theory}},}\ }\href
  {\doibase 10.1103/PhysRevB.69.155330} {\bibfield  {journal} {\bibinfo
  {journal} {Phys. Rev. B}\ }\textbf {\bibinfo {volume} {69}},\ \bibinfo
  {pages} {155330} (\bibinfo {year} {2004}{\natexlab{a}})}\BibitemShut
  {NoStop}%
\bibitem [{\citenamefont {Paaske}\ \emph
  {et~al.}(2004{\natexlab{b}})\citenamefont {Paaske}, \citenamefont {Rosch},
  \citenamefont {Kroha},\ and\ \citenamefont {W{\"o}lfle}}]{Paaske2004b}%
  \BibitemOpen
  \bibfield  {author} {\bibinfo {author} {\bibfnamefont {J.}~\bibnamefont
  {Paaske}}, \bibinfo {author} {\bibfnamefont {A.}~\bibnamefont {Rosch}},
  \bibinfo {author} {\bibfnamefont {J.}~\bibnamefont {Kroha}}, \ and\ \bibinfo
  {author} {\bibfnamefont {P.}~\bibnamefont {W{\"o}lfle}},\ }\bibfield  {title}
  {\enquote {\bibinfo {title} {{Nonequilibrium transport through a Kondo dot:
  Decoherence effects}},}\ }\href {\doibase 10.1103/PhysRevB.70.155301}
  {\bibfield  {journal} {\bibinfo  {journal} {Phys. Rev. B}\ }\textbf {\bibinfo
  {volume} {70}},\ \bibinfo {pages} {155301} (\bibinfo {year}
  {2004}{\natexlab{b}})}\BibitemShut {NoStop}%
\bibitem [{\citenamefont {Rosch}\ \emph {et~al.}(2005)\citenamefont {Rosch},
  \citenamefont {Paaske}, \citenamefont {Kroha},\ and\ \citenamefont
  {W{\"o}lfle}}]{Rosch2005}%
  \BibitemOpen
  \bibfield  {author} {\bibinfo {author} {\bibfnamefont {A.}~\bibnamefont
  {Rosch}}, \bibinfo {author} {\bibfnamefont {J.}~\bibnamefont {Paaske}},
  \bibinfo {author} {\bibfnamefont {J.}~\bibnamefont {Kroha}}, \ and\ \bibinfo
  {author} {\bibfnamefont {P.}~\bibnamefont {W{\"o}lfle}},\ }\bibfield  {title}
  {\enquote {\bibinfo {title} {{The Kondo Effect in Non-Equilibrium Quantum
  Dots: Perturbative Renormalization Group}},}\ }\href {\doibase
  10.1143/JPSJ.74.118} {\bibfield  {journal} {\bibinfo  {journal} {J. Phys.
  Soc. Japan}\ }\textbf {\bibinfo {volume} {74}},\ \bibinfo {pages} {118--126}
  (\bibinfo {year} {2005})}\BibitemShut {NoStop}%
\bibitem [{\citenamefont {Karrasch}\ \emph {et~al.}(2006)\citenamefont
  {Karrasch}, \citenamefont {Enss},\ and\ \citenamefont
  {Meden}}]{Karrasch2006}%
  \BibitemOpen
  \bibfield  {author} {\bibinfo {author} {\bibfnamefont {C.}~\bibnamefont
  {Karrasch}}, \bibinfo {author} {\bibfnamefont {T.}~\bibnamefont {Enss}}, \
  and\ \bibinfo {author} {\bibfnamefont {V.}~\bibnamefont {Meden}},\ }\bibfield
   {title} {\enquote {\bibinfo {title} {{Functional renormalization group
  approach to transport through correlated quantum dots}},}\ }\href {\doibase
  10.1103/PhysRevB.73.235337} {\bibfield  {journal} {\bibinfo  {journal} {Phys.
  Rev. B}\ }\textbf {\bibinfo {volume} {73}},\ \bibinfo {pages} {235337}
  (\bibinfo {year} {2006})}\BibitemShut {NoStop}%
\bibitem [{\citenamefont {Pletyukhov}\ and\ \citenamefont
  {Schoeller}(2012)}]{Pletyukhov2012}%
  \BibitemOpen
  \bibfield  {author} {\bibinfo {author} {\bibfnamefont {M.}~\bibnamefont
  {Pletyukhov}}\ and\ \bibinfo {author} {\bibfnamefont {H.}~\bibnamefont
  {Schoeller}},\ }\bibfield  {title} {\enquote {\bibinfo {title}
  {{Nonequilibrium Kondo model: Crossover from weak to strong coupling}},}\
  }\href {\doibase 10.1103/PhysRevLett.108.260601} {\bibfield  {journal}
  {\bibinfo  {journal} {Phys. Rev. Lett.}\ }\textbf {\bibinfo {volume} {108}},\
  \bibinfo {pages} {260601} (\bibinfo {year} {2012})}\BibitemShut {NoStop}%
\bibitem [{\citenamefont {Nestmann}\ and\ \citenamefont
  {Nestmann}(2021)}]{Nestmann2021}%
  \BibitemOpen
  \bibfield  {author} {\bibinfo {author} {\bibfnamefont {K.}~\bibnamefont
  {Nestmann}}\ and\ \bibinfo {author} {\bibfnamefont {M. R.}~\bibnamefont
  {Wegewijs}},\ }\bibfield  {title} {\enquote {\bibinfo {title}
  {{Renormalization group for open quantum systems using environment temperature as flow parameter}},}\
  } {\bibfield  {journal}
  {\bibinfo  {journal} {arXiv:2111.07320}\ } (\bibinfo {year} {2021})}\BibitemShut {NoStop}%
\bibitem [{\citenamefont {Kamenev}(2005)}]{Kamenev2005}%
  \BibitemOpen
  \bibfield  {author} {\bibinfo {author} {\bibfnamefont {A.}~\bibnamefont
  {Kamenev}},\ }in\ \href@noop {} {\emph {\bibinfo {booktitle} {{Nanoscopic
  Quantum Transport, Proceedings of the Les Houches Summer School of
  Theoretical Physics, 2004}}}},\ Vol.~\bibinfo {volume} {81}\ (\bibinfo
  {publisher} {Elsevier, Amsterdam},\ \bibinfo {year} {2005})\ pp.\ \bibinfo
  {pages} {177--246}\BibitemShut {NoStop}%
\bibitem [{\citenamefont {Altland}\ and\ \citenamefont
  {Simons}(2010)}]{Altland2010}%
  \BibitemOpen
  \bibfield  {author} {\bibinfo {author} {\bibfnamefont {A.}~\bibnamefont
  {Altland}}\ and\ \bibinfo {author} {\bibfnamefont {B.~D.}\ \bibnamefont
  {Simons}},\ }\href@noop {} {\emph {\bibinfo {title} {{Condensed Matter Field
  Theory}}}}\ (\bibinfo  {publisher} {Cambridge University Press},\ \bibinfo
  {year} {2010})\BibitemShut {NoStop}%
\bibitem [{\citenamefont {Bock}\ \emph {et~al.}(2016)\citenamefont {Bock},
  \citenamefont {Liluashvili},\ and\ \citenamefont {Gasenzer}}]{Bock2016}%
  \BibitemOpen
  \bibfield  {author} {\bibinfo {author} {\bibfnamefont {S.}~\bibnamefont
  {Bock}}, \bibinfo {author} {\bibfnamefont {A.}~\bibnamefont {Liluashvili}}, \
  and\ \bibinfo {author} {\bibfnamefont {T.}~\bibnamefont {Gasenzer}},\
  }\bibfield  {title} {\enquote {\bibinfo {title} {{Buildup of the Kondo effect from real-time effective action for the Anderson impurity model}},}\ }\href {\doibase
  10.1103/PhysRevB.94.045108} {\bibfield  {journal} {\bibinfo  {journal} {Phys.
  Rev. B}\ }\textbf {\bibinfo {volume} {94}},\ \bibinfo {pages} {045108}
  (\bibinfo {year} {2016})}\BibitemShut {NoStop}%
\bibitem [{\citenamefont {Weiss}(2012, 4th ed.)}]{Weiss2012}%
  \BibitemOpen
  \bibfield  {author} {\bibinfo {author} {\bibfnamefont {U.}~\bibnamefont
  {Weiss}},\ }\href@noop {} {\emph {\bibinfo {title} {{Quantum Dissipative
  Systems}}}}\ (\bibinfo  {publisher} {World Scientific, Singapore},\ \bibinfo
  {year} {2012, 4th ed.})\BibitemShut {NoStop}%
\bibitem [{\citenamefont {Jin}\ \emph {et~al.}(2010)\citenamefont {Jin},
  \citenamefont {Tu}, \citenamefont {Zhang},\ and\ \citenamefont
  {Yan}}]{Jin2010}%
  \BibitemOpen
  \bibfield  {author} {\bibinfo {author} {\bibfnamefont {J.}~\bibnamefont
  {Jin}}, \bibinfo {author} {\bibfnamefont {M.~W.-Y.}\ \bibnamefont {Tu}},
  \bibinfo {author} {\bibfnamefont {W.-M.}\ \bibnamefont {Zhang}}, \ and\
  \bibinfo {author} {\bibfnamefont {Y.~J.}\ \bibnamefont {Yan}},\ }\bibfield
  {title} {\enquote {\bibinfo {title} {{Non-equilibrium quantum theory for
  nanodevices based on the Feynman--Vernon influence functional}},}\ }\href
  {http://stacks.iop.org/1367-2630/12/i=8/a=083013} {\bibfield  {journal}
  {\bibinfo  {journal} {New J. Phys.}\ }\textbf {\bibinfo {volume} {12}},\
  \bibinfo {pages} {083013} (\bibinfo {year} {2010})}\BibitemShut {NoStop}%
\bibitem [{\citenamefont {Altland}\ and\ \citenamefont
  {Egger}(2009)}]{Altland2009}%
  \BibitemOpen
  \bibfield  {author} {\bibinfo {author} {\bibfnamefont {A.}~\bibnamefont
  {Altland}}\ and\ \bibinfo {author} {\bibfnamefont {R.}~\bibnamefont
  {Egger}},\ }\bibfield  {title} {\enquote {\bibinfo {title} {{Nonequilibrium
  Dephasing in Coulomb Blockaded Quantum Dots}},}\ }\href {\doibase
  10.1103/PhysRevLett.102.026805} {\bibfield  {journal} {\bibinfo  {journal}
  {Phys. Rev. Lett.}\ }\textbf {\bibinfo {volume} {102}},\ \bibinfo {pages}
  {026805} (\bibinfo {year} {2009})}\BibitemShut {NoStop}%
\bibitem [{\citenamefont {Wingreen}\ and\ \citenamefont
  {Meir}(1994)}]{Wingreen1994}%
  \BibitemOpen
  \bibfield  {author} {\bibinfo {author} {\bibfnamefont {N.~S.}\ \bibnamefont
  {Wingreen}}\ and\ \bibinfo {author} {\bibfnamefont {Y.}~\bibnamefont
  {Meir}},\ }\bibfield  {title} {\enquote {\bibinfo {title} {{Anderson model
  out of equilibrium: Noncrossing-approximation approach to transport through a
  quantum dot}},}\ }\href {\doibase 10.1103/PhysRevB.49.11040} {\bibfield
  {journal} {\bibinfo  {journal} {Phys. Rev. B}\ }\textbf {\bibinfo {volume}
  {49}},\ \bibinfo {pages} {11040--11052} (\bibinfo {year} {1994})}\BibitemShut
  {NoStop}%
\bibitem [{\citenamefont {Smirnov}\ and\ \citenamefont
  {Grifoni}(2013)}]{Smirnov2013}%
  \BibitemOpen
  \bibfield  {author} {\bibinfo {author} {\bibfnamefont {S.}~\bibnamefont
  {Smirnov}}\ and\ \bibinfo {author} {\bibfnamefont {M.}~\bibnamefont
  {Grifoni}},\ }\bibfield  {title} {\enquote {\bibinfo {title} {{Nonequilibrium
  Kondo transport through a quantum dot in a magnetic field}},}\ }\href
  {\doibase 10.1088/1367-2630/15/7/073047} {\bibfield  {journal} {\bibinfo
  {journal} {New J. Phys.}\ }\textbf {\bibinfo {volume} {15}},\ \bibinfo
  {pages} {073047} (\bibinfo {year} {2013})}\BibitemShut {NoStop}%
\bibitem [{\citenamefont {Schmid}\ \emph {et~al.}(2015)\citenamefont {Schmid},
  \citenamefont {Smirnov}, \citenamefont {Marga{\'n}ska}, \citenamefont
  {Dirnaichner}, \citenamefont {Stiller}, \citenamefont {Grifoni},
  \citenamefont {H{\"u}ttel},\ and\ \citenamefont {Strunk}}]{Schmid2015}%
  \BibitemOpen
  \bibfield  {author} {\bibinfo {author} {\bibfnamefont {D.~R.}\ \bibnamefont
  {Schmid}}, \bibinfo {author} {\bibfnamefont {S.}~\bibnamefont {Smirnov}},
  \bibinfo {author} {\bibfnamefont {M.}~\bibnamefont {Marga{\'n}ska}}, \bibinfo
  {author} {\bibfnamefont {A.}~\bibnamefont {Dirnaichner}}, \bibinfo {author}
  {\bibfnamefont {P.~L.}\ \bibnamefont {Stiller}}, \bibinfo {author}
  {\bibfnamefont {M.}~\bibnamefont {Grifoni}}, \bibinfo {author} {\bibfnamefont
  {A.~K.}\ \bibnamefont {H{\"u}ttel}}, \ and\ \bibinfo {author} {\bibfnamefont
  {C.}~\bibnamefont {Strunk}},\ }\bibfield  {title} {\enquote {\bibinfo {title}
  {{Broken SU(4) symmetry in a Kondo-correlated carbon nanotube}},}\ }\href
  {http://link.aps.org/doi/10.1103/PhysRevB.91.155435} {\bibfield  {journal}
  {\bibinfo  {journal} {Phys. Rev. B}\ }\textbf {\bibinfo {volume} {91}},\
  \bibinfo {pages} {155435} (\bibinfo {year} {2015})}\BibitemShut {NoStop}%
\bibitem [{\citenamefont {Niklas}\ \emph {et~al.}(2016)\citenamefont {Niklas},
  \citenamefont {Smirnov}, \citenamefont {Mantelli}, \citenamefont
  {Marga{\'n}ska}, \citenamefont {Nguyen}, \citenamefont {Wernsdorfer},
  \citenamefont {Cleuziou},\ and\ \citenamefont {Grifoni}}]{Niklas2016}%
  \BibitemOpen
  \bibfield  {author} {\bibinfo {author} {\bibfnamefont {M.}~\bibnamefont
  {Niklas}}, \bibinfo {author} {\bibfnamefont {S}~\bibnamefont {Smirnov}},
  \bibinfo {author} {\bibfnamefont {D.}~\bibnamefont {Mantelli}}, \bibinfo
  {author} {\bibfnamefont {M.}~\bibnamefont {Marga{\'n}ska}}, \bibinfo {author}
  {\bibfnamefont {N.-V.}\ \bibnamefont {Nguyen}}, \bibinfo {author}
  {\bibfnamefont {W.}~\bibnamefont {Wernsdorfer}}, \bibinfo {author}
  {\bibfnamefont {J.-P.}\ \bibnamefont {Cleuziou}}, \ and\ \bibinfo {author}
  {\bibfnamefont {M.}~\bibnamefont {Grifoni}},\ }\bibfield  {title} {\enquote
  {\bibinfo {title} {{Blocking transport resonances via Kondo many-body
  entanglement in quantum dots}},}\ }\href {\doibase 10.1038/ncomms12442}
  {\bibfield  {journal} {\bibinfo  {journal} {Nat. Commun.}\ }\textbf {\bibinfo
  {volume} {7}},\ \bibinfo {pages} {12442} (\bibinfo {year}
  {2016})}\BibitemShut {NoStop}%
\bibitem [{\citenamefont {Lahiri}\ \emph {et~al.}(2020)\citenamefont {Lahiri},
  \citenamefont {Hata}, \citenamefont {Smirnov}, \citenamefont {Ferrier},
  \citenamefont {Arakawa}, \citenamefont {Niklas}, \citenamefont {Marganska},
  \citenamefont {Kobayashi},\ and\ \citenamefont {Grifoni}}]{Lahiri2020}%
  \BibitemOpen
  \bibfield  {author} {\bibinfo {author} {\bibfnamefont {A.}~\bibnamefont
  {Lahiri}}, \bibinfo {author} {\bibfnamefont {T.}~\bibnamefont {Hata}},
  \bibinfo {author} {\bibfnamefont {S.}~\bibnamefont {Smirnov}}, \bibinfo
  {author} {\bibfnamefont {M.}~\bibnamefont {Ferrier}}, \bibinfo {author}
  {\bibfnamefont {T.}~\bibnamefont {Arakawa}}, \bibinfo {author} {\bibfnamefont
  {M.}~\bibnamefont {Niklas}}, \bibinfo {author} {\bibfnamefont
  {M.}~\bibnamefont {Marganska}}, \bibinfo {author} {\bibfnamefont
  {K.}~\bibnamefont {Kobayashi}}, \ and\ \bibinfo {author} {\bibfnamefont
  {M.}~\bibnamefont {Grifoni}},\ }\bibfield  {title} {\enquote {\bibinfo
  {title} {{Unraveling a concealed resonance by multiple Kondo transitions in a
  quantum dot}},}\ }\href {\doibase 10.1103/PhysRevB.101.041102} {\bibfield
  {journal} {\bibinfo  {journal} {Phys. Rev. B}\ }\textbf {\bibinfo {volume}
  {101}},\ \bibinfo {pages} {041102} (\bibinfo {year} {2020})}\BibitemShut
  {NoStop}%
\bibitem [{\citenamefont {Negele}\ and\ \citenamefont
  {Orland}(1988)}]{Negele1988}%
  \BibitemOpen
  \bibfield  {author} {\bibinfo {author} {\bibfnamefont {J.~W.}\ \bibnamefont
  {Negele}}\ and\ \bibinfo {author} {\bibfnamefont {H.}~\bibnamefont
  {Orland}},\ }\href@noop {} {\emph {\bibinfo {title} {{Quantum Many-Particle
  Systems}}}}\ (\bibinfo  {publisher} {Addison-Wesley Publishing Company,
  Redwood City, California},\ \bibinfo {year} {1988})\BibitemShut {NoStop}%
\bibitem [{\citenamefont {Cahill}\ and\ \citenamefont
  {Glauber}(1999)}]{Cahill1999}%
  \BibitemOpen
  \bibfield  {author} {\bibinfo {author} {\bibfnamefont {K.~E.}\ \bibnamefont
  {Cahill}}\ and\ \bibinfo {author} {\bibfnamefont {R.~J.}\ \bibnamefont
  {Glauber}},\ }\bibfield  {title} {\enquote {\bibinfo {title} {{Density
  operators for fermions}},}\ }\href {\doibase 10.1103/PhysRevA.59.1538}
  {\bibfield  {journal} {\bibinfo  {journal} {Phys. Rev. A}\ }\textbf {\bibinfo
  {volume} {59}},\ \bibinfo {pages} {1538--1555} (\bibinfo {year}
  {1999})}\BibitemShut {NoStop}%
\bibitem [{\citenamefont {Tu}\ and\ \citenamefont {Zhang}(2008)}]{Tu2008}%
  \BibitemOpen
  \bibfield  {author} {\bibinfo {author} {\bibfnamefont {M.~W.-Y.}\
  \bibnamefont {Tu}}\ and\ \bibinfo {author} {\bibfnamefont {W.-M.}\
  \bibnamefont {Zhang}},\ }\bibfield  {title} {\enquote {\bibinfo {title}
  {{Non-Markovian decoherence theory for a double-dot charge qubit}},}\ }\href
  {http://link.aps.org/doi/10.1103/PhysRevB.78.235311} {\bibfield  {journal}
  {\bibinfo  {journal} {Phys. Rev. B}\ }\textbf {\bibinfo {volume} {78}},\
  \bibinfo {pages} {235311} (\bibinfo {year} {2008})}\BibitemShut {NoStop}%
\bibitem [{\citenamefont {{W.-M. Zhang}}\ \emph {et~al.}(2012)\citenamefont
  {{W.-M. Zhang}}, \citenamefont {{P.-Y. Lo}}, \citenamefont {{H.-N. Xiong}},
  \citenamefont {{M. W.-Y. Tu}},\ and\ \citenamefont {{F. Nori}}}]{Zhang2012}%
  \BibitemOpen
  \bibfield  {author} {\bibinfo {author} {\bibnamefont {{W.-M. Zhang}}},
  \bibinfo {author} {\bibnamefont {{P.-Y. Lo}}}, \bibinfo {author}
  {\bibnamefont {{H.-N. Xiong}}}, \bibinfo {author} {\bibnamefont {{M. W.-Y.
  Tu}}}, \ and\ \bibinfo {author} {\bibnamefont {{F. Nori}}},\ }\bibfield
  {title} {\enquote {\bibinfo {title} {{General Non-Markovian Dynamics of Open
  Quantum Systems}},}\ }\href {\doibase 10.1103/PhysRevLett.109.170402}
  {\bibfield  {journal} {\bibinfo  {journal} {Phys. Rev. Lett.}\ }\textbf
  {\bibinfo {volume} {109}},\ \bibinfo {pages} {170402} (\bibinfo {year}
  {2012})}\BibitemShut {NoStop}%
\bibitem [{\citenamefont {Yang}\ and\ \citenamefont {Zhang}(2016)}]{Yang2016}%
  \BibitemOpen
  \bibfield  {author} {\bibinfo {author} {\bibfnamefont {P.-Y.}\ \bibnamefont
  {Yang}}\ and\ \bibinfo {author} {\bibfnamefont {W.-M.}\ \bibnamefont
  {Zhang}},\ }\bibfield  {title} {\enquote {\bibinfo {title} {{Master equation
  approach to transient quantum transport in nanostructures}},}\ }\href
  {\doibase 10.1007/s11467-016-0640-z} {\bibfield  {journal} {\bibinfo
  {journal} {Front. Phys.}\ }\textbf {\bibinfo {volume} {12}},\ \bibinfo
  {pages} {127204} (\bibinfo {year} {2016})}\BibitemShut {NoStop}%
\bibitem [{\citenamefont {Grabert}\ and\ \citenamefont {{Devoret
  (Eds.)}}(1992)}]{Grabert1992}%
  \BibitemOpen
  \bibfield  {author} {\bibinfo {author} {\bibfnamefont {H.}~\bibnamefont
  {Grabert}}\ and\ \bibinfo {author} {\bibfnamefont {M.~H.}\ \bibnamefont
  {{Devoret (Eds.)}}},\ }\href@noop {} {\emph {\bibinfo {title} {{Single Charge
  Tunneling}}}},\ {Nato Science Series B}\ (\bibinfo  {publisher} {Springer},\
  \bibinfo {year} {1992})\BibitemShut {NoStop}%
\bibitem [{\citenamefont {Meir}\ \emph {et~al.}(1993)\citenamefont {Meir},
  \citenamefont {Wingreen},\ and\ \citenamefont {Lee}}]{Meir1993}%
  \BibitemOpen
  \bibfield  {author} {\bibinfo {author} {\bibfnamefont {Y.}~\bibnamefont
  {Meir}}, \bibinfo {author} {\bibfnamefont {N.~S.}\ \bibnamefont {Wingreen}},
  \ and\ \bibinfo {author} {\bibfnamefont {P.~A.}\ \bibnamefont {Lee}},\
  }\bibfield  {title} {\enquote {\bibinfo {title} {{Low-temperature transport
  through a quantum dot: The Anderson model out of equilibrium}},}\ }\href
  {\doibase 10.1103/PhysRevLett.70.2601} {\bibfield  {journal} {\bibinfo
  {journal} {Phys. Rev. Lett.}\ }\textbf {\bibinfo {volume} {70}},\ \bibinfo
  {pages} {2601--2604} (\bibinfo {year} {1993})}\BibitemShut {NoStop}%
\bibitem [{\citenamefont {Leggett}\ \emph {et~al.}(1987)\citenamefont
  {Leggett}, \citenamefont {Chakravarty}, \citenamefont {Dorsey}, \citenamefont
  {Fisher}, \citenamefont {Garg},\ and\ \citenamefont {Zwerger}}]{Leggett1987}%
  \BibitemOpen
  \bibfield  {author} {\bibinfo {author} {\bibfnamefont {A.~J.}\ \bibnamefont
  {Leggett}}, \bibinfo {author} {\bibfnamefont {S.}~\bibnamefont
  {Chakravarty}}, \bibinfo {author} {\bibfnamefont {A.~T.}\ \bibnamefont
  {Dorsey}}, \bibinfo {author} {\bibfnamefont {Matthew P.~A.}\ \bibnamefont
  {Fisher}}, \bibinfo {author} {\bibfnamefont {Anupam}\ \bibnamefont {Garg}}, \
  and\ \bibinfo {author} {\bibfnamefont {W.}~\bibnamefont {Zwerger}},\
  }\bibfield  {title} {\enquote {\bibinfo {title} {{Dynamics of the dissipative
  two-state system}},}\ }\href {\doibase 10.1103/RevModPhys.59.1} {\bibfield
  {journal} {\bibinfo  {journal} {Rev. Mod. Phys.}\ }\textbf {\bibinfo {volume}
  {59}},\ \bibinfo {pages} {1--85} (\bibinfo {year} {1987})}\BibitemShut
  {NoStop}%
\bibitem [{\citenamefont {Hanson}\ \emph {et~al.}(2007)\citenamefont {Hanson},
  \citenamefont {Kouwenhoven}, \citenamefont {Petta}, \citenamefont {Tarucha},\
  and\ \citenamefont {Vandersypen}}]{Hanson2007}%
  \BibitemOpen
  \bibfield  {author} {\bibinfo {author} {\bibfnamefont {R.}~\bibnamefont
  {Hanson}}, \bibinfo {author} {\bibfnamefont {L.~P.}\ \bibnamefont
  {Kouwenhoven}}, \bibinfo {author} {\bibfnamefont {R.~J.}\ \bibnamefont
  {Petta}}, \bibinfo {author} {\bibfnamefont {S.}~\bibnamefont {Tarucha}}, \
  and\ \bibinfo {author} {\bibfnamefont {K.~L.~M.}\ \bibnamefont
  {Vandersypen}},\ }\bibfield  {title} {\enquote {\bibinfo {title} {{Spins in
  few-electron quantum dots}},}\ }\href {\doibase 10.1103/RevModPhys.79.1217}
  {\bibfield  {journal} {\bibinfo  {journal} {Rev. Mod. Phys.}\ }\textbf
  {\bibinfo {volume} {79}},\ \bibinfo {pages} {1217--1265} (\bibinfo {year}
  {2007})}\BibitemShut {NoStop}%
\bibitem [{\citenamefont {Laird}\ \emph {et~al.}(2015)\citenamefont {Laird},
  \citenamefont {Kuemmeth}, \citenamefont {Steele}, \citenamefont
  {Grove-Rasmussen}, \citenamefont {Nyg{\aa}rd}, \citenamefont {Flensberg},\
  and\ \citenamefont {Kouwenhoven}}]{Laird2015}%
  \BibitemOpen
  \bibfield  {author} {\bibinfo {author} {\bibfnamefont {A.~E.}\ \bibnamefont
  {Laird}}, \bibinfo {author} {\bibfnamefont {F.}~\bibnamefont {Kuemmeth}},
  \bibinfo {author} {\bibfnamefont {G.~A.}\ \bibnamefont {Steele}}, \bibinfo
  {author} {\bibfnamefont {K.}~\bibnamefont {Grove-Rasmussen}}, \bibinfo
  {author} {\bibfnamefont {J.}~\bibnamefont {Nyg{\aa}rd}}, \bibinfo {author}
  {\bibfnamefont {K.}~\bibnamefont {Flensberg}}, \ and\ \bibinfo {author}
  {\bibfnamefont {L.~P.}\ \bibnamefont {Kouwenhoven}},\ }\bibfield  {title}
  {\enquote {\bibinfo {title} {{Quantum transport in carbon nanotubes}},}\
  }\href {\doibase 10.1103/RevModPhys.87.703} {\bibfield  {journal} {\bibinfo
  {journal} {Rev. Mod. Phys.}\ }\textbf {\bibinfo {volume} {87}},\ \bibinfo
  {pages} {703--764} (\bibinfo {year} {2015})}\BibitemShut {NoStop}%
\bibitem [{\citenamefont {Grifoni}\ and\ \citenamefont
  {H{\"a}nggi}(1998)}]{Grifoni1998}%
  \BibitemOpen
  \bibfield  {author} {\bibinfo {author} {\bibfnamefont {M.}~\bibnamefont
  {Grifoni}}\ and\ \bibinfo {author} {\bibfnamefont {P.}~\bibnamefont
  {H{\"a}nggi}},\ }\bibfield  {title} {\enquote {\bibinfo {title} {{Driven
  quantum tunneling}},}\ }\href {\doibase 10.1016/S0370-1573(98)00022-2}
  {\bibfield  {journal} {\bibinfo  {journal} {Phys. Rep.}\ }\textbf {\bibinfo
  {volume} {304}},\ \bibinfo {pages} {229--354} (\bibinfo {year}
  {1998})}\BibitemShut {NoStop}%
\bibitem [{\citenamefont {Schiro}\ and\ \citenamefont
  {Scarlatella}(2019)}]{Schiro2019}%
  \BibitemOpen
  \bibfield  {author} {\bibinfo {author} {\bibfnamefont {M.}~\bibnamefont
  {Schiro}}\ and\ \bibinfo {author} {\bibfnamefont {O.}~\bibnamefont
  {Scarlatella}},\ }\bibfield  {title} {\enquote {\bibinfo {title} {{Quantum
  impurity models coupled to Markovian and non-Markovian baths}},}\ }\href
  {\doibase 10.1063/1.5100157} {\bibfield  {journal} {\bibinfo  {journal} {J.
  Chem. Phys.}\ }\textbf {\bibinfo {volume} {151}},\ \bibinfo {pages} {044102}
  (\bibinfo {year} {2019})}\BibitemShut {NoStop}%
\bibitem [{\citenamefont {K{\"o}nig}\ and\ \citenamefont
  {Martinek}(2003)}]{Koenig2003}%
  \BibitemOpen
  \bibfield  {author} {\bibinfo {author} {\bibfnamefont {J.}~\bibnamefont
  {K{\"o}nig}}\ and\ \bibinfo {author} {\bibfnamefont {J.}~\bibnamefont
  {Martinek}},\ }\bibfield  {title} {\enquote {\bibinfo {title}
  {{Interaction-Driven Spin Precession in Quantum-Dot Spin Valves}},}\ }\href
  {\doibase 10.1103/PhysRevLett.90.166602} {\bibfield  {journal} {\bibinfo
  {journal} {Phys. Rev. Lett.}\ }\textbf {\bibinfo {volume} {90}},\ \bibinfo
  {pages} {166602} (\bibinfo {year} {2003})}\BibitemShut {NoStop}%
\bibitem [{\citenamefont {Braun}\ \emph {et~al.}(2006)\citenamefont {Braun},
  \citenamefont {K{\"o}nig},\ and\ \citenamefont {Martinek}}]{Braun2006}%
  \BibitemOpen
  \bibfield  {author} {\bibinfo {author} {\bibfnamefont {M.}~\bibnamefont
  {Braun}}, \bibinfo {author} {\bibfnamefont {J.}~\bibnamefont {K{\"o}nig}}, \
  and\ \bibinfo {author} {\bibfnamefont {J.}~\bibnamefont {Martinek}},\
  }\bibfield  {title} {\enquote {\bibinfo {title} {{Frequency-dependent current
  noise through quantum-dot spin valves}},}\ }\href {\doibase
  10.1103/PhysRevB.74.075328} {\bibfield  {journal} {\bibinfo  {journal} {Phys.
  Rev. B}\ }\textbf {\bibinfo {volume} {74}},\ \bibinfo {pages} {075328}
  (\bibinfo {year} {2006})}\BibitemShut {NoStop}%
\bibitem [{\citenamefont {Hornberger}\ \emph {et~al.}(2008)\citenamefont
  {Hornberger}, \citenamefont {Koller}, \citenamefont {Begemann}, \citenamefont
  {Donarini},\ and\ \citenamefont {Grifoni}}]{Hornberger2008}%
  \BibitemOpen
  \bibfield  {author} {\bibinfo {author} {\bibfnamefont {R.}~\bibnamefont
  {Hornberger}}, \bibinfo {author} {\bibfnamefont {S.}~\bibnamefont {Koller}},
  \bibinfo {author} {\bibfnamefont {G.}~\bibnamefont {Begemann}}, \bibinfo
  {author} {\bibfnamefont {A.}~\bibnamefont {Donarini}}, \ and\ \bibinfo
  {author} {\bibfnamefont {M.}~\bibnamefont {Grifoni}},\ }\bibfield  {title}
  {\enquote {\bibinfo {title} {{Transport through a double-quantum-dot system
  with noncollinearly polarized leads}},}\ }\href {\doibase
  10.1103/PhysRevB.77.245313} {\bibfield  {journal} {\bibinfo  {journal} {Phys.
  Rev. B}\ }\textbf {\bibinfo {volume} {77}},\ \bibinfo {pages} {245313}
  (\bibinfo {year} {2008})}\BibitemShut {NoStop}%
\bibitem [{\citenamefont {Rezaei}\ \emph {et~al.}(2020)\citenamefont {Rezaei},
  \citenamefont {Hussein}, \citenamefont {Kamra},\ and\ \citenamefont
  {Belzig}}]{Rezaei2020}%
  \BibitemOpen
  \bibfield  {author} {\bibinfo {author} {\bibfnamefont {A.}~\bibnamefont
  {Rezaei}}, \bibinfo {author} {\bibfnamefont {R.}~\bibnamefont {Hussein}},
  \bibinfo {author} {\bibfnamefont {A.}~\bibnamefont {Kamra}}, \ and\ \bibinfo
  {author} {\bibfnamefont {W.}~\bibnamefont {Belzig}},\ }\bibfield  {title}
  {\enquote {\bibinfo {title} {{Phase-controlled spin and charge currents in a
  superconductor-ferromagnet hybrid}},}\ }\href {\doibase
  10.1103/PhysRevResearch.2.033336} {\bibfield  {journal} {\bibinfo  {journal}
  {Phys. Rev. Research}\ }\textbf {\bibinfo {volume} {2}},\ \bibinfo {pages}
  {033336} (\bibinfo {year} {2020})}\BibitemShut {NoStop}%
\bibitem [{\citenamefont {Rohrmeier}\ and\ \citenamefont
  {Donarini}(2021)}]{Rohrmeier2021}%
  \BibitemOpen
  \bibfield  {author} {\bibinfo {author} {\bibfnamefont {C.}~\bibnamefont
  {Rohrmeier}}\ and\ \bibinfo {author} {\bibfnamefont {A.}~\bibnamefont
  {Donarini}},\ }\bibfield  {title} {\enquote {\bibinfo {title} {{Pseudospin
  resonances reveal synthetic spin-orbit interaction}},}\ }\href@noop {}
  {\bibfield  {journal} {\bibinfo  {journal} {arXiv:2008.09857}\ } (\bibinfo
  {year} {2021})}\BibitemShut {NoStop}%
\bibitem [{\citenamefont {Darau}\ \emph {et~al.}(2009)\citenamefont {Darau},
  \citenamefont {Begemann}, \citenamefont {Donarini},\ and\ \citenamefont
  {Grifoni}}]{Darau2009}%
  \BibitemOpen
  \bibfield  {author} {\bibinfo {author} {\bibfnamefont {D.}~\bibnamefont
  {Darau}}, \bibinfo {author} {\bibfnamefont {G.}~\bibnamefont {Begemann}},
  \bibinfo {author} {\bibfnamefont {A.}~\bibnamefont {Donarini}}, \ and\
  \bibinfo {author} {\bibfnamefont {M.}~\bibnamefont {Grifoni}},\ }\bibfield
  {title} {\enquote {\bibinfo {title} {{Interference effects on the transport
  characteristics of a benzene single-electron transistor}},}\ }\href {\doibase
  10.1103/PhysRevB.79.235404} {\bibfield  {journal} {\bibinfo  {journal} {Phys.
  Rev. B}\ }\textbf {\bibinfo {volume} {79}},\ \bibinfo {pages} {235404}
  (\bibinfo {year} {2009})}\BibitemShut {NoStop}%
\bibitem [{\citenamefont {Donarini}\ \emph {et~al.}(2019)\citenamefont
  {Donarini}, \citenamefont {Niklas}, \citenamefont {Schafberger},
  \citenamefont {Paradiso}, \citenamefont {Strunk},\ and\ \citenamefont
  {Grifoni}}]{Donarini2019}%
  \BibitemOpen
  \bibfield  {author} {\bibinfo {author} {\bibfnamefont {A.}~\bibnamefont
  {Donarini}}, \bibinfo {author} {\bibfnamefont {M.}~\bibnamefont {Niklas}},
  \bibinfo {author} {\bibfnamefont {M.}~\bibnamefont {Schafberger}}, \bibinfo
  {author} {\bibfnamefont {N.}~\bibnamefont {Paradiso}}, \bibinfo {author}
  {\bibfnamefont {C.}~\bibnamefont {Strunk}}, \ and\ \bibinfo {author}
  {\bibfnamefont {M.}~\bibnamefont {Grifoni}},\ }\bibfield  {title} {\enquote
  {\bibinfo {title} {{Coherent population trapping by dark state formation in a
  carbon nanotube quantum dot}},}\ }\href {\doibase 10.1038/s41467-018-08112-x}
  {\bibfield  {journal} {\bibinfo  {journal} {Nat. Commun.}\ }\textbf {\bibinfo
  {volume} {10}},\ \bibinfo {pages} {381} (\bibinfo {year} {2019})}\BibitemShut
  {NoStop}%
\bibitem [{Note1()}]{Note1}%
  \BibitemOpen
  \bibinfo {note} {This integral can be solved by noting that the denominator
  in the integral splits as $\left \protect \{[(\epsilon -\epsilon
  _0)-{\protect \rm i}\Gamma /2]^{-1}-[(\epsilon -\epsilon _0)+{\protect \rm
  i}\Gamma /2]^{-1}\right \protect \}/{\protect \rm i}\Gamma $ and by applying
  Eq.~\protect \textup {\hbox {\mathsurround \z@ \protect \normalfont
  (\ignorespaces \ref {Isummary}\unskip \@@italiccorr )}}.}\BibitemShut {Stop}%
\bibitem []{Bickers1987}%
  \BibitemOpen
  \bibfield  {author} {\bibinfo {author} {\bibfnamefont {N. E.}~\bibnamefont
  {Bickers}}}, \bibfield  {title} {\enquote {\bibinfo {title} {{Review of techniques in the large-$N$ expansion for dilute magnetic alloys}},}\ }\href {\doibase 10.1103/RevModPhys.59.845} {\bibfield
  {journal} {\bibinfo  {journal} {Rev. Mod. Phys.}\ }\textbf {\bibinfo {volume}
  {59}},\ \bibinfo {pages} {845} (\bibinfo {year} {1987})}\BibitemShut
  {NoStop}%
\bibitem [{\citenamefont {{J. K{\"o}nig}}(1995)}]{KoenigDiploma1995}%
  \BibitemOpen
  \bibfield  {author} {\bibinfo {author} {\bibnamefont {{J. K{\"o}nig}}},\
  }\href@noop {} {Master's thesis},\ \bibinfo  {school} {Universit{\"a}t
  Karlsruhe} (\bibinfo {year} {1995})\BibitemShut {NoStop}%
\bibitem [{\citenamefont {Koller}\ \emph {et~al.}(2012)\citenamefont {Koller},
  \citenamefont {Grifoni},\ and\ \citenamefont {Paaske}}]{Koller2012}%
  \BibitemOpen
  \bibfield  {author} {\bibinfo {author} {\bibfnamefont {S.}~\bibnamefont
  {Koller}}, \bibinfo {author} {\bibfnamefont {M.}~\bibnamefont {Grifoni}}, \
  and\ \bibinfo {author} {\bibfnamefont {J.}~\bibnamefont {Paaske}},\
  }\bibfield  {title} {\enquote {\bibinfo {title} {{Sources of negative
  tunneling magnetoresistance in multilevel quantum dots with ferromagnetic
  contacts}},}\ }\href {\doibase 10.1103/PhysRevB.85.045313} {\bibfield
  {journal} {\bibinfo  {journal} {Phys. Rev. B}\ }\textbf {\bibinfo {volume}
  {85}},\ \bibinfo {pages} {045313} (\bibinfo {year} {2012})}\BibitemShut
  {NoStop}%
\bibitem [{\citenamefont {Dirnaichner}\ \emph {et~al.}(2015)\citenamefont
  {Dirnaichner}, \citenamefont {Grifoni}, \citenamefont {Pr{\"u}fling},
  \citenamefont {Steininger}, \citenamefont {H{\"u}ttel},\ and\ \citenamefont
  {Strunk}}]{Dirnaichner2015}%
  \BibitemOpen
  \bibfield  {author} {\bibinfo {author} {\bibfnamefont {A.}~\bibnamefont
  {Dirnaichner}}, \bibinfo {author} {\bibfnamefont {M.}~\bibnamefont
  {Grifoni}}, \bibinfo {author} {\bibfnamefont {A.}~\bibnamefont
  {Pr{\"u}fling}}, \bibinfo {author} {\bibfnamefont {D.}~\bibnamefont
  {Steininger}}, \bibinfo {author} {\bibfnamefont {A.~K.}\ \bibnamefont
  {H{\"u}ttel}}, \ and\ \bibinfo {author} {\bibfnamefont {C.}~\bibnamefont
  {Strunk}},\ }\bibfield  {title} {\enquote {\bibinfo {title} {{Transport
  across a carbon nanotube quantum dot contacted with ferromagnetic leads:
  Experiment and nonperturbative modeling}},}\ }\href {\doibase
  10.1103/PhysRevB.91.195402} {\bibfield  {journal} {\bibinfo  {journal} {Phys.
  Rev. B}\ }\textbf {\bibinfo {volume} {91}},\ \bibinfo {pages} {195402}
  (\bibinfo {year} {2015})}\BibitemShut {NoStop}%
\bibitem [{\citenamefont {Grabert}\ \emph {et~al.}(1988)\citenamefont
  {Grabert}, \citenamefont {Schramm},\ and\ \citenamefont {Ingold}}]{Grab88}%
  \BibitemOpen
  \bibfield  {author} {\bibinfo {author} {\bibfnamefont {H.}\ \bibnamefont
  {Grabert}}, \bibinfo {author} {\bibfnamefont {P.}\ \bibnamefont
  {Schramm}}, \ and\ \bibinfo {author} {\bibfnamefont {G.-L.}\
  \bibnamefont {Ingold}},\ }\bibfield  {title} {\enquote {\bibinfo {title}
  {{Quantum Brownian motion: the functional integral approach}},}\ }\href
  {\doibase 10.1016/0370-1573(88)90023-3} {\bibfield  {journal} {\bibinfo
  {journal} {Phys. Rep.}\ }\textbf {\bibinfo {volume} {168}},\ \bibinfo {pages}
  {115--207} (\bibinfo {year} {1988})}\BibitemShut {NoStop}%
\bibitem [{\citenamefont {Nilsson}\ \emph {et~al.}(2010)\citenamefont
  {Nilsson}, \citenamefont {Karlstr{\"o}m}, \citenamefont {Larsson},
  \citenamefont {Caroff}, \citenamefont {Pedersen}, \citenamefont {Samuelson},
  \citenamefont {Wacker}, \citenamefont {Wernersson},\ and\ \citenamefont
  {Xu}}]{Nilsson2010}%
  \BibitemOpen
  \bibfield  {author} {\bibinfo {author} {\bibfnamefont {H.~A.}\ \bibnamefont
  {Nilsson}}, \bibinfo {author} {\bibfnamefont {O.}~\bibnamefont
  {Karlstr{\"o}m}}, \bibinfo {author} {\bibfnamefont {M.}~\bibnamefont
  {Larsson}}, \bibinfo {author} {\bibfnamefont {P.}~\bibnamefont {Caroff}},
  \bibinfo {author} {\bibfnamefont {J.~N.}\ \bibnamefont {Pedersen}}, \bibinfo
  {author} {\bibfnamefont {L.}~\bibnamefont {Samuelson}}, \bibinfo {author}
  {\bibfnamefont {A.}~\bibnamefont {Wacker}}, \bibinfo {author} {\bibfnamefont
  {L.~E.}\ \bibnamefont {Wernersson}}, \ and\ \bibinfo {author} {\bibfnamefont
  {H.~Q.}\ \bibnamefont {Xu}},\ }\bibfield  {title} {\enquote {\bibinfo {title}
  {Correlation-induced conductance suppression at level degeneracy in a quantum
  dot},}\ }\href {\doibase 10.1103/PhysRevLett.104.186804} {\bibfield
  {journal} {\bibinfo  {journal} {Phys. Rev. Lett.}\ }\textbf {\bibinfo
  {volume} {104}},\ \bibinfo {pages} {186804} (\bibinfo {year}
  {2010})}\BibitemShut {NoStop}%
\bibitem [{\citenamefont {Kohler}\ \emph {et~al.}(2004)\citenamefont {Kohler},
  \citenamefont {Camalet}, \citenamefont {Strass}, \citenamefont {Lehmann},
  \citenamefont {Ingold},\ and\ \citenamefont {H{\"a}nggi}}]{Kohler2004}%
  \BibitemOpen
  \bibfield  {author} {\bibinfo {author} {\bibfnamefont {S.}~\bibnamefont
  {Kohler}}, \bibinfo {author} {\bibfnamefont {S.}~\bibnamefont {Camalet}},
  \bibinfo {author} {\bibfnamefont {M.}~\bibnamefont {Strass}}, \bibinfo
  {author} {\bibfnamefont {J.}~\bibnamefont {Lehmann}}, \bibinfo {author}
  {\bibfnamefont {G.-L.}\ \bibnamefont {Ingold}}, \ and\ \bibinfo {author}
  {\bibfnamefont {P.}~\bibnamefont {H{\"a}nggi}},\ }\bibfield  {title}
  {\enquote {\bibinfo {title} {{Charge transport through a molecule driven by a
  high-frequency field}},}\ }\href {\doibase 10.1016/j.chemphys.2003.09.023}
  {\bibfield  {journal} {\bibinfo  {journal} {Chem. Phys.}\ }\textbf {\bibinfo
  {volume} {296}},\ \bibinfo {pages} {243--249} (\bibinfo {year}
  {2004})}\BibitemShut {NoStop}%
\bibitem [{\citenamefont {Smirnov}\ and\ \citenamefont
  {Grifoni}(2013)}]{Smirnov2013PRB}%
  \BibitemOpen
  \bibfield  {author} {\bibinfo {author} {\bibfnamefont {S.}~\bibnamefont
  {Smirnov}}\ and\ \bibinfo {author} {\bibfnamefont {M.}~\bibnamefont
  {Grifoni}},\ }\bibfield  {title} {\enquote {\bibinfo {title} {{Keldysh effective action theory for universal physics in spin-$\frac{1}{2}$ Kondo dots}},}\ }\href
  {\doibase 10.1103/PhysRevB.87.121302} {\bibfield  {journal} {\bibinfo
  {journal} {Phys. Rev. B}\ }\textbf {\bibinfo {volume} {87}},\ \bibinfo
  {pages} {121302(R)} (\bibinfo {year} {2013})}\BibitemShut {NoStop}%
\bibitem {Wolfram_digamma}%
  \BibitemOpen
  \bibfield  {author} {\bibinfo {author} {\bibfnamefont {E.  W.}~\bibnamefont
  {Weisstein}},\ }\bibfield  {title} {\enquote {\bibinfo {title} {{Digamma Function." From MathWorld -- A Wolfram Web Resource}},}\ }\href
  { https://mathworld.wolfram.com/DigammaFunction.html}
  {\bibfield  {journal} {\bibinfo
  {journal} {https://mathworld.wolfram.com/DigammaFunction.html}}}\BibitemShut {NoStop}%

\end{thebibliography}
%
%

\end{document}